\documentclass[10pt]{book}

\usepackage[top=2.8cm, bottom=2.8cm, left=2.4cm, right=2.4cm]{geometry}

\usepackage[utf8]{inputenc}  
\usepackage[T1]{fontenc}     
\usepackage[french]{babel}  
\usepackage{amsmath,amssymb,lmodern} 
\usepackage{graphicx}
\usepackage{diagbox}
\usepackage{hyperref}
\usepackage{color}
\usepackage{lipsum}
\usepackage{dsfont}
\usepackage{braket}
\usepackage{slashed}
\usepackage{bm}
\usepackage{pdflscape}
\usepackage{lscape}
\usepackage{afterpage}
\usepackage{bbm}
\usepackage{young}
\usepackage{anyfontsize}
\usepackage[toc,page]{appendix}

\usepackage{array}
\newcolumntype{L}[1]{>{\raggedright\let\newline\\\arraybackslash\hspace{0pt}}m{#1}}
\newcolumntype{C}[1]{>{\centering\let\newline\\\arraybackslash\hspace{0pt}}m{#1}}
\newcolumntype{R}[1]{>{\raggedleft\let\newline\\\arraybackslash\hspace{0pt}}m{#1}}

\usepackage{fancyhdr}

\usepackage[small]{caption2}

\usepackage[nottoc,notlot,notlof]{tocbibind}

\newcommand{\dd}{\mbox{d}}

\renewcommand{\a}{{\color{red}a\color{black}}}
\renewcommand{\b}{{\color{red}b\color{black}}}
\newcommand{\EAc}{{\color{red}c\color{black}}}
\renewcommand{\d}{{\color{red}d\color{black}}}
\newcommand{\e}{{\color{red}e\color{black}}}
\newcommand{\f}{{\color{red}f\color{black}}}
\newcommand{\g}{{\color{red}g\color{black}}}

\newcommand{\al}{{\color{red}d\color{black}}}
\newcommand{\be}{{\color{red}e\color{black}}}
\newcommand{\ga}{{\color{red}\gamma\color{black}}}
\newcommand{\au}{{\color{red}a_1\color{black}}}
\newcommand{\am}{{\color{red}a_m\color{black}}}
\newcommand{\aN}{{\color{red}a_N\color{black}}}

\usepackage{xcolor}
\definecolor{blue}{rgb}{0.19,0.64,0.54}
\definecolor{reddish}{rgb}{0.65, 0.2, 0.2}
\definecolor{red}{rgb}{0.7,0.3,0.3}
\definecolor{darkgreen}{rgb}{0.2,0.7,0.3}
\definecolor{darkblue}{rgb}{0.3,0.40,0.48}
\definecolor{gray}{rgb}{.8,.8,.8}

\title{Au-delà des modèles standards en cosmologie}
\author{Erwan Allys}
\date{\today}

\begin{document}
%
\thispagestyle{empty}

\begin{figure}[h!]
\begin{center}
\begin{minipage}[c]{0.30\linewidth}
\includegraphics[height=1.4cm]{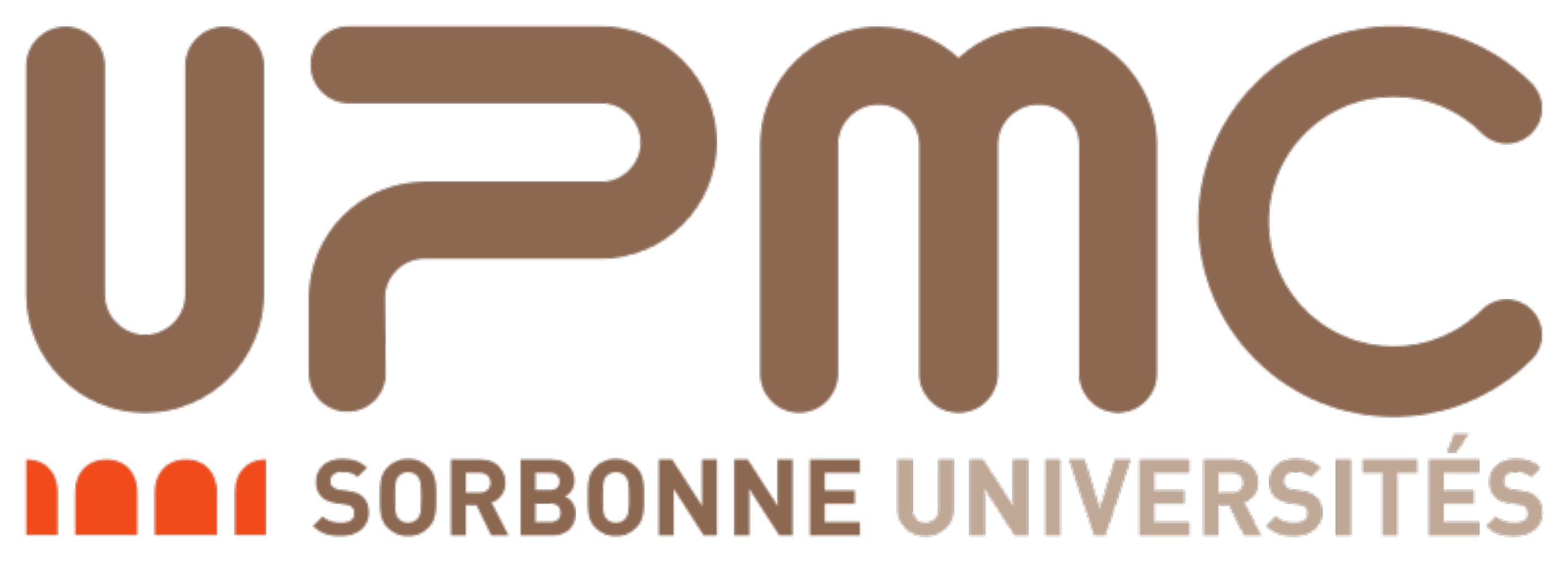}
\hfill
\end{minipage}
\hspace{0.9cm}
\begin{minipage}[c]{0.30\linewidth}
\hspace{0.75cm}
\includegraphics[height=2.6cm]{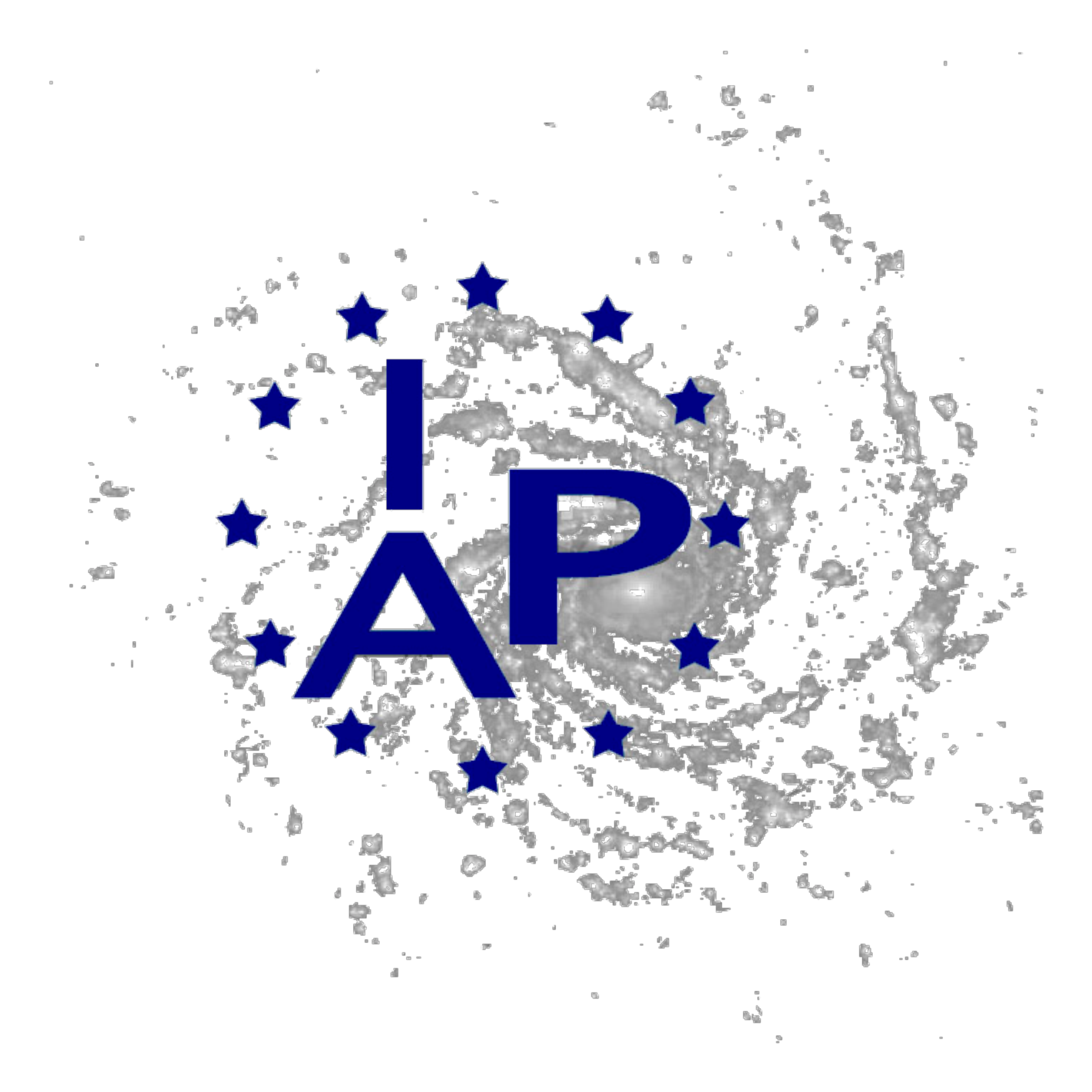}
\end{minipage}
\hspace{0.1cm}
\begin{minipage}[c]{0.30\linewidth}
\hfill
\includegraphics[height=2.6cm]{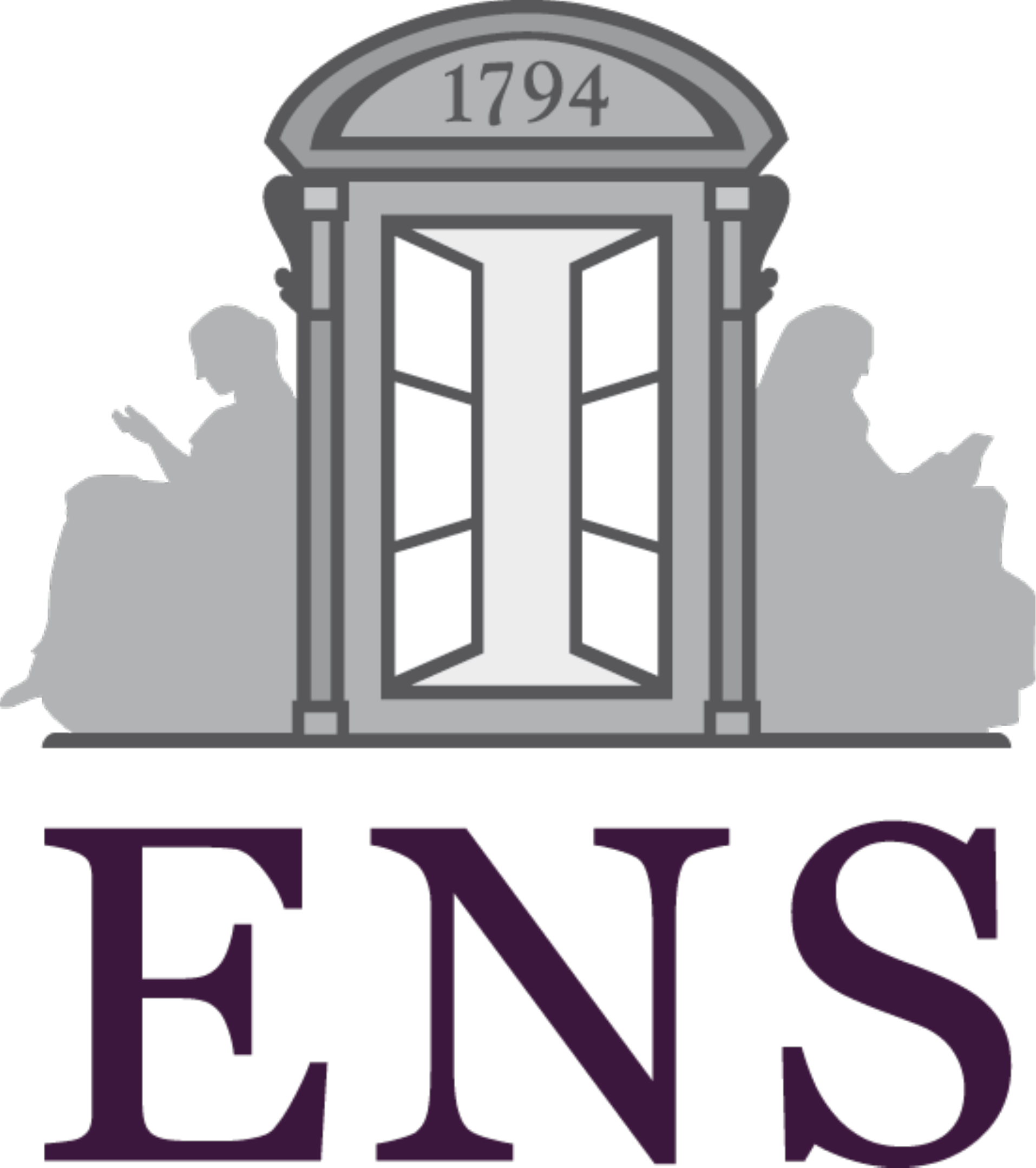}
\end{minipage}
\end{center}
\end{figure}

\begin{center}

{\Large
{\bf Université Pierre et Marie Curie - Paris VI}\\
}
\vspace{0.2cm}
{\large
École Doctorale de Physique en Île-de-France
}

\vspace{1.25cm}

\textbf{{\LARGE THÈSE DE DOCTORAT}}\\
\vspace{0.25cm}
{\Large Spécialité~: \textbf{Physique Théorique}}

\vspace{0.8cm}

{\large réalisée à }\\
\vspace{0.1cm}
{\Large l'Institut d'Astrophysique de Paris}

\vspace{1.6cm}

\rule{0.6\linewidth}{.5pt}

{\LARGE {\bf Au-delà des modèles standards
\\ \vspace{0.25cm} en cosmologie}}

\rule{0.6\linewidth}{.5pt}

\vspace{0.8cm}

{\large présentée par }\\
\vspace{0.15cm} {\Large Erwan {\sc Allys}}
  
\vspace{1.75cm} 

{\large et soutenue publiquement le 9 juin 2017}\\
\vspace{0.5cm}
{\large devant le jury composé de} \\
\vspace{0.2cm}

\begin{large}
\begin{tabular}{r@{\ }ll}
  & M$^{\text{me}}$ Claudia {\sc de Rham} & Rapporteuse\\
  & M. David {\sc Langlois} & Rapporteur \\
  & M. Robert {\sc Brandenberger } & Examinateur  \\
  & M. Philippe {\sc Brax} & Examinateur  \\
  & M. Michael {\sc Joyce} & Examinateur  \\
  & M$^{\text{me}}$ Danièle {\sc Steer} & Examinatrice  \\
  & M. Patrick {\sc Peter} & Directeur de thèse  \\
  & M. Gilles {\sc Esposito-Farèse} & Membre invité  \\
\end{tabular}

\end{large}
\end{center}

\newpage
\thispagestyle{empty}
~ 

\newpage
\thispagestyle{empty}

~ 
\vspace{7cm}
\begin{flushright}
\textit{“We are never prepared for what we expect.” \\
James A. Michener}
\end{flushright}

\vfill

\newpage
\thispagestyle{empty}

\chapter*{Remerciements\markboth{Remerciements}{Remerciements}}
\addcontentsline{toc}{chapter}{\protect\numberline{}Remerciements} 

\vspace{2cm}

\noindent
Bien qu'identifiée comme une réalisation individuelle, une thèse est avant tout une réussite collective, permise par l'échange pendant trois années entre un thésard et un environnement universitaire toujours très stimulant. J'ai personnellement la chance d'avoir eu pendant ma thèse des interactions particulièrement agréables et profitables avec cet environnement, que cela soit du point de vue de la recherche, de l'enseignement, ou bien extra-professionnel. Je sais que ce n'est pas forcément le cas pour tous les thésards, et je suis donc éminemment reconnaissant envers tous ceux qui ont permis que mon travail se déroule dans de si bonnes conditions.

\vspace{0.1cm}

Mais remerciements vont tout particulièrement à Patrick Peter, mon directeur de thèse, qui a eu la patience de m'apprendre progressivement le métier de chercheur, avec toute la rigueur qui lui est associée. Je lui suis reconnaissant pour sa disponibilité de chaque instant, ainsi que pour la grande liberté qu'il m'a laissée dès le début de ma thèse, et la confiance qui allait de pair. Je remercie également le $\mathcal{G}\mathbb{R}\epsilon\mathbb{C}\mathcal{O}$ pour son accueil, notamment Guillaume Faye pour sa gestion et sa disponibilité, et plus spécifiquement Luc Blanchet, Cédric Deffayet, Gilles Esposito-Farèse, Jérôme Martin et Jean-Philippe Uzan, auprès de qui j'ai beaucoup appris. Je remercie aussi François Bouchet, Alain Riazuelo, Christophe Ringeval, et Mairi Sakellariadou pour les discussions que nous avons pu avoir en conférence, en colloque, ou ailleurs, ainsi que Marie-Christine Angonin pour sa disponibilité lorsque je lui faisais part de mes nombreuses interrogations.

\vspace{0.1cm}

Je souhaite bien sûr exprimer toute ma gratitude envers les deux rapporteurs de cette thèse, David Langlois et Claudia de Rham, qui ont pris le temps nécessaire pour l'évaluer -- je suis bien conscient que ma prose n'est malheureusement pas toujours aussi fluide que je le souhaiterais... --, ainsi qu'envers les examinateurs et invités, Robert Brandenberger, Philippe Brax, Michael Joyce, Danièle Steer, ainsi que Gilles Esposito-Farèse. Je suis particulièrement honoré de voir mon travail jugé par des scientifiques d'une si grande qualité. Je remercie également mes collaborateurs colombiens, Andres et Juan-Pablo, et tout particulièrement Yeinzon, qui ont tout fait pour m'accueillir dans les meilleurs conditions possibles, et avec qui il a été très agréable de travailler. 

\vspace{0.1cm}

Je remercie les doctorants de l'Institut d'Astrophysique de Paris, auprès de qui j'ai énormément appris. Je pense tout particulièrement à Clément, Jean-Baptiste, Laura, Pierre et Vivien qui m'ont guidé scientifiquement tout au long de ma thèse, ainsi qu'à Caterina, Christoffel, Laura (encore), Tanguy, Thomas et Oscar avec qui j'ai partagé mon bureau dans une belle atmosphère d'échange et de convivialité. Je ne sais également pas ce qu'aurait été cette thèse sans la présence de Julia (on a été dans la même classe, tu te souviens ?) et de Alba et Mélanie (auxquelles j'ai évidemment immédiatement plu !). Je pense aussi à Federico, Jesse, Tilman, ainsi qu'à Giulia et Benoît, avec qui les discussions ont toujours été intéressantes. Je remercie par ailleurs Madeleine Roux-Merlin et tout particulièrement Christophe Gobet, dont le grand engagement permet de travailler dans de si bonnes conditions à l'IAP.

\vspace{0.1cm}

Cette thèse a également été très stimulante pour moi grâce à la chance que j'ai eue de pouvoir enseigner, qui plus est dans d'excellentes conditions. Je remercie grandement pour cela Jean-Michel Raimond et François Levrier, qui m'ont fait confiance pour enseigner au centre de préparation à l'agrégation de Montrouge, où j'ai eu le plaisir d'enseigner à des étudiants toujours agréables et intéressants, notamment avec Alexis, Antoine, Arnaud, Benoît, Jérémy, Kenneth, et Matthieu. Cela fut possible grâce à la grande qualité du travail d'Eric Guineveu, envers qui je suis très reconnaissant. J'ai aussi passé de très agréables moments en encadrant l'équipe de l'ENS pour l'IPT (je remercie Frédéric Chevy pour sa confiance), et en organisant différentes éditions du FPT et de l'IPT, avec Andréane, Arnaud, Charlie, Cyrille, Daniel, Maxime et Marguerite.

\vspace{0.1cm}

Une part non négligeable de ce manuscrit a vu le jour lors d'innombrables discussions avec des amis physiciens et mathématiciens non spécialistes, et je remercie tout particulièrement Anne, Damien \og binôme \fg{}, Jemi et Olivier, Paul, ainsi que les doctorants qui m'ont permis de tester différentes présentations originales lors des fameux YMCA. Je suis également reconnaissant envers tous ceux qui ont donné de leur temps pour relire ce manuscrit, en premier lieu Patrick, mais aussi Amandine, Delphine, François, Julia, Mélanie, Tanguy, Oscar et Vivien, ainsi qu'Yves et mon père. J'ai une pensée émue pour mes amis de toujours, les dieux de la forêt. Leur foi inébranlable en ma réussite scientifique a indubitablement contribué à faire de moi ce que je suis devenu. Plus spécialement pour cette thèse, je pense tout particulièrement à Matthieu qui a lu des livres de physique rien que pour pouvoir en discuter avec moi, et bien sûr à Florian qui m'a supporté deux ans en collocation, et qui m'a accueilli en Guadeloupe pendant la fin de la rédaction de ma thèse pour que je puisse le faire les pieds dans l'eau ! 

\vspace{0.1cm}

Mes derniers remerciement vont vers ceux qui m'ont enseigné la Physique pendant mes études, ainsi que vers tous ceux qui se sont donné la peine d'écrire des livres scientifiques d'une réelle qualité, faisant le travail de fond indispensable à toute transmission du savoir dans le temps. Enfin, et bien qu'ils sachent ce que je leur dois, j'ai une profonde reconnaissance pour ma famille, Auberie et les longues discussions scientifiques qu'elle endure pendant nos différents repas, Josselin pour sa grande aide en informatique et son accueil spontané pendant mes périodes de doute, et bien évidemment pour mes parents, sans qui rien n'aurait été possible.

\vspace{0.8cm}

\begin{flushright}
Paris, mai 2017
\end{flushright}

\setcounter{tocdepth}{1}
\tableofcontents


\chapter*{Introduction\markboth{Introduction}{Introduction}} 
\addcontentsline{toc}{chapter}{\protect\numberline{}Introduction} 
\label{IntroThese}

\vspace{0.75cm}
\noindent
Pourrons-nous aller au-delà des modèles standards actuels de la physique ? Pourrons-nous décrire la gravitation au-delà de la relativité générale, ainsi que la structure fondamentale de la matière et les interactions électromagnétiques et nucléaires au-delà du Modèle Standard de la physique des particules ? Motivées par le succès croissant des vérifications expérimentales et observationnelles de ces théories physiques, qui ont justement conféré à celles-ci le statut de modèles standards, ces questions viennent effectivement à l'esprit. Après la détection du boson de Higgs au LHC en 2012 ainsi que celle d'ondes gravitationnelles par des méthodes interférométriques en 2015, et alors qu'il semble difficile de trouver une simple mesure remettant directement en cause ces modèles standards, ce questionnement vient avec d'autant plus de force.

Une telle situation n'est cependant pas sans rappeler les années 1900, ou certains physiciens considéraient que l'avenir de la physique n'était plus à la découverte de lois fondamentales, mais seulement à l'amélioration des précisions expérimentales. À la même époque, Lord Kelvin identifiait néanmoins deux \emph{nuages noirs} qui assombrissaient l'édifice de la physique classique, à savoir l'impossibilité de calculer la capacité calorifique de certains solides, ainsi que la non-invariance des équations de Maxwell sous les transformations de Galilée qui amenait à introduire la notion d'éther. Et il n'a fallut que quelques années pour que ces \emph{nuages noirs} amènent des révolutions dans les théories physiques, à savoir la physique quantique et la relativité restreinte.

À l'heure actuelle, et malgré les succès expérimentaux de la relativité générale et du Modèle Standard de la physique des particules, force est de constater que ces modèles ne sont eux aussi pas dépourvus de \emph{nuages noirs}, et tout particulièrement dans le domaine de la cosmologie. Avec l'avènement de la cosmologie moderne et l'amélioration constante de la précision des observations cosmologiques, la nécessité d'inclure la matière noire et l'énergie noire dans les modèles cosmologiques s'impose en effet à présent, sans que ces deux ingrédients ne soient justifiés par les modèles standards actuels. Une compréhension complète des observations cosmologiques nécessitera donc une remise en question d'au moins un de ces modèles standards. Plutôt que de questionner si nous pourrons aller au-delà des modèles standards, et comme cela apparaît comme une nécessité, il est donc plutôt légitime de se demander comment cela peut être fait.

Plus qu'un simple marqueur de notre ignorance, la cosmologie apparait comme un formidable laboratoire pour tester de nouvelles théories. L'histoire temporelle de l'univers étant aussi une histoire thermique, les phénomènes en jeu peu après le big-bang peuvent en effet avoir laissé des traces encore observables actuellement, permettant de sonder des échelles d'énergie beaucoup trop hautes pour être accessibles sur Terre, allant \emph{a priori} jusqu'à l'énergie de Planck $E_{\text{{\sc p}}}\simeq 10^{19}$ GeV. À l'opposé du spectre énergétique, la possibilité d'observer des phénomènes jusqu'à des tailles caractéristiques proches du rayon de Hubble permettent de sonder des échelles d'énergie également inaccessibles sur Terre, allant jusqu'à $E_{H_0} \simeq 10^{-33}$ eV. Une étape nécessaire dans l'investigation de toute théorie au-delà des modèles standards est donc d'étudier leur phénoménologie dans un cadre cosmologique. 

En accord avec ces motivations, les travaux effectués pendant ma thèse, et que je vais présenter ici, ont été consacrés au développement de théories au-delà des modèles standards actuels, ainsi qu'à l'étude de certaines de leurs conséquences dans un cadre cosmologique. J'ai étudié dans un premier temps les conséquences cosmologiques des théories de grande unification -- des modèles de physique des particules qui unifient les interactions du Modèle Standard dans un seul groupe de jauge --, et plus précisément la formation de défauts topologiques stables, des cordes cosmiques, lors de leur brisure spontanée de symétrie dans l'univers primordial. J'ai montré comment étudier la structure microscopique de ces défauts dans le cadre d'une implémentation réaliste des théories de grandes unification, ce qui implique notamment la condensation dans ces cordes de tous les champs de Higgs participant à la brisure de symétrie. J'ai étudié dans un deuxième temps la construction de différentes théories de gravité modifiée de Galiléons, et j'ai notamment contribué à construire la théorie la plus générale pour des multi-Galiléons scalaires ainsi que des Galiléons vectoriels. J'ai aussi participé au développement et à l'application d'une procédure systématique de recherche de Lagrangiens à des théories de multi-Galiléons vectoriels.

Une présentation complète des différents aspects des théories de grande unification et de gravité modifiée ainsi que leurs conséquences cosmologiques irait bien au-delà d'un simple manuscrit de thèse. Aussi a-t-il été nécessaire de faire des choix sur les sujets discutés, et de mettre l'accent sur certains thèmes plutôt que d'autres. Étant donné que ma thèse a principalement consisté à étudier la structure des théories de grande unification comme de gravité modifiée, j'ai choisi de suivre dans ce document un fil directeur basé sur la construction de théories physiques.

Dans la partie~\ref{ChapterModelBuilding}, j'introduis donc en les justifiant les concepts centraux dans la construction de modèles physiques. Je discute par exemple pourquoi les outils de la théorie des groupes sont particulièrement bien adaptés pour décrire les notions de symétries, ainsi que le fait de construire une théorie à partir d'une action. Je présente également comment on peut faire émerger naturellement le concept de jauge de la distinction entre les représentations unitaires du groupe de Lorentz inhomogène et les représentations covariantes de Lorentz.

Dans la partie~\ref{ChapterTheoriesDeJauges}, je présente les théories de jauge et le mécanisme de Higgs, en m'appuyant sur les notions développées dans la première partie. Afin de simplifier au maximum la description de ces théories dans le cadre des théories de grande unification, je fais cette discussion à l'aide des notions de racines et de poids, qui permettent de mettre en avant la structure des groupes de jauge, et d'identifier simplement les nombres quantiques des particules ainsi que l'action des transformations de jauge sur celles-ci. J'introduis pour cela au début de cette partie les notions de racines et de poids, sans entrer dans leurs fondements mathématiques, mais plutôt en montrant comment on peut les utiliser de façon pratique.

Dans la partie~\ref{ChapterSMAndBeyond}, je présente le Modèle Standard de la physique des particules et les théories de grande unification, du point de vue de la construction de modèles physiques. Je discute par exemple dans ce cadre l'importance de la matrice CKM en tant que validation $\emph{a posteriori}$ du caractère fondamental des théories de jauge et du mécanisme de Higgs de brisure spontanée de symétrie. Je présente ensuite les motivations et la construction des théories de grande unification, en utilisant notamment le formalisme des racines et des poids pour simplifier la discussion. Je discute finalement des propriétés des modèles actuels de grande unification qui seront utilisés dans la partie suivante.

Dans la partie~\ref{ChapterRealisticStrings}, j'introduis les défauts topologiques à partir de l'examen des propriétés topologiques des configurations de vide après brisure spontanée de symétrie. Je justifie ensuite la formation des cordes cosmiques dans l'univers primordial, et discute son caractère universel ainsi que le lien fort entre la structure microscopique des cordes cosmiques et leurs propriétés macroscopiques. Je présente alors mes articles sur les structures réalistes de cordes cosmiques prenant en compte une implémentation complète des théories de grande unification. Ceux-ci montrent notamment qu'il est nécessaire de prendre en compte dans la structure des cordes l'intégralité des champs de Higgs contribuant au schéma de brisure de symétrie, et décrivent les structures de cordes qui en résulte.

La partie~\ref{ChapterGalileons} donne une introduction aux théories de gravité modifiée, et en particulier aux théories de Galiléons scalaires. Je commence par introduire la relativité générale, en discutant notamment ses différentes approches théoriques. Je présente ensuite les motivations menant aux théories de gravité modifiée, ainsi que les difficultés inhérentes à la construction de telles théories. Je discute ensuite les modèles de Galiléons en tant que théorie de gravité modifiée, avant de discuter de leur importance en tant que théorie tenseur/scalaire, en lien avec le théorème Horndeski/Galiléons. Je présente finalement mon article sur les multi-Galiléons scalaires, qui introduit une nouvelle classe de Lagrangiens de multi-Galiléons scalaires et examine ses propriétés en détail, avec une application à des multi-Galiléons dans des représentations fondamentales et adjointes de groupes de symétrie globale SO($N$) et SU($N$). 

La partie~\ref{ChapterGalVec} présente finalement les articles écrits en collaboration avec J.-P. Beltran, P. Peter et Y. Rodriguez sur les théories de Galiléons et multi-Galiléons vectoriels, qui ont notamment contribué à obtenir la théorie la plus générale pour des Galiléons vectoriels, et dans lesquels nous avons également construit de façon exhaustive les premiers termes de modèles de multi-Galiléons vectoriels dans la représentation adjointe d'une groupe de symétrie globale SU(2). Ces articles sont introduits en détails dans le chapitre~\ref{PartIntroProca}.

L'annexe~\ref{ChapterTdG} de ce manuscrit regroupe finalement certains éléments de théorie des groupes. Cette annexe contient les principaux résultats mathématiques utilisés tout au long du manuscrit, qui y sont rassemblés afin de ne pas saturer la discussion du corps de texte. Elle ne se veut pas une présentation détaillée, mais plutôt une exposition concise de ces différents résultats mathématiques, qui puisse servir de glossaire si nécessaire.

\section*{Articles publiés}
\addcontentsline{toc}{section}{Articles publiés}

\noindent
Les articles qui ont été publiés suite au travail effectué pendant ma thèse sont les suivants~:

\begin{center}
~ 
\hfill
\begin{minipage}[c]{0.90\linewidth}
\begin{itemize}
\item[\textbf{1.}] E.~Allys, \textit{Bosonic condensates in realistic supersymmetric GUT cosmic strings}, \href{http://iopscience.iop.org/article/10.1088/1475-7516/2016/04/009/meta}{\textbf{JCAP 1604 (2016) 009}}, \href{http://arxiv.org/abs/1505.07888}{arXiv:1505.07888 [gr-qc]}.
\vspace{0.5cm}
\item[\textbf{2.}] E.~Allys, P.~Peter and Y.~Rodriguez, \textit{Generalized Proca action for an Abelian vector field}, \href{http://iopscience.iop.org/article/10.1088/1475-7516/2016/02/004/meta}{\textbf{JCAP 1602 (2016) 004}}, \href{http://arxiv.org/abs/1511.03101}{arXiv:1511.03101 [hep-th]}.
\vspace{0.5cm}
\item[\textbf{3.}] E.~Allys, \textit{Bosonic structure of realistic SO(10) supersymmetric cosmic strings}, \href{http://journals.aps.org/prd/abstract/10.1103/PhysRevD.93.105021}{\textbf{Phys.\~Rev.\ D 93 (2016) 105021}}, \href{http://arxiv.org/abs/1512.02029}{arXiv:1512.02029 [gr-qc]}.
\vspace{0.5cm}
\item[\textbf{4.}] E.~Allys, J.~P.~Beltran Almeida, P.~Peter and Y.~Rodriguez, \textit{On the $4D$ generalized Proca action for an Abelian vector field}, \href{http://iopscience.iop.org/article/10.1088/1475-7516/2016/09/026/meta}{\textbf{JCAP 1609 (2016) 026}}, \href{http://arxiv.org/abs/1605.08355}{arXiv:1605.08355 [hep-th]}.
\vspace{0.5cm}
\item[\textbf{5.}] E.~Allys, P.~Peter and Y.~Rodriguez, \textit{Generalized SU(2) Proca Theory}, \href{http://journals.aps.org/prd/abstract/10.1103/PhysRevD.94.084041}{\textbf{Phys.\ Rev.\ D 94 (2016) 084041}}, \href{http://arxiv.org/abs/1609.05870}{arXiv:1609.05870 [hep-th]}.
\vspace{0.5cm}
\item[\textbf{6.}] E.~Allys, \textit{New terms for scalar multi-galileon models and application to SO(N) and SU(N) group representations}, \href{https://journals.aps.org/prd/abstract/10.1103/PhysRevD.95.064051}{\textbf{Phys.\ Rev.\ D 95 (2017) 064051}}, \href{http://arxiv.org/abs/1612.01972}{arXiv:1612.01972 [hep-th]}.
\end{itemize}
\end{minipage}
\hfill
~
\end{center}

\noindent
Ils sont reproduits dans les chapitres~\ref{PartArticleCordes1},~\ref{PartArticleCorde2},~\ref{PartArticleMultigal},~\ref{PartArticleProca1},~\ref{PartArticleProca2} et~\ref{PartArticleProcaSU2}.

\section*{Conventions et figures}
\addcontentsline{toc}{section}{Conventions et figures}

\begin{itemize}
\item Sauf lorsque le contraire est précisé explicitement, les équations sont données dans un système d'unités naturelles dans lequel $\hbar=c=k_{_{\text{B}}}=1$. 
\item La métrique de Minkowski $\eta_{\mu\nu}$ est diagonale, et vaut $\eta_{11}=\eta_{22}=\eta_{33} = 1$ et $\eta_{00}=-1$, pour un intervalle de distance ${\text{d}s}^2 = -{\text{d}t}^2 + {\text{d}\vec x} ^2$. Sa généralisation en espace courbe est $g_{\mu\nu}$ de signature positive.
\item Les matrices $\sigma_i$ pour $i=1,\cdots,3$ désignent les matrices de Pauli qui vérifient $\left[\frac{\sigma_i}{2},\frac{\sigma_j}{2}\right] = i \epsilon_{ijk}\frac{\sigma_j}{2}$, auxquelles on ajoute $\sigma_0 = \mathds{1}$. 
\item Les matrices $\gamma_\mu$ pour $\mu=0,\cdots,3$ sont les matrices de Dirac, qui vérifient $\left\{\gamma_\mu,\gamma_\nu\right\} =  2 \eta_{\mu\nu}$. On définit aussi conventionnellement $\gamma_5 = i \gamma_0 \gamma_1 \gamma_2 \gamma_3$.\\
\end{itemize}

\noindent
Toutes les figures ont été faites par mes soins, sauf mention contraire dans leur légende. Elles ont été réalisées avec une combinaison de GeoGebra, GIMP, Inkscape, \LaTeX, Mathematica, mjograph, ainsi que l'outil de capture d'écran Mac.


\part{Symétries, théorie des groupes, construction de modèles physiques}
\label{ChapterModelBuilding}
\chapter{Théorie des groupes et construction de modèles physiques}
\section{Pourquoi la théorie des groupes ?}
\label{IntroGroupes}

\noindent
L'application de la théorie des groupes à la physique n'a commencé à être systématisée qu'à partir du vingtième siècle. D'une part, elle permet d'avoir une compréhension particulièrement profonde de la description des systèmes physiques et des lois régissant leurs comportements. D'autre part, elle donne des contraintes très fortes sur les théories qu'il est possible d'écrire, ainsi que sur quelles lois de conservations celles-ci doivent vérifier.

L'apparition d'une structure de groupe est en fait naturelle en physique. En effet, supposons qu'un ensemble de transformations laissent invariant un système physique, décrivant ainsi les propriétés de symétrie de ce système. Cet ensemble contient \emph{a priori} l'identité (puisqu'elle laisse le système inchangé), et une combinaison successive de ces transformations décrit toujours une invariance du système. D'autre part, étant donné une transformation qui laisse le système physique invariant, on s'attend à ce que la transformation inverse soit bien définie, et laisse elle aussi le système inchangé. Ces conditions sont suffisantes pour pouvoir décrire ces transformations avec les outils de la représentation des groupes, et donc de décrire la symétrie en question via une structure de groupe. En retour, la structure de groupe implique un certain nombre de résultats importants sur le système.

Les deux premières conséquences au fait qu'un système physique soit invariant sous l'action d'un groupe de symétrie sont :
\begin{itemize}
\item[i)] Les différentes propriétés de ce système physique ne peuvent être décrites que via des représentations de ce groupe de symétrie.
\item[ii)] Les lois régissant ce système physique ne peuvent relier que des termes ou des produits de termes se trouvant dans une même représentation du groupe de symétrie.
\end{itemize}
La premier point se comprend bien par le fait que si un groupe de symétrie peut agir sur un système physique, alors celui-ci est décrit par des représentations de ce groupe de symétrie (qui peuvent être triviales). Le deuxième point exprime qu'on ne peut égaler que des quantités qui se transforment similairement sous l'action d'un groupe, car on obtiendrait sinon une équation différente après chaque transformation censée laisser le système invariant (comme en égalant une composante d'une quantité vectorielle avec une quantité scalaire). Une conséquence importante est que pour écrire des équations dans une représentation donnée, on pourra identifier quels produits d'autres représentations contiennent la représentation recherchée. Ce point traduit par exemple que les seules dérivées spatiales utilisables en physique classique sont celles qui produisent bien des grandeurs par exemple scalaire ou vectorielles sous les rotations de l'espace, à savoir les gradients, divergences, rotationnels, etc. ; ou qu'on ne doit de même n'utiliser que la dérivée quadrivectorielle $\partial_\mu$ comme opérateur différentiel dans le formalisme covariant de la relativité restreinte. 

Ces résultats limitent et encadrent considérablement toute description d'un système physique dès que l'on sait sous l'action de quels groupes de symétrie il est invariant. L'ajout de seulement quelques conditions supplémentaires est souvent suffisant pour identifier de manière univoque un seul modèle physique possible, ou bien une classe réduite de modèles. Ces propriétés, formulées de façon plus précise dans la section~\ref{SymPartWigner}, sont à présent à la base de la construction de modèles en physique moderne. Par la suite, et sauf mention contraire, le terme de représentation désignera des représentations irréductibles, les représentation réductibles pouvant être décomposées en leur composantes irréductibles.

\section{Exemple de l'électrodynamique classique}
\label{PartElectromagClassique}

\noindent
Avant de discuter formellement les points évoqués précédemment, on peut les illustrer en les appliquant à la construction \emph{a posteriori} de l'électrodynamique classique (non formulée comme une théorie de jauge). Les symétries du modèle sont alors le groupe des rotations spatiales SO(3), ainsi que les opérations conjugaison de charge $C$, de parité $P$, et renversement du temps $T$. Pour SO(3), les représentations peuvent être indiquées par leur dimension. Pour les autres symétries (chacune associée à un groupe $\mathbb{Z}_2$), on indiquera comment se transforment les différentes grandeurs physiques par $\pm1$ selon qu'elles changent de signe ou pas sous ces transformations. Les représentations des différentes grandeurs physiques, et de certaines de leurs dérivées sont récapitulées dans le tableau~\ref{RepMaxwell}. On notera que les grandeurs $\vec r$ et $q$ décrivent les propriétés d'une particule donnée, alors que les autres grandeurs sont des champs.

\begin{table}[h!]
\begin{center}
\hspace{0.5cm}
\begin{minipage}[t]{.49\textwidth}
\renewcommand{\arraystretch}{1.4}
\begin{tabular}{|c|c|c|}
\hline
Nom & Grandeur & SO$(3)\times C\times P\times T$ \\
\hline
Champ électrique & $\displaystyle{\vec E}$ & $\displaystyle{(\mathbf{3},-1,-1,1)}$ \\
\hline
Champ magnétique & $\displaystyle{\vec B}$ & $\displaystyle{(\mathbf{3},-1,1,-1)}$ \\
\hline
Densité de charge & $\displaystyle{\rho}$ & $\displaystyle{(\mathbf{1},-1,1,1)}$ \\
\hline
Densité de courant & $\displaystyle{\vec j}$ & $\displaystyle{(\mathbf{3},-1,-1,-1)}$ \\
\hline
Position & $\displaystyle{\vec r}$ & $\displaystyle{(\mathbf{3},1,-1,1)}$ \\
\hline
Charge & $q$ & $\displaystyle{(\mathbf{1},-1,1,1)}$ \\
\hline
\end{tabular}
\end{minipage}
\hspace{1cm}
\begin{minipage}[t]{.35\textwidth}
\renewcommand{\arraystretch}{1.4}
\begin{tabular}{|c|c|}
\hline
Grandeur & SO$(3)\times C\times P\times T$ \\
\hline
$\displaystyle{\partial_t \vec E}$ & $\displaystyle{(\mathbf{3},-1,-1,-1)}$ \\
\hline
$\displaystyle{\partial_t \vec B}$ & $\displaystyle{(\mathbf{3},-1,1,1)}$ \\
\hline
$\displaystyle{\vec\nabla \cdot \vec E}$ & $\displaystyle{(\mathbf{1},-1,1,1)}$ \\
\hline
$\displaystyle{\vec\nabla \cdot \vec B}$ & $\displaystyle{(\mathbf{1},-1,-1,-1)}$ \\
\hline
$\displaystyle{\vec\nabla \wedge \vec E}$ & $\displaystyle{(\mathbf{3},-1,1,1)}$ \\
\hline
$\displaystyle{\vec\nabla \wedge  \vec B}$ & $\displaystyle{(\mathbf{3},-1,-1,-1)}$ \\
\hline
\end{tabular}
\end{minipage}
\end{center}
\caption{Représentations de certaines grandeurs de l'électromagnétisme classique}
\label{RepMaxwell}
\end{table}

L'écriture d'équations dans des représentations données est très contrainte. Postulons que l'on cherche des équations de champs pour l'électromagnétisme, avec la condition qu'elles sont du premier ordre en les champs, linéaires, et sourcées par les densités de charge et de courant. Les seules équations possibles sont alors les suivantes, où l'on indique à chaque fois la représentation dans laquelle est l'équation sous SO$(3)\times C\times P\times T$.
\begin{align}
 \vec \nabla \cdot \vec E &= \alpha_{_1} \rho, & ~~~~~~~~~~~~  (\mathbf{1},-1,1,1)\\
 \vec \nabla \cdot \vec B &=0, &~~~~~~~~~~~~ (\mathbf{1},-1,-1,-1)\\
 \vec\nabla \wedge \vec E &=\alpha_{_2} \partial_t \vec B, &~~~~~~~~~~~ (\mathbf{3},-1,1,1)\\
 \vec\nabla \wedge  \vec B &=\alpha_{_3} \vec j + \alpha_{_4} \partial_t \vec E. &~~~~~~~~~~~ (\mathbf{3},-1,-1,-1)
\end{align}
Les $\alpha_i$ sont des constantes réelles qui ne rentrent pas en compte dans les produits de représentations\footnote{Ces constantes ne sont en général pas arbitraires, mais des puissances des différentes constantes fondamentales associées aux lois étudiées. Dans le cas de l'électromagnétisme du vide, cela donne des constantes de la forme $\alpha_i\propto {\epsilon_0}^{n_i} {\mu_0}^{mi}$.}. On reconnait aisément la forme des équations de Maxwell. Cette écriture montre aussi qu'une densité de charge qui sourcerait le champ magnétique (une densité de monopôles magnétiques) devrait être décrite par un scalaire de parité négative, ce qui n'existe pas en physique classique.

On peut de même essayer de construire la force qui s'exerce sur une particule dans un champ électromagnétique, en supposant qu'elle est linéaire en les champs électromagnétiques. Une force est dans la même représentation que le vecteur accélération, à savoir $(\mathbf{3},1,-1,1)$ sous SO$(3)\times C\times P\times T$, une dérivation par rapport au temps rajoutant simplement un signe à la façon dont la grandeur physique se transforme sous renversement du temps. Les champs électromagnétiques n'étant pas invariants sous conjugaison de charge, il est nécessaire d'introduire une grandeur physique caractérisant la particule et n'étant pas invariante sous cette opération de symétrie. Ce rôle est joué par sa charge électrique, permettant d'écrire une force électrique en $\vec F_E=q \vec E$. Il n'est par contre pas possible d'utiliser directement le champ magnétique, et il est nécessaire de construire une force magnétique à l'aide d'un produit de représentations. Sans introduire d'autres grandeurs que celles décrivant la particule, et restant linéaire vis-a-vis du champ, la seule possibilité est $\vec F_B = q \vec v \wedge \vec B$. On retrouve bien la force de Lorentz.

Cette construction \emph{a posteriori} est avant tout didactique. Cependant, elle montre bien qu'une fois déterminées les symétries d'un système, ainsi que les représentations sous ces symétries des différentes grandeurs physiques le décrivant, la nécessité de n'égaler que des grandeurs dans la même représentation donne des contraintes très fortes sur le système. Nous reviendrons dans la section~\ref{TheorieFonda&Effect} sur les hypothèses supplémentaires qui ont été prises, dans le cadre d'une discussion plus générale sur la construction de théories physiques.

\section{Théorème de Wigner et construction d'une action}
\label{SymPartWigner}

\noindent
Dans cette section, nous décrivons de façon plus formelle les deux points de la section~\ref{IntroGroupes}. Cette formulation mathématique ainsi que les résultats associés sont éminemment importants dans la cadre de la construction de modèles en physique théorique, puisqu'ils posent la base de la description mathématique des objets physiques, ainsi que des équations régissant leur évolution. Sauf précision contraire, nous nous plaçons à présent dans le cadre de la théorie quantique des champs.

Dans une théorie quantique, les objets physiques sont décrits par des rayons vecteurs normés, repérés par des vecteurs (ou kets) $\ket{\Psi}$ d'un espace de Hilbert $\mathcal{H}$ et définis à une phase près. Les observables sont définies par des opérateurs hermitiens $A$, dont les valeurs propres $a_n\in \mathbb{R}$ sont les seuls résultats possibles des processus de mesures associés à ces observables, avec comme probabilité $P_n \equiv P(a_n) \equiv {|\braket{\Phi_n|\Psi}|}^2$, notant $\ket{\Phi_n}$ le vecteur propre de $A$ associé à la valeur propre $a_n$ tel que $A \ket{\Phi_n} = a_n \ket{\Phi_n}$. Dans ce cadre, les transformations de symétrie $T$ sont définies comme les transformations s'appliquant à tous les rayons vecteurs et qui ne changent pas les probabilités de mesure, \emph{i.e.} telles que les nouveaux états $\ket{\Psi^\prime}$ et $\ket{\Phi^\prime_n}$ après transformation vérifient ${|\braket{\Phi^\prime_n|\Psi^\prime}|}^2= {|\braket{\Phi_n|\Psi}|}^2$.

Le théorème de Wigner~\cite{Wigner:1931,Weinberg:1995mt}, démontré au début des années 30, énonce sous ces seules hypothèses que les transformations de symétrie sont représentées sur l'espace de Hilbert par des opérateurs soit linéaires et unitaires, soit antilinéaires et antiunitaires. Les transformations qui peuvent être reliées continument à l'identité doivent être représentées par un opérateur linéaire et unitaire, le cas antilinéaire et antiunitaire étant principalement lié au renversement du temps. Le fait que les représentations doivent être unitaires n'impose pas de n'utiliser que des groupes unitaires dans leur représentation de définition, tant que les représentations qu'on utilise sont bel et bien unitaires. Cela permet cependant d'exclure les groupes qui n'ont pas de représentations unitaires, comme c'est le cas par exemple de SO(11)~\cite{PeterUzan}.

Comme discuté dans la section~\ref{IntroGroupes}, les transformations de symétrie forment un groupe, puisqu'une composition de transformations qui ne changent pas les probabilités garde cette propriété, et qu'on peut définir les transformations inverses et identité. Les transformations de symétrie sont donc décrites sur l'espace de Hilbert par des représentations linéaires de ces groupes de symétrie\footnote{Ces représentations peuvent \emph{a priori} être projectives car les états sont repérés par des rayons vecteurs. Cependant, et dans le cadre des groupes de Lie, on peut se ramener à des représentations non projectives si les charges centrales apparaissant dans les relations de structure de l'algèbre de Lie peuvent être absorbées dans les générateurs, et si le groupe est simplement connexe~\cite{Weinberg:1995mt}. Pour un groupe non simplement connexe, on pourra considérer son recouvrement universel, qui aura lui des représentations non projectives.}. Dans le cas des groupes continus (et donc des représentations linéaires et unitaires), la notion de groupe de Lie apparaît naturellement si on suppose que ces groupes peuvent être décrits par un nombre fini de paramètres réels.

En théorie des champs, on construit alors les modèles théoriques à partir d'une action. Cette action est une fonctionnelle qui s'écrit sous la forme 
\begin{equation}
\label{DefAction}
S\left(\tau_1,\tau_2,[\Phi^a]\right) \equiv \int^{\tau_1}_{\tau_2} \text{d}^4 x ~ \mathcal{L}\left(\Phi^a, \partial_\mu \Phi^a,\cdots \right), 
\end{equation} 
où $\mathcal{L}\left(\Phi^a, \partial_\mu \Phi^a, \cdots\right)$ est la densité locale de Lagrangien, et où $\Phi^a$ désigne un ensemble de champs dans des représentations du groupe complet de symétrie. On a ici présupposé qu'on étudiait une théorie invariante par translation, et que le Lagrangien ne dépendait donc pas explicitement de $x^\mu$. En général, on demande que cette action construite à partir d'un Lagrangien soit réelle, donne des équations classiques du mouvement au maximum du second ordre (nous y reviendrons dans la section~\ref{TheorieFonda&Effect}), et soit invariante sous les transformation du groupe de Poincaré~\cite{Ramond:1981pw}. Les groupes de transformation qui laissent l'action invariante sont les groupes de symétrie du modèle étudié, qui peuvent décrire des symétries aussi bien globales que locales. En d'autres termes, l'action doit être un singlet de toutes les symétries de la théorie physique étudiée, y compris le groupe de Poincaré. Cette propriété de l'action implique que toutes les équations du mouvement sont bien dans une représentation donnée, car celles-ci sont obtenues en variant l'action par rapport à un champ, ce qui donne une équation dans la représentation conjuguée de ce champ. Cette définition des symétries comme les transformations qui laissent invariante l'action d'un système physique sera celle qui sera utilisée par la suite. Elle est équivalente à la définition liée à l'invariance de toutes les mesures physiques, mais plus simple à utiliser.

La description d'un système par une action est plus restrictive que la simple donnée d'un ensemble d'équations différentielles décrivant l'évolution de ses différents paramètres, car toutes les équations différentielles doivent découler d'une même action. Le formalisme de l'action donne cependant un cadre naturel à la description de l'évolution d'un système possédant des lois de conservation\footnote{Ce n'est donc pas étonnant si il est difficile de décrire dans le cadre d'une action des équations impliquant des phénomènes dissipatifs, comme les équations de Navier-Stokes.}. En effet, lorsqu'on décrit un système de particules possédant des quantités conservées, la diminution de cette quantité conservée pour une particule devra être compensée par l'augmentation de cette même quantité pour d'autres particules, impliquant un couplage entre les équations d'évolution des différentes particules. C'est ce couplage entre les différentes équations que permet l'action, en plus de faire un lien direct entre les symétries et les quantités conservées via le théorème de Noether (voir la section~\ref{PartNoether}). En d'autre termes, la description d'une théorie à l'aide d'une action revient à exprimer les couplages à l'aide de vertex d'interaction. Les lois de conservation apparaissent alors au niveau des vertex, par exemple par la conservation des nombres quantiques sur chaque vertex pour les théories quantifiées. Et les liens entre les différentes équations régissant les évolutions des particules sont liés au fait que celles-ci interagissent par les mêmes vertex : si un électron peut réagir en émettant ou absorbant un photon, alors un photon peut être absorbé par un électron ou un positron, et se transformer en une paire électron-positron, mais ne peut pas par exemple se transformer en un ou plusieurs électrons. Cette dernière considération sur les équations d'évolution de l'électron puis du photon montre bien le lien direct entre les équations d'évolution et les lois de conservation.

Le théorème de Wigner et la description d'une théorie par son action sont des éléments centraux pour construire des modèles physiques. L'approche moderne, plus systématique, consiste plutôt à construire des théories à partir du haut, c'est à dire en postulant initialement des groupes de symétrie et les représentations associées, et en étudiant les théories physiques qu'il est alors possible d'écrire. L'approche historique a plutôt consisté à construire des théories par le bas, étudiant \emph{a posteriori} les symétries des modèles étudiés et en tirant les conséquences mathématiques. 

\section{Symétries et quantités conservées, théorème de Noether}
\label{PartNoether}

\noindent
Une conséquence très forte des symétries d'un système physique est la présence de quantités conservées indépendantes lors de l'évolution temporelle de celui-ci. Ce résultat est décrit par le théorème de Noether~\cite{Noether:1918zz}, formulé en 1918, qui énonce que lorsqu'une action est invariante sous les transformations d'un groupe de symétrie continu, il existe autant de quantités conservées dans le temps que de transformations infinitésimales associées au groupe de symétrie, soit la dimension du groupe de symétrie. En pratique, lorsqu'une loi de conservation apparaît dans une théorie, on cherche systématiquement la symétrie continue associée, et réciproquement.

Explicitons ce théorème, valide pour une théorie décrite par une action, à partir des notations de la formule~\eqref{DefAction}~\cite{Ramond:1981pw,Weinberg:1995mt,PeterUzan}. On peut écrire la transformation infinitésimale du champ $\Phi^a$ associée à un groupe de symétrie continu de dimension $N$, et décrite par un jeu de paramètres $\delta \theta^\alpha$ pour $\alpha=1,\cdots,N$ -- voir le complément~\ref{DefTransfoGroupes} pour plus de détails --, comme
\begin{equation}
\label{EqTransfoChamps}
\delta \Phi^a = i \delta \theta^\alpha (T_\alpha)^a{}_b\Phi^b.
\end{equation}
Cette transformation infinitésimale laisse l'action invariante, ce qui signifie que la variation du Lagrangien est une dérivée totale~:
\begin{equation}
\delta \mathcal{L}= \delta \theta^\alpha  \partial_\mu \Lambda^\mu_\alpha.
\end{equation}
Reliant ces deux variations, et utilisant une intégration par partie ainsi que les équations du mouvement, on obtient $N$ courants conservés, à savoir
\begin{equation}
j^\mu_\alpha = \Lambda^\mu_\alpha - i \frac{\partial \mathcal{L}}{\partial\left(\partial_\mu \Phi^a \right)} (T_\alpha)^a{}_b\Phi^b,
\end{equation}
tels que 
\begin{equation}
\partial_\mu j^\mu_\alpha =0, 
\end{equation}
formulation standard d'une loi de conservation en théorie des champs. Les charges conservées sont obtenues en intégrant la composante temporelle des courants sur l'espace, 
\begin{equation}
Q_\alpha(t) = \lim\limits_{V \rightarrow +\infty} \int_V \text{d}^3 x j^0_\alpha (t,\vec{x}),
\end{equation}
et vérifient bien
\begin{equation}
\frac{\text{d}Q_\alpha}{\text{d}t} = 0,
\end{equation}
ce qui se montre en utilisant le théorème de Green-Ostrogradski. En fait, le terme en $\Lambda^\mu_\alpha$ n'est non nul que si la symétrie correspond à une transformation générale des coordonnées en plus du champ, et il alors possible d'exprimer $\Lambda^\mu_i$ en fonction de cette transformation\footnote{L'écriture complète de l'équation~\eqref{EqTransfoChamps} est alors, pour une transformation infinitésimale, et en notant avec des primes les grandeurs après la transformation,
\begin{equation}
\Phi^{\prime a} (x^\prime_\mu) = \Phi^a (x_\mu) +i \delta\theta^\alpha(x) (T_\alpha)^a{}_b\Phi^b(x_\mu).
\end{equation}
Et pour une transformation de coordonnées $\delta x_\mu = \delta\theta^\alpha f_{\alpha\mu}$ (où l'on rappelle que $\alpha$ est un indice lié au groupe de symétrie), on a~\cite{PeterUzan}
\begin{equation}
\Lambda^\mu_\alpha = \left(\frac{\partial \mathcal{L}}{\partial(\partial_\mu \Phi^a)}\partial^\nu \Phi^a - g^{\mu\nu}\right)f_{\alpha\nu}.
\end{equation}}. Ce n'est pas le cas pour les symétries dites internes, qui laissent invariant le Lagrangien et pas seulement l'action, et où l'on a alors simplement
\begin{equation}
j^\mu_\alpha = - i \frac{\partial \mathcal{L}}{\partial\left(\partial_\mu \Phi^a \right)} (T_\alpha)^a{}_b\Phi^b. 
\end{equation}

On peut considérer comme premier exemple le cas du Lagrangien associé à un spineur de Weyl (voir la section~\ref{PartSpineurs} pour plus de détails), qui s'écrit pour un spineur gauche 
\begin{equation}
\mathcal{L_\psi} =  \psi^\dagger_L \sigma^\mu \partial_\mu \psi_L.
\end{equation}
Ce Lagrangien est invariant sous la transformation globale
\begin{equation}
\psi_L\rightarrow e^{i\theta}\psi_L,
\end{equation}
lié à un groupe de symétrie abélien U(1). On en déduit la conservation du courant
\begin{equation}
j^\mu = i \psi^\dagger_L \sigma^\mu \psi_L,
\end{equation}
qui a comme charge conservée 
\begin{equation}
Q = i \int \text{d}^3 x \psi^\dagger_L \psi_L,
\end{equation}
ce qui correspond à la conservation du nombre de particules. Dans le cas ou plusieurs particules sont chargées sous le même groupe abélien, il est nécessaire d'introduire une charge sous les transformations U(1) pour chacun des champs. Par exemple pour un Lagrangien 
\begin{equation}
\mathcal{L_{\psi,\chi}} =  \psi^\dagger_L \sigma^\mu \partial_\mu \psi_L + \chi^\dagger_L \sigma^\mu \partial_\mu \chi_L,
\end{equation}
associé à une symétrie globale
\begin{equation}
\left\{
\begin{array}{l}
\psi_L\rightarrow e^{iq_\psi\theta}\psi_L,\\
\chi_L\rightarrow e^{iq_\chi\theta}\chi_L,
\end{array}
\right.
\end{equation}
la charge conservée est
\begin{equation}
Q = i \int \text{d}^3 x \left( q_\psi \psi^\dagger_L \psi_L + q_\chi \chi^\dagger_L \chi_L\right).
\end{equation}
Ceci montre la conservation du nombre de particules pondéré par les charges de celles-ci, ou en d'autres mots la conservation de la charge totale des particules.

Les calculs sont exactement similaires pour une symétrie locale associée à des fermions, le couplage avec le champ de jauge n'entrant pas dans le courant car n'impliquant pas de dérivées. Dans le cas général, les quantités conservées ne seront cependant pas toujours associées à des charges. Pour les symétries associées à des groupes de Lie non abéliens, il y a par exemple autant de charges conservées que le rang du groupe de symétrie, puisque ces charges sont associées à des symétries U(1) qui commutent. Les autres quantités conservées sont alors liées à la structure des transformations de symétries. Nous y reviendrons plus en détails dans la section~\ref{PartJaugesNonAbeliennes}.

La discussion sur le lien entre les symétries et les quantités conservées a été faite dans un cadre classique. En pratique, il arrive que les lois de conservation liées à des symétries ne soient plus vérifiées après quantification des théories. Ces violations des lois de conservation après quantification,  rendant généralement caduques les modèles où elles apparaissent, sont nommées anomalies. Il est à noter que dans certaines théories de jauge, plusieurs anomalies distinctes peuvent s'annuler entre elles pour un contenu en matière particulier ; c'est notamment le cas du Modèle Standard de la Physique des particules, comme nous le discuterons ultérieurement.

\section{Quantification et renormalisation}
\label{PartQuantification}

\noindent
Bien que le théorème de Wigner concerne les grandeurs quantiques, la construction de l'action d'une théorie physique a pour l'instant été discutée au niveau classique. Le passage de la description classique à la description quantique nécessite donc un processus dit de quantification. Plusieurs processus équivalents de quantification existent, chacun d'eux mettant en avant certaines propriétés particulières des théories (unitarité, covariance, etc.). À la méthode historique de la quantification canonique, consistant à promouvoir les observables physiques en opérateurs quantiques et à introduire les relations de commutations dites canoniques entre variables conjuguées, on préfère à présent l'approche moderne de la quantification par l'intégrale de chemin~\cite{VanVleck:1928zz,Dirac:1933xn,Feynman:1948ur}. Nous présentons ici les résultats de base de cette quantification sans nous attacher aux méthodes calculatoires associées et au formalisme mathématique sous-jacent.

Dans la quantification par l'intégrale de chemin\footnote{Cette quantification est initialement écrite à partir de la formulation Hamiltonienne de la théorie, et certaines précautions doivent être prises avant d'arriver à la forme écrite en fonction du Lagrangien, notamment lorsque la théorie contient des degrés de liberté fermioniques ou de jauge.}, la probabilité pour un champ $\phi$ de passer d'un état $\ket{\tilde{\phi}}$ à un temps $t$ à un état $\ket{\tilde{\phi}^\prime}$ à un temps $t^\prime$ est donnée par 
\begin{equation}
\label{EqPathIntegral}
\displaystyle{
\braket{\tilde\phi^\prime,t^\prime|\tilde\phi,t} = \int_{\phi(t,\vec x) = \tilde{\phi}(\vec x)}^{\phi(t^\prime,\vec x)=\tilde{\phi}^\prime(\vec{x})} \mathcal{D}\phi ~ e^{\frac{i}{\hbar} \int_{t}^{t^\prime}\text{d}^4 x\mathcal{L}(\phi,\partial_\mu \phi)} 
= \int_{\phi(t,\vec x) = \tilde{\phi}(\vec x)}^{\phi(t^\prime,\vec x)=\tilde{\phi}^\prime(\vec{x})} \mathcal{D}\phi ~ e^{\frac{i}{\hbar} S(t,t^\prime,[\phi])},
}
\end{equation}
où l'on a exceptionnellement fait apparaître le facteur $\hbar$. La première intégrale est calculée sur toutes les configurations classiques du champ $\phi$ telles que celui-ci vaut $\tilde{\phi}$ en $t$ et $\tilde{\phi}^\prime$ en $t^\prime$. La mesure d'intégration $\mathcal{D}\phi$ contient un certain nombre de normalisations que nous ne préciserons pas. Chacune des configurations classiques apparaît avec un facteur $e^{\frac{i}{\hbar} S(t,t^\prime,[\phi])}$, qui est un terme oscillatoire. Il est cependant possible d'identifier ce terme à un facteur de Boltzmann après passage en temps euclidien $\tau=i t$ (ce qui correspond à une rotation dite de Wick). Il est possible de rajouter un terme de source classique $J$, apparaissant sous la forme 
\begin{equation}
\label{EqPathIntegralSource}
\displaystyle{
\braket{\tilde\phi^\prime,t^\prime|\tilde\phi,t}_J = \int_{\phi(t,\vec x) = \tilde{\phi}(\vec x)}^{\phi(t^\prime,\vec x)=\tilde{\phi}^\prime(\vec{x})} \mathcal{D}\phi ~ e^{\frac{i}{\hbar} \int_{t}^{t^\prime}\text{d}^4 x\left[\mathcal{L}(\phi,\partial_\mu \phi)+J \phi\right]}, 
}
\end{equation}
ce qui est notamment très utile à des fins calculatoires\footnote{Les termes apparaissant dans les calculs de théorie quantique des champs sont principalement exprimés à l'aide des fonctions de Green, dont le calcul est donc central. Notant $\ket{\Omega}$ les vides asymptotiques à $t=\pm\infty$ et utilisant les notations des équations~\eqref{EqPathIntegralOperateurs} et~\eqref{EqPathIntegral}, ces fonctions de Green s'écrivent (pour une théorie à un champ scalaire)
\begin{equation}
\label{EqGreenFunction}
\displaystyle{
G^{(N)}(x_1,\cdots,x_N) = 
\bra{\Omega} T\left\{ \mathbf{\Phi}(x_1) \cdots \mathbf{\Phi}(x_N)  \right\}\ket{\Omega}
 = \int_{\phi(-\infty,\vec x) = \Omega}^{\phi(+\infty,\vec x)=\Omega} \mathcal{D}\phi ~ \phi(x_1) \cdots \phi(x_N) ~ e^{i S(t,t^\prime,[\phi])}.
}
\end{equation}
où l'on a utilisé des notations concises pour les coordonnées d'espace-temps. Introduisant comme fonction génératrice l'amplitude de transition des vides asymptotiques en présence d'une source extérieure $J$
\begin{equation}
W[J]=\braket{\Omega|\Omega}_J = \int_{\phi(-\infty,\vec x) = \Omega}^{\phi(+\infty,\vec x)=\Omega} \mathcal{D}\phi ~ e^{i \int_{t}^{t^\prime}\text{d}^4 x\left[\mathcal{L}(\phi,\partial_\mu \phi)+J \phi\right]}, 
\end{equation}
les fonctions de Greens s'obtiennent alors via
\begin{equation}
G^{(N)}(x_1,\cdots,x_N)=\frac{1}{i^N} \frac{\delta}{\delta J(x_1)} \cdots \frac{\delta}{\delta J(x_N)} W[J] \Big|_{J=0}. 
\end{equation}
}. Il est aussi possible de calculer les éléments matriciels d'opérateurs locaux $\mathcal{O}_A$ à des points d'espace-temps $(t_A,\vec{x}_A)$ via
\begin{multline}
\label{EqPathIntegralOperateurs}
\displaystyle{
\bra{\phi^\prime,t^\prime}T\left\{ \mathcal{O}_A[\mathbf{\Phi}(t_A,\vec{x}_A)]\mathcal{O}_B[\mathbf{\Phi}(t_B,\vec{x}_B)]  \cdots \right\}\ket{\phi,t}
}
\\ \displaystyle{
 = \int_{\phi(t,\vec x) = \tilde{\phi}(\vec x)}^{\phi(t^\prime,\vec x)=\tilde{\phi}^\prime(\vec{x})} \mathcal{D}\phi ~ \mathcal{O}_A[\phi(t_A,\vec{x}_A)]\mathcal{O}_B[\phi(t_B,\vec{x}_B)]  \cdots ~ e^{\frac{i}{\hbar} \int_{t}^{t^\prime}\text{d}^4 x\mathcal{L}(\phi,\partial_\mu \phi)}, 
}
\end{multline}
où $\mathbf{\Phi}$ désigne le champ quantique associé au champ classique $\phi$, et où $T$ désigne le produit chronologique. Ces considérations données ici pour un unique champ scalaire se généralisent à tout type de champ.

Dans la limite classique $\hbar\rightarrow 0$, les variations de phase pour des trajectoires adjacentes sont très importantes, et toutes les contributions vont s'annuler par interférence sauf pour les configurations où l'action est stationnaire. On retrouve donc bien les trajectoires classiques comme les extrema de l'action. Dans le cas général, toutes les configurations possibles pour les champs (aussi appelés les trajectoires des champs) jouent un rôle dans l'obtention des résultats de mesure, rendant les calculs analytiques impraticables. Une approche perturbative est donc privilégiée quand elle est possible, la plus simple d'entre elle consistant à faire des développement en puissance de $\hbar$ (ce qui permet par exemple de calculer une action classique effective prenant en compte les premières contributions quantiques). En physique des particules, il est commode de considérer la propagation de particules \og libres \fg{} dont l'interaction est traitée ensuite de façon perturbative. C'est par exemple le cas pour l'électrodynamique quantique, dont la constante de couplage vaut $\alpha\simeq \frac{1}{137}$. Le calcul de l'intégrale de champ consiste alors à calculer ordre par ordre tous les diagrammes de Feynman correspondant à la réaction étudiée, ceux-ci décrivant des particules libres réagissant via des vertex d'interactions. 

\begin{figure}[h!]
\begin{center}
\includegraphics[scale=0.6]{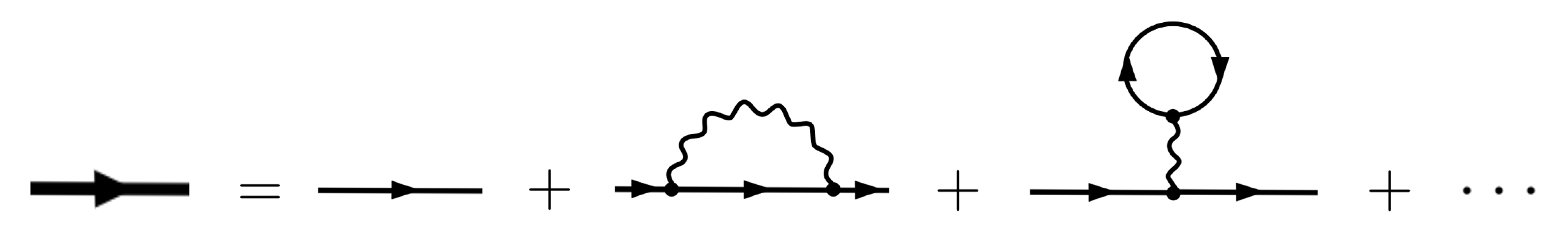}
\vspace{-0.5cm}
\end{center}
 \caption{Calcul du propagateur habillé d'un électron à partir des diagrammes de Feynman écrits en fonction des grandeurs nues.}
 \label{RenormalisationElectron}
\end{figure}

La notion de particule \og libre \fg{} est cependant une notion idéale difficile à définir, l'auto-interaction d'une particule avec elle-même étant par exemple toujours présente\footnote{Dans le cadre de l'électrodynamique quantique, traiter la notion de particule \og libre \fg{} revient à ignorer l'interaction électromagnétique.}. Pour prendre en compte ces auto-interactions et décrire des particules réelles, on introduit des grandeurs dites habillées, en opposition aux grandeurs nues liées aux particules libres idéales (ce sont ces grandeurs nues qui apparaissent dans le Lagrangien). La masse habillée d'un électron prend alors en compte l'auto-interaction d'un électron avec lui-même lors de sa propagation, et se calcule en sommant les contributions de tous les diagrammes de Feynman écrits à partir des grandeurs nues de la théorie, comme décrit dans la figure~\ref{RenormalisationElectron}. Une théorie décrite par ses grandeurs habillées est dite renormalisée, la renormalisation devant être effectuée pour les masses des particules comme pour les constantes de couplages des interactions~\cite{Weinberg:1996kr}.

Une caractéristique particulière des théories quantiques est que la valeur des paramètres habillés d'une théorie dépend de l'échelle d'énergie considérée (ou, de manière équivalente, des longueurs caractéristiques considérées). C'est une conséquence des procédés de régularisation utilisés pour renormaliser une théorie. En effet, les calculs de masses habillées comme celui présenté dans la figure~\ref{RenormalisationElectron} donnent en général des résultats infinis, et il est nécessaire d'introduire un procédé dit de régularisation ainsi que des contre-termes infinis pour obtenir des résultats finis. Ces procédés peuvent être étudiés de manière non-perturbative, mais ils sont généralement considérés ordre par ordre. Les régularisations font alors apparaître une dépendance explicite des grandeurs renormalisées en l'échelle d'énergie considérée. En conséquence, les masses des particules et les couplages associés aux interactions ont des valeurs qui dépendent de l'échelle d'énergie à laquelle ils sont mesurés.

\begin{figure}[h!]
\begin{center}
\includegraphics[scale=0.6]{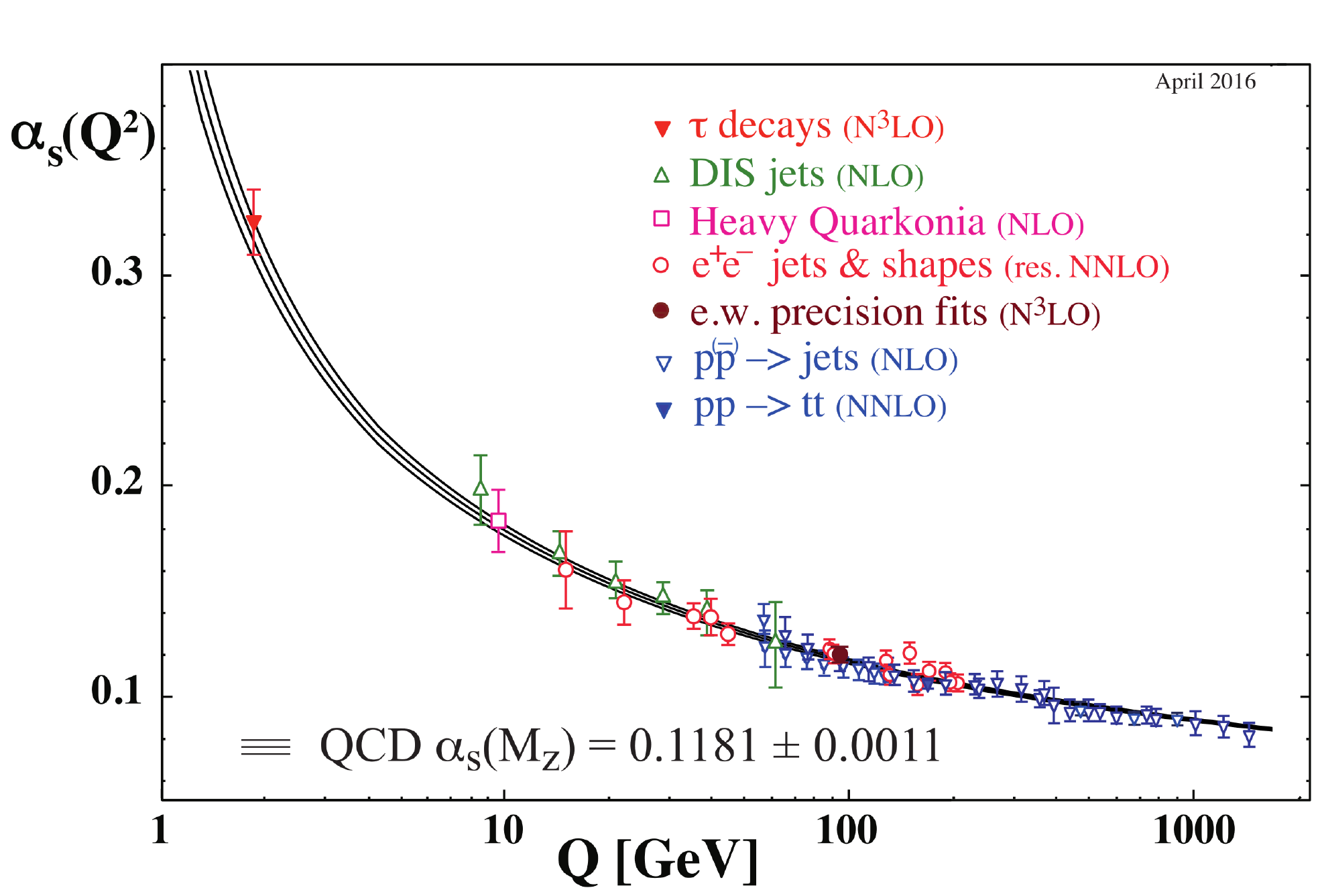}
\vspace{-0.5cm}
\end{center}
 \caption{Mesures de la constante de couplages des interactions fortes $\alpha_s$ pour différentes échelles d'énergie $Q$, tiré de~\cite{PDG2016}. Pour chacune de ces expériences, l'ordre des calculs perturbatifs effectués pour extraire $\alpha_s$ est indiqué, NLO signifiant \og Next-to-leading-order \fg{}, et ainsi de suite.}
 \label{AlphaQCD}
\end{figure}

Ces résultats, conséquences directes du processus de quantification des théories des champs, sont bien observés expérimentalement. L'exemple de la renormalisation de la constante de couplage des interactions fortes en fonction de l'énergie est donné figure~\ref{AlphaQCD}. Historiquement, l'apparition d'infinis dans les processus de renormalisation et l'utilisation de régularisation pour les gérer a été considéré comme un problème central des théories quantiques des champs. Il est cependant intéressant de noter que sans l'apparition d'infinis et l'introduction de régularisations pour obtenir des résultats finis, la renormalisation des grandeurs physiques ne dépendrait pas de l'échelle d'énergie, ce qui contredirait les observations expérimentales. Dans ce cadre, l'apparition de tels infinis semble donc nécessaire pour écrire une théorie quantique cohérente. 

En pratique, la possibilité de renormaliser une théorie n'est pas garantie. Pour certaines théories, dite non-renormalisables, il est nécessaire d'introduire une quantité infinie de contre-termes, ce qui rend impossible l'obtention de résultats finis au dessus de certaines échelles d'énergie. Les termes Lagrangiens ayant des comportements renormalisables ou non-renormalisables sont classifiés\footnote{Le critère introduit ici permet de déterminer les termes qui sont \og power counting renormalizable \fg{}. Il peut être possible d'obtenir des théories incluant des termes ne validant pas ces conditions mais qui sont renormalisables suite à l'annulation des contributions non renormalisables individuellement. De telles théories sont cependant très contraintes.}. On les écrits pour cela sous la forme
\begin{equation}
S \ni \int \text{d}^4 x ~ g_i \phi \psi \chi \cdots,
\end{equation}
sans spécifier les spins des différents champs (on rappelle que l'action est adimensionnée, et que $\text{d}^4 x$ est de dimension $-4$, les dimensions étant comptées via $[L]^{-1}=[T]^{-1}=[M]=1$). Dans le cas où l'on ne considère que des scalaires, des spin $\frac12$, et des vecteurs sans masse -- de dimensions respectives 1, $\frac32$ et 1 --, les termes renormalisables sont alors ceux qui ont un pré-facteur de dimension positive ou nulle. Exprimé différemment, les termes renormalisables sont des produits de champs de dimension inférieure ou égale à 4. Ainsi, des termes comme $e \psi \slashed A \psi$, $F_{\mu\nu}F^{\mu\nu}$ ou $\lambda \phi^4$ sont renormalisables, mais pas $\frac{1}{m} \psi \phi^2 \psi$, $\frac{1}{m^2}\psi\psi\psi\psi$ ou $\frac1{m^2} \phi^6$. Ces résultats, valables en dimension 4, se généralisent aisément pour des espaces de dimension différente.

\section{Théories fondamentales et effectives}
\label{TheorieFonda&Effect}

\noindent
Il est à présent admis que les théories physiques fondamentales doivent être décrites dans un formalisme quantique. Dans le cadre de la théorie quantique des champs, seuls les modèles ne comportant que des termes renormalisables peuvent être décrits à toutes les échelles d'énergie. Cela ne signifie cependant pas que l'inclusion de termes non-renormalisables dans une théorie lui fait perdre son pouvoir prédictif à toutes les échelles d'énergie. Les constantes de couplage des termes non renormalisables étant dimensionnées, elles font en effet apparaître naturellement une échelle d'énergie caractéristique. Notant 
\begin{equation}
g_i = \frac{\tilde{g}_i}{M^{|d_i|}},
\end{equation}
la constante de couplage d'un terme non renormalisable de dimension $d_i<0$, on peut alors montrer que les fonctions de Green à l'ordre $N$ pour des échelles d'énergie $\Lambda$ se comportent après régularisation en~\cite{Weinberg:1996kr}
\begin{equation}
G^{(N)}(\Lambda) \simeq (\tilde{g_i})^N \left(\frac{\Lambda}{M} \right)^{|d_i|N}.
\end{equation}
Ces fonctions de Green impliquent des résultats divergents pour des échelles d'énergie plus grandes que l'échelle caractéristique $M$, mais elles sont également rapidement supprimées pour des échelles d'énergie petites devant $M$. Les termes non-renormalisables ne sont donc pas problématiques à toutes les échelles d'énergie, mais sont au contraire négligeables en dessous de certaines énergies caractéristiques.

Cette propriété implique une séparation des échelles d'énergie : étudiant un modèle à une énergie donnée, les termes pertinents du point de vue phénoménologique seront généralement seulement les termes renormalisables. Si on considère que seuls les termes renormalisables sont fondamentaux du point de vue de la description quantique d'une théorie, il est donc possible d'écrire une théorie fondamentale valable à certaines échelles d'énergie sans préjuger de sa validité à toutes les échelles d'énergie. Ainsi, et même si la description à basse énergie fait apparaître des termes non-renormalisables à haute énergie, ceux-ci sont supprimés en loi de puissance à basse énergie et peuvent donc y être ignorés. C'est cette séparation des échelles d'énergie qui justifie la recherche de théories fondamentales, dans un domaine de validité donné, par l'investigation de modèles ne comportant que des termes renormalisables. 

Le Modèle Standard de la physique des particules peut ainsi être considéré comme une théorie fondamentale, dont on considère à l'heure actuelle qu'elle est valable jusqu'à des énergies maximales de l'ordre de $10^{16}$ GeV -- voir la section~\ref{PartConclusionSM} --. Dans ce cadre, il est possible d'introduire la physique des particules et ses outils théoriques de façon plutôt naturelle si on part du principe qu'on cherche à décrire une théorie fondamentale et donc renormalisable. C'est ce que l'on fera par la suite en discutant la construction et les propriétés des théories de jauge et du mécanisme de Higgs de brisure spontanée de symétrie dans la partie~\ref{ChapterTheoriesDeJauges}, qui serviront à construire le Modèle Standard et les théories de grande unification dans la partie~\ref{ChapterSMAndBeyond}.

L'utilisation de théories ne comportant que des termes renormalisables n'est cependant pas toujours possible, ou du moins pas nécessairement adaptée au phénomène que l'on souhaite décrire. L'interaction gravitationnelle, dont aucune description quantique satisfaisante n'est admise à l'heure actuelle, en est un exemple. L'étude pratique des interactions fortes en est un autre, l'obtention de résultats comparables avec l'expérience étant beaucoup plus aisé à partir de théories effectives qu'avec l'utilisation de la seule chromodynamique quantique. D'autres hypothèses que la renormalisabilité sont alors introduites afin de limiter les termes qu'il est possible d'écrire, la première d'entre elle étant de n'utiliser que des termes n'impliquant pas d'instabilités. Cette approche effective est notamment utilisée pour construire des théories de gravité modifiée, ce que nous développerons dans les parties~\ref{ChapterGalileons} et~\ref{ChapterGalVec}.

Sans prétendre donner une classification précise des théories fondamentales et effectives\footnote{Le Modèle Standard de la physique des particules peut par exemple être considéré comme une théorie effective du fait qu'il ne décrit pas la physique à très haute énergie.}, ces dernières sont nécessaires lorsque l'état de vide d'une théorie nécessite une description compliquée, ou n'est pas bien compris à un niveau fondamental. Les champs apparaissant dans de telles théories seront liés à des excitations autour d'un état du vide dont la description exacte n'est pas forcément connue. Ces excitations ne seront alors pas identifiées à des particules élémentaires, comme on aura tendance à le faire pour les théories fondamentales. C'est par exemple le cas pour l'interaction gravitationnelle -- où la description du vide quantique de la théorie à un niveau fondamental pose problème --, ou pour les interactions fortes -- où les états réels sont très difficiles à décrire aux énergies où le confinement a lieu, voir la section~\ref{PartRenormalisationPhenomSM}.

On peut conclure cette section en appliquant les méthodes des théories fondamentales et effectives à la construction de l'électrodynamique classique. Construisant cette théorie à partir de la théorie des groupes, il avait alors été choisi de ne garder pour les équations du mouvement que des termes linéaires en les dérivées premières des champs électromagnétiques et en les sources. Écrivant une théorie renormalisable pour un champ vecteur sans masse couplé à de la matière -- décrite par des fermions --, le Lagrangien ne peut contenir que des termes de la forme $\partial A \partial A$, ou $A \psi \psi$, où $\psi$ est le champ fermionique décrivant la matière. Différenciant ces termes par rapport au champ vecteur, on obtient des termes en $\partial\partial A$ et en $\psi \psi$, ce qui se ramène aux précédentes hypothèses de linéarité menant aux équations de Maxwell. On peut aussi chercher à construire cette théorie d'un point de vue effectif. Par exemple, l'article~\cite{Deffayet:2013tca} montre qu'en espace plat, l'extension la plus générale d'une théorie de jauge abélienne issue d'une action invariante sous les transformations de jauge et de Lorentz, et impliquant un Lagrangien et des équations du mouvement au maximum de degré deux en le champ de jauge -- \emph{i.e.} sans instabilité d'Ostrodraski --, a des équations du mouvement au maximum linéaire en les dérivées secondes de ce champ. Cette approche effective ramène donc également aux équations de Maxwell pour des équations écrites en fonctions des champs électromagnétiques.

\section{Conclusion}

\noindent
On a discuté dans ce chapitre les principes sur lesquels reposent la construction moderne de théories physiques. Les symétries jouent un rôle fondamental dans cette construction, puisqu'elles restreignent considérablement les grandeurs physiques qu'il est possible d'écrire, ainsi que les équations qui peuvent relier ces grandeurs. Les symétries, liées à des quantités conservées, justifient également la description des théories physiques par une action, celle-ci permettant de relier directement les symétries continues aux invariances grâce au théorème de Noether. Ces outils sont alors utilisés pour rechercher des théories que l'on peut quantifier, et qui sont donc renormalisables, ou des théories effectives, décrivant les degrés de liberté pertinents pour l'étude d'une phénoménologie.

Dans cette discussion, il a été nécessaire d'introduire des champs pour décrire les grandeurs physiques. Ces champs existent dans l'espace de Minkowski, dont les différentes symétries sont contenues dans le groupe de Poincaré. Il est alors possible d'appliquer les outils que l'on vient de développer à ces symétries pour étudier la structure de ces champs. Il convient alors de faire la distinction entre les représentations unitaires du groupe de Poincaré, symétrie globale permettant de décrire les états physiques liés au concept de particule, et les représentations covariantes du groupe de Lorentz, nécessaire pour écrire un Lagrangien. De cette distinction émerge le concept de symétries de jauge. 

\chapter{Symétries globales, particules, degrés de liberté physiques et de jauge}
\label{SecSymGlob}
\section{Symétries globales, notion de particules}

\noindent
Les symétries globales sont les premières à avoir été comprises en physique. Ces symétries sont liées à des transformations similaires en tout point de l'espace-temps et qui laissent invariante l'action des systèmes physiques étudiés. Elles peuvent agir sur les champs comme sur les coordonnées de l'espace-temps. Les exemples les plus courants en sont les transformations du groupe de Poincaré, du groupe de Lorentz\footnote{Dans tout ce document, et sauf indication contraire, nous désignons par groupe de Lorentz non pas le groupe de Lorentz complet (associé aux transformations qui laissent invariante la contraction de deux quadrivecteurs), mais sa composante propre et orthochrone, regroupant les transformations de déterminant positif et laissant invariant le signe de la composante temporelle des quadrivecteurs. Ce sous-groupe est la composante connexe de l'identité du groupe complet, et forme donc un groupe de Lie, correspondant à SO(3,1). Cette distinction est importante, car elle permet de séparer les transformations liées aux boosts et aux rotations spatiales des opérations de parité et de renversement du temps, contenus dans le groupe de Lorentz complet. Par exemple, on dit souvent que le modèle standard est invariant sous les transformations de Lorentz, ce qui est incorrect si on considère la forme complète qui contient $P$ et $T$.}, les permutations entre particules identiques, ou bien les conjugaison de charge $C$, opération de parité $P$, et renversement du temps $T$. 

Ces symétries étant identiques en tout point de l'espace-temps, elles peuvent être considérées comme liées à une redéfinition des grandeurs physiques : on peut décrire exactement la même physique avant et après une telle transformation. L'exemple le plus commun est le cas de la description d'une position dans un espace tri-dimensionnel doté d'une symétrie globale par rotation décrite par le groupe SO(3). On peut y décrire un vecteur par ses coordonnées cartésiennes $\vec r = (x,y,z)$, sans que ces coordonnées aient un sens intrinsèque, car toute rotation du système permet une description équivalente via un nouveau jeu $(x^\prime,y^\prime,z^\prime)$, où les nouvelles coordonnées sont décrites par une combinaison linéaire des anciennes. En conséquence, on ne peut pas identifier de façon univoque des grandeurs reliées par des transformations de symétrie globale, ces grandeurs nécessitant une convention pour être différenciées par un détecteur, et il faut les considérer comme différentes composantes de la description d'un même objet. Par exemple, quand on divise un ensemble de particules identiques en celles qui ont un moment magnétique sur une direction donnée de $+\frac12$ et $-\frac12$, n'aurait-on pas pu choisir sans perdre de généralité les signes opposés, de même qu'une autre direction de mesure ?

Cela nous amène à associer les états à une particule, que nous appellerons plus simplement particules, aux représentations irréductibles du groupe des symétries globales du système étudié. Conventionnellement, on les lie souvent uniquement aux représentations irréductibles du groupe de Lorentz inhomogène $\mathbb{R}^4\rtimes$SO(3,1)~\cite{Weinberg:1995mt}, produit semi-direct du groupe des translation de l'espace temps et de Lorentz\footnote{C'est un produit semi-direct car ayant effectué une translation de vecteur $a^\mu$, une transformation de Lorentz ultérieure va être aussi appliquée sur cette contribution aux nouvelles coordonnées.}. Cependant, suivant la discussion précédente, il n'y a pas de raison d'ignorer d'autres symétries globales associées au système physique en question. En pratique, nous partirons du groupe de Lorentz inhomogène, et identifierons ensuite comme différents états internes d'une même particule les représentations irréductibles de ce groupe reliées entre elles par d'autres symétries globales. De plus, et même si nous allons appliquer par la suite ces considérations principalement aux particules fondamentales du Modèle Standard, elles sont aussi valides pour des particules non fondamentales, comme des protons ou des molécules.

Les symétries que l'on considère usuellement pour construire le Modèle Standard de la physique des particules sont seulement les transformations du groupe de Lorentz inhomogène. Le Modèle Standard a cependant d'autres symétries globales, comme la transformation simultanée sous $C$, $P$, et $T$, notée $CPT$. L'invariance d'une théorie sous $CPT$ peut être démontrée à partir de son invariance sous les transformations de Lorentz. C'est le théorème $CPT$, démontré dans les années 1950~\cite{CPT}. Cela implique notamment qu'une violation de Lorentz peut être déduite d'une observation de la violation de la symétrie $CPT$~\cite{Greenberg:2002uu,Chaichian:2011fc}. Les transformations $C$, $P$ et $T$ ne sont pas des symétries du Modèle Standard, ainsi que toute combinaison de deux de ces transformations. La violation de $P$ par les interactions faibles a été questionnée puis testée dès les années 50~\cite{Parite1,Parite2,Parite3}, et la violation de $CP$ a été observée dans les années 60~\cite{CPviolation}. 

Certaines interactions du Modèle Standard contiennent cependant des symétries supplémentaires à celles du modèle complet, comme les interactions électromagnétiques et fortes qui sont invariantes sous transformation de parité. Toutes les particules soumises seulement à ces interactions n'interagiront donc que par des vertex similaires pour des états reliés par parité, permettant d'identifier ces états comme états internes d'une même particule. C'est par exemple le cas des états de polarisations circulaires gauche et droite du photon.

\section{Représentations unitaires, degrés de liberté physiques, spin}
\label{PartRepUnitaires}

\noindent
En conséquence de la discussion précédente, les représentations irréductibles du groupe de Lorentz inhomogène jouent un rôle très important dans la construction de modèles physiques. Pour se conformer au théorème de Wigner, on identifie les particules aux représentations irréductibles unitaires. Cela permet notamment d'identifier le nombre de composantes associées à chacune de ces représentations, qui sont associées aux degrés de liberté physiques. 

Pour le groupe des translations, ces états à une particule sont repérés par leur valeur propre pour l'opérateur $P^\mu$, leur quadri-impulsion $p^\mu$. On peut alors isoler l'action du groupe des translations sur ces états dans un facteur de phase $e^{-i a_\mu p^\mu}$ pour les translations $x^\mu \rightarrow x^\mu + a^\mu$. L'invariant relativiste associé est $p^\mu p_\mu$, qui vaut $-m^2$ pour les particules massives de masse $m$ et 0 pour les particules sans masse (on ne discute pas ici de la possibilité d'avoir des tachyons de masse carré négative). Les états internes sont alors décrits par les représentations irréductibles du groupe d'invariance de ce quadrivecteur\footnote{Une façon plus rigoureuse de décrire ces représentations irréductibles passe par l'introduction des deux opérateurs de Casimir associés à $\mathbb{R}^4\rtimes$SO(3,1), qui sont les contractions avec eux-même du vecteur impulsion et du vecteur de Pauli-Lubanski, \begin{equation}
W^\mu = \frac12 \epsilon^{\mu\nu\rho\sigma}P_\nu J_{\rho\sigma},
\end{equation}avec $J_{\mu\nu}$ le tenseur décrivant les boosts et les rotations spatiales. Les valeurs propres associées redonnent effectivement la masse des particules, ainsi que leur spin ou hélicité selon que celles-ci ont une masse ou pas.}. Il est plus simple de visualiser le groupe d'invariance en effectuant préalablement une transformation de Lorentz pour mettre le quadrivecteur impulsion sous une forme standard. On obtient les résultats suivants :
\begin{itemize}
\item Pour les particules de masse non nulle, il existe un référentiel dans lequel on peut mettre le vecteur impulsion sous la forme $p^\mu = (m,0,0,0)$. Le groupe d'invariance associé est donc le groupe SO(3) des rotations spatiales. Les représentations irréductibles sont décrites par leur spin $s$, qui prend des valeurs demi-entières ou entières, la projection de spin sur un axe prenant des valeurs $s_3 = -s, -s+1, \cdots , s-1, s$. Les états sont donc décrits par leur masse $m$, leur spin $s$, une impulsion spatiale $P_i$, et leur projection de spin $s_3$. Les particules massives de spin $s$ ont donc $2s+1$ degrés de liberté.
\item Pour les particules de masse nulle, il existe un référentiel dans lequel on peut mettre le vecteur impulsion sous la forme $p^\mu = (-\kappa,\kappa, 0, 0)$. Le groupe d'invariance associé est ISO(2), composé des rotations autour de l'axe de propagation, et des translations sur les axes perpendiculaires, l'impulsion en question décrivant une onde plane d'extension spatiale infinie. Des représentations continues sont possibles, mais ne semble pas réalisées dans la nature. Les représentations restantes sont caractérisées par leur hélicité, associée à la valeur propre de $\vec J$ projetée sur l'axe de propagation, et égale à $\pm s$ avec $s$ le spin de la représentations. Les deux états d'hélicités opposées ne sont pas reliés par une transformation de Lorentz, mais s'échangent sous opération de parité\footnote{Ils ne sont donc identifiés \emph{a priori} comme états internes d'une même particule que dans le cadre de théories invariantes sous parité (comme les deux états de polarisation du photons pour l'électromagnétisme), et pas dans le cas général.}.
\end{itemize}

Ces représentations unitaires sont très utiles pour comptabiliser le nombre de degrés de liberté nécessaires pour décrire chaque type de particule, que nous appellerons aussi degrés de liberté physique. Elles ne sont cependant pas écrites dans un formalisme covariant, puisque définies vis-à-vis du groupe d'invariance de la quadri-impulsion $p^\mu$ des particules, ce qui les rend difficilement utilisables pour écrire une théorie invariante sous les transformations de Lorentz, ce qui est nécessaire pour écrire un Lagrangien. Par exemple, la définition du spin d'une particule est explicitement liée aux rotations spatiales seulement. Il est donc nécessaire d'écrire les champs dans un formalisme covariant lié aux représentations du groupe de Lorentz, qui ne seront pas unitaires, et de voir comment ces représentations sont liées aux représentations unitaires associées aux particules.

\section{Représentations covariantes, degrés de liberté de jauge}
\label{PartRepCovariante}

\noindent
Pour décrire les représentations irréductibles du groupe de Lorentz, il est commode de se ramener à un produit de groupes simples compacts, en utilisant l'identité SO(3,1)$\simeq$SU(2)$\times$SU(2) pour son algèbre complexifiée. En effet, les relations de structure reliant les générateurs des rotations $J_i$ et des boosts $K_i$ étant 
\begin{equation}
\begin{array}{l}
\vspace{0,1cm}  \left[J_i,J_j \right] = i \epsilon_{ijk}J_k,\\
\vspace{0,1cm}  \left[J_i,K_j \right] = i \epsilon_{ijk}K_k,\\
\left[K_i,K_j \right] = -i \epsilon_{ijk}J_k,
\end{array}
\end{equation}
l'identification s'obtient à partir des générateurs
\begin{equation}
\begin{array}{l}
\vspace{0,1cm}  N_i^{L} = \frac12\left(J_i + i K_i \right),\\
N_i^{R} = \frac12\left(J_i - i K_i \right),
\end{array}
\end{equation}
qui ont bien des relations de commutations correspondant à deux SU(2) découplés~:
\begin{equation}
\begin{array}{l}
\vspace{0,1cm}  \left[N_i^{L},N_j^{L}\right] = i \epsilon_{ijk}N_k^{L},\\
\vspace{0,1cm}  \left[N_i^{R},N_j^{R}\right] = i \epsilon_{ijk}N_k^{R},\\
\left[N_i^{L},N_j^{R}\right] = 0.
\end{array}
\end{equation}
On peut alors décrire les représentations de SO(3,1) par les représentations associées à chaque SU(2), utilisant les représentations de spin, indicées par des entiers ou demi-entiers $j\geq 0$ (les spins ne sont pas ici directement équivalents au spin \og physique \fg{}, qui est celui associé aux rotations spatiales). On décrit donc les représentations de SO(3,1) sous la forme $(j_1,j_2)$, les $j_i$ décrivant la représentation sous chaque SU(2), et les produits de représentations s'obtiennent selon les règles standards de composition des spins pour chaque SU(2). Les deux groupes SU(2) apparaissant dans cette décomposition, et qui s'échangent sous opération de parité, sont notés $L$ et $R$ car leur représentations fondamentales de spin $j=\frac12$ sont associées aux spineurs de Weyl de chiralités gauche et droite. Dans le cadre d'une théorie invariante sous opération de parité, ces deux groupes ne pourraient pas être identifiés de manière univoque. Cependant, les interactions du Modèle Standard n'ayant pas cette symétrie, ils sont effectivement distincts. Par convention, on considère que les fermions interagissant via les interactions faibles sont de chiralité gauche, les autres étant de chiralité droite.

Les représentations reliées aux particules élémentaires en physique des particules sont~:
\begin{itemize}
\item $(0,0)$ est la représentation de spin zéro, elle peut être scalaire ou pseudo-scalaire selon son comportement sous l'opération de parité.
\item $(\frac12,0)$ correspond aux spineurs de Weyl de chiralité gauche, et $(0,\frac12)$ aux spineurs de Weyl de chiralité droite. Ils ont deux composantes complexes, et se transforment l'un en l'autre sous la conjugaison complexe et la parité.
\item $(\frac12,0) \oplus (0,\frac12)$ correspond aux spineurs de Dirac quand les spineurs gauches et droits sont indépendants, et aux spineurs de Majorana quand les spineurs gauches et droits sont complexes conjugués l'un de l'autre. Le spineur de Dirac a donc quatre composantes complexes, alors que le spineur de Majorana est réel et peut être décrit par deux composantes complexes.
\item $(\frac12,\frac12) = (\frac12,0) \otimes (0,\frac12)$ est la représentation de spin 1 qui correspond aux 4-vecteurs.
\item $(1,1)$ est la représentation de spin 2 qui correspond aux 4-tenseurs de rang 2 symétriques sans trace\footnote{Ce type de tenseur a 9 degrés de liberté. En relativité générale, la métrique $g_{\mu\nu}$ a dix degrés de liberté car elle correspond à une représentation réductible, somme de la représentation $(1,1)$ et $(0,0)$. La composante scalaire de $g_{\mu\nu}$ s'obtient en prenant sa trace.}.
\end{itemize}
 Seules les représentations $(j,j)$ ou $(j_1,j_2) \oplus(j_2,j_1)$ peuvent être réelles, les autres représentations étant complexes.

Nous avons ici fait le lien avec les représentations de spin 1, de spin 2, etc. Cependant, les représentations du groupe de Lorentz, qui sont non-unitaires, contiennent en général plus qu'une représentation unitaire liée à un type de particules, d'une façon imposée par leur structure. Pour obtenir le spin \og physique \fg{} des représentations unitaires qu'elles contiennent, lié aux rotations spatiales, il suffit d'utiliser l'équation $J_3 = N_3^L + N_3^R$, qui donne que la représentation $(j_1,j_2)$ contient les représentations sous les rotations associées aux spins $j=|j_1-j_2|, |j_1-j_2| + 1, \cdots, j_1+j_2$. Ainsi, la représentation quadrivecteur comporte bien un vecteur spatial et un scalaire spatial.

La notion de particule étant liée aux représentations unitaires, on voit que le fait de travailler avec un formalisme covariant va imposer certaines structures particulières. Un quadrivecteur est par exemple la façon la plus simple de décrire une particule de spin 1. Mais ce quadrivecteur décrit \emph{a priori} la propagation de jusqu'à quatre degrés de liberté, et comporte un scalaire en plus du spin 1. Dans ces cas-là, même si certains degrés de liberté ne se propageront pas à cause de contraintes dans les modèles, il reste la possibilité d'avoir propagation de degrés de liberté non physiques. Ces degrés de liberté ne sont pas accessibles via des processus de mesure, et correspondent en conséquence à des redondances dans la description des systèmes, associées à des symétries -- puisque leur valeur peut changer sans avoir un impact sur les processus de mesure -- continues et locales -- puisque les composantes non physiques sont des composantes d'un champ et donc des fonctions continues de l'espace-temps. Le concept de symétrie locale, décrivant ces degrés de liberté dits de jauge, émerge donc naturellement de cette construction. 

Les scalaires et les spineurs n'incluent pas de degrés de liberté de jauge. En effet, la représentation triviale est la même pour tous les groupes, et les particules de spin $\frac12$, qu'elles soient massives ou non, propagent deux degrés de liberté, ce qui est aussi le cas des spineurs de Weyl gauches ou droits. La possibilité de degrés de liberté de jauge apparaît par contre naturellement pour les spins 1 et 2 décrits par des vecteurs et des tenseurs symétriques de rang 2. Le fait que ces champs soient massifs ou non est important, car il y aura dans le deuxième cas propagation de moins de degrés de liberté physiques pour autant de composantes dans les représentations covariantes, ce qui libérera plus de degrés de liberté qui pourront être associés à des symétries locales. En pratique, les champs vectoriels contiennent un degré de liberté de jauge lorsqu'ils décrivent des particules de spin 1 sans masse, et aucun degré de liberté de jauge lorsqu'ils décrivent des particules de spin 1 massives\footnote{On raisonne ici pour les Lagrangiens ne comportant qu'un terme cinétique standard, de même pour le terme de masse. Pour les Lagrangiens impliquant d'autres termes, une étude Hamiltonienne est nécessaire pour étudier les différents degrés de liberté.}. 

On note finalement qu'écrits à partir de quadrivecteurs, et dans le cas où ils n'ont pas de masses, les deux d'hélicité $\pm1$ d'une particule de spin 1 sont reliés notamment par la transformation $CPT$. Cela implique qu'on peut les identifier comme deux états internes d'une même particule. Ce sera le cas notamment pour les deux états d'hélicité des bosons de jauge associés aux interactions faibles, même si cette interaction n'est pas invariante sous la parité.

\section{Spineurs de Weyl, Dirac et Majorana, masse des fermions}
\label{PartSpineurs}

\noindent
L'introduction des représentations du groupe de Lorentz fait aussi apparaître une distinction entre deux particules de spin $\frac12$ différentes, les spineurs de Weyl de chiralités gauches et droites. Dans une théorie invariante sous l'opération de parité, il est impossible de définir de manière univoque les différentes chiralités. Cependant, et comme l'invariance sous transformaiton de parité est brisée par les interactions faibles, on peut les différencier en identifiant les spineurs gauches à ceux qui sont couplés à l'interaction faible. Nous dénoterons ces spineurs à deux composants $\psi_L$ et $\psi_R$, dans les représentations $(\frac12,0)$ et $(0,\frac12)$ du groupe de Lorentz. Leurs composantes sont des variables de Grassmann anticommutantes.

Dans la littérature, on introduit souvent les théories spinorielles via les spineurs de Dirac $\Psi_{\hspace{-0.06cm}_D}$ à quatre composantes, dans la représentation $(\frac12,0)\oplus (0,\frac12)$ du groupe de Lorentz. Pour une théorie invariante sous parité, c'est la première représentation irréductible du groupe global de symétrie qu'il est possible d'écrire avec des spineurs, et il est donc légitime de l'utiliser comme point de départ. Cependant, pour une théorie qui n'est pas invariante sous parité, comme le Modèle Standard, il est \emph{a priori} suffisant d'introduire les spineurs de Weyl pour décrire les particules élémentaires. Nous préférerons cette approche plus fondamentale, plutôt que de travailler dès le début avec des grandeurs invariantes sous parité, et d'introduire ensuite en complément les interactions faibles n'agissant que sur des sous-ensembles de ces grandeurs\footnote{Lorsque les interactions faibles sont introduites à partir de spineurs de Dirac, les projecteurs $P_L = \frac12 (1+\gamma_5)$ et $P_R = \frac12 (1-\gamma_5)$ sont utilisés pour écrire des termes spécifiques aux spineurs de Weyl de chiralités gauches et droites. Ces projecteurs vérifient  
\begin{equation}
P_L  \Psi_{\hspace{-0.06cm}_D} = 
\left(
\begin{array}{c}
\psi_L\\
0
\end{array}
\right),
~~~~ \text{et} ~~~~
P_R  \Psi_{\hspace{-0.06cm}_D} = 
\left(
\begin{array}{c}
0\\
\psi_R
\end{array}
\right),
\end{equation}
où l'on utilise les notations de l'équation~\eqref{DefSpineurDirac}. Cela permet par exemple d'écrire des couplages avec les bosons $W$ dans les dérivées covariantes qui ne concernent que les spineurs de chiralité gauche. Nous éviterons cependant cette description par des spineurs de Dirac, qui peut être particulièrement trompeuse et contre-intuitive lorsque les deux spineurs de Weyl qu'ils contiennent ne sont pas dans une même représentation du groupe de jauge.}.

Les spineurs de Weyl ont des propriétés plus restrictives que les spineurs de Dirac. En effet, une théorie décrivant un spineur de Weyl sans interaction avec d'autres champs ne peut pas comporter de terme de masse, car les contributions quadratiques en ces spineurs décrivant des termes cinétiques et de masse ne sont pas obtenues via les mêmes produits de représentations. Nous ferons ce raisonnement pour un spineur de Weyl gauche, sans perdre de généralité.
Le terme cinétique, est construit en obtenant un terme quadrivectoriel quadratique en $\psi_L$  qu'on contracte avec une dérivée quadrivectorielle pour former un scalaire. Ce quadrivecteur étant obtenu via le produit $(\frac12,\frac12) \in (\frac12,0)\otimes(0,\frac12)$, il nécessite l'introduction du complexe conjugué du spineur de Weyl, et s'écrit
\begin{equation}
\label{LagSpinL}
\mathcal{L}_{\text{cin},L}= \frac12 \psi_L^\dagger \sigma^\mu \overleftrightarrow{\partial}_\mu\psi_L = \frac12 \psi_L^\dagger \sigma^\mu \partial_\mu\psi_L - \frac12  \partial_\mu \psi_L^\dagger \sigma^\mu\psi_L,
\end{equation}
le même terme avec un spineur d'hélicité droite s'obtenant en remplaçant $\sigma^\mu$ par $\bar\sigma^\mu$, vérifiant $\bar\sigma^0=\sigma^0=\mathds{1}$ et $\bar\sigma^i = -\sigma^i$.
\begin{equation}
\label{LagSpinR}
\mathcal{L}_{\text{cin},R}= \frac12 \psi_R^\dagger \bar\sigma^\mu \overleftrightarrow{\partial}_\mu\psi_R = \frac12 \psi_R^\dagger \bar \sigma^\mu \partial_\mu\psi_R - \frac12  \partial_\mu \psi_R^\dagger \bar \sigma^\mu\psi_R.
\end{equation} 
 Le terme de masse, quant à lui, est obtenu en construisant un scalaire via $(0,0)\in (\frac12,0) \otimes (\frac12,0)$, donnant 
\begin{equation}
\mathcal{L}_m= m \psi_L^T \sigma^2 \psi_L.
\end{equation}Il n'est donc pas possible d'écrire un Lagrangien décrivant un spineur de Weyl massif sans perdre la conservation du nombre de particules, ce qu'on souhaite \emph{a priori} imposer pour la dynamique d'une particule sans interaction. 

Il est par contre possible d'écrire le Lagrangien d'un spineur de Dirac massif, décrivant en fait deux spineurs de Weyl de chiralités différentes en interaction. Ce spineur s'écrit
\begin{equation}
\label{DefSpineurDirac}
\Psi_{\hspace{-0.06cm}_D} =
\left(
\begin{array}{l}
\psi_L\\
\psi_R
\end{array}
\right),
\end{equation}
et on introduit aussi son adjoint de Pauli, défini comme $\bar{\Psi}_{\hspace{-0.06cm}_D} = \Psi_{\hspace{-0.06cm}_D}^\dagger \gamma^0$. Les différentes contributions au Lagrangien s'écrivent alors
\begin{equation}
\mathcal{L}_{m,D}= m \bar{\Psi}_{\hspace{-0.06cm}_D}  \Psi_{\hspace{-0.06cm}_D} =m \left( \psi_R^\dagger \psi_L + \psi_L^\dagger \psi_R\right),
\end{equation}
et
\begin{equation}
\label{LagSpinD}
\mathcal{L}_{\text{cin},D}= \frac12 \bar{\Psi}_{\hspace{-0.06cm}_D} \gamma^\mu \overleftrightarrow{\partial}_\mu \Psi_{\hspace{-0.06cm}_D}
= \frac12\left( \psi_L^\dagger \sigma^\mu \overleftrightarrow{\partial}_\mu\psi_L + \psi_R^\dagger \bar \sigma^\mu \overleftrightarrow{\partial}_\mu\psi_R \right),
\end{equation}
qui forment bien un Lagrangien conservant le nombre total de particules. Ce Lagrangien, qui peut aussi être considéré comme décrivant le couplage entre deux spin $\frac12$ de chiralités différentes, exprime que la propagation massive de particules de spin $\frac12$ nécessite une oscillation pendant la propagation entre une particule de chiralité gauche et une particule de chiralité droite. 

Dans le cadre de la physique des particules, les interactions sont décrites par un certain nombre de symétries locales internes, auxquelles sont associées les nombres quantiques de chaque particule. La possibilité d'écrire un Lagrangien de Dirac nécessite alors que les particules associées aux deux spineurs de Weyl de chiralités différentes aient les mêmes nombres quantiques sous les symétries internes, puisque le Lagrangien doit être un singlet sous les transformations associées à ces symétries. C'est une hypothèse très forte, qui n'a pas de raison d'être vérifiée dans le cas général, et qui ne l'est d'ailleurs pas dans le Modèle Standard. Dans une théorie non invariante sous opération de parité, il faut donc décrire les particules de spin $\frac12$ par des spineurs de Weyl sans termes de masse.

Pour finir cette section de façon exhaustive, il faut mentionner le cas particulier où il est possible d'écrire un terme de masse pour un spineur de Weyl. La non-conservation du nombre de particules n'est en effet interdite que pour les particules chargées sous des symétries internes, car elle implique alors la violation des lois de conservation des nombres quantiques associées à ces symétries. Mais une telle non-conservation est possible dans le cas d'une particule qui n'est chargée sous aucune symétrie interne, et qui est donc sa propre anti-particule. C'est le cas d'un spineur de Majorana, qui peut être décrit soit simplement pas un spineur de Weyl, soit par un spineur à quatre composantes rassemblant un spineur de Weyl et son complexe conjugué. Ce type de spineur peut par exemple décrire les neutrinos de chiralité droite, qui ne sont chargés sous aucun groupe de jauge du Modèle Standard.

Finalement, on voit que de façon très contre-intuitive, la discussion précédente implique d'identifier comme deux particules différentes les électrons de chiralités gauche et droite. Certes, ils partagent un terme de masse à basse énergie, et la propagation des deux particules dans le vide résulte alors d'une oscillation entre deux électrons de chiralités gauche et droite. Cependant, ce couplage n'est permis que via le champ de Higgs, et il n'y a aucune raison de considérer ces deux électrons comme différents états internes d'une même particule. C'est pour cela qu'ils sont considérés de façon indépendante dans la construction du Modèle Standard. À haute énergie, les fermions sont donc décrits par des spineurs de Weyl sans masses qui interagissent via les symétries locales (en échangeant des bosons de jauge) et via des couplages de Yukawa avec le champ de Higgs.

Par la suite, et dans un but de simplicité, les Lagrangiens décrivant les termes cinétiques des fermions de Weyl seront écrits avec une dérivée agissant à sa droite seulement et sans facteur $\frac12$. Cette écriture est équivalente à celle donnée précédemment à une dérivée totale près\footnote{Le Lagrangien sous cette nouvelle forme n'est d'ailleurs pas forcément réel, mais c'est simplement parce qu'on a rajouté une dérivée totale complexe à un Lagrangien initialement réel.}. On contractera aussi les dérivées avec les matrices de Pauli, introduisant les notations $\slashed \partial_L = \sigma^\mu \partial_\mu$ et $\slashed \partial_R = \bar \sigma^\mu \partial_\mu$ pour les spineurs de Weyl gauches et droits respectivement. On étend également cette notation aux Lagrangiens impliquant des spineurs de Dirac, introduisant l'opérateur $\slashed \partial_D$.

\section{Conclusion}

\noindent
J'ai essayé de présenter ici dans un cadre cohérent les outils utilisés couramment pour construire des modèles physiques. Cette discussion a permis de mettre en avant les hypothèses sous-jacentes à ces outils. Bien que ces hypothèses correspondent à des principes physiques très généraux, il est cependant important d'identifier quand elles ne sont pas respectées, car la discussion de ce chapitre est alors à reconsidérer. Un exemple commun consiste à travailler avec des champs d'un espace à plus de trois dimensions spatiales. Tous les résultats découlant de l'utilisation des représentations décrites dans les sections~\ref{PartRepUnitaires} et~\ref{PartRepCovariante} -- soit une bonne partie des résultats de base de théorie des champs -- doivent alors être reconsidérés. Ce sera par exemple le cas du théorème de spin-statistiques~\cite{fierz1939relativistische,pauli1940connection}, prouvant la symétrie des fonctions d'ondes sous l'échange des particules de spin entier -- les bosons --, et leur antisymétrie sous l'échange des particules de spin demi-entier -- les fermions --~: dans un espace à deux dimensions spatiales, le spin des particules n'est en effet pas quantifié, et les particules peuvent être régies par des statistiques intermédiaires entre les statistiques de Bose-Einstein et de Fermi-Dirac~\cite{Wilczek:1982wy}.


\part{Théories de jauge : interactions et brisures spontanées de symétrie}
\label{ChapterTheoriesDeJauges}

\chapter{Algèbres de Lie simples, représentations et classifications}
\label{PartRacinesPoids}
\section{Introduction}

\noindent
Les groupes étudiés dans la partie~\ref{ChapterModelBuilding} étaient de taille réduite. Pour étudier les théories de jauge, et notamment les théories de grande unification, il est cependant nécessaire de décrire des groupes de taille plus importante, qui sont compacts. L'introduction d'outils performants pour décrire la structure des groupes de Lie compacts de grandes dimensions, ainsi que leurs représentations, est donc souhaitable. Nous ferons par la suite appel au formalisme des racines et des poids, présenté dans ce chapitre. Ce formalisme est particulièrement efficace pour décrire la structure des théories de jauge, identifier les nombres quantiques des particules dans les représentations du groupe de jauge, et décrire l'action des générateurs du groupe de jauge sur ses représentations. Il est cependant moins efficaces pour décrire les calculs en théories des champs, où la construction des représentations d'un groupe par produits tensoriels de ses représentations fondamentales est souvent plus commode.

Pour alléger la présentation, l'accent est mis sur l'utilisation des racines et des poids pour construire les représentations d'un groupe de jauge plutôt que sur leur fondement mathématique. Une grande partie des résultats sera donc simplement énoncée, et non démontrée. On pourra également se référer à l'annexe~\ref{ChapterTdG}, synthétisant les principales définitions et propriétés de théorie des groupes.


\section{Préliminaire, étude de SU(2)}
\label{PartSU2Ini}

\noindent
Commençons par étudier l'algèbre simple la plus réduite, à savoir l'algèbre de SU(2), de dimension 3 (on rappelle que U(1) n'est pas simple). Celle-ci comporte trois générateurs, les $T_i$ pour $i=1,2,3$, qui vérifient les règles de commutation
\begin{equation}
\left[T_i,T_j \right] = i \epsilon_{ijk} T_k.
\end{equation}
Comme le groupe est compact et simple, on peut écrire ces générateurs sous une forme hermitienne. Son opérateur de Casimir quadratique est 
\begin{equation}
\label{EqDefCasimir}
C_2 \equiv \left( T_1\right)^2 + \left( T_2\right)^2 + \left( T_3\right)^2,
\end{equation}
qui commute avec tous les générateurs de l'algèbre : 
\begin{equation}
\left[C_2,T_i\right]=0.
\end{equation}
Pour étudier les représentations, il est commode d'introduire les opérateurs d'échelle
\begin{equation}
\label{EqDefOpEchelle}
T_\pm \equiv T_1 \pm i  T_2,
\end{equation}
qui vérifient les relations de commutation 
\begin{equation}
\label{EqCommutationTpm}
\left[T_+,T_- \right] = 2 T_3, ~~~~ \left[T_3,T_\pm \right]= \pm T_\pm.
\end{equation}
Ces opérateurs d'échelle ne sont pas hermitiens, mais vérifient $(T_\pm)^\dagger =T_\mp$.

Identifions à présent les représentations irréductibles. Souhaitant décrire les différents éléments d'un espace de représentation par leur valeurs propres vis-à-vis d'un ensemble complet d'opérateurs qui commutent, on commence par déterminer le nombre maximal d'opérateurs d'un tel ensemble. Sans perdre de généralité, on choisira $C_2$ et $T_3$, qu'on peut alors simultanément diagonaliser dans la représentation étudiée. Comme l'opérateur de Casimir commute avec tous les autres opérateurs, tous les éléments d'une représentation donnée auront la même valeur propre pour cet opérateur, qu'on notera $c$ et dont on se servira pour identifier la représentation. On note $m$ la valeur propre associée à l'opérateur $T_3$, et on peut donc dénoter $\ket{c,m}$ les différents états de la représentation, qui vérifient
\begin{equation}
C_2 \ket{c,m} = c \ket{c,m}, ~~~~ T_3 \ket{c,m} = m \ket{c,m}.
\end{equation}
Les valeurs propres sont réelles car ce sont celles d'opérateurs hermitiens. De plus, la valeur propre de $T_3$ est bornée, puisque la valeur propre de $C_2$ est positive et finie [voir l'équation~\eqref{EqDefCasimir}].

Partant d'un élément de la représentation, on peut en obtenir d'autres éléments en faisant agir les opérateurs d'échelle $T_\pm$ sur le premier élément. Utilisant les relations de commutations~\eqref{EqCommutationTpm}, on obtient directement que 
\begin{equation}
T_+ \ket{c,m}  \propto \ket{c,m+1}.
\end{equation}
Continuant à appliquer cet opérateur, on doit à un moment obtenir zéro car la valeur de $m$ est bornée. On a donc une valeur propre maximum de $T_3$ telle que
\begin{equation}
T_+ \ket{c,j}=0, ~~~~ T_3 \ket{c,j} =j \ket{c,j}. 
\end{equation}
Cette valeur propre $j$ est appelée poids le plus haut de la représentation. On peut de façon similaire se déplacer dans la représentation par l'action de l'opérateur $T_-$, et on définit le poids le plus bas $k$ tel que
\begin{equation}
T_- \ket{c,k}=0, ~~~~ T_3 \ket{c,k} =k \ket{c,k}. 
\end{equation}
Un peu d'algèbre avec les relations de commutations permet alors de montrer que $k=-j$, $c=j(j+1)$ et que $j$ est entier ou demi-entier puisque la représentation contient $2j +1$ éléments. Un point important est que le lien entre la valeur propre de l'opérateur de Casimir sur la représentation et le poids le plus haut permet d'identifier une représentation par son poids le plus haut, et donc de décrire ces éléments par les notations plus communes $\ket{j,m}$.

\section{Poids et racines}
\label{Poids&Racines}

\noindent
Pour discuter les représentations irréductibles des algèbres de Lie simples, il est nécessaire de généraliser les méthodes utilisées pour SU(2). Pour commencer, on a vu que les représentations irréductibles peuvent être identifiées par les valeurs propres des opérateurs de Casimir, constantes sur une représentation donnée. L'algèbre de SU(2) étant de rang $1$ et ne possédant donc qu'un seul opérateur de Casimir, un seul nombre était nécessaire. Mais pour des algèbres plus élaborées, $n$ valeurs propres sont nécessaires, avec $n$ le nombre d'opérateurs de Casimir, égal au rang de l'algèbre (un seul de ces opérateurs est quadratique, les ordres des autres opérateurs de Casimir étant par ailleurs classifiés pour les groupes simples~\cite{Ramond:2010zz}). La description des représentations est donc $n$-dimensionnelle : on identifiera une représentation irréductible par $n$ indices, et il en faudra aussi $n$ pour décrire les éléments de cette représentation. 

Dans le cas d'une algèbre de rang $n$, on s'attend à ce que $n$ opérateurs jouent le rôle que $T_3$ avait dans SU(2), permettant d'identifier les différents éléments de l'espace vectoriel de représentation via leurs valeurs propres pour ces opérateurs. Ces opérateurs correspondent aux différents éléments d'une sous-algèbre de Cartan de l'algèbre de Lie étudiée\footnote{Cette sous-algèbre est choisie arbitrairement et sera conservée par la suite sans perdre de généralité. Tous les résultats obtenus en choisissant une autre sous-algèbre sont équivalents, et peuvent être reliés entre eux par des automorphismes. Voir par exemple le complément~\ref{RepEquivSU3}, qui utilise des notions introduites par la suite.}, qui est de dimension $n$, et seront notés $T_3^{(i)}$ avec $i\in\{1,\cdots,n\}$. Ces opérateurs commutent entre eux et peuvent être simultanément diagonalisés dans toute représentation de l'algèbre, ce qui légitime leur utilisation. On nomme poids d'un élément de l'espace vectoriel de représentation ses valeurs propres vis à vis de ces $n$ opérateurs. Il est alors commode de décrire les éléments de l'espace de représentation par leurs poids, qui forment des vecteurs $\boldsymbol\lambda$ réels -- les opérateurs étant hermitiens -- de dimension $n$. Les notant par des kets $\ket{\boldsymbol\lambda}$, ceux-ci vérifient donc
\begin{equation}
T_3^{(i)} \ket{\lambda_1,\cdots,\lambda_n} = \lambda_i \ket{\lambda_1,\cdots,\lambda_n}.
\end{equation}

Reste alors à obtenir l'action des éléments restants de l'algèbre, \emph{i.e.} qui ne sont pas dans la sous-algèbre de Cartan, sur l'espace de représentation. En pratique, ces éléments vont jouer le rôle d'opérateur d'échelle, comme $T_\pm$ dans le cas de SU(2), et vont soit annihiler les poids soit les transformer en un autre poids~\cite{Slansky:1981yr}. Plus précisément, et dans le cas où ces opérateurs donnent un résultat non nul, ils sont chacun associés à une unique translation dans l'espace des poids. Ce résultat est obtenu en utilisant les relations de commutations entre les opérateurs $T_3^{(i)}$ de la sous-algèbre de Cartan et les autres opérateurs, après diagonalisation simultanée des opérateurs de la sous-algèbre de Cartan sur l'algèbre de Lie elle-même. Prenons un tel opérateur $T_p$, il vérifie des relations de commutations de la forme
\begin{equation}
\left[ T_3^{(i)},T_p\right] = \alpha_p^i T_p.
\end{equation}
On identifie les $\alpha_p^i$ aux $n$ composantes d'un vecteur à $n$ dimensions, $\boldsymbol\alpha_p$. Il est alors plus commode de repérer les différents opérateurs directement par ces vecteurs, soit de les écrire $T_{\boldsymbol\alpha_p}$ avec 
\begin{equation}
\left[ T_3^{(i)},T_{\boldsymbol\alpha_p}\right] = \alpha_p^i T_{\boldsymbol\alpha_p}.
\end{equation}
En pratique, cela identifie toutes les composantes de l'algèbre de manière unique pour les algèbres de Lie simples, sauf pour le cas $\boldsymbol\alpha_p = 0$, qui est dégénéré $n$ fois et correspond aux éléments de la sous-algèbre de Cartan. Utilisant ces relations de commutation, on peut obtenir l'action des opérateurs associés à ces différents éléments sur l'espace vectoriel de représentation. Partant d'un poids $\ket{\boldsymbol\lambda}$, on identifie les composantes de $\ket{\boldsymbol\lambda^\prime}=T_{\boldsymbol\alpha_p}\ket{\boldsymbol\lambda}$ via (on rappelle que les $T_3^{(i)}$ sont diagonalisables sur toute représentation)
\begin{equation}
T_3^{(i)} \ket{\boldsymbol\lambda^\prime} = \left(\lambda_i + \alpha_p^i \right) \ket{\boldsymbol\lambda^\prime}, 
\end{equation}
qui donne
\begin{equation}
\lambda^\prime_i = \lambda_i + \alpha_p^i.
\end{equation}
Soit ce poids ne fait pas partie de l'espace vectoriel de représentation, et la solution est alors $\ket{\boldsymbol\lambda^\prime}=0$, soit l'action de $T_{\boldsymbol\alpha_p}$ implique bien une translation du vecteur $\boldsymbol\alpha_p$ dans l'espace des poids, la normalisation et la phase restant à déterminer. Cela justifie qu'on peut décrire ces opérateurs comme des opérateurs d'échelle associés à une translation dans l'espace poids.

Les vecteurs $\boldsymbol\alpha_p$ décrivant l'action des éléments de l'algèbre sont nommés racines. Celles-ci ne dépendent pas du choix initial de la sous-algèbre de Cartan, et sont classifiées pour tous les groupes de Lie. Toutes ces racines apparaissent par paires impliquant des translations opposées. En effet, on montre aisément que $T_{-\boldsymbol\alpha_p} = T_{\boldsymbol\alpha_p}^\dagger$. Cela permet de classer les racines en racines positives et négatives, en définissant comme racines positives toutes les racines dans un même demi-espace arbitraire, les racines négatives étant contenues dans le demi-espace complémentaire. Finalement, les racines étant des vecteurs à $n$ dimensions, il est possible d'en choisir $n$ indépendantes qui forment une base de cet espace, qui sont appelées racines simples, et qu'on peut prendre positives. Elles sont indicées de $\boldsymbol\alpha_1$ à $\boldsymbol\alpha_n$. Ces racines simples sont définies de manière unique si elles sont choisies telles que toutes les racines positives peuvent s'écrire comme une combinaison linéaire à coefficients positifs des racines simples -- ce qui est toujours possible pour les groupes simples~\cite{Slansky:1981yr}. 

\section{Description et classifications des représentations irréductibles}

\noindent
L'introduction des racines simples permet de décrire les représentations irréductibles d'une façon similaire à ce qui a été fait dans le cas de SU(2), rajoutant simplement le fait que l'espace des poids et des racines est de dimension $n$, le rang de l'algèbre étudiée. On a déjà vu que la valeur propre des opérateurs de Casimir était constante sur une représentation irréductible, car ces opérateurs commutent avec tous les éléments de l'algèbre. Or, l'opérateur de Casimir quadratique étant proportionnel à la somme des carrés des opérateurs associés à tous les éléments de l'algèbre, on obtient comme dans le cas de SU(2) que les valeurs propres des poids sous les $T_3^{(i)}$ doivent être bornées. Cela signifie qu'appliquant successivement les opérateurs d'élévation associés aux racines simples, le résultat sera forcément nul à un moment. 

On peut donc définir le poids le plus haut, unique pour une représentation irréductible donnée, tel que ce poids est annihilé sous l'action de tous les opérateurs d'élévation associés aux racines simples (et en fait à toutes les racines positives). Appliquant ensuite successivement les opérateurs d'abaissement sur ce poids le plus haut, on peut générer tous les poids et donc tous les éléments de l'espace vectoriel de représentation. Plusieurs chaînes sont \emph{a priori} nécessaires à partir du poids le plus haut, si on n'utilise que les opérateurs d'abaissement. Similairement au cas de SU(2), on décrit de cette façon l'intégralité de l'espace de représentation, et on connait l'effet des opérateurs associés à tous les éléments de l'algèbre de Lie. Les seules inconnues encore à déterminer sont la normalisation et la phase associées aux différentes translations dans l'espace des poids. On notera qu'on génère souvent l'espace vectoriel de représentation à partir du poids le plus haut en appliquant les opérateurs d'abaissement associées aux racines simples seulement. Cependant, les translations dans l'espace des poids associées aux générateurs restants étant des combinaisons linéaires de celles associées aux racines simples, la description via les racines simples suffit bien à décrire toute l'action de l'algèbre.

On pourrait identifier les représentations irréductibles par les valeurs des opérateurs de Casimir sur celles-ci. Cependant, on a vu dans le cas de SU(2) qu'il était plus commode de les décrire par les valeurs propres pour $T_3$ du poids le plus haut, c'est à dire l'élément annihilé par l'opérateur d'élévation $T_+$. On procède de même pour les représentations irréductibles de toutes les algèbres de Lie simples, la seule différence étant que le poids le plus haut est indicé par ses valeurs propres associées aux $n$ éléments de la sous-algèbre de Cartan $T_3^{(i)}$. L'exhaustivité de cette description pour les algèbres simples a été montrée en 1957 par Dynkin, qui a prouvé que toute représentation irréductible possédait un poids le plus haut, et que celui-ci la définissait de manière unique \cite{Dynkin:1957um}. La description de tous les poids les plus hauts possibles d'une algèbre de Lie simple, les poids dominants, permet alors une classification de toutes les représentations de cette algèbre.

\begin{table}[h!]
\begin{center}
\hspace{1.9cm}
\begin{minipage}[t]{.4\textwidth}
\begin{tabular}{|c|c|}
\hline
$ \begin{array}{c}
\text{Indices}\\
\text{de Dynkin}
\end{array}
$ & Dimension \\
\hline
$(1 0)$ & $\mathbf{3}$ \\
\hline
$(2 0)$ & $\mathbf{6}$ \\
\hline
$(1 1)$ & $\mathbf{8}$ \\
\hline
$(3 0)$ & $\mathbf{10}$ \\
\hline
$(2 1)$ & $\mathbf{15}$ \\
\hline
$(4 0)$ & $\mathbf{15}^\prime$ \\
\hline
$(0 5)$ & $\mathbf{21}$ \\
\hline
\end{tabular}
\end{minipage}
\hspace{0.5cm}
\begin{minipage}[t]{.4\textwidth}
\begin{tabular}{|c|c|}
\hline
$ \begin{array}{c}
\text{Indices}\\
\text{de Dynkin}
\end{array}
$ & Dimension \\
\hline
$(1 0 0)$ & $\mathbf{4}$ \\
\hline
$(0 1 0)$ & $\mathbf{6}$ \\
\hline
$(2 0 0)$ & $\mathbf{10}$ \\
\hline
$(1 0 1)$ & $\mathbf{15}$ \\
\hline
$(0 1 1)$ & $\mathbf{20}$ \\
\hline
$(0 2 0)$ & $\mathbf{20}^\prime$ \\
\hline
$(0 0 3)$ & $\mathbf{20}^{\prime\prime}$ \\
\hline
\end{tabular}
\end{minipage}
\hspace{1cm}
\end{center}
\caption{Indices de Dynkin des premières représentations de SU(3) (à gauche) et SU(4) (à droite).}
\label{TableRepSU3Et4}
\end{table}

Pour cela, il est avantageux de travailler dans la base dite de Dynkin, dans laquelle il est toujours possible de se placer, et où tous les poids et racines ont des composantes entières. Dans cette base, Dynkin a montré que tous les poids dominants peuvent être associés à ceux qui n'ont que des composantes positives ou nulles \cite{Dynkin:1957um}. On peut donc identifier de manière unique toutes les représentations irréductibles d'une algèbre simple de rang $n$ par un élément de $\mathbb{N}^n$. Les $n$ entiers décrivant les coordonnées du poids le plus haut d'une représentation irréductible sont appelés les indices de Dynkin d'une représentation, et communément utilisés pour désigner les différentes représentations irréductibles. Rassemblant ces indices dans une liste entre parenthèses, on ne sépare souvent pas les indices par des virgules avant que ceux-ci ne dépassent 9. Le cas des représentations de basse dimension de SU(3) et SU(4) sont données en exemple dans le tableau~\ref{TableRepSU3Et4}~\cite{Slansky:1981yr}. Dans la suite, on désignera les représentations par leurs indices de Dynkin ou par leur dimension. On voit cependant qu'il peut y avoir plusieurs représentations irréductibles de même dimension pour une algèbre donnée.

Pour l'étude pratique des différentes représentations irréductibles d'une algèbre de Lie simple, il n'est bien sûr pas nécessaire de recalculer les coordonnées des racines simples en fonction des relations de commutation des générateurs. Les racines simples sont en effet classifiées pour chaque algèbre de Lie (en fait, les racines simples d'une algèbre de Lie la définissent intégralement, comme on le verra dans la section~\ref{PartClassificationAlgebres}), et il est facile de les obtenir dans la base de Dynkin où elles ont des coordonnées entières. On peut les trouver par exemple dans la référence~\cite{Slansky:1981yr}. D'autre part, la matrice de Cartan, utilisée pour décrire les différentes algèbres de Lie, contient toute l'information sur les racines simples de celles-ci. En effet, cette matrice $n\times n$ a pour lignes les $n$ racines simples positives écrites dans la base de Dynkin. Ces racines simples obtenues, on peut décrire l'intégralité des représentations irréductibles des algèbres de Lie simple utilisant les propriétés décrites ci-dessus. Et la connaissance de l'action des racines dans l'espace des poids contient toute l'information nécessaire pour avoir l'action des différents éléments du groupe sur l'espace vectoriel de représentation.

\section{Exemples de SU(3) et SO(10)}
\label{SU3&SO10}

\noindent
Appliquons pour commencer le formalisme précédent au cas de SU(2), déjà étudié dans la section~\ref{PartSU2Ini}. Les poids dans la base de Dynkin sont alors le double des moments magnétiques, qui sont bien des entiers. Le couple de racines simples non nulles correspond à $\pm2$ (en temps que vecteurs à une dimension). Conventionnellement, les représentations de SU(2) sont décrites par leur représentation de spin décrite précédemment, comme on fera par la suite, plutôt que dans la base de Dynkin. Pour aller plus loin, nous prenons à présent l'exemple de représentations de SU(3) et SO(10), qui seront utiles par la suite dans l'étude des interactions fortes et des théories de grande unification. 

\begin{figure}[h!]
\begin{center}
\includegraphics[scale=1.3]{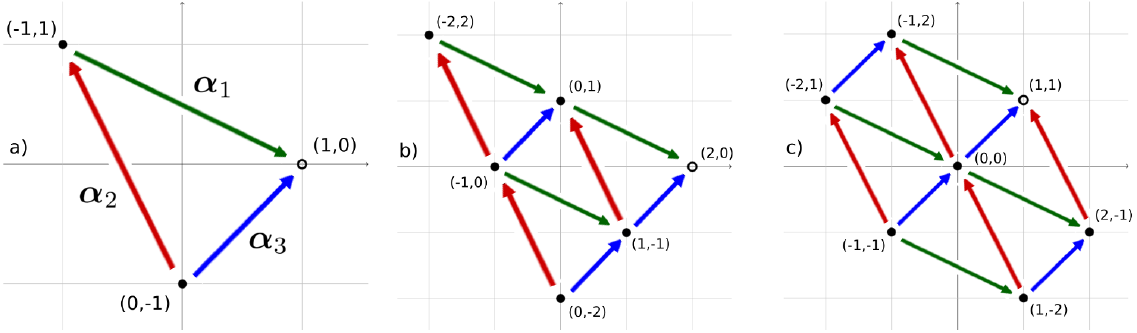}
\end{center}
 \caption{Diagrammes des poids de plusieurs représentations de SU(3), pour les représentations de dimensions 3 (fig. a), 6 (fig. b) et 8 (fig. c). Le poids le plus haut est représenté par un cercle. Les translations associées aux racines positives sont représentées par des vecteurs. Les deux racines simples $\boldsymbol\alpha_1$ et $\boldsymbol\alpha_2$ sont respectivement en vert et en rouge. La troisième racine, $\boldsymbol\alpha_3 = \boldsymbol\alpha_1 + \boldsymbol\alpha_2$ est représentée en bleu. Le point $(0,0)$ du diagramme des poids de la représentation de dimension 8 est dégénéré, et correspond à deux poids distincts.}
 \label{DiagPoidsSU3}
\end{figure}

L'algèbre de SU(3) étant de rang 2, l'espace des poids associé est de dimension 2. Les racines simples positives dans la base de Dynkin sont~\cite{Slansky:1981yr}
\begin{equation}
\begin{array}{l}
\boldsymbol\alpha_1 = (2,-1),\\
\boldsymbol\alpha_2 = (-1,2).
\end{array}
\end{equation}
La troisième racine positive est
\begin{equation}
\boldsymbol\alpha_3 = \boldsymbol\alpha_1 + \boldsymbol\alpha_2 = (1,1).
\end{equation}
Les représentations de dimension 3, 6 et 8 sont données figure~\ref{DiagPoidsSU3}, où l'on retrouve les poids les plus hauts donnés dans le tableau~\ref{TableRepSU3Et4}. On identifie bien les poids les plus hauts par le fait que ce sont les poids de la représentation sur lesquels on termine forcément lorsqu'on se déplace suivant les racines simples positives (et en fait, toutes les racines positives). La représentation de dimension 8 de SU(3) est sa représentation adjointe, agissant sur l'algèbre elle-même. Le poids $(0,0)$ de cette représentation est donc double, correspondant aux deux éléments de la sous-algèbre de Cartan de SU(3)\footnote{Comme les deux opérateurs de cette sous-algèbre commutent entre eux, ils ont des valeurs propres nulles l'un par rapport à l'autre.}. Dans la littérature, les diagrammes des racines et des poids sont parfois donnés dans d'autres bases que celle de Dynkin, mais les différentes descriptions sont bien sûr équivalentes (voir le complément~\ref{RepEquivSU3}).

\begin{figure}[h!]
\begin{center}
\includegraphics[scale=1]{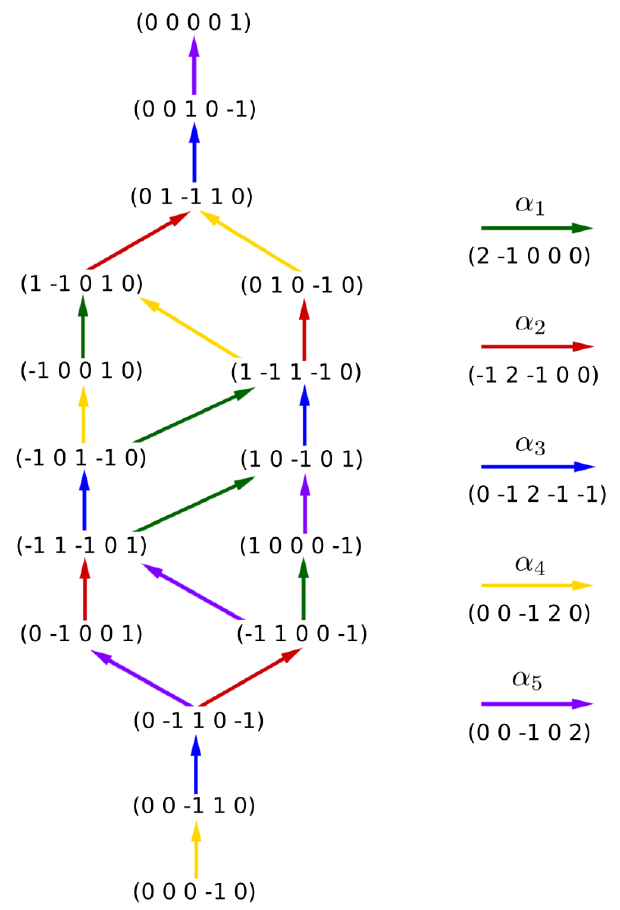}
\end{center}
 \caption{Poids de la représentation de dimension 16 et des racines simples positives de SO(10). Les translations correspondantes aux différentes racines simples sont également indiquées.}
 \label{DiagPoidsS010}
\end{figure}

Pour obtenir l'étendue exacte de la représentation en partant du poids le plus haut, il suffit de construire les points extrémaux, c'est à dire qui forment l'enveloppe des poids d'une représentation, et de remplir ensuite si nécessaire la figure obtenue à l'aide des différentes racines. Partant du poids le plus haut, les points extrémaux s'obtiennent grâce à la propriété que dans la base de Dynkin, leurs coordonnées donnent l'opposé algébrique du nombre de fois qu'on peut leur appliquer les opérateurs d'échelle liées aux racines simples avant d'obtenir un résultat nul~\cite{Slansky:1981yr}. Prenant par exemple un poids extrémal $\ket{\boldsymbol\lambda} = (a_1,\cdots,a_n)$ dans une algèbre de rang $n$, on a que $(T_{-\boldsymbol\alpha_i})^{a_i} \neq 0$ mais que $(T_{-\boldsymbol\alpha_i})^{a_i+1}= 0$, quelque soit $i$ entre $1$ et $n$. D'autre part, on a aussi que $(T_{-\boldsymbol\alpha_i})^{- \frac{a_i}{|a_i|}} = 0$, puisque le poids est un extrémal. Appliquant les opérateurs d'échelle ces nombres donnés de fois, on construit l'enveloppe du graphe de la représentation. Prenons l'exemple de la représentation de dimension 6 donné dans la figure~\ref{DiagPoidsSU3}. Le poids le plus haut en est $(2,0)$. On peut donc lui soustraire deux fois  $\boldsymbol\alpha_1$, et on obtient le poids $(-2,2)$. On peut ajouter à ce nouveau poids deux fois $\boldsymbol\alpha_1$ et revenir au poids initial, et on peut aussi lui soustraire deux fois $\boldsymbol\alpha_2$, et on obtient le poids $(0,-2)$ qui complète l'enveloppe du graphique. 

Pour des groupes de rang supérieur ou égal à 3, la représentation graphique de l'espace des poids perd en intérêt. On représente alors les liens entre les différents poids, en partant du poids le plus haut de la représentation étudiée. L'exemple de la représentation de dimension 16 de SO(10) est donné dans la figure~\ref{DiagPoidsS010}. Les racines simples positives de SO(10), qui est de rang 5, sont \cite{Slansky:1981yr}
\begin{equation}
\begin{array}{l}
\boldsymbol\alpha_1 = (2,-1,0,0,0),\\
\boldsymbol\alpha_2 = (-1,2,-1,0,0),\\
\boldsymbol\alpha_3 = (0,-1,2,-1,-1)\\
\boldsymbol\alpha_4 = (0,0,-1,2,0)\\
\boldsymbol\alpha_5 = (0,0,-1,0,2).
\end{array}
\end{equation}
On vérifie aisément que tous les poids de la représentation en forment l'enveloppe, et que le poids de composantes $(0 0 0 0 1)$ est bien le poids le plus haut.

La représentation de dimension 16 de SO(10) n'a aucun poids dégénéré, et vérifie la propriété que si $\ket{\boldsymbol\lambda}$ est un poids, alors $\ket{-\boldsymbol\lambda}$ ne l'est pas. Ce sont des propriétés générales des représentations complexes~\cite{Slansky:1981yr}. Ces représentations complexes ont par ailleurs toujours un poids le plus haut qui n'est pas invariant sous l'opération
\begin{equation}
\label{InversionPoids}
\ket{\lambda_1,\lambda_2,\cdots,\lambda_{n-1},\lambda_n} \rightarrow \ket{\lambda_n,\lambda_{n-1},\cdots,\lambda_2,\lambda_1}.
\end{equation}
Cette opération permet au contraire d'obtenir le poids le plus haut de leur représentation conjuguée. Cela permet notamment de compléter le tableau~\ref{TableRepSU3Et4}, la représentation obtenue en appliquant l'opération~\eqref{InversionPoids} au poids les plus haut d'une première représentation étant soit équivalente soit complexe conjuguée de la première représentation. En corollaire, les représentations complexes ne contiennent jamais le poids identiquement nul $\ket{0,\cdots,0}$, ou tout poids correspondant à une combinaison linéaire de racines (puisque cela impliquerait la présence du poids nul dans la représentation).

\section{Classification des algèbres de Lie}
\label{PartClassificationAlgebres}

\noindent
Un des grands succès de l'étude de la théorie des groupes de Lie a été de classifier toutes les algèbres de Lie simples possibles. Cette classification est appelée classification de Killing-Cartan, du nom de ses deux principaux contributeurs, et a été achevée au début du vingtième siècle. Cette classification permet de décrire de façon systématique les groupes de Lie compacts d'intérêt en physique. Les théorèmes de Cartan permettent en effet d'écrire une algèbre semi-simple comme un produit direct d'algèbres simples, voir l'annexe~\ref{TheorèmeCartan}. Pour les algèbres non semi-simples, il est possible d'isoler leurs idéaux invariants liés à des groupes abéliens U(1). On peut alors les écrire sous la forme $G\simeq \left\{ G/[U(1)]^p\right\} \times [U(1)]^p$, le groupe $G/[U(1)]^p$ étant semi-simple et pouvant donc être décrit par un produit d'algèbres simples. On peut alors écrire de façon systématique les groupes de Lie compacts sous la forme $G_1\times \cdots \times G_N \times [U(1)]^p$, où les $G_i$ sont associés à des algèbres simples. Des produits directs avec des groupes finis peuvent aussi être considérés de la même façon.

\begin{table}[h!]
\begin{center}
\renewcommand{\arraystretch}{1.4}
\begin{tabular}{|l|c|l|}
\hline
Groupe & Dimension & Nom\\
\hline
$A_n(n\geq 1)=$ SU($n+1$)  & $n(n+2)$ & Spécial Unitaire \\
\hline
$B_n(n\geq2)=$ SO($2n+1$) &  $n(2n+1)$ & Spécial Orthogonal \\
\hline
$C_n(n\geq3)=$ Sp($2n$) & $n(2n+1)$ & Symplectique \\
\hline
$D_n(n\geq4)=$ SO($2n$) &  $n(2n-1)$ & Spécial Orthogonal \\
\hline
\end{tabular}
\hspace{1cm}
\end{center}
\caption{Familles infinies de la classification des algèbres de Lie simples}
\label{CartanClassification}
\end{table}

Passons à la description des algèbres simples. \emph{A priori}, il n'est pas nécessaire de considérer les représentations d'une algèbre pour la décrire. En effet, la description de celle-ci se fait via les crochets de Lie entre ses différents éléments, qui correspondent à la donnée des constantes de structure. Cependant, ces crochets de Lie correspondent aussi à la représentation de l'algèbre sur elle-même, à savoir à sa représentation adjointe. On peut donc décrire une algèbre via les outils introduits précédemment pour décrire les représentations d'une algèbre, et la donnée des racines est équivalente à la donnée des constantes de structure. L'étude de la structure d'une algèbre correspond donc à l'étude de la géométrie de son système des racines, et plus particulièrement à l'étude de celle de ses racines simples. Or, cette géométrie n'est pas arbitraire, car elle doit être compatible avec les identités de Jacobi, le critère de simplicité, et l'antisymétrie des crochets de Lie. Cela fixe notamment les angles et les rapports de longueur entre les différentes racines simples de façon très contraignante, restreignant les possibilités de structures d'algèbres simples, et permettant de les dénombrer et de les classifier~\cite{Dynkin:1957um,Fuchs:1997jv,Georgi:1999wka,Ramond:2010zz}.

Lorsqu'on liste les différentes algèbres de Lie simples, il est souvent commode de les décrire par les groupes simplement connexes associés, ce qu'on fera ici. Les algèbres simples se classent en quatre familles infinies dénombrables résumées dans le tableau ~\ref{CartanClassification}, ainsi que cinq algèbres exceptionnelles nommées $G_2$, $F_4$, $E_6$, $E_7$ et $E_8$. Le rang des ces algèbres est indiqué par $n$ pour les familles infinies, et par le chiffre en indice pour les algèbres exceptionnelles. 

Finissons cette section en notant que comme on l'a vu, l'étude des relations de commutation de l'algèbre et donc de ses racines  revient à étudier la représentation adjointe de celle-ci. C'est pourquoi cette représentation a la particularité d'avoir les racines de l'algèbre comme poids, permettant de visualiser directement l'espace des racines. On peut le voir par exemple pour la représentation de dimension 8 de SU(3), dans la figure~\ref{DiagPoidsSU3}. Ce résultat s'obtient du fait que le poids identiquement nul apparaît $n$ fois, pour chaque élément de la sous-algèbre de Cartan, et qu'appliquer les opérateurs d'échelle associés à toutes les racines à partir d'un de ces poids nuls va donner l'intégralité des racines comme poids.

\chapter{Théories de jauge et interactions}
\label{PartTheoriesJauge}
\section{Introduction}

\noindent
La nécessité du concept de symétrie de jauge a été discutée dans le chapitre~\ref{SecSymGlob}, apparaissant naturellement de la distinction entre les représentations unitaires du groupe de Poincaré, nécessaires pour décrire les états physiques, et les représentations covariantes du groupe de Lorentz, nécessaires pour écrire un Lagrangien. Les théories de jauge sont construites ici en suivant cette observation, et en essayant d'identifier dans les différentes transformations des champs les degrés de liberté physiques et de jauge.

Les théories de jauge ainsi construites décrivent une interaction entre particules, liée aux transformations du groupe de symétrie de jauge. Les notions de racines et de poids peuvent alors être introduites pour permettre une description compacte des différentes transformations de symétrie. Les bosons de jauge s'identifient à des racines, et les particules sont groupées dans des représentations irréductibles, dont les poids jouent le rôle de nombres quantiques conservés. C'est ce formalisme qui est développé ici, appliqué aux interactions entre fermions et bosons de jauge, puis aux auto-interactions entre bosons de jauge.

\section{Théories de jauge abéliennes}
\label{AbelianGaugeTheory}

\noindent
La nécessité d'introduire des théories de jauge apparaît lorsqu'on tente de construire un terme de couplage entre un spineur de Weyl $\psi$ et un champ vectoriel sans masse $A_\mu$. En effet, on a vu dans la section~\ref{PartRepCovariante} qu'un des degrés de liberté de $A_\mu$ était un degré de liberté de jauge, associé à une symétrie locale (et continue). Pour qu'un couplage soit possible avec des spineurs, il faut donc que la modification du terme de jauge du champ vecteur corresponde à une transformation de symétrie sur les spineurs, permettant d'écrire un terme de couplage invariant sous la symétrie en question au niveau du Lagrangien. Dans le cas contraire, la variation du degré de liberté de jauge de $A_\mu$ ne peut être annulée en aucune manière, et mène nécessairement à un terme Lagrangien non invariant de jauge. Partant du Lagrangien cinétique pour un spineur de Weyl,
\begin{equation}
\mathcal{L}_{\text{Cin},\psi} = i \psi^\dagger \slashed \partial \psi,
\end{equation}
cette considération amène naturellement aux théories de jauge.

Nous commençons ici par le cas le plus simple, à savoir une symétrie locale associée à un groupe abélien U(1). Comme ce groupe ne possède que des représentations irréductibles de dimension 1, les spineurs ne se transforment sous l'action de celui-ci que par un terme de phase. En repérant les transformations par un paramètre $\theta(x)$, on obtient, 
\begin{equation}
\label{TransfoU1}
\psi \rightarrow e^{-iq_\psi \theta(x)} \psi.
\end{equation}
Cette transformation ne laisse pas invariant le Lagrangien cinétique des spineurs, puisqu'un terme en $+ \psi^\dagger q_\psi (\slashed \partial \theta) \psi$ apparaît, qui dépend explicitement de la transformation de jauge. On peut cependant compenser l'apparition de ce terme en rajoutant un couplage entre le champ vecteur et le spineur par l'introduction d'une dérivée covariante pour le champ $\psi$ de la forme
\begin{equation}
\label{DevCov}
D_{\mu} \equiv \partial_\mu - ig q_\psi A_\mu(x),
\end{equation} 
et une loi de transformation pour le champ vectoriel
\begin{equation}
\label{TransfoU1A}
A_\mu(x) \rightarrow A_\mu(x) - \frac1g \partial_\mu \theta(x),
\end{equation}
où $g$ est une constante de couplage arbitraire. Cette transformation pour le champ vectoriel modifie sa composante longitudinale, qui peut être identifié avec le degré de liberté de jauge, comme nous le discutons ci-dessous. On peut alors introduire un terme cinétique n'affectant que les degrés de liberté physiques du champ vecteur, et donc invariant de jauge, sous la forme 
\begin{equation}
\mathcal{L}_{\text{Cin},A} =-\frac14 F_{\mu\nu} F^{\mu\nu},
\end{equation}
avec $F_{\mu\nu}$ le tenseur de Faraday défini par
\begin{equation}
F_{\mu\nu} =\partial_\mu A_\nu - \partial_\nu A_\mu.
\end{equation}
Ce tenseur est bien invariant de jauge.

On voit donc qu'effectuer une transformation de jauge consiste à faire évoluer le degré de liberté de jauge du champ vecteur tout en effectuant une transformation de symétrie sur les spineurs, transformation repérée par l'évolution du terme de jauge du vecteur. Cette construction nécessite d'avoir un champ qui propage des degrés de liberté non physiques. En effet, à partir d'une transformation locale de la forme donnée en \eqref{TransfoU1}, une dérivée du champ spineur fait apparaître hors de l'exponentielle le paramètre $\theta(x)$, associé à la transformation de symétrie et n'ayant pas de sens physique en soi. La seule façon de le faire disparaitre implique un couplage avec un champ donc les composantes peuvent être modifiées en accord avec l'évolution du paramètre $\theta(x)$, ce qui implique que ce champ possède des degrés de liberté de jauge. C'est ce couplage qui permet d'écrire des dérivées covariantes, c'est à dire se transformant comme le champ spinoriel lui-même, donnant dans une transformation de symétrie
\begin{equation}
D_{\mu} \psi \rightarrow (D_{\mu} \psi )^\prime =  e^{-iq_\psi\theta (x)}  D_{\mu} \psi.
\end{equation}
La présence d'un champ qui propage des degrés de liberté non physiques est donc nécessaire pour écrire un Lagrangien invariant sous des symétries locales. Dans le cas présent, cela implique que $A_\mu$ ne doit pas avoir de masse, car il ne propagerait alors pas de degré de liberté de jauge, et ne pourrait pas servir à former des dérivées covariantes.

Les lois de transformation du champ vectoriel sont mieux comprises dans ce cadre. En effet, observant l'équation~\eqref{TransfoU1A} donnant l'évolution de $A_\mu$ sous l'action de la symétrie U(1) étudiée, on voit que ce champ n'est pas dans une représentation du groupe de symétrie, ce qui semble contraire au théorème de Wigner. Le problème est résolu en notant que le théorème de Wigner décrit les degrés de liberté physiques, liés aux représentations irréductibles unitaires du groupe de symétrie complet. Or on a vu dans le chapitre~\ref{SecSymGlob} que $A_\mu$ n'était pas dans une telle représentation. En fait, les deux degrés de liberté physiques, liés aux deux états de polarisation du photon contenu dans le champ vecteur, ne sont pas affectés par la transformation de symétrie, et sont donc dans sa représentation triviale\footnote{Dans le vide, les états de polarisation du photon sont transverses. Effectuant une transformée de Fourier du problème, on voit que la transformation de jauge du champ vecteur est lié à sa composante longitudinale, puisque se modifiant sous la forme $ k_\mu \theta (x)$. Les degrés de liberté physiques ne sont donc pas affectés.}. Le fait d'être dans une représentation non triviale du groupe de symétrie U(1) impliquant d'être chargé sous celui-ci, le boson vecteur associé à cette symétrie n'est pas lui-même couplé à l'interaction qu'il médiate. On note aussi que l'introduction de la dérivée covariante peut être liée à la nécessité d'avoir une dérivée du champ spineur qui ne contienne que des degrés de liberté physiques, ce qui revient à dire qu'elle doit se transformer dans une représentation donnée du groupe de symétrie, ce qui est bien le cas ici.

En récapitulatif, le Lagrangien d'une théorie de jauge abélienne s'écrit sous la forme 
\begin{equation}
\mathcal{L}_{U(1)} = i \psi^\dagger \left( \slashed \partial - i g q_\psi \slashed A \right) \psi -\frac14 F_{\mu\nu} F^{\mu\nu},
\end{equation}
où il est nécessaire d'introduire pour dérivées quadrivectorielles les notations $\slashed A_L = \sigma^\mu A_\mu$ et $\slashed A_R = \bar \sigma^\mu A_\mu$, de façon similaire aux dérivées des spineurs de Weyl gauches et droits respectivement. Ce Lagrangien est invariant sous la transformation de jauge 
\begin{equation}
\displaystyle{
\left\{
\begin{array}{l}
\psi \rightarrow e^{-iq_\psi \theta(x)} \psi,\\
A_\mu(x) \rightarrow A_\mu(x) - \frac1g \partial_\mu \theta(x).
\end{array}
\right.
}
\end{equation}
La loi de conservation associée à cette symétrie est la conservation de la charge des particules, notées $q_\psi$. Cela se montre via le théorème de Noether de la même manière que pour une symétrie globale U(1) dans la section~\ref{PartNoether}, vu qu'il n'y a pas de termes supplémentaires en la dérivée du spineur. Chaque champ a une charge associée à l'interaction avec le champ de jauge, caractérisant l'intensité de son couplage à cette interaction. On a aussi introduit dans les équations~\eqref{DevCov} et~\eqref{TransfoU1A} un terme de couplage additionnel $g$, caractérisant la force de l'interaction. Dans une théorie classique, un tel couplage peut être réabsorbé dans les charges, et est donc un paramètre superflu. Dans le cadre d'une théorie quantique, la renormalisation des couplages va faire varier l'intensité de l'interaction en fonction de l'échelle d'énergie considérée, comme discuté dans la section~\ref{PartQuantification}. L'ajout de ce couplage $g$ permet alors d'isoler ces variations, plutôt que de faire varier les charges avec l'énergie, le rapport entre les différentes charges restant fixé.

Avant de discuter plus avant les notions d'interactions via les symétries locales, nous généralisons d'abord les théories de jauge au cas d'un groupe de symétrie non-abélien. Le cas de l'étude d'une théorie des jauge associée à un champ scalaire est quant à lui décrit dans le complément~\ref{TheorieJaugeScalaire}.

\section{Théories de jauge non-abéliennes}
\label{PartJaugesNonAbeliennes}

\noindent
Nous nous intéressons à présent à des symétries locales décrites par un groupe de symétrie non-abélien de dimension $N$. Nous nous restreignons pour l'instant au cas d'un groupe simple. Comme discuté dans la section~\ref{PartClassificationAlgebres}, on s'attend à ce que tout groupe de symétrie puisse être écrit comme un produit de groupes simples et de groupes U(1), ce qu'on considérera dans la section~\ref{ProduitGroupesJauge}. En physique des particules, les groupes utilisés sont principalement le groupe spécial unitaire SU et le groupe spécial orthogonal SO. Comme les transformations associées au groupe de symétrie agissent sur le champ spinoriel, celui-ci est nécessairement dans une représentation de ce groupe. Nous repérons les éléments de cette représentation par l'ajout d'un indice interne au champ (\emph{i.e.} non lié à l'espace-temps), l'écrivant $\psi_a$ avec $a=1,\cdots,d$ pour une représentation de dimension $d$. La transformation d'un champ spinoriel sous cette symétrie est alors décrite par un ensemble de $N$ paramètres $\boldsymbol\theta = \{\theta^\alpha(x)\}$ avec $\alpha=1,\cdots,N$, et s'écrit 
\begin{equation}
\psi_a \rightarrow \psi^\prime_a= U(\boldsymbol\theta)_a^b\psi_b = (e^{-i \theta_\alpha T^\alpha})_a^b \psi_b,
\end{equation}
avec $T^\alpha$ les générateurs de l'algèbre du groupe dans la représentation en question (se référer au complément~\ref{DefTransfoGroupes} pour une présentation exhaustive). Par la suite, et quand il n'y aura pas d'ambiguïté, on utilisera aussi la notation simplifiée $U$ pour décrire l'action du groupe, laissant implicites les indices internes ainsi que la dépendance en $\boldsymbol\theta$.

Le champ $(\psi_a)^\dagger = (\psi^\dagger)^a$ est dans la représentation conjuguée de celle du champ $\psi_a$, et il est possible de contracter ensemble les indices internes de ces deux champs pour former un singlet du groupe. Cependant, un terme cinétique standard pour ces champs, de la forme $ i (\psi^\dagger)^a \slashed \partial \psi_a$ n'est pas invariant sous les transformations du groupe de symétrie. En effet, la dérivée du champ spinoriel se transforme en
\begin{equation}
\label{PsiNonAbel}
\partial_\mu \psi_a \rightarrow (\partial_\mu \psi_a)^\prime = U \left(\partial_\mu \psi_a\right) +  \left(\partial_\mu U\right) \psi_a, 
\end{equation}
et n'est donc pas dans une représentation donnée du groupe de jauge. Comme précédemment, on introduit des champs vectoriels sans masse dont la variation des degrés de liberté de jauge est liée aux transformations du groupe de symétrie. Chaque vecteur ayant un degré de liberté de jauge, et les transformations du groupe étant décrit par $N$ paramètres réels, il est nécessaire d'introduire $N$ champs de jauge, que l'on note $A_\mu^\alpha$. On construit alors une dérivée covariante de la forme 
\begin{equation}
\label{DevCovNonAbel}
D_\mu \psi_a = \partial_\mu \psi_a  - i g A_\mu^\alpha (T_\alpha)_a^b \psi_b,
\end{equation}
avec la transformation suivante pour les champs vecteurs :
\begin{equation}
\label{ANonAbel}
A_\mu^\alpha T_\alpha \rightarrow (A^\prime)_\mu^\alpha T_\alpha= A_\mu ^\alpha U  T_\alpha U^{-1}  - \frac{i}{g} (\partial_\mu U) U^{-1},
\end{equation}
où l'on a laissé implicites les indices liés à la représentation des spineurs. Cette transformation s'écrit pour une transformation infinitésimale
\begin{equation}
\label{ANonAbelInf}
\delta A_\mu^\alpha  =  f_{\beta\gamma}{}^{\alpha}\delta \theta^\beta A_\mu^\gamma - \frac1g \partial_\mu \delta \theta ^\alpha,
\end{equation}
où les $f_{\beta\gamma}{}^{\alpha}$ sont les constantes de structure du groupe. Utilisant les équations~\eqref{PsiNonAbel} et~\eqref{ANonAbel}, on montre que la dérivée covariante du champ se transforme comme le champ sous l'action du groupe de symétrie, ce qui permet d'écrire un terme Lagrangien cinétique pour $\psi_a$ de la forme 
\begin{equation}
\mathcal{L}_{\text{Cin},\psi_a} = i (\psi^\dagger)^a \left[\slashed \partial\delta_a^b -  i g \slashed A^\alpha (T_\alpha)_a^b \right] \psi_b,
\end{equation}
qui est bien invariant sous les transformations de jauge.

Comme dans le cas d'une théorie de jauge abélienne, il est intéressant de se questionner sur la transformation du champ vecteur sous l'action du groupe de symétrie. On différenciera pour cela les deux termes de la transformation des champs de jauge, écrits en~\eqref{ANonAbel} pour une transformation finie et en~\eqref{ANonAbelInf} pour une transformation infinitésimale. Comme dans le cas abélien, on identifie le terme en $ - \frac{i}{g} (\partial_\mu U) U^{-1}$ ou de façon équivalentes $\frac1g \partial_\mu \delta \theta^\alpha$ comme la modification des degrés de liberté de jauge du champ, effectuée simultanément à la transformation de symétrie. Cette modification, qui affecte indépendamment chaque champ $A_\mu^\alpha$, est reliée à la paramétrisation par $\theta^\alpha$ de l'action du générateur $T_\alpha$ qui lui est associée dans la transformation totale de symétrie. C'est l'évolution du terme de jauge, qu'on retrouve dans le cas abélien. L'autre contribution à la transformation du champ, $ A_\mu ^\alpha U T_\alpha U^{-1} $, ou $ f_{\beta\gamma}{}^{\alpha}\delta \theta^\beta A_\mu^\gamma$ pour l'écriture infinitésimale, agit par contre sur les indices internes. On reconnait cette transformation comme celle associée à la représentation adjointe du groupe de symétrie. S'appliquant sur l'intégralité des composantes spatio-temporelles des champs $A_\mu^\alpha$, elle agit aussi bien sur leurs degrés de liberté physiques que sur leurs degrés de liberté de jauge. Ainsi, les degrés de liberté physiques des différents champs vectoriels sont reliés entre eux et se transforment dans la représentation adjointe du groupe de symétrie de jauge. Ce résultat n'est pas surprenant, car comme discuté précédemment, les degrés de liberté physiques doivent être dans une représentation du groupe de symétrie ; avoir autant de champs que la dimension du groupe mène naturellement à la représentation adjointe.

Dans le cas d'un groupe de symétrie non-abélien, la construction de dérivées covariantes pour les spineurs nécessite donc, comme attendu en généralisant le cas abélien, l'introduction d'autant de champs vectoriels que de dimensions du groupe de symétrie. Cependant, cette construction implique également que les composantes physiques des champs vectoriels se transforment elles-mêmes sous l'action du groupe de symétrie, dans la représentation adjointe de celui-ci. Cette propriété a des conséquences phénoménologiques importantes, car elle signifie que les champs vectoriels propageant l'interaction associée à la symétrie locale sont eux-mêmes sujets à cette interaction. Cela complexifie aussi l'introduction d'un terme cinétique pour les champs $A_\mu^a$, sachant que celui-ci doit être aussi invariant de jauge.

Une façon simple d'introduire le tenseur de Faraday non-abélien consiste à considérer la commutation des dérivées covariantes. En effet, en construisant un tenseur en combinant des termes dans une représentation donnée du groupe de symétrie, on s'assure qu'il sera lui aussi dans une représentation du groupe de symétrie. Et les opérateurs de dérivées covariantes sont dans la représentation adjointe du groupe de symétrie, se transformant en $D_\mu \rightarrow U D_\mu U^{-1}$. En conséquence, le tenseur construit à partir de ces dérivées ne contiendra pas de degrés de liberté de jauge, et sera donc adapté pour décrire la dynamique des degrés de liberté physique. On a
\begin{equation}
\frac{i}{g}[D_\mu,D_\nu] = F_{\mu\nu}^\alpha T_\alpha = \partial_\mu A_\nu^\alpha T_\alpha - \partial_\nu A_\mu^\alpha T_\alpha - i g [A_\mu^\beta T_\beta , A_\nu^\gamma T_\gamma ],
\end{equation}
où l'on a laissé implicites les indices de la représentation. On identifie donc
\begin{equation}
\label{FaradayNonAbel}
F_{\mu\nu}^\alpha = \partial_\mu A_\nu^\alpha - \partial_\nu A_\mu^\alpha + g f_{\beta \gamma}{}^\alpha A_\mu^\beta A_\nu^\gamma,
\end{equation}
ce tenseur de Faraday se transformant dans la représentation adjointe du groupe de symétrie, à savoir
\begin{equation}
\delta F_{\mu\nu}^\alpha = f_{\beta\gamma}{}^\alpha \delta\theta^\beta F_{\mu\nu}^\gamma,
\end{equation}
pour une transformation infinitésimale.

Le Lagrangien complet d'une théorie de jauge non abélienne pour un spineur donne finalement 
\begin{equation}
\mathcal{L}_{\text{NA}} = i (\psi^\dagger)^a \left[\slashed \partial\delta_a^b -  i g \slashed A^\alpha (T_\alpha)_a^b \right] \psi_b - \frac14 F_{\mu\nu}^\alpha  F^{\mu\nu}_\alpha,
\end{equation}
où l'on a inclus le terme cinétique pour le champ vectoriel. Ce dernier terme est obtenu en formant un singlet du groupe de symétrie interne à partir de $F_{\mu\nu}^\alpha$ qui est dans sa représentation adjointe. Il n'y a pas besoin de contracter ce tenseur avec son complexe conjugué car la représentation adjointe d'une algèbre simple est toujours réelle~\cite{Slansky:1981yr}. Ce Lagrangien est bien invariant sous la transformation infinitésimale
\begin{equation}
\displaystyle{
\left\{
\begin{array}{l}
\delta \psi_a =  -i \delta\theta_\alpha (T^\alpha)_a^b \psi_b,\\
\delta A_\mu^\alpha  =  f_{\beta\gamma}{}^{\alpha}\delta \theta^\beta A_\mu^\gamma - \frac1g \partial_\mu \delta \theta ^\alpha.
\end{array}
\right.
}
\end{equation}
Il est possible d'écrire ce Lagrangien en faisant apparaître les différents couplages résultant de l'écriture de termes invariants de jauge,
\begin{equation}
\label{LagNonAbelFull}
\mathcal{L}_{\text{NA}} = i (\psi^\dagger)^a \slashed \partial \psi_a + g \slashed A^\alpha  (\psi^\dagger)^a (T_\alpha)_a^b  \psi_b - \partial_{[\mu} A_{\nu]}^\alpha \partial^{[\mu} A^{\nu]}_\alpha + g \partial^{[\mu} A^{\nu]}_\alpha f_{\beta\gamma}{}^\alpha A_\mu^\beta A_\nu^\gamma  + g^2 f_{\beta\gamma}{}^\alpha f^{\delta\epsilon}{}_\alpha A_\mu^\beta A_\nu^\gamma A^\mu_\delta A^\nu_\epsilon,
\end{equation}
où les crochets indiquent une antisymétrisation normalisée\footnote{Cette antisymétrisation s'écrit par exemple
\begin{equation}
\partial_{[\mu} A_{\nu]}^\alpha = \frac12 \left(\partial_\mu A_\nu^\alpha - \partial_\nu A_\mu^\alpha \right).
\end{equation}}.
Ce sont les termes de couplage de ce Lagrangien que nous allons étudier plus en détails par la suite.

\section{Interactions via les symétries de jauge}
\label{PartIntSymDeJauge}

\noindent
Nous avons précédemment détaillé la construction d'un Lagrangien invariant sous une transformation de symétrie locale pour des fermions. Cette construction nécessite l'introduction d'un ensemble de champs vectoriel associés au groupe de symétrie en question, et implique un certain nombre de couplages entre le champ vectoriel et les spineurs et d'auto-couplage du champ vectoriel avec lui-même. Nous discutons ici ces couplages plus en détails.

Dans le cas d'une symétrie globale, c'était la même opération de symétrie qui avait lieu en tout point de l'espace, et on pouvait relier cette symétrie à une redéfinition des différentes grandeurs physiques. Ce n'est plus possible pour une symétrie locale, où une opération de symétrie différente est opérée en chaque point de l'espace. Les indices internes $a$ d'un spineur distinguent alors des états décrivant des particules différentes. C'est aussi le cas pour les indices de groupe $\alpha$ des champs vectoriels. Ainsi, pour les interactions électrofaibles, l'électron et le neutrino électronique de chiralités gauches font partie d'une même représentation d'un groupe de symétrie locale SU(2). Pour les interactions fortes, décrites par un groupe de symétrie locale SU(3), les trois états de couleur (rouge, vert, et bleu) d'un même quark font aussi partie d'une même représentation. Les opérations de symétrie reliant différentes particules étant locales, elles peuvent en un point de l'espace-temps transformer une particule en une autre. Le groupe abélien U(1) n'ayant que des représentations irréductibles de dimension 1, les transformations de symétrie n'affectent que la phase d'une particule donnée. Par la suite, on discutera le cas d'un groupe simple (\emph{a priori} différent de U(1), qui ne l'est pas), et qui contient donc des représentations de dimension $d \geq 1$, regroupant plusieurs particules différentes.

\begin{figure}[h!]
\begin{center}
\includegraphics[scale=1.]{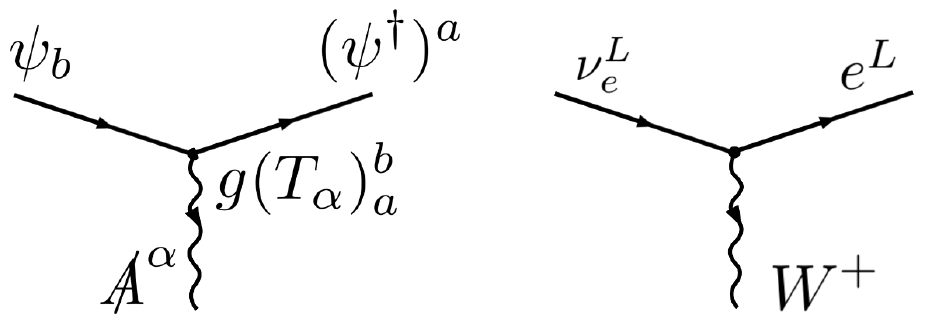}
\vspace{-0.4cm}
\end{center}
 \caption{Gauche : Vertex correspondant au terme Lagrangien donné dans l'équation~\eqref{VertexCouplage}. Droite : Vertex des interactions électrofaibles, transformant un neutrino électronique en électron, avec émission d'un boson $W$.}
 \label{VertexeW}
\end{figure}

Les transformations locales sont décrites par les vertex de couplage entre les champs, imposés par ces symétries. Si un spineur se transforme sous une symétrie de jauge non abélienne via
\begin{equation}
\label{TransfoPsiNonAbel}
\delta \psi_a =  -i \delta\theta_\alpha (T^\alpha)_a^b \psi_b,
\end{equation}
alors on a vu dans la section précédente que la nécessité d'écrire une action invariante de jauge impose un couplage dans le Lagrangien de la forme
\begin{equation}
\label{VertexCouplage}
\mathcal{L}_{A\psi\psi}= g \slashed A^\alpha  (\psi^\dagger)^a (T_\alpha)_a^b  \psi_b,
\end{equation}
reliant le spineur et le champ de jauge. Pour un indice $\alpha$ donné, ce terme de couplage décrit l'annihilation\footnote{On rappelle que les opérateurs de champs $\psi_a$ décrivent l'annihilation d'un fermion en un point de l'espace-temps, ou bien la création de l'anti-particule associée. De même, un opérateur $(\psi^\dagger)^a$ décrit la création d'un fermion, ou bien l'annihilation de l'anti-particule associée. La discussion est faite par la suite en termes de création et d'annihilation de particules, mais on peut les remplacer par des annihilations et créations d'antiparticules, respectivement.} d'une particule de type $\psi_b$ et la création d'une particule de type $\psi_a$ en même temps que l'émission d'un boson de jauge $A_\mu^\alpha$, lorsque les deux fermions sont reliés entre eux par la matrice de représentation $(T_\alpha)_a^b$. Ce vertex décrit donc le fait d'appliquer à un fermion et en un point de l'espace-temps la transformation associée à un des générateur du groupe de symétrie en émettant le boson de jauge correspondant, comme décrit dans la figure~\ref{VertexeW}. Dans le cadre des interactions faibles, par exemple, un neutrino électronique peut se transformer en électron (tous deux de chiralité gauche) en émettant le boson de jauge associé à cette transformation, comme décrit par le vertex de la figure~\ref{VertexeW}.

Les bosons de jauge jouent dans ce cadre un rôle très particulier. Émis en fonction des transformations de jauge qui sont effectuées, ils véhiculent l'information de l'opération de symétrie locale qui a été réalisée en un point de l'espace-temps : capter sur un détecteur une particule liée à $A_\mu^\alpha$ signifie qu'une transformation correspondante à $T_\alpha$ a été effectuée dans la direction d'où il provient. On identifiera donc par la suite les champ de jauge aux transformations qu'ils décrivent. Ce rôle est aussi visible sur la transformation de symétrie de ces bosons de jauge. Celle-ci contient un terme en $\frac1g \partial_\mu \delta \theta ^\alpha \in \delta A_\mu^\alpha $, qui en théorie des champ lie la création et la propagations de ces bosons aux modifications de \og l'état local de symétrie \fg{}, à savoir le paramètre $\theta(x)$. Il est d'ailleurs intéressant que ce terme de transformation contienne la dérivée de $\theta(x)$, et pas ce paramètre lui-même, car seules ses modifications ont un sens en soi, une fois un état initial donné. Le fait d'effectuer des transformations de symétrie locale nécessite donc d'émettre un boson de jauge pour propager \og l'état local de symétrie \fg{}. Cette propagation est par ailleurs permise par le fait que ces bosons sont sans masse, et donc de portée infinie ; nous y reviendrons en étudiant les brisures spontanées de symétrie dans la section~\ref{PhenomSymBrisées}. 

Dans la construction que l'on a suivie, les spineurs ne peuvent interagir qu'en échangeant des bosons de jauge, et donc en effectuant des transformations locales de symétries, qui impliquent parfois de changer de type de particule. C'est pourquoi on relie les groupes de symétrie locale à la notion d'interaction en physique des particules. Les champ de jauge étant identifiés aux différents générateurs du groupe de symétrie, on a vu qu'ils sont dans la représentation adjointe de celui-ci. C'est cette représentation particulière qui permet de coupler le champ de jauge avec toutes les représentations de la matière. En effet, le produit d'une représentation d'un groupe simple et de sa représentation conjuguée contient toujours la représentation adjointe de ce groupe~\cite{Slansky:1981yr}. Et la représentation adjointe étant réelle, le produit de celle-ci avec elle-même contient toujours l'identité. Ainsi, quelle que soit la représentation des différents champ spinoriels décrivant la matière, $\psi_a$ ou $\chi_{\tilde{a}}$ par exemple, il est toujours possible de créer des couplages $g \slashed A^\alpha  (\psi^\dagger)^a (T_\alpha)_a^b  \psi_b$ ou $g \slashed A^\alpha  (\chi^\dagger)^{\tilde{a}} (\tilde{T}_\alpha)_{\tilde{a}}^{\tilde{b}}  \chi_{\tilde{b}}$, les différents champ interagissant toujours de la même manière avec le champ de jauge. Les théories de jauge ont donc un avantage important d'un point de vue de la construction de modèles physiques, car elles permettent de coupler de façon naturelle des spineurs dans n'importe quelle représentation, indépendamment du fait qu'il soit possible d'obtenir des singlets à partir de ces spineurs seulement.

\section{Lien avec les poids et racines, application à SU(2) et SU(3)}
\label{NonAbelRacines&Poids}

\noindent
Pour décrire la physique des particules par des théories de jauge, il est important de déterminer pour une représentation donnée quelles particules sont reliées entre elles par les transformations de symétrie. Pour cela, le formalisme des racines et des poids explicité dans le chapitre~\ref{PartRacinesPoids} est particulièrement efficace. Prenant un certain nombres de particules rassemblées dans une représentation irréductible $\psi_a$ du groupe de jauge de rang $n$, on peut identifier chacune d'elles à un poids $\ket{\boldsymbol\lambda_a}= \ket{\lambda_{a_1},\cdots,\lambda_{a_n}}$. Les différentes composantes des poids sont les nombres quantiques de chaque particule pour le groupe de symétrie locale (on considère ici une théorie de jauge vis-à-vis d'un groupe simple, le cas des produits de groupes sera discuté dans la section~\ref{ProduitGroupesJauge}). Dans cette base, chaque représentant de l'algèbre du groupe de symétrie $(T^\alpha)_a^b$, et donc chaque champ de jauge associé $A_\mu^\alpha$, correspond à une racine. Ainsi, une particule $\psi_a$ ne peut se transformer localement en une particule $\psi_b$ en émettant un champ $A_\mu^\alpha$ que si la racine associée à $A_\mu^\alpha$ génère une translation du poids associé à $\psi_a$ au poids associé à $\psi_b$. On retrouve le caractère particulier de la représentation adjointe permettant de coupler les champ de jauge à tous les fermions par le fait que les racines peuvent agir sur tous les espaces de poids.

Associer chaque particule d'une représentation irréductible à un poids et chaque champ de jauge à une racine implique des choix de base particuliers, de la représentation comme des générateurs de l'algèbre. Les choix alternatifs de bases ne correspondent cependant qu'à des descriptions équivalentes des différents objets physiques, pour la plupart moins avantageuses. En effet, les champ de jauge étant associés à des racines, ils ne peuvent que transformer un poids en un autre, et donc une particule en une autre, ce qui est très adapté pour décrire des vertex de la forme~\eqref{VertexCouplage}\footnote{Faisant un combinaison linéaire de deux champs de jauge transformant une particule $\psi_{a_1}$ en des particules $\psi_{a_2}$ et $\psi_{a_3}$, les interactions avec les champ de jauge correspondraient alors à une superposition de couplage~\eqref{VertexCouplage} écrits en termes des particules $\psi_{a_i}$.}. Si un champ de jauge est lié à la transformation d'une particule $\psi_{a_1}$ en une particule $\psi_{a_2}$, alors il existe un champ de jauge, lié à la racine opposée, qui effectue la transformation inverse. Les champ de jauge liés aux racines nulles jouent un rôle particulier. En effet, les opérateurs liés à ces racines ne modifient pas le poids qu'ils affectent, et interagissent avec une particule sans la modifier. Le couplage associé à un tel vertex est proportionnel au poids de cette particule vis à vis de ces champ de jauge. Ainsi, un poids $\ket{\boldsymbol\lambda} = \ket{\lambda_1,\cdots,\lambda_n}$ a un couplage proportionnel à $\lambda_i$ avec la racine nulle liée à l'élément $T_3^{(i)}$ de l'algèbre de son groupe de symétrie. Les coordonnées des poids jouent alors le rôle de charges sous les théories de jauge abélienne U(1) générées par ces opérateurs $T_3^{(i)}$. 

Passons à l'identification des quantités conservées dans les théories de jauge non-abéliennes. On a vu que les composantes des poids des particules peuvent être identifiées à des charges sous les symétries U(1) associées aux différents éléments de la sous-algèbre de Cartan du groupe de symétrie, qui commutent entre eux. Le théorème de Noether implique donc que toutes ces grandeurs sont conservées. Cela légitime \emph{a posteriori} l'identification des composantes des poids comme nombres quantiques des particules, et le choix fait d'avoir utilisé l'espace des poids pour décrire les particules (voire aussi la note de bas de page \ref{footnote}). Dans les transformations de symétrie des fermions décrits par les termes de couplage~\eqref{VertexCouplage}, les nombres quantiques des bosons de jauge correspondent donc à la différence des nombres quantiques des fermions incidents et émergents du vertex. Les bosons de jauge associés aux racines nulles ont donc des nombres quantiques identiquement nuls. On a pour l'instant montré la conservation de $n$ grandeurs pour un groupe de rang $n$, qui comporte en général bien plus de dimensions, et donc de transformations infinitésimales. Ce sont en fait les seules quantités conservées que l'on peut faire apparaître en terme de charges, \emph{i.e.} de nombres quantiques\footnote{\label{footnote}La sous-algèbre de Cartan décrit en effet le nombre maximal d'opérateurs qui commutent entre eux dans l'algèbre de Lie du groupe de symétrie. On pourrait essayer de considérer les charges associées à des générateurs supplémentaires à cette sous-algèbre, mais cela impliquerait une diagonalisation sur une nouvelle base de l'espace de représentation composée d'éléments qui ne seraient plus vecteurs propres des opérateurs de la sous-algèbre de Cartan (on ne peut simultanément diagonaliser que des opérateurs qui commutent entre eux). En fait, la description des charges conservées vis-à-vis des éléments de la sous-algèbre de Cartan est simplement la formulation en théorie des groupes de la méthode en mécanique quantique qui consiste à prendre comme nombres quantiques des différents états leurs valeurs propres par rapport à un ensemble complet d'observables qui commutent entre eux, après diagonalisation simultanée de l'espace de Hilbert. C'est ce qu'on a fait en se plaçant dans l'espace des poids, ce qui explique l'apparition naturelle des nombres quantiques dans cet espace.}. Les autres quantités conservées sont contenues dans la structure de l'action des opérateurs associés aux champ de jauge sur les différentes particules. Par exemple dans le fait qu'un champ de jauge n'est associé qu'à une seule translation donnée dans l'espace des poids, \emph{i.e.} à une racine.

Appliquons cette construction au groupe SU(2), qui joue un rôle très important en physique car c'est le seul groupe simple possédant des représentations à bidimensionnelles\footnote{Cette propriété implique notamment que toutes les symétries continues sur des espaces à deux dimensions -- et donc sur des systèmes quantiques à deux états --, sont \emph{a priori} décrites par la représentation de dimension 2 d'un groupe SU(2), dont les transformations sont décrites par les matrices de Pauli.}. On se placera dans la description de spin des représentations de ce groupe, et pas dans la base de Dynkin (voir la construction associée dans la section~\ref{PartSU2Ini}). Les éléments des représentations sont donc décrits par des poids $\ket{m}_j = \ket{j,m}$, où le spin total $j$ décrit la représentation, et sa projection $m$ sur un axe distingue les différents éléments d'une représentation donnée. Les grandeurs $j$ et $m$ sont entières ou demi-entières dans la représentation de spin, et $m$ prend des valeurs séparées d'une unité entre $-j$ et $+j$. Les deux racines non nulles de SU(2) décrivent des translations dans l'espace des poids (ici de dimension 1) de $\pm 1$. On retrouve que les opérateurs d'élévation et d'abaissement $T_{\pm}$ transforment un état $\ket{m}_j$ en des états $\ket{m+1}_j$ et $\ket{m-1}_j$, et que l'opérateur $T_3$ est diagonal sur la représentation avec $m$ comme valeur propre pour $\ket{m}_j$.

\begin{figure}[h!]
\begin{center}
\includegraphics[scale=1.]{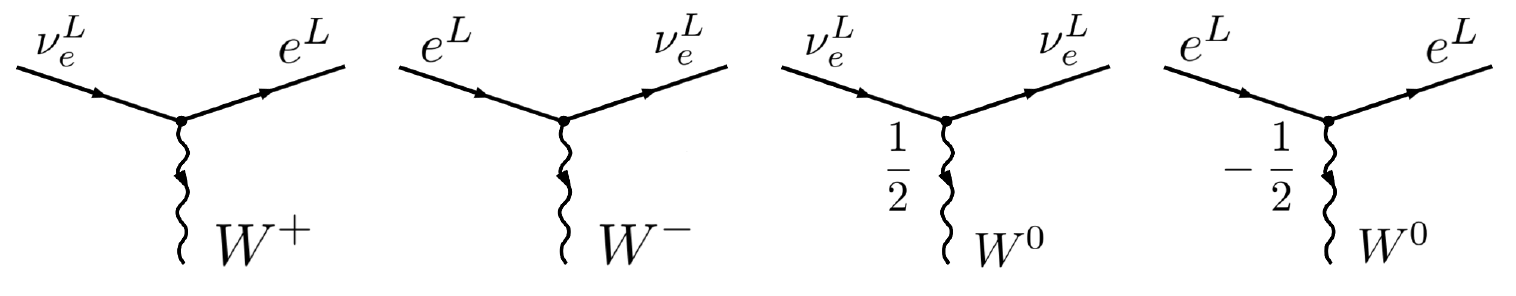}
\vspace{-0.6cm}
\end{center}
 \caption{Ensemble des vertex associés au secteur SU(2)$_L$ des interactions électrofaibles pour les spineurs décrivant l'électron et le neutrino électronique de chiralités droites.}
 \label{VertexFulleW}
\end{figure}

Pour les interactions électrofaibles, les neutrinos électroniques et électrons de chiralités gauches sont dans un doublet d'une symétrie locale SU(2) nommée SU(2)$_L$, de même que les quarks up et down de chiralités gauches
\begin{equation}
\label{DoubletsChirL_SM}
\boldsymbol\psi_{\text{leptons}}^L = 
\left(
\begin{array}{l}
\nu_e^L\\
e^L
\end{array}
\right),
~~~~ ~~~~ 
\boldsymbol\psi_{\text{quarks}}^L = 
\left(
\begin{array}{l}
u^L\\
d^L
\end{array}
\right),
\end{equation}
où l'on décrits directement les spineurs de Weyl par les symboles associés aux particules, précisant simplement leur chiralité par un exposant. Afin de différencier les nombres de spin liés aux représentations de SU(2)$_L$ du spin sous les rotations, on les nomme nombres quantiques d'isospin faible. Pour SU(2)$_L$ des interactions électrofaibles, on note $W^{\pm}$ les champ de jauge associés à $T_\pm$, et $W^0$ le champ de jauge associé à $T_3$. Suivant la discussion précédente, on peut alors récapituler l'action de ces champ de jauge sur le doublet $\psi_{\text{leptons}}^L$ par les vertex de la figure~\ref{VertexFulleW}. Pour le couplage avec le boson $W^0$, on a indiqué la charge pour le générateur associé, ce boson pris isolément générant une théorie de jauge abélienne.

Dans cette construction, les bosons $W^+$ et $W^-$ sont complexes conjugués l'un de l'autre. Cela se voit notamment par le fait que l'hermitien conjugué de l'opérateur lié à une racine donne un opérateur lié à la racine opposé, comme discuté dans la section~\ref{Poids&Racines}. Les opérateurs d'échelle ont en effet été construits à partir de combinaisons linéaires complexes des opérateurs hermitiens définissant la structure de l'algèbre de SU(2)$_L$, voir l'équation~\eqref{EqDefOpEchelle}. Formant un champ complexe, ils sont chargés sous (au moins) un U(1), ici celui associé à l'opérateur $W^0$. On retrouve cette structure pour tous les groupes simples. Les champ de jauge liés aux opérateurs d'échelle portant la différence de charge (ici d'isospin faible) entre les particules qu'ils relient, ils seront toujours chargés sous au moins un des éléments de la sous-algèbre de Cartan du groupe de Lie. Revenant au cas de SU(2), les bosons $W^\pm$ portent une charge d'isospin faible $\pm 1$. Les bosons de jauge $A_\mu^\alpha$ sont en effet dans la représentation adjointe du groupe de jauge, ici celle de dimension 3 de SU(2)$_L$, formant un triplet d'isospins respectifs $-1$, $0$, et $+1$ pour $W^-$, $W^0$, et $W^+$ (on parle ici de charges d'isospin faible, le lien avec la charge électrique sera fait plus tard). On peut alors vérifier la conservation de l'isospin sur les vertex de la figure~\ref{VertexFulleW}. On note qu'étant émis, les bosons de jauge emportent la différence de charge entre la particule initiale et la particule finale de la transformation qui leur est associée. Ils portent donc la charge opposée à la translation qu'ils décrivent dans l'espace des poids, et il faut associer $W^-$ à $T_+$ et $W^+$ à $T_-$. Cette convention est inversée si on écrit les vertex avec des bosons de jauge incidents.

\begin{figure}[h!]
\begin{center}
\includegraphics[scale=1.1]{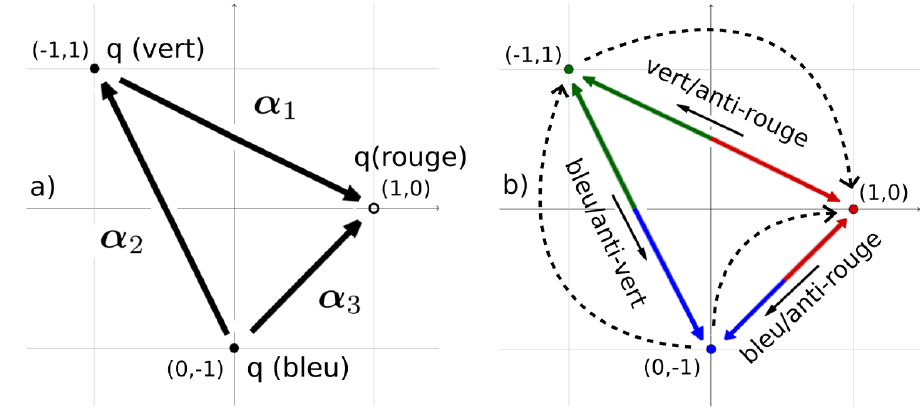}
\vspace{-0.6cm}
\end{center}
 \caption{Diagrammes des poids de la représentation de dimension 3 de SU(3)$_C$ dans la base de Dynkin. Dans la figure a), on indique les différentes translations liées aux trois racines positives $\boldsymbol\alpha_i$ de SU(3)$_C$, voir la section~\ref{Poids&Racines}. Dans la figure b), on utilise la description en termes de charges de couleur des bosons émis en effectuant les transformations indiquées par les flèches en pointillés.}
 \label{DiagPoidsQuarks}
\end{figure}

Intéressons nous à présent aux interactions fortes, et donc à la chromodynamique quantique (QCD). Ces interactions ne brisent pas la parité\footnote{La prise en compte des états propres de masse associés aux quarks, discuté dans la section~\ref{PartCKMMatrix}, brise par contre la parité.}, et agissent donc de la même façon sur les états de chiralités gauche et droite, qui peuvent être considérés indépendamment avec des spineurs de Weyl, ou associés en un spineur de Dirac. On omettra donc par la suite les indices $L$ et $R$, et le type de spineur qu'on considère ne sera pas précisé. Chaque quark est dans une représentation de dimension 3 du groupe de symétrie locale SU(3)$_C$ des interactions fortes. On notera par exemple leur spineurs $u^\alpha$, $d^\alpha$, et $s^\alpha$ avec $\alpha\in\{1,2,3\}$, ici pour les quarks up, down, et strange respectivement. L'espace des poids de cette représentation est donné dans la figure~\ref{DiagPoidsQuarks}. On y visualise l'action des bosons de jauge liés aux racines positives non-nulles, les transformations inverses liées aux racines négatives non-nulles étant aussi permises. Les deux bosons de jauge restants, liés aux opérateurs $T_3^{(1)}$ et $T_3^{(2)}$ de la sous-algèbre de Cartan de SU(3)$_C$, ont des interactions avec les différents états internes qui correspondent à une théorie de jauge abélienne, avec un couplage pour chaque état interne des quarks proportionnel à la composante de leur poids associé à ces $T_3^{(i)}$. Ces transformations décrivent l'intégralité des interactions possibles entre les quarks et les gluons.

\begin{figure}[h!]
\begin{center}
\includegraphics[scale=1.2]{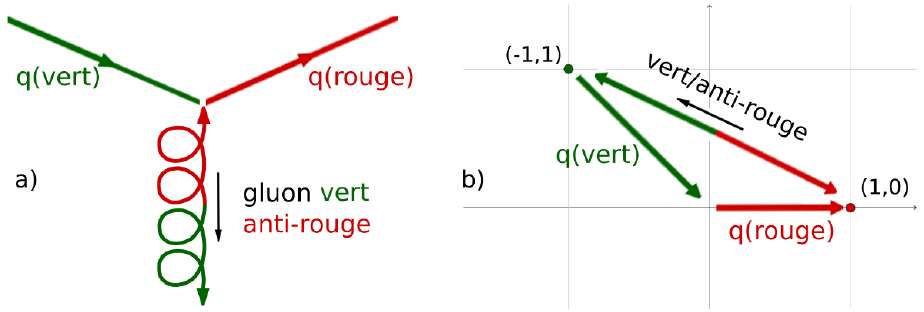}
\vspace{-0.6cm}
\end{center}
 \caption{La figure a) est un vertex de la chromodynamique quantique décrivant la transformation d'un quark vert en un quark rouge, avec émission d'un gluon vert/anti-rouge. La figure b) décrit la même transformation dans le diagramme des poids de SU(3)$_C$, comptant négativement et positivement les charges des particules entrantes et sortantes du vertex, respectivement.}
 \label{VertexQCD}
\end{figure}

Pour décrire ces interactions, il est commode d'introduire la notion de charge de couleur. Ces charges sont introduites en associant aux trois états internes des quarks une couleur donnée, soit dans le cas présent et dans la base de Dynkin $(1,0)$ pour la charge rouge, $(-1,1)$ pour la charge verte, et $(0,-1)$ pour la charge bleue. Ces triplets sont par exemple notés
\begin{equation}
u = 
\left(
\begin{array}{l}
u^r\\
u^v\\
u^b
\end{array}
\right),
~~~~ ~~~~ 
d = 
\left(
\begin{array}{l}
d^r\\
d^v\\
d^b
\end{array}
\right),
\end{equation}
pour les trois états de couleurs des quarks up et down\footnote{Il est intéressant de noter que comme ces états quarks ont été découverts tardivement, notamment à cause de la propriété de confinement des interactions forte, ils portent le même nom. Ce sont pourtant \emph{a priori} des particules différentes du point de vue de la construction de modèles physiques.}. Le complément~\ref{RepEquivSU3} donne des détails supplémentaires sur ce choix de base, et sur des choix de base alternatives rencontrés dans la littérature. Cette nomenclature permet une description simple des différents quarks et gluons, par la donnée des charges de couleurs qu'ils véhiculent, la conservation des poids donnant la conservation des charges de couleur. La figure~\ref{VertexQCD} présente un exemple de vertex en QCD avec les dénominations des particules par les charges de couleurs, ainsi que la visualisation du même vertex dans l'espace de poids de SU(3)$_C$, faisant apparaître explicitement la conservation des nombres quantiques. On retrouve dans cette figure qu'on peut considérer de façon équivalente un gluon vert/anti-rouge émergent du vertex, ou son anti-particule incidente, un gluon rouge/anti-vert. Seul le sens de la flèche noire identifiant la direction de propagation du gluon ainsi que le nom de celui-ci changera\footnote{Il faudra alors compter négativement les nombres quantiques du gluon rouge/anti-vert pour vérifier les lois de conservation, car celui-ci sera incident au vertex.}. Pour les vertex liés aux opérateurs $T_3^{(1)}$ et $T_3^{(2)}$, on peut les décrire de la même façon en introduisant des gluons par exemple rouge/anti-rouge, et en listant leurs couplages avec chacune des états de couleur des quarks. Par exemple, en lisant les poids des différents états de couleur dans la figure~\ref{DiagPoidsQuarks}, on voit que l'opérateur $T_3^{(1)}$ correspond à la différence du gluon rouge/anti-rouge et du gluon vert/anti-vert.

\begin{figure}[h!]
\begin{center}
\includegraphics[scale=1.0]{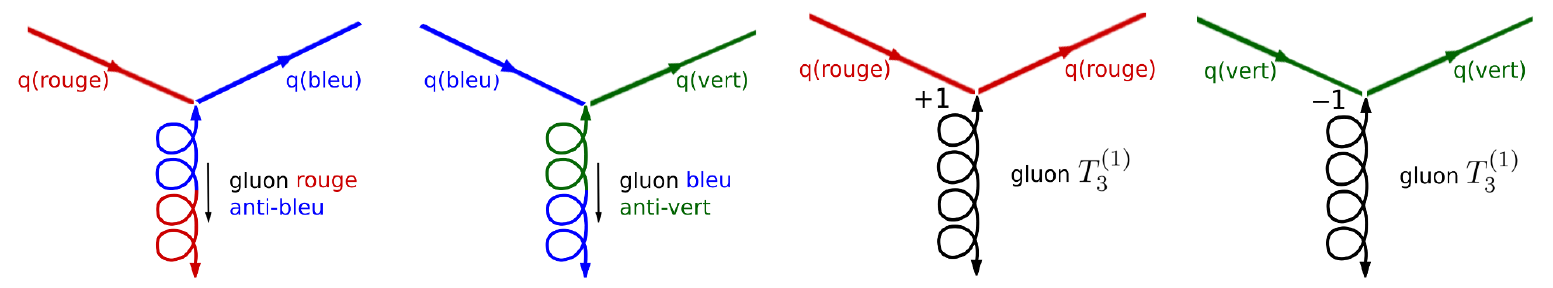}
\vspace{-0.6cm}
\end{center}
 \caption{Exemples de vertex de transformations de symétrie locale des quarks. Pour les vertex liés à l'élément $T_3^{(1)}$ de la sous-algèbre de Cartan de SU(3)$_C$, on a indiqué la charge des fermions pour ces générateurs (qui se lit par exemple dans la figure~\ref{DiagPoidsQuarks}).}
 \label{VertexQCDFermions}
\end{figure}

La description par les charges de couleur, bien que commode pour décrire les états internes des quarks et les gluons, ne doit pas faire oublier qu'une description correcte des nombres quantiques des interactions fortes est intrinsèquement bidimensionnelle. On ne peut par exemple pas identifier dans SU(3)$_C$ trois symétries U(1) associées à chacune des charges de couleur et qui commutent. Des vertex liés à des transformations de symétrie locale de quarks sont donnés dans la figure~\ref{VertexQCDFermions} ; les gluons non chargés y sont associés aux racines nulles, plutôt que décrits par des couleurs. La description des gluons comme des paires quark/anti-quark est aussi à prendre avec précaution. D'une part, elle n'explique pas la structure exacte des transformations associées aux gluons, car neuf combinaisons indépendantes de paires peuvent ainsi être formées (dont trois liées à des symétries U(1) qui ne sont pas indépendantes, comme discuté plus haut). D'autre part, cette description cache le fait que les opérateurs liés aux gluons peuvent agir sur toute représentation non-triviale du groupe de jauge, et pas seulement les quarks (même si  seuls les quarks semblent être dans une telle représentation à basse énergie, à cause des propriétés de confinement de la QCD discutées dans la section~\ref{PartRenormalisationPhenomSM}). Il faut donc \emph{a priori} considérer les gluons comme des particules élémentaires à part entière.

\section{Interactions entre bosons de jauge}
\label{PartIntBosonsDeJauge}

\noindent
En conséquence de la construction des théories de jauge non-abéliennes dans la section~\ref{PartJaugesNonAbeliennes}, on a vu que les champs de jauge devaient être dans une représentation non triviale du groupe de jauge. Cela implique que les champ de jauge sont eux-même sujets à l'interaction dont ils sont les médiateurs. Ces couplages apparaissent dans la construction du terme cinétique pour le champ de jauge, car le tenseur de Faraday non abélien défini par l'équation~\eqref{FaradayNonAbel} contient un terme ne décrivant pas la dérivée du champ vecteur. Les deux termes d'auto-couplage apparaissent dans le Lagrangien de l'équation~\eqref{LagNonAbelFull}, et sont de la forme
\begin{equation}
\mathcal{L}_{\text{auto-couplage}} =  g \partial^{[\mu} A^{\nu]}_\alpha f_{\beta\gamma}{}^\alpha A_\mu^\beta A_\nu^\gamma  + g^2 f_{\beta\gamma}{}^\alpha f^{\delta\epsilon}{}_\alpha A_\mu^\beta A_\nu^\gamma A^\mu_\delta A^\nu_\epsilon.
\end{equation}

Le terme cubique correspond à des interactions quasiment similaires à celles décrites pour les fermions, aux contractions des indices d'espace-temps près. La contribution de la transformation de $A_\mu^\alpha$ qui correspond à une représentation du groupe de symétrie est (voir la section~\ref{PartJaugesNonAbeliennes})
\begin{equation}
\label{TransfoAAdj}
\delta_{\text{adj}} A_\mu^\alpha  =  f_{\beta\gamma}{}^{\alpha}\delta \theta^\beta A_\mu^\gamma = -i \delta^\alpha (T^A_\alpha)_\beta^\gamma  A_\mu^\beta,
\end{equation}
en identifiant $(T^A_\alpha)_\beta^\gamma = i f_{\alpha \beta}{}^\gamma$. On reconnait la représentation adjointe du groupe de jauge, avec notamment le fait que les indices de la représentation coïncident avec les indices du groupe. Le terme de couplage
\begin{equation}
\label{AutoCouplageAAA}
\mathcal{L}_{\text{auto-couplage,}A^3} = g \partial^{[\mu} A^{\nu]}_\alpha f_{\beta\gamma}{}^\alpha A_\mu^\beta A_\nu^\gamma = -i g \partial^{[\mu} A^{\nu]\alpha} A_\mu^\beta (T^A_\alpha)_\beta^\gamma A_{\nu\gamma},
\end{equation}
est donc la correspondance pour le champ vectoriel\footnote{En fait, les transformations et couplages dans la formulation avec des constantes de structure des équations~\eqref{TransfoAAdj} et~\eqref{AutoCouplageAAA} sont donnés dans les représentations de définition des groupes de Lie, où les matrices des opérateurs de la représentation adjointe sont écrites directement à partir des constantes de structure elles-même. Les formulations avec les représentants $(T^A_\alpha)$ des éléments de l'algèbre de Lie dans la représentation adjointe sont par contre intrinsèques, et ne présupposent pas une base particulière pour les champ de jauge. On peut donc les utiliser dans le cadre de la description des représentations par le formalisme des racines et des poids.} au terme de couplage avec le champ spinoriel donné dans l'équation~\eqref{VertexCouplage}. La seule différence est la nécessité d'introduire une dérivée du champ vectoriel médiateur de l'interaction, à la place d'une contraction avec une matrice de Pauli, afin de construire un scalaire. Tous les raisonnements faits dans le cas du couplage avec des spineurs sont inchangés, et un vertex pour cette interaction correspond à l'application de $T^A_\alpha$ sur le champ de jauge en un point de l'espace-temps simultanément à l'émission du champ $\partial^{[\mu} A^{\nu]\alpha}$. En pratique, les vertex cubiques sont généralement symétrisés sur les trois bosons de jauge incidents, afin de ne pas avoir à identifier dans les calculs l'impulsion d'un seul des bosons de jauge.

Le terme quadratique est de même lié à la contribution dans la transformation de $A_\mu^\alpha$ qui correspond à la représentation adjointe. En effet, on peut l'écrire 
\begin{equation}
\label{AutoCouplageAAAA}
\mathcal{L}_{\text{auto-couplage,}A^4} = g^2 f_{\beta\gamma}{}^\alpha f^{\delta\epsilon}{}_\alpha A_\mu^\beta A_\nu^\gamma A^\mu_\delta A^\nu_\epsilon = - g^2A_\mu^\beta (T_A^\alpha)_\beta^\gamma A^\nu_\gamma   \cdot A^\mu_\delta (T^A_\alpha)^\delta_\epsilon A_\nu^\epsilon,
\end{equation}
qui correspond au couplage de deux paires de champ liés entre eux par un même opérateur du groupe de symétrie $T_\alpha$. Il est en fait souvent aisé de décrire les différents couplages quartiques (voire cubiques) en listant ceux qui vérifient les lois de conservation des nombres quantiques. La formulation par les transformations de symétrie du groupe est cependant primordiale, notamment pour décrire les couplages impliquant les champ de jauge liés aux racines nulles.

\begin{figure}[h!]
\begin{center}
\includegraphics[scale=1.2]{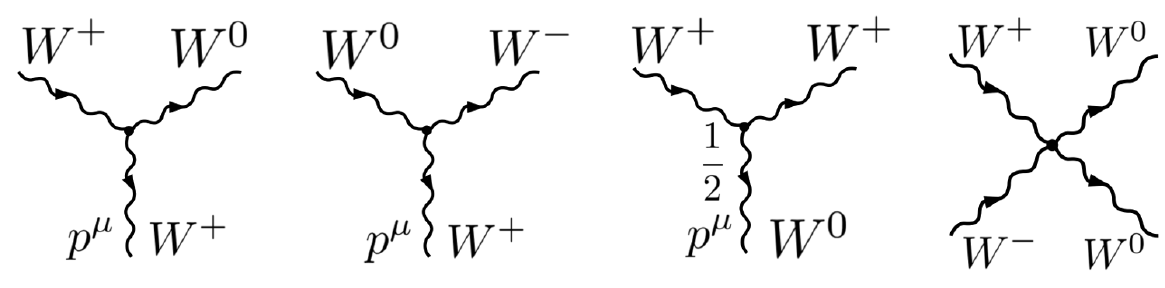}
\vspace{-0.6cm}
\end{center}
 \caption{Exemples de vertex d'interactions entre bosons de jauge associés au secteur SU(2)$_L$ des interactions électrofaibles. Les deux vertex de gauches sont les transformations possibles associées à $T_-$, et donc à $W^+$. Dans le troisième vertex, associé à la transformation $T_3$, on a noté la charge d'isospin sous la symétrie U(1) associée. Les charges d'isospin sont omises dans le vertex quartique.}
 \label{VertexJaugeeW}
\end{figure}

Pour les interactions faibles, la représentation adjointe de SU(2)$_L$ est de dimension 3, et les champ de jauge sont dans un triplet
\begin{equation}
\left(
\begin{array}{l}
W^+\\
W^0\\
W^-
\end{array}
\right),
\end{equation}
d'isospins respectifs $-1$, $0$ et $+1$. Les actions des différents champ de jauge pour les vertex cubiques sont donc simples à lister, en rappelant que les bosons de jauge transportent les nombres quantiques opposés de la transformation qu'ils impliquent, et que $W^+$ et $W^-$ correspondent donc respectivement aux opérateurs $T_-$ et $T_+$. L'émission d'un boson $W^+$ correspond à la transformation de $W^+$ en $W^0$, ou de $W^0$ en $W^-$, et l'émission d'un boson $W^-$ correspond à la transformation de $W^-$ en $W^0$, ou de $W^0$ en $W^+$. Le boson $W^0$ se couple avec $W^+$ et $W^-$ avec des charges respectives $+1$ et $-1$. On procède similairement pour les vertex à quatre champ, suivant les considérations présentées plus haut. Des exemples de tels vertex sont donnés dans la figure~\ref{VertexJaugeeW}, où l'on indique avec un terme en $p^\mu$ le champ vecteur qui contient une dérivée. 

\begin{figure}[h!]
\begin{center}
\includegraphics[scale=1.35]{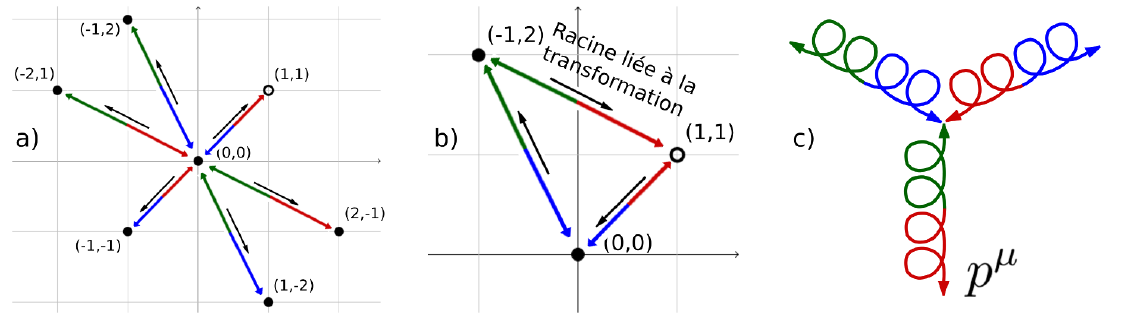}
\vspace{-0.6cm}
\end{center}
 \caption{La figure a) est le diagramme des poids de la représentation adjointe de SU(3)$_C$, où l'on a indiqué les charges de couleurs. Les figures b) et c) décrivent une transformation de symétrie possible pour les champ de jauge, correspondant à une interaction cubique. La figure b) décrit cette transformation dans le diagramme des poids, et la figure c) en terme de vertex d'interaction.}
 \label{TransfoGluonPoids}
\end{figure}

\begin{figure}[h!]
\begin{center}
\includegraphics[scale=1.0]{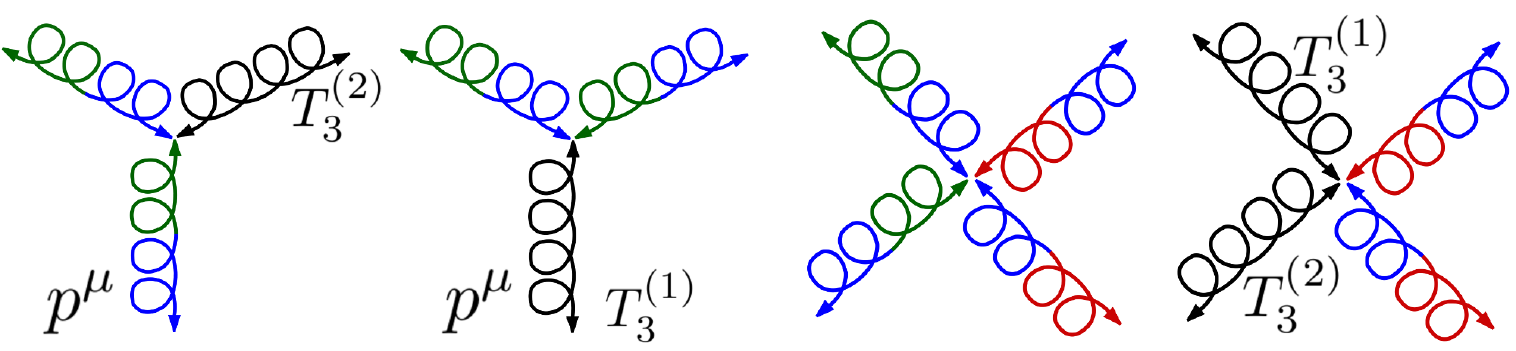}
\vspace{-0.6cm}
\end{center}
 \caption{Exemples de vertex d'auto-interaction cubiques et quartiques pour les gluons. Pour les gluons liés aux éléments de la sous-algèbre de Cartan de SU(3)$_C$, on n'a pas indiqué les charges associées.}
 \label{VertexGluons}
\end{figure}

Le groupe de jauge SU(3)$_C$ des interactions fortes étant de rang 2, le diagramme des poids associé à la représentation adjointe, et donc aux bosons de jauge, est de dimension 2. Ce diagramme des poids est donné dans la figure~\ref{TransfoGluonPoids}. On y a aussi représenté une transformation de symétrie possible pour les gluons, avec le vertex cubique associé. Les champs étant complexes, on n'a pas précisé sur le vertex le sens de propagation des particules, vu qu'on peut prendre de façon équivalente leur anti-particule se propageant dans la direction opposée. Sur le diagramme des poids, on a par contre indiqué les gluons émergeant du vertex. On vérifie bien la conservation des charges des interactions fortes, via les poids dans le diagramme associé, et via les couleurs dans le vertex. La figure~\ref{VertexGluons} donne également d'autres exemples de vertex d'auto-interaction pour les gluons. Comme précédemment, si on a pris les notations de charges de couleurs pour les gluons associés aux racines non-nulles, on a dénoté les gluons liés aux racines nulles par l'élément de la sous-algèbre de Cartan qui leur est associé. La vérification de la conservation des charges de couleur est immédiate sur ces vertex. 

Pour les groupes de rang $n\geq 2$ comme  SU(3)$_C$, une subtilité supplémentaire apparaît pour les membres de la sous-algèbre de Cartan, qui ont les mêmes coordonnées nulles sur le diagramme des poids. Ainsi, partant du poids $(-1,2)$ de la représentation adjointe de la figure~\ref{TransfoGluonPoids} et effectuant la transformation associée à l'émission d'un gluon vert/anti-bleu, le diagramme des poids ne permet pas d'obtenir l'état final obtenu. Il est pour cela nécessaire de se ramener aux relations de commutations entre les différents générateurs du groupe, décrivant les transformations de la représentation adjointe. On a déjà vu dans la section~\ref{Poids&Racines} la relation
\begin{equation}
\label{EqComRacines}
\left[ T_3^{(i)},T_{\boldsymbol\alpha_p}\right] = \alpha_p^i T_{\boldsymbol\alpha_p},
\end{equation}
décrivant l'action des générateurs de la sous-algèbre de Cartan sur les générateurs liés aux racines. On peut également calculer que 
\begin{equation}
\label{EqComRacinesOpposées}
\left[ T_{\boldsymbol\alpha_p},T_{-\boldsymbol\alpha_p}\right] = \alpha_p^i T_{3}^{(i)}.
\end{equation}
Cette équation décrit donc que partant du poids $(-1,2)$ et effectuant la transformation associée à l'émission d'un gluon vert/bleu, le poids obtenu sera la combinaison linéaire $[-(0,0)_3^{(1)} + 2 (0,0)_3^{(2)}]$, où l'on a repéré les poids nuls par les éléments de la sous-algèbre de Cartan auxquels ils correspondent. De tels poids sont en fait associés aux générateurs de la sous-algèbre de Cartan qui forment une sous-algèbre SU(2) avec les générateurs d'une racine $\boldsymbol \alpha_p$. En effet, définissant
\begin{equation}
\label{DefTSU2}
T_3^{(\boldsymbol\alpha_p)}=\alpha_p^i T_3^{(i)},
\end{equation} 
on obtient directement 
\begin{equation}
\label{DefComTSU2}
\left[ T_{\boldsymbol\alpha_p},T_{-\boldsymbol\alpha_p}\right] = T_3^{(\boldsymbol\alpha_p)}.
\end{equation}
Cela montre, avec l'équation~\eqref{EqComRacines}, que l'ensemble formé des générateurs $\{ T_3^{(\boldsymbol\alpha_p)} ,T_{\boldsymbol\alpha_p} ,T_{-\boldsymbol\alpha_p}\}$ décrit bien une sous-algèbre SU(2) du groupe complet.

Dans tous les cas, il n'est pas possible d'avoir des vertex cubiques impliquant plus d'un champ de jauge lié à un élément $T_3^{(i)}$ de la sous-algèbre de Cartan du groupe de symétrie, et un vertex quartique impliquant plus de deux tels champs. Les champs liés aux éléments de la sous-algèbre de Cartan ne portant aucune charge pour le groupe, un vertex cubique comportant deux de ces champs violerait automatiquement la conservation des charges du groupe. Pour les vertex cubiques et quartiques impliquant respectivement trois ou quatre éléments $T_3^{(i)}$, cela impliquerait [voir les équations~\eqref{AutoCouplageAAA} et~\eqref{AutoCouplageAAAA}] que deux tels champs sont reliés entre eux par une transformation du groupe de symétrie, ce qui n'est pas le cas. 

\section{Produit de groupes de jauge}
\label{ProduitGroupesJauge}

\noindent
Nous avons pour l'instant considéré les théories de jauge associées à un groupe de symétrie simple ou abélien. Les théories de jauge peuvent cependant aussi être liées à des symétries décrites par des produits directs de groupes simples ou abélien. En fait, c'est même la forme générale qu'on attend pour des groupes de symétrie continues, voir la section~\ref{PartClassificationAlgebres}. Du point de vue de la terminologie, on identifie une interaction à un groupe de symétrie, ou à un produit restreint de groupes de symétrie, comme SU(2)$_L\times$U(1)$_Y$ pour les interactions électrofaibles, ou SU(3)$_C$ pour les interactions fortes. Mais la symétrie complète d'un modèle contenant ces différentes interactions est le produit direct des groupes associés (les définitions et propriétés des produits directs et de leurs représentations sont données dans l'annexe~\ref{ProduitsGroupes}).

Les représentations de produits directs de groupes se décrivent par les produits des représentations de chaque groupe. Partant de deux groupes $G_1$ et $G_2$ portant respectivement des représentations de dimensions $d_1$ et $d_2$, notées $\psi_1^{a}$ et $\psi_2^b$ pour $a=1,\cdots,d_1$ et $b=1,\cdots,d_2$, alors le groupe $G_1\times G_2$ porte une représentation de dimension $d=d_1\cdot d_2$ notée $\psi_{1\times2}^{a,b}$. Dans cette écriture, $G_1$ agira uniquement sur l'exposant $a$ et de la même façon que pour la représentation $\psi_1^{a}$, de même pour $G_2$. Pour les groupes abéliens U(1), la donnée d'une représentation revient à la donnée de la charge de chaque particule sous l'action du groupe. La représentation $\psi_{1\times2}$ est irréductible si $\psi_1^{a}$ et $\psi_2^{b}$ le sont. Cette formulation des représentations de produits de groupes permet d'expliciter simplement leurs propriétés.

Pour commencer, chaque élément de la représentation de $G_1$ porte une représentation complète de $G_2$, et réciproquement. Cette contrainte est très forte, puisqu'on ne pourra assembler dans des représentations de $G_1$ que des ensembles de particules se transformant dans une même représentation de $G_2$. Ainsi, on ne peut par exemple rassembler dans une même représentation de SU(2)$_L$ que des particules dans une représentation triviale de SU(3)$_C$ (comme les électrons et neutrinos électroniques de chiralités gauches), ou bien des particules dans une même représentation de dimension 3 de SU(3)$_C$ (comme les quarks up et down). Une autre propriété est que les éléments de $G_1$ n'agissant que sur l'exposant $a$ de $\psi_{1\times 2}^{a,b}$, ils relient toujours deux états de mêmes nombres quantiques sous $G_2$, et sont donc non chargés sous ce dernier groupe. Dans un produit de groupe, les bosons de jauge liés à un groupe $G_i$ sont ainsi uniquement dans sa représentation adjointe, et  dans la représentation triviale des groupes $G_j$ pour $j\neq i$. Ils ne sont donc sujets qu'à l'interaction du groupe auquel ils sont associés. En particulier, les bosons de jauge de $G_i$ ne peuvent être chargés que pour les sous-groupes abéliens générés par les éléments de la sous-algèbre de Cartan de $G_i$.

Les quarks forment un exemple de particules dans la représentation de plusieurs groupes distincts. On a en effet vu que les quarks up et down de chiralité gauche étaient dans une représentation de dimension 2 de SU(2)$_L$. Mais chacun de ces états est aussi dans une représentation de dimension 3 de SU(3)$_C$, associée à leurs états de couleurs. Prenant en compte ces deux groupes de symétrie, ces quarks s'assemblent dans une seule représentation de dimension 6 de SU(2)$_L\times$SU(3)$_C$, qu'on peut noter $\psi^{L,a,\alpha}_{\text{quarks}}$, l'exposant $a$ prenant deux valeurs distinguant le quark up du quark down, et l'exposant $\alpha$ prenant trois valeurs désignant la charge de couleur. On a par exemple
\begin{equation}
\psi^{L,u,\alpha}_{\text{quarks}}=
\left(
\begin{array}{l}
u^{L,r}\\
u^{L,g}\\
u^{L,b}
\end{array}
\right),
\end{equation}
et
\begin{equation}
\psi^{L,a,r}_{\text{quarks}}=
\left(
\begin{array}{l}
u^{L,r}\\
d^{L,r}
\end{array}
\right).
\end{equation}
Les transformations de chacun de ces groupes n'agissent que sur les indices associés à leurs représentations. Ainsi, les gluons peuvent faire changer d'état de couleur un quark up ou down, et un boson $W$ peut changer un quark up d'une couleur en un quark down de la même couleur, par exemple. Mais ces deux secteurs n'interagissent pas entre eux. Le Modèle Standard de la physique des particules contient aussi une symétrie U(1) liée à l'hypercharge, ou hypercharge faible. En conséquence, $\psi^{L,a,\alpha}_{\text{quarks}}$ peut porter une hypercharge, qui sera la même pour toutes les particules contenues dans cette représentation.

On peut faire de même pour les quarks up et down de chiralité droite, qui sont tous deux des singlets de SU(2)$_L$ mais des triplets de couleurs, $u^R_\alpha$ et $d^R_\alpha$, et portent une hypercharge. Ou avec les électrons et neutrinos électroniques de chiralité gauche, qui sont dans un doublet de SU(2)$_L$ et dans un singlet de SU(3)$_C$ (ils ne portent pas de charges de couleurs), $\psi^{L^,a}_{\text{leptons}}$. Les électrons et neutrinos électroniques de chiralité droite sont quant à eux des singlets de SU(2)$_L\times$SU(3)$_C$, $e^R$ et $\nu_e^R$, et ne peuvent porter qu'une hypercharge. 

Les notations utilisées pour décrire les représentations de produits de groupes consiste à juxtaposer les représentations sous chaque groupe apparaissant dans le produit. Les représentations des groupes non abélien sont communément décrites par leur dimension notée en gras, et les représentations des groupes abélien par leur charge. Ainsi, $\psi^{L,a,\alpha}_{\text{quarks}}$ est dans la représentation $(\mathbf{3},\mathbf{2},\frac16)$ sous SU(3)$_C\times$SU(2)$_L\times$U(1)$_Y$. Les dérivées covariantes des jauge de champs dans la représentation d'un produit de groupes de jauge sont obtenues en sommant les contributions liées aux champs de jauge des différents groupes de symétrie ; se rappelant que l'action des opérateurs de chaque groupe ne joue que sur la représentation restreinte qui leur est associée. Ainsi, la dérivée covariante de $\psi^{L,a,\alpha}_{\text{quarks}}$ s'écrit 
\begin{equation}
D_\mu \psi^{L,a,\alpha}_{\text{quarks}} = \left[\partial_\mu \delta^a_b \delta^\alpha_\beta - i g_Y y^L_\text{quarks} A_\mu ^Y \delta^a_b \delta^\alpha_\beta - ig_L  A_{\mu,L}^{\alpha_L}(T_{\alpha_L})^a_b \delta^\alpha_\beta - i g_{_C} A_{\mu,C}^{\alpha_{_C}}(T_{\alpha_{_C}})^\alpha_\beta \delta^a_b\right]\psi^{L,b,\beta}_{\text{quarks}},
\end{equation}
où les indices $\alpha_L\in[1,2,3]$ et $\alpha_C\in[1,\cdots,8]$ identifient les différents générateurs de l'algèbre de SU(2)$_L$ et SU(3)$_L$. Cette formule est un peu lourde écrite avec tous ses indices explicitement, mais est conceptuellement équivalente à l'équation~\eqref{DevCovNonAbel} après identification des composantes simples du groupe de symétrie complet.

Finalement, on peut aussi s'interroger sur quelle est la symétrie liée à l'électromagnétisme dans la théorie de jauge SU(2)$_L\times$U(1)$_Y$ des interactions électrofaibles. La solution la plus simple serait d'identifier U(1)$_Y$ à l'électromagnétisme. L'expérience montre cependant que les bosons $W^\pm$ liés aux racines non-nulles de SU(2)$_L$ portent une charge électrique. Cela implique que U(1)$_Y$ n'est pas la symétrie de l'électromagnétisme, car $W^\pm$ ne pourraient alors pas porter une charge électrique. L'électromagnétisme n'est pas lié non plus à la symétrie U(1) générée par l'élément $T_3$ de l'algèbre de SU(2)$_L$, car le neutrino électronique de chiralité gauche serait alors chargé électriquement (il a une charge d'isospin faible $+\frac12$). Il faut donc chercher le U(1) de l'électromagnétisme dans une combinaison des deux symétries abéliennes contenues dans SU(2)$_L\times$U(1)$_Y$, voir la section~\ref{RepChampsHiggs}.

\chapter{Brisures spontanées de symétries}
\label{PartSSB}
\section{Introduction}

\noindent
Bien que les théories de jauge décrivent particulièrement bien certaines interactions fondamentales, comme les interactions électromagnétiques et fortes, elles ne sont pas suffisantes pour expliquer la phénoménologie de la physique des particules. Elles ne décrivent par exemple pas les interactions faibles, véhiculées par des champs vectoriels massifs. Nécessitant de grouper les particules dans des représentations du groupe de jauge, elles interdisent aussi d'écrire des termes de masse pour les spineurs de Weyl décrivant la matière (cf la section~\ref{PartSpineurs}).

Ces problèmes sont résolus par le mécanisme de Brout-Englert-Higgs-Guralnik–Hagen–Kibble~\cite{PhysRevLett.13.321,PhysRevLett.13.508,PhysRevLett.13.585}, aussi appelé mécanisme de Brout-Englert-Higgs ou simplement de Higgs,
qui décrit la brisure spontanée d'une symétrie de jauge lorsqu'un champ scalaire -- appelé champ de Higgs -- acquiert une valeur moyenne dans le vide non-nulle. Après avoir explicité ce mécanisme dans le modèle de Higgs abélien, il est ensuite décrit dans un cadre général, en prenant notamment l'exemple de la brisure de la symétrie électrofaible du Modèle Standard.

\section{Modèle de Higgs abélien et brisure spontanée de symétrie}
\label{PartSSBHiggsAbelien}

\begin{figure}[h!]
\begin{center}
\includegraphics[scale=0.75]{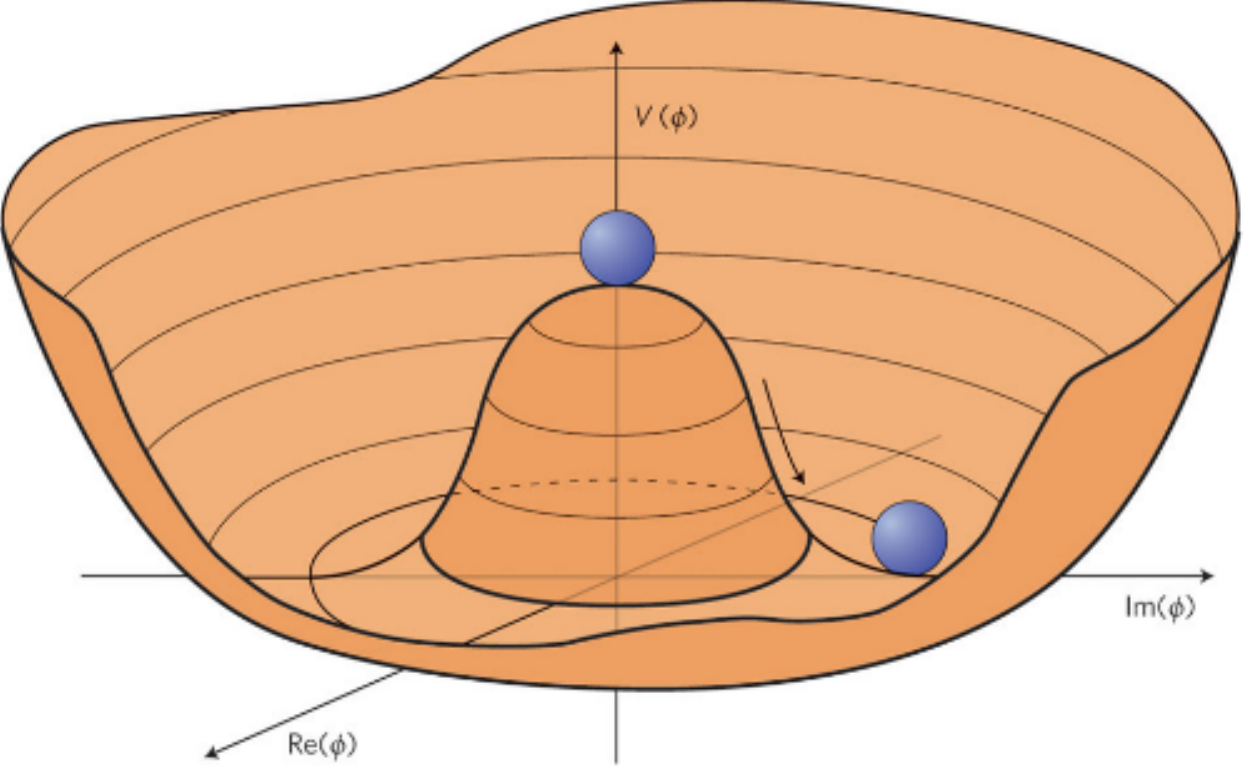}
\vspace{-0.5cm}
\end{center}
 \caption{Potentiel du champ de Higgs dans le modèle de Higgs abélien, issu de~\cite{FigureHiggs}. Les billes indiquent deux extrema du potentiel : le maximum local pour $\phi=0$, qui est instable, et un minimum global pour $|\phi|=\eta$, pour une phase arbitraire.}
 \label{HiggsPotential}
\end{figure}

\noindent
Le modèle de Higgs abélien est l'un des modèles les plus simples permettant la réalisation d'une brisure spontanée de symétrie. Nous l'utiliserons donc pour faire une première présentation du mécanisme de Higgs, avant de décrire ce mécanisme sous sa forme la plus générale dans la section suivante. Ce modèle décrit un champ scalaire complexe $\phi$, nommé champ de Higgs, chargé sous une symétrie de jauge abélienne U(1). La charge du champ scalaire sous les transformations de jauge est notée $q_\phi$. Le Lagrangien de ce modèle, présenté dans la section~\ref{TheorieJaugeScalaire} s'écrit 
\begin{equation}
\label{LagHiggsAbelienBulk}
\mathcal{L}_{\text{Higgs abélien}} = - (D_\mu \phi)^* (D^\mu \phi) - \frac14 F_{\mu\nu} F^{\mu\nu} - V(|\phi|^2),
\end{equation}
avec 
\begin{equation}
V(|\phi|^2) = \lambda(|\phi|^2 - \eta^2)^2.
\end{equation}
Afin d'avoir un mécanisme de brisure spontanée de symétrie, on prendra $\lambda>0$ et $\eta>0$. Un tel potentiel est représenté figure~\ref{HiggsPotential}. Le Lagrangien de l'équation~\eqref{LagHiggsAbelienBulk} est bien invariant sous la transformation de jauge
\begin{equation}
\label{TransfoJaugeHiggsAbelien}
\displaystyle
\left\{
\begin{array}{l}
\phi \rightarrow e^{-iq_\phi \theta(x)} \phi,\\
A_\mu(x) \rightarrow A_\mu(x) - \frac1g \partial_\mu \theta(x).
\end{array}
\right.
\end{equation}

Le mécanisme de Higgs permet d'avoir à basse énergie un champ scalaire dont l'état de vide ne vérifie pas les propriétés de symétrie du modèle complet. Cette brisure de symétrie a lieu lors du passage d'un état de haute énergie, où l'état du vide du champ correspond à sa valeur nulle, à l'état le plus stable énergétiquement, qui brise la symétrie du modèle. Le caractère dynamique de cette transition de phase, qui a lieu lorsque l'énergie caractéristique d'excitation thermique est de l'ordre de la profondeur du puits de potentiel du champ scalaire, est discuté dans la section~\ref{PartPhaseTransition}. Le Hamiltonien associé au Lagrangien de l'équation~\eqref{LagHiggsAbelienBulk} s'écrivant 
\begin{equation}
\mathcal{H}_{\text{Higgs abélien}} = \frac12 \dot\phi^2 + \frac12 (\vec\nabla \phi)^2 + V(|\phi|^2) ,
\end{equation}
les états d'énergie minimale sont stationnaires, \emph{i.e.} vérifient $\dot\phi=0$ et $\vec\nabla\phi=0$, et correspondent à un minimum du potentiel. Les minima du potentiel sont les états sur le cercle $|\phi|=\eta$. Le champ prend donc comme état du vide un point localisé sur ce cercle, impliquant une brisure de la symétrie U(1) du modèle, cette symétrie correspondant à la variation de la phase du champ. On nomme valeur moyenne dans le vide, ou {\sc vev} pour \og vacuum expectation value \fg{}, la valeur du champ définissant l'état du vide. Elle correspond ici à $\bra{0}\phi\ket{0} = \braket{\phi} = \eta e^{iq\theta_0}$, avec $\ket{0}$ l'état vide, et $\theta_0$ l'angle définissant la direction pointée par le champ scalaire. 

Pour analyser les conséquences physiques de la brisure spontanée de symétrie, il est nécessaire d'étudier les excitations fondamentales autour de la {\sc vev} non nulle du champ scalaire. En effet, même avec une telle {\sc vev}, le modèle écrit en fonction des champs $\phi$ et $A_\mu$ est toujours invariant sous les transformations de jauge de l'équation~\eqref{TransfoJaugeHiggsAbelien}. Cependant, ce n'est plus le cas si on le décrit avec les excitations des champs autour de la {\sc vev} du champ de Higgs ; et les notions de particules étant uniquement liées à ces excitations, c'est sur ces dernières qu'il faut étudier les symétries réelles du modèle. Cela explique le lien important entre l'état de vide d'un modèle physique et ses symétries (voir aussi la note de bas de page~\ref{Footnote2} page suivante). Pour étudier les excitations autour du vide, on développe le champ de Higgs autour de sa {\sc vev} en
\begin{equation}
\label{DevPhiVEV}
\phi(x) = \left[\eta + \frac{h(x)}{\sqrt{2}}\right] e^{i q_\phi \tilde{\theta}(x)},
\end{equation}
où $h$ et $\tilde\theta$ sont deux champs réels (on a précisé leur dépendance spatio-temporelle pour éviter toute confusion). Écrits en fonctions de ces champs, les termes cinétique et potentiel deviennent respectivement
\begin{equation}
\label{DevCinVEV}
( D_\mu \phi )^* (D^\mu \phi) = \frac12 (\partial_\mu h) (\partial^\mu h) + \Big(\eta + \frac{h}{\sqrt{2}} \Big)^2\Big(q_\phi \partial_\mu \tilde{\theta} - g q_\phi A_\mu \Big)\Big(q_\phi \partial^\mu \tilde{\theta} - g q_\phi A^\mu \Big),
\end{equation}
et 
\begin{equation}
\label{DevPotVEV}
V(h) = \frac12 {m_h}^2 h^2 + \sqrt{2}\lambda\eta h^3 + \frac{\lambda}{4} h^4,
\end{equation}
où l'on a introduit $m_h = 2 \sqrt{\lambda} \eta$. 

Le champ $\tilde\theta$ n'apparaît pas dans le potentiel, et correspond donc à un champ de masse nulle. Il est associé aux rotations autour du puits de potentiel, correspondant à des trajectoires de potentiel constant. Les équations~\eqref{DevPhiVEV} et~\eqref{DevCinVEV} montrent que ses variations correspondent exactement aux transformations de jauge du modèle décrites dans l'équation~\eqref{TransfoJaugeHiggsAbelien}. Cette excitation du champ scalaire ne peut donc pas être physique, puisque cela impliquerait que le degré de liberté de jauge du modèle initial le serait\footnote{\label{Footnote2}On peut légitimement s'interroger sur le fait que l'apparition du boson de Goldstone ne soit liée qu'à l'écriture du champ scalaire en coordonnées polaires comme dans l'équation~\eqref{DevPhiVEV}, la composante angulaire du champ décrivant alors les transformations de jauge. Il est cependant important de bien considérer les particules d'une théorie comme les excitations fondamentales à partir de l'état définissant son vide. Avant la brisure spontanée de symétrie, le champ de Higgs a un potentiel effectif avec une masse positive (voir la section~\ref{PartPhaseTransition}), et un minimum stable pour l'état $\phi=0$ qui définit le vide. Les excitations autour de cet état ne peuvent pas être développées en coordonnées polaires, et correspondent bien à deux particules massives. Réciproquement, il est nécessaire pour pouvoir écrire les excitations du champ scalaire en coordonnées polaires (et donc pour identifier une de ses composantes aux transformations de jauge) de les développer autour d'un vide vérifiant $|\phi|\neq0$, vide qui brise explicitement la symétrie de jauge.}. C'est un boson de Goldstone (ou de Nambu-Goldstone), qui est une réminiscence de la symétrie de jauge dans la description du modèle. Il est possible de faire une transformation de jauge pour réabsorber ce boson dans le choix de jauge.

Le Lagrangien du modèle en fonction des autres excitations autour de l'état de vide s'écrit alors
\begin{equation}
\mathcal{L}_{\text{{\sc vev}}} = -\frac12 (\partial_\mu h) (\partial^\mu h) - \frac14 F_{\mu\nu} F^{\mu\nu} - V_{\text{{\sc vev}}}(h,A_\mu),  
\end{equation}
avec
\begin{equation}
V_{\text{{\sc vev}}}(h,A_\mu) = \frac12 {m_h}^2 h^2 + \sqrt{2}\lambda\eta h^3 + \frac{\lambda}{4} h^4 + \frac12 {m_A}^2 A_\mu A^\mu + \frac12 (gq_\phi)^2 h^2 A_\mu A^\mu, 
\end{equation}
où l'on a identifié $m_A = \sqrt{2}gq_\phi \eta$. Ce Lagrangien correspond à un couplage entre un champ scalaire réel et un champ vectoriel massif. Les excitations du champ vectoriel autour du vide brisant la symétrie U(1) sont donc massives. Elles ne contiennent plus le degré de liberté lié aux transformations de jauge, mais propagent un mode longitudinal. Ce degré de liberté de jauge est porté à la place par le boson de Goldstone. Le champ vectoriel a donc absorbé un des degrés de liberté du champ scalaire. L'excitation de masse non nulle du champ de Higgs correspond aux déplacements radiaux du champ de part et d'autre du minimum de la cuvette ; elle est décrite par un champ scalaire réel, et propage bien un seul degré de liberté. C'est cette excitation particulière du champ de Higgs qui peut être observée expérimentalement, et qu'on peut identifier au boson de Higgs d'un point de vue phénoménologique.

Pour mieux comprendre l'apparition du boson de Goldstone, on peut s'interroger sur la description de la direction angulaire prise par la valeur moyenne dans le vide du champ de Higgs. Le modèle initial étant invariant sous les transformations de la symétrie U(1), il est en effet impossible de décrire une direction particulière d'un point de vue intrinsèque, tant que cette symétrie n'est pas brisée. Cette ambigüité dans la description est contenue dans le champ $\tilde\theta$ servant à décrire la direction angulaire (locale) de la {\sc vev} du champ scalaire dans l'équation~\eqref{DevPhiVEV}. Ce champ de Goldstone, lié à la symétrie initiale du modèle, quantifie donc les redéfinitions possibles de l'état de vide en chaque point. Une fois la brisure de symétrie effectuée, il est par contre possible de décrire les champs vis-à-vis de la direction de cette brisure, ce qui est le cas pour le champ $h$, aligné avec la {\sc vev} du champ de Higgs. Nous reviendrons sur ces concepts après avoir donné une description plus générale du mécanisme de Higgs.

\section{Mécanisme de Brout-Englert-Higgs}
\label{PartBEH}

\noindent
Nous décrivons à présent le mécanisme de Brout-Englert-Higgs dans le cas d'une théorie de jauge non-abélienne basée sur un groupe de symétrie simple $G$ de dimension $N$. Le cas d'une théorie de jauge basée sur un produit de groupes simples et de composantes U(1), comme détaillé dans la section~\ref{ProduitGroupesJauge}, se traite de façon similaire. Nous utilisons les notations introduites dans le complément~\ref{TheorieJaugeScalaire}, prenant un champ scalaire hermitien\footnote{Ce raisonnement peut s'appliquer ensuite à des champs complexes en les écrivant comme la somme en deux champs hermitiens.} $\Phi_a$ dans une représentation de dimension $d$ du groupe $G$. Le Lagrangien du modèle est
\begin{equation}
\label{TheorieJaugeNonAbelScalaireBulk}
\mathcal{L}_{\Phi_a}= - (D_\mu \Phi_a)^\dagger (D^\mu \Phi_a) - V(\Phi_a) -\frac14 F_{\mu\nu}^\alpha F_\alpha^{\mu\nu}.
\end{equation}
Il est invariant sous les transformations de jauge infinitésimales
\begin{equation}
\left\{
\begin{array}{l}
\delta \Phi_a (x) = -i \delta\theta_\alpha \left(T^\alpha\right)_a^b \Phi_b(x),\vspace{0.1cm}\\
\delta A_\mu^\alpha  =  f_{\beta\gamma}{}^{\alpha}\delta \theta^\beta A_\mu^\gamma - \frac1g \partial_\mu \delta \theta ^\alpha.
\end{array}
\right.
\end{equation}
Le potentiel $V(\Phi_a)$ peut contenir des termes quadratiques à quartiques en le champ scalaire formant des singlets de $G$ (voir le complément~\ref{TheorieJaugeScalaire}), et nous supposons qu'il est minimum pour des valeurs non nulles de $\Phi_a$. Cet ensemble de minima correspond souvent à une norme fixée pour le champ scalaire, en plus d'autres contraintes possibles. Dans les états de basse énergie, le champ scalaire va prendre une valeur moyenne dans le vide non nulle correspondant à un de ces minima, et briser spontanément la symétrie du modèle sous le groupe $G$. C'est le mécanisme de Higgs.

L'état du vide à basse énergie permet de déterminer la symétrie résiduelle après brisure de symétrie. On note cette la valeur moyenne dans le vide du champ
\begin{equation}
\boldsymbol v = \bra{0}\boldsymbol\Phi\ket{0} = 
\left(
\begin{array}{c}
v_1\\
\vdots \vspace{0.03cm}\\
v_d
\end{array}
 \right),
\end{equation}
avec $v_a=\bra{0}\Phi_a\ket{0}$. Elle correspond à un minimum du potentiel et doit vérifier
\begin{equation}
\label{MinPotVev}
\left.\frac{\partial V}{\partial \Phi_a}\right|_{\boldsymbol\Phi=\boldsymbol v}=0.
\end{equation}
Le groupe de symétrie résiduel après la brisure spontanée de symétrie est noté $H$, de dimension $M$. Comme les transformations de $H$ laissent invariant l'état du vide, les $M$ générateurs de ce groupe annihilent cet état~: 
\begin{equation}
\label{GenH}
(T^\alpha_{_H})^b_a \Phi_b = 0.
\end{equation}
Comme ces éléments décrivant un ensemble de transformations laissant invariant un système, ils génèrent bien un groupe (voir la discussion de la section~\ref{IntroGroupes}), qui est nécessairement un sous-groupe du groupe de symétrie initial $G$. Les $N-M$ autres générateurs de $G$ vérifient
\begin{equation}
\label{GenG-H}
(T^{\tilde{\alpha}}_{\text{br}})^b_a \Phi_b \neq 0.
\end{equation}
Ils ne décrivent plus une symétrie du modèle puisqu'ils ne laissent pas invariant son état fondamental. Par exemple, un vecteur de norme fixée brise la symétrie SO(3) d'un espace tridimensionnel. Comme il est invariant sous les rotations sur son plan transverse, mais pas sous les autres rotations de l'espace, il implique une brisure spontanée de symétrie SO(3)$\rightarrow$SO(2). Pour décrire les états physique après la brisure spontanée de symétrie, il est alors nécessaire d'étudier les excitations fondamentales autour du vide décrit par la {\sc vev} $\boldsymbol v$. On décrit ces excitations du champ scalaire par les champs
\begin{equation}
\tilde{\phi}_a = \Phi_a - v_a,
\end{equation}
qui ont une valeur moyenne dans le vide nulle.

Comme dans le cas abélien, une brisure spontanée de symétrie est associée à l'apparition de bosons de Goldstone. De façon générale, le théorème de Nambu-Goldstone~\cite{Goldstone:1961eq,Nambu:1961fr,Nambu:1961tp,Goldstone:1962es} énonce que toute brisure spontanée d'une symétrie locale implique l'apparition d'autant de bosons de Goldstone que de générateurs de la symétrie brisée. Pour identifier ces particules, on peut étudier le terme de masse des excitations du champ scalaire,
\begin{equation}
\mathcal{L}_{m_{\tilde{\phi}}}= - \frac{\mu^2_{ab}}{2}\tilde\phi^a \tilde\phi^b,
\end{equation}
où l'on a défini
\begin{equation}
\displaystyle{
\mu^2_{ab} = \left. \frac{\partial V}{\partial \tilde\phi^a \partial\tilde\phi^b}\right|_{\boldsymbol\Phi = \boldsymbol v}.
}
\end{equation}
Le potentiel étant un singlet du groupe de jauge, il doit être invariant lors d'une transformation infinitésimale de symétrie, qu'on peut paramétrer par des variables $\delta \theta_\alpha$. On a donc
\begin{equation}
\delta V (\Phi_a) = \frac{\partial V}{\partial\Phi_a}\delta \Phi_a = [-i \delta \theta_\alpha (T^\alpha)_a^b \Phi_b] \frac{\partial V}{\partial\Phi_a} = 0.
\end{equation}
Ce résultat étant vrai quel que soit $\delta \theta_\alpha$, on a 
\begin{equation}
 \frac{\partial V}{\partial\Phi_a} (T^\alpha)_a^b \Phi_b = 0.
\end{equation}
Dérivant cette expression par rapport à $\Phi_a$, et l'évaluant pour $\boldsymbol\Phi = \boldsymbol v$ en utilisant l'équation~\eqref{MinPotVev} et la définition de $\mu^2_{ab} $, on obtient
\begin{equation}
\label{SpectreMassesNulles}
\mu^2_{ab} (T^\alpha)^b_c v^c =0.
\end{equation}
Connaissant l'action des générateurs de $G$ sur la {\sc vev} définissant l'état de vide, donnée dans les équations~\eqref{GenH} et~\eqref{GenG-H}, on en déduit que $\mu^2_{ab}$ a $N-M$ vecteurs propres de valeur propre nulle. On peut montrer que ceux-ci sont linéairement indépendants~\cite{Abers:1973qs}, ce qui implique que le spectre de masse des excitations du champ de Higgs autour du vide défini par $\boldsymbol v$ contient bien $N-M$ particules de masse nulle.

Ces excitations de masse nulle, notées $\tilde\theta_{\tilde{\alpha}}(x)$ pour $\tilde{\alpha}=1,\cdots,(M-N)$, correspondent aux bosons de Goldstone apparaissant après la brisure de symétrie. Les ayant identifiés, on peut décrire l'intégralité des excitations du champ de Higgs autour de sa {\sc vev} non nulle comme~\cite{Kibble:1967sv} 
\begin{equation}
\label{DevPhiNonAbel}
\boldsymbol \Phi = 
\text{exp}\Big[-i \tilde\theta_{\tilde\alpha}(x) T^{\tilde{\alpha}}_{\text{brisé}}\Big]
\left(
\begin{array}{c}
\vspace{-0.12cm}v_1 + h_1(x)\vspace{-0.0cm}\\
\vdots \\
v_p+h_p(x)\\
\vspace{-0.12cm}0\\
\vdots \vspace{0.03cm}\\
0
\end{array}
\right),
\end{equation}
où $h_1 (x)$ à $h_p (x)$, avec $p=d-(M-N)$, correspondent aux excitations massives du champ de Higgs. Dans cette formule, généralisant l'équation \eqref{DevPhiVEV} du modèle de Higgs abélien, on retrouve que les champs associés aux bosons de Goldstone paramètrent les transformations de symétries associées aux générateurs de $G$ brisés par la {\sc vev} non nulle du champ scalaire\footnote{On peut le démontrer à partir de l'équation~\eqref{SpectreMassesNulles}, qui montre que les excitations de masse nulle du champ scalaire se font dans les directions $(T_{\text{brisé}}^{\tilde\alpha})_a^b v_b$, correspondant à une transformation infinitésimale de symétrie générée par $T_{\text{brisé}}^{\tilde\alpha}$. Ces transformations seront donc paramétrées par la valeur $\tilde\theta_{\tilde\alpha}$ des bosons de Goldstone associés.}. Ils n'ont donc pas de sens physique, puisqu'ils correspondent aux degrés de liberté de jauge des champs vectoriels avant la brisure de symétrie, et peuvent être annulés par une transformation de jauge des champs $\Phi_a$ comme dans le cas abélien. Leurs degrés de liberté sont en fait absorbés par les champs de jauge associés aux générateurs de la symétrie brisé, qui acquièrent une masse. 

Comme dans le cas abélien, on peut faire un lien entre les bosons de Goldstone et l'impossibilité de décrire de façon intrinsèque une direction pour la {\sc vev} non-nulle brisant la symétrie. Le modèle initial étant invariant sous les transformations de jauge, on peut en effet effectuer une transformation de jauge sur le champ scalaire, et donc modifier la direction de la {\sc vev}, sans modification sur les observables physiques. Ces transformations de jauge s'écrivent
\begin{equation}
\label{RotationVEV}
\boldsymbol v \rightarrow \boldsymbol v^\prime = \text{exp}\Big[-i \theta_{\alpha}(x) T^{\alpha}\Big] \boldsymbol v = \text{exp}\Big[-i \theta_{\tilde\alpha}(x) T^{\tilde{\alpha}}_{\text{brisé}}\Big] \boldsymbol v,
\end{equation}
l'écriture impliquant seulement les générateurs de la symétrie brisée étant possible car les autres générateurs laissent invariant l'état du vide. Ces transformations s'identifient bien à celles paramétrées par les bosons de Goldstone. 

Lorsqu'on effectue une transformation de jauge sur la {\sc vev} décrivant la brisure de symétrie, le groupe d'invariance $H$ de celle-ci est inchangé. Si l'on l'avait identifié en terme de générateurs de $G$, il est par contre nécessaire d'effectuer aussi une transformation sur ceux-ci pour obtenir les générateurs de $H$ après la transformation de jauge. Étant donné que les générateurs de $G$ se transforment dans la représentation adjointe de $G$, une transformation de jauge appliquée aux générateurs $T_H^\beta$ annulant la {\sc vev} $\boldsymbol v$ donne les nouveaux générateurs~:
\begin{equation}
T_{H^\prime}^\beta = \text{exp}\Big[-i \theta_{\alpha}(x) T^{\alpha}\Big] T_H^\beta \text{exp}\Big[i \theta_{\alpha}(x) T^{\alpha}\Big].
\end{equation}
Ces générateurs laissent bien invariante la {\sc vev} $\boldsymbol v^\prime$ après la transformation de jauge~:
\begin{equation}
T_{H^\prime}^\beta \boldsymbol v^\prime = \text{exp}\Big[-i \theta_{\alpha}(x) T^{\alpha}\Big] T_H^\beta \text{exp}\Big[i \theta_{\alpha}(x) T^{\alpha}\Big] \text{exp}\Big[-i \theta_{\alpha}(x) T^{\alpha}\Big] \boldsymbol v  = \text{exp}\Big[-i \theta_{\alpha}(x) T^{\alpha}\Big] T_H^\beta \boldsymbol v = 0.
\end{equation}
Les groupes $H$ et $H^\prime$ ont nécessairement la même structure, puisque les transformations de jauge ne changent pas les propriétés physiques du système\footnote{Par exemple, la brisure de symétrie SO(3)$\rightarrow$SO(2) par un vecteur avait été décrite de façon intrinsèque après l'équation $\eqref{GenG-H}$. Supposons à présent que la {\sc vev} non nulle du vecteur soit dirigée sur l'axe $z$, invariante sous les rotations générées par $T_z$. Il est possible de faire une rotation initiale du système pour que la {\sc vev} soit dirigée sur $z^\prime = R(\theta_1,\theta_2,\theta_3) z$ (où les $\theta_i$ désignent par exemple les angles d'Euler décrivant la rotation). Elle ne sera alors plus invariante sous les rotations générées par $T_z$, mais sous celles générées par $T_{z^\prime} = R(\theta_1,\theta_2,\theta_3) T_z R^{-1}(\theta_1,\theta_2,\theta_3)$.}. On retrouve donc finalement que $H$ doit être un sous-groupe invariant de $G$, ce qui permet de décrire la symétrie brisée par le groupe $G/H$. Ce résultat montre bien que quelle que soit la façon dont on décrit la {\sc vev} du champ de Higgs, à savoir à des transformations de jauge près, la brisure de symétrie qui aura lieu sera toujours la même.

On peut par contre définir des grandeurs par rapport à la direction de la {\sc vev} après brisure de symétrie, même si la direction de celle-ci n'a pas de sens intrinsèque. C'est le cas des champs $h_i(x)$ dans l'équation~\eqref{DevPhiNonAbel}, toujours alignés dans la direction de la {\sc vev}. Les transformations de jauge ne changent en effet que la description des grandeurs physiques, mais pas les relations de structure entre elles. On l'a vu précédemment pour les générateurs du groupe $H$, et de la même façon, s'il n'est pas possible de définir intrinsèquement une direction particulière dans un espace invariant sous rotation (par exemple), on peut définir des orientations relativement à cette première direction.

\section{Détermination du groupe de symétrie résiduel, brisure électrofaible}
\label{RepChampsHiggs}

\noindent
Dans l'explication du mécanisme de Higgs pour les théories de jauge non-abéliennes dans la section précédente, nous n'avons pas précisé comment identifier en pratique le groupe d'invariance. Une fois les contraintes sur la {\sc vev} non nulle du champ de Higgs obtenues à partir du potentiel, une méthode systématique consiste à faire une transformation de jauge sur cette {\sc vev} pour l'exprimer sous une forme simple, puis à évaluer quels générateurs annulent cette {\sc vev} et ne sont pas brisés, formant le groupe de symétrie résiduel. Il faut aussi vérifier que ces générateurs forment bien une algèbre, et ne génèrent pas de nouveaux termes dans leurs relations de commutation. Ces considérations sont faites ici à l'aide des opérateurs liés aux racines, mais une discussion similaire est possible avec les opérateurs hermitiens liés aux constantes de structure.

L'identification des générateurs annulant la {\sc vev} est aisée pour les éléments de la sous-algèbre de Cartan. Il suffit en effet de chercher pour quelles combinaisons des générateurs $T_3^{(i)}$ la {\sc vev} n'est pas chargée. Cette donnée permet de connaitre le rang de la symétrie résiduelle, ce qui implique notamment que les représentations ne contenant pas de poids identiquement nul, comme les représentations complexes, diminuent forcément le rang de l'algèbre d'au moins une unité. De plus, cela permet d'éliminer tous les générateurs liés aux racines chargées sous les $T_3^{(i)}$ de la symétrie brisée, car ils feraient apparaître ces mêmes $T_3^{(i)}$ dans leurs relations de commutations entre racines opposées, voir l'équation~\eqref{EqComRacinesOpposées}, et ne permettraient pas d'écrire une algèbre avec des relations fermées. Les symétries résiduelles sont donc décrites par les  combinaisons de racines qui sont non chargées sous les $T_3^{(i)}$ annulant la {\sc vev}\footnote{Ce sont leur angles et leurs rapports de longueurs qui vont caractériser le groupe de symétrie résiduel, ce qui revient au procédé utilisé lors de la classification des différentes algèbres simples dans la section~\ref{CartanClassification}.}. 

Prenons l'exemple de la brisure de symétrie électrofaible SU(2)$_L\times$U(1)$_Y$ en la théorie abélienne décrivant l'électromagnétisme basée sur une théorie de jauge U(1)$_Q$. Le champ de Higgs $\Phi$ est dans une représentation complexe de dimension 2 de SU(2)$_L$, et porte une hypercharge\footnote{Nous reviendrons sur cette définition de l'hypercharge dans la section~\ref{PartJaugesSM}.} $y_\Phi=\frac12$. On le représente communément sous la forme
\begin{equation}
\label{DefPhiSM}
\Phi =
\left(
\begin{array}{l}
\phi^+\\
\phi^0
\end{array}
\right).
\end{equation}
Les deux composantes portent la même hypercharge $y_\Phi=\frac12$ [puisqu'ils dans la même représentation de SU(2)$_L$], et les charges des champs sous $T_3^L$ de SU(2)$_L$ sont $-\frac12$ pour $\phi^0$ et $+\frac12$ pour $\phi^+$. Le terme potentiel quartique le plus général pour ce champ de Higgs est de la forme
\begin{equation}
V(\Phi) = \lambda\left[(\Phi^\dagger \Phi) - \eta^2\right]^2,
\end{equation}
similaire au cas abélien discuté dans la section~\ref{TheorieJaugeScalaire}

À basse énergie, les coefficients $\lambda$ et $\eta$ sont strictement positifs, et le champ de Higgs prend une {\sc vev} non nulle vérifiant $(\Phi^\dagger \Phi) = \eta^2$. Effectuant une transformation de jauge SU(2)$_L$ puis U(1)$_Y$, il est alors possible d'écrire cette {\sc vev} sous la forme
\begin{equation}
\label{VevHiggsSM}
\boldsymbol v = 
\left(
\begin{array}{l}
0\\
v_0
\end{array}
\right),
\end{equation}
où $v_0$ est réel. Cette {\sc vev} porte une charge sous $T_3^L$ de SU(2)$_L$ et $T_3^Y$ de U(1)$_Y$, et brise donc ces deux groupes [les générateurs d'échelle associés à SU(2)$_L$ sont brisés puisque chargés sous $T_3^L$, ou de manière équivalente puisque $T^+$ n'annule pas la {\sc vev}]. Cette {\sc vev} ne brise cependant pas le groupe U(1)$_Q$ = U(1)$_L$+U(1)$_Y$, où U(1)$_L$ est le sous-groupe abélien généré par $T_3^L$. Pour voir cela, on peut écrire ces transformations de jauge de U(1)$_L$ et U(1)$_Y$, de paramètres respectifs $\theta_L$ et $\theta_Y$, sur $v_0$
\begin{equation}
\label{TransfoVevSM}
v_0 \rightarrow v_0^\prime = e^{-i(-\frac{1}{2} )\theta_L} e^{-i (+\frac{1}{2} ) \theta_Y} v_0.
\end{equation}
Il est alors clair que le U(1)$_Q$ de générateur
\begin{equation}
\label{ChargeQSU2}
T_3^Q = T_3^L + T_3^Y
\end{equation}
laisse invariant cette {\sc vev}, puisque la charge de celle-ci pour ce générateur est nulle : les transformations associées correspondent à celles décrites dans l'équation~\eqref{TransfoVevSM} avec $\theta_L = \theta_Y$. Ce champ de Higgs cause donc la brisure de symétrie électrofaible
\begin{equation}
\text{SU(2)}_L \times \text{U(1)}_Y \overset{\braket{\Phi}}{\relbar\joinrel\relbar\joinrel\longrightarrow} \text{U(1)}_Q.
\end{equation}
À titre d'autre exemple spécifique, la description de la brisure d'une symétrie SU(3) par des champs de Higgs dans des représentations de dimension 3 et 8 est donnée dans la section~\ref{BrisureSU3}.

Il n'est pas nécessaire d'étudier en détails les générateurs brisés par un champ de Higgs pour obtenir la symétrie résiduelle après brisure spontanée de symétrie. Dans le cas de la brisure électrofaible qu'on vient d'étudier, la condition $(\Phi^\dagger \Phi) = \eta^2$ est par exemple suffisante pour déterminer la brisure de symétrie. Dans le cas général, la donnée de la représentation du champ de Higgs causant la brisure spontanée de symétrie permet de lister les différentes schémas de brisure possibles. Deux propriétés sont importantes pour cette description. Pour commencer, un champ de Higgs dans une représentation non triviale d'un groupe et qui acquiert une {\sc vev} non nulle brisera toujours ce groupe. Dans le cas contraire, les poids associés à cette {\sc vev} ne seraient en effet reliés à aucun autre poids par une racine non-nulle, ce qui définit la représentation triviale. À l'opposé, un champ de Higgs dans la représentation triviale d'un groupe ne peut pas briser celui-ci, puisque les générateurs du groupe l'annulent toujours, par définition.

Une méthode systématique pour obtenir simplement les schémas de brisure d'un groupe $G$ brisé par un champ de Higgs dans une représentation donnée consiste donc à étudier les règles de branchement de ce champ sur les sous-groupes $H=H_1\times \cdots \times H_p$ de $G$, et d'identifier pour lesquels de ces sous-groupes le champ de Higgs peut se transformer comme un singlet. Prenons l'exemple de la brisure de SU(3) par des champ de Higgs dans des représentations de dimension 3 et 8, présenté dans la section~\ref{BrisureSU3}. On peut écrire les règles de branchement de ces deux champs sur SU(2)$\times$U(1), notant en gras la représentation des dimensions, et entre parenthèses la charge sous la symétrie U(1) (voir la section~\ref{ProduitGroupesJauge}) :
\begin{align}
\text{SU(3)} &\supset \text{SU(2)}\times\text{U(1)} \nonumber\\
\mathbf{3} &= \mathbf{1}(-2)+\mathbf{2}(1),\\
\mathbf{8} &= \mathbf{1}(0)+\mathbf{2}(3) + \mathbf{2}(-3)+\mathbf{3}(0).
\end{align}
Ces règles de branchement peuvent se visualiser sur l'espace des poids, comme indiqué en~\ref{BrisureSU3}. On y lit que la représentation de dimension 3 peut briser SU(3) et SU(2) mais pas en SU(2)$\times$U(1), le singlet sous SU(2) étant chargé sous U(1). De même, on voit que la brisure de SU(3) en SU(2) par la représentation de dimension 8 laisse aussi le U(1) non brisé. Cette méthode est particulièrement utile pour les groupes de symétrie de grande taille, leur différentes règles de branchement étant classifiées~\cite{Slansky:1981yr}, où pouvant être obtenues par ordinateur~\cite{Feger:2012bs}. Dans les décompositions des représentations d'un groupe $G$ sur un sous-groupe $H$, les racines non-nulles contenues dans $G$ mais pas dans $H$ peuvent relier différentes représentations restreintes. L'action de ces racines décrit des transformations de jauge du groupe de symétrie brisé, et contient notamment les transformations liées aux bosons de Goldstone. 


\section{Masse des particules, application à la brisure électrofaible}
\label{PartMasseFermionsHiggs}

\noindent
On a vu dans le cas abélien qu'après une brisure spontanée de symétrie par le mécanisme de Higgs, les excitations du champ vectoriel autour de la {\sc vev} non nulle du champ de Higgs étaient massives. Le degré de liberté supplémentaire par rapport au cas d'un champ de jauge sans masse correspondait alors à celui du boson de Goldstone du champ de Higgs, qui s'identifiait au paramètre décrivant les transformations de jauge avant la brisure de symétrie. On retrouve cette propriété dans le cas d'une théorie de jauge non-abélienne : les champs vecteurs acquérant une masse sont ceux correspondant aux générateurs de la symétrie brisée, et les degrés de liberté supplémentaires sont récupérés sur les excitations du champ de Higgs correspondant à des bosons de Goldstone.

Le terme de masse des champs vecteurs après la brisure de symétrie provient du terme cinétique du champ scalaire, qui s'écrit
\begin{equation}
\mathcal{L}_{\text{Cin.},\Phi_a} =  - (D_\mu \Phi_a)^\dagger (D^\mu \Phi_a) 
= \left[\partial^\mu \delta_b^a + i g {A^\mu_\alpha}^\dagger ({T^\alpha}^\dagger)_b^a\right]\Phi^b  \left[\partial^\mu \delta_a^c -i g A^\mu_\beta (T^\beta)_a^c\right]\Phi_c,
\end{equation}
pour un champ scalaire hermitien. Cette formule est écrite pour des champs de jauge et les générateurs associés qui peuvent être complexes, comme c'est le cas pour les champs de jauge associés aux racines non nulles dans la formulation liée aux racines et aux poids. Les générateurs et les champs de jauge pouvant être mis dans une base hermitienne, cette formule est généralement écrite dans la littérature en supposant ces propriétés. Après développement autour de la {\sc vev} non nulle du champ de Higgs donnée par l'équation~\eqref{DevPhiNonAbel}, les termes de masse des champs vectoriels correspondent à
\begin{equation}
\mathcal{L}_{m_A} = (\mu_A^2)^{\alpha\beta} A^\mu_\alpha A_{\mu\beta} =  g^2 (A^\mu_\alpha T^\alpha \boldsymbol v)^\dagger ( A_{\mu\beta} T^\beta \boldsymbol v) .
\end{equation}
Utilisant les équations~\eqref{GenH} et~\eqref{GenG-H}, on obtient alors que les champs vectoriels associés aux générateurs de la symétrie résiduelle restent sans masse, alors que les ceux qui sont associés aux symétries brisées sont massifs, comme attendu.

Ces formules permettent d'identifier quels bosons de jauge acquièrent une masse lors de la brisure de symétrie électrofaible. L'action des différents générateurs de SU(2)$_L\times$U(1)$_Y$ sur la {\sc vev} non nulle du champ de Higgs est
\begin{equation}
\displaystyle{
T^L_+ \left( \begin{array}{c} \hspace{-0.15cm}0\\ \hspace{-0.15cm} v_0  \end{array} \hspace{-0.15cm}\right) =\left( \begin{array}{c} \hspace{-0.15cm}v_0\\ \hspace{-0.15cm} 0  \end{array} \hspace{-0.15cm}\right),
\hspace{0.5cm}
T^L_- \left( \begin{array}{c} \hspace{-0.15cm}0\\ \hspace{-0.15cm} v_0  \end{array} \hspace{-0.15cm}\right)=\left( \begin{array}{c} \hspace{-0.15cm}0\\ \hspace{-0.15cm} 0  \end{array} \hspace{-0.15cm}\right),
\hspace{0.5cm}
T_3^L \left( \begin{array}{c} \hspace{-0.15cm}0\\ \hspace{-0.15cm} v_0  \end{array} \hspace{-0.15cm}\right)= -\frac12 \left( \begin{array}{c} \hspace{-0.15cm}0\\ \hspace{-0.15cm} v_0  \end{array} \hspace{-0.15cm}\right),
\hspace{0.5cm}
T_3^Y \left( \begin{array}{c} \hspace{-0.15cm}0\\ \hspace{-0.15cm} v_0  \end{array} \hspace{-0.15cm}\right)= +\frac12 \left( \begin{array}{c} \hspace{-0.15cm}0\\ \hspace{-0.15cm} v_0  \end{array} \hspace{-0.15cm}\right),
}
\end{equation}
On obtient donc un terme de masse
\begin{equation}
\displaystyle{
\mathcal{L}_{m_W,m_Z} = g_L^2 v_0^2 (W^{+\mu})^\dagger W^+_\mu +\frac14 g_L^2 v_0^2 W^{0\mu}W_\mu^0 +\frac14 g_Y^2 v_0^2 A_Y^\mu A_{Y\mu} = M^2_W W^{-\mu} W^+_\mu + \frac12 M^2_Z Z^\mu Z_\mu, 
}
\end{equation}
où l'on a identifié le boson de jauge 
\begin{equation}
Z_\mu =  \frac{g_Y A_{Y\mu} - g_L W^0_\mu}{\sqrt{g_L^2 + g_y^2}},
\end{equation}
associé au générateur 
\begin{equation}
T^Z_3 = -T_3^L + T_3^Y,
\end{equation}
et les masses
\begin{equation}
\displaystyle{
\left\{
\begin{array}{l}
M_W^2 = g_L^2 v_0^2,\vspace{0.2cm}\\
M_Z^2 = (g_L^2 + g_Y^2)\frac{v_0^2}{2}.
\end{array}
\right.
}
\end{equation}
Finalement, on identifie le bosons de jauge restant sans masse après la brisure de symétrie comme
\begin{equation}
A_{Q\mu} = \frac{g_Y A_{Y\mu} + g_L W^0_\mu}{\sqrt{g_L^2 + g_y^2}}.
\end{equation}
Il est associé au générateur $T_3^Q$ de l'électromagnétisme défini dans l'équation~\eqref{ChargeQSU2}, et correspond donc au photon.

Le mécanisme de brisure de symétrie est communément décrit par l'angle de Weinberg, ou angle de la brisure électrofaible, défini par
\begin{equation}
\label{AngleWeinberg}
\tan\theta_{W} = \frac{g_Y}{g_L}.
\end{equation}
Cet angle permet de quantifier la contribution relative des interactions liées à SU(2)$_L$ et à U(1)$_Y$ dans les interactions électromagnétiques résiduelles à basse énergie. Utilisant la définition de l'angle de Weinberg, on a en effet 
\begin{equation}
A_{Q\mu} = \cos \theta_W A_{Y\mu} + \sin \theta_W W^0_\mu.
\end{equation}
Expérimentalement, cet angle peut être obtenu en comparant les masses des bosons $Z$ et $W$, puisque 
\begin{equation}
M_Z^2 = \frac{M_W^2}{2 \cos^2 \theta_W.}
\end{equation}
L'angle de Weinberg a une valeur mesurée de $\sin^2 \theta_W \simeq 0.23$~\cite{PDG2016}.

Une autre conséquence de la {\sc vev} non nulle du champ de Higgs est l'apparition de termes de masse pour les fermions après la brisure de symétrie. Ces termes proviennent des couplages de Yukawa $\psi\Phi\psi$ dans le Lagrangien, reliant les spineurs et le champ de Higgs, après développement du champ de Higgs autour de sa {\sc vev} non nulle. La principale restriction de ce type de couplage est de former un singlet. Comme la {\sc vev} non nulle du champ de Higgs n'est chargée que sous les symétries brisées, elle ne donnera des masses qu'aux fermions interagissant avec les symétries brisées, sauf cas d'espèce\footnote{Il faudrait alors qu'un des champs spinoriels $\psi_a$ soit dans la représentation conjuguée du champ de Higgs sous le secteur brisé, et que le deuxième champ spinoriel $\psi_b$, non chargé sous la symétrie brisée soit dans la représentation conjuguée de $\psi_a$ pour la symétrie résiduelle.}. Ainsi, dans le modèle standard, les particules acquérant une masse après la brisure de symétrie électrofaible sont chargées sous SU(2)$_L\times$U(1)$_Y$. Il n'y a aucune raison \emph{a priori} pour que les couplages de Yukawa n'impliquent que des spineurs de chiralités différentes, même si cela a lieu pour le Modèle Standard.

Prenons l'exemple de la brisure de symétrie du Modèle Standard, détaillée dans la section précédente. La relation de l'équation~\eqref{ChargeQSU2} permet d'obtenir \emph{a posteriori} les hypercharges des différents spineurs, connaissant les charges électriques des particules qu'ils contiennent. Ainsi, le doublet $\psi^L_{\text{leptons}}$ porte une hypercharge $y^L_{\text{leptons}}=-\frac12$ et le singlet $e^R$ a hypercharge de $y_e^R = -1$. Le champ de Higgs du Modèle Standard étant un doublet de SU(2)$_L$ et portant une hypercharge $y_\Phi = \frac12$, il est possible d'écrire un couplage de Yukawa
\begin{equation}
\mathcal{L}_{\text{Yukawa},e} = \lambda_e {(\psi^L_{\text{leptons}})^\dagger}^a \Phi_a e^R + \text{h.c.} = \lambda_e \left[(\nu_e^L)^\dagger \phi^+ + (e^L)^\dagger \phi_0\right] e^R + \lambda_e (e^R)^\dagger\left[(\phi^+)^* \nu_e^L + (\phi^0)^* e^L \right],
\end{equation}
où l'on a utilisé les notations définies dans les équations~\eqref{DoubletsChirL_SM} et~\eqref{DefPhiSM}, et où $\lambda_e$ est une constante de couplage qu'on peut prendre réelle. Après la brisure spontanée de symétrie, développant ce couplage autour de la {\sc vev} non nulle du champ de Higgs donnée dans l'équation~\eqref{VevHiggsSM}, on obtient
\begin{equation}
\mathcal{L}_{m_e} = \lambda_e v \left[ (e^L)^\dagger e^R + (e^R)^\dagger e^L \right].
\end{equation}
Ce Lagrangien décrit bien un terme de masse de Dirac couplant les électrons de chiralités gauche et droite, pour une masse $m_e =  \lambda_e v$. La {\sc vev} non nulle du champ de Higgs du Modèle Standard permet d'écrire similairement un terme de masse pour les quarks up et down.

\section{Phénoménologie d'une symétrie brisée}
\label{PhenomSymBrisées}

\noindent
Les interactions liées à des symétries qui ont été brisées ne peuvent plus être décrites comme dans le chapitre~\ref{PartTheoriesJauge}. En effet, les bosons de jauge qui ont acquis une masse ne se propagent plus à basse énergie, se désintégrant en des particules plus légères. On peut considérer que les champs de jauge ayant acquis une masse ne propagent plus l'état local de symétrie $\theta^\alpha$ pour les transformations qui leurs sont associées, ce qui est en accord avec l'identification faite précédemment entre les générateurs de la symétrie brisée et les bosons vectoriels devenus massifs. 

\begin{figure}[h!]
\begin{center}
\includegraphics[scale=1.2]{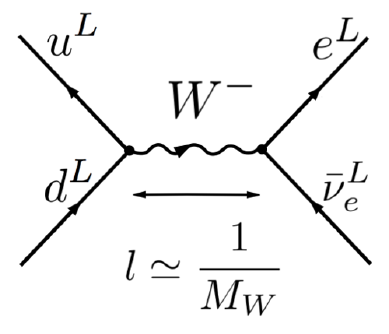}
\vspace{-0.5cm}
\end{center}
 \caption{Exemple de transformation liée aux interactions faibles, après la brisure de symétrie du Modèle Standard. Le boson $W$ virtuel ne se propage que sur une courte distance, et se désintègre ensuite en émettant un électron et un anti-neutrino.}
 \label{VertexWbrise}
\end{figure}

À des échelles d'énergie petites devant celle de brisure de symétrie (et donc devant la masse des bosons de jauge), les fermions ne sont pas assez énergétiques pour pouvoir émettre un boson de jauge massif lors d'une transformation de symétrie. Il y a cependant la possibilité d'émettre un boson de jauge virtuel se propageant pendant un temps très court, qui va ensuite se désintégrer. Le temps de vie caractéristique de ce boson de jauge, comme sa distance de propagation, est alors inversement proportionnel à sa masse. Pour les interactions faibles, avec des masses de l'ordre de 100 MeV pour les bosons $Z$ et $W$, on obtient une distance caractéristique de propagation de l'ordre de 
\begin{equation}
l_{W,Z}\simeq \frac{1}{M_{W,Z}} \simeq 10^{-14}m,
\end{equation}
et des temps de vie de l'ordre de 
\begin{equation}
\tau_{W,Z} \simeq \frac{1}{M_{W,Z}}  \simeq 10^{-22}s.
\end{equation}
Un exemple de telle transformation de symétrie est donné dans la figure~\ref{VertexWbrise}.

\begin{figure}[h!]
\begin{center}
\includegraphics[scale=0.65]{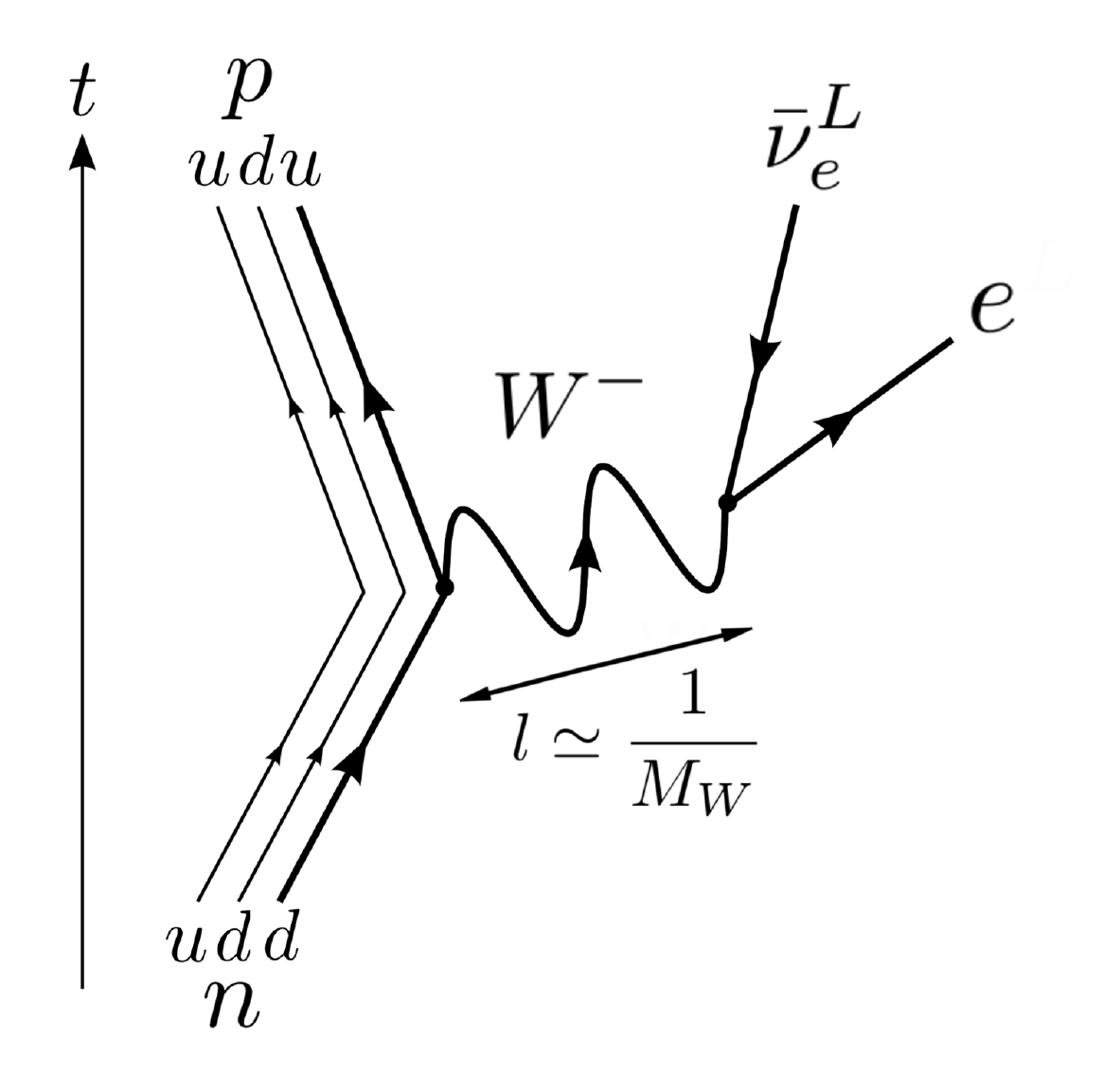}
\vspace{-0.5cm}
\end{center}
 \caption{Description de la désintégration $\beta^-$ d'un neutron en proton, par l'émission d'un boson virtuel $W$ qui se désintègre en un électron et un anti-neutrino (modification d'une figure issue de wikipédia). Les chiralités des quarks et de l'électron n'ont pas été indiquées, ces particules ayant des termes de masse après brisure de symétrie causant une oscillation entre les deux états de chiralité lors de leur propagation.}
 \label{NeutronDecayW}
\end{figure}

Pour des échelles d'énergie petites devant la masse des bosons de jauge (et donc de longueur grande devant sa distance caractéristique de propagation), ce type de transformation peut être décrit par un vertex impliquant 4 fermions. Au niveau du Lagrangien, ces couplages sont décrits par des termes effectifs de la forme
\begin{equation}
\mathcal{L}_{\text{eff}} = G_F \left[ (d^L)^c \gamma_\mu u^L \right]\left[ (\nu_e^L)^c \gamma^\mu e^L\right]+\text{h.c.},
\end{equation} 
où l'on a pris l'exemple du vertex effectif associé au processus décrit dans la figure~\ref{VertexWbrise}.
Ces vertex, liés aux interactions effectives dites de Fermi, ont été une des premières manières de décrire les interactions faibles~\cite{Fermi:1933jpa}. Sans spécifier un vertex effectif en particulier, le couplage $G_F$ apparaissant pour les interactions faibles est de la forme  
\begin{equation}
G_F \simeq \frac{g^2}{M^2_{W,Z}},
\end{equation}
où $g$ est la constante de couplage de la symétrie de jauge. Pour des échelles d'énergie $\mu$, ce terme implique des couplages proportionnels à $(\mu/M_{W,Z})^2$, qui sont effectivement négligeables dès lors que $\mu\ll M_{W,Z}$. Pour des échelles d'énergie grandes devant $M_{W,Z}$, la description par un vertex à quatre fermions n'est plus valable, car les interactions électrofaibles ne sont pas brisées, et les bosons associés n'ont pas de masse. Les vertex à 4 fermions étant non-renormalisables, ces résultats sont en accord avec la discussion de la section~\ref{TheorieFonda&Effect} sur l'apparition possible de termes non renormalisables à basse énergie comme résidus de théories renormalisables à haute énergie. 

Les interactions liées aux symétries brisées ne disparaissent donc pas à basse énergie, mais deviennent rapidement négligeables par rapport aux interactions associées à des symétries non brisées. La façon la plus simple d'en observer la phénoménologie est donc d'étudier des réactions qui ne sont possibles que par les interactions associées aux symétries brisées. Pour les interactions faibles, c'est par exemple le cas de la désintégration $\beta^{-}$, représentée figure~\ref{NeutronDecayW}, et qui correspond à une transformation associée au boson $W$. Un ordre de grandeur du temps de vie associé à la désintégration d'une particule non-relativiste de masse $m$ est alors
\begin{equation}
\tau_m^{-1} \simeq G_F^2 m^5 \simeq \frac{g^4  m^5}{M_{W,Z}^4}.
\end{equation}
On retrouve bien que de tels processus sont supprimés en loi de puissance pour des énergies petites devant celle de la brisure de symétrie.

\section{Conclusion}

\noindent
Les discussions de cette partie sur les théories de jauge et le mécanisme de Higgs de brisure spontanée de symétrie ont été menées indépendamment des symétries et des champs qu'ils décrivent. Construire des modèles physiques pour décrire la phénoménologie expérimentale revient alors à choisir des symétries de jauge, et à placer les particules de ces modèles dans des représentations de ces symétries. Cette construction est discutée pour le Modèle Standard et les théories de grande unification dans la partie~\ref{ChapterSMAndBeyond}. Par ailleurs, la brisure spontanée d'une symétrie locale par un champ de Higgs peut mener à la formation de structures stables par les configurations de vide du champ de Higgs, des défauts topologiques. La formation de tels défauts est discutée dans la partie~\ref{ChapterRealisticStrings}.

\chapter{Annexes à la deuxième partie}

\section{Représentations et transformations de symétries des champs}
\label{DefTransfoGroupes}

\noindent
On donne ici un bref récapitulatif de la représentation des groupes de Lie appliquée à la théorie des champs. Cela vient en complément de l'annexe~\ref{ChapterTdG} sur la théorie des groupes, et permet de fixer les notations. On considère un groupe de Lie $G$ de dimension $N$ ayant une algèbre de Lie avec $N$ générateurs $t_\alpha$ vérifiant 
\begin{equation}
\left[t_\alpha,t_\beta\right] = i f_{\alpha\beta}{}^\gamma t_\gamma,
\end{equation}
avec $f_{\alpha\beta}{}^\gamma$ les constantes de structure du groupe. Une représentation du groupe se fait sur un espace de dimension $d$ décrit par un champ $\Phi_a$ avec $a\in\left[1,\cdots,d\right]$. Ce champ peut être scalaire, spinoriel, vectoriel, ou dans toute autre représentation vis-à-vis des transformations de Lorentz, et on laisse les indices associés implicites. On se restreint pour commencer aux symétries internes, n'ayant pas d'effets sur les coordonnées de l'espace-temps. 

L'action du groupe sur le champ est décrite à partir de ses $N$ générateurs $\left(T_\alpha\right)_a^b$, associés aux éléments de l'algèbre de Lie $t_\alpha$. Ils génèrent les transformations infinitésimales 
\begin{equation}
\left[t_\alpha,\Phi_a \right] = - \left(T_\alpha\right)_a^b \Phi_b,
\end{equation}
et vérifient la même algèbre que les $t_\alpha$,
\begin{equation}
\left[T_\alpha,T_\beta\right] = i f_{\alpha\beta}{}^\gamma T_\gamma.
\end{equation}
La représentation du groupe sur le champ s'obtient en exponentiant ces générateurs. Les transformation finies sont décrites par $N$ paramètres réels $\theta_\alpha$, tandis que les transformations infinitésimales sont décrites par des paramètres $\delta\theta_\alpha$. Selon que ces paramètres sont ou non des fonctions de l'espace-temps, la symétrie décrite est locale ou globale. L'action des éléments du groupe associés à ces paramètres donne
\begin{equation}
\label{TransfoJaugeScalaire}
\Phi_a(x) \rightarrow \Phi^\prime_a(x) = (e^{-i \theta_\alpha T^\alpha})_a^b \Phi_b(x),
\end{equation}
pour une transformation finie, et
\begin{equation}
\label{TransfoJaugeInfScalaire}
\delta \Phi_a (x) = -i \delta\theta_\alpha \left(T^\alpha\right)_a^b \Phi_b(x),
\end{equation}
pour une transformation infinitésimale.

Sous l'action d'un groupe abélien U(1), toutes les représentations irréductibles sont de dimension 1, et aucun indice interne n'est nécessaire pour les décrire. Celles-ci sont seulement décrites par un réel associé à la charge du champ sous cette transformation, notée $q_\phi$ pour un champ $\phi$. L'action du groupe se réduit alors à 
\begin{equation}
\label{AbTransfoJaugeScalaire}
\phi(x) \rightarrow \phi^\prime_a(x) = e^{-i q_\phi \theta} \phi(x),
\end{equation}
pour une transformation finie, et 
\begin{equation}
\label{AbTransfoJaugeInfScalaire}
\delta \phi (x) = -i q_\phi \delta \theta \phi(x),
\end{equation}
pour une transformation infinitésimale. La charge $q_\phi$, qui peut être absorbée dans une redéfinition de $\theta$ pour un champ donné, est nécessaire dès que l'on considère l'action du groupe sur  des champs différents, qu'il faut \emph{a priori} décrire avec des charges distinctes.

Lorsqu'on considère une symétrie liée aux transformations de l'espace-temps, comme les translations ou les transformations de Lorentz, les formules sont un peu modifiées. L'action du groupe peut avoir lieu notamment sur les indices liés au spin des particules : les indices spinoriels, quadrivectoriels, etc. D'autre part, la transformation a aussi lieu sur les coordonnées repérant l'espace-temps, donnant des formules de la forme
\begin{equation}
\Phi_a(x) \rightarrow \Phi^\prime_a(x^\prime) = (e^{-i \theta_\alpha T^\alpha})_a^b \Phi_b(x^\prime),
\end{equation}
où
\begin{equation}
x_\mu \rightarrow x^\prime_\mu = (e^{-i \theta_\alpha L^\alpha})_\mu^\nu x_\nu,
\end{equation}
avec $L^\alpha$ les représentants de l'algèbre de Lie dans la représentation associés aux quadrivecteurs coordonnées.

\section{Théorie de jauge pour un champ scalaire, modèle de Higgs abélien}
\label{TheorieJaugeScalaire}

\noindent
Nous décrivons dans cette section les théories de jauge associées à un champ scalaire, pour un groupe de Lie simple. Nous utilisons pour cela les notations introduites dans la section~\ref{DefTransfoGroupes} qui définit les différentes lois de transformation pour un champ scalaire dans une représentation d'un tel groupe. Les résultats sont obtenus en suivant le raisonnement effectué dans la section~\ref{PartJaugesNonAbeliennes} pour construire une théorie de jauge pour un spineur de Weyl, et en utilisant les résultats sur les champs de jauge donnés dans cette section.

Le Lagrangien d'un champ scalaire\footnote{On a pris ici un champ scalaire complexe. Celui-ci peut cependant aussi être réel. Dans ce cas-là, le Lagrangien associé s'écrira simplement\begin{equation}
\mathcal{L}_{\Phi,\text{réel}}= -\frac12 \partial_\mu \Phi \partial^\mu \Phi - V(\Phi).
\end{equation}
} s'écrit
\begin{equation}
\mathcal{L}_{\Phi}= - (\partial_\mu \Phi)^\dagger (\partial^\mu \Phi) - V(\Phi),
\end{equation}
avec $V(\Phi)\in \mathbb{R}$, la théorie étant renormalisable tant que le potentiel n'inclut pas des termes de degré plus élevé que quartique en le champ scalaire. Considérant à présent des transformations de jauge pour le champ scalaire de la forme donnée dans l'équation~\eqref{TransfoJaugeScalaire}  [ou l'équation~\eqref{TransfoJaugeInfScalaire} pour leur formulation infinitésimale], la théorie de jauge associée est
\begin{equation}
\label{TheorieJaugeNonAbelScalaire}
\mathcal{L}_{\Phi_a}= - (D_\mu \Phi_a)^\dagger (D^\mu \Phi_a) - V(\Phi_a) -\frac14 F_{\mu\nu}^\alpha F_\alpha^{\mu\nu}.
\end{equation}
Dans ce Lagrangien, les dérivées covariantes de jauge sont définies par
\begin{equation}
D^\mu \Phi_a = \left[\partial^\mu \delta_a^b -i g A^\mu_\alpha (T^\alpha)_a^b\right]\Phi_b,
\end{equation}
et sont dans la même représentation que $\Phi_a$, en considérant les transformations~\eqref{ANonAbel} pour le champ de jauge [ou l'équation~\eqref{ANonAbelInf} pour leur formulation infinitésimale]. Pour le potentiel $V(\Phi_a)$, on ne considère que des contractions de $\Phi_a$ formant des singlets du groupe de jauge, et qui sont donc invariantes sous les transformations de jauge. Le Lagrangien de l'équation~\eqref{TheorieJaugeNonAbelScalaire} est invariant sous les transformations de jauge infinitésimales
\begin{equation}
\displaystyle{
\left\{
\begin{array}{l}
\delta \Phi_a (x) = -i \delta\theta_\alpha \left(T^\alpha\right)_a^b \Phi_b(x),\vspace{0.1cm}\\
\delta A_\mu^\alpha  =  f_{\beta\gamma}{}^{\alpha}\delta \theta^\beta A_\mu^\gamma - \frac1g \partial_\mu \delta \theta ^\alpha.
\end{array}
\right.
}
\end{equation}

Dans cette construction, les seules limitations pour le potentiel $V(\Phi_a)$ sont de contenir au maximum des termes quartiques en le champ scalaire, et de n'inclure que des contractions se transformant comme des singlets du groupe de jauge. La contraction d'un champ dans une représentation quelconque avec un champ dans sa représentation conjuguée contenant toujours un singlet, le potentiel contient toujours les termes $(\Phi_a)^\dagger \Phi_a = (\Phi^\dagger)^a \Phi_a$ et $[(\Phi^\dagger)^a \Phi_a]^2$. En fonction de la représentation du champ $\Phi_a$, il peut être possible d'écrire des termes cubiques, ou des termes quartiques différant de $[(\Phi^\dagger)^a \Phi_a]^2$. C'est pourquoi on évitera l'écriture $V[(\Phi^\dagger)^a \Phi_a]$, incorrecte dans le cas général. Pour une théorie incluant plusieurs champs scalaires, il sera aussi nécessaire de considérer dans le potentiel les contractions entres les différents champs qui peuvent former des singlets.

La théorie de jauge abélienne associée à un champ scalaire complexe est nommée modèle de Higgs abélien. Cette théorie est la plus simple pour décrire le mécanisme de Higgs (cf le chapitre~\ref{PartSSB}), d'où son nom. Le champ scalaire $\phi$ de charge $q_\phi\neq 0$ se transforme sous les transformations de jauge U(1) comme décrit dans les équations~\eqref{AbTransfoJaugeScalaire} et~\eqref{AbTransfoJaugeInfScalaire}. Le Lagrangien du modèle de Higgs abélien s'écrit
\begin{equation}
\label{LagHiggsAbelien}
\mathcal{L}_{\text{Higgs abélien}} = - ( D_\mu \phi )^* (D^\mu \phi) - \frac14 F_{\mu\nu} F^{\mu\nu} - V(|\phi|^2),
\end{equation}
avec 
\begin{equation}
D_\mu\phi = \partial_\mu \phi - i g q_\phi A_\mu.
\end{equation}
Ce Lagrangien est invariant sous les transformations de jauge infinitésimales 
\begin{equation}
\displaystyle{
\left\{
\begin{array}{l}
\delta \phi (x) = -i q_\phi \delta \theta \phi(x),\vspace{0.1cm}\\
\delta A_\mu  =  - \frac1g \partial_\mu \theta(x).
\end{array}
\right.
}
\end{equation}
Le potentiel dans ce Lagrangien est écrit comme une fonction de $|\phi|^2=\phi^*\phi$, qui est le seul terme qui peut apparaître dans les singlets du groupe U(1) formés à partir de $\phi$ seulement (pour les modèles incluant plusieurs champs scalaires, d'autres singlets peuvent être formés si les charges sont commensurables). Une théorie renormalisable ne comportant que des termes au maximum quartiques en les champs scalaires, le potentiel général pour le modèle de Higgs abélien s'écrit donc
\begin{equation}
V_{\text{Higgs abélien}}(|\phi|^2) = {m_\phi}^2|\phi|^2+ \lambda |\phi|^4 ~(+\eta^4)= \lambda(|\phi|^2 - \eta^2)^2,
\end{equation}
où $m_\phi$, $\lambda$ et $\eta$ sont des paramètres constants arbitraires apparaissant dans les deux formes canoniques du potentiel, ${m_\phi}^2$, $\lambda$ et $\eta^2$ pouvant être pris réels. Pour que la théorie décrite par le Lagrangien de l'équation~\eqref{LagHiggsAbelien} soit viable, il faut cependant que le potentiel soit borné par le bas, nécessitant de prendre $\lambda\geq 0$ (avec ${m_\phi}^2\geq 0$ dans le cas où $\lambda=0$). C'est le cas ${m_\phi}^2<0 \Longleftrightarrow \eta^2 > 0$ qui est d'intérêt pour l'étude du mécanisme de Higgs décrit dans le chapitre~\ref{PartSSB}, le champ scalaire développant alors une instabilité tachyonique pour la valeur $\phi=0$.

\section{Représentations équivalentes liées aux charges de couleurs}
\label{RepEquivSU3}

\noindent
Dans la section~\ref{NonAbelRacines&Poids}, les états de couleur des quarks sont reperés dans l'espace des poids de SU(3)$_C$ dans la base de Dynkin~\cite{Slansky:1981yr}. On trouve dans la littérature d'autres façons équivalentes de représenter ces états de couleurs. L'algèbre du groupe SU(3) est souvent décrite par sa représentation de définition impliquant les matrices $3\times3$ de Gell-Mann $\lambda_\alpha$ vérifiant les relations de commutations décrites par les constantes de structure~\cite{Ramond:2010zz},
\begin{equation}
\displaystyle{
\left[\frac{\lambda_\alpha}{2},\frac{\lambda_\beta}{2}\right] = if_{\alpha\beta}{}^\gamma \frac{\lambda_\gamma}{2},
}
\end{equation}
où les $f_{\alpha\beta}{}^\gamma$ sont les constantes de structure de SU(3). Ces matrices hermitiennes sont l'équivalent pour SU(3) des matrices de Pauli $\sigma_\alpha$ de SU(2). Une façon commode de les représenter est de prendre comme base de l'espace sur lequel elles agissent des états de couleur $\ket{rouge}$, $\ket{vert}$, et $\ket{bleu}$ notés respectivement $\boldsymbol r$, $\boldsymbol v$ et $\boldsymbol b$. On note les bras associés $\bar{\boldsymbol r}$, $\bar{\boldsymbol v}$ et $\bar{\boldsymbol b}$. Dans cette base, les matrices de Gell-Mann s'écrivent 
\begin{equation}
\label{GellMann}
\begin{array}{l}
\lambda_1 = (\boldsymbol r\bar{\boldsymbol v}+\boldsymbol v\bar{\boldsymbol r}),\\
\lambda_4 = (\boldsymbol r\bar{\boldsymbol b}+\boldsymbol b\bar{\boldsymbol r}),\\
\lambda_6 = (\boldsymbol v\bar{\boldsymbol b}+\boldsymbol b\bar{\boldsymbol v}),\\
\lambda_3 = (\boldsymbol r\bar{\boldsymbol r}-\boldsymbol v\bar{\boldsymbol v}),
\end{array}
~~~~
\begin{array}{l}
\lambda_2 = -i (\boldsymbol r\bar{\boldsymbol v}-\boldsymbol v\bar{\boldsymbol r}),\\
\lambda_5 = -i (\boldsymbol r\bar{\boldsymbol b}-\boldsymbol b\bar{\boldsymbol r}),\\
\lambda_7 = -i (\boldsymbol v\bar{\boldsymbol b}-\boldsymbol b\bar{\boldsymbol v}),\\
\lambda_8 = (\boldsymbol r\bar{\boldsymbol r}+\boldsymbol v\bar{\boldsymbol v}-2\boldsymbol b\bar{\boldsymbol b} ).
\end{array}
\end{equation}
En fait, décrire ces matrices par leur action sur les états de couleurs les identifie directement aux gluons via leur interactions avec les trois états internes des quarks.

Un premier choix pour passer sur l'espace des poids et racines consiste à prendre comme sous-algèbre de Cartan les générateurs diagonaux apparaissant dans la formulation de SU(3) avec les matrices de Gell-Mann
\begin{equation}
\label{RacinesSU3_2}
\begin{array}{l}
T_{3,GM}^{(1)}=\lambda_3 = (\boldsymbol r\bar{\boldsymbol r}-\boldsymbol v\bar{\boldsymbol v}),\\
T_{3,GM}^{(2)} =  \lambda_8 = (\boldsymbol r\bar{\boldsymbol r}+\boldsymbol v\bar{\boldsymbol v}-2\boldsymbol b\bar{\boldsymbol b} ).
\end{array}
\end{equation}
Les opérateurs liés aux racines sont alors
\begin{equation}
\label{RacinesSU3}
\displaystyle{
\begin{array}{l}
T_{\boldsymbol r\bar{\boldsymbol v}}=\frac12 (\lambda_1 - i \lambda_2)= \boldsymbol r\bar{\boldsymbol v} ,\vspace{0.1cm}\\
T_{\boldsymbol r\bar{\boldsymbol b}}=\frac12 (\lambda_4 - i \lambda_5)= \boldsymbol r\bar{\boldsymbol b} ,\vspace{0.1cm}\\
T_{\boldsymbol v\bar{\boldsymbol b}}=\frac12 (\lambda_6 - i \lambda_7)= \boldsymbol v\bar{\boldsymbol b} ,
\end{array}
~~~~
\begin{array}{l}
T_{-\boldsymbol r\bar{\boldsymbol v}}=(T_{\boldsymbol r\bar{\boldsymbol v}})^\dagger=T_{\boldsymbol v\bar{\boldsymbol r}}=\frac12 (\lambda_1 + i \lambda_2)= \boldsymbol v\bar{\boldsymbol r},\vspace{0.1cm}\\
T_{-\boldsymbol r\bar{\boldsymbol b}}=(T_{\boldsymbol r\bar{\boldsymbol b}})^\dagger=T_{\boldsymbol b\bar{\boldsymbol r}}=\frac12 (\lambda_4 + i \lambda_5)= \boldsymbol b\bar{\boldsymbol r},\vspace{0.1cm}\\
T_{-\boldsymbol v\bar{\boldsymbol b}}=(T_{\boldsymbol v\bar{\boldsymbol c}})^\dagger=T_{\boldsymbol b\bar{\boldsymbol v}}=\frac12 (\lambda_6 + i \lambda_7)= \boldsymbol b\bar{\boldsymbol v},
\end{array}
}
\end{equation}
On retrouve les différentes propriétés des racines discutées dans la section~\ref{Poids&Racines}, ainsi qu'une similitude forte avec la construction donnée dans la section~\ref{PartSU2Ini} des opérateurs d'échelle de SU(2) à partir de la représentation de définition -- qui sont écrits avec les matrices de Pauli et dont les relations de commutation donnent les constantes de structure du groupe. Les relations de commutations des éléments de le nouvelle base de l'algèbre de SU(3)  donnée par les équations~\eqref{RacinesSU3_2} et~\eqref{RacinesSU3} ne sont par contre plus décrites par les constantes de structure du groupe.

\begin{figure}[h!]
\begin{center}
\includegraphics[scale=1.2]{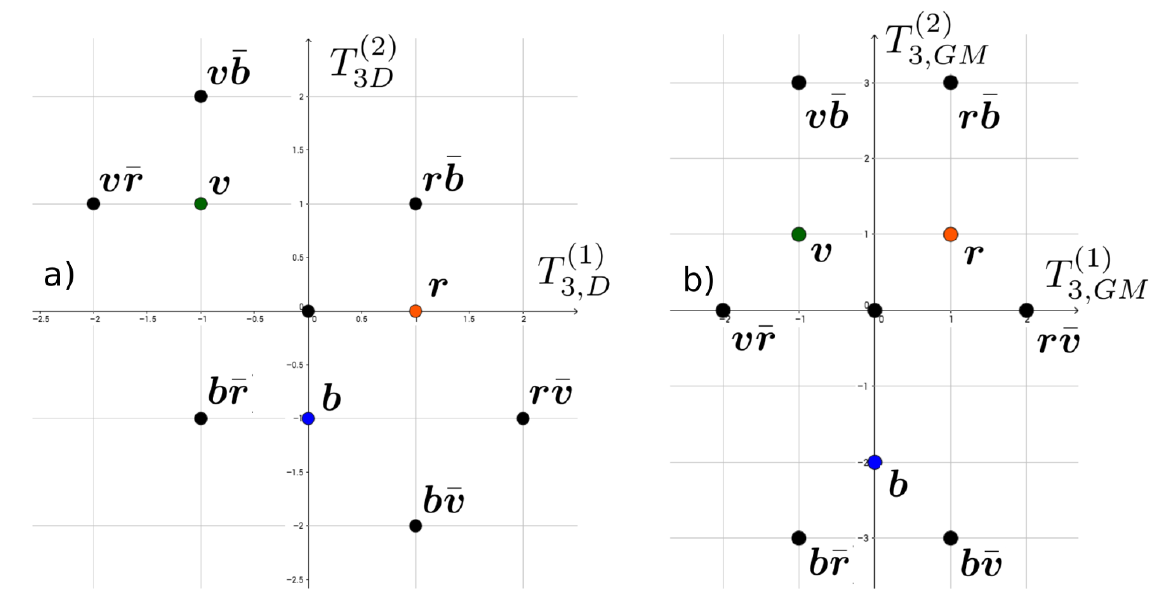}
\vspace{-0.4cm}
\end{center}
 \caption{Diagrammes des poids de SU(3)$_C$ associés aux états de couleurs des quarks (représentation de dimension 3) et aux gluons (représentation adjointe de dimension 8). Dans la figure a), on a pris les nombres quantiques correspondants à la base de Dynkin. Dans la figure b), on a pris les nombres quantiques correspondant à la sous-algèbre de Cartan définie par les matrices de Gell-Mann. Dans les deux cas, le poids de coordonnées nulles est double, et correspond aux deux éléments de la sous-algèbre de Cartan.}
 \label{SU3_GM_D}
\end{figure}

Le choix des éléments de la sous-algèbre de Cartan est arbitraire, et sert à définir les coordonnées de poids des éléments des différentes représentations, et donc leur nombres quantiques. Ainsi, dans cette base, les nombres quantiques des états de quarks sont 
\begin{equation}
\begin{array}{l}
\boldsymbol r = \ket{1,1}_{GM},\\
\boldsymbol v = \ket{-1,1}_{GM},\\
\boldsymbol b = \ket{0,-2}_{GM}.
\end{array}
\end{equation}
En décrivant les gluons, et donc les opérateurs de l'algèbre, par la différence des états de couleurs qu'ils impliquent, on obtient
\begin{equation}
\displaystyle{
\begin{array}{l}
\boldsymbol r\bar{\boldsymbol v}=\ket{2,0}_{GM},\\
\boldsymbol r\bar{\boldsymbol b}=\ket{1,3}_{GM},\\
\boldsymbol v\bar{\boldsymbol b}=\ket{-1,3}_{GM},
\end{array}
~~~~
\begin{array}{l}
\boldsymbol v\bar{\boldsymbol r}=\ket{-2,0}_{GM},\\
\boldsymbol b\bar{\boldsymbol r}=\ket{-1,-3}_{GM},\\
\boldsymbol b\bar{\boldsymbol v}=\ket{1,-3}_{GM}.
\end{array}
}
\end{equation}
Ces vecteurs, en complément des poids nuls associés à $T_{3,GM}^{(1)}$ et $T_{3,GM}^{(2)}$,  décrivent l'espace des poids de la représentation adjointe de SU(3). Le diagramme des poids associé à ces deux représentations pour la sous-algèbre de Cartan liée aux matrices de Gell-Mann est représentée dans la figure~\ref{SU3_GM_D}.

Pour se placer dans la base de Dynkin \cite{Slansky:1981yr}, il faut modifier les opérateurs de la sous-algèbre de Cartan, et prendre 
\begin{equation}
\begin{array}{l}
T_{3,D}^{(1)} = T_{3,GM}^{(1)} = (\boldsymbol r\bar{\boldsymbol r}-\boldsymbol v\bar{\boldsymbol v}),\vspace{0.1cm}\\
T_{3D}^{(2)} =  \frac12 (T_{3,GM}^{(2)} -  T_{3,GM}^{(1)}) = (\boldsymbol v\bar{\boldsymbol v}-\boldsymbol b\bar{\boldsymbol b} ).
\end{array}
\end{equation}
Ce nouveau choix de sous-algèbre de Cartan ne change en rien l'action des différents opérateurs liés aux racines donnée dans l'équation~\eqref{RacinesSU3}. Les poids, et donc les nombres quantiques, des différents quarks et gluons sont par contre modifiés. On a dans cette base
\begin{equation}
\begin{array}{l}
\boldsymbol r = \ket{1,0}_D,\\
\boldsymbol v = \ket{-1,1}_D,\\
\boldsymbol b = \ket{0,-1}_D,
\end{array}
\end{equation}
pour les états de couleurs des quarks, et 
\begin{equation}
\displaystyle{
\begin{array}{l}
\boldsymbol r\bar{\boldsymbol v}=\ket{2,-1}_D,\\
\boldsymbol r\bar{\boldsymbol b}=\ket{1,1}_D,\\
\boldsymbol v\bar{\boldsymbol b}=\ket{-1,2}_D,
\end{array}
~~~~
\begin{array}{l}
\boldsymbol v\bar{\boldsymbol r}=\ket{-2,1}_D,\\
\boldsymbol b\bar{\boldsymbol r}=\ket{-1,-1}_D,\\
\boldsymbol b\bar{\boldsymbol v}=\ket{1,-2}_D,
\end{array}
}
\end{equation}
pour les gluons associés à des racines non-nulles. Les diagrammes de ces deux représentations repérés vis-à-vis de la sous-algèbre de Cartan liée aux indices de Dynkin est représenté dans la figure~\ref{SU3_GM_D}. Ce sont ces nombres quantiques qu'on utilise tout au long de ce document, par exemple dans les figures~\ref{DiagPoidsQuarks} et~\ref{TransfoGluonPoids}.

Cette discussion montre bien que des choix différents de sous-algèbre de Cartan ne modifient que la description des différentes représentations, mais donnent des résultats complètement équivalents. Les opérateurs de cette sous-algèbre permettent de définir les poids des représentations, et passer d'une sous-algèbre à une autre en redéfinissant par une combinaison linéaire les éléments de la sous-algèbre de Cartan correspond seulement à un changement linéaire de base dans les espaces de représentations.

\section{Brisure de SU(3) par ses représentations de dimension 3 et 8}
\label{BrisureSU3}

\begin{figure}[h!]
\begin{center}
\includegraphics[scale=1.2]{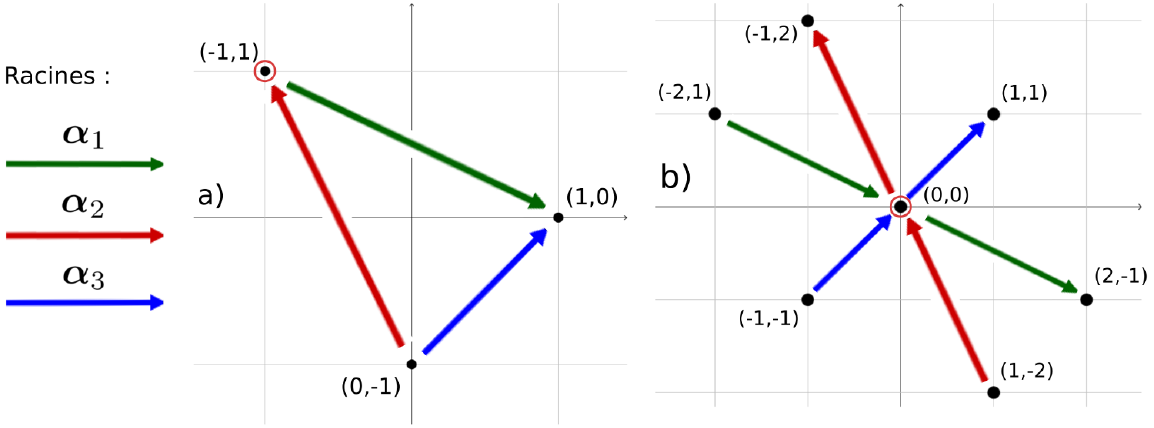}
\vspace{-0.4cm}
\end{center}
 \caption{Diagramme des poids des représentations de dimension 3 (figure a) et 8 (figure b) de SU(3), où l'on a indiqué certaines translations liées aux racines. Sur chaque figure, on a aussi entouré en rouge le poids associé à la {\sc vev} non-nulle pour les brisures de symétrie décrites dans la présente section.}
 \label{FigBrisureSU3}
\end{figure}

\noindent
Nous explicitons dans cette section des brisures possibles de SU(3) par ses représentations de dimension 3 et 8. Ces exemples sont indépendant de la chromodynamique quantique. Les espaces des poids de ces représentations et les racines de SU(3) sont représentés sur la figure~\ref{FigBrisureSU3}, dans la base de Dynkin.

Commençons par le cas d'une représentation de dimension 3 de SU(3). Sans perdre de généralité, on peut supposer que la {\sc vev} du champ de Higgs correspond à l'état de nombres quantiques $(-1,1)$, et identifier sur cette {\sc vev} les générateurs de la symétrie non brisée. Bien que $T_3^{(1)}$ et $T_3^{(2)}$ n'annulent pas la {\sc vev}, la combinaison linéaire $T_3^{(1)} + T_3^{(2)}$ l'annule, et correspond donc à un générateur de la symétrie résiduelle. Sur l'espace des poids, on lit aussi que les générateurs associés à $\boldsymbol\alpha_1$ et $\boldsymbol\alpha_2$ n'annulent pas la {\sc vev}, la transformant en un autre poids de la représentation. Le générateur associé à $\boldsymbol\alpha_3$ annule par contre la {\sc vev}, et correspond également à un générateur de la symétrie résiduelle. Reste à vérifier que les générateurs non brisés forment une algèbre. Le générateur de la sous-algèbre de Cartan annulant la {\sc vev} correspond à
\begin{equation}
\label{DefT3Alpha3}
T_3^{(\boldsymbol\alpha_3)} = T_3^{(1)} + T_3^{(2)},
\end{equation}
qui forme bien une algèbre fermée avec $T_{\pm\boldsymbol\alpha_3}$ [voir les équations~\eqref{DefTSU2} et~\eqref{DefComTSU2}]. Ainsi, la {\sc vev} non nulle correspondant au poids $(-1,1)$ de la représentation de dimension 3 cause une brisure spontanée de symétrie de SU(3) vers le SU(2) de symétrie résiduelle généré par $\{T_3^{(\boldsymbol\alpha_3)},T_{\boldsymbol\alpha_3},T_{-\boldsymbol\alpha_3}  \}$ :
\begin{equation}
\text{SU(3)} \overset{\mathbf{3}}{\relbar\joinrel\relbar\joinrel\longrightarrow} \text{SU(2)}.
\end{equation}

On peut faire de même pour la représentation de dimension 8 de SU(3). Pour décrire la {\sc vev}, on écrit la sous-algèbre de Cartan à l'aide de l'opérateur $T_{\boldsymbol\alpha_3}$ défini dans l'équation~\eqref{DefT3Alpha3}, et l'opérateur orthogonal
\begin{equation}
T_3^{(\boldsymbol\alpha^\perp_3)} = T_3^{(1)}- T_3^{(2)}.
\end{equation}
Ce dernier opérateur vérifie
\begin{equation}
\label{EqBrisureSU3}
\left[T_3^{(\boldsymbol\alpha^\perp_3)} , T_{\boldsymbol\alpha_3}  \right] =0.
\end{equation}
On peut alors étudier la configuration où la {\sc vev} du champ de Higgs correspond à $(0,0)_3^{(\boldsymbol\alpha^\perp_3)}$, où l'on décrit le poids nul en l'identifiant à un opérateur de la sous-algèbre de Cartan. L'équation~\eqref{EqBrisureSU3} donne alors que l'action des racines $\pm\boldsymbol \alpha_3$ annule le vide, mais pas celle des autres racines. La {\sc vev} ayant des nombres quantiques nuls, car correspondant à un poids $(0,0)$, elle est aussi annulée par les générateurs $T_3^{(\boldsymbol\alpha_3)} $ et $T_3^{(\boldsymbol\alpha^\perp_3)} $. Ainsi, le groupe de symétrie résiduel correspond à SU(2)$\times$U(1), où le SU(2) est généré par $\{T_3^{(\boldsymbol\alpha_3)},T_{\boldsymbol\alpha_3},T_{-\boldsymbol\alpha_3}  \}$ et le U(1) par $T_3^{(\boldsymbol\alpha^\perp_3)}$ :
\begin{equation}
\text{SU(3)} \overset{\mathbf{8}}{\relbar\joinrel\relbar\joinrel\longrightarrow} \text{SU(2)}\times\text{U(1)}.
\end{equation}

Dans ces deux cas, il est possible d'identifier les règles de branchement entre les représentations du groupe de symétrie initial et résiduel par l'observation du diagramme des poids. On fait cette identification sur le sous-groupe maximal SU(2)$\times$U(1), où SU(2) est identifié aux générateurs $\{T_3^{(\boldsymbol\alpha_3)},T_{\boldsymbol\alpha_3},T_{-\boldsymbol\alpha_3}  \}$ et U(1) à $T_3^{(\boldsymbol\alpha^\perp_3)}$. Les représentations de SU(2) s'identifient comme les éléments liés entre eux par la racine $\boldsymbol\alpha_3$. La charge sous la symétrie U(1), identique pour toute la représentation, s'obtient alors en prenant la charge pour $T_3^{(\boldsymbol\alpha^\perp_3)}$. Pour la représentation de dimension 3, on identifie $\mathbf{1}(-2)$ et $\mathbf{2}(1)$ aux ensembles de poids $\{(-1,1)\}$ et $\{(0,-1),(1,0)\}$. Pour la représentation de dimension 8, on identifie $\mathbf{1}(0)$, $\mathbf{2}(3)$, $\mathbf{2}(-3)$ et $\mathbf{3}(0)$ aux ensembles de poids $\{(0,0)_3^{\boldsymbol\alpha_3^\perp}\}$, $\{(1,-2),(2,-1)\}$, $\{(-2,1),(-1,2)\}$ et $\{(-1,-1), (0,0)_3^{\boldsymbol\alpha_3},(1,1)\}$.



\part{Physique des particules : le Modèle Standard, et au delà}
\label{ChapterSMAndBeyond}

\chapter{Le Modèle Standard de la physique des particules}
\section{Introduction}

\noindent
Le Modèle Standard de la physique des particules décrit dans un formalisme unifié les interactions électromagnétiques, fortes, et faibles. Sa description repose sur les théories de jauge et sur le mécanisme de Higgs de brisure spontanée de symétrie, tous deux décrits dans la partie~\ref{ChapterTheoriesDeJauges}. La brisure spontanée de symétrie du Modèle Standard correspond à la brisure des interactions électrofaibles, dont la symétrie résiduelle correspond aux interactions électromagnétiques. Le développement du Modèle Standard a commencé dès les années 1930 avec les premières tentatives infructueuses de quantifier les interactions électromagnétiques, et s'est conclu au début des années 1970 avec la compréhension des propriétés de confinement et de liberté asymptotique de la chromodynamique quantique.

Il y a beaucoup de façon de décrire le Modèle Standard, du point de vue le plus théorique au plus phénoménologique. On prendra ici une approche liée à la construction de modèles physiques, afin de discuter ensuite dans les meilleures conditions possibles les théories de grande unification. On mettra notamment l'accent sur la place des théories de jauge et du mécanisme de Higgs dans le Modèle Standard. Les principales considérations théoriques ont été introduites dans les chapitres précédents, et ont été exemplifiées sur le Modèle Standard lui-même. La présentation du présent chapitre sera donc plutôt une description du contenu en champs et des couplages du Modèle Standard, ainsi que de son statut actuel.

\section{Modèle Standard avant la brisure électrofaible}
\label{PartJaugesSM}

\noindent
Avant la brisure spontanée de symétrie du Modèle Standard par le champ de Higgs électrofaible, les interactions du Modèle Standard sont décrites par une théorie de jauge basée sur le groupe de symétrie
\begin{equation}
G_{\text{SM}} = \text{SU(3)}_C \times \text{SU(2)}_L \times \text{U(1)}_Y.
\end{equation}
Le groupe SU(3)$_C$ décrit les interactions fortes dans le cadre de la chromodynamique quantique (QCD). Les bosons de jauge associés à ce groupe sont nommés gluons, et les représentations de SU(3)$_C$ sont décrites dans le présent document par leurs poids dans la base de Dynkin, comme cela a été le cas dans la partie~\ref{ChapterTheoriesDeJauges}. Le groupe SU(2)$_L\times$U(1)$_Y$ décrit les interactions électrofaibles avant la brisure de symétrie du même nom. Les bosons associés à ce groupe sont les trois bosons $W$ de SU(2)$_L$ et le boson $A_\mu^Y$ associé au générateur de U(1)$_Y$. Les charges d'isospin faible sont décrites dans la représentation conventionnelle de spin de SU(2), les isospins prenant des valeurs entières et demi-entières. Les hypercharges faibles sont normalisées de façon à ce que
\begin{equation}
\label{ChargeQSU2Chap3}
T_3^Q = T_3^L + T_3^Y,
\end{equation}
ce qui permet d'obtenir très simplement la charge électrique d'une particule à partir de son isospin faible et de son hypercharge. C'est la formule de Gell-Mann–Nishijima~ \cite{Nakano:1953zz,Gell-Mann:1956iqa}. Dans la littérature, cette relation est souvent écrite avec un facteur $\frac12$ devant $T_3^Y$, que nous omettrons dans ce document par soucis de simplicité (il suffit pour cela de redéfinir le paramètre de la transformation de symétrie de U(1)$_Y$ et d'inclure ce facteur dans la constante de couplage associée à ce groupe). 

\begin{figure}[h!]
\begin{center}
\includegraphics[scale=1.78]{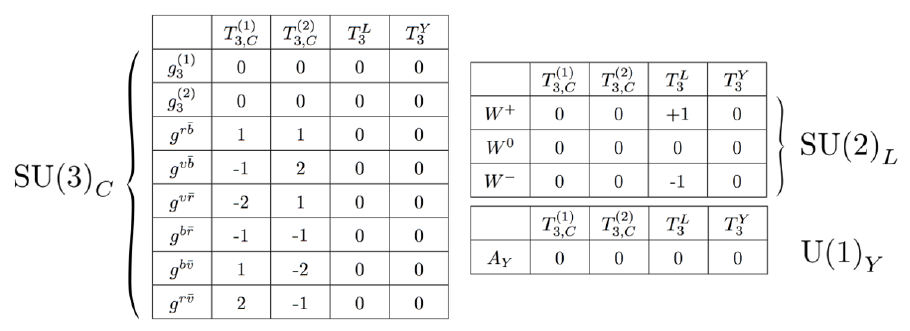}
\vspace{-0.6cm}
\end{center}
 \caption{Noms et nombres quantiques des bosons de jauge du Modèle Standard avant la brisure électrofaible. Les gluons $g$ sont dénommés soit par les indices de l'élément de la sous-algèbre de Cartan de SU(3)$_C$ auxquels ils sont associés, soit par leur charge de couleur. Les bosons $W$ sont désignés par leur charge d'isospin faible.}
 \label{TableJaugeSM}
\end{figure}

Les bosons de jauge jouent le rôle de médiateurs des différentes interactions du Modèle Standard. Les degrés de liberté physiques des bosons de SU(3)$_C$ et SU(2)$_L$ sont dans les représentations adjointes de ces groupes, et le boson $A_\mu^Y$ associé au générateur de U(1)$_Y$ est dans une représentation triviale. Ces bosons sont sans masse, et ne sont pas chargés sous les groupes associés aux interactions dont ils ne sont pas les médiateurs, comme discuté dans la section~\ref{ProduitGroupesJauge}. On identifie ces bosons aux différentes racines de chaque groupe, les bosons associés aux racines nulles ayant tous leurs nombres quantiques identiquement nuls. Les noms et nombres quantiques des différents bosons de jauge du Modèle Standard sont récapitulés dans la figure~\ref{TableJaugeSM}.

\begin{figure}[h!]
\begin{center}
\includegraphics[scale=1.25]{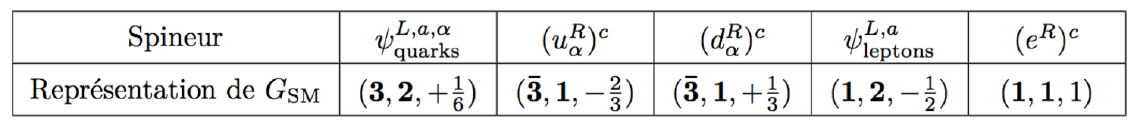}
\vspace{-0.6cm}
\end{center}
 \caption{Représentations contenant chaque génération de fermions, repérées ici pour la première génération. Les particules de chiralité droite sont décrites par leur anti-particule de chiralité gauche.}
 \label{TableRepSM}
\end{figure}

Les fermions du Modèle Standard forment trois générations identiques, liées à des échelles de masse croissantes après brisure de symétrie, et rassemblant chacune 15 particules. Nous discutons pour commencer la première génération de fermions, associée aux masses les plus faibles. Dans l'optique de grouper tout ou partie des particules contenues dans une génération de fermions dans une représentation des théories de grande unification, comme discuté dans le chapitre~\ref{PartGUT}, nous décrivons dès maintenant tous les fermions avec des spineurs de Weyl de chiralité gauche. Les fermions de chiralité droite sont donc décrits par les anti-particules qui leurs sont associées, de chiralité gauche. Les fermions de chaque génération se rassemblent dans cinq représentations de $G_{\text{SM}}$, résumées dans la figure~\ref{TableRepSM}. Les représentations sont décrites en donnant successivement les dimensions des représentations\footnote{Une distinction est parfois faite dans la littérature entre la représentation $\mathbf{2}$ de SU(2) et sa représentation complexe conjugée $\mathbf{\bar{2}}$. Bien que ne pouvant pas être directement identifiées du point de vue calculatoire, ces deux représentations sont cependant équivalentes, la $\mathbf{2}$ de SU(2) étant une représentation pseudo-réelle. Il est par exemple toujours possible de former un singlet à partir de deux quelconques de ces représentations, même si l'écriture détaillée de la contraction associée peut varier. Aussi, et dans tout ce document, nous ne distinguerons pas ces représentations.} sous SU(3)$_C$, SU(2)$_L$, et la charge sous U(1)$_Y$.

\begin{figure}[h!]
\begin{center}
\includegraphics[scale=1.35]{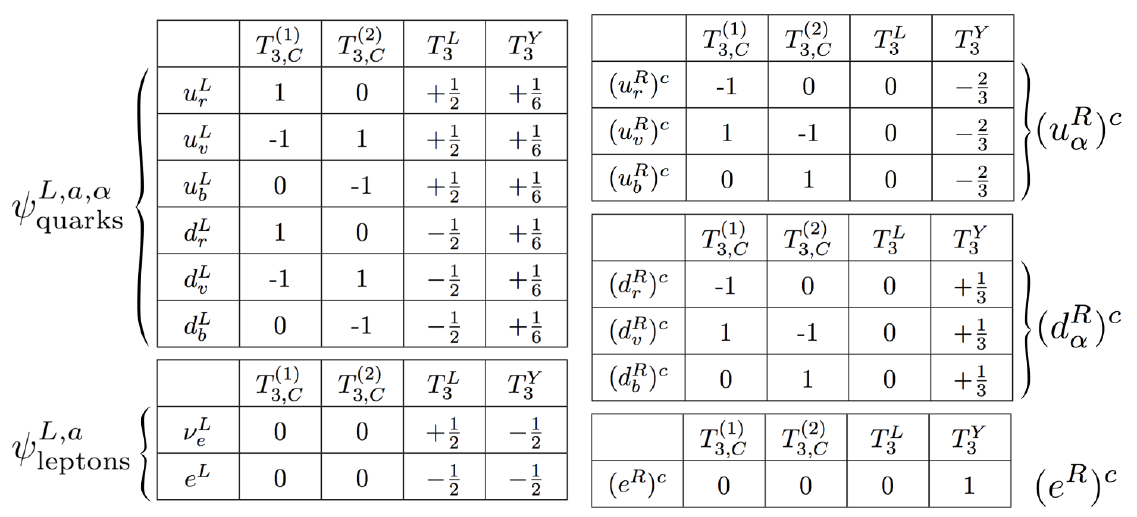}
\vspace{-0.6cm}
\end{center}
 \caption{Fermions de la première génération du Modèle Standard, groupés par représentations irréductibles. On a indiqué le spineur associé à chacune de ces représentations. Les particules de chiralité droite sont décrites par leur anti-particule, de chiralité gauche.}
 \label{TableFermionsSM}
\end{figure}

Les fermions soumis aux interactions fortes sont les quarks. Ils sont rassemblés dans une représentation ($\mathbf{3},\mathbf{2},+\frac16$) pour les quarks de chiralité gauche, notée $\psi^{L,a,\alpha}_{\text{quarks}}$, et dans deux représentations ($\mathbf{\bar 3},\mathbf{1},-\frac23$) et ($\mathbf{\bar 3},\mathbf{1},+\frac13$) pour les antiparticules associées aux quarks de chiralité droite, $(u^R_\alpha)^c$ et $(d^R_\alpha)^c$. Les leptons, non soumis aux interactions fortes, sont dans une représentation $(\mathbf{1},\mathbf{2},-\frac12)$ pour ceux de chiralité gauche, notée $\psi^{L,a}_{\text{leptons}}$, et dans une représentation singlet d'hypercharge $+1$ pour l'anti-particule associée à l'électron de chiralité droite, $(e^R)^c$. Les assignations des différentes particules dans ces représentations sont données dans la figure~\ref{TableFermionsSM}, où l'on a regroupé les fermions par représentation irréductibles. 

\begin{figure}[h!]
\begin{center}
\includegraphics[scale=1.75]{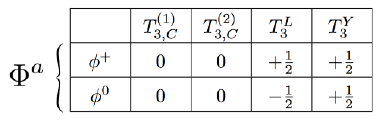}
\vspace{-0.6cm}
\end{center}
 \caption{Nombres quantiques des différentes composantes du champ de Higgs du Modèle Standard.}
 \label{TableChampHiggs}
\end{figure}

Les deux autres générations de fermions sont identiques du point de vue des représentations de $G_{\text{SM}}$ dans lesquelles ils se regroupent. Les quarks "up" et "down" ($u$ et $d$) deviennent "charm" et "strange" ($c$ et $s$) dans la deuxième génération, et "top" et "bottom" ($t$ et $b$) dans la troisième génération\footnote{Ceux-ci sont parfois (rarement) appelés "truth" et "beauty".}. Les électrons et neutrinos électroniques ($e$ et $\nu_e$) deviennent des muons et des neutrinos muoniques ($\mu$ et $\nu_\mu$) dans la deuxième génération, et des taus et neutrinos tauiques ($\tau$ et $\nu_\tau$) dans la troisième génération. Finalement, le Modèle Standard contient un champ scalaire, le champ de Higgs $\Phi^a$, dans une représentation $(\mathbf{1},\mathbf{2},+\frac12)$ de $G_{\text{SM}}$. Les nombres quantiques de ses différentes composantes sont récapitulés dans la figure~\ref{TableChampHiggs}.

Suivant les considérations développées dans les sections~\ref{PartIntSymDeJauge} à~\ref{ProduitGroupesJauge}, les tableaux récapitulant les nombres quantiques des particules du Modèle Standard permettent d'obtenir leur vertex d'interaction. Prenons l'exemple des vertex cubiques impliquant des fermions. Les transformations associées aux racines non-nulles ont lieu au sein d'une même représentation, et s'identifient par les translations dans chaque représentation qui sont permises par les bosons de jauge liés à ces racines vis-à-vis de la conservation des charges du Modèle Standard. Pour les vertex associés aux bosons de jauge de charge nulle, associés à des éléments de la sous-algèbre de Cartan de $G_{\text{SM}}$, ils ont des interactions se ramenant à celles d'une théorie de jauge abélienne pour la charge qui leur est associée [par exemple la charge sous $T_{3,C}^{(1)}$ pour le gluon $g_3^{(1)}$]. Les vertex quartiques se construisent similairement.

Les couplages de Yukawa, qui relient le champ de Higgs à deux champs spinoriels, peuvent aussi s'obtenir de cette façon. La première étape consiste à identifier deux représentations de fermions qui peuvent former un singlet avec le champ de Higgs dans un terme cubique. Le champ de Higgs étant dans la représentation $(\mathbf{1},\mathbf{2},+\frac12)$ de $G_{\text{SM}}$, la lecture de la figure~\ref{TableRepSM} permet d'identifier rapidement quelles représentations peuvent former un tel singlet. Celles-ci devront être dans des représentations $\mathbf{3}$ et $\mathbf{\bar 3}$ ou bien $\mathbf{1}$ de SU(3)$_C$, $\mathbf{1}$ et $\mathbf{2}$ de SU(2)$_C$, et avoir une somme d'hypercharges égale à $\pm\frac12$ selon qu'on considère le champ de Higgs ou bien son anti-particule. Ces deux représentations fermioniques identifiées, la conservation des charges du Modèle Standard permet d'identifier les fermions de chaque représentation qui sont reliés par une des composantes du champ de Higgs.

Les couplages de Yukawa permis par le champ de Higgs sont très différents des couplages avec les bosons de jauge, car ils relient des fermions dans des représentations différentes. Les fermions des différentes générations se regroupant dans des représentations identiques de $G_{\text{SM}}$, deux représentations couplées par le champ de Higgs le sont alors quelle que soit la génération des fermions de chacune de ces représentations. Le champ de Higgs permet donc un couplage inter-générationnel au niveau du Lagrangien. Nous verrons dans les parties~\ref{PartMasseFermions} et~\ref{PartCKMMatrix} les conséquences de ces couplages au niveau des états propres de masse des fermions.

\begin{figure}[h!]
\begin{center}
\includegraphics[scale=1.6]{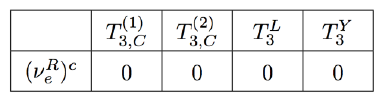}
\vspace{-0.6cm}
\end{center}
 \caption{Nombres quantiques de l'anti-particule associée au neutrino électronique de chiralité droite.}
 \label{TableNeutrino}
\end{figure}

Pour conclure cette section, on note que les fermions de chaque génération ne contiennent pas de neutrinos de chiralité droite. Leur présence en complément des neutrinos de chiralité gauche pourrait sembler nécessaire, mais ce n'est pas \emph{a priori} le cas dans une théorie dont la parité n'est pas une symétrie, comme le Modèle Standard. Ces neutrinos peuvent cependant être rajoutés au Modèle Standard, et sont alors dans la représentation triviale de $G_{\text{SM}}$, de nombres quantiques identiquement nuls comme indiqués dans la figure~\ref{TableNeutrino}. 

Cette représentation particulière implique qu'aucun couplage n'est possible entre le neutrino de chiralité droite et les autres particules du Modèle Standard. Ce neutrino étant dans la représentation triviale du groupe de jauge, il ne se couple avec aucun boson de jauge du Modèle Standard. De même, cette représentation ne permet pas de construire un couplage de Yukawa avec le champ de Higgs, qui demanderait que des fermions ou anti-fermions soient dans la même représentation que le champ de Higgs lui-même, ce qui n'est pas le cas. Cependant, et comme le neutrino de chiralité droite n'est chargé sous aucune symétrie interne, il peut avoir un terme de masse de Majorana, comme décrit dans la section~\ref{PartSpineurs}, sans contrainte théorique particulière sur la valeur de cette masse.

Cette donnée exhaustive du contenu en champ permet d'écrire le Lagrangien complet du Modèle Standard. Les couplages entre les champs de jauge et les spineurs proviennent des dérivées covariantes de jauge, décrites dans le chapitre~\ref{PartTheoriesJauge}. Le potentiel du champ de Higgs et les couplages de Yukawa avec les spineurs sont décrits dans le chapitre~\ref{PartSSB}. Dans chaque cas, les champs sont normalisés par le fait que leur terme cinétique initial ne contient aucun pré-facteur. Nous ne donnerons pas ici le Lagrangien explicite, qui présente peu d'intérêt pratique, et peut se trouver aisément dans la littérature~\cite{Langacker:1980js,Donoghue:1992dd,PDG2016}.

\section{Modèle Standard après la brisure électrofaible}

\noindent
La symétrie de jauge complète du Modèle Standard est brisée spontanément à basse énergie par le mécanisme de Higgs en un groupe résiduel comprenant SU(3)$_C$ de la chromodynamique quantique et U(1)$_Q$ de l'électrodynamique quantique :
\begin{equation}
G_{\text{SM}} = \text{SU(3)}_C \times \text{SU(2)}_L \times \text{U(1)}_Y \overset{\braket{\Phi^a}}{\relbar\joinrel\relbar\joinrel\longrightarrow} G_{\text{SM}}^{^{\text{{\small br}}}} =  \text{SU(3)}_C \times \text{U(1)}_Q.
\end{equation}
Ce mécanisme est décrit dans le chapitre~\ref{PartSSB}, le cas de la brisure électrofaible étant spécifiquement étudiée dans la section~\ref{RepChampsHiggs}. Le champ de Higgs du modèle électrofaible prend alors une valeur moyenne dans le vide ({\sc vev}) de $\simeq 246$~GeV, qui est aussi l'échelle caractéristique de brisure de symétrie, et la masse de l'excitation résiduelle du champ scalaire est $\simeq 125$~GeV~\cite{PDG2016}. La théorie de jauge associée à U(1)$_Q$ décrit les interactions électromagnétiques, tandis que les interactions associées aux bosons de jauge de SU(2)$_L\times$U(1)$_Y$ ayant acquis une masse correspondent aux interactions faibles. Les contributions respectives des interactions de SU(2)$_L$ et U(1)$_Y$ dans les interactions électromagnétiques sont décrites par l'angle de Weinberg, défini dans l'équation~\eqref{AngleWeinberg}, qui caractérise la brisure de symétrie électrofaible.

\begin{figure}[h!]
\begin{center}
\includegraphics[scale=1.4]{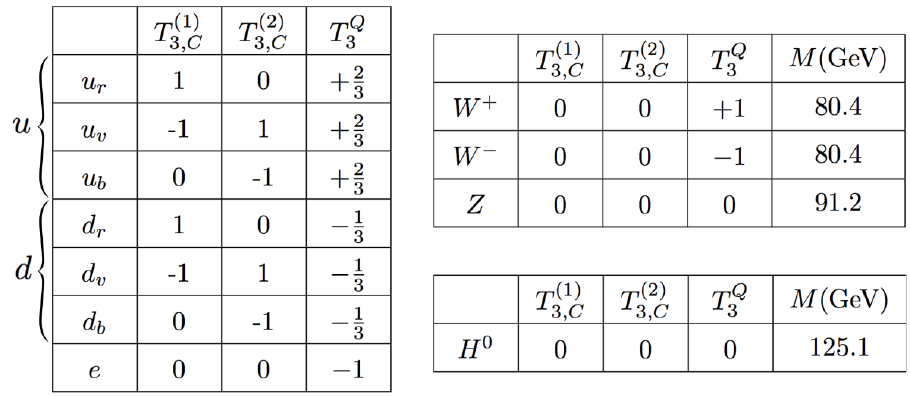}
\vspace{-0.6cm}
\end{center}
 \caption{Nombres quantiques des particules dans des représentations du groupe de symétrie du Modèle Standard après la brisure électrofaible, SU(3)$_C \times$U(1)$_Q$. Les tables rassemblent les fermions de première génération, les bosons de jauge du Modèle Standard ayant acquis une masse, et le boson de Higgs. Les représentations irréductibles des fermions sont identifiées par des accolades.}
 \label{TableParticulesSMBrise}
\end{figure}

Le groupe de symétrie de jauge résiduel après la brisure électrofaible est SU(3)$_C\times$U(1)$_Q$. Il convient donc de décrire les interactions associées vis-à-vis des représentations irréductibles de ce groupe de jauge. Ces interactions ne violant pas la parité, les spineurs de Weyl peuvent être rassemblés par paires de chiralités opposées, qui sont dans les mêmes représentations du groupe de jauge. L'utilisation de spineurs de Dirac, rassemblant de telles paires, est donc adaptée (voir la discussion de la section~\ref{PartSpineurs}). Ces spineurs de Dirac ont un terme de masse reliant leurs composantes de chiralité droite et gauche, de par leur couplage avec la {\sc vev} non-nulle du champ de Higgs, comme présenté dans la section~\ref{PartMasseFermionsHiggs}. Les nombres quantiques de ces spineurs sont donnés figure~\ref{TableParticulesSMBrise}, pour les fermions de la première génération, ainsi que ceux des bosons de jauge massifs après la brisure de symétrie, et du boson de Higgs (l'excitation de masse non nulle du champ de Higgs après brisure de symétrie).

Pour ces particules, la description des interactions de jauge liées à SU(3)$_C\times$U(1)$_Q$ se fait comme détaillé dans le chapitre~\ref{PartTheoriesJauge}. Les nombres quantiques des gluons pour SU(3)$_C$, donnés dans la figure~\ref{TableJaugeSM}, sont inchangés. Les gluons ne sont pas chargés sous U(1)$_Q$, comme attendu (voir la discussion de la section~\ref{ProduitGroupesJauge}). Ils étaient en effet non chargés sous SU(2)$_L\times$U(1)$_Q$, et ne peuvent donc pas être chargés sous un de ses sous-groupe. De façon similaire, on peut montrer que le champ de jauge associé au photon 
ne porte aucune charge sous SU(3)$_C\times$U(1)$_Q$, ce qui est normal pour le générateur d'un groupe abélien. 

\begin{figure}[h!]
\begin{center}
\includegraphics[scale=1.3]{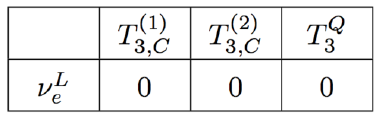}
\vspace{-0.6cm}
\end{center}
 \caption{Nombres quantiques du neutrino électronique de chiralité gauche sous SU(3)$_C\times$U(1)$_Y$.}
 \label{TableNeutrinoSMBrise}
\end{figure}

Avec la description des particules dans les représentations de SU(3)$_C\times$U(1)$_Q$, il n'est plus possible de décrire les interactions faibles, contenues dans les interactions électrofaibles. En effet, il est nécessaire pour cela de se ramener à la description de la section précédente, impliquant les représentations de $G_\text{SM}$. Ces interactions ont cependant des contributions très faibles par rapport à celles du groupe non brisé à des échelles d'énergies petites devant l'échelle de brisure de symétrie, et ne sont dominantes que pour les interactions non permises par la symétrie résiduelle, comme les désintégrations $\beta$. Les effets des interactions faibles correspondent alors à ceux d'une symétrie brisée, discutés dans la section~\ref{PhenomSymBrisées}. Le neutrino de chiralité gauche, décrit dans la figure~\ref{TableNeutrinoSMBrise}, n'est pas chargé sous SU(3)$_C\times$U(1)$_Q$, et n'interagit donc avec aucun champ de jauge après la brisure de symétrie. En fait, il peut ne pas être décrit par une représentation de ce groupe, étant donné qu'il ne peut pas être décrit à l'aide d'un spineur de Dirac en l'absence d'un neutrino de chiralité droite. Il a alors une position similaire pour SU(3)$_C\times$U(1)$_Q$ à celle qu'a le neutrino de chiralité droite pour $G_\text{SM}$. Il interagit cependant par les interactions faibles.

\section{Masse des fermions et états propres de masse}
\label{PartMasseFermions}

\noindent
Les fermions donnés dans la figure~\ref{TableParticulesSMBrise} sont définis vis-à-vis des représentations de SU(3)$_C\times$U(1)$_Q$. Ce sont les états propres d'interaction. On a cependant vu dans la section~\ref{PartJaugesSM} que le champ de Higgs permettait des termes de couplage dans une même génération comme entre générations. Cette propriété se retrouve au niveau des termes de masse après la brisure de symétrie électrofaible. L'électron de chiralité gauche possède ainsi un terme de masse le couplant à l'électron de chiralité droite, mais aussi aux muons et taus de chiralité droite, ces considérations étant faites vis-a-vis des états propres d'interaction. Regroupant sous la notation $\boldsymbol\ell^i = (e,\mu,\tau)$ les trois leptons acquérant une masse dans la brisure de symétrie, le terme de masse le plus général qu'il est nécessaire d'écrire est
\begin{equation}
\mathcal{L}_{m_l} = 
\boldsymbol{\bar\ell}_L^i \mathbf{M}^\ell_{ij} \boldsymbol\ell_R^j + \text{h.c.},
\end{equation}
où $\mathbf{M}^\ell$ est la matrice $3\times3$ de masse des leptons. Cette matrice n'est pas symétrique. En effet, les termes de masse $\bar{e}_L \cdot \mu_R$ et $\bar{\mu}_L \cdot e_R$ n'ont par exemple pas de raison d'avoir le même couplage avec le champ de Higgs (on a vu que dans la construction du Modèle Standard, les fermions de chiralités gauche et droite sont des particules \emph{a priori} distinctes), ce qui se répercute sur leur terme de masse après brisure de symétrie. 

Il est cependant possible de diagonaliser cette matrice de masse en effectuant deux transformations unitaires différentes sur les leptons de chiralités gauche et droite (les matrices de passage entre deux bases d'états quantiques doivent être unitaires afin de conserver la densité totale de probabilité). Ces matrices $\mathbf{S}_L^\ell$ et $\mathbf{S}_R^\ell$ permettent d'écrire les leptons par leurs états propre de masse, notés avec un prime :
\begin{equation}
\displaystyle{
{\boldsymbol\ell^\prime}_L^i = (\mathbf{S}_L^\ell)^i_j \boldsymbol\ell_L^j, ~~~~~ ~~~~{\boldsymbol\ell^\prime}_R^i  = (\mathbf{S}_R^\ell)^i_j \boldsymbol\ell_R^j.
}
\end{equation}
Ces transformations unitaires diagonalisant la matrice de masse, la matrice
\begin{equation}
{\mathbf{M}^\prime}^\ell = \mathbf{S}_L^\ell \mathbf{M}^\ell (\mathbf{S}_R^\ell)^\dagger,
\end{equation}
s'écrit
\begin{equation}
{\mathbf{M}^\prime}^\ell  = \text{Diag}(m_e,m_\mu,m_\tau),
\end{equation}
et le Lagrangien exprimé en fonction des états propres de masse donne
\begin{align}
\displaystyle{
\mathcal{L}_{m_\ell}} &\displaystyle{= \boldsymbol{\bar{\ell}}_L \mathbf{M}^\ell \boldsymbol\ell_R + \text{h.c.} = \boldsymbol{\bar{\ell}}_L (\mathbf{S}_L^\ell)^\dagger  \mathbf{S}_L^\ell \mathbf{M}^\ell (\mathbf{S}_R^\ell)^\dagger \mathbf{S}_R^\ell\boldsymbol\ell_R^j + \text{h.c.}\nonumber}\\
&\vspace{01.5cm}\displaystyle{=  \bar{e}_L^\prime m_e e_R^\prime + \bar{\mu}_L^\prime m_\mu \mu_R^\prime + \bar{\tau}_L^\prime m_\tau \tau_R^\prime + \text{h.c.}\nonumber}\\
&\vspace{0.5cm}\displaystyle{=  \bar{e}^\prime m_e e^\prime + \bar{\mu}^\prime m_\mu \mu^\prime + \bar{\tau}^\prime m_\tau \tau^\prime,\nonumber }
\end{align}
où la dernière ligne est écrite avec les spineurs de Dirac associés aux états propres de masse. 

\begin{figure}[h!]
\begin{center}
\includegraphics[scale=1.6]{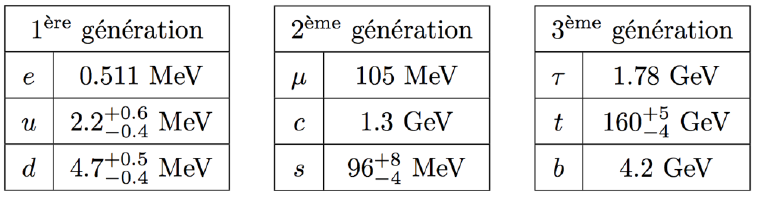}
\vspace{-0.6cm}
\end{center}
 \caption{Masses des états propres de masse des leptons et quarks du Modèle Standard après la brisure de symétrie électrofaible. Les barres d'erreurs sont omises lorsqu'elles apparaissent sur des décimales ultérieures à celles données.}
 \label{TableFermionsSMBrise}
\end{figure}

On peut procéder de façon similaire pour les quarks, sans prendre en compte leur état de couleur. Le champ de Higgs étant en effet un singlet de SU(3)$_C$, des quarks ne différant que par leur charge de couleur acquièrent des masses exactement semblables. En fait, et comme les termes de masse proviennent de la brisure de la symétrie électrofaible effectuée par un champ de Higgs chargé seulement sous cette interaction, les états propres d'interaction et de masse ne peuvent être différenciés que vis-à-vis des interactions de cette symétrie brisée : une dénomination plus précise serait donc "états propres d'interaction électrofaible". Laissant implicites les états de couleurs, il est donc possible d'introduire les états propres de masse des quarks~:
\begin{equation}
\displaystyle{
{\mathbf{d}^\prime}_L^i = (\mathbf{S}_L^d)^i_j \mathbf{d}_L^j, ~~~~~ ~~~~{\mathbf{d}^\prime}_R^i  = (\mathbf{S}_R^d)^i_j \mathbf{d}_R^j, ~~~~~ ~~~~ {\mathbf{u}^\prime}_L^i = (\mathbf{S}_L^u)^i_j \mathbf{u}_L^j, ~~~~~ ~~~~{\mathbf{u}^\prime}_R^i  = (\mathbf{S}_R^u)^i_j \mathbf{u}_R^j, 
}
\end{equation}
où l'on a généralisé les notations utilisées pour les leptons. Ces états propres donnent des termes de masse diagonaux, 
\begin{equation}
\displaystyle{
\mathcal{L}_{m_q} = \bar{u}^\prime m_u u^\prime + \bar{d}^\prime m_d d^\prime 
+ \bar{c}^\prime m_c c^\prime + \bar{s}^\prime m_s s^\prime 
+ \bar{t}^\prime m_t t^\prime + \bar{b}^\prime m_b b^\prime 
},
\end{equation}
où l'on utilise directement les spineurs de Dirac associés aux quarks. Les masses des états propres des fermions sont récapitulées dans la figure~\ref{TableFermionsSMBrise}. On voit notamment que les masses des particules augmentent avec les générations (historiquement, cette nomenclature provient simplement du fait que les particules les plus légères ont été observées en premier, la première génération étant d'ailleurs la seule stable à nos échelles d'énergie).

Les résultats expérimentaux sont exprimés en fonction des états propres de masse. Pour permettre des comparaisons avec la théorie, il est donc commode de décrire celle-ci en fonction de ces états propres de masse, et non pas des états propres d'interaction. Pour les leptons, la différence entre les deux bases d'états propres n'a en fait pas d'importance, et il est possible de travailler directement avec les états propres de masse. Cela est dû à la forme des interactions auxquelles ils sont sujets, qui ne mélangent pas de leptons massifs. Le couplage des leptons massifs avec $A_\mu^Y$ a par exemple la forme (introduisant les hypercharges données figure~\ref{TableFermionsSM}) :
\begin{equation}
\displaystyle{\mathcal{L}_{\ell,Y}= g_Y \slashed A ^Y \left( -\frac12 \boldsymbol{\bar{\ell}}_L  \boldsymbol\ell_L +  \boldsymbol{\bar{\ell}}_R  \boldsymbol\ell_R  \right),}
\end{equation}
qui est inchangée en passant dans la base des états propres de masse
\begin{align}
\displaystyle{\mathcal{L}_{\ell,Y}}&\displaystyle{= g_Y \slashed A^Y \left[ -\frac12 \boldsymbol{\bar{\ell}}_L  (\mathbf{S}_L^\ell)^\dagger  \mathbf{S}_L^\ell  \boldsymbol\ell_L +\boldsymbol{\bar{\ell}}_R  (\mathbf{S}_R^\ell)^\dagger  \mathbf{S}_R^\ell \boldsymbol\ell_R  \right],}\nonumber\\
&\displaystyle{=  g_y \slashed A ^Y \left( -\frac12 \boldsymbol{\bar{\ell}}^\prime_L  \boldsymbol\ell^\prime_L +  \boldsymbol{\bar{\ell}}^\prime_R  \boldsymbol\ell^\prime_R  \right),}
\end{align}
Le résultat est le même pour le couplage avec $W^0$, et donc aussi pour les interactions liées aux bosons $Z$ et $A_\mu^Q$. La situation est par contre plus compliquée pour les transformations liées aux bosons $W^\pm$. Rassemblant les neutrinos des différentes générations avec la notation $\boldsymbol\nu_L^i = (\nu_e,\nu_\mu,\nu_\tau)$, ces termes de couplage donnent
\begin{equation}
\mathcal{L}_{\ell,W} = g_L \left(\slashed W^-  \boldsymbol{\bar{\ell}}_L \boldsymbol\nu_L + \slashed W^+ \bar{\boldsymbol\nu}_L \boldsymbol\ell_L\right).
\end{equation}
Le passage aux états propres de masse des fermions ne laisse pas invariant ce couplage, car il est linéaire en $ \boldsymbol\ell_L$. Cependant, en définissant les états propres de masse des neutrinos (supposés sans masse dans le Modèle Standard) par la même matrice que celle définissant les états propres de masse de $ \boldsymbol\ell_L$ par rapport à leur état d'interaction, on obtient un terme de couplage avec $W^\pm$ identique pour les états propres de masse et d'interaction. En effet, en définissant 
\begin{equation}
{\mathbf{\nu}^\prime}_L^{i} = {\mathbf{S}^\ell_L}^i_j \boldsymbol\nu_L^j,
\end{equation}
le Lagrangien de couplage en fonction des états propres de masse devient
\begin{equation}
\mathcal{L}_{\ell,W} = g_L \left(\slashed W^- \boldsymbol{\bar{\ell}}^\prime_L \boldsymbol\nu^\prime_L + \slashed W^+ \bar{\boldsymbol\nu}^\prime_L  \boldsymbol\ell^\prime_L\right),
\end{equation}
qui a la même forme que pour les états propres d'interaction. Il est donc possible d'identifier \emph{a posteriori} les états propres de masse comme états propres d'interaction.

\section{Mélange entre les états propres des quarks, matrice CKM}
\label{PartCKMMatrix}

\noindent
On a vu qu'il était possible pour les leptons d'identifier les états propres de masse et d'interaction, puisqu'aucune transformation liée à une racine non nulle ne couple des particules massives distinctes. En effet, un terme de couplage $\boldsymbol\ell_L \slashed A^\alpha \boldsymbol\ell_R$ prendrait une forme différente écrit en fonction des états propres de masse, les transformations unitaires $\mathbf{S}_L^\ell$ et $\mathbf{S}_R^\ell$ étant distinctes. Or un tel terme existe pour les quarks de chiralité gauche, $\mathbf{u}_L$ et $\mathbf{d}_L$, qui sont reliés entre eux par les bosons $W^\pm$ de l'interaction électrofaible. Les interactions liées aux gluons ne jouent par contre aucun rôle dans le lien entre les différents états propres, car elles ne mélangent entre elles que des particules partageant un même terme de masse -- puisque les états propres d'interaction sont des états propres d'interaction électrofaible. Nous n'indiquerons donc pas par la suite les indices de couleur.

Le terme non invariant par passage des états propres d'interaction à ceux de masse est
\begin{equation}
\mathcal{L}_{\text{quarks},W} =  g_L \left(\slashed W^- \mathbf{\bar{d}}_L  \mathbf{u}_L + \slashed W^+  \mathbf{\bar{u}}_L\mathbf{d}_L\right).
\end{equation}
En effet, l'introduction des états propres de masse donne
\begin{equation}
\mathcal{L}_{\text{quarks},W} =  g_L \left[\slashed W^- \mathbf{\bar{d}}^\prime_L (\mathbf{S}^d_L)^\dagger  \mathbf{S}^u_L \mathbf{u}^\prime_L + \slashed W^+ \mathbf{\bar{u}}^\prime_L (\mathbf{S}^u_L) ^\dagger  \mathbf{S}^d_L \mathbf{d}^\prime_L\right],
\end{equation}
qui ne peut pas se ramener à la forme précédente, $\mathbf{S}^d_L$ et $\mathbf{S}^d_L$ étant \emph{a priori} différents. La description du lien entre les états propres de masse et d'interaction peut cependant se faire à l'aide d'une seule matrice unitaire. Partant des états propres de masse, on peut en effet définir \emph{a posteriori} les états propres d'interaction de telle façon à identifier les deux états propres de $\mathbf{u}_L$. Ces nouveaux états propres d'interaction sont,
\begin{equation}
\mathbf{u}^{\prime\prime}_L = \mathbf{u}^\prime_L, ~~~~~ \mathbf{d}^{\prime\prime}_L = \mathbf{S}^u_L (\mathbf{S}^d_L)^\dagger \mathbf{d}^\prime_L.
\end{equation}
qui donnent un couplage 
\begin{equation}
\mathcal{L}_{\text{quarks},W} =  g_L \left(\slashed W^- \mathbf{\bar{d}}^\prime_L  \mathbf{u}^\prime_L + \slashed W^+  \mathbf{\bar{u}}^\prime_L\mathbf{d}^\prime_L\right) =   g_L \left(\slashed W^- \mathbf{\bar{d}}^{\prime\prime}_L  \mathbf{u}^{\prime\prime}_L + \slashed W^+  \mathbf{\bar{u}}^{\prime\prime}_L\mathbf{d}^{\prime\prime}_L\right).
\end{equation}
La transformation entre les états propres de masse et d'interaction des quarks peut donc être décrite par la matrice
\begin{equation}
\mathbf{V}= \mathbf{S}^u_L (\mathbf{S}^d_L)^\dagger,
\end{equation}
qui est unitaire puisqu'elle s'écrit comme un produit de matrices unitaires.

Cette matrice est nommée matrice de Cabibbo lorsque deux générations de quarks sont considérées~\cite{Cabibbo:1963yz}, et matrice de Kobayashi-Maskawa lorsque trois générations de quarks sont considérées~\cite{Kobayashi:1973fv}, ou plus généralement matrice CKM (pour Cabibbo-Kobayashi-Maskawa). 
Cette matrices étant unitaire, elle peut être décrite par un certain nombre d'angles et de phases. Certaines phases peuvent cependant être absorbées dans des redéfinitions des champs fermioniques. La matrice de Cabibbo peut alors s'écrire en fonction d'un seul angle $\theta_C$, l'angle de Cabibbo, sous la forme
\begin{equation}
V = 
\left(
\begin{matrix}
   \cos \theta_C & \sin \theta_C  \\
   -\sin \theta_C & \cos\theta_C
\end{matrix}
\right).
\end{equation}
La matrice CKM pour trois générations de quarks s'écrit en fonction de trois angles $\theta_i$ et d'une phase $\delta$~:
\begin{equation}
V = 
\left(
\begin{matrix}
   c_1 & -s_1 c_3 & -s_1 s_3 \\
   s_1 c_2 & c_1 c_2 c_3 - s_2 s_3 e^{i\delta} & c_1c_2s_3+s_2c_3 e^{i\delta} \\
   s_1 s_2 & c_1 s_2 c_3 + c_2 s_3 e^{i\delta} & c_1 s_2 s_3 - c_2 c_3 e^{i\delta}
\end{matrix}
\right),
\end{equation}
où l'on a utilisé les notations condensées $c_i=\cos \theta_i$ et $s_i = \sin \theta_i$ pour $i=1,2,3$~\cite{Chau:1984fp}. Les amplitudes de ces coefficients sont à présent déterminées avec précision~\cite{PDG2016}:
\begin{equation}
V = 
\left(
\begin{matrix}
   0.97434^{+0.00011}_{-0.00012} & 0.22506 \pm 0.00050 & 0.00357 \pm 0.00015 \\
   0.22492\pm0.00050 & 0.97351\pm 0.00013 & 0.0411 \pm 0.0013 \\
0.00875^{+0.00032}_{-0.00033} &0.0403\pm 0.0013 &  0.99915 \pm 0.00005
\end{matrix}
\right).
\end{equation}
La présence d'une phase non-nulle cause une violation de la symétrie $CP$. C'en est d'ailleurs la seule cause admise actuellement dans le cadre du Modèle Standard\footnote{Dans le cadre des interactions fortes, il est permis d'écrire un terme Lagrangien de la forme 
\begin{equation}
\label{EqDefThetaCP}
\mathcal{L}_{CP} = \theta_{_{CP}} \frac{g_C^2}{32 \pi^2}F_{\mu\nu}^a \tilde{F}^{\mu\nu}_a,
\end{equation}
qui brise explicitement la symétrie $CP$. Il peut être écrit comme une dérivée totale, mais les termes de bords associés ne peuvent pas être considérés comme nuls à cause d"effets non-perturbatifs~\cite{Jackiw:1976pf,Callan:1976je,Weinberg:1977ma,Wilczek:1977pj}. À l'heure actuelle, un tel paramètre de brisure de symétrie n'a pas été mesuré, et est très contraint par les mesures de moment dipolaire électrique du neutron~:  $\theta_{_{CP}}\leq  10^{-10}$~\cite{Baker:2006ts,PDG2016}. D'un point de vue théorique, il n'a cependant aucune raison d'être nul. Le fait de déterminer si ce paramètre est nul ou pas constitue le "strong-$CP$ problem", encore non résolu. Un tel problème n'apparaît pas pour une théorie de jauge basée sur le groupe SU(2)~\cite{Donoghue:1992dd}.}.

Interprété vis-à-vis des états propres de masse, cette matrice de mélange implique que les transformations de symétrie liées aux bosons $W^\pm$ relient des particules de générations différentes. Après quantification, les coefficients anti-diagonaux de la matrice CKM impliquent que les bosons $W^\pm$ peuvent par exemple transformer des quarks $u$ en quarks $s$ ou $b$. Interprété vis-à-vis des états propres d'interaction, les transformations associées aux bosons $W^\pm$ ne vont pas modifier les générations des quarks, mais les termes non-diagonaux de la matrice de masse vont impliquer des oscillations de génération des quarks lors de leur propagation. Ces oscillations ne sont cependant pas accessibles aux échelles d'énergies où les quarks sont confinés et interagissent fortement avec des gluons (c'est le contraire pour les oscillations de génération des neutrinos pendant leur propagation, ceux-ci n'interagissant que faiblement et se propageant sur des grandes distances).

L'existence de la matrice CKM -- et donc la présence de couplages de Yukawa entre le champ de Higgs et toutes les représentations fermioniques qui le permettent -- est un marqueur clair de la pertinence de la description des particules par des représentations des groupes de symétrie. La notion de représentation a en effet été introduite afin de décrire les couplages avec les champs vectoriels, couplages qui ont la particularité de ne mélanger que des fermions d'une même représentation irréductible. D'aucun pourrait cependant penser qu'elles ne sont pertinentes que pour décrire ce type d'interaction. Or, suite à l'introduction du mécanisme de Higgs pour donner une masse aux fermions, la reproduction à l'identique des représentations fermioniques sur plusieurs générations implique des couplages inter-générationnels liés à la description des fermions par des représentations de $G_{\text{SM}}$, couplages qui sont effectivement mesurés. Cela renforce l'idée que la description des interactions par des symétries de jauge ainsi que le mécanisme de brisure spontanée de symétrie par l'introduction d'un champ de Higgs ont effectivement un caractère fondamental, puisque la matrice CKM est une conséquence de l'application coïncidente de ces deux descriptions.

Au-delà des spécificités techniques de la matrice CKM, qui ne nous servira pas particulièrement par la suite, cette matrice justifie l'introduction de couplages dans le Lagrangien entre toutes les représentations qui permettent de créer des singlets. Dans le cadre du Modèle Standard, construit à partir de termes renormalisables, cette propriété n'est visible que sur les couplages de Yukawa, liés à la matrice CKM. Cela conforte l'idée qu'en construisant des théories renormalisables, il est toujours nécessaire de considérer tous les couplages qu'il est possible d'écrire du point de vue de la théorie des groupes. Les constantes de couplage de ces interactions sont alors \emph{a priori} complexes, prenant en compte le fait qu'ils doivent assurer une évolution unitaire, comme on le montre par l'existence d'une phase dans la matrice CKM. 

Le fait de considérer tous les couplages possibles donne également des informations sur la possibilité d'inclure une quatrième génération de fermions dans le Modèle Standard. En effet, en présence d'une quatrième (au moins) génération de fermions, la matrice CKM ne serait qu'un sous-bloc d'une matrice unitaire, et n'aurait par conséquent aucune raison d'être elle-même unitaire. Or, la matrice CKM vérifie particulièrement bien les conditions d'unitarité, à savoir $V^\dagger V = V V^\dagger = \mathds{1}$, ce qui implique notamment que la somme des carrés des amplitudes des composantes d'une même ligne ou colonne vaut 1. Les mesures donnent par exemple
\begin{equation}
\begin{array}{l}
|V_{11}|^2 + |V_{12}|^2 + |V_{13}|^2 = 1.00000 \pm 0.00031,\\
|V_{21}|^2 + |V_{22}|^2 + |V_{23}|^2 =  1.00000 \pm 0.00036,\\
|V_{31}|^2 + |V_{32}|^2 + |V_{33}|^2 =  1.00000 \pm 0.00015,
\end{array}
\end{equation}
les écarts à l'unité apparaissant dans le calcul sont au minimum à l'ordre $10^{-6}$ (l'annulation des décimales jusqu'à cette ordre n'est cependant pas représentative, puisqu'au delà des incertitudes expérimentales). Ces vérifications \emph{a posteriori} sur les résultats expérimentaux confortent encore la pertinence de la description théorique utilisée jusqu'à présent, en plus d'être une indication claire de l'existence de seulement trois générations de fermions.

\section{Renormalisation des constantes de couplage, phénoménologie du Modèle Standard}
\label{PartRenormalisationPhenomSM}

\noindent
Nous avons pour l'instant détaillé les particules élémentaires du Modèle Standard. Or, si les leptons apparaissant dans cette classification sont bien observés expérimentalement, ce n'est pas le cas des quarks. Au contraire, une zoologie importante de particules subissant l'interaction forte est observée. L'explication de cette phénoménologie est contenue dans la variation en fonction de l'échelle d'énergie des constantes de couplage associées aux différentes interactions. Ces variations, liées à la renormalisation des théories quantiques des champs discutée dans la section~\ref{PartQuantification}, peuvent être étudiées de façon perturbative. À l'ordre dominant, les variations de ces constantes de couplage sont logarithmiques. La charge électrique, constante de couplage de l'électrodynamique quantique, varie par exemple comme~\cite{Donoghue:1992dd}
\begin{equation}
e^2 (\mu_2) = e^2 (\mu_1) + \frac{e_0^4}{12 \pi^2} \ln \frac{\mu_2^2}{\mu_1^2},
\end{equation}
où $e_0$ est la valeur nue de la charge électrique apparaissant dans le Lagrangien de l'électrodynamique quantique, et $\mu_1$ et $\mu_2$ deux échelles d'énergie. Dans la pratique, on préfère décrire les variations des constantes de couplages par les "$\beta$-fonctions" associées (aussi appelées fonctions de Callan-Symanzik)
\begin{equation}
\beta = \mu_R \frac{\partial g}{\partial \mu_R} = \frac{\partial g}{\partial \ln \mu_R},
\end{equation}
qui ont l'avantage de ne pas dépendre explicitement de l'échelle d'énergie considérée. 
Par exemple, pour l'électrodynamique quantique, cette fonction vaut~\cite{Gell-Mann:1956iqa}
\begin{equation}
\beta(e) = \frac{e^3}{12 \pi^2} + \mathcal{O}(e^5).
\end{equation}
On peut de même calculer la $\beta$-fonction de la chromodynamique\footnote{Pour une théorie de jauge non-abélienne impliquant $n_f$ spineurs de Dirac dans une représentation $R_f$, la $\beta$-fonction vaut à l'ordre dominant~\cite{Peskin:1995ev}
\begin{equation}
\label{EqCallanSymanzik}
\beta(g)=-\frac{g^3}{16\pi^2}\left[\frac{11}{3}C_2(R_{\text{adj}})-\frac{4}{3}n_f C_2(R_f)\right],
\end{equation}
où $C_2$ correspond à la valeur de l'opérateur de Casimir quadratique sur la représentation de chaque champ (voir la discussion de la section~\ref{Poids&Racines} sur la possibilité de classifier les représentations par la valeur de l'opérateur de Casimir quadratique sur celles-ci). La première contribution est celle des bosons de jauge, qui sont dans la représentation adjointe du groupe de jauge. La valeur du Casimir quadratique de SU($N$) valant $N$ dans sa représentation adjointe, et $\frac12$ dans sa représentation fondamentale, on retrouve l'équation~\eqref{BetaQCD} pour une théorie de jauge basée sur le groupe SU(3)$_C$ et comprenant 6 spineurs de Dirac.
}, qui vaut
\begin{equation}
\label{BetaQCD}
\beta(g_{_C}) = -\frac{g_{_C}^3}{4\pi^2}\left(\frac{11}{4} - \frac{2}{3}n_f\right) + \mathcal{O}(g_{_C}^5),
\end{equation}
avec $n_f$ le nombre de fermions décrits par leur spineurs de Dirac, et ici dans le cas où la masse des quarks est négligeable~\cite{Politzer:1973fx,Gross:1973id}.

Ces $\beta$-fonctions décrivant l'évolution en fonction de l'échelle d'énergie des constantes de couplage sont très utiles pour décrire la phénoménologie des interactions associées. Celles-ci décrivent en effet les interactions effectives prenant en compte les fluctuations quantiques de la théorie et les interactions avec les particules virtuelles pouvant apparaître. Ainsi, l'intensité du couplage électromagnétique augmente avec l'énergie, et diminue avec la distance séparant deux charges. Cela peut s'interpréter par le fait que les paires virtuelles électron-positron -- ou plus généralement de tout type de fermion chargé électriquement -- qui apparaissent entre deux particules chargées vont se polariser, et créer un effet d'écran entre ces particules. Une plus grande distance entre deux particules autorise l'interaction avec plus de paires virtuelles, et un écrantage plus important. La $\beta$-fonction de la chromodynamique quantique comporte elle-aussi une contribution positive, liée aux quarks qui peuvent causer un effet similaire d'écrantage des charges de couleur. Il apparaît cependant aussi une contribution négative, liée aux bosons de jauge qui interagissent entre eux dans une théorie de jauge non-abélienne. Ces bosons de jauge, dont la contribution est dominante en présence de seulement trois générations de fermions pour le Modèle Standard, impliquent alors un mécanisme d'anti-écrantage des  interactions fortes, dont l'intensité augmente avec la distance.

Le mécanisme d'anti-écrantage croissant avec la distance décrit par la $\beta$-fonction de la QCD explique la propriété de confinement des interactions fortes.
En effet, lors de la séparation de deux particules portant des charges de couleur non-nulles, l'interaction causée par l'apparition et l'interaction d'un grand nombre de gluons virtuels entre ces particules va augmenter considérablement la force de l'interaction forte, empêchant la séparation des particules, ou menant à l'apparition d'une paire quark-antiquark. En conséquence, seuls des états liés de quarks, de charges de couleur nulles, peuvent être observés à nos échelles d'énergie, à l'opposé des leptons qui ne forment pas d'états liés. Ces états liés, formant les hadrons, sont nommés mésons pour ceux formés de deux quarks, et baryons pour ceux formés de trois quarks. Réciproquement, la $\beta$-fonction de la QCD implique que pour des échelles d'énergie importantes, et donc sur des petites échelles de distance, les interactions fortes vont devenir très faibles. Cette propriété est nommée liberté asymptotique des interactions fortes. Cela implique que la QCD peut être décrite perturbativement dans le cadre de la liberté asymptotique, mais qu'une approche non-perturbative doit être menée pour expliquer la structure des hadrons à partir de quarks, avec l'utilisation notamment de méthodes numériques liées à l'étude de la QCD sur réseau, ou l'utilisation de théories effectives des champs.

\begin{figure}[h!]
\begin{center}
\includegraphics[scale=1.35]{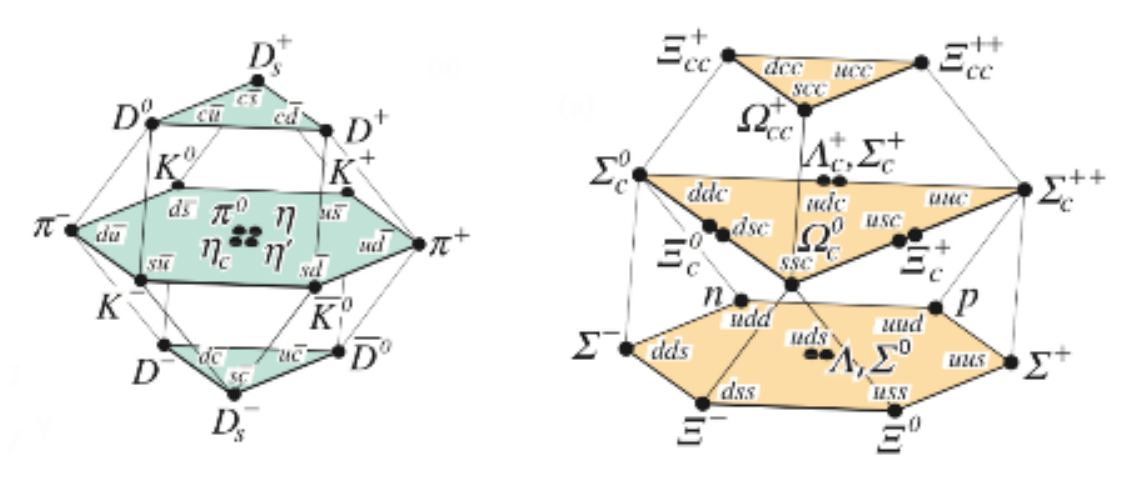}
\vspace{-0.6cm}
\end{center}
 \caption{Localisation de certains hadrons dans les représentations de la symétrie SU(4) de saveurs des quarks, tiré de~\cite{PDG2016}. La figure de gauche est le 16-tuplet de mésons pseudo-scalaires, et la figure de droite le 20-tuplet de baryons contenant un octet de SU(3). On peut identifier sur les plans horizontaux les représentations de SU(3) données dans la figure~\ref{DiagPoidsSU3}.}
 \label{QuarksSaveurs}
\end{figure}

Sans entrer dans les détails, on notera que les outils de la théorie des groupes peuvent être utilisés pour décrire de façon performante les hadrons, observant que les quarks apparaissent de façon exactement similaire dans le Modèle Standard si on omet leur masse et leurs interactions via SU(2)$_L\times$U(1)$_Y$. Or, l'interaction forte est bien dominante dans la description de ces quarks, puisque les interactions électromagnétiques et faible ont des contributions respectives 60 à $10^5$ fois plus faible à l'échelle des quarks. Il est par exemple possible de rassembler les quarks des deux premières générations dans une représentation de dimension 4 d'un groupe de symétrie global SU(4), nommée symétrie de saveur\footnote{C'est une symétrie explicitement brisée, puisque qu'elle ne laisse pas invariant le Lagrangien du Modèle Standard. Cependant, elle permet une bonne description dans la limite où les termes qui la brisent, à savoir ici les interactions électrofaibles et les masses des quarks, peuvent être considérés comme des perturbations.}. Les mésons et baryons formés à partir de ces quarks se placent alors dans les représentations de SU(4) apparaissant dans les produits des représentations des quarks. La figure~\ref{QuarksSaveurs} donne deux exemples de positionnement de hadrons dans de telles représentations. Il est important de bien différencier les représentations des symétries de jauge considérées jusqu'à présent et les représentations des groupes de symétries globales liés aux saveurs, qui décrivent des concepts complètement différents.

Les interactions entre hadrons (comme les protons et neutrons) ne sont pas décrites par des échanges de gluons car ces fermions ont des charges de couleur identiquement nulles suite aux propriétés de confinement des interactions fortes. Les interactions fortes sont alors véhiculées par des hadrons, principalement des mésons. Ainsi, alors que les interactions entre les quarks $uud$ et $udd$ formant les protons et neutrons sont véhiculées par les gluons, les interactions entre neutrons et protons sont  associées à des échanges de mésons, principalement des pions. L'étude des interactions résiduelles de la chromodynamique quantique aux échelles atomiques est du ressort de la physique nucléaire, traitée principalement par des théories effectives ; nous ne la décrirons pas plus avant. Concernant les interactions faibles, elles peuvent être traitées comme des perturbations aux échelles des particules élémentaires et des hadrons, étant alors de $10^4$ à $10^7$ fois plus faible que l'électromagnétisme. La phénoménologie de ces interactions, associées à une symétrie brisée, est décrite dans la section~\ref{PhenomSymBrisées}.

\begin{figure}[h!]
\begin{center}
\includegraphics[scale=1.5]{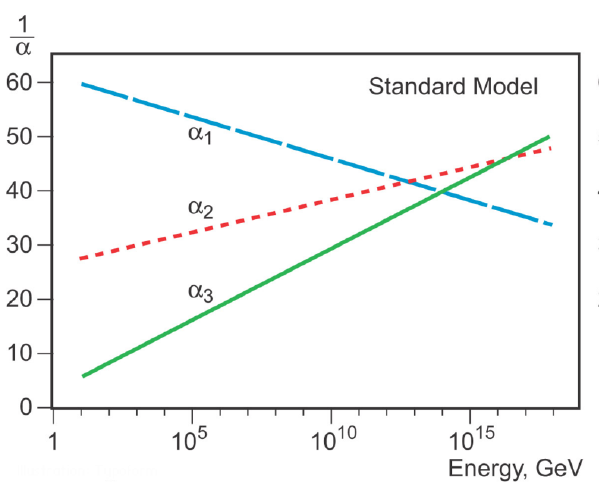}
\vspace{-0.6cm}
\end{center}
 \caption{Renormalisation des constantes de couplages associées aux trois composantes simples ou abéliennes de $G_{\text{SM}}$, issue de~\cite{Nobel}. Les constantes de couplage $\alpha_1$, $\alpha_2$ et $\alpha_3$ sont associées respectivement à U(1)$_Y$, SU(2)$_L$ et SU(3)$_C$.}
 \label{AlphaRunning}
\end{figure}

À haute énergie, lorsque la théorie électrofaible n'est pas brisée, il est nécessaire de considérer la renormalisation des trois constantes de couplages associées aux trois composantes de $G_{\text{SM}}$. La constante de couplage de U(1)$_Y$ augmente alors avec l'échelle d'énergie, alors que les constantes de SU(2)$_L$ et SU(3)$_C$ diminuent. Cette évolution, représentée figure~\ref{AlphaRunning}, reste valide tant qu'aucune nouvelle physique supplémentaire au Modèle Standard n'apparaît.

\section{Succès et défauts du Modèle Standard}
\label{PartConclusionSM}

\begin{figure}[h!]
\begin{center}
\includegraphics[scale=1.5]{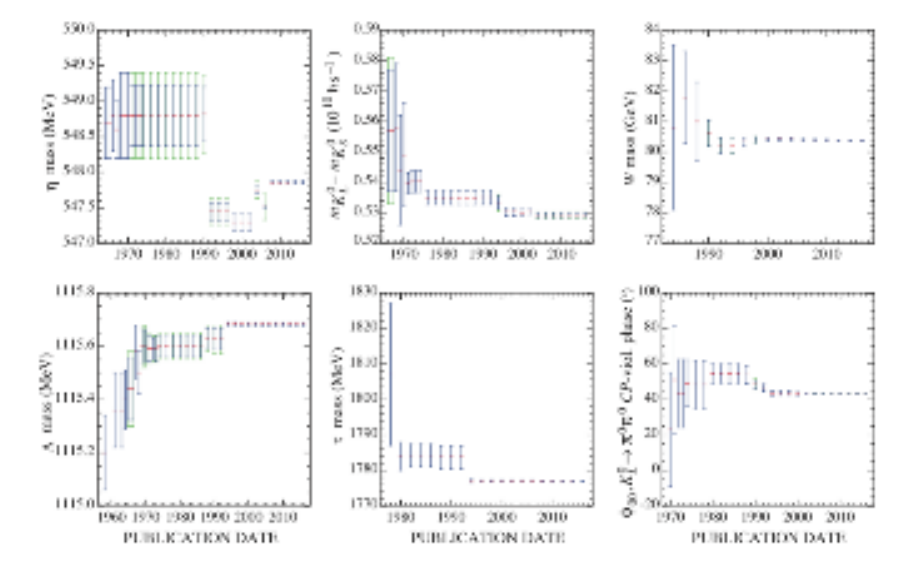}
\vspace{-0.6cm}
\end{center}
 \caption{Évolution au cours du temps de la précision de certaines mesures expérimentales liées au Modèle Standard, tiré de~\cite{PDG2016}.}
 \label{PrecisionSM}
\end{figure}

\noindent
Malgré ses quelques défauts que nous allons détailler ci-dessous, le Modèle Standard représente un important succès de la physique moderne. Expérimentalement, il a été conforté par la prédiction et la détection du quark top en 1995~\cite{Abachi:1994td,Abe:1995hr}, du neutrino tauique en 2000~\cite{Kodama:2000mp}, et du boson de Higgs en 2012~\cite{Aad:2012tfa,Chatrchyan:2012xdj}. Il est également vérifié expérimentalement pour la plupart de ses prédictions avec un grand degré de précision, allant jusqu'à dix chiffres significatifs. Par exemple, le calcul du moment magnétique anomal de l'électron donne~\cite{Aoyama:2012wj,Aoyama:2014sxa}
\begin{equation}
{\displaystyle a_{e}^{\text{SM}}=0.001\;159\;652\;181\;643(\pm 764)},
\end{equation}
alors que les données expérimentales les plus récentes donnent~\cite{Hanneke:2010au,PDG2016}
\begin{equation}
 {\displaystyle a_{e}^{\text{exp}}=0.001\;159\;652\;180\;91(\pm 26)},
\end{equation}
les valeurs étant compatibles aux barres d'erreur près, et la précision étant de l'ordre du milliardième.
À l'heure actuelle, et alors que la précision des résultats expérimentaux augmente significativement avec le temps (voir figure~\ref{PrecisionSM}), il n'y a pas de signal remettant définitivement en cause les prédictions du Modèle Standard dans son domaine d'application (laissant à part la masse des neutrinos), c'est à dire ayant des incompatibilités statistiques à plus de 5 écarts-types. Étant donné que des centaines d'expériences comportant un traitement statistique ont été et sont encore comparées avec ses prédictions, il est normal qu'à un instant donné certaines d'entre elles dévient légèrement des prédictions théoriques, avant d'y revenir suite à l'acquisition de donnée supplémentaires\footnote{Le contraire indiquerait plutôt une surestimation des erreurs des différents résultats d'expériences.} ; il est donc nécessaire de traiter avec beaucoup de précaution les déviations du Modèle Standard observées dans les expériences.

On notera cependant le problème persistant du résultat des mesures du moment magnétique anomale du muon. Les calculs du Modèle Standard donnent en effet une valeur de \cite{PDG2016,Blum:2013xva}
\begin{equation}
{\displaystyle a_{\mu}^{\text{SM}}=0.001\;165\;918\;03(\pm 49)},
\end{equation}
pour des résultats expérimentaux menant à \cite{Bennett:2006fi,PDG2016}
\begin{equation}
 {\displaystyle a_{\mu}^{\text{exp}}=0.001\;165\;920\;91(\pm 63)},
\end{equation}
soit un écart à 3.6 écarts-types. Une augmentation de la précision expérimentale de cette mesure est prévue pour 2018. D'autre part, des résultats de l'expérience BaBar obtenus en 2012, sont aussi en désaccord avec les prédictions du Modèle Standard, une désintégration particulière étant observée plus fréquemment qu'attendu~\cite{Lees:2012xj}. Ces résultats, ayant un écart de 3.4 écarts-types avec les prédictions du Modèle Standard, ont été confortés par des mesures aux LHC en 2015, amenant le désaccord théorie-expérience à 3.6 écarts-types~\cite{Aaij:2015yra}.

Du point de vue théorique, le Modèle Standard de la physique des particules unifie dans un seul cadre théorique la description de trois des quatre interactions fondamentales. De plus, on a vu dans ce document comment la construction du Modèle Standard apparaît de façon plutôt naturelle si on la dérive à partir d'une action se basant sur des symétries formulées dans le cadre de la théorie des groupes et n'utilisant que des termes renormalisables, ces différents concepts restreignant considérablement les modèles théoriques qu'il est possible d'écrire. Cela nous mène à croire que ces concepts théoriques décrivent certaines propriétés fondamentales de la physique, et ne se résument pas uniquement à une description effective des phénomènes étudiés (voir notamment la discussion de la section~\ref{PartCKMMatrix}). Cela est aussi un sérieux encouragement à étudier les théories au-delà du Modèle Standard en tout premier lieu avec les mêmes outils que ceux qui se sont montrés aussi fructueux pour décrire les échelles d'énergie jusqu'à la centaine de GeV.

En plus des résultats discutés dans les paragraphes précédents, les neutrinos représentent un cas particulier de désaccord entre les prédictions du Modèle Standard et les observations expérimentales. En effet, le phénomène d'oscillation de neutrinos, à savoir le changement de type de neutrinos lors de leur propagation, n'est pas décrit dans le cadre du Modèle Standard tel que présenté dans cette section. Ce phénomène est très bien observé par différentes expériences, que ce soit pour les neutrinos solaires~\cite{Anselmann:1992um,Fukuda:1996sz,Cleveland:1998nv} ou atmosphériques~\cite{Eguchi:2002dm}, et ceux produits en réacteurs~\cite{Fukuda:1998mi} ou dans des accélérateurs~\cite{Ahn:2006zza,Michael:2006rx}. Le phénomène d'oscillation de neutrinos~\cite{Pontecorvo:1957cp,Pontecorvo:1957qd,Maki:1962mu} est une conséquence directe de la distinction entre les états propres de masse et d'interaction de ces particules, comme développé dans les sections~\ref{PartMasseFermions} et~\ref{PartCKMMatrix} ; il ne peut cependant pas s'expliquer dans le cadre du Modèle Standard, qui n'associe pas de termes de masse aux neutrinos. La méthode la plus simple pour ajouter une masse aux neutrinos consiste en l'ajout d'un couplage entre les neutrinos de chiralité gauche du Modèle Standard, et des neutrinos stériles de chiralité droite. Ces derniers neutrinos pouvant supporter un terme de masse de Majorana lié à des échelles de haute énergie, un mécanisme de see-saw~\cite{Minkowski:1977sc,Mohapatra:1979ia,GellMann:1980vs} (mécanisme à bascule) est alors possible, expliquant la faiblesse des masses des neutrinos de chiralité gauche, de l'ordre de l'eV.

Bien que cela ne soit pas contenu dans le Modèle Standard tel qu'il a été présenté dans ce chapitre, il est donc aisé d'ajouter une masse aux neutrinos par la prise en compte d'une matrice de masse faisant le lien entre les états propres de masse et d'interaction des neutrinos. Cette matrice est nommée matrice de Pontecorvo–Maki–Nakagawa–Sakata, ou matrice PMNS~\cite{Pontecorvo:1957cp,Pontecorvo:1957qd,Maki:1962mu}. Dans le cas où les neutrinos de chiralité droite n'ont pas de terme de masse de Majorana, la matrice PMNS est décrite par trois angles et une phase, comme la matrice CKM présentée dans la section~\ref{PartCKMMatrix}. Dans le cas où l'on considère des masses de Majorana pour les neutrinos stériles, la matrice PMNS doit être décrite par deux phases additionnelles. Selon les appellations, le Modèle Standard inclut ou n'inclut pas les masses des neutrinos, et on ne considère en tout cas pas ces termes comme une remise en question du Modèle Standard initial, mais plutôt comme son extension naturelle.

Le Modèle Standard possède aussi des défauts du point de vue théorique. Sans prétendre introduire une distinction exacte, ces défauts se classent principalement en deux catégories. Dans une première classe, et admettant la description théorique du Modèle Standard (notamment les théories de jauge et le mécanisme de Higgs), on peut se demander s'il est possible d'expliquer le contenu en symétries et en champs de ce modèle, ainsi que la valeur de ses différents paramètres. Pourquoi le groupe de jauge est-il SU(3)$_C\times$SU(2)$_L\times$U(1)~? Y a-t-il une explication à la répartition particulière des fermions dans les représentations du groupe de jauge~? Pourquoi les fermions apparaissent dans trois générations identiques, qui plus est de masses croissantes~? Sans prétendre répondre au questionnement absolu de pourquoi l'électron possède la masse et la charge mesurées, il peut être possible de faire un lien entre les propriétés des différentes particules du Modèle Standard, de la même façon que le Modèle Standard fait un lien entre les nombres quantiques des particules dans une même représentation irréductible. Et le contenu en champ du Modèle Standard contient pour cela des indications fortes d'une relation entre les différentes représentations des fermions. Citons le fait que la charge électrique soit quantifiée et commensurable entre les différentes représentations, ou que la somme des hypercharges $Y$ ainsi que celle de leurs cubes $Y^3$ s'annulent sur chaque génération de fermions du Modèle Standard, permettant notamment d'écrire une théorie libre d'anomalies quantiques\footnote{Après quantification, certaines lois de conservation valides pour la théorie classique peuvent ne plus être vérifiées. Cette violation des lois de conservation au niveau quantique est nommée anomalie. En pratique, des anomalies apparaissant pour un sous-ensemble de particules d'une théorie peuvent s'annuler pour la théorie complète. C'est le cas du Modèle Standard, où le calcul de l'intégralité des diagrammes de Feynman triangulaires associés à certains processus mène à une annulation des anomalies. Cela est permis par la structure des représentations utilisées dans le Modèle Standard, notamment le fait que la somme des valeurs de $\text{Tr}\left( T_\alpha \left\{T_\beta,T_\gamma\right\} \right)$ sur l'ensemble des représentations fermioniques s'annule pour tous les générateurs du Modèle Standard. Effectuant une somme pour $\beta$ et $\gamma$ sur les générateurs de SU(3)$_C$ ou SU(2)$_L$ vérifiant $\sum T_\beta T_\gamma = \frac12 \delta_{\beta\gamma}$, et prenant $\alpha=Y$, on obtient par exemple que la somme des hypercharges faibles de tous les fermions du Modèle Standard doit être nulle, ce qui se vérifie sur la première génération en prenant les données de la figure~\ref{TableFermionsSM}. De même, prenant $\alpha=\beta=\gamma=Y$, on obtient que la somme des cubes des hypercharges faibles de tous les fermions du Modèle Standard s'annule aussi, ce qu'on vérifie également~! On peut vérifier des résultats similaires pour les trois autres nombres quantiques du Modèle Standard, ceux-ci étant en fait vérifiés pour chaque représentation individuelle.}. On peut de même se questionner sur l'indépendance des 19 paramètres libres du Modèle Standard [3 masses pour les leptons, 6 pour les quarks, 3 couplages de jauge, trois angles et une phase pour la matrice CKM, une masse et un couplage pour le champ de Higgs, et l'angle $\theta_{CP}$ de la QCD défini dans l'équation~\eqref{EqDefThetaCP}], auxquels il faut ajouter les 6 paramètres de la matrice PMNS. Ces points sont traités dans le cadre des théories de grande unification, ou GUT (pour "Grand Unified Theories"), discutés dans le chapitre~\ref{PartGUT}.

D'autres défauts du Modèle Standard questionnent plutôt les fondements théoriques de celui-ci. Le premier est le problème de la hiérarchie: le fait que la masse du champ de Higgs doive être renormalisée au premier ordre en carré de l'énergie \emph{a priori} jusqu'à l'échelle de Planck, demandant un ajustement très important de sa masse nue (à $10^{-17}$ près). Ce problème, provenant de la grande différence entre les échelles d'énergie liées à la gravité (l'énergie de Planck) et à la physique des particules (la centaine de GeV) apparaît seulement pour le champ de Higgs\footnote{Les champs de jauge et les fermions sont en quelque sorte "protégés" par les symétries, étant sans masse au-delà de l'échelle électrofaible.}, et questionne sa pertinence pour décrire une théorie physique. Vient aussi la difficulté d'incorporer la gravité dans le cadre de la théorie quantique des champs, la Relativité Générale étant non-renormalisable, et les problèmes liés à l'énergie du vide quantique comme source de la constante cosmologique (la valeur calculée dans le Modèle Standard étant de l'ordre de $10^{50}$ fois plus grande que celle obtenue via la cosmologie observationnelle~\cite{Martin:2012bt,Burgess:2013ara}). Ces problèmes sont traités dans le cadre de théories remettant en cause certains fondements de la construction du Modèle Standard. Mentionnons notamment la supersymétrie~\cite{Martin:1997ns,Weinberg:2000cr,Binetruy:2006ad}, qui résout entre autre le problème de la hiérarchie, et son extension locale nommée supergravité qui incorpore la gravitation~\cite{deWit:2002vz}. D'autres théories, comme la théorie des cordes~\cite{Polchinski:1998rq,Polchinski:1998rr,Zwiebach:2004tj} ou la gravité quantique à boucles~\cite{Rovelli:2014ssa}, tentent quant à elles de décrire la gravité dans un cadre quantique. Nous ne décrirons pas ces théories dans le présent document.

Notons finalement d'autres problèmes ne rentrant pas forcément dans une des deux catégories précédentes, à savoir le problème de la matière noire, l'asymétrie matière-antimatière observée (qui n'est pas expliquée par la violation de $CP$ trop faible dans le Modèle Standard), ou le "strong $CP$-problem" expliqué après l'équation~\eqref{EqDefThetaCP}.

\chapter{Théories de Grande Unification}
\label{PartGUT}

\section{Motivations et concepts}

\noindent
Les théories de grande unification, ou GUT, tentent d'unifier le Modèle Standard dans un nombre réduit de groupes de jauge et de leurs représentations. Elles utilisent les outils théoriques discutés dans la partie~\ref{ChapterTheoriesDeJauges}, qui ont fait leur preuve dans le cadre du Modèle Standard. Cette approche est justifiée par la structure particulière des représentations de fermions du Modèle Standard. En effet, et comme discuté dans le chapitre précédente, il n'y a par exemple aucune raison \emph{a priori} pour que les hypercharges faibles des différentes représentations de fermions soit commensurables, ce qui sous-tend un lien entre les différentes représentations. De même, le fait que les sommes des valeurs de $T_Y$ ou de $T_Y^3$ soient nulles sur chaque génération de fermions semble indiquer que les représentations d'une même génération forment une structure dans leur ensemble, ce qui apparaît naturellement si chaque génération est décrite par une seule représentation.

Le point de départ de la construction de GUTs est donc le groupement de plusieurs représentations fermioniques du Modèle Standard dans un petit nombre de représentations, une seule dans l'idéal. Cette construction n'est pas possible sans étendre le groupe de symétrie $G_{\text{SM}}$, puisque des représentations de plus grande dimension que celles déjà utilisées introduiraient des nombres quantiques ne rentrant pas dans les représentations du Modèle Standard (les valeurs extrêmes des poids d'une représentation augmentant avec la dimension de celle-ci). Il est donc nécessaire d'introduire un groupe de jauge étendu, $G_{\text{GUT}}$, ayant $G_{\text{SM}}$ comme sous-groupe. La brisure de ce groupe de jauge par un champ de Higgs fournit alors un mécanisme naturel pour décomposer les représentations de $G_{\text{GUT}}$ sur les représentations de $G_{\text{SM}}$. Le Modèle Standard a alors le même statut vis-à-vis de la théorie de grande unification que le Modèle Standard après brisure électrofaible par rapport au Modèle Standard complet.

\begin{figure}[h!]
\begin{center}
\includegraphics[scale=1.1]{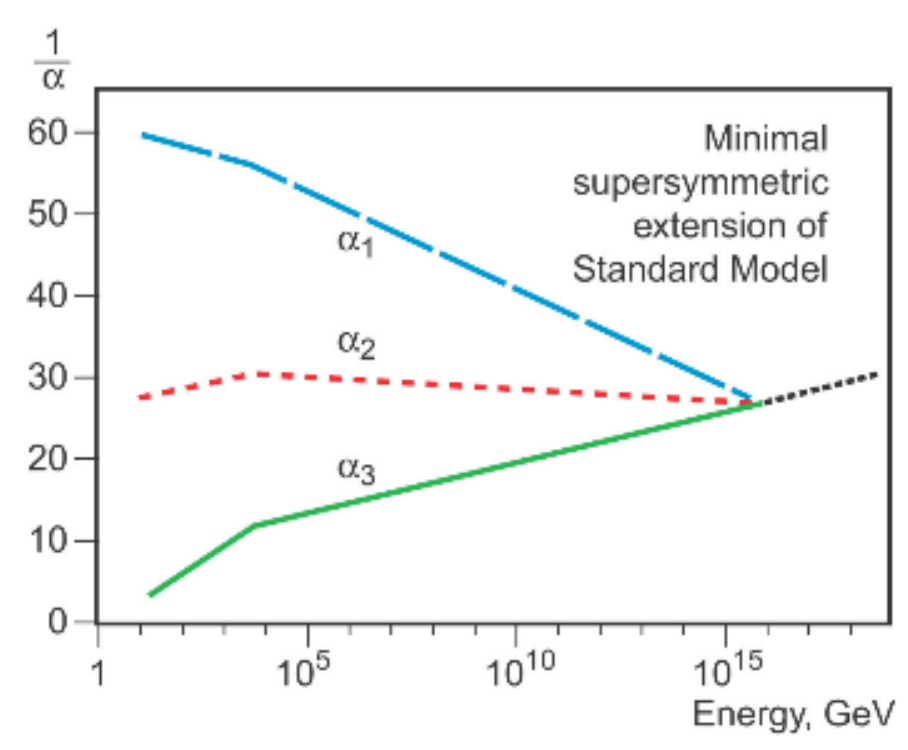}
\vspace{-0.6cm}
\end{center}
 \caption{Renormalisation des constantes de couplage associées aux trois composantes simples ou abéliennes de $G_{\text{SM}}$, dans le cadre de l'extension supersymétrique minimale du Modèle Standard, issue de~\cite{Nobel}. Les constantes de couplage $\alpha_1$, $\alpha_2$ et $\alpha_3$ sont associées respectivement à U(1)$_Y$, SU(2)$_L$ et SU(3)$_C$.}
 \label{AlphaRunningSUSY}
\end{figure}

C'est cette unification des représentations fermioniques comme des groupes de jauge qui octroie un pouvoir prédictif important vis-à-vis du Modèle Standard. En effet, une seule constante de couplage est suffisante pour décrire l'action d'un groupe de jauge simple. De même, très peu de couplages de Yukawa différents peuvent être écrits si les fermions sont rassemblés dans une ou deux représentations. Les différentes constantes du Modèle Standard apparaissent alors dans les règles de branchement numérique -- les coefficients de Clebsch-Gordan -- des différentes composantes de $G_{\text{SM}}$ dans $G_{\text{GUT}}$ comme des représentations fermioniques du Modèle Standard dans celles des théories de grande unification. Les théories de grande unification sont donc plus contraintes théoriquement que le Modèle Standard, mais ont pour cette raison un pouvoir prédictif beaucoup plus important. Les grandeurs prédites pour le Modèle Standard ne sont cependant pas celles obtenues directement par les règles de branchement, puisqu'il il est nécessaire de les renormaliser préalablement des échelles du GUT jusqu'à celles du Modèle Standard. 

Sans faire de calculs détaillés pour une théorie de grande unification donnée, il est possible d'estimer l'échelle d'énergie de brisure des théories de grande unification à $10^{15-16}$ GeV. En effet, l'étude de la renormalisation des constantes de couplage du Modèle Standard à haute énergie (voir la section~\ref{PartRenormalisationPhenomSM}) indique une convergence de ces couplages vers ces échelles d'énergie. Cette convergence est d'ailleurs grandement améliorée dans le cadre de l'extension supersymétrique minimale de Modèle Standard, comme observé dans la figure~\ref{AlphaRunningSUSY}. Ces valeurs sont donc couramment utilisées pour obtenir des ordres de grandeurs sans spécifier une théorie de grande unification en particulier.

La construction de théories de grande unification se fait en deux étapes. Dans un premier temps, il est nécessaire d'identifier un groupe de jauge de grande unification ainsi que ses représentations qui peuvent contenir les fermions d'une génération. Cette étape, très contrainte par les propriétés des groupes compacts, est discutée dans la section~\ref{PartGroupeUnification}. Le plongement des représentations fermioniques du Modèle Standard dans la représentation de dimension 16 de SO(10) est présenté dans la section~\ref{PartRep16SO10}, et les décompositions de cette représentation sur les sous-algèbres maximales de SO(10) sont discutées dans le chapitre~\ref{PartRep16SousAlgebre}. Dans un deuxième temps, et une fois un groupe de jauge et une représentation fermionique choisis, les théories de grande unification sont construites en explicitant leur contenu en champs de Higgs. Ces champs de Higgs doivent causer la brisures des symétries de grande unification et électrofaible, et permettre de retrouver le Modèle Standard à basse énergie. Cette construction théorique est discutée dans le cas d'un groupe d'unification SO(10) dans la section~\ref{PartGUTHiggsSector}, suivie par une conclusion sur le statut actuel des théories de grande unification.

\section{Groupe d'unification et désintégration du proton}
\label{PartGroupeUnification}

\noindent
Les différents groupes compacts et leurs représentations étant classifiés -- voir le chapitre~\ref{PartRacinesPoids} --, l'investigation des possibles groupes de grande unification et des représentations fermioniques associées peut être faite de façon systématique. De plus, et gardant les hypothèses utilisées pour construire le Modèle Standard, un certain nombre de conditions sur les groupes d'unification potentiels permet de restreindre le champ d'investigation.
\begin{itemize}
\item[i)] Le groupe de grande unification devant contenir $G_{\text{SM}}$ comme sous-groupe~: il doit être au moins de rang $n_{\text{GUT}}\geq 4$, et contenir un sous-groupe SU(3).
\item[ii)] La renormalisation des constantes de couplage à partir des données de basse énergie semble indiquer une convergence vers une seule valeur $g_{\text{GUT}}$. Les groupes associés à une seule constante de couplage sont soit simples, soit des produits symétrisés d'un groupe simple, comme 
$\left[\text{SU}(N)\right]^3$, où les constantes de couplage des trois groupes peuvent être supposées identiques.
\item[iii)] Pour écrire une théorie quantique viable, il est nécessaire de ne considérer que des représentations unitaires. Cela exclut les groupes de symétrie n'ayant que des représentations non-unitaires, comme SO(11)~\cite{PeterUzan}.
\item[iv)] Le Modèle Standard introduit des symétries chirales, agissant différemment sur les fermions de chiralité droite et gauche. Cela est permis par l'utilisation de représentations complexes, différenciant explicitement les états d'hélicités différentes\footnote{Lorsque ce n'est pas le cas, un fermion et son anti-particule sont dans la même représentation du groupe de symétrie interne. Il est alors toujours possible de décrire des phénomène chiraux, mais cela demande l'introduction de champs supplémentaires pour briser la symétrie chirale~\cite{Georgi:1979md}. Cela n'est pas le cas dans le Modèle Standard du fait notamment de la présence d'un groupe de symétrie abélien, dont les représentations non-triviales sont complexes.}. Il semble donc naturel de grouper les fermions dans des représentations complexes du groupe de jauge~\cite{PhysRevLett.32.438,GellMann:1976pg,GellMann:1980vs}. Les seuls groupes simples possédant de telles représentations sont SU($N$) pour $N\geq3$, SO(4$N$+2), et $E_6$~\cite{Metha1,Metha2}.
\item[v)] Les fermions doivent être groupés dans des représentations garantissant une annulation des anomalies quantiques. C'est le cas pour toutes les représentations de SO($N$) pour $N\geq7$ et $E_6$~\cite{Gursey:1975ki,Okubo:1977sc}. Ce n'est par contre pas le cas pour l'intégralité des représentations de SU($N$), qui doivent pour certaines d'entre elles être groupées par représentations de sommes d'anomalies nulles -- celles-ci étant classifiées~\cite{Okubo:1977sc,Banks:1976yg}.
\end{itemize}  

Toutes ces conditions limitent fortement les groupes de symétrie de grande unification qu'il est possible d'écrire. Pour $n_{\text{GUT}}=4$, les seules possibilités compatibles avec les conditions précédentes sont SU(5) et SU(3)$\times$SU(3). Il est cependant impossible d'écrire simplement un GUT basé sur SU(3)$\times$SU(3)~\cite{Langacker:1980js}, restreignant les possibilités au seul SU(5). Pour $n_{\text{GUT}}=5$, les seules possibilités sont SU(6) et SO(10). Il est donc légitime de tenter de construire en priorité des théories de grande unification basées sur ces groupes. Considérant les représentations de ces différents groupes, un plongement naturel d'une génération de fermions du Modèle Standard ne peut se faire que pour SU(5)~\cite{Georgi:1974sy}, dans ses représentations $\mathbf{5}^*$ et $\mathbf{10}$, et pour SO(10)~\cite{Georgi:1974sy,Fritzsch:1974nn}, dans sa représentation $\mathbf{16}$ qui inclut le neutrino droit. Ces deux groupes sont par conséquent les candidats naturels pour construire des théories de grande unification. Ces plongements, n'étant absolument pas garantis \emph{a priori}, forment de fait une justification très forte à la construction de théories de grande unification. 

Étant donné qu'il n'y a qu'une seule façon de positionner les fermions du Modèle Standard dans ces représentations de SU(5) et SO(10), les interactions de jauge associées à ces deux groupes sont définies de manière univoque. Les bosons de jauge associés à des racines contenues dans ces théories de grande unification mais pas dans le Modèle Standard acquièrent une masse à haute énergie. Aux énergies du Modèle Standard, ils correspondent à des symétries brisées, comme discuté dans la section~\ref{PhenomSymBrisées}. Ces interactions étant supprimées quadratiquement en l'échelle d'énergie\footnote{Dans une théorie non supersymétrique. Le cas supersymétrique est discuté dans la section~\ref{PartGUTHiggsSector}.}, elles sont alors complètement négligeables par rapport aux symétries non brisées. Elles restent cependant potentiellement observables sur des processus interdits par ailleurs, comme des désintégrations de particules stables vis-à-vis des interactions du Modèle Standard, comme les électrons et les protons. Comme une désintégration doit émettre des particules de masse totale inférieure à celle de la particule qui se désintègre, cela réduit les possibilités à la désintégration du proton (à moins de ne considérer la désintégration d'un électron en seulement des neutrinos). Comme les théories de jauge basées sur SU(5) et SO(10) incluent des transformations de symétrie transformant des quarks en leptons qui permettent la désintégration du proton en un méson et un lepton, elles peuvent être contraintes par l'observation de cette désintégration.

À l'heure actuelle, les processus de désintégration du proton n'ont pas été observés et seules des bornes inférieures de temps de vie sont obtenues expérimentalement. Les résultats actuels donnent des temps de vie supérieurs à $10^{33}-10^{34}$ années, dépendant du processus de désintégration considéré~\cite{Takhistov:2016eqm,PDG2016}. Ces processus sont très contraignants pour les théories de grande unification. Les prédictions des théories de grande unification raisonnables\footnote{C'est-à-dire sans rajout artificiel d'un trop grand nombre de termes faisant perdre tout pouvoir prédictif.} basées sur SU(5) étant incompatibles avec ces observations, celles-ci sont à présent exclues par les observations~\cite{Murayama:2001ur}. Les théories basées sur SO(10) et sa représentation fermionique de dimension 16 apparaissent alors comme les plus prometteuses à l'heure actuelle.

\section{GUT SO(10) : secteur fermionique et interactions de jauge}
\label{PartRep16SO10}

\noindent
Les théories de grande unification basées sur SO(10) regroupent dans une seule représentation, de dimension 16, l'intégralité d'une génération de fermions du Modèle Standard, y compris un hypothétique neutrino de chiralité droite~\cite{Georgi:1974sy,Fritzsch:1974nn}. Dans cette section, nous décrivons les représentations des groupes simples dans la base de Dynkin, comme discuté dans le chapitre~\ref{PartRacinesPoids}. La représentation $\mathbf{16}$ de SO(10) dans cette base est donnée dans la figure~\ref{DiagPoidsS010}. Les particules de chaque génération de fermions sont placées dans ce diagramme des poids comme dans la figure~\ref{RepSO10ToSM}. Dans cette figure, on a aussi indiqué les poids des racines simples positives, et les bosons de jauge du Modèle Standard qui leurs sont associés le cas échéant.

\afterpage{
\begin{landscape}
\begin{figure}[t!]
\begin{center}
\includegraphics[scale=0.97]{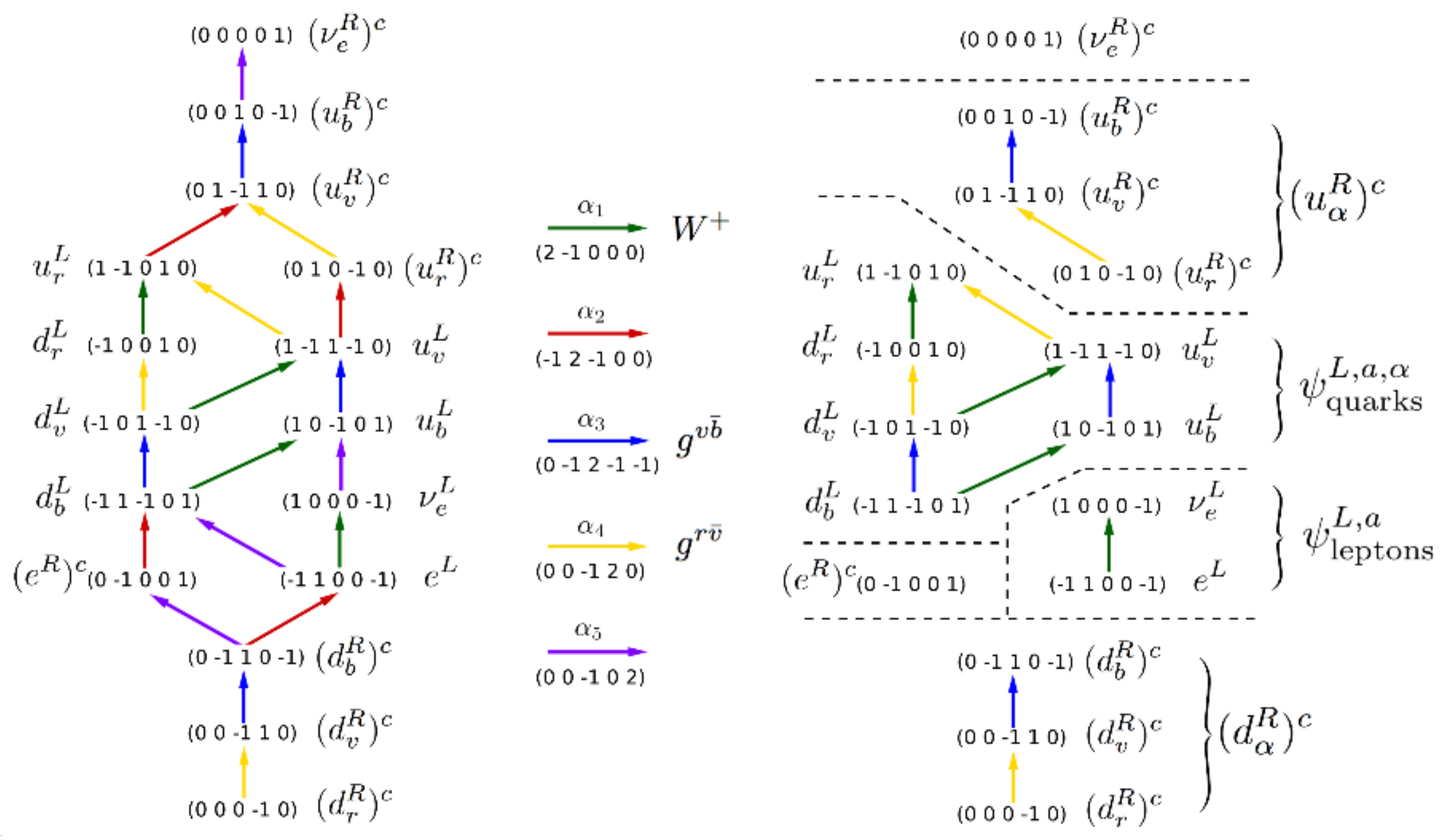}
 \caption{Gauche : Placement des fermions de la première génération du Modèle Standard dans la représentation de dimension 16 de SO(10). Droite : Représentations du Modèle Standard issues de la représentation de dimension 16 de SO(10), décrites avec les nombres quantiques de SO(10).}
 \label{RepSO10ToSM}
 \end{center}
\end{figure}
\end{landscape}
}

Les poids de chaque particule donnés dans la figure~\ref{RepSO10ToSM} correspondent aux nombres quantiques dans SO(10). Les éléments de la sous-algèbre de Cartan de $G_{\text{SM}}$ et de SO(10) sont reliés de la façon suivante~:
\begin{equation}
\label{EqDefT3L}
\begin{array}{l}
\displaystyle{T_{3,C}^{(1)} = T_3^{(4)}}\vspace{0.2cm},\\
\displaystyle{T_{3,C}^{(2)}= T_3^{(3)}}\vspace{0.2cm},\\
\displaystyle{T_3^L=\frac12 T_3^{(1)}}\vspace{0.2cm},\\
\displaystyle{T_3^Y = -\frac12 T_3^{(1)} - T_3^{(2)} - \frac23 T_3^{(3)} - \frac13 T_3^{(4)}},
\end{array}
\end{equation}
où l'on a nommé $T_3^{(i)}$ pour $i=1,\cdots,4$ les quatre premiers générateurs de la sous-algèbre de SO(10). Ces équations permettent de passer des nombres quantiques de SO(10) à ceux du Modèle Standard donnés dans la figure~\ref{TableFermionsSM}. On reconnait de même les nombres quantiques des bosons $W^+$, $g^{v\bar b}$ et $g^{r\bar v}$ pour les racines simples $\boldsymbol \alpha_1$, $\boldsymbol \alpha_3$ et $\boldsymbol \alpha_4$. Les deux autres racines simples, $\boldsymbol \alpha_2$ et $\boldsymbol \alpha_5$, ne correspondent pas à des bosons de jauge du Modèle Standard.


Seules les translations associées aux racines simples sont indiquées sur la figure~\ref{RepSO10ToSM}. Les translations liées aux autres racines non nulles de SO(10) correspondent également à des transformations de symétrie. Le groupe SO(10) étant de dimension 45 et de rang 5, il possède 15$[=(45-5)/2-5]$ racines positives en plus de ses racines simples $\boldsymbol \alpha_1$ à $\boldsymbol \alpha_5$, qui correspondent à des combinaisons linéaires positives de ces racines simples. L'ensemble des racines positives est donné dans l'équation~\eqref{RacinesSO10}, où elles sont obtenues en partant de $(01000)$ le poids le plus haut de la représentation adjointe de SO(10), les racines simples apparaissant sur la dernière ligne.
\begin{equation}
\label{RacinesSO10}
\begin{array}{c}
(01000)\\
(1\hspace{-0.075cm}-\hspace{-0.075cm}1100)\\
(\hspace{-0.075cm}-\hspace{-0.075cm}10100)(10\hspace{-0.075cm}-\hspace{-0.075cm}111)\\
(1001\hspace{-0.075cm}-\hspace{-0.075cm}1)(100\hspace{-0.075cm}-\hspace{-0.075cm}11)(\hspace{-0.075cm}-\hspace{-0.075cm}11\hspace{-0.075cm}-\hspace{-0.075cm}111)\\
(101\hspace{-0.075cm}-\hspace{-0.075cm}1\hspace{-0.075cm}-\hspace{-0.075cm}1)(\hspace{-0.075cm}-\hspace{-0.075cm}1101\hspace{-0.075cm}-\hspace{-0.075cm}1)(\hspace{-0.075cm}-\hspace{-0.075cm}110\hspace{-0.075cm}-\hspace{-0.075cm}11)(0\hspace{-0.075cm}-\hspace{-0.075cm}1011)\\
(11\hspace{-0.075cm}-\hspace{-0.075cm}100)(\hspace{-0.075cm}-\hspace{-0.075cm}111\hspace{-0.075cm}-\hspace{-0.075cm}1\hspace{-0.075cm}-\hspace{-0.075cm}1)(0\hspace{-0.075cm}-\hspace{-0.075cm}111\hspace{-0.075cm}-\hspace{-0.075cm}1)(0\hspace{-0.075cm}-\hspace{-0.075cm}11\hspace{-0.075cm}-\hspace{-0.075cm}11)\\
(2\hspace{-0.075cm}-\hspace{-0.075cm}1000)(\hspace{-0.075cm}-\hspace{-0.075cm}12\hspace{-0.075cm}-\hspace{-0.075cm}100)(0\hspace{-0.075cm}-\hspace{-0.075cm}12\hspace{-0.075cm}-\hspace{-0.075cm}1\hspace{-0.075cm}-\hspace{-0.075cm}1)(00\hspace{-0.075cm}-\hspace{-0.075cm}120)(00\hspace{-0.075cm}-\hspace{-0.075cm}102)
\end{array}
\end{equation}
Chacune de ces racines est associée à un boson de jauge complexe de SO(10), et l'équation précédente permet donc de décrire l'intégralité des transformations de SO(10) sur les fermions. La racine associée au gluon $g^{r\bar b}$ est ainsi $\boldsymbol\alpha_{r\bar b} = \boldsymbol\alpha_3 + \boldsymbol\alpha_4 = (0	\hspace{-0.075cm}-\hspace{-0.075cm}111\hspace{-0.075cm}-\hspace{-0.075cm}1)$, et on peut vérifier qu'elle n'est pas chargée sous $T_3^L$ ou $T_3^Y$. La donnée des nombres quantiques des racines permet également de décrire les interactions entre les bosons de jauge.

\begin{table}[h!]
\begin{center}
\renewcommand{\arraystretch}{1.4}
\begin{tabular}{|c|C{0.80cm}|C{0.80cm}|C{0.80cm}|C{0.80cm}|C{0.80cm}|C{0.80cm}|}
\hline
~ &$T_{3,C}^{(1)}$&$T_{3,C}^{(2)}$ & $T_3^L$ & $T_3^Y$ & $T_3^{_{B\text{-}\hspace{-0.03cm}L}}$ & $T_3^{X}$ \\
\hline
$\boldsymbol \alpha_1$ ($W^+$) & 0 & 0 & 1 & 0 & 0 & 0 \\
\hline
$\boldsymbol \alpha_2$ & 0 & -1 & $-\frac12$ & $-\frac56$& $-\frac23$ & 0 \\
\hline
$\boldsymbol \alpha_3$ ($g^{r\bar v}$) & -1 & 2 & 0 & 0 & 0 & 0 \\
\hline
$\boldsymbol \alpha_4$ ($g^{v\bar b}$) & 2 & -1 & 0 & 0 & 0 & 0 \\
\hline
$\boldsymbol \alpha_5$ &0 & -1 & 0 & $+\frac23$ & $+\frac43$ & -4 \\
\hline
\end{tabular}
\end{center}
\caption{Nombres quantiques des racines simples positives de SO(10) exprimés en fonction des nombres quantiques du Modèle Standard ainsi que sous U(1)$_{B-L}$ et U(1)$_X$ (voir la section~\ref{PartSO10ToSU5}).}
\label{TableNQRacinesSO10}
\end{table}

L'algèbre de SO(10) étant de rang 5, il est possible d'identifier un cinquième nombre quantique conservé en plus de ceux du Modèle Standard. On associe ce cinquième nombre quantique au générateur
\begin{equation}
\displaystyle{T_{3}^{B-L} = \frac23 T_3^{(3)} + \frac13 T_3^{(4)}+  T_3^{(5)}}.
\end{equation}
Il correspond à la différence entre le nombre baryonique $B$ et leptonique $L$, et vaut $\frac13$ pour les quarks et $-1$ pour les leptons. Les nombres quantiques du Modèle Standard et $B-L$ sont indiqués pour les racines simples positives dans le tableau~\ref{TableNQRacinesSO10}. On voit que les bosons de jauge du Modèle Standard ne sont pas chargés sous U(1)$_{B-L}$, ce qui est lié au fait qu'ils ne transforment que des quarks en quarks ou des leptons en leptons. Au contraire, les bosons de jauge associés à $\boldsymbol\alpha_2$ et $\boldsymbol\alpha_5$, parfois appelés bosons $X$ ou bosons $X$ et $Y$, transforment des quarks en leptons. Ils portent également des charges de couleur et d'isospin faible, mélangeant les différentes représentations du Modèle Standard.

Après la brisure de SO(10) en $G_{\text{SM}}$, le générateur $T_3^{B-L}$ ne décrit plus une symétrie puisque le rang du groupe de jauge diminue d'une unité. En conséquence, tous les bosons de jauge chargés sous U(1)$_{B-L}$ vont acquérir une masse, comme discuté dans la section~\ref{RepChampsHiggs}. Les représentations fermioniques du Modèle Standard apparaissent alors sur le diagramme des poids de SO(10), en ne considérant que les racines simples $\boldsymbol\alpha_1$, $\boldsymbol\alpha_2$ et $\boldsymbol\alpha_3$ décrivant les bosons de jauge de nombres quantiques non nuls du Modèle Standard. Cet enchâssement des représentations du Modèle Standard dans la représentation fermionique de grande unification est explicité dans la figure~\ref{RepSO10ToSM}. Comme aucune des racines des $G_{\text{SM}}$ n'est chargée sous U(1)$_{B-L}$, on retrouve que les représentations du Modèle Standard ne groupent que des quarks ou des leptons.

Les décompositions de la représentation fermionique de SO(10) sur ses sous-algèbres maximales $\text{SU(5)}\times\text{U(1)}_X$ et SU(4)$_C\times$SU(2)$_L\times$SU(2)$_R$ (le groupe de Pati-Salam) sont données dans le chapitre~\ref{PartRep16SousAlgebre}. Ces décompositions permettent une compréhension plus approfondie de la structure de la représentation $\mathbf{16}$ de SO(10). De plus, le tableau~\ref{TableNQSO10Full} récapitulant les différents nombres quantiques d'une génération de fermions se trouve à la fin de cette annexe, page~\pageref{TableNQSO10Full}.

\section{GUT SO(10) : secteur de Higgs}
\label{PartGUTHiggsSector}

\noindent
Une fois un groupe d'unification donné, celui-ci permettant le rassemblement des fermions du Modèle Standard dans un nombre réduit de représentations, la construction de théories de grande unification revient à décrire leur secteur de Higgs. Nous discutons ici cette construction dans le cadre d'une théorie de grande unification basée sur SO(10). Cette discussion fait apparaître des contraintes fortes afin de retrouver le Modèle Standard, mais montre aussi qu'un certain nombre de caractéristiques intéressantes découlent de la structure des représentations de SO(10), celles-ci permettant naturellement de mettre en place certaines propriétés attendues d'une théorie de grande unification. On se place par la suite dans un cadre supersymétrique.

D'un point de vue de la construction de modèles, il est \emph{a priori} nécessaire de considérer tous les couplages possibles entre les champs de Higgs et les fermions, ainsi qu'entre les champs de Higgs eux-même. On s'attend également à ce que le nombre de couplages augmente quadratiquement avec le nombre de champs de Higgs. Dans ce cadre, les modèles ayant le plus grand pouvoir prédictif sont ceux qui contiennent le plus petit nombre de champs de Higgs, et sont donc minimaux. Les deux rôles des champs de Higgs sont de briser SO(10) en $G_{\text{SM}}$ et de donner leur masse aux fermions. On va considérer successivement ces deux points, et les contraintes associées en ce qui concerne les champs de Higgs. On raisonnera au niveau des représentations de SO(10) ou bien de leurs décompositions sur ses sous-algèbres maximales détaillées dans le chapitre~\ref{PartRep16SousAlgebre}, ce qui simplifie souvent la discussion.

Puisque les rangs de $G_{\text{SM}}$ et SO(10) sont respectivement 4 et 5, la brisure de symétrie du groupe de grande unification implique une diminution du rang du groupe de symétrie d'une unité. Il est pour cela nécessaire d'avoir des champs de Higgs dans des représentations complexes\footnote{Lors de la décomposition d'une représentation réelle sur un sous-groupe de symétrie, les champs complexes apparaissent avec leur complexes conjugués~\cite{Slansky:1981yr}. Identifiant un schéma de brisure de la forme $G\rightarrow G^\prime$ avec $G\supset G^\prime \times \text{U(1)}_{G^\prime}$, un singlet de $G^\prime$ contenu dans une représentation réelle de $G$ aura donc toujours une charge nulle sous U(1)$_{G^\prime}$, ce qui empêchera la brisure de symétrie de la composante abélienne.}. En ne considérant que les représentations de dimensions par trop élevées, cela implique d'utiliser des champs de Higgs dans les représentations $\mathbf{16}$, $\mathbf{126}$ ou $\mathbf{144}$.

Lorsque nous avons évoqué la désintégration du proton dans la section~\ref{PartGroupeUnification}, seules les désintégrations via des générateurs de symétrie brisée ont été considérées. Dans des théories supersymétriques, il est aussi nécessaire de considérer les couplages entre les fermions et leurs partenaires supersymétrique, comme des couplages de Yukawa de la forme $\mathbf{16}\times\mathbf{16}\times\mathbf{16}$ pouvant relier des quarks et des leptons et causer la désintégration du proton. Ces couplages étant renormalisables, ils ne sont pas négligeables à basse énergie, et sont donc incompatibles avec les bornes inférieures de temps de vie du proton. Ils peuvent cependant être interdits dans des théories invariantes sous une symétrie nommée $R$-parité~\cite{Farrar:1978xj}, ou parité de matière~\cite{Ellis:1981tv,Dimopoulos:1981dw}. Cette symétrie, décrite par un groupe $\mathbb{Z}_2$, change le signe des fermions et sfermions mais laisse invariants les champs de Higgs et leurs partenaires supersymétriques, interdisant les couplages de Yukawa entre fermions et sfermions, mais les autorisant entre fermions et champs de Higgs.

Pouvant sembler arbitraire à première vue, la R-parité est en fait une symétrie résiduelle de certain schémas de brisure de symétrie de SO(10). La $R$-parité peut en effet être identifiée à un sous-groupe non brisé de U(1)$_{B-L}$. En pratique, il a été montré que ce sous-groupe de symétrie est conservé à basse énergie lorsque la brisure de symétrie est effectuée par des champs de Higgs dans certaines représentations particulières~\cite{Martin:1992mq}. Ces représentations sont par exemple les $\mathbf{10}$, $\mathbf{45}$, $\mathbf{54}$, $\mathbf{120}$, $\mathbf{126}$, $\mathbf{210}$ et leurs représentations conjuguées. Les autres représentations, comme par exemple les $\mathbf{16}$, $\mathbf{144}$ et leurs conjuguées, n'ont pas cette propriété. Ces dernières représentations sont donc exclues dans un cadre supersymétrique. En conséquence, les schémas de brisure de SO(10) dans un cadre supersymétrique contiennent généralement la brisure de U(1)$_{B-L}$ par des champs de Higgs dans les représentations $\mathbf{126}$ et $\mathbf{\overline{126}}$, cette brisure laissant le $\mathbb{Z}_2$ associé à la $R$-parité comme symétrie résiduelle\footnote{La représentation de plus petite dimension permettant ensuite cette brisure de symétrie tout en conservant la $R$-parité est la $\mathbf{2772}$~\cite{Slansky:1981yr,Martin:1992mq}.}.

Seules quelques représentations de SO(10) peuvent former un couplage avec la matière, puisque celles-ci doivent pouvoir former un couplage de Yukawa singlet avec la représentation $\mathbf{16}$. Ces représentations s'identifient dans le produit tensoriel de la représentation fermionique avec elle-même donnant~:
\begin{equation}
\label{EqHiggsMasseSO10}
\mathbf{16} \times \mathbf{16} = \mathbf{10}_s + \mathbf{120}_a + \mathbf{126}_s,
\end{equation}
où l'indice $s$ ou $a$ indique si la représentation apparaît dans le produit symétrique ou antisymétrique des deux représentations. La première tâche est d'identifier le champ de Higgs électrofaible, qui doit être exprimé dans une représentation de SO(10). L'étude complète des règles de branchement montre que seul un champ de Higgs dans la représentation $\mathbf{10}$ de SO(10) reproduit bien les couplages du champ de Higgs électrofaible, ce qui l'identifie de manière unique~\cite{Barbieri:1979ag}. 

On a vu précédemment que les schémas de brisure réalistes contiennent généralement un champ de Higgs dans les représentations $\mathbf{126}$ et $\mathbf{\overline{126}}$, brisant U(1)$_{B-L}$. Ces champs forment des couplages de Yukawa avec la représentation fermionique et peuvent donc donner des masses de grande unification à certains fermions, ce qui pourrait être \emph{a priori} très problématique. Cependant, la valeur moyenne dans le vide non nulle de la représentation $\mathbf{\overline{126}}$ brisant U(1)$_{B-L}$ est dans la représentation $\mathbf{1}(10)$ de SU(5)$\times$U(1)$_X$, et ne se couple pas avec les fermions du Modèle Standard qui sont dans des représentations non-triviales de SU(5) (voir la section~\ref{PartSO10ToSU5} pour plus de détail). Un couplage est par contre possible avec le neutrino de chiralité droite, dans la représentation $\mathbf{1}(-5)$ de SU(5)$\times$U(1)$_X$, qui acquiert donc une masse aux énergies de grande unification. Cela rajoute un intérêt supplémentaire à l'utilisation des représentations $\mathbf{126}$ et $\mathbf{\overline{126}}$ pour briser U(1)$_{B-L}$, octroyant un terme de masse à haute énergie au neutrino droit, permettant par là même un mécanisme de balançoire expliquant la phénoménologie des neutrinos à basse énergie. Ce terme de masse pour le neutrino droit relie deux spineurs de même chiralité. Après brisure de SO(10), et pour un groupe de jauge comme SU(5) ou $G_{\text{SM}}$ pour lequel le neutrino droit n'aura que des nombres quantiques nuls, il s'identifiera donc à un terme de masse de Majorana, comme attendu du point de vue du Modèle Standard.

On peut finalement s'interroger sur le fait d'avoir rassemblé les fermions d'une même génération du Modèle Standard dans la représentation $\mathbf{16}$ de SO(10). Les trois générations ayant la même structure de groupe, il aurait en effet été possible de mélanger les différentes générations du Modèle Standard dans trois représentations fermioniques de SO(10). Cependant, comme il est possible d'écrire un seul couplage entre une représentation fermionique et le champ de Higgs électrofaible de SO(10), chaque représentation fermionique va être associée à une même échelle de masse après renormalisation jusqu'au Modèle Standard. Ainsi, plutôt que de répondre à la question de pourquoi chaque génération de fermions est contenue dans une même représentation de SO(10), la construction d'un GUT SO(10) explique naturellement pourquoi trois structures identiques de particules vont apparaître avec une hiérarchie de masse, structures qui seront identifiées comme les générations du Modèle Standard.

\section{Statut actuel}
\label{PartGUTStatutActuel}

\noindent
Dans le cas de SO(10), un grand nombre de schémas de brisure impliquant des champs de Higgs dans des représentations $\mathbf{10}$, $\mathbf{126}$ et $\mathbf{\overline{126}}$ ont été proposés dès la fin des années 70 après le développement initial des théories de grande unification~\cite{Chanowitz:1977ye,Georgi:1978fu,Machacek:1979tx,Georgi:1979dq,Georgi:1979ga,Mohapatra:1979nn,Rajpoot:1980xy,delAguila:1980qag}. Ils contiennent des champs de Higgs additionnels parmi les représentations $\mathbf{45}$, $\mathbf{54}$, $\mathbf{120}$ et $\mathbf{210}$. Les modèles minimaux les plus réalistes à l'heure actuelle sont supersymétriques et brisent SO(10) jusqu'au Modèle Standard par une combinaison de champs de Higgs dans les représentations $\mathbf{126}$, $\overline{\mathbf{126}}$ et $\mathbf{210}$ de SO(10)~\cite{Aulakh:2003kg,Goh:2003hf,Fukuyama:2004xs,Bajc:2004xe,Aulakh:2004hm,Cacciapaglia:2013tga,Aulakh:2015sha}, un champ de Higgs dans la représentation $\mathbf{10}$ de SO(10) effectuant ensuite la brisure électrofaible. Les schémas de brisure associés ne passent pas par SU(5) -- évitant ainsi les défauts inhérents à ce type d'unification --, mais par le groupe de Pati-Salam, discuté dans le complément~\ref{PartSO10ToPS}. Les deux schémas de brisure possibles sont 
\begin{equation}
SO(10) \rightarrow \text{SU3}_C\times \text{SU2}_L \times\text{SU2}_R\times \text{U(1)}_{B-L} \rightarrow G_{\text{SM}} \times \mathbb{Z}_2,
\end{equation}
ou 
\begin{equation}
SO(10) \rightarrow \text{SU3}_C\times \text{SU2}_L \times\text{U(1)}_R \times\text{U(1)}_{B-L} \rightarrow G_{\text{SM}} \times \mathbb{Z}_2,
\end{equation}
le groupe $\mathbb{Z}_2$ résiduel désignant la $R$-parité. La brisure de symétrie sera étudiée en détails dans le chapitre~\ref{PartArticleCorde2}. 

Ces modèles de GUTs supersymétriques basés sur SO(10) sont généralement nommés MSGUT pour "Minimal Supersymmetric Grand Unified Theory". Ils ont toutes les caractéristiques attendues d'un GUT~: ils sont compatibles avec toute la phénoménologie du Modèle Standard y compris la désintégration du proton, expliquent la structure de son secteur fermionique et les échelles de masse différentes des générations de fermions, décrivent le spectre de masse des fermions avec un nombre réduit de paramètres, prédisent une masse au neutrino droit et un mécanisme de see-saw, décrivent la brisure de supersymétrie, etc.. De plus, la description des modèles les plus minimaux demande 26 paramètres réels~\cite{Aulakh:2003kg}, soit le même nombre que pour décrire l'extension supersymétrique minimale du Modèle Standard, alors même qu'ils décrivent aussi la phénoménologie aux échelles d'énergie $E_{\text{GUT}}$. Ils représentent donc des modèles théoriques performants pour décrire l'unification des interactions du Modèle Standard à haute énergie.

Finalement, et en plus de leur statut de théorie de physique des particules à haute énergie, les théories de grande unification apparaissent souvent comme des limites de "basse énergie" des théories de cordes -- comme dans les théories hétérotiques ou dans le "flux landscape" de ces théories -- et des théories d'unification construites dans un espace à plus de quatre dimensions comme les orbifolds GUTs~\cite{PDG2016}. L'étude des GUTs semble donc être une étape déterminante de l'exploration des théories de haute énergie, justifiant la recherche de tous signaux observationnels qui pourraient permettre de les sonder. Or, un test expérimental des GUTs autre que l'étude de la désintégration du proton semble inaccessible sur Terre. Cela justifie la recherche de cordes cosmiques, défauts topologiques formés par la brisure de symétrie des GUTs dans l'univers primordial, et qui pourrait permettre d'observer les conséquences directes de théories de grande unification, comme discuté dans la partie~\ref{ChapterRealisticStrings}.

\chapter{Annexes à la troisième partie}
\label{PartRep16SousAlgebre}


\section{Décomposition de SO(10) sur SU(5)$\times$U(1)$_X$}
\label{PartSO10ToSU5}

\noindent
Dans cette section, on décompose SO(10) et sa représentation $\mathbf{16}$ sur sa sous-algèbre maximale SU(5)$\times$U(1), en identifiant SU(5) par le fait qu'il contient intégralement $G_{\text{SM}}$. Cela permet notamment de faire le lien entre les théories de grande unification basées sur le groupe de jauge SU(5), nommées théories de Georgi-Glashow~\cite{Georgi:1974sy}, et celles basées sur SO(10). Le U(1) apparaissant dans ce produit de groupes est appelé U(1)$_X$, défini par le générateur
\begin{equation}
\label{EqDefT3X}
\displaystyle{T_3^X = -2 T_3^{(1)}-4 T_3^{(2)} - 6 T_3^{(3)} - 3 T_3^{(4)} - 5 T_3^{(5)}.}
\end{equation}
Le groupe SO(10) étant de rang 5, cet élément de sa sous-algèbre de Cartan n'est pas indépendant de ceux précédemment définis, et on vérifie que
\begin{equation}
\displaystyle{T_3^X = 4 T_3^Y - 5 T_3^{B-L}.}
\end{equation}
La représentation de dimension 16 de SO(10) se décompose sur SU(5)$\times$U(1)$_X$ comme
\begin{equation}
\label{Eq16SO10ToSU5}
\mathbf{16} = \mathbf{1}(-5) + \mathbf{\bar{5}}(3) + \mathbf{10}(-1),
\end{equation}
où l'on a noté la charge sous U(1)$_X$ entre parenthèses. 

\afterpage{
\begin{landscape}
\begin{figure}[t!]
\begin{center}
\includegraphics[scale=0.97]{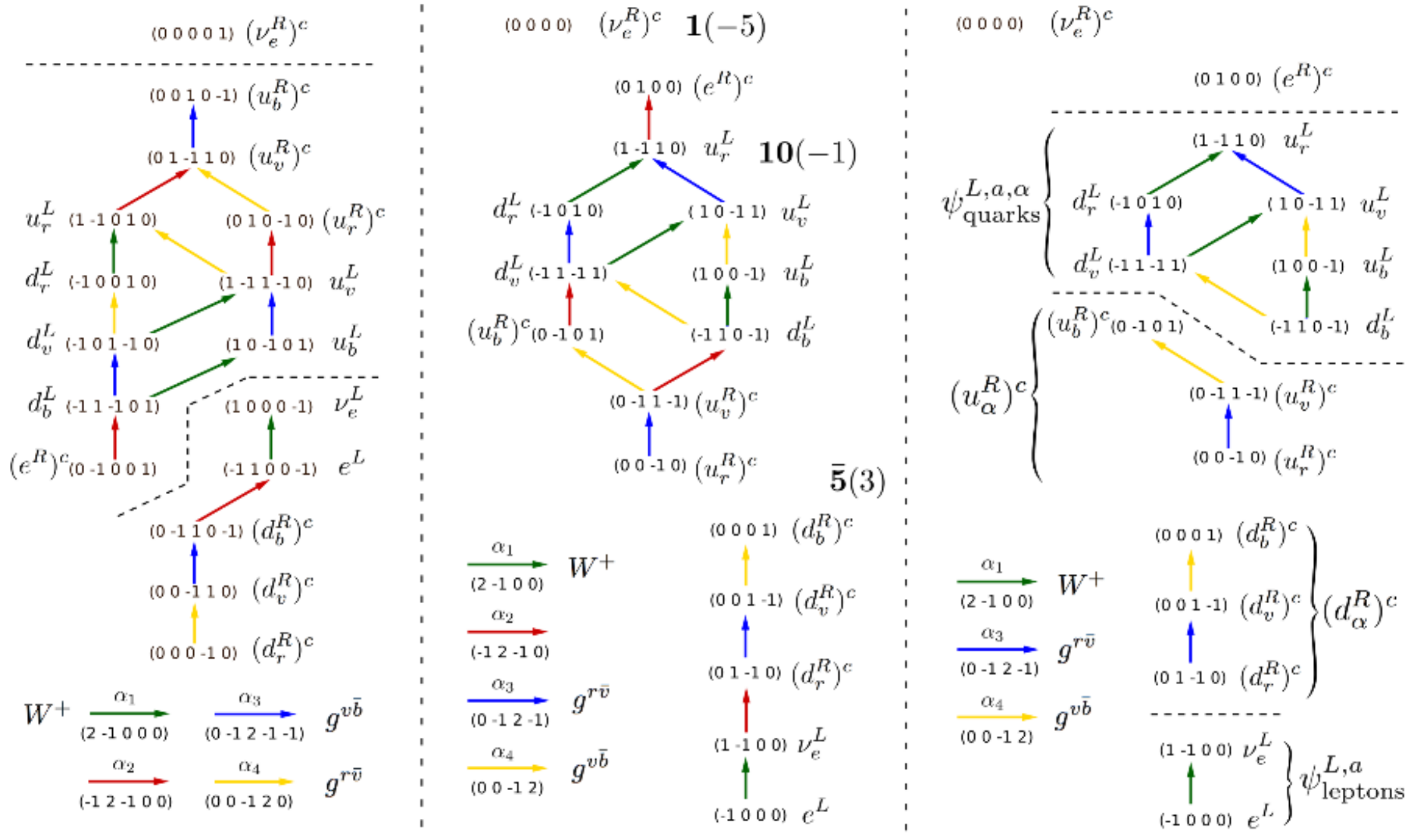}
 \caption{Gauche : Description des représentations de SU(5) issues de la représentation de dimension 16 de SO(10) avec les nombres quantiques de SO(10). Centre : Les mêmes représentations avec les nombres quantiques de SU(5). Droite : Représentations du Modèle Standard issues de ces représentations, avec les nombres quantiques de SU(5).}
 \label{RepSO10ToSU5}
 \end{center}
\end{figure}
\end{landscape}
}

\begin{table}[h!]
\begin{center}
\renewcommand{\arraystretch}{1.4}
\begin{tabular}{|c|C{0.80cm}|C{0.80cm}|C{0.80cm}|C{0.80cm}|C{0.80cm}|}
\hline
  & $T_{3,C}^{(1)}$ & $T_{3,C}^{(2)}$ & $T_3^L$ & $T_3^Y$ &   $T_3^Q$    \\ \hline
$\boldsymbol \alpha_1$ ($W^+$)  & 0  & 0  & 1    & 0 & 1  \\ \hline
$\boldsymbol \alpha_2$ &  -1 & 0  & $-\frac12$ & $\frac56$ & $\frac13$ \\ \hline
$\boldsymbol \alpha_3$ ($g^{v\bar b}$) & 2  & -1 & 0    & 0   & 0  \\ \hline
$\boldsymbol \alpha_4$ ($g^{r\bar v}$) & -1 & 2  & 0    & 0   & 0  \\ \hline
\end{tabular}
\end{center}
\caption{Nombres quantiques des racines simples positives de SU(5) exprimés en fonction des nombres quantiques du Modèle Standard.}
\label{TableNQRacinesSU5}
\end{table}

Parmi les racines simples de SO(10) données dans la table~\ref{TableNQRacinesSO10}, seule $\boldsymbol\alpha_5$ est chargée sous U(1)$_X$ et n'est pas une racine de SU(5). Elle fait en effet le lien entre des particules de charges sous U(1)$_X$ différentes, appartenant donc à différentes représentations de SU(5). Le plongement des représentations de SU(5) dans la représentation de dimension 16 de SO(10) s'obtient alors en ne considérant que les racines simples $\boldsymbol\alpha_1$ à $\boldsymbol\alpha_4$. Cette décomposition est décrite dans la figure~\ref{RepSO10ToSU5}, où les représentations de SU(5) sont exprimées en fonction des nombres quantiques de SO(10) puis de SU(5). La matrice de projection de SO(10) dans SU(5)$\times$U(1)$_X$ s'écrit à partir de l'équation~\eqref{EqDefT3X} et de 
\begin{equation}
\begin{array}{l}
\displaystyle{\tilde{T}_3^{(1)}= T_3^{(1)}}\vspace{0.2cm},\\
\displaystyle{\tilde{T}_3^{(2)}= -T_3^{(1)}  -T_3^{(2)}  -T_3^{(3)}  -T_3^{(4)}}\vspace{0.2cm},\\
\displaystyle{\tilde{T}_3^{(3)}= T_3^{(4)}}\vspace{0.2cm},\\
\displaystyle{\tilde{T}_3^{(4)}= T_3^{(3)}}\vspace{0.2cm},\\
\end{array}
\end{equation}
où l'on a dénoté avec des tildes les générateurs de SU(5) pour les différencier de ceux de SO(10) [les nombres quantiques de SU(5) sont indiqués dans la base de Dynkin].

Les racines simples de SU(5) ne s'identifient pas directement avec celles de SO(10) dans la base de Dynkin. Ce sont cependant bien des racines de SO(10), et les trois racines non-nulles indépendantes du Modèle Standard s'identifient à trois des quatre racines simples de SU(5). Cette dernière propriété permet d'identifier la position des représentations du Modèle Standard dans celles de SU(5), comme montré dans la figure~\ref{RepSO10ToSU5}. Le lien entre les éléments des sous-algèbres de Cartan du Modèle Standard et de SU(5) est 
\begin{equation}
\begin{array}{l}
\displaystyle{T_{3,C}^{(1)} = \tilde{T}_3^{(3)}}\vspace{0.2cm},\\
\displaystyle{T_{3,C}^{(2)}= \tilde{T}_3^{(4)}}\vspace{0.2cm},\\
\displaystyle{T_3^L=\frac12 \tilde{T}_3^{(1)}}\vspace{0.2cm},\\
\displaystyle{T_3^Y = \frac12 \tilde{T}_3^{(1)}+\tilde{T}_3^{(2)}+\frac23 \tilde{T}_3^{(3)} +\frac13 \tilde{T}_3^{(4)}}\vspace{0.2cm}.
\end{array}
\end{equation}
Il n'est pas possible d'identifier la charge $B-L$ comme un nombre quantique de SU(5) qui est de rang 4. Cependant, les représentations $\mathbf{\bar{5}}$ et $\mathbf{10}$ étant contenues dans une même représentation de SO(10), les interactions de jauge associées à ces deux représentations conservent cette charge. Les nombres quantiques des racines simples de SU(5) vis-à-vis du Modèle Standard sont donnés dans le tableau~\ref{TableNQRacinesSU5}. On retrouve le boson $W^+$ et deux gluons portant des charges de couleur, en plus d'un boson de jauge reliant les différents secteurs du Modèle Standard.

\section{Décomposition de SO(10) sur le groupe de Pati-Salam}
\label{PartSO10ToPS}

\afterpage{
\begin{landscape}
\begin{figure}[t!]
\begin{center}
\includegraphics[scale=0.97]{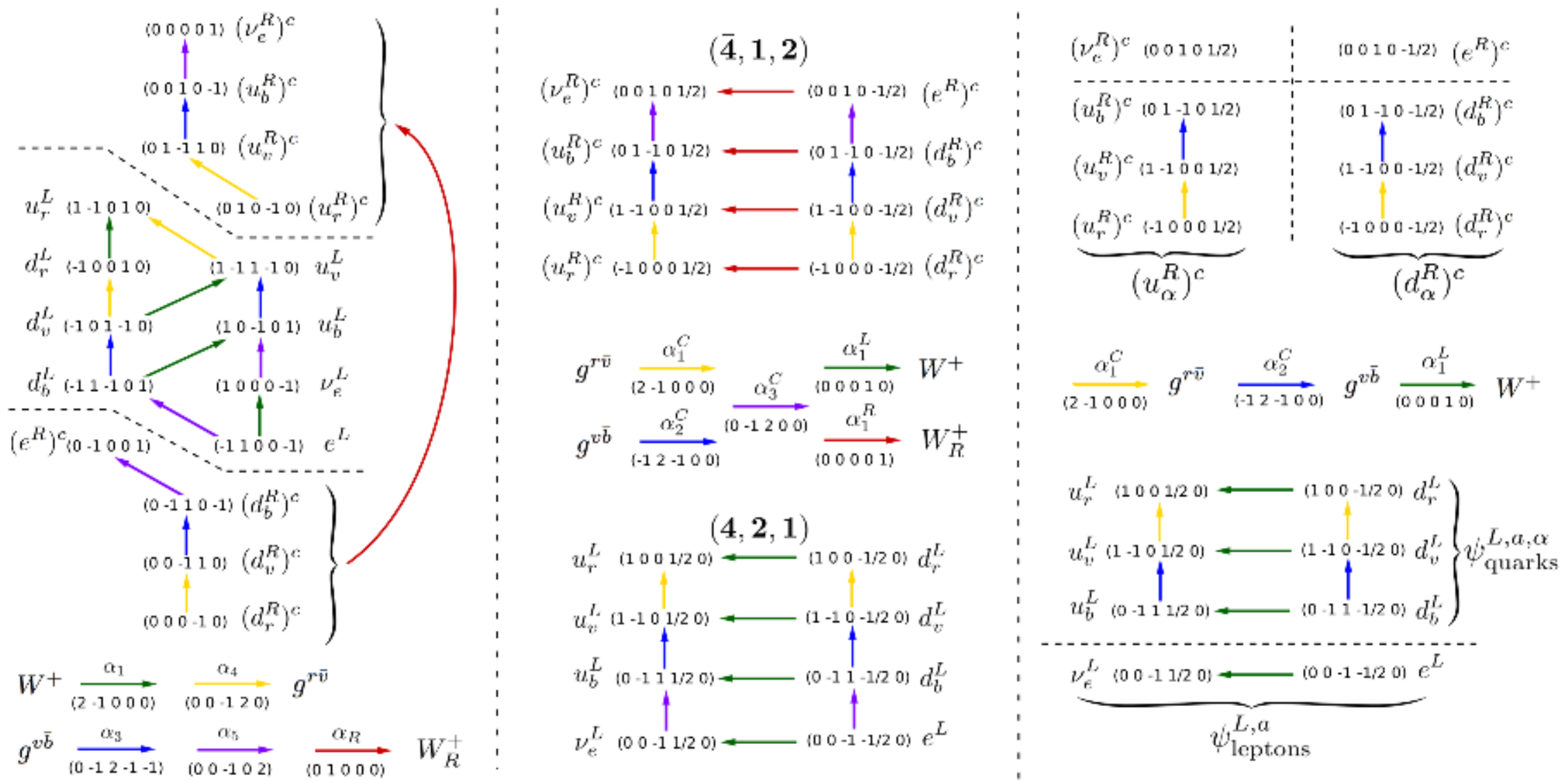}
 \caption{Gauche : Description des représentations du groupe de Pati-Salam issues de la représentation de dimension 16 de SO(10) avec les nombres quantiques de SO(10). Centre : Les mêmes représentations avec les nombres quantiques du groupe de Pati-Salam. Droite : Représentations du Modèle Standard issues de ces représentations, avec les nombres quantiques du groupe de Pati-Salam.}
 \label{RepSO10ToPS}
 \end{center}
\end{figure}
\end{landscape}
}

\noindent
Il est aussi possible de décomposer SO(10) sur sa sous-algèbre maximale SU(4)$_C\times$SU(2)$_L\times$SU(2)$_R$. C'est le groupe de Pati-Salam, qui fut proposé historiquement comme groupe d'unification partielle~\cite{Pati:1974yy}. Le groupe SU(2)$_L$ est celui du Modèle Standard, et SU(4)$_C$ contient SU(3)$_C$ comme sous-groupe. Le groupe SU(2)$_R$ reproduit la structure de SU(2)$_L$ sur les fermions de chiralité droite ; il décrit donc une interaction similaire aux interactions faibles sur ces fermions. L'isospin faible droit s'identifie par 
\begin{equation}
\label{EqDefT3R}
T_3^R = T_3^{(1)} + 2 T_3^{(2)} + 2 T_3^{(3)} + T_3^{(4)}  + T_3^{(5)}, 
\end{equation}
et la racine positive associée à l'équivalent du boson $W^+$ pour SU(2)$_L$ est
\begin{equation}
\boldsymbol\alpha_R  = \boldsymbol\alpha_1 + \boldsymbol\alpha_2 + 2 \boldsymbol\alpha_3 + \boldsymbol\alpha_4 + \boldsymbol\alpha_5.
\end{equation}
En plus des définitions de $T_3^L$ et $T_3^R$ données dans les équations~\eqref{EqDefT3L} et~\eqref{EqDefT3R}, la matrice de projection de SO(10) dans le groupe de Pati-Salam est donnée par 
\begin{equation}
\begin{array}{l}
\displaystyle{\tilde{T}_{3,C}^{(1)} = \tilde{T}_3^{(3)}}\vspace{0.2cm},\\
\displaystyle{\tilde{T}_{3,C}^{(2)}= \tilde{T}_3^{(4)}}\vspace{0.2cm},\\
\displaystyle{\tilde{T}_{3,C}^{(3)}= \tilde{T}_3^{(5)}}\vspace{0.2cm},\\
\end{array}
\end{equation}
où l'on a noté $\tilde{T}_{3,C}^{(i)}$ les éléments de la sous-algèbre de Cartan de SU(4)$_C$. Les nombres quantiques de SU(4)$_C$ sont indiqués dans la base de Dynkin. Les nombres quantiques associés à SU(2)$_L$ et SU(2)$_R$ sont indiqués dans les représentations de spin, ce qui explique l'apparition de poids demi-entiers\footnote{Les nombres quantiques de SU(2)$_L$ dans la base de Dynkin sont le double de ceux dans les représentations de spin.}.

La représentation $\mathbf{16}$ de SO(10) se décompose sur le groupe de Pati-Salam en
\begin{equation}
\mathbf{16} = (\mathbf{4},\mathbf{2},\mathbf{1}) + (\mathbf{\bar 4},\mathbf{1},\mathbf{2}).
\end{equation}
Pour identifier le plongement de ces représentations dans la représentation initiale de SO(10), on peut observer que la racine $\alpha_2$ relie forcément des représentations de SU(2)$_L$ différentes, puisque qu'elle change le caractère entier ou demi-entier de l'isospin faible. D'autre part, des éléments reliés par $\boldsymbol\alpha_R$ sont dans la même représentation de SU(2)$_R$, et ne sont donc pas dans une même représentations du Modèle Standard, ce qui permet de conclure l'identification. Ce plongement est explicité dans la figure~\ref{RepSO10ToPS}, où l'on a précisé les particules reliées par $\boldsymbol\alpha_R$ dans le diagramme ne faisant apparaître initialement que les racines simples de SO(10).

La figure~\ref{RepSO10ToPS} fait apparaître différentes propriétés de l'unification du Modèle Standard dans le groupe de Pati-Salam. D'une part, ce groupe restaure la symétrie entre les fermions de chiralité gauche et droite du Modèle Standard, nommée \og left-right symmetry \fg{}. Cette symétrie se retrouve notamment dans l'écriture de la charge électrique en fonction des charges du groupe de Pati-Salam~:
\begin{equation}
T_3^Q = \frac16 \tilde{T}_{3,C}^{(1)} + \frac13 \tilde{T}_{3,C}^{(2)} + \frac12 \tilde{T}_{3,C}^{(3)} + T_3^R + T_3^L.
\end{equation}
Ce groupe fait également apparaître les leptons comme une quatrième couleur, parfois associée à une charge de couleur violette. Les relations entre les charges du groupe de Pati-Salam et les charges du Modèle Standard sont relativement simples~:
\begin{equation}
\begin{array}{l}
\displaystyle{T_{3,C}^{(1)} = \tilde{T}_{3,C}^{(1)}}\vspace{0.2cm},\\
\displaystyle{T_{3,C}^{(2)}=  \tilde{T}_{3,C}^{(2)}}\vspace{0.2cm},\\
\displaystyle{T_3^Y = \frac16 \tilde{T}_{3,C}^{(1)} + \frac13 \tilde{T}_{3,C}^{(2)} + \frac12 \tilde{T}_{3,C}^{(3)} + T_3^R }\vspace{0.2cm},
\end{array}
\end{equation}
ce qui implique une certaine facilité d'utilisation du groupe de Pati-Salam.

\afterpage{
\begin{landscape}
\begin{figure}[b!]
\begin{center}
\includegraphics[scale=1.1]{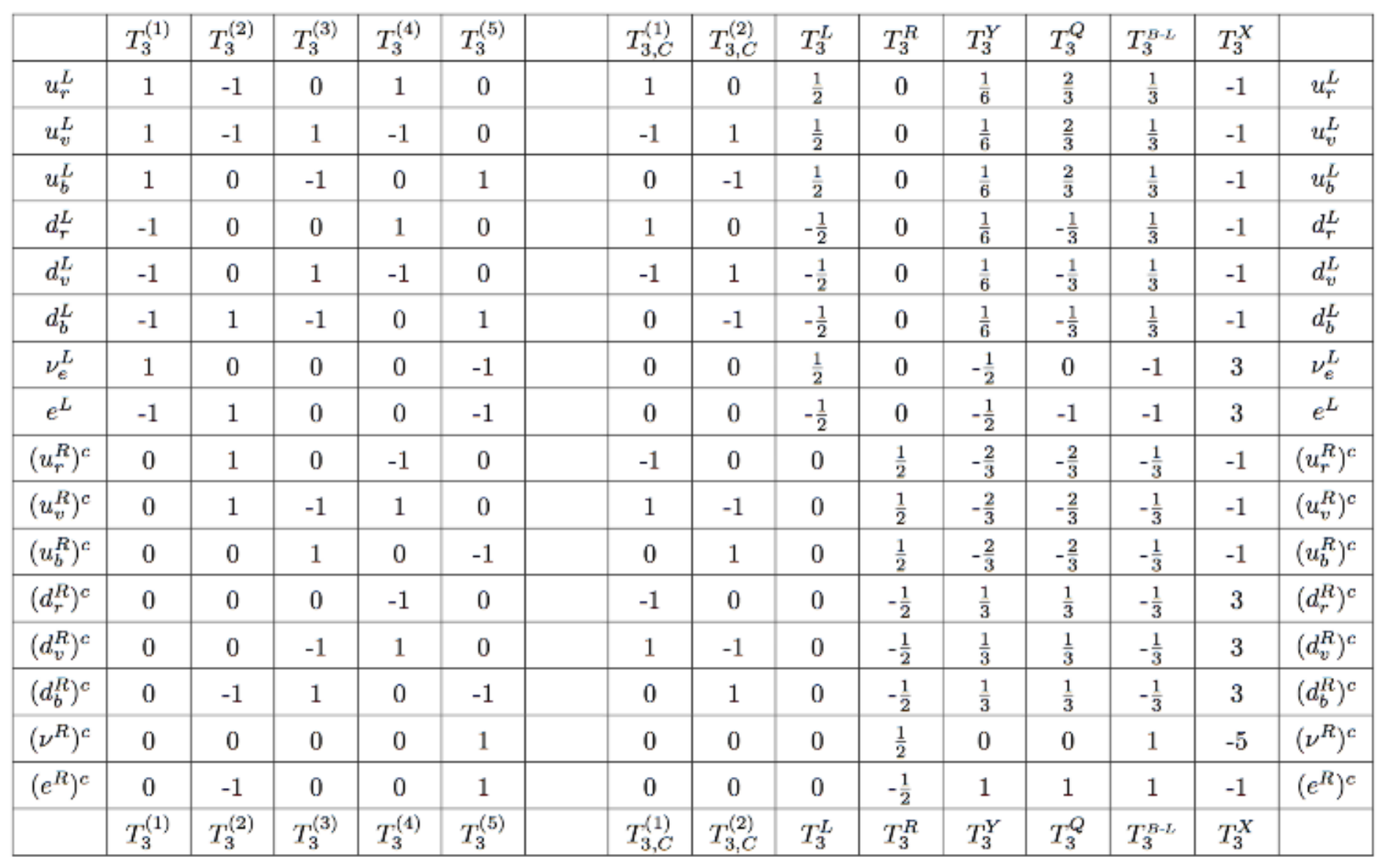}
 \caption{Nombres quantiques des fermions de la première génération du Modèle Standard lorsqu'ils sont placés dans la représentation de dimension 16 de SO(10). Sont aussi indiquées les charges vis-à-vis du Modèle Standard, ainsi que les charges d'isospin faible droit et associées à U(1)$_{B-L}$ et U(1)$_X$.}
 \label{TableNQSO10Full}
 \end{center}
\end{figure}
\end{landscape}
}

\part{Cordes cosmiques réalistes}
\label{ChapterRealisticStrings}
\chapter{Défauts topologiques et univers primordial}
\section{Introduction}

\noindent
La formation de défauts topologiques est une conséquence générique des transitions de phase impliquant des brisures spontanées de symétrie. Ce sont des configurations de vide non triviales apparaissant après la brisure de symétrie, formant des structures topologiquement stables et ayant des propriétés physiques localisées (énergie, champ magnétique, etc.). Ce sont des structures non perturbatives, correspondant à des solitons en théorie des champs. De tels types de défauts sont notamment formés lors de la brisure spontanée de symétries de jauge par un champ de Higgs, comme cela apparaît en physique des particules.

Pour étudier ces défauts topologiques, il est nécessaire d'étudier les différentes configurations possibles du vide après la brisure de symétrie. Ces configurations ont un statut non trivial dans le cadre des théories de jauge, puisqu'il est possible de redéfinir la direction de brisure de symétrie en chaque point à l'aide d'une transformation de jauge. Il est cependant possible d'identifier des structures invariantes de jauge dans les configurations de vide, liées aux propriétés topologiques de l'espace formé par l'ensemble des directions possibles de brisure de symétrie. C'est la présence de telles structures invariantes qui est associée à la formation de défauts topologiques. Une étude de la dynamique des transitions de phase permet alors de prédire la densité de formations de tels défauts après une brisure spontanée de symétrie.

Après avoir examiné de façon générale la structure et la formation des défauts topologiques, on discutera la formation de tels défauts dans le cadre de la brisure spontanée des théories de grande unification dans l'univers primordial. On se focalisera plus particulièrement sur les défauts linéaires, ou cordes cosmiques, évoquées pour la première fois par Kibble en 1976~\cite{Kibble:1976sj,Kibble:1980mv}, et notamment sur leurs conséquences cosmologiques et observationnelles. L'accent sera mis sur le lien entre leur structure microscopique et leurs propriétés macroscopiques. Nous ne discuterons pas d'autres types de structures linéaires pouvant apparaître en cosmologie, comme des cordes globales, semi-locales~\cite{Achucarro:1999it}, de Alice~\cite{Schwarz:1982ec,Schwarz:1982zt}, ou même des supercordes fondamentales~\cite{Copeland:2011dx}.

\section{Modèle de Higgs abélien et cordes de Nielsen-Olesen}
\label{PartCordesHiggsAbelien}

\noindent
Avant de faire une discussion générale sur les défauts topologiques, nous considérons le cas particulier des défauts topologiques linéaires, appelés généralement lignes de vortex ou cordes cosmiques. Nous prendrons l'exemple de la formation de tels défauts lors de la brisure spontanée de symétrie du modèle de Higgs abélien. Ce modèle, présenté dans le complément~\ref{TheorieJaugeScalaire}, décrit un champ de Higgs scalaire complexe $\phi$, ayant une charge $q$ sous une symétrie de jauge U(1). Son Lagrangien s'écrit
\begin{equation}
\mathcal{L}_{\text{Higgs abélien}} = - ( D_\mu \phi )^* (D^\mu \phi) - \frac14 F_{\mu\nu} F^{\mu\nu} - \lambda(|\phi|^2 - \eta^2)^2,
\end{equation}
les constantes réelles $\lambda$ et $\eta$ étant prises strictement positives. La brisure de la symétrie locale U(1) par l'acquisition en chaque point de l'espace d'une {\sc vev} non nulle $\braket{\phi}$ pour le champ scalaire a été décrite dans la section~\ref{PartSSBHiggsAbelien}. Pour le modèle de Higgs abélien, les {\sc vev}s sont de la forme 
\begin{equation}
\label{EqVevHiggsAbelien}
\braket{\phi(x)} = \eta e^{i \theta_0(x)}.
\end{equation}
Les angles $\theta_0(x)$ indiquent une direction de brisure de symétrie en chaque point de l'espace. Le champ scalaire étant continu, la fonction $\theta_0$ doit l'être également. Dans cette section, l'état de vide défini par le champ scalaire et les défauts topologiques associés sont étudiés dans un cadre statique ; le cas dynamique sera discuté ultérieurement.

Dans le cadre des théories de jauge, la signification de la direction de brisure de symétrie n'est pas forcément univoque. On a en effet vu dans la section~\ref{PartSSBHiggsAbelien} qu'il était difficile de décrire de façon absolue une telle direction, le modèle étant invariant sous les transformations de jauge. On peut par exemple se questionner sur la possibilité de transformer continument toute configuration de vide en une configuration où $\theta_0$ a une même valeur en tout point de l'espace. Pour discuter cette possibilité, on introduit une grandeur $N$ associée à toute courbe fermée et orientée $\Gamma$, nommée nombre d'enroulement (winding number). Cette grandeur est définie pour chaque courbe fermée comme la circulation du gradient de $\theta_0$ sur celle-ci,
\begin{equation}
N(\Gamma) = \frac{1}{2\pi} \oint_\Gamma \vec \nabla \theta_0 \cdot \text{d}\vec l~.
\end{equation}
La phase étant une fonction continue, $N(\Gamma)$ caractérise le nombre algébrique de rotations que fait $\theta_0$ sur la courbe orientée $\Gamma$ avant de revenir à sa valeur initiale. C'est donc un entier relatif : $N(\Gamma)\in\mathbb{Z}$. 

\begin{figure}[h!]
\begin{center}
\includegraphics[scale=1]{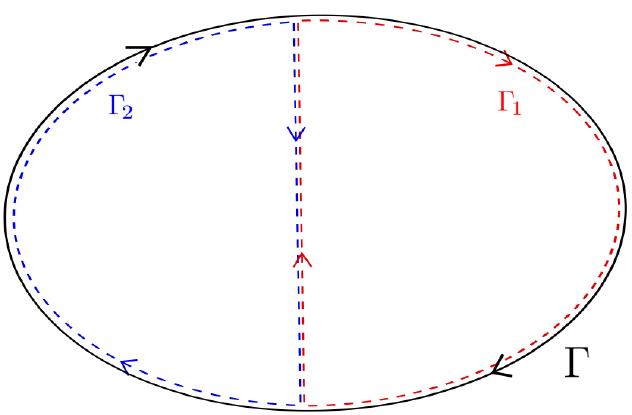}
\end{center}
 \caption{Décomposition d'un chemin $\Gamma$ en deux sous-chemins $\Gamma_{\hspace{-0.05cm}1}$ et $\Gamma_{\hspace{-0.05cm}2}$.}
 \label{CompositionChemins}
\end{figure}

Ces nombres d'enroulement ont la particularité d'être invariants sous les transformations de jauge continues. On peut le voir par exemple pour les transformations de jauge infinitésimales
\begin{equation}
\theta_0(x) \rightarrow \theta_0(x) + \delta \theta_0 (x),
\end{equation}
où $\delta \theta_0 (x)$ est une fonction continue. Ces transformations ne modifient pas les nombres d'enroulement de la configuration. Si c'était le cas, le champ $\delta\theta_0(x)$ aurait des nombres d'enroulement non-nuls, ce qui n'est pas possible sans que ce champ ne prenne des valeurs non infinitésimales. En effet, prenant un chemin avec un nombre d'enroulement non-nul, et intégrant la variation de $\delta \theta_0$ à partir d'un point $x_1$ où $\delta \theta_0(x_1)$ est infinitésimal, il y aurait par exemple au moins un point $x_2$ où l'on aurait $\delta \theta_0 (x_2) = \delta \theta_0(x_1) + \pi$, ce qui est incompatible avec le fait que $\delta \theta_0$ soit infinitésimal. Les transformations de jauge continues s'obtenant par l'intégration de transformations infinitésimales, elles conservent donc aussi les nombres d'enroulement\footnote{Il est toujours possible de considérer des transformations de jauge non continues, qui peuvent par exemple amener la phase $\theta_0$ dans la jauge unitaire $\tilde\theta_0 = 0$ quelque soit la configuration initiale de $\theta_0$. Ces transformations ne sont cependant pas incompatibles avec la présence de défauts topologiques. Les transformations de jauge affectant aussi les champs de jauge $A_\mu$, les défauts topologiques se manifesteront dans la jauge unitaire comme des singularités non-physiques des champs de jauge~\cite{Rajantie:2001ps}. Dans le cas du modèle de Higgs abélien, les transformations de jauge menant à la jauge unitaire contiendront par exemple des composantes du type $2\pi \delta(x)$ permettant de \og casser \fg{} les enroulements de $\theta_0$, et qui seront liées à des singularités du champ vecteur.}. Les nombres d'enroulement étant invariants de jauge, ils décrivent des propriétés physiques de l'état du vide après brisure de symétrie. Des nombres d'enroulement non-nuls vont en effet mener à des structures de vide non-triviales ayant une énergie supérieure à la configuration où la phase $\theta_0$ est constante sur l'espace. Ces configurations particulières sont des défauts topologiques, qui sont des lignes de vortex dans le cas du modèle de Higgs abélien. Dans un cadre cosmologique, ces lignes de vortex sont appelées cordes cosmiques. C'est l'appellation que l'on prendra par la suite.

Étudions une configuration de vide contenant un chemin $\Gamma$ avec un nombre d'enroulement non nul\footnote{En tout rigueur, il faut partir d'un chemin dont des petites déformations ne changent pas le nombre d'enroulement -- et ne considérer de même que ce type de chemins durant toute la démonstration --, ce qui signifie que la singularité recherchée n'est pas située sur celui-ci. Cela permet d'éviter les contre-exemples triviaux où l'on divise un chemin de nombre d'enroulement nul en deux chemins de nombres d'enroulement opposés. C'est ce qu'on requiert généralement en demandant que les configurations de vide \og loin de la corde \fg{} aient un nombre d'enroulement non-nul.}. On peut alors montrer qu'en au moins un point de chaque surface portée par $\Gamma$, la norme du champ de Higgs doit s'annuler. Pour comprendre cela, il est possible de diviser successivement le chemin $\Gamma$ en sous-chemins, comme représenté sur la figure~\ref{CompositionChemins}. Dans cette figure, les branches verticales des sous-chemins $\Gamma_{\hspace{-0.05cm}1}$ et $\Gamma_{\hspace{-0.05cm}2}$ ont été séparées à des fins de lisibilités, mais elles sont supposées confondues. Une intégration sur les deux sous-chemins donne alors
\begin{equation}
N(\Gamma_{\hspace{-0.05cm}1}) + N(\Gamma_{\hspace{-0.05cm}2}) = N(\Gamma),
\end{equation}
ce qui implique que le nombre d'enroulement d'au moins un des deux sous-chemins est non nul si le nombre d'enroulement de $\Gamma$ est non nul. Continuant cette décomposition autant que nécessaire, on obtient alors des chemins de nombres d'enroulement non nuls ayant des dimensions linéaires tendant vers zéro. Pour un tel chemin $\gamma$ de dimension caractéristique $l_\gamma$, la contribution cinétique à l'énergie provenant des variations de $\theta_0$ est de l'ordre
\begin{equation}
K_{\theta_0} \simeq \eta^2 \nabla_\mu \theta_0 \nabla^\mu \theta_0 \simeq \eta^2 \left[\frac{  N(\gamma)}{l_\gamma}\right]^2,
\end{equation}
où l'on a considéré seulement la contribution de la valeur moyenne dans le vide, que l'on a supposée être de la forme donnée dans l'équation~\ref{EqVevHiggsAbelien}. Cette contribution diverge pour $l_\gamma\rightarrow 0$, ce qui n'est pas physique. La seule solution pour éviter une telle divergence est alors que la norme du champ de Higgs s'annule en un point au centre des chemins de nombre d'enroulement non nul. En effectuant des raisonnements similaires après avoir effectué des translations verticales, on montre la présence d'un objet linéaire au centre duquel la norme du champ de Higgs s'annule : c'est une corde cosmique. Cet objet possède une énergie localisée -- puisque le potentiel a une valeur non nulle lorsque le champ de Higgs s'annule -- et n'est donc pas simplement une singularité dans la description du champ.

\begin{figure}[t!]
\begin{center}
\includegraphics[scale=1]{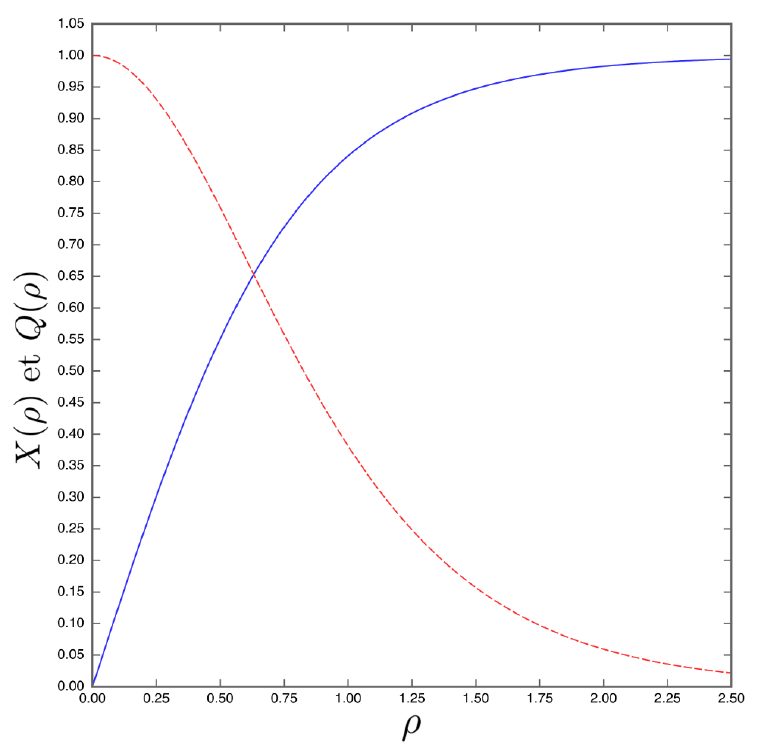}
\end{center}
 \caption{Structure d'une corde de Nielsen-Olesen du modèle de Higgs abélien, pour $q=1$ et $\lambda=1$. La ligne continue bleue correspond à l'amplitude du champ de Higgs, et la ligne pointillée rouge à celle du champ $Q$.}
 \label{AbelianHiggsStructure}
\end{figure}

Un modèle simple pour décrire ce type de défauts est la solution de Nielsen-Olesen~\cite{Nielsen:1973cs}, décrivant également les structures de tube de flux d'Abrikosov dans le cadre de la superconductivité~\cite{Abrikosov:1956sx}. Cette solution statique, dans un jeu de coordonnées cylindriques ($r,\alpha,z$), suppose qu'on peut décrire la corde suffisamment proche pour négliger sa courbure. La position du centre de la corde, c'est à dire de l'ensemble des points où le champ de Higgs s'annule, est alors repérée par $r=0$. La solution de Nielsen-Olesen est postulée à symétrie cylindrique et invariante par translation sur l'axe vertical. Le champ de Higgs est noté $\phi = \varphi e^{i\theta}$, avec $\varphi$ sa norme. Étudiant une corde de nombre d'enroulement $N$ (aussi appelé indice de Pontryagin), la phase du champ de Higgs autour de la corde est alors simplement 
\begin{equation}
\theta = N \alpha.
\end{equation}
Loin de la corde, cette solution redonne un état de vide de la forme $\phi = \eta e^{i N \alpha}$. Tenant compte des symétries de la solution, le champ de jauge n'a qu'une composante non nulle $A_\alpha$, et la norme du champ de Higgs $\varphi$ comme le champ de jauge sont des fonctions de $r$ uniquement. Plutôt que de travailler avec $A_\alpha (r)$, il est commode d'introduire la grandeur $Q(r)=N + q A_\alpha (r)$, qui s'annule loin de la corde\footnote{La solution loin de la corde minimise l'énergie, et doit donc notamment annuler la partie spatiale des dérivées covariantes. Or, la contribution de cette dérivée covariante loin de la corde et sur l'indice $\alpha$ est $D_\alpha \phi \propto (N + q A_\alpha) \phi \propto Q \phi$.}. Les équations du mouvement régissant ces fonctions sont alors
\begin{equation}
\frac{\text{d}^2 X}{\text{d}\rho^2} + \frac{1}{\rho}\frac{\text{d}X}{\text{d}\rho} = \frac{X Q^2}{\rho^2} +\frac14 \left(X^2 -1\right) X,
\end{equation}
et
\begin{equation}
\frac{\text{d}^2 Q}{\text{d}^2 \rho^2} -\frac{1}{\rho}\frac{\text{d}Q}{\text{d}\rho} = \tilde{q}^2 X^2 Q,
\end{equation}
où l'on a introduit les variables adimensionnées $X=\varphi/\eta $, $\rho = \sqrt{8\lambda} \eta r$, et $\tilde{q}^2 = q^2/(4\lambda)$, qui est le seul paramètre dont dépend la solution. La solution associée à une corde cosmique correspond aux conditions aux limites
\begin{equation}
\left\{
\begin{array}{l}
X(0)=0,\\
Q(0)=N,
\end{array}
\right.
~~~~ ~~~~
\left\{
\begin{array}{l}
X(\infty)=1,\\
Q(\infty)=0.
\end{array}
\right.
\end{equation}
Une résolution numérique pour $N=1$ est donnée dans la figure~\ref{AbelianHiggsStructure}. Ce type de structure ne peut pas être décrite de façon perturbative. Nous reviendrons sur ses propriétés dans la section~\ref{PartStringsMicroToMacro}.

\section{Topologie du vide et défauts topologiques}
\label{PartTopologie}

\begin{figure}[t!]
\begin{center}
\includegraphics[scale=1]{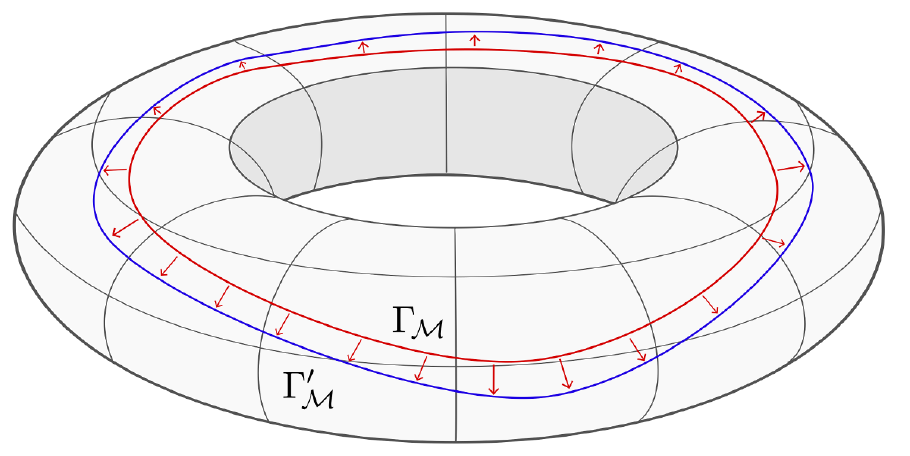}
\end{center}
 \caption{Déplacement d'un chemin $\Gamma_{\hspace{-0.05cm}\mathcal{M}}$ lors d'une transformation de jauge, sur une variété de vide $\mathcal{M}$ torique.} 
 \label{DeformationTore}
\end{figure}

\noindent
La description des cordes cosmiques pour le modèle de Higgs abélien peut être généralisée à des brisures de symétrie plus générales, qu'on écrira sous la forme
\begin{equation}
\label{EqBrisureGtoH}
G \overset{\braket{\mathbf{\Phi}}}{\relbar\joinrel\relbar\joinrel\longrightarrow} H,
\end{equation}
$G$ et $H$ étant les groupes de symétrie initial et résiduel, et $\mathbf{\Phi}$ le champ de Higgs réalisant la brisure spontanée de symétrie. Pour identifier si des défauts topologiques linéaires peuvent exister, il est toujours possible de partir de l'étude de l'évolution de la direction de brisure de symétrie $\braket{\mathbf{\Phi}}$ sur un chemin fermé $\Gamma$. La donnée de $\braket{\mathbf{\Phi}}$ sur ce chemin $\Gamma$ correspond alors à une courbe fermée $\Gamma_{\mathcal{M}}$ sur la variété $\mathcal{M}$ définie par l'ensemble des états de vide associés à la brisure de symétrie étudiée, soit les différentes \og directions de brisure de symétrie \fg{}. Dans le modèle de Higgs abélien, cette variété est le cercle situé au fond de la cuvette de potentiel : $\mathcal{M}_{HA}\simeq S^1$ (où l'on note de façon systématique $S^d$ la $d$-sphère\footnote{Cette $d$-sphère est une surface à $d$ dimension formant une sphère dans un espace à $d+1$ dimension ; $S^0$ correspond à deux points distincts, $S^1$ à un cercle, et $S^2$ à la sphère d'un espace tri-dimensionnel.}). Les transformations de jauge continues modifiant les valeurs de $\braket{\mathbf{\Phi}}$ sur le chemin $\Gamma$ correspondent alors à un déplacement de la courbe fermée $\Gamma_{\mathcal{M}}$ associée à ce chemin sur la variété $\mathcal{M}$. Un exemple de transformation de jauge modifiant un chemin $\Gamma_{\hspace{-0.05cm}\mathcal{M}}$ en un chemin $\Gamma^\prime_{\hspace{-0.05cm}\mathcal{M}}$ sur une variété $\mathcal{M}$ torique est représenté dans la figure~\ref{DeformationTore}.

Pour que des cordes cosmiques puissent se former après la brisure de symétrie décrite équation~\eqref{EqBrisureGtoH}, il faut qu'il puisse exister des configurations où la valeur moyenne dans le vide du champ de Higgs sur un chemin $\Gamma$ ne puisse pas être ramenée à une valeur constante par une transformation de jauge. Cela revient à questionner la possibilité d'avoir des chemins $\Gamma_{\mathcal{M}}$ non contractibles sur la variété $\mathcal{M}$, soit à étudier le premier groupe d'homotopie $\pi_1(\mathcal{M})$ de cette variété (les définitions et propriétés du premier groupe d'homotopie sont données dans l'annexe~\ref{AnnexePartTopologie}). Si ce groupe est trivial, il est toujours possible d'effectuer une transformation de jauge ramenant à une constante la valeur du champ de Higgs sur un chemin $\Gamma$, et il n'est pas possible de former des cordes cosmiques. Si $\pi_1(\mathcal{M})$ est non trivial, il est possible de former des cordes cosmiques. 

Indépendamment des transformations de jauge, les boucles $\Gamma_{\hspace{-0.05cm}\mathcal{M}}$ associées à un chemin $\Gamma$ seront alors toujours associées à un même élément de $\pi_1(\mathcal{M})$. Ce sont ces éléments du premier groupe d'homotopie qui généralisent la notion de nombre d'enroulement du modèle de Higgs abélien ; l'invariance de jauge sur l'espace réel étant reliée à la notion d'invariance topologique sur $\mathcal{M}$. Dans le cas du modèle de Higgs abélien, le premier groupe d'homotopie d'un cercle est bien $\mathbb{Z}$, les entiers relatifs s'identifiant au nombre d'enroulement défini dans la section précédente. Il est également possible d'introduire ces grandeurs comme des intégrales sur les coordonnées de la variété $\mathcal{M}$, ces coordonnées repérant la \og direction de brisure de symétrie \fg{}. La composition des chemins $\Gamma_{\hspace{-0.05cm}i}$ sur l'espace réel correspond alors à la composition des chemins $\Gamma_{\hspace{-0.05cm}\mathcal{M},i}$ sur la variété $\mathcal{M}$, cette dernière composition s'identifiant à l'opération du groupe $\pi_1(\mathcal{M})$.

\begin{figure}[h!]
\begin{center}
\includegraphics[scale=1.3]{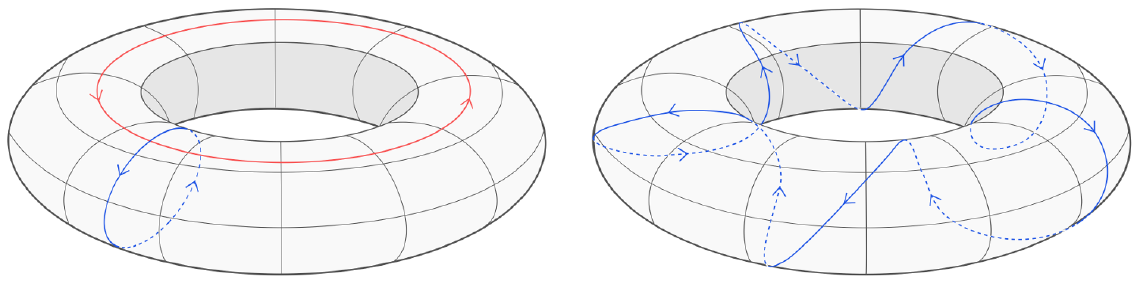}
\end{center}
 \caption{Différentes configurations de boucles fermées sur une variété correspondante au vide d'une brisure de symétrie vérifiant $G/H \simeq \text{U(1)}\times \text{U(1)}$. Dans la figure de gauche, les boucles rouge et bleue ont des nombres d'enroulement respectifs de (1,0) et (0,1). Dans la figure de droite, la boucle a un nombre d'enroulement de (1,5).}
 \label{ToreHomotopie}
\end{figure}

Une propriété importante du mécanisme de Higgs est qu'il est possible d'identifier la variété $\mathcal{M}$ formée par les différentes valeurs moyennes dans le vide possibles pour le champ de Higgs avec le groupe quotient $H/G$. Pour faire cette identification, on remarque pour commencer que les différents vides définissant $\mathcal{M}$ sont reliés par des transformations de $G$. De plus, comme les transformations de $H$ laissent invariantes les valeurs moyennes dans le vide -- par définition même du groupe de symétrie résiduel --, on peut factoriser les transformations de $G$ par celles de son sous-groupe invariant $H$, et ne considérer que les transformations de $G/H$. Cela implique que la variété $\mathcal{M}$ est dans une représentation de $G/H$. Cependant, il a été montré dans la section~\ref{PartBEH} qu'il est possible de repérer la direction de brisure de symétrie du champ de Higgs -- et donc tous les éléments de $\mathcal{M}$ -- à l'aide de $\left[\text{dim}(G) - \text{dim} (H)\right]$ bosons de Goldstone, qui sont des champs scalaires réels. Cela signifie que $\mathcal{M}$ est une variété de dimension $\left[\text{dim}(G) - \text{dim} (H)\right]$. Cette dimension étant aussi celle de $G/H$, la représentation de $G/H$ sur $\mathcal{M}$ est donc la représentation adjointe de $G/H$ sur lui-même, et il est possible d'identifier les deux structures : $G/H \simeq \mathcal{M}$.

Ces résultats permettent d'étudier la possibilité de formation de cordes cosmiques lors d'une brisure spontanée de symétrie par la simple considération des groupes de symétrie initial et résiduel, à savoir que des cordes cosmiques peuvent se former si et seulement si $\pi_1(G/H)$ n'est pas trivial. On identifie par exemple la structure de vide de forme torique donnée dans la figure~\ref{DeformationTore} à une brisure de symétrie vérifiant $G/H \simeq \text{U(1)} \times \text{U(1)}$. Le groupe d'homotopie de $\text{U(1)} \times \text{U(1)}$ étant $\mathbb{Z} \times \mathbb{Z}$, cette brisure de symétrie peut former des cordes. D'autre part, les nombres d'enroulement associés à ces cordes seront décrits par deux entiers relatifs $[N_1(\Gamma),N_2(\Gamma)]$, des configurations de nombres d'enroulement (1,0) et (0,1) étant représentées dans la figure~\ref{ToreHomotopie}. Ces deux configurations correspondent à des cordes cosmiques similaires à celles du modèle de Higgs abélien, et elle n'ont chacune un flux magnétique que pour les champs de jauge associés à un des U(1) brisés. Des configurations du champ plus élaborées sont aussi possibles, comme la configuration de nombre d'enroulement (1,5) représentée figure~\ref{ToreHomotopie}. Cette dernière configuration correspond à des cordes cosmiques différentes de celles apparaissant dans le modèle de Higgs abélien, qui ont notamment un flux magnétique pour les champs de jauge associés aux deux U(1) brisés.

\begin{figure}[h!]
\begin{center}
\includegraphics[scale=0.75]{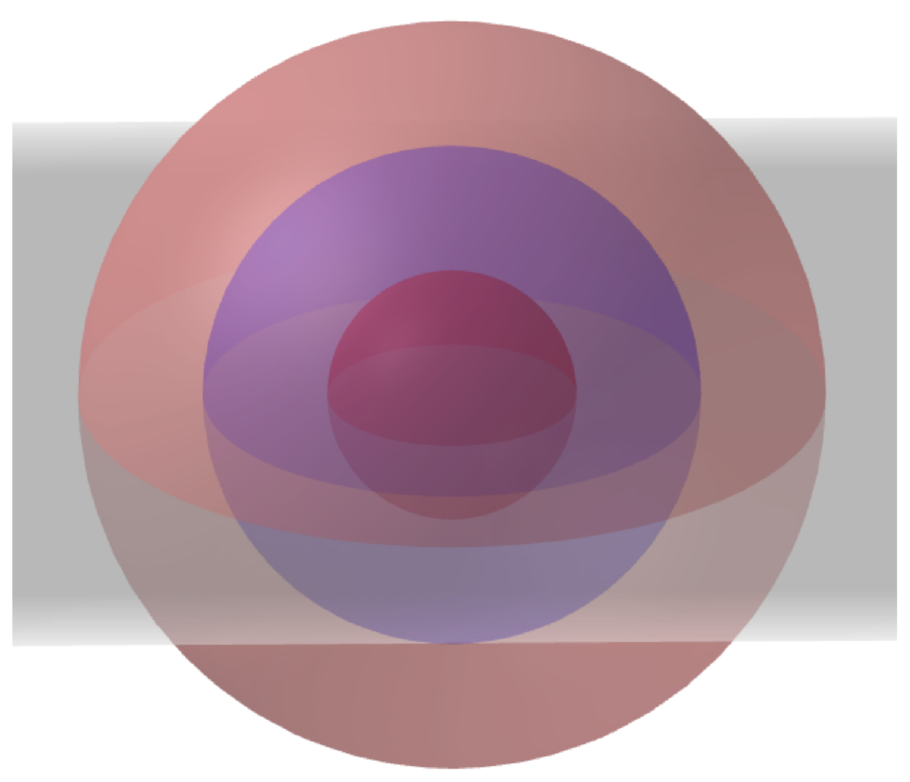}
\end{center}
 \caption{Illustration d'une variété de second groupe d'homotopie non trivial. La variété correspond au volume entre les deux sphères rouges. Il n'est pas possible de déformer continument la sphère bleue en un point tout en restant dans cette variété. Le plan gris a simplement vocation à améliorer la lisibilité du schéma.}
 \label{SecondGroupeHomotopie}
\end{figure}

Cette investigation de la présence de défauts topologiques linéaires pour une configuration de vide peut être étendue à d'autres types de défauts topologiques. Dans un espace tri-dimensionnel, les autres types de défauts topologiques sont les monopôles et les murs de domaines -- aussi appelées membranes. Les monopôles sont liés à des configurations du champ de Higgs sur une surface fermée (comme une sphère) qui ne peuvent pas être ramenées à une valeur constante de $\braket{\Phi}$ par une transformation de jauge. Ces configurations correspondent à des surfaces fermées de $\mathcal{M}$ qui ne peuvent pas être déformées continument en un point, et ne peuvent exister que si le second groupe d'homotopie $\pi_2$ de $\mathcal{M}$ est non-trivial. Les configurations sont alors repérées par les éléments de $\pi_2(\mathcal{M})$ qui leurs sont associés, et qui sont invariants sous les transformations de jauge.
Un exemple heuristique d'une telle situation est donnée dans la figure~\ref{SecondGroupeHomotopie}, où la variété $\mathcal{M}$ correspond au volume entre les deux sphères rouges et où les valeurs du champ sur une surface donnée de l'espace correspondent à la sphère bleu. Il est évident qu'il n'est pas possible de déformer continument la sphère bleue en un point tout en restant dans la variété. On peut alors montrer que la norme du champ de Higgs doit s'annuler en au moins un point de la sphère, autour duquel se situera un défaut topologique ponctuel, nommé monopôle. Le premier groupe simple dont la brisure mène à la formation de monopôles est SU(2), lorsqu'il est brisé en un groupe résiduel abélien U(1).

\begin{figure}[h]
\begin{center}
\includegraphics[scale=1.2]{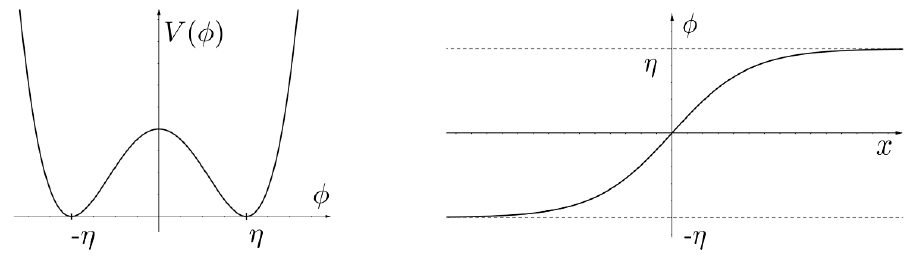}
\end{center}
 \caption{Gauche : Potentiel d'un champ de Higgs scalaire réel pouvant mener à l'apparition de murs de domaines après la brisure spontanée de la symétrie $\mathbb{Z}_2$ du potentiel. Droite : Configurations types du champ de Higgs après la brisure de symétrie, sur une ligne traversant un mur de domaine. Le mur est ici localisé autour de $x=0$, où le champ de Higgs s'annule, et on retrouve loin du mur les valeurs moyennes dans le vide correspondant aux minima de la cuvette de potentiel.}
 \label{CourbesDomainWall}
\end{figure}

Les murs de domaines apparaissent quand des états de vides $\braket{\Phi_1}$ et $\braket{\Phi_2}$ en deux points de l'espace appartiennent à des parties non connexes de l'ensemble des minima du potentiel. Ces différents domaines du potentiel sont généralement reliés par les composantes finies du groupe de symétrie résiduel, et ne sont en tout cas pas reliés par des composantes continues. Or, les groupes finis ne correspondent pas à des symétries de jauge, puisqu'ils n'ont pas de paramètres continus pouvant être associés à des transformations de jauge -- \emph{i.e.} au degré de liberté de jauge d'un champ de jauge ou à un boson de Goldstone après brisure de symétrie. Les groupes finis sont donc traités comme des symétries globales. Cela signifie qu'il n'existe pas de transformation de jauge pouvant relier $\braket{\Phi_1}$ à $\braket{\Phi_2}$, qui sont fondamentalement dans des états de vide différents. La continuité du champ de Higgs impose alors qu'il prenne des valeurs ne correspondant pas à l'état de vide sur les courbes spatiales reliant deux points où le champ prend de telles valeurs $\braket{\Phi_1}$ à $\braket{\Phi_2}$. Les zones de vrai vide sont donc séparées par des zones frontières, formant les murs de domaine. L'exemple simple d'un champ scalaire réel dans un potentiel $V(\phi)=\lambda(\phi^2 - \eta^2)$ est donné figure~\ref{CourbesDomainWall}. On y voit que pour passer d'un état de vide à l'autre, le champ scalaire est obligé de passer par une zone frontière où il prend une valeur nulle\footnote{La valeur du champ de Higgs au centre du mur correspond plus généralement à un point selle du potentiel.}.

La discussion de cette section a été effectuée pour un espace à trois dimensions spatiales. Elle se généralise aisément au cas d'un espace à $D$ dimensions~\cite{Rajantie:2001ps}. L'étude précédente revient à considérer les valeurs du champ de Higgs sur des sphères $S^n$, où $n$ prend des valeurs entre 0 et $D-1$, et donc à associer à ces sphères des contours de même dimension dans la variété $\mathcal{M}\simeq G/H$ des différents vides possibles. Les cas où ces contours ne sont pas contractibles continument en un point indiquent la présence de défauts topologiques. La possibilité de tels défauts est alors encodée dans les $D-1$ premiers groupes d'homotopie de $G/H$, un contour non contractible de dimension $n$ étant associé à un défaut de dimension $d=D-(n+1)$ (un monopôle, ponctuel, étant considéré de dimension nulle). Les défauts topologiques pouvant exister dans des espaces à 3 et 4 dimensions spatiales sont listés dans le tableau~\ref{TableRecapDefautsTopologiques}. On y note l'apparition de défauts topologiques tri-dimensionnels, nommés textures, dans les espaces à 4 dimensions spatiales.

\begin{table}[h!]
\begin{center}
\renewcommand{\arraystretch}{1.4}
\begin{tabular}{|c|c|c|c|}
\hline
\multicolumn{2}{|c|}{Topologie} & Espace 3D & Espace 4D   \\
\hline
$S^0$ & $\pi_0$ & mur de domaine (2$d$) & texture (3$d$) \\
\hline
$S^1$ & $\pi_1$ & corde cosmique (1$d$) & mur de domaine (2$d$)\\
\hline
$S^2$ & $\pi_2$ & monopôle (0$d$) &  corde cosmique (1$d$) \\
\hline
$S^3$ & $\pi_3$ &  & monopôle (0$d$) \\
\hline
\end{tabular}
\end{center}
\caption{Récapitulatif des différents défauts topologiques pouvant exister dans des espaces comportant respectivement 3 et 4 dimension spatiales. La colonne topologie indique les contours dans l'espace réel sur lesquels des configurations de champ non-triviales mènent à l'existence de défauts, ainsi que le groupe d'homotopie de $\mathcal{M}$ lié à la possibilité de telles configurations non triviales. Les dimensions des différents défauts topologiques sont indiquées.}
\label{TableRecapDefautsTopologiques}
\end{table}

\section{Transitions de phase et formation de défauts}
\label{PartPhaseTransition}

\noindent
Dans les sections précédentes, les configurations de vide liées à la présence de défauts topologiques ont été étudiées, ainsi que la possibilité pour une brisure spontanée de symétrie $G \rightarrow H$ donnée de former différents types de défauts. Aucune considération n'a cependant été faite sur l'apparition de ces configurations, liées à l'acquisition par un champ de Higgs de valeurs moyennes dans le vide non nulles lors d'une transition de phase. Il est nécessaire pour cela d'étudier la dynamique de la transition de phase elle-même. Les brisures de symétrie étant spontanées, il est évidemment impossible de prédire les configurations de vide qui se forment pendant de telles transitions ; il est par contre possible de déterminer leurs propriétés statistiques.

Pour décrire la dynamique des transitions de phase, il est nécessaire de considérer l'évolution des champs non plus dans le vide mais dans un bain thermique. Ce bain thermique, pouvant être considéré comme le vide à température non-nulle, est une superposition du vide à température nulle considéré jusqu'à présent et de ses différents états excités. Décrivant par exemple un modèle ne contenant qu'un seul champ $\Psi$ dans le formalisme grand canonique~\cite{Birrell:1982ix}, les état $\ket{\Psi_i}$ d'énergie totale $E_i$ et contenant $n_i$ excitations fondamentales de modes de vibrations apparaissent avec la probabilité
\begin{equation}
\rho_\alpha \propto \exp \left[{-\frac{1}{T} \left(E_i - \mu n_i\right)}\right],
\end{equation}
avec $\mu$ le potentiel chimique de ce champ. Le vide thermique est alors de la forme
\begin{equation}
\ket{\Omega}_T = \frac{1}{Z} \sum_i \ket{\Psi_i}  \exp \left[{-\frac{1}{T} \left(E_i - \mu n_i\right)}\right],
\end{equation}
avec $Z$ la fonction de partition grand canonique permettant de normaliser les probabilités. Les calculs des fonctions de Green définies équation~\eqref{EqGreenFunction} doivent alors être effectués avec le vide thermique $\ket{\Omega}_T$ à la place du vide de température nulle $\ket{\Omega}$.

Ces effets thermiques peuvent être traités comme des corrections au potentiel de température nulle, qui s'identifie avec celui apparaissant dans le Lagrangien. Pour un champ de Higgs $\phi$, le potentiel effectif prenant en compte les corrections thermiques est de la forme~\cite{Rajantie:2001ps,PeterUzan},
\begin{equation}
V_{\text{eff}}(\phi,T) = \left.V(\phi)\right|_{T=0} + \Delta V(\phi,T).
\end{equation}
C'est l'évolution de ce potentiel effectif en fonction de la température, et particulièrement de ses minima, qui va régir la transition de phase. Pour le modèle de Higgs abélien considéré dans la section~\ref{PartCordesHiggsAbelien}, le potentiel a ainsi un seul minimum en $\phi=0$ à haute température, et un ensemble de minima vérifiant $|\phi|=\eta$ à basse température. Dans tous les cas, la transition de phase a lieu à la température critique $T_{\text{c}}$ où l'état $\phi=0$ cesse d'être un minimum global du potentiel, l'échelle d'énergie associée à $T_{\text{c}}$ étant de l'ordre de grandeur de la profondeur de la cuvette de potentiel avant prise en compte de l'interaction avec le bain thermique. 

\begin{figure}[h]
\begin{center}
\includegraphics[scale=1.3]{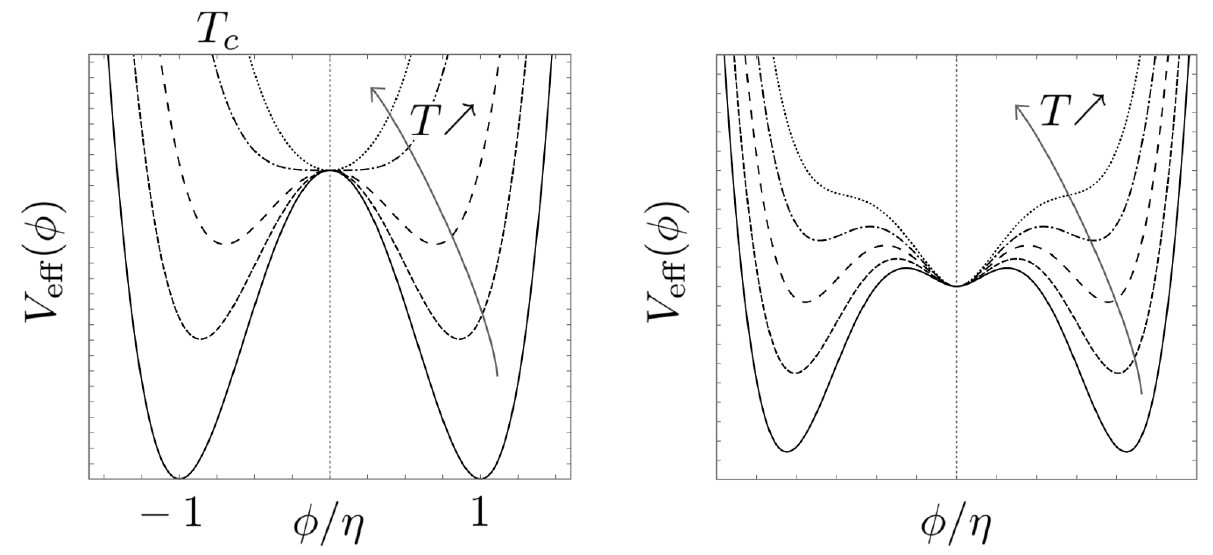}
\end{center}
 \caption{Gauche : Évolution du potentiel effectif en fonction de la température pour la transition de phase du modèle de Higgs abélien, qui est du second ordre. Le potentiel à la température critique est indiqué, et le potentiel en trait plein est celui de température nulle. Droite : Détails de l'évolution du potentiel effectif en fonction de la température pour une transition de phase du premier ordre.}
 \label{PhaseTransition}
\end{figure}

De manière générale, les transitions de phase se classent en transitions du premier et du deuxième ordre, dont l'évolution caractéristique des potentiels en fonction de la température est donnée figure~\ref{PhaseTransition}. Pour les transitions de phase du premier ordre, le minimum global du potentiel ne passe pas continument de $\phi=0$ à haute énergie à des valeurs non-nulles après la transition de phase, et la {\sc vev} du champ de Higgs doit traverser une barrière de potentiel pour accéder à son état de basse température, les états avant et après brisure étant clairement distingués. Pour les transitions du second ordre, aussi appelées transitions continues, le minimum du potentiel passe continument de $\phi=0$ à des valeurs non-nulles, la transition de phase ayant lieu progressivement. Par la suite, nous décrirons les transitions de phase d'un point de vue phénoménologique ; toutes les grandeurs liées aux transitions peuvent cependant être calculées explicitement à partir des paramètres lagrangiens d'un modèle.

Dans les transitions de phase du premier ordre, la phase symétrique reste métastable un peu en dessous de la température critique, et le champ doit localement franchir une barrière de potentiel pour  acquérir une valeur correspondant à un minimum global du potentiel. Ce processus de nucléation a lieu localement, par effet tunnel ou suite à des fluctuations thermiques, et crée une zone de \og vrai vide \fg{} dans le vide métastable. Une bulle de vrai vide ainsi formée va alors grossir jusqu'à rencontrer d'autres bulles similaires de telle façon que le vrai vide remplisse tout l'espace. Les {\sc vev}s des champs de Higgs étant choisies indépendamment dans chaque bulle de vrai vide, la mise en contact de ces bulles va permettre des configurations de champs menant à la formation de défauts comme discuté dans les sections précédentes. En pratique, et pourvu que la topologie de $G/H$ le permette, les rencontres impliquant deux bulles peuvent mener à la formation de murs de domaine~\cite{Zeldovich:1974uw}, celles impliquant trois bulles de cordes cosmiques~\cite{Prokopec:1991ab,Vachaspati:1984dz}, et celles impliquant quatre bulles des monopôles\footnote{Voir aussi~\cite{Hindmarsh:1993av,Borrill:1995gu,Copeland:1996jz} pour une discussion de la possibilité d'effectuer des transformations de jauge dans chacune des bulles de vide avant leur rencontre.}. Il est donc possible d'évaluer la densité moyenne de défauts topologiques formés après la transition de phase à partir de la distance moyenne entre deux bulles de vide, qui peut être là encore calculée à partir des paramètres microscopiques du modèle étudié.

Les transitions de phase du deuxième ordre ne contiennent pas de mécanisme de nucléation, et ont lieu progressivement en tout point de l'espace. Il est cependant possible de faire apparaître une longueur de corrélation $\xi$ au delà de laquelle les {\sc vev}s du champ de Higgs ne sont pas corrélées. Des domaines de tailles caractéristique $D > \xi$ peuvent alors jouer le rôle des bulles de vrai vide dans les raisonnements effectués pour les transitions du premier ordre~\cite{Zeldovich:1974uw,Kibble:1976sj}. Un problème inhérent à cette description est que le calcul à l'équilibre de la longueur de corrélation d'une transition de phase continue montre que celle-ci diverge à la transition. Cependant, et lorsqu'une transition de phase a lieu pendant un temps fini $t_Q$ -- appelé temps de quench -- imposé par des paramètres extérieurs (comme l'expansion de l'univers en cosmologie), il est nécessaire de prendre en compte le temps d'établissement de l'équilibre $t_0$. La prise en compte de ce temps de relaxation décrit le mécanisme dit de Kibble-Zurek, et permet de calculer les longueurs de corrélation effectives apparaissant dans les transitions de phase du second ordre~\cite{Zurek:1985qw,Zurek:1993ek}. 

Les concepts théoriques décrits ci-dessus on pu être testés en laboratoire sur des systèmes variés. La formation de défauts topologiques lors de la brisure de symétries globales a par exemple été étudié pour des cristaux liquides~\cite{Chuang:1991zz,Bowick:1992rz} et des superfluides~\cite{hendry1994generation,bauerle1996laboratory,ruutu1996vortex,dodd1998nonappearance}. Des études similaires dans le cadre de la brisure spontanée de symétries locales ont été faites dans des milieux supraconducteurs~\cite{Carmi:2000zz,PhysRevB.67.104506,PhysRevB.74.144513,PhysRevB.77.054509}, des cristaux ioniques et ferroélectriques~\cite{1367-2630-12-11-115003,PhysRevLett.108.167603,PikaKeller,LinWang} et dans des condensats de Bose-Einstein~\cite{Lamporesi,PhysRevLett.113.065302,Tylutki2015,PhysRevLett.115.170402,PhysRevA.94.023628}.
Toutes ces études confirment et appuient les modèles théoriques utilisés, rendant particulièrement robuste la prédiction de la formation de défauts topologiques lors de transitions de phase avec brisure spontanées de symétrie, et notamment l'évaluation de la densité de défauts formés.

\section{Conséquences cosmologiques des théories de grande unification}

\noindent
Les théories de grande unification reposent sur un mécanisme de brisure spontanée de symétrie permettant de retrouver à basse énergie les symétries locales du Modèle Standard de la physique des particules. Il est alors légitime de questionner la possible formation de défauts topologiques lors de la transition de phase associée à cette brisure de symétrie.
Une telle transition de phase doit partir d'un état de très haute température, de l'ordre de l'échelle d'énergie d'unification $10^{16}\text{ GeV} \simeq 10^{28}$ K, où la symétrie de grande unification est réalisée. Bien que ces températures ne semblent pas atteignables sur Terre ou dans des systèmes astrophysiques, on s'attend à ce qu'elles soient réalisées dans la phase primordiale\footnote{Dans les modèles cosmologiques de big-bang chaud, la brisure de symétrie de grande unification se produit lorsque l'univers a entre $10^{-39}$ et $10^{-37}$s. En comparaisant, la brisure de symétrie électrofaible a lieu de l'ordre de $10^{-11}$s après le big-bang~\cite{Hindmarsh:1994re}.} de l'univers~\cite{PeterUzan}. Or, des défauts topologiques formés suite à la brisure de symétrie des théories de grande unification dans l'univers primordial seraient stables et donc encore présents aujourd'hui, et permettraient alors une observation directe de la phénoménologie des théories de grande unification.

Des calculs d'ordres de grandeur montrent cependant qu'il n'est pas possible d'accommoder l'évolution observée de l'univers avec la présence de murs de domaines~\cite{Zeldovich:1974uw} et de monopôles~\cite{PhysRevLett.43.1365} formés à des échelles de grande unification. Des murs de domaines formés à des échelles d'énergie $E_{\text{GUT}}\simeq 10^{16}$ GeV ont par exemple une masse surfacique de l'ordre de $U_{\text{mur}}\simeq 10^{52} \text{kg}.\text{m}^{-2}$. Si un seul de ces murs traverse l'univers observable, la densité d'énergie qu'il apporte à l'univers est de l'ordre de $\rho_{\text{mur}} \simeq 10^{26}  \text{kg}.\text{m}^{-3}$, soit $10^{52}$ fois la densité d'énergie observée à l'heure actuelle. La façon la plus simple d'éviter ce problème manifeste est de ne considérer que des schémas de brisure de théorie de grande unification ne formant pas de murs de domaine.

On peut montrer de même que les monopôles ne sont pas compatibles avec les observations cosmologiques actuelles. Cependant, et contrairement aux murs de domaines, les monopôles sont une conséquence \emph{a priori} inévitable des théories de grande unification basées sur des groupes simples. En effet, pour une brisure de symétrie $G \rightarrow H$ associée à un vide $\mathcal{M}\simeq G/H$, on peut montrer l'implication~\cite{nakahara2003geometry}
\begin{equation}
\label{EqPropTopologie}
\pi_n(G) \simeq \pi_{n-1}(G) \simeq \text{Id} ~~~~ \Rightarrow ~~~~ \pi_n(\mathcal{M}) \simeq \pi_{n-1}(H).
\end{equation}
Les théories de grande unification étant basées sur des groupes simples vérifiant $\pi_2(G_{\text{GUT}}) \simeq \pi_{1}(G_{\text{GUT}}) \simeq \text{Id}$, et le groupe de symétrie du Modèle Standard ayant une composante U(1)$_Y$ de premier groupe d'homotopie non trivial, le vide $\mathcal{M}$ brisant les théories de grande unification jusqu'au modèle standard vérifie alors $\pi_2(\mathcal{M})\neq\text{Id}$ et doit contenir des configurations associées à des monopôles. Cet incompatibilité manifeste entre les prédictions des théories de grande unification et les observations cosmologiques a historiquement été nommée le \og problème des monopôles \fg{}~\cite{ZELDOVICH1978239,PhysRevLett.43.1365}. La solution communément admise à ce problème réside dans la présence d'une phase d'inflation cosmologique primordiale, qui dilue suffisamment les monopôles préalablement formés pour que leurs influence cosmologique ultérieure soit négligeable~\cite{PeterUzan}.

Les considérations précédentes amènent à rechercher des schémas de brisure de théories de grandes unifications qui sont \og réalistes \fg{} vis-à-vis des observations cosmologiques. Ces schémas de brisures comportent généralement plusieurs étapes~:
\begin{equation}
G_{\text{GUT}} \rightarrow \cdots \rightarrow G_{\text{SM}}.
\end{equation}
Un point important pour l'étude de la validité de ces schémas de brisure consiste à identifier à quelle étape du schéma de brisure a lieu l'inflation. Lorsqu'on met l'accent sur le fait que la phase d'inflation est corrélée avec les différentes transitions de phase cosmologiques\footnote{Si on suppose que l'inflaton est relié à la physique des particules, il est alors naturel que les mécanismes menant à l'inflation soient reliés à des transitions de phases de la physique des particules~\cite{Mazumdar:2010sa}.}, on peut notamment supposer dans un cadre d'inflation hybride que l'inflation commence après la formation des monopôles, et se termine via une nouvelle transition de phase. Une investigation systématique des schémas réalistes\footnote{Si on ne considère que les théories supersymétriques laissant invariante la $R$-parité à basse énergie, comme discuté dans la section~\ref{PartGUTHiggsSector}, la propriété~\eqref{EqPropTopologie} implique alors que des cordes cosmiques sont systématiquement formées à au moins une étape du schéma de brisure. Ce raisonnement n'exclut cependant pas la possibilité que les cordes cosmiques ne se forment que pendant l'étape de brisure formant des monopôles par exemple, et qu'elles soient subséquemment diluées par l'inflation. D'où la nécessité de l'investigation de la référence~\cite{Jeannerot:2003qv}.} du point de vue de la cosmologie comme de la physique des particules a montré que la formation de cordes topologiquement stables lors de la dernière étape de brisure de symétrie mettant fin à l'inflation était inévitable pour des théories d'unification basées sur SO(10) et $E_6$~\cite{Jeannerot:2003qv}. Cela semble indiquer que la formation de cordes cosmiques stables, et pouvant donc être observées jusqu'à aujourd'hui, est une conséquence inévitable des théories de grande unification. Ce sont ces cordes cosmiques que nous allons étudier par la suite. Étant des reliquats de phénomènes liés aux énergies de grande unification, leur masse caractéristique est de l'ordre de $10^{21}$ kg.m$^{-1}$, et leur épaisseur caractéristique de l'ordre de $10^{-30}$ m.

\begin{figure}[h!]
\begin{center}
\includegraphics[scale=1.5]{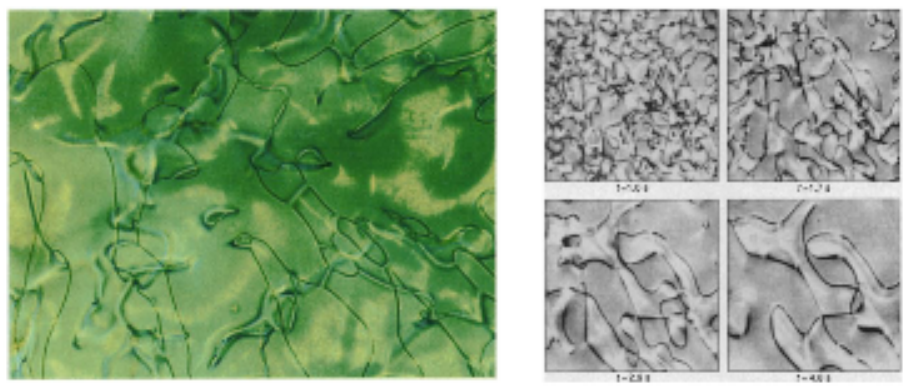}
\end{center}
 \caption{Gauche : Photographie de défauts topologiques linéaires formés après une baisse rapide de température dans une couche mince de cristal liquide nématique. La largeur de la région photographiée est 790 $\mu$m. Droite : Visualisation de l'évolution d'un réseau de défauts topologiques linéaires dans un couche mince de cristal liquide. Les boucles ont tendance à se dissiper en se contractant, menant à des configurations principalement composées de défauts \og infinis \fg{}. La largeur de la région photographiée est 360 $\mu$m. Les deux images sont issues de~\cite{Chuang:1991zz}.}
 \label{RealDefects}
\end{figure}

Les cordes cosmiques formées au moment de la brisure spontanée de symétrie des théories de grande unification forment un réseau cosmologique. La longueur de corrélation de la transition de phase accompagnant la brisure de symétrie donne accès aux propriétés statistiques de la répartition spatiales des cordes, permettant de prédire des distributions types de cordes dans l'univers~\cite{Vachaspati:1984dz}. De tels réseaux de cordes cosmiques contiennent à la fois des boucles et des cordes infinies, c'est à dire qui traversent un volume de Hubble. Pour connaître les propriétés spatiales du réseau de cordes à des temps ultérieurs, il est nécessaire de considérer son évolution dans un cadre cosmologique, impliquant l'expansion de l'univers, la dynamique propre de chaque corde, ainsi que les possibilités pour des cordes de se croiser en échangeant leurs extrémités, ou d'émettre de l'énergie sous forme de rayonnement ou de particules. Une telle évolution d'un réseau de défauts topologiques linéaires n'est pas unique à la cosmologie, et peut s'observer notamment dans différents cristaux liquide~\cite{Chuang:1990hf,Chuang:1991zz}, voir figure~\ref{RealDefects}.

L'évolution temporelle d'un réseau cosmologique de cordes a été étudié de façon analytique et numérique. Les premiers modèles analytiques reposaient sur un seul paramètre~\cite{Kibble:1976sj,Kibble:1980mv}, et ils ont été ensuite raffinés pour inclure jusqu'à trois paramètres~\cite{Copeland:1991kz,Austin:1993rg,Martins:1996jp,Martins:2000cs}, voir aussi~\cite{Lorenz:2010sm} pour une étude analytique récente. Ces modèles sont en bon accord avec les simulations numériques, les premières d'entre elles ayant été faites avec des cordes de Nambu-Goto -- voir la section suivante -- dont la structure microscopique est ignorée~\cite{Bennett:1987vf,Albrecht:1989mk,Allen:1990tv}, les simulations plus récentes~\cite{Vincent:1997cx,Martins:2005es,Ringeval:2005kr,BlancoPillado:2011dq,Hindmarsh:2017qff} introduisant également une structure microscopique décrite par le modèle de Higgs abélien. Ces analyses mettent en avant un phénomène de mise à l'échelle (\og scaling \fg{}), impliquant que les configurations du réseau de cordes sont très peu dépendantes pour des temps longs des conditions initiales comme des paramètres exacts des modèles utilisés. On note cependant que ces raisonnements ne sont \emph{a priori} valables que lors de l'utilisation des modèles de Higgs abélien et des cordes de Nambu-Goto.

\begin{figure}[h]
\begin{center}
\includegraphics[scale=1]{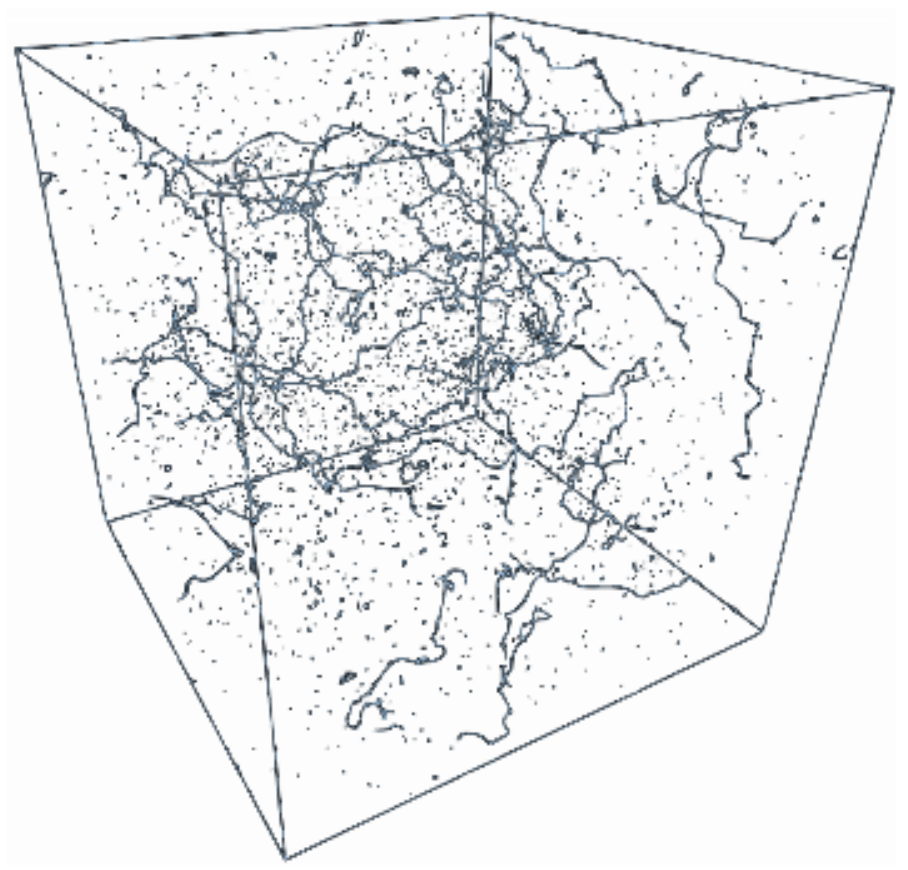}
\end{center}
 \caption{Etat final de l'évolution d'un réseau cosmologique de cordes cosmiques. On observe quelques cordes infinies et des boucles de toutes tailles. Modification d'une figure réalisée par C. Ringeval à l'aide d'un programme de simulation numérique de F. Bouchet et D. Bennett, issue de~\cite{PeterUzan}}
 \label{ReseauCosmologique}
\end{figure}

\section{Cordes cosmiques : de la structure microscopique aux propriétés macroscopiques}
\label{PartStringsMicroToMacro}

\noindent
Une part importante des possibilités d'observation des cordes repose sur leurs effets gravitationnels, liés principalement à leur masse par unité de longueur. De plus, toutes les observations cosmologiques reposent sur les résultats décrits précédemment sur la formation et l'évolution d'un réseau cosmologique de cordes. Or, la plupart de ces résultats sont principalement évalués à partir du modèle de Higgs abélien, sans nécessairement faire de lien avec les théories de grande unification dont la brisure de symétrie mène à la formation de cordes. Il est alors légitime de se questionner sur le bien-fondé d'un tel modèle pour décrire la structure microscopique des cordes cosmiques. D'autant plus que les cordes étant des solitons et ayant une structure essentiellement non-linéaires, leurs propriétés macroscopiques peuvent être considérablement affectées par une modification de leur structure microscopique.

Dans la section~\ref{PartCordesHiggsAbelien}, la description des cordes cosmiques dans le modèle de Higgs abélien a fait apparaître un unique paramètre $q^2/\lambda$ dont dépend la structure des cordes. Si ce paramètre vérifie $q^2/\lambda<1$, une corde de nombre d'enroulement $N$ est plus stable que $N$ cordes de nombre d'enroulement $1$, et la stabilité des cordes augmente avec leur nombre d'enroulement~\cite{Bogomolny:1976tp,Jacobs:1978ch}. Cela implique que des cordes cosmiques se croisant peuvent fusionner en une seule corde~\cite{Salmi:2007ah}, donnant des jonctions dites \og en fermeture éclair \fg{}. Au contraire, dans le cas où $q^2/\lambda>1$, les cordes de nombres d'enroulement plus grands que 1 sont instables, et la probabilité que deux cordes intercommutent lorsqu'elles se croisent est de de l'ordre de l'unité~\cite{Shellard:1987bv,matzner1988interaction}. Ces deux comportements sont associés aux vortex de type I et de type II dans les superconducteurs~\cite{tilley1990superfluidity}, voir figure~\ref{FigTypeCordes}, et on suppose généralement que les cordes cosmiques sont de type II, sans que cela soit forcément justifié par un lien avec la physique des particules.

\begin{figure}[h]
\begin{center}
\includegraphics[scale=0.8]{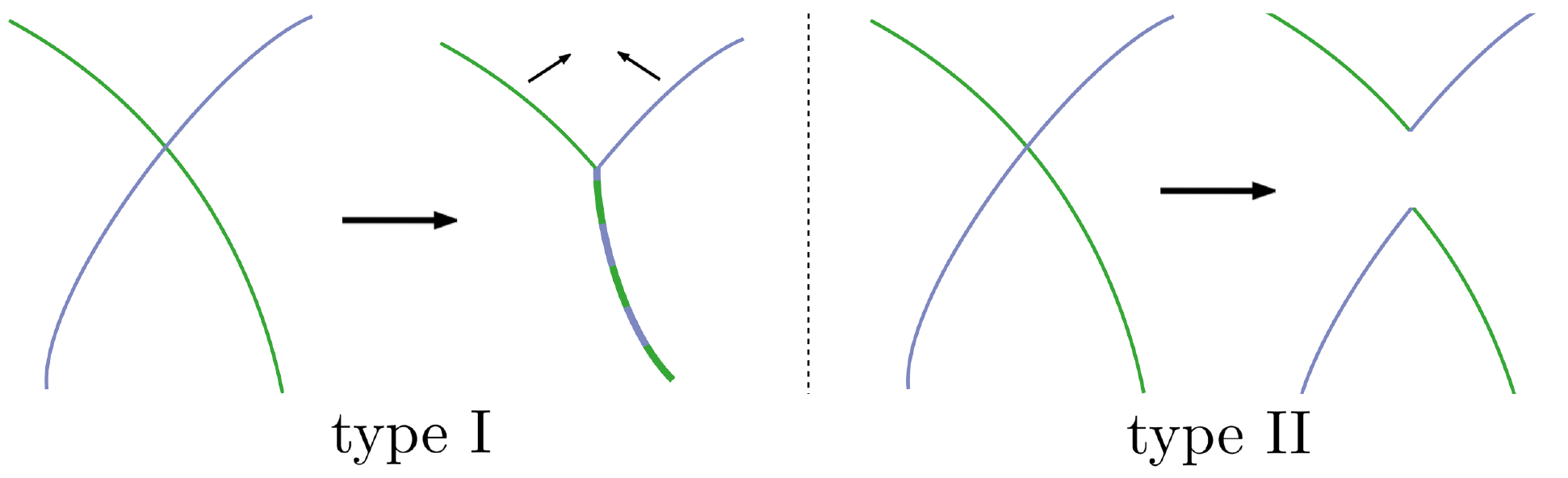}
\end{center}
 \caption{Illustration de vortex de type I (avec des jonctions en fermeture éclair) et de type II (avec des jonctions en intercommutation).}
 \label{ReseauCosmologique}
\end{figure}

Dans le modèle de Higgs abélien, la densité d'énergie décroît exponentiellement à grande distance de la corde. Les défauts topologiques linéaires se formant dans ces théories sont donc localisés, et on peut les caractériser par leur énergie par unité de longueur $\mu$, qui ne peut être que de la forme
\begin{equation}
\mu = \pi \eta^2 f(\lambda/e^2),
\end{equation}
où $f$ est une fonction à calculer dans le modèle microscopique sous jacent. On peut montrer analytiquement que $f(1)=1$~\cite{bogomol1976stability}, et les calculs numériques montrent que $f$ varie logarithmiquement en $e^2/\lambda$~\cite{Jacobs:1978ch,Hill:1987qx} (voir aussi les résultats numériques de la figure~\ref{CS2Graph_eta=0}, où $U_0$ désigne la masse par unité de longueur et où $\kappa$ peut s'identifier à $\lambda/q^2$). Ces résultats dans le cadre du modèle de Higgs abélien tendent à justifier l'approximation $\mu \simeq \eta^2$ communément admise. Cependant, et dans le cas où des champs additionnels sont nécessaires pour décrire la structure microscopique des cordes, la possibilité de former des structures non-perturbatives plus stables énergétiquement n'est pas exclue.

Dans la solution de Nielsen-Olesen, les champs formant les cordes n'ont pas de dépendance en les composantes $z$ ou $t$, dans un jeu de coordonnées cylindriques. Cela implique notamment qu'elles sont invariantes sous les transformations de Lorentz le long de leur direction, et leur tension est alors égale à leur énergie par unité de longueur. Négliger la structure microscopique des cordes en gardant cette équation d'état revient à étudier des cordes dites de Nambu-Goto~\cite{nambu1970symmetries,goto1971relativistic}, qui décrivent macroscopiquement des objets dont les dimensions radiales sont négligeables. Cependant, en présence de symétries locales et de champs additionnels à ceux nécessaire à leur formation, les cordes peuvent être parcourues par des courants, souvent de type supraconducteurs~\cite{Witten:1984eb}. D'un point de vue macroscopique, ces cordes doivent alors être décrites par une équation d'état non triviale reliant leur tension et leur énergie par unité de longueur~\cite{Peter:1992dw,Peter:1992nz,Peter:1992ta,Carter:1994hn,Carter:1996rn,Ringeval:2000kz,Peter:2000sw,Ringeval:2001xd}. Une implication forte de ces courants est notamment la possibilité de stabiliser des boucles, produisant des configurations nommées vortons~\cite{Davis:1988ij}, pouvant être incompatibles avec les observations cosmologiques actuelles~\cite{PeterUzan,Brandenberger:1996zp}.

Il est aussi possible de former des cordes en brisant des symétries de jauge non-abéliennes, et pas seulement des symétries U(1) comme dans le modèle de Higgs abélien. Partant d'un groupe simple comme SU(2) ou SU(3), connexes et de premier groupe d'homotopie trivial, cela nécessite d'avoir un groupe de symétrie résiduel ayant une composante discrète [voir l'implication de l'équation~\eqref{EqPropTopologie}], comme pour une brisure SU(2)$\rightarrow \mathbb{Z}_3$. De telles cordes, dites non-abéliennes, sont associées à des générateurs du groupe de symétrie brisé. Lorsque deux cordes associées à des générateurs $T_1$ et $T_2$ se croisent, elles se retrouvent alors reliées entre elles par une troisième corde, de générateur proportionnel à $[T_1,T_2]$~\cite{PeterUzan}. Les réseaux cosmologiques de cordes non-abéliennes se comportent différemment des réseaux formés de cordes abéliennes ; selon les cas, ils peuvent contenir beaucoup plus de cordes, avoir une dépendance très forte dans la configuration initiale du réseau, ou même avoir un comportement proche de celui d'un solide~\cite{Spergel:1996ai,McGraw:1997nx,Bucher:1998mh}.

Ces quelques exemples montrent que la modification de la structure microscopique des cordes cosmiques peuvent avoir des conséquences très importantes sur leurs propriétés macroscopiques. Ces modifications sont liées à des changements non perturbatifs dans les structures et les propriétés des cordes, et ne peuvent pas être prédites par un simple calcul d'ordre de grandeur. C'est tout particulièrement le cas en présence de symétries et de champs additionnels à ceux suffisants pour former la corde, ce qui apparaît naturellement dans les schémas de brisure de théories de grande unification discutés dans le chapitre~\ref{PartGUT}. Il est donc désirable d'étudier la structure microscopique des cordes cosmiques à partir de descriptions réalistes de théories de grande unification, afin d'établir sur une base solide les propriétés macroscopiques des cordes cosmiques. 

\section{Observation des cordes cosmiques}

\noindent
Les possibilités d'observation des cordes cosmiques peuvent être cosmologiques ou astrophysiques. Les observations cosmologiques sont liées à la structure du réseau de cordes cosmologique. Leurs traitement quantitatifs sont donc dépendantes de l'exactitude des modèles utilisés pour décrire la formation et l'évolution de ce réseau de cordes. À l'opposé, les observations astrophysiques sont liées à des phénomènes n'impliquant qu'une seule corde cosmique, et ne dépendent pas de la répartition des cordes dans l'univers. À l'heure actuelle, aucun signal observationnel n'a pu être associé de façon certaine à des cordes cosmiques, et la plupart des observations impliquent donc uniquement des bornes inférieures et supérieures à l'énergie par unité de longueur des cordes.

Les principales observations cosmologiques des cordes cosmiques consistent en l'empreinte que le réseau de cordes a pu laisser dans le fond diffus cosmologique, ou CMB pour \og Cosmic Microwave Background \fg{}, principalement dû à l'effet Kaiser-Strebbins~\cite{Kaiser:1984iv,Gott:1984ef,Bouchet:1988hh}. Aucune observation claire n'émerge de l'étude du spectre de puissance du CMB~\cite{Bouchet:2000hd,Bevis:2007gh,Ringeval:2010ca,Urrestilla:2011gr,Ade:2013xla}, la contribution des cordes étant au maximum de $3\%$ du total, et le signal étant dominé par la contribution inflationnaire. Il est également possible d'étudier le sillage laissé par les cordes cosmiques~\cite{1986MNRAS.222P..27R,Charlton:1987qy,Stebbins:1987cy,Hara:1987zm}, en étudiant notamment les propriétés topologiques du CMB via l'introduction de fonctionnelles de Minkowski~\cite{Mecke:1994ax,Schmalzing:1997aj,Novikov:1998st,Hikage:2006fe}. Des méthodes de détection directe des discontinuités locales dues aux cordes, ou \og edges \fg{}, ont été étudiées~\cite{Amsel:2007ki,Danos:2008fq}, ainsi que les distorsion spatiales du CMB dues à l'émission d'ondes électromagnétiques par les cordes~\cite{Tashiro:2012nb}. Les empreintes des cordes pourront aussi être aussi observées sur la polarisation du CMB~\cite{Danos:2010gx,Lizarraga:2014eaa,Moss:2014cra,Lazanu:2014eya,Lizarraga:2014xza}, ou sur le rayonnement de fond  à 21 cm~\cite{Khatri:2008kb,Brandenberger:2010hn,Berndsen:2010xc}. Finalement, il est également possible d'observer la contribution du réseau de cordes sur le fond diffus gravitationnel, par l'étude du signal de pulsars binaires, fournissant actuellement la meilleure contrainte à l'énergie des cordes~\cite{Jenet:2006sv,Blanco-Pillado:2013qja}. 

En plus de leur contribution au fond diffus gravitationnel, provenant principalement de l'oscillation de boucles, les cordes cosmiques peuvent émettre des jets gravitationnels au niveau de leur points de rebroussement ou \og cusps \fg{}~\cite{Damour:2001bk,Damour:2004kw}. Ces émissions intenses pourraient être détectées par LIGO ou VIRGO dans le futur. Dû à leur masse très importante, les cordes cosmiques peuvent causer des effets observables de lentillage~\cite{Vilenkin:1984ea,Garriga:1994sy,Mack:2007ae} et de micro-lentillage gravitationnel~\cite{Pshirkov:2009vb,Tuntsov:2010fu,Bloomfield:2013jka}. Etant donné qu'elles contiennent en leur coeur des champs à des énergies de grande unification, les cordes cosmiques peuvent être également la source de rayon cosmiques de très haute énergie~\cite{Vilenkin:1986zz,Bhattacharjee:1989vu,Bonazzola:1997tk,Bhattacharjee:1998qc,Berezinsky:2009xf,Long:2014lxa}. Dans certains cas, l'importance observationnelle de ces rayons est inversement proportionnelle à l'énergie des cordes, et la recherche de rayons cosmiques donne donc à l'heure actuelle une borne inférieure à l'énergie des cordes. Pour finir, les cordes cosmiques peuvent également émettre des jets intenses d'ondes électromagnétiques. Alors que de tels jets dans le domaine visible~\cite{Garriga:1989bx,JonesSmith:2009ti,Steer:2010jk} semblent difficilement observables~\cite{Vachaspati:2015cma}, certains signaux liés à des jets radios, ou \og radio burst \fg{}, émis par des cordes superconductrices~\cite{Vachaspati:2008su,Cai:2012zd} semblent compatibles avec des observations récentes~\cite{Yu:2014gea}.

\section{Conclusion, cordes cosmiques réalistes}

\noindent
On a vu que la formation de défauts topologiques lors d'une transition de phase brisant une théorie de jauge par le mécanisme de Higgs est une prédiction particulièrement générale. Cette prédiction a en effet pu être étudiée quantitativement sur différents systèmes de physique de la matière condensée, et repose sur des concepts théoriques -- théorie de jauge et mécanisme de Higgs -- ayant largement fait leur preuve par ailleurs, notamment dans le cadre du Modèle Standard de la physique des particules.
Si les symétries de grande unification sont une façon correcte de décrire la physique des particules à haute énergie, il semble donc important d'étudier les défauts topologiques se formant lors de la brisure à haute énergie de ces symétries dans un cadre cosmologique, et plus particulièrement les cordes cosmiques, qui apparaissent comme une conséquence inévitable des schémas de brisure réalistes du point de vue de la physique des particules comme de la cosmologie.

On a également vu que les propriétés macroscopiques des cordes cosmiques dépendent considérablement de leur structure microscopique, car elles ont des structures de soliton qui doivent être décrites de façon essentiellement non perturbative. La justification de la structure microscopique des cordes à partir des théories de grande unification dont la brisure mène à leur formation permet d'établir sur une base solide leurs propriétés macroscopiques, et la phénoménologie qui en découle. C'est ce travail que je présente dans cette partie.

Des premiers travaux dans cette direction ont été faits dans les articles~\cite{Ma:1992ky,Davis:1996sp}, dans lesquels j'ai étudié les cordes se formant lors d'une étape de brisure de symétrie de théories de grande unification. Les travaux que je présente par la suite étudient les cordes se formant à la fin d'un schéma complet et réaliste de brisure de symétrie de grande unification. Une propriété importante des schémas complets de brisure réside dans le fait que la valeur des {\sc vev}s des champs de Higgs participant aux premières étapes de brisure de symétrie continuent à être modifiées aux étapes ultérieures. Décrivant un schéma de brisure de la forme
 \begin{equation}
G_{\text{GUT}}\overset{\braket{\Phi_1}}{\relbar\joinrel\relbar\joinrel\longrightarrow} G_1  \overset{\braket{\Phi_2}}{\relbar\joinrel\relbar\joinrel\longrightarrow} G_2  \overset{\braket{\Phi_3}}{\relbar\joinrel\relbar\joinrel\longrightarrow} G_{\text{SM}},
\end{equation}
les $\braket{\Phi_i}$ étant les champs de Higgs implémentant les étapes de brisure de symétrie, cela signifie que les champs $\Phi_1$ et $\Phi_2$ jouent aussi un rôle dans la structure microscopique des cordes formées lors de la brisure de symétrie implémentée par $\Phi_3$ et donnant le Modèle Standard. Les cordes ainsi formées contiennent alors plus de champs que dans les modèles ne prenant en compte qu'une étape de brisure de symétrie, ce qui peut affecter considérablement leurs propriétés macroscopiques, comme discuté dans la section~\ref{PartStringsMicroToMacro}. Cette approche permet également d'identifier les champs apparaissant dans la structure microscopique des cordes cosmiques vis-à-vis des théories de grande unification étudiées ; il est notamment possible de connaitre les différentes charges des champs additionnels interagissant avec la corde, et donc d'étudier par exemple les possibles courants pouvant se former.  

Le premier article sur le sujet~\cite{Allys:2015yda}, reproduit dans le chapitre~\ref{PartArticleCordes1}, étudie la formation de cordes cosmiques à la fin d'un schéma réaliste de brisure de symétrie d'une théorie de grande unification, sans spécifier la théorie en particulier. Cette étude montre que tous les champs de Higgs contribuant au schéma de brisure de symétrie sont nécessaires pour décrire la structure microscopique de ces cordes. La structure minimale pour décrire de tels objets est examinée dans le cas des cordes abéliennes, faisant apparaître les composantes des champs de Higgs singlets du Modèle Standard, \emph{i.e.} qui se transforment trivialement sous l'action des symétries du Modèle Standard. Deux grands types de structures sont identifiées, et un ansatz est donné pour décrire chacune d'elles. Je montre aussi comment normaliser les champs de grande unification de telle façon qu'il soit possible de retrouver le modèle de Higgs abélien dans une limite de découplage. Cette normalisation permet finalement une étude perturbative de l'énergie par unité de longueur des structures réalistes de cordes.

Dans un deuxième article sur le sujet~\cite{Allys:2015kge}, reproduit dans le chapitre~\ref{PartArticleCorde2}, j'effectue l'étude discutée précédemment pour une théorie donnée de grande unification basée sur le groupe SO(10) (le \og Minimal Supersymmetric GUT \fg{} décrit dans la section~\ref{PartGUTStatutActuel}). L'implémentation du schéma de brisure par les champs de Higgs est discutée en détails, et la structure microscopique minimale des cordes cosmiques est obtenue. Cette structure est décrite avec les champs normalisés de telle façon que le modèle de Higgs abélien est retrouvé dans une limite de découplage. J'ai calculé pour cela les règles de branchement numériques -- les coefficients de Clebsch-Gordan -- entre les différents singlets du Modèle Standard apparaissant dans les champs de Higgs. J'ai ensuite effectué un ensemble de résolutions numériques pour obtenir la structure microscopique des cordes cosmiques. Cette étude numérique permet notamment de confirmer les études perturbatives de l'article de le chapitre~\ref{PartArticleCordes1}.

\chapter[Bosonic condensates in realistic supersymmetric GUT cosmic strings (article)]{Bosonic condensates in realistic supersymmetric GUT cosmic strings}
\label{PartArticleCordes1}
%

\begin{figure}[h!]
\begin{center}
\includegraphics[scale=1.1]{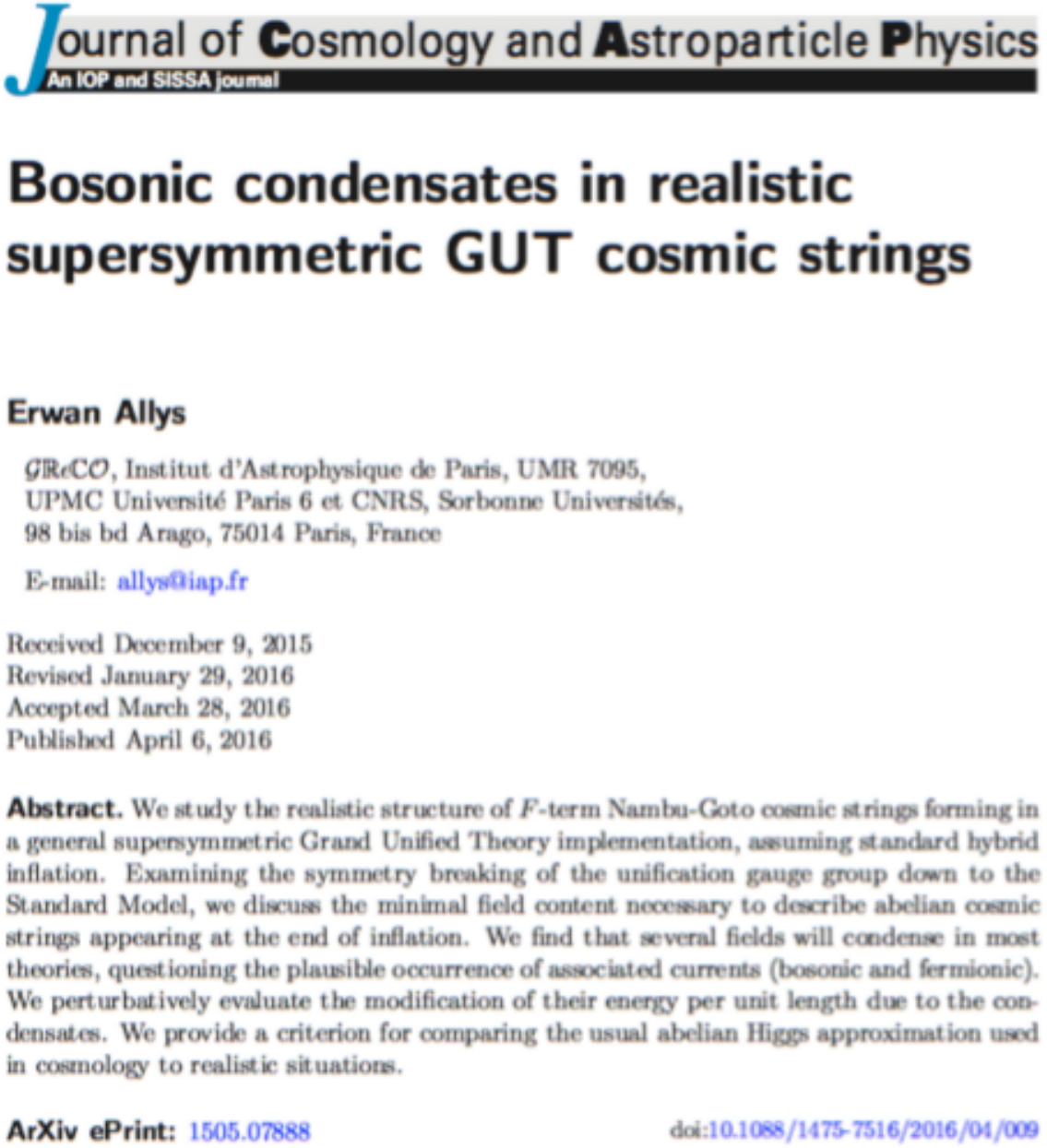}
\end{center}
\end{figure}


\section{Introduction}

Decade long improvements in experimental data led to the conclusion that only those Grand Unified Theories (GUT) involving some amount of Supersymmetry (SUSY) were acceptable \cite{Aulakh:2002zr,Fukuyama:2004ps,Fukuyama:2004xs,Aulakh:2004hm,Bajc:2004xe,Aulakh:2005mw,Bajc:2005qe,Aulakh:2003kg,Raby:2011jt}. The vacuum structure of these theories implies that they should have
produced topological defects during their successive steps of Spontaneous Symmetry Breaking (SSB),
including monopoles and cosmic strings \cite{Kibble:1980mv,Hindmarsh:1994re}, a phase of cosmic inflation being then necessary to dilute the former.
If we consider furthermore a $F-$term hybrid inflation scenario~\cite{Copeland:1994vg,Linde:1993cn,Dvali:1994ms,Lyth:1998xn,Kyae:2005vg,Mazumdar:2010sa,Davis:1999tk},
most of the SSB schemes lead to the formation of cosmic strings at the end of inflation \cite{Jeannerot:2003qv}.
So, constraining the string energy per unit length, e.g. through Cosmic Microwave Background (CMB) observations
\cite{Bouchet:2000hd,Bevis:2007gh,Ringeval:2010ca,Ade:2013xla,Brandenberger:2013tr} provides a general way to constrain GUT themselves. 

The structure of cosmic strings forming at the end of inflation has already been studied in details, see Refs.~\cite{Kibble:1982ae,Aryal:1987sn,Ma:1992ky,Davis:1996sp,Davis:1997bs,Ferreira:2002mg,Morris:1997ua,Davis:1997ny}. These works considered models where only the minimal field content necessary to form a string was introduced, and where the scale of formation of strings is the only dimensionful parameter. The aim of the present paper is to investigate the structure of cosmic strings in a realistic GUT context, \emph{i.e.} starting from the complete GUT field content, and analyzing how the Higgs fields implementing the SSB scheme form the strings. This is done in $F$-terms models, focusing only on the bosonic part of the supermultiplets.

Considering a general SUSY GUT, we identify the minimal field content sufficient to describe the realistic string structure. This minimal structure involves all the fields which take non vanishing Vacuum Expectations Values (VEVs) at the end of the SSB scheme, and thus are singlet of the Standard Model (SM). The energy par unit length will be modified by this condensation of several Higgs fields in the core of the string. The additional fields also give natural candidates to carry bosonic currents \cite{Witten:1984eb,Peter:1992dw,Peter:1992nz,Peter:1993tm,Morris:1995wd}, and we expect their superpartner to carry zero-modes fermionic currents \cite{Witten:1984eb,Jackiw:1981ee,Weinberg:1981eu,Davis:1995kk,Ringeval:2000kz,Peter:2000sw}. Furthermore, the intercommutation process \cite{Shellard:1988ki,Laguna:1990it,Matzner:1988ky,Moriarty:1988qs,Moriarty:1988em,Shellard:1987bv} and the cusps evaporations \cite{Srednicki:1986xg,Brandenberger:1986vj,Gill:1994ic,Olum:1998ag,Bhattacharjee:1989vu} can be qualitatively modified due to this extra structure. The modification of any of these properties can have a major impact on the cosmological consequences of cosmic strings. In this paper, we give a complete description of the realistic microscopic structure of cosmic strings, and we perform a perturbative study of the modification of their energy per unit length from standard toy models. This gives a first step for the study of the other phenomena mentioned above.

For this purpose, we give an ansatz and boundary conditions for such a minimal structure, in the case of Nambu-Goto abelian strings. The conventions and normalizations chosen are such that if all the fields were to decouple from the string-forming Higgs, one would recover the standard abelian Higgs model. Two different classes of strings are discussed, only depending on how the Higgs fields of the GUT implement the SSB scheme, referred to as single and many-field strings. We perturbatively evaluate the modification of the energy per unit length from standard toy models due to this complex structure, taking into account the numerical factors appearing when one writes the complete formulation of the GUT.

In Sec.~\ref{CS1GeneralModel}, we specify the SUSY GUT studied while setting the notation, and briefly describe its SSB scheme in parallel with the inflationary process. In Sec.~\ref{CS1PartAbelianStrings}, using properties of the GUT reviewed in the first section, we present the abelian cosmic strings considered, discuss their minimal structure, and distinguish the two categories of strings. Finally, in Secs.~\ref{CS1SingleFieldStrings} and~\ref{CS1ManyFieldsStrings}, we propose ansätze and perform a perturbative study of their properties.


\section{Theoretical framework}
\label{CS1GeneralModel}

\subsection{Field content and superpotential}
\label{CS1Superpotential}

We consider a general SUSY GUT associated with a gauge group $G$. Usual
examples are based on either SO(10) or SU(6). The spontaneous symmetry
breaking down to the SM will be implemented in
the context of a $F$-term hybrid inflation
\cite{Copeland:1994vg,Linde:1993cn,Dvali:1994ms,Lyth:1998xn,Kyae:2005vg,Mazumdar:2010sa,Davis:1999tk}. The field content of the theory includes a set of chiral supermultiplets
and a gauge supermultiplet associated with the generators of $G$.
We restrict ourselves to the bosonic sector of the
model, and so only write down the scalar part of the chiral
supermultiplets. As we study a $F$-term theory, we assume the
$D$-terms are identically zero,
with no Fayet-Iliopoulos terms~\cite{Martin:1997ns}. It will add some constraints to the fields.

In order to implement hybrid inflation, it is a necessity to have at
least two fields $ \mathbf{\Sigma}$ and $\overline{ \mathbf{\Sigma}}$ in complex conjugate
representations, which have a coupling term with the inflaton in the
superpotential \cite{Copeland:1994vg,Linde:1993cn,Dvali:1994ms}. The inflaton is assumed to be a chiral supermultiplet
of scalar component $S$, singlet under $G$. In order to reach the SM
symmetry and have a phase of inflation washing away monopoles, one
needs to have more than one SSB step; this means other fields
should be present. The other chiral supermultiplets are
denoted $\mathbf{\Phi}_I$ for the fields in real representations, and $\mathbf{\Phi}_i$
and $\mathbf{\Phi}_{\bar{\imath}}$ for the fields in complex representations.

We assume the most general superpotential taking into
account all the above chiral supermultiplets, supplemented by
a specific term for the inflaton $S$ to implement hybrid inflation. It is,
assuming explicit summation on all repeated indices, \cite{Martin:1997ns}
\begin{align}
\label{CS1EqSuperpotential}
W=&~ m_\Sigma \mathbf{\Sigma} \overline{ \mathbf{\Sigma}} + \frac{1}{2} m_{IJ} \mathbf{\Phi}_I \mathbf{\Phi}_J + m_{i\bar{\jmath}} \mathbf{\Phi}_i \mathbf{\Phi}_{\bar{\jmath}} 
+ \eta_x \mathbf{\Sigma} \overline{ \mathbf{\Sigma}} \mathbf{\Phi}_x \nonumber \\
& ~~ + \beta_{xy} \mathbf{\Sigma} \mathbf{\Phi}_x \mathbf{\Phi}_y + \bar{\beta}_{xy} \overline{ \mathbf{\Sigma}} \mathbf{\Phi}_x \mathbf{\Phi}_y 
+ \frac{1}{3}\lambda_{xyz} \mathbf{\Phi}_x \mathbf{\Phi}_y \mathbf{\Phi}_z + \kappa S ( \mathbf{\Sigma}\overline{ \mathbf{\Sigma}} - M^2),
\end{align}
the last term actually implementing hybrid inflation. In this
equation, the label $x$, $y$ and $z$ can be either $I$, $i$ or
$\bar{\imath}$. All the coefficients which appear in addition to the fields
are complex constants, and $\beta_{xy}$, $\bar{\beta}_{xy}$ and
$\lambda_{xyz}$ are totally symmetric in their indices. The constants
$\kappa$, $M$, $m_\Sigma$ and the diagonal elements of the mass
matrices can be set real and positive by redefinition of the phases of
the fields. Depending on the explicit choices of representations, the
coefficients $m_{IJ}$, $m_{i\bar{\jmath}}$, $\eta_x$, $\beta_{xy} $,
$\bar{\beta}_{xy} $ and $\lambda_{xyz}$ will be non-zero only when they allow to build gauge singlets. We did not include terms like
$\mathbf{\Sigma}\mathbf{\Sigma}\mathbf{\Phi}$ and
$\mathbf{\overline{\Sigma}}\mathbf{\overline{\Sigma}}\mathbf{\Phi}$ for the sake of simplicity; their contribution to the macroscopic structure of the string is shortly discussed in Sec. \ref{CS1ManyFieldsModifEnergy}, and found similar to that of terms like $\mathbf{\Sigma}\mathbf{\overline{\Sigma}}\mathbf{\Phi}$.

In this work, we chose to consider separately the GUT and inflation parts, being as realistic as possible for the former and setting up the simplest possible model for the latter. This is the approach found also e.g. in \cite{Cacciapaglia:2013tga}. This simple inflation term is sufficient to reproduce the standard inflation phenomenology. Besides, different kinds of terms such as $S^2$, $S^3$ or $S \mathbf{\Phi}^2$ generate mass or quartic terms for the inflaton in the scalar potential and can thus spoil the inflation. They must therefore be considered with care \cite{Cacciapaglia:2013tga}.

In the example of a SO(10) GUT, the fields $ \mathbf{\Sigma}$ and $\overline{\mathbf{\Sigma}}$
are often taken to transform as the $\mathbf{126}$ and
$\mathbf{\overline{126}}$ representations, which are the lowest dimensional complex conjugate
representations which are safe, \emph{i.e.} permitting $R$-parity conservation at low energy to ensure proton stability
\cite{Martin:1992mq}. The other fields used to implement the SSB can
be for example a $\mathbf{210}$ and a $\mathbf{10}$
\cite{Aulakh:2003kg}, or two $\mathbf{54}$, two $\mathbf{45}$, and two
$\mathbf{10}$ representations \cite{Jeannerot:1995yn}. The whole expression of a SO(10) model, with the normalizations and conventions of the present paper, is given in Ref.~\cite{Allys:2015kge}.


\subsection{Lagrangian of the bosonic sector}

Let us write down the general form for the Lagrangian of
the bosonic sector, while setting the notation. In
the following, we take the signature of the metric to be $+2$, and
label with latin indices $a$, $b$, $\dots$ the generators of G, with gauge coupling constant
$g$. The kinetic part of the Lagrangian is thus (with implicit summations on $i$, $\bar{i}$ and $I$)
\begin{align}
\label{CS1Kinetic}
 K =&- (D_\mu \mathbf{\Phi}_{\bar{\imath}})^\dagger (D^\mu \mathbf{\Phi}_{\bar{\imath}})
 - ( D_ \mu\mathbf{\Phi}_i)^\dagger( D^\mu\mathbf{\Phi}_i) 
 - (D_\mu\mathbf{\Phi}_I)^\dagger(D^\mu\mathbf{\Phi}_I) \nonumber \\
 & ~~ -(D_\mu \overline{ \mathbf{\Sigma}})^\dagger (D^\mu \overline{ \mathbf{\Sigma}}) 
 - ( D_\mu \mathbf{\Sigma})^\dagger  ( D^\mu \mathbf{\Sigma}) 
  - (\nabla_\mu S)^* (\nabla^\mu S) -\frac{1}{4}F_{\mu \nu}^a F^{a \mu \nu}, ~~~~  ~~~~  ~~~~ 
\end{align}
with
\begin{equation}
\label{CS1CovDev}
D_\mu X=(\nabla_\mu - i g A_\mu^a \tau^a_X)X,
\end{equation}
$\tau^a_X$ being the relevant operators encoding the action of the
generator labeled by $a$ on the field $X$. From now on, we will
denote by $X$ a generic scalar field when unspecified, 
\emph{i.e.} $X\in \{ \mathbf{\Sigma}, \overline{ \mathbf{\Sigma}}, \mathbf{\Phi}_I, \mathbf{\Phi}_i, \mathbf{\Phi}_{\bar{\imath}},S\}$.
The strength tensor is defined in the usual way,
\begin{equation}
F_{\mu\nu}^a=\nabla_\mu A_{\nu}^a-\nabla_\nu A_{\mu}^a+g
f^a{}_{bc}A_\mu^b A_\nu^c,
\end{equation}
with $f^a{}_{bc}$ the structure constants of $G$.

The potential term is constructed from the $F$-term, obtained by
taking the derivative of the superpotential with respect to the chiral
supermultiplets
\begin{equation}
F_X=\frac{\partial W}{\partial X}.
\end{equation}
This yields the terms
\begin{equation}
\label{CS1F-terms}
\begin{array}{l}
\mathbf{F}_{\Sigma}=\overline{ \mathbf{\Sigma}} (m_\Sigma +\eta_x \mathbf{\Phi}_x + \kappa S)
+\beta_{xy} \mathbf{\Phi}_x \mathbf{\Phi}_y,\\ 
\mathbf{F}_{\bar{\Sigma}}= \mathbf{\Sigma}(m_\Sigma +\eta_x
\mathbf{\Phi}_x + \kappa S) +\bar{\beta}_{xy} \mathbf{\Phi}_x \mathbf{\Phi}_y,\\
 \mathbf{F}_{\Phi_I}=m_{IJ}\mathbf{\Phi}_J + \eta_I
\overline{ \mathbf{\Sigma}} \mathbf{\Sigma} + 2 \beta_{Iy} \mathbf{\Sigma} \mathbf{\Phi}_y +
2 \bar{\beta}_{Iy}\overline{ \mathbf{\Sigma}}\mathbf{\Phi}_y + \lambda_{Iyz}\mathbf{\Phi}_y
\mathbf{\Phi}_z,\\
 \mathbf{F}_{\Phi_i}=m_{i\bar{\jmath}}\mathbf{\Phi}_{\bar{\jmath}} + \eta_i
\overline{ \mathbf{\Sigma}} \mathbf{\Sigma} + 2 \beta_{iy} \mathbf{\Sigma} \mathbf{\Phi}_y +
2 \bar{\beta}_{iy}\overline{ \mathbf{\Sigma}}\mathbf{\Phi}_y + \lambda_{iyz}\mathbf{\Phi}_y
\mathbf{\Phi}_z,\\ 
\mathbf{F}_{\Phi_{\bar{\imath}}}=m_{j\bar{\imath}}\mathbf{\Phi}_j + \eta_{\bar{\imath}}
\overline{ \mathbf{\Sigma}} \mathbf{\Sigma} + 2 \beta_{\bar{\imath}y} \mathbf{\Sigma} \mathbf{\Phi}_y + 2 \bar{\beta}_{\bar{\imath}y}\overline{ \mathbf{\Sigma}}\mathbf{\Phi}_y +
\lambda_{\bar{\imath}yz}\mathbf{\Phi}_y \mathbf{\Phi}_z,
\\ F_{S}=\kappa
(\overline{ \mathbf{\Sigma}} \mathbf{\Sigma}-M^2),
\end{array}
\end{equation}
$F_X$ being in the conjugate representation of $X$. These $F$-terms are discussed in more details in Sec.~\ref{CS1LagEOM}. The scalar
potential is finally obtained from these terms through
\begin{equation}
\label{CS1potential}
V=\sum_X \mathbf{F}_X^\dagger \mathbf{F}_X\equiv\sum_X V_X,
\end{equation}
where $V_X>0$ and $V>0$, and where we use additional bold symbols for the $F$-terms to remind that they are not singlet of the gauge group in general. Note that we did not include in $V$ the $D$-term contribution, since these terms identically vanish in a $F$-term scenario, and thus play no role in the dynamical study of the fields. However, they are indeed taken into account by imposing some constraints which are discussed in Sec.~\ref{CS1PartAnsatzBCSingle} and~\ref{CS1IntroManyFields}.

Finally, the full Lagrangian density is derived from these two
terms
\begin{equation}
\label{CS1L=K-V}
\mathcal{L}=K-V,
\end{equation}
\emph{i.e.} by adding Eqs. (\ref{CS1Kinetic}) and (\ref{CS1potential}).


\subsection{Hybrid inflation and SSB scheme}
\label{CS1HI&SSB}

It has been shown in \cite{Jeannerot:2003qv} that the formation of cosmic strings at the end of an hybrid inflation phase is essentially unavoidable during the final stage of a SUSY GUT symmetry breaking. 
It is this category of strings that we discuss in the present paper. In such models, all the monopoles which may have been produced in a previous phase are washed out during inflation.

The matter content we introduced previously implements
the SSB scheme down to the SM in at least two steps :
\begin{equation}
G
\overset{\langle\Phi_x\rangle}{\relbar\joinrel\relbar\joinrel\cdots\joinrel\longrightarrow}
G' \overset{\langle\Sigma\rangle\langle\Phi_{x'}\rangle}
{\relbar\joinrel\relbar\joinrel\relbar\joinrel\longrightarrow}
G_{\text{SM}} \times \mathbb{Z}_2.
\end{equation}
In many cases, the symmetry
that is broken at the last step contains U(1)$_{B-L}$, but since there are no
general constraints about it, it is better to leave it arbitrary.

All the non vanishing VEVs after the end of inflation have to be
singlet under the SM gauge group, otherwise the vacuum would
have non vanishing quantum numbers under this symmetry.
We also assume there is no symmetry restoration, \emph{i.e.} all fields acquiring a non-zero VEV at a given stage keep it non vanishing at later stages. So, we can restrict the study of the SSB scheme to SM singlets only.

At the onset of inflation, we can assume an initially very large
value for the inflaton $S$ in comparison with all the other
fields, as is expected for chaotic inflation. To minimize $V_\Sigma
\sim |\kappa \overline{ \mathbf{\Sigma}} S|^2$ and $V_{\bar{\Sigma}} \sim |\kappa
 \mathbf{\Sigma} S|^2$, the fields $ \mathbf{\Sigma}$ and $\overline{ \mathbf{\Sigma}}$ must take a
vanishing VEV. 
The terms $\beta_{xy} \mathbf{\Phi}_x \mathbf{\Phi}_y$ must be in the same
representation as $\overline{ \mathbf{\Sigma}}$ for the terms $\beta_{xy} \mathbf{\Phi}_x \mathbf{\Phi}_y \mathbf{\Sigma}$ to be scalars. So, if they take a VEV before the end of inflation, all the symmetries
broken by $\overline{ \mathbf{\Sigma}}$ after inflation would already be broken at this current step, which is contrary to our assumptions. Thus, they cannot take a non-zero VEV before the end of inflation. 
This property and
the same reasoning for $\bar{\beta}_{xy} \mathbf{\Phi}_x \mathbf{\Phi}_y$ show that
$V_\Sigma=0$ and $V_{\bar{\Sigma}}=0$ before the end of inflation.
This also ensures that all the terms implying the inflaton $S$ 
identically vanish in the potential at tree level.

Finally, we
can assume that at the onset of inflation, the VEVs verify that all
the potential terms except $V_S$ are zero; it is indeed the
global minimum for the potential taking into account the constraint 
on $V_S$. Different field configurations give this minimal value, 
including that with all the fields having a vanishing value\footnote{We assume that other configurations can exist, since the potential conditions 
give fourth order polynomial equations in the fields.}. 
So, it gives several sets of solutions
\begin{equation}
\label{CS1VEV0}
\{\langle \mathbf{\Sigma}_{\scriptscriptstyle{(-)}}\rangle=0,\langle\overline{ \mathbf{\Sigma}}_{\scriptscriptstyle{(-)}}\rangle=0,\langle \mathbf{\Phi}_{x,{\scriptscriptstyle{(-)}}}\rangle\}_{\delta},
\end{equation}
the index $(-)$ meaning that we consider the set of VEVs before the
end of inflation, while $\delta$ labels the different sets
themselves.

At the end of inflation, the fields reach a global
minimum for the potential, meaning all $F$-terms contributions must independently
vanish. Setting $V_S=0$ then implies
\begin{equation}
\label{CS1SigmaInf}
\langle \mathbf{\Sigma}\rangle\langle\overline{ \mathbf{\Sigma}}\rangle=M^2,
\end{equation}
and the fields $ \mathbf{\Sigma}$ and $\overline{ \mathbf{\Sigma}}$ take non zero VEVs as expected.
The vanishing conditions of all the other potential terms finally
give several sets of solutions
\begin{equation}
\label{CS1VEV1}
\{\langle \mathbf{\Sigma}_{\scriptscriptstyle{(+)}}\rangle,\langle\overline{ \mathbf{\Sigma}}_{\scriptscriptstyle{(+)}}\rangle,S_{\scriptscriptstyle{(+)}},\langle\mathbf{\Phi}_{x,{\scriptscriptstyle{(+)}}}\rangle\}_{\delta^\prime},
\end{equation} 
the index $(+)$ denoting that we consider the set of VEVs after the end
of inflation, while $\delta^\prime$ labels the different sets of solution.

During inflation, the potential for the inflaton yields $V=V_0 + \text{quant.corr.}$, where $V_0=\kappa^2 M^4$ is the value of the potential at tree-level. Due to the quantum corrections, the value of the inflaton field will then slowly roll, until it meets its critical value, thus ending inflation (see Ref.\cite{Cacciapaglia:2013tga} for an explicit example). 

As we study strings in models which are relevant in a particle physics and cosmological point of view, we assume that the superpotential and field content used imply at least one appropriate SSB scheme. For these schemes, the set of VEVs at the end of inflation defines the SM gauge group, no harmful topological defects are produced, and the stability of the inflationary valley is ensured (see the associated discussion\footnote{It was also
argued in Ref.~\cite{Cacciapaglia:2013tga} that one should
verify that the set of VEVs before the end of inflation is close in
field space to only one set of VEVs after the end of inflation, since otherwise the possibility that the fields take different sets of VEVs at the end of
inflation could create domain
walls.} in Ref.~\cite{Cacciapaglia:2013tga}). We consider from now on such a SSB scheme and the associated non vanishing VEVs, those being described as in Eqs.~(\ref{CS1VEV0})
and (\ref{CS1VEV1}) without considering the labels $\delta$ and $\delta '$
anymore, \emph{i.e.} explicitly assuming a specific set in each ensemble.

Such a SSB scheme in parallel with the inflationary process in a given SO(10) GUT is described in Ref.~\cite{Allys:2015kge}.


\subsection{Description with the restricted representations}
\label{CS1RestrictedRep}

As we work with large dimensional representations, it is useful to consider
their branching rules to simplify the description of these fields. Indeed, we only need to work with fields behaving as singlets under the SM
gauge group to describe the whole SSB scheme. For instance, the
\textbf{210} representation of SO(10) contains three such restricted
representations, included in its 
(\textbf{1},\textbf{1},\textbf{1}), (\textbf{1},\textbf{1},\textbf{15}) 
and (\textbf{1},\textbf{3},\textbf{15}) representations of its decomposition 
under the Pati-Salam group [SU(2)$_L\times$SU(2)$_R\times$SU(4)] \cite{Slansky:1981yr}.
The \textbf{126} and $\mathbf{\overline{126}}$ representations only have one such sub-representation, contained respectively
in their (\textbf{1},\textbf{3},\textbf{10}) and (\textbf{1},\textbf{3},$\mathbf{\overline{10}}$)
restricted representations under Pati-Salam.

Let us consider such a VEV singlet. For
a given field $\mathbf{\Phi}_x$, we ascribe an index $\alpha$ to describe the
different sub-representations transforming trivially under the SM, and write the associated VEVs as
\begin{equation}
\label{CS1defsubmultiplet}
\langle \mathbf{\Phi}_{x,\alpha} \rangle = \phi_{x,\alpha} \left(x^\mu\right) {\langle \mathbf{\Phi}_{x,\alpha}\rangle}_0,
\end{equation}
without implicit summation, and where $\phi_{x,\alpha}$ is a complex function of space-time and ${\langle
 \mathbf{\Phi}_{x,\alpha}\rangle}_0$ a constant normalized vector in representation
space (${\langle
 \mathbf{\Phi}_{x,\alpha}\rangle}_0^\dagger{\langle
 \mathbf{\Phi}_{x,\alpha}\rangle}_0=1$). Using these notations, the complete part of the field which is singlet under the SM
can be written as the combination
\begin{equation}
\label{CS1SubMultiplet}
\langle \mathbf{\Phi}_x \rangle = \sum_\alpha \phi_{x,\alpha}(x^{\mu}) {\langle \mathbf{\Phi}_{x,\alpha}\rangle}_0.
\end{equation}
This procedure reduces the fields description to only a few complex
functions.

\section{Abelian cosmic strings}
\label{CS1PartAbelianStrings}
\subsection{Strings studied}

Let us focus on the strings created at the last step of
SSB of the GUT, lowering the rank of the gauge group by one unit and ending
the inflation phase. Several kinds of strings can
appear, depending on the quotient group $H\sim G^\prime / G_{\text{SM}}$. As the SSB lowers 
the rank of the group, $H$ must contain at least one U(1) subgroup, and we will focus on
the Nambu-Goto abelian strings which form at this step, associated with this abelian
generator. When $H$ is larger than U(1), non abelian strings
could also form \cite{Aryal:1987sn,Ma:1992ky,Davis:1996sp}. However, and since we want to constraint all GUTs, we restrict attention to the minimal U(1) case, any other kind of strings tightening the constraints. We will denote
$\text{U}(1)_{\text{str}}$ this particular subgroup and
$\tau^{\text{str}}$ the associated generator.

These strings cannot be connected to monopoles. As shown in Ref. \cite{Vilenkin:1982hm}, the strings are stable with respect to breaking into monopoles when the scale of formation of these monopoles is higher than the scale of formation of the strings. It is the case here, as we assumed that only strings form at the SSB considered. So, these strings could connect only to pre-existing monopoles, and those have, by construction, already been washed away during the inflation phase.

In what follows, we use a set of cylindrical coordinates $(r,\theta,z,t)$ based on the location 
of the string, and taken to be locally aligned along the $z$-axis at $r=0$. 
We also focus on strings with fields functions
of $r$ and $\theta$ only, and
consider uniquely the bosonic structure of the string. It means that we will not consider in this paper any currents in the core of the string. This possibility will however be discussed in the following section.

\subsection{Minimal structure}
\label{CS1PartMinimalStructure}

We have to determine which sub-representations are sufficient
to describe the structure of the string. On the one hand, the fields take at infinity the non vanishing VEVs defining the SM symmetry, so all the restricted representations non singlet under this symmetry take vanishing values far from the string. On the other hand, the sub representations non singlet under the SM appear at least in a quadratic form in the potential. Indeed, a potential term containing only one such field would be charged under the SM. Both these results imply that an ansatz where all the sub-representations charged under the SM take an identically vanishing value is solution of the equations of motion with the boundary conditions at infinity.

To understand in another way this ansatz, one can consider the static configuration which minimizes the potential at the center of the string. Such a configuration is given by the set of non vanishing VEVs singlet under the SM defining $G^\prime$ in the SSB scheme, see Sec.~\ref{CS1HI&SSB}. These VEVs are not charged under U(1)$_{\text{str}}$, which would otherwise already be broken at this step. Using arguments discussed e.g. in Ref.~\cite{Witten:1984eb}, we expect fields in the string to take intermediate values between the configurations which minimize the potential at the center of the string and at infinity, depending on the competition between kinetic and potential terms. Then, as both these configurations only contain singlets of the SM, one expects an ansatz where all the other fields take an identically vanishing value. 

This particular ansatz, which we consider in the following, is what we define as the minimal structure. As the configurations which minimize the potential at the center of the string and at infinity are \emph{a priori} different since they are solutions of two different quartic polynomial equations, all the fields taking non vanishing values at the last step of SSB condense in the string, and thus have to be taken into account. Note that a complete study of the stability of such an ansatz should be considered in each given model.

The minimal structure for cosmic strings, where several bosonic fields condense in the string, should modify most of the properties of these objects. Some of these properties are described below. For instance, it becomes mandatory to describe the microscopic structure of the strings by taking into account the condensation of these additional Higgs fields. This then permits to evaluate the modification of the energy per unit length due to this complex structure. As most of the observational constraints on this macroscopic parameter are rather stringent, every sizable modification of it in a given model could rule out this model.

Several other properties of cosmic strings will be modified by the fact that the actual structure is more involved than that of the toy models usually considered. First, the condensation of several Higgs fields in the core of the string gives natural candidates to carry bosonic currents \cite{Witten:1984eb,Peter:1992dw,Peter:1992nz,Peter:1993tm,Morris:1995wd}. Also, and as we work in a SUSY framework, we can expect the superpartners of these Higgs fields to build fermionic currents via their zero modes \cite{Witten:1984eb,Jackiw:1981ee,Weinberg:1981eu,Davis:1995kk,Ringeval:2000kz,Peter:2000sw}. This more complex structure can also qualitatively modify the intercommutation process \cite{Shellard:1988ki,Laguna:1990it,Matzner:1988ky,Moriarty:1988qs,Moriarty:1988em,Shellard:1987bv}, which has a tremendous impact on the temporal evolution of the cosmological string network, and thus on the consequences on the CMB \cite{Kibble:1976sj,Kibble:1980mv,Copeland:1991kz,Austin:1993rg,Martins:1996jp,Martins:2000cs}. Furthermore, the link made between the fields forming the string and the particle physics model used allows a more detailed examination of the cusps evaporation phenomenon \cite{Srednicki:1986xg,Brandenberger:1986vj,Gill:1994ic,Olum:1998ag,Bhattacharjee:1989vu}. A modification of each of these properties can have major consequences on the cosmological implications of cosmic strings.


\subsection{Equation of state, toy model limit}
\label{CS1ToyModel}

Since we explicitly assume currentless strings, nothing in the configuration we are 
interested in can depend on the internal string worldsheet coordinates,
here locally $z$ and $t$. We have 
\begin{equation}
T^\mu_\nu=-2g^{\mu \alpha}\frac{\delta \mathcal{L}}{\delta
 g^{\alpha\nu}} + \delta^\mu_\nu \mathcal{L},
\end{equation}
yielding $T^{tt}=-T^{zz}=\mathcal{L}$. Then
\begin{equation}
\label{CS1Nambu-Goto}
{\displaystyle U = 2\pi \int r\text{d}r~ T^{tt}=-2\pi \int r \text{d}r~ T^{zz}=T},
\end{equation}
\emph{i.e.} the Nambu-Goto equation of state, Lorentz-invariant along the worldsheet. 
Thus, the only parameter of interest is the energy per unit length $U$ defined in Eq.~(\ref{CS1Nambu-Goto}). This parameter, frequently denoted by $\mu$ in the cosmology literature, is directly constrained from, e.g., CMB observations: from Planck and WMAP data, there is a constraint of $G\mu/c^2 < 3.2\times 10^{-7}$ at $95\%$ confidence level for the abelian Higgs model \cite{Ade:2013xla}. Thus, every modification of $U$ due to the realistic structure of the strings will have to be compared to the already stringent observational constraints, and could rule out the associated model.

To translate the standard abelian Higgs model to a $F$-term SUSY formalism, three fields are necessary, $\Sigma$, $\overline{\Sigma}$ and $S$, the first two fields having opposite U(1) charges, the last one being uncharged. The superpotential is 
\begin{equation}
W=\kappa S \left( \overline{\Sigma} \Sigma -M^2\right).
\end{equation}
Assuming that $S$ identically vanishes, it yields as expected the standard U(1) string model for $\mathbf{\Sigma}$ and $\overline{\mathbf{\Sigma}}$ (see for example \cite{Davis:1997bs,Aryal:1987sn,Peter:1992dw}), with\footnote{Note that if one is more familiar with the model containing a single kinetic term of the form $(D_\mu \overline{\Sigma})(D^\mu \Sigma)$, usual for non supersymmetric theories, a link between both models can be easily performed. For this, it is sufficient to introduce new variables of the form $\tilde{\Sigma}=\sqrt{2}\Sigma$, $\tilde{M}=\sqrt{2}M$, and $\tilde{\kappa}=\kappa/2$. Indeed, one sees that the superpotential in term of the new variables remains the same, while the factor 2 of the kinetic term will disappear. This is only valid in a $F$-term scenario, where $\Sigma$ and $\overline{\Sigma}$ are complex conjugate due to the $D$-term condition (see \ref{CS1PartAnsatzBCSingle}).}.
\begin{equation}
\mathcal{L}=-(D_\mu \overline{\Sigma})^\dagger(D^\mu \overline{\Sigma}) -(D_\mu \Sigma)^\dagger (D^\mu \Sigma)-\frac{1}{4} F_{\mu\nu}F^{\mu\nu}-\kappa^2 \left| \Sigma \overline{\Sigma}-M^2 \right| ^2.
\end{equation}

We recover explicitly this limit from the realistic model when all the parameters but $\kappa$ and $M$ go to zero, and considering only $\text{U}(1)_{\text{str}}$ defined in the previous part. Note that it is the case due to the previous normalizations and conventions, see Sec.~\ref{CS1SingleFieldStrings} for the details. This well defined limit must be verified when using results from the abelian Higgs model, since important GUT numerical factors can appear. For this well known problem, we have $\Sigma \sim \overline{\Sigma} \sim M$, a characteristic radius for $\Sigma$ of $(\kappa M)^{-1}$, and a characteristic energy per unit length $U_0 \sim M^2$ \cite{Aryal:1987sn,Peter:1992dw,Hindmarsh:1994re,Allys:2015kge}.

\subsection{Two classes of strings}
\label{CS1PartTwoClasses}

We assume for the entire paper that $ \mathbf{\Sigma}$ and $\overline{\mathbf{\Sigma}}$ only have one sub-representation singlet of the SM, charged under U(1)$_{\text{str}}$. If it was not the case, we could treat the additional representations, charged or not under U(1)$_{\text{str}}$, in the same manner than the additional fields $\mathbf{\Phi}$. 

We distinguish two classes of strings. For the first kind, no other fields are charged under U(1)$_{\text{str}}$. We call these 
strings single-field strings, since only one field (and its conjugate) is directly coupled 
to the gauge field. They are discussed in Sec. \ref{CS1SingleFieldStrings}. 
For the second case, other fields can be charged under 
$\text{U}(1)_{\text{str}}$, and we naturally call them many-field strings, see Sec. \ref{CS1ManyFieldsStrings}. This property only depends on the field content of the GUT and how it permits to implement the SSB scheme.

In the case of a single-field string, no $\beta$-couplings, e.g. in $\mathbf{\Sigma \Phi \Phi}$, can appear in the superpotential between the restricted representations singlet under the SM, since such terms would be charged under $\text{U}(1)_{\text{str}}$. It is not anymore the case with a many-field string.


\section{Single-field strings}
\label{CS1SingleFieldStrings}

\subsection{Ansatz and boundary conditions}
\label{CS1PartAnsatzBCSingle}

According to our definition of single-field strings, we consider the case where
only the restricted representations of $\mathbf{\Sigma}$ and $\overline{ \mathbf{\Sigma}}$ are charged under $\text{U}(1)_{\text{str}}$. We normalize their charges to $q_\Sigma = 1$ and
$q_{\bar{\Sigma}}=-1$, the charges being defined by identifying $\tau_X=q_X
\text{Id}_X$ for an abelian generator in Eq.~(\ref{CS1CovDev}).
To describe the different sub-representations singlet under the SM, we use the decomposition of Eq. (\ref{CS1SubMultiplet}), which gives
\begin{equation}
\label{CS1VEVSigma}
\langle \mathbf{\Sigma}(r,\theta) \rangle = \sigma(r,\theta) {\langle \mathbf{\Sigma} \rangle}_0,
\end{equation}
and
\begin{equation}
\label{CS1VEVSigmab}
\langle \overline{ \mathbf{\Sigma}}(r,\theta) \rangle = \bar{\sigma} (r,\theta){\langle
 \overline{ \mathbf{\Sigma}} \rangle}_0,
\end{equation}
with ${\langle \mathbf{\Sigma} \rangle}_0^\dagger{\langle
 \mathbf{\Sigma} \rangle}_0=1$ and ${\langle \overline{ \mathbf{\Sigma}} \rangle}_0={\langle
 \mathbf{\Sigma} \rangle}_0^\dagger$.
 
The $D$-term condition associated with $\tau^{\text{str}}$, \emph{i.e.} \cite{Martin:1997ns}
\begin{equation}
\label{CS1D-term}
D^{\text{str}}=-g\sum_X (X^\dagger \tau^{\text{str}}_X X)=0,
\end{equation}
ensures that $\sigma$ and $\bar{\sigma}$ have the same norm. In addition, as the phase of the inflaton $S$ have been rephased in order to make $M$ real, it ensures that the global minimum of the potential is reached when $\boldsymbol\Sigma \overline{\boldsymbol\Sigma}=M^2 \in \mathbb{R}$. Both these results impose that $\overline{\boldsymbol\Sigma}=\boldsymbol\Sigma^\dagger$, which finally gives $\overline{\sigma}=\sigma^*$.

 The string itself is defined through
\cite{Kibble:1976sj,Hindmarsh:1994re}
\begin{equation}
\label{CS1null_center}
\langle \mathbf{\Sigma} \rangle_ {(r=0)} = \langle \overline{ \mathbf{\Sigma}} \rangle_
 {(r=0)} = 0.
\end{equation}
Thus, we can introduce an ansatz similar to the Nielsen-Olesen solution, with integer winding number $n$ : \cite{Aryal:1987sn,Peter:1992dw,Ma:1992ky,Hindmarsh:1994re}
\begin{equation}
\label{CS1Antsatz}
\begin{array}{l}
\sigma=f(r)\text{e}^{in\theta},\\
\phi_{x,\alpha}=\phi_{x,\alpha}(r),\\
S=S(r),\\
A_\mu=A_\theta^{\text{str}}(r) \tau^{\text{str}} \delta_\mu^\theta,
\end{array}
\end{equation}
where $f(r)$ and $A_\theta^{\text{str}}(r)$ are real, $\phi_{x,\alpha}(r)$ and $S(r)$ being complex. At infinity, we have 
\begin{equation}
\label{CS1BCdebut}
\begin{array}{l}
\displaystyle{\lim_{r \to \infty} f(r)=M},\\
\displaystyle{\lim_{r \to \infty} A_\theta^{\text{str}}(r)=\frac{n}{g}},
\end{array}
\end{equation}
\emph{i.e.} the value of $f$ ensures Eq.~(\ref{CS1SigmaInf}) holds, while
the gauge field cancels $D_\mu \mathbf{\Sigma}$. 
After generalizing the notation of Eq. (\ref{CS1VEV1}), the boundary conditions for the other fields satisfy
\begin{equation}
\begin{array}{l}
\displaystyle{\lim_{r \to \infty} \phi_{x,\alpha}(r)=\phi_{x,\alpha,{\scriptscriptstyle{(+)}}},}\\
\displaystyle{\lim_{r \to \infty} S(r)=S_{\scriptscriptstyle{(+)}}}.
\end{array}
\end{equation}
At the center of the string,
\begin{equation}
f(0)=0, ~~~~~~~ \text{and} ~~~~~~~ A_\theta^{\text{str}}(0)=0,
\end{equation}
while cylindrical symmetry imposes
\begin{equation}
\label{CS1BCfin}
\displaystyle{\frac{\text{d}\phi_{x,\alpha}}{\text{d} r}(0)=0},
 ~~~~~~~ \text{and} ~~~~~~~
\displaystyle{\frac{\text{d} S}{\text{d} r}(0)=0}.
\end{equation}


\subsection{Lagrangian and equations of motion}
\label{CS1LagEOM}

With the ansatz of the previous section, we can simplify the model tremendously. The kinetic term yields (with implicit summations on the representations singlet of the SM)
\begin{align}
\label{CS1KSingleField}
K =& - 2 \left|(\nabla_\mu-ig A_\mu^{\text{str}})\sigma\right|^{2}
-(\nabla_\mu \phi_{\bar{\imath},\bar{\alpha}})^*(\nabla^\mu \phi_{\bar{\imath},\bar{\alpha}})
-(\nabla_\mu \phi_{i,\alpha})^*(\nabla^\mu \phi_{i,\alpha}) \nonumber\\
 &~~ -(\nabla_\mu \phi_{I,\alpha})^*(\nabla^\mu \phi_{I,\alpha})
 -\left| (\nabla_\mu S)\right|^2 - \frac{1}{4}F_{\mu \nu}^{\text{str}} F^{\mu \nu \, \text{str} },
\end{align}
where $F_{\mu \nu}^{\text{str} } = \nabla_\mu A_\nu^{\text{str}} - \nabla_\nu A_\mu^{\text{str}}$.
We label $(\bar{\imath},\bar{\alpha})$ the restricted representation
complex conjugate to $(i,\alpha)$.
No scalar products
between vectors in representation spaces are present due to the
normalization choice of Sec.~\ref{CS1HI&SSB}. There is no cross-terms since singlet quadratic terms can only be built from products of two conjugate representations.

Writing down the potential is a bit trickier, as it contains high order terms
whose derivation w.r.t. the fields need to be done with care.
We add a subscript to the VEV indicating in which
representation is the product we consider,
${\langle X Y\rangle}_{Z}$ denoting the field in the
representation of $Z$ coming from the contraction between $X$ and $Y$. 

In the SO(10) case, 
the \textbf{126} and
$\mathbf{\overline{126}}$ representations are fifth-rank anti-symmetric respectively
self-dual\footnote{Self-duality being here defined by $\Sigma_{ijklm}=\frac{i}{5!}\epsilon_{ijklmabcde}\Sigma_{abcde}$.} and anti-self-dual
tensors $\Sigma_{ijklm}$ and $\bar{\Sigma}_{ijklm}$, while the
\textbf{210} is a fourth-rank anti-symmetric tensor
$\Phi_{ijkl}$. The singlet which can be formed with these fields is
$\bar{\Sigma}_{ijklm}\Sigma_{ijkno}\Phi_{lmno}$. Differentiating
with respect to $\bar{ \mathbf{\Sigma}}$, we obtain $\langle\mathbf{\Sigma}\mathbf{\Phi}\rangle_{\Sigma}$ in the same
representation as $ \mathbf{\Sigma}$, which is $\frac{1}{2}\left(\Phi_{[ij|\alpha\beta}\Sigma_{\alpha\beta |klm]}+\frac{i}{5!}\epsilon_{ijklmabcde}\Phi_{ab\alpha\beta}\Sigma_{\alpha\beta cde}\right)$
\cite{Slansky:1981yr,Allys:2015kge}, totally antisymmetric and self-dual as expected.

With these notations and with implicit summation on all the indices but e.g. $I$ when considering $V_{\Phi_I}$, the scalar potential becomes (noting that there is no $\beta$-coupling, as explained in Sec.~\ref{CS1PartTwoClasses})
\begin{equation}
\label{CS1V_S}
V_S=\kappa^2 (\sigma \sigma^*-M^2)^2,
\end{equation}
for the inflaton part,
\begin{align}
V_\Sigma=& m_\Sigma^{2} \sigma\sigma^* + \kappa^2 S S^* \sigma
\sigma^* +|\eta_x|^2|{\langle
 \overline{ \mathbf{\Sigma}}\mathbf{\Phi}_{x,\alpha}\rangle}_{\bar{\Sigma},0}|^2\sigma\sigma^*\phi_{x,\alpha}\phi_{x,\alpha}^* +m_\Sigma \kappa \sigma\sigma^* S^* +
\text{h.c.} \nonumber\\ 
& ~~ + m_\Sigma \eta_x^*{\langle \overline{ \mathbf{\Sigma}}\rangle}_0{\langle
 \overline{ \mathbf{\Sigma}}\mathbf{\Phi}_{x,\alpha}\rangle}_{\bar{\Sigma},0}^\dagger
\sigma\sigma^*\phi_{x,\alpha} +\text{h.c.} + \eta_x \kappa {\langle
 \overline{ \mathbf{\Sigma}}\mathbf{\Phi}_{x,\alpha}\rangle}_{\bar{\Sigma},0}{\langle
 \overline{ \mathbf{\Sigma}}\rangle}_0^\dagger \sigma \sigma^*S^*\phi_{x,\alpha} +
\text{h.c.},
\end{align}
and
\begin{equation}
V_{\bar{\Sigma}}=V_\Sigma ( \mathbf{\Sigma} \longleftrightarrow \overline{\mathbf{\Sigma}} ),
\end{equation}
for the string forming field part, and 
\begin{align}
\label{CS1V_Phi}
V_{\Phi_I}=&
|m_{IJ}|^2\phi_{J,\alpha}\phi_{J,\alpha}^*+|\eta_I|^2|\langle \mathbf{\Sigma}
\overline{ \mathbf{\Sigma}}\rangle _{\Phi_I,
 0}|^{2}(\sigma\sigma^*)^2 +|\lambda_{Ixy}|^2|{\langle\mathbf{\Phi}_{x,\alpha}\mathbf{\Phi}_{y,\beta}\rangle}_{\Phi_I,0}|^2\phi_{x,\alpha}\phi_{x,\alpha}^*\phi_{y,\beta}\phi_{y,\beta}^*\nonumber\\
& +m_{IJ}\eta_I^*{\langle\mathbf{\Phi}_{J,\alpha}\rangle}_0{\langle
 \mathbf{\Sigma} \overline{ \mathbf{\Sigma}}\rangle}_{\Phi_{J,\alpha},0}^\dagger \sigma
\sigma^* \phi_{J,\alpha}+\text{h.c.} +
m_{IJ}\lambda_{Ixy}^*{\langle\mathbf{\Phi}_{J,\gamma}\rangle}_0{\langle\mathbf{\Phi}_{x,\alpha}\mathbf{\Phi}_{y,\beta}\rangle}_{\Phi_{J,\gamma},0}^\dagger
\phi_{J,\gamma}\phi_{x,\alpha}^*\phi_{y,\beta}^* + \text{h.c.} \nonumber\\ 
& ~~ + \eta_I \lambda_{Ixy}^*{\langle \mathbf{\Sigma}
 \overline{ \mathbf{\Sigma}}\rangle}_{\Phi_I,0}{\langle\mathbf{\Phi}_{x,\alpha}\mathbf{\Phi}_{y\beta}\rangle}_{\Phi_I,0}^
\dagger \sigma \sigma^* \phi_{x,\alpha}^*\phi_{y,\beta}^* +
\text{h.c.},
\end{align}
\begin{equation}
V_{\Phi_i}=V_{\Phi_I} ( I \longleftrightarrow i, J \longleftrightarrow \bar{\jmath}),
\end{equation}
\begin{equation}
V_{\Phi_{\bar{\imath}}}=V_{\Phi_I} ( I \longleftrightarrow \bar{\imath}, J \longleftrightarrow j),
\end{equation}
for the other fields. An example of such potential with the conventions used in this paper can be found in Ref.~\cite{Allys:2015kge}.

The Lagrangian is
\begin{equation}
\label{CS1Lagrangian}
\mathcal{L}= K - V_S - V_\Sigma - V_{\bar{\Sigma}} -
\sum_{x=I,i,\bar{\imath}}V_{\Phi_x},
\end{equation}
with the kinetic term given in Eq. (\ref{CS1KSingleField}). Note that we recover the abelian Higgs model with no additional numerical factor in the limit described in Sec.~\ref{CS1ToyModel}, due to the conventions used.

Finally, using the ansatz given in Eq. (\ref{CS1Antsatz}), we obtain
the following equations of motion
\begin{equation}
\label{CS1EOM}
\begin{array}{l}
\displaystyle{2\left(f''+\frac{f'}{r}\right)=\frac{fQ^2}{r^2}+\frac{1}{2}\frac{\partial
 V}{\partial f}},\\
  \displaystyle{\phi_{I,\alpha}''+\frac{\phi_{I,\alpha}'}{r}=\frac{\partial
 V}{\partial \phi_{I,\alpha}^*}},\\
  \displaystyle{\phi_{i,\alpha}''+\frac{\phi_{i,\alpha}'}{r}=\frac{\partial
 V}{\partial \phi_{i,\alpha}^*}},\\ 
 \displaystyle{\phi_{\bar{\imath},\bar{\alpha}}''+\frac{\phi_{\bar{\imath},\bar{\alpha}}'}{r}=\frac{\partial V}{\partial \phi_{\bar{\imath},\bar{\alpha}}^*}},\\
  \displaystyle{S''+\frac{S'}{r}=\frac{\partial V}{\partial S^*}},\\ 
  \displaystyle{Q''-\frac{Q'}{r}=2 g^2 f^2Q},
\end{array}
\end{equation}
where a prime means a derivative with respect to the radial
coordinate ($'\equiv \text{d}/\text{d}r$). We also introduced the field
\begin{equation}
\label{CS1defQ}
Q(r)=n-gA_\theta^{\text{str}}(r),
\end{equation}
which is a real field function of $r$ only with for boundary
conditions
\begin{equation}
\label{CS1CLQ}
Q(0)=n, ~~~~~~~ \text{and} ~~~~~~~ \displaystyle{\lim_{r \to \infty} Q(r)=0}.
\end{equation}

Eq. (\ref{CS1EOM}) with associated boundary conditions can only be solved once the actual theory is implemented, thus giving the relevant and necessary coefficients in the potential.

\subsection{Modification of the energy per unit length}
\label{CS1OdGsingle}

We now evaluate the influence of the extra fields contribution to the energy per unit length, when a perturbative study of the condensation of these extra fields in the core of string is possible.

We take into account the possibility to work with high dimensional representations, of characteristic dimension $N$, as is often the case in GUTs. Let us remind that we chose the vectors which define the VEV directions to be normalized, \emph{i.e.}
$\langle \mathbf{\Phi}_{x,\alpha}\rangle_0 \langle
\mathbf{\Phi}_{x,\alpha}\rangle_0^\dagger=1$ and $\langle \mathbf{\Sigma} \rangle_0
\langle \mathbf{\Sigma}\rangle_0^\dagger=1$. The cubic and
quartic contraction between these VEV directions must be smaller than
$1$ due to the Cauchy-Schwarz inequality. They can be approximately
estimated\footnote{For 
this purpose, we describe the VEV directions by vectors of $N$ 
components of value $1/\sqrt{N}$ (in order to be normalized), and the vectors 
formed from two different VEV directions by $N$ component of values $1/N$. 
A scalar product of two vectors giving $N$ times the product of their components, 
we found the results used.}
 to be of order $1/\sqrt{N}$ for the cubic terms, for example
${\langle\mathbf{\Phi}_{x,\alpha}\mathbf{\Phi}_{y,\beta}\rangle}_{\Phi_{z,\gamma},0}
\langle \mathbf{\Phi}_{z,\gamma}\rangle _0^\dagger$, and of order $1/N$ for the
quartic terms, as ${\langle \mathbf{\Sigma}
 \overline{ \mathbf{\Sigma}}\rangle}_{\Phi_I,0}{\langle\mathbf{\Phi}_{x,\alpha}\mathbf{\Phi}_{y\beta}\rangle}_{\Phi_I,0}^
\dagger $.

Important remarks must be done at this step. On the one hand, we consider here a dynamical case, and not only a static and uniform configuration, where it is sufficient to independently ensure that the different $F$-terms vanish (e.g. when studying the SSB scheme). Then, the different multiplicities of the $F$-term components must be taken into account. On the other hand, the choice of conventions in the definition of the superpotential and the kinetic part of the Lagrangian (where we included here no multiplicative factor) and of normalizations for the constant vectors in the representation space (see Sec.~\ref{CS1RestrictedRep}) can considerably affect the formulation of the model. 

A complete model using the conventions of the present paper is described in Ref.~\cite{Allys:2015kge}. In this SO(10) case, taking the VEV directions for the singlets of the SM as given in
Ref. \cite{Bajc:2004xe,Aulakh:2003kg,Fukuyama:2004xs}, and after normalization, we find cubic
coefficients like $1/(10\sqrt{2})$ and $1/(6\sqrt{6})$, and quartic
coefficients like $1/54$ and $1/164$. This is roughly in agreement with
what was expected with a characteristic dimension of order $100$ for
the representation, and sufficient to do a first approximation. Note that Ref.~\cite{Allys:2015kge} also shows that the different formulations are indeed equivalent.

Let us evaluate a rough order of magnitude for the potential, and for that purpose consider a generic field $\phi$, without specifying
indices. From Eq. (\ref{CS1V_Phi}), we have
\begin{equation}
V \simeq m^2\phi^2+ \frac{\eta^2 \sigma^4}{N}+ \frac{\lambda^2 \phi^4}{N} 
+\frac{m \eta \phi \sigma^2}{\sqrt{N}}+ \frac{m \lambda \phi^3}{\sqrt{N}}+\frac{\eta \lambda \sigma^2 \phi^2}{N}.
\end{equation}
To evaluate the contributions of $\phi$ to the energy, we consider its characteristic scale of variation due to the presence of the string. For this purpose, we estimate the value of $\phi$ at the center of the string and at infinity by taking the values which minimize the potential for $\sigma=0$ and $\sigma=M$, that we write respectively $\phi_0$ and $\phi_0 + \phi_1$. We obtain $\phi_0 \sim (\sqrt{N} m)/\lambda$, and $\phi_1 \sim(\eta M^2)/(\sqrt{N} m)$ for the perturbation parameter, leading to
\begin{equation}
\frac{\phi_1}{\phi_0}\sim \frac{\lambda \eta M^2}{N m^2}.
\end{equation}
Since $N\gg 1$, $M\leq m$, and one has $\lambda$ and $\eta \leq 1$ in most of the cases, a first approach by considering the modifications of $\phi$ as a perturbation is often relevant. 

When $\phi_1 \ll \phi_0$, we can evaluate the maximal modification of the energy per unit length due to the condensation of $\phi$ in the string. Close to the configuration with $\phi_0$, which is a minimum of the potential, the additional potential term is
\begin{equation}
\label{CS1dU1}
\delta U \simeq \int r \text{d}r ~ m^2 \phi_1^2\sim \frac{\eta^2 M^2}{N \kappa^2 },
\end{equation}
which finally gives, since $U_0 \sim M^2$,
\begin{equation}
\label{CS1dU1/U}
\frac{\delta U}{U_0} \sim \frac{\eta^2}{N \kappa^2}.
\end{equation}
This result gives a criterion to estimate if the toy model description of the cosmic strings is relevant from a macroscopic point of view. Note that when $\phi_1$ can not be treated as a perturbation or when $\delta U \ll U$ is not verified, the estimate (\ref{CS1dU1}) of $\delta U$ is meaningless, and a complete calculation must be done.

Up to now, we did not consider the inflaton $S$. A careful examination however shows that we can estimate the modifications of $S$ due to presence of the string to be of order $\phi_1$, and that the contribution to the energy per unit length of this field is of the same order or lower than the contribution of $\phi_1$. It can be understood by the fact that $S$ has no characteristic scale, which is necessary for it in order to play the role of the inflaton.



\section{Many-field strings}
\label{CS1ManyFieldsStrings}

\subsection{Ansatz and boundary conditions}
\label{CS1IntroManyFields}

We now turn to the case of a many-field string assuming the minimal structure ansatz, \emph{i.e.} with only the restricted representations singlet under the SM taking non vanishing values. We denote by $\tilde{\mathbf{\Phi}}_{x,\alpha}$ the sub-representations charged under U(1)$_{\text{str}}$, using as before the notations of Sec.~\ref{CS1RestrictedRep} for the different restricted representations. In the case of a many-field string, such representations appear in the GUT field content, and the superpotential contains some $\beta$-couplings. As U(1)$_{\text{str}}$ is not broken before the last step of SSB, see Sec.~\ref{CS1HI&SSB}, these particular fields must have vanishing VEVs before the end of inflation, yielding $\langle \tilde{\mathbf{\Phi}}_{x,\alpha}\rangle =0$.

An example of such a field can be the \textbf{16} representation of
SO(10), with a $\beta$ coupling through
\textbf{16}$\times$\textbf{16}$\times\mathbf{\overline{126}}$. Its
SM singlet is contained in its
($\mathbf{1},\mathbf{2},\mathbf{4}$) restricted
representation under the Pati-Salam group
[SU(2)$_L\times$SU(2)$_R$ $\times$SU(4)]. Then, the $\beta$-coupling can be constructed from this previous sub-representation and the ($\mathbf{1},\mathbf{3},\overline{\mathbf{10}}$) representation under Pati-Salam contained in the $\overline{\mathbf{126}}$ representation of SO(10). Decomposing SU(4) to SU(3)$_C\times$U(1)$_{B-L}$, the restricted representation of this additional field is
charged under the U(1)$_{B-L}$ which can play the role of
U$(1)_\text{str}$ \cite{Slansky:1981yr,Jeannerot:2003qv}. A similar coupling can be formed with the $\overline{\mathbf{16}}$ and $\mathbf{126}$ representations.

More than one field and its conjugate are now charged
under U$(1)_\text{str}$, so we have to specify these charges. We
call them $q_{x,\alpha}$ and $q_\Sigma=-q_{\bar{\Sigma}}$. 
As we can multiply these charges (which can be fractional) by a common factor without loss of
generality, we choose the smallest set of charges where they all are
integers, i.e where a rotation of $2\pi$ in the U(1) group is the smallest one which reduces to identity.
It will be convenient in what follows to define the winding number.

The $D$-term condition given in Eq. (\ref{CS1D-term}) should now include all the fields that are charged under
U$(1)_\text{str}$. This condition is \emph{a priori} not sufficient to ensure that
$\sigma^{*}=\bar{\sigma}$ and similar relations for all the
fields charged under this abelian symmetry, \emph{i.e.}
$\tilde{\phi}_{x,\alpha}^{*}=\tilde{\phi}_{\bar{x},\bar{\alpha}}$, with $\tilde{\mathbf{\Phi}}_{\bar{x},\bar{\alpha}}$ in the conjugate representation of $\tilde{\mathbf{\Phi}}_{x,\alpha}$. 
We will however assume it from now on, since this is a solution of $D_{\text{str}}=0$, and so a global minimum of the associated potential.

The same topological arguments than above imply that all the fields which
are charged under U$(1)_\text{str}$ vanish at the center of
the string \cite{Hindmarsh:1994re}:
\begin{equation}
\left\{
\begin{array}{l}
\langle \mathbf{\Sigma} \rangle_ {(r=0)} = \langle \overline{ \mathbf{\Sigma}} \rangle_
 {(r=0)} = 0,\\ \langle \tilde{\mathbf{\Phi}}_{x,\alpha}\rangle_
 {(r=0)}=0.
\end{array}
\right.
\end{equation}

We consider the following ansatz for an abelian cosmic string associated with the
generator $\tau^{\text{str}}$,
\begin{equation}
\label{CS1Antsatz2}
\begin{array}{l}
\sigma=f_\sigma (r)\text{e}^{iq_\Sigma n\theta},\\
\tilde{\phi}_{x,\alpha}=f_{x,\alpha}(r)\text{e}^{iq_{x,\alpha} n(\theta-\theta_{x,\alpha})},\\
\phi_{x,\alpha}=\phi_{x,\alpha}(r),\\
S=S(r),\\
A_\mu=A_\theta^{\text{str}}(r) \tau^{\text{str}} \delta_\mu^\theta.
\end{array}
\end{equation}
In these equations, the $f_X$ and $A_\theta^{\text{str}}$ are real functions, and
$\theta_{x,\alpha}$ are reals constants. The possibility to freely
define an origin for the coordinate $\theta$ permits to put no initial
phase to the field $\sigma$. The integer $n$ is well identified with the
winding number, taking into account the choice of normalization we did
for the charges.
We also have the following boundary conditions, similarly to the single-field string case, 
\begin{equation}
\label{CS1CLinf}
\begin{array}{l}
\displaystyle{\lim_{r \to \infty} f(r)=M,}\\
\displaystyle{\lim_{r \to \infty} f_{x,\alpha}(r)=|\tilde{\phi}_{x,\alpha,{\scriptscriptstyle{(+)}}}|,}\\
\displaystyle{\lim_{r \to \infty} \phi_{x,\alpha}(r)=\phi_{x,\alpha,{\scriptscriptstyle{(+)}}},}\\
\displaystyle{\lim_{r \to \infty} S(r)=S_{\scriptscriptstyle{(+)}},}\\
\displaystyle{\lim_{r \to \infty} A_\theta^{\text{str}}(r)=\frac{n}{g}},
\end{array}
\end{equation}
at infinity, and 
\begin{equation}
\begin{array}{l}
f(0)=0,~~~~~~
\tilde{\phi}_{x,\alpha}(0)=0, ~~~~~
 \displaystyle{A_\theta^{\text{str}}(0)=0},\\
\displaystyle{\frac{\text{d}\phi_{x,\alpha}}{\text{d} r}(0)=0},
 ~~~~~~~ \text{and} ~~~~~~~
\displaystyle{\frac{\text{d} S}{\text{d} r}(0)=0},
\end{array}
\end{equation}
at the center of the string.

At infinity, our ansatz simply gives an absolute minimum for the
potential on which we applied a local U$(1)_\text{str}$ transformation. It is the standard restated result for a cosmic strings,
but here with several fields charged under the string forming U(1). 
It shows that the angles $\theta_{x,\alpha}$ are not freely
chosen, but are those which give the absolute minimum of the potential and thus
define the SM symmetry at infinity, in addition to the limits
given in Eq. (\ref{CS1CLinf}).


\subsection{Modification of the energy per unit length}
\label{CS1ManyFieldsModifEnergy}

We now turn to the modification of the energy per unit length from standard toy models due to the condensation of the additional fields in the core of the string, with a perturbative approach. As in the case of a single-field string, we work with a generic field $\phi$, without considering its indices anymore. 

The potential term gives
\begin{align}
V \simeq& ~ m^2\phi^2+ \frac{\eta^2 \sigma^4}{N} + \frac{\beta^2 \sigma^2 \phi^2}{N}+ \frac{\lambda^2 \phi^4}{N} 
+\frac{m \eta \phi \sigma^2}{\sqrt{N}} + \frac{m \beta \sigma \phi^2}{\sqrt{N}} \nonumber\\
& ~~ + \frac{m \lambda \phi^3}{\sqrt{N}}+\frac{\eta \beta \sigma^3 \phi}{N}+\frac{\eta \lambda \sigma^2 \phi^2}{N} + \frac{\beta \lambda \sigma \phi^3}{N}.
\end{align}
Introducing $\phi_0 \sim (\sqrt{N} m)/\lambda$, two perturbation scales appears, $\phi_1 \sim(\eta M^2)/(\sqrt{N} m)$ and $\phi_1^\prime \sim (\beta M)/\lambda$. So, in order to make a perturbative study of the string, the small parameters we have to consider are 
\begin{equation}
\frac{\phi_1}{\phi_0}\sim \frac{\lambda \eta M^2}{N m^2} ~~~~ \text{and} ~~~~\frac{\phi_1^\prime}{\phi_0}\sim \frac{\beta M}{\sqrt{N} m}.
\end{equation}
If a perturbative study is possible, we then estimate the maximal modification of $U$ due to the second term to be of order 
\begin{equation}
\delta U ^\prime \simeq \int r \text{d}r ~ m^2 \phi_1^2 \sim \frac{\beta^2 m^2}{\kappa^2 \lambda^2},
\end{equation}
which gives
\begin{equation}
\label{CS1dU2/U}
\frac{\delta U^\prime}{U_0} \sim \frac{\beta ^2 m^2}{\kappa^2 \lambda^2 M^2}.
\end{equation}
We see that even in a model where the coupling constants are smaller than $1$, we cannot in general consider that the modification of the energy per unit length of the strings can be treated as a perturbation. It means that in the case of a many-field string, it is necessary to do a complete study of the microscopic structure of the string, and also that the toy model approximation is in general not valid.

Finally, we see that the simplification of Sec. \ref{CS1Superpotential} consisting in
omitting terms like $\mathbf{\Sigma}\mathbf{\Sigma}\mathbf{\Phi}$ and
$\mathbf{\overline{\Sigma}}\mathbf{\overline{\Sigma}}\mathbf{\Phi}$ can be justified 
\emph{a posteriori}. Indeed, these terms must be studied in the case of 
a many-field string, as the field $\mathbf{\Phi}$ appearing in this cubic term 
has to be charged under U(1)$_{\text{str}}$ in order to provide a singlet term.
However, its contribution to the energy will be similar to a standard $\mathbf{\Sigma}
\mathbf{\overline{\Sigma}}\mathbf{\Phi}$ term, which is already taken into account here.


\section{Conclusions and discussions}

Let us summarize by stating the different results obtained in this paper, focusing on the abelian strings forming at the last step of the SSB scheme, at the end of inflation. We worked in the framework of a SUSY GUT with hybrid inflation, but some results are more general, and we state for each point which hypotheses are actually necessary.
\begin{itemize}

\item[i)] The minimal structure model for realistic cosmic strings contains at least all the fields taking non vanishing VEVs at the end of the SSB down to the SM, which all condense in the core of the strings. These fields are singlets of the SM. This property neither depends on the GUT nor on the inflationary process.

\item[ii)] It is possible to distinguish two classes of strings, the single-field strings where only one field and its conjugate are charged under U(1)$_\text{str}$, and the many-field strings. The way the field content performs the SSB scheme is sufficient to determine which kind of strings form, independently of having a SUSY model or of the inflationary process. Different couplings appear in each model, which may give different phenomenologies for the strings.

\item[iii)] For each class of strings and with a SUSY GUT and hybrid inflation, we gave an ansatz and the associated boundary conditions to describe the minimal structure of the strings, completely defining it from a mathematical point of view. These strings are singlet of the SM. The fields take intermediate values in the core of the string between the minimum of the potential defining the SM at infinity and the configuration taken by the fields before the end of inflation. This ansatz can easily be generalized for other models.

\item[iv)] Going from the GUT description to the minimal structure ansatz, important numerical factors appear and must be taken into account. How these factors appear depends on the conventions and normalizations used. A special care has to be taken in SUSY models, where these factors are omitted when studying the static configurations which minimize the potential, \emph{i.e.} with all $F$-terms independently vanishing. We emphasized how to normalize the GUT in order to recover the abelian Higgs model in the limit where the couplings between the string-forming Higgs and the other fields go to zero.

\item[v)] In the case of a SUSY GUT with hybrid inflation, we performed perturbative estimates of the modification of the energy per unit length with respect to standard toy models, which are given in Eqs.~(\ref{CS1dU1/U}) and~(\ref{CS1dU2/U}). These results are very different in the case of a single-field string, where the modifications are sizable in a high-coupling limit, and in the case of a many-field string, where the modifications are always important and require a complete computation of each model.

\end{itemize}

These results represent a first step of a more thorough investigation of cosmic strings taking into account their realistic structure. A complete study of the microscopic structure and the energy per unit length of such strings in a given SO(10) model has been performed in parallel, including numerical solutions, and can be found in Ref.~\cite{Allys:2015kge}. As discussed in Sec.~\ref{CS1PartMinimalStructure}, several properties of the strings can also be modified in this framework and have major cosmological consequences. Indeed, the Higgs fields condensing in the core of the string could carry bosonic currents \cite{Witten:1984eb,Peter:1992dw,Peter:1992nz,Peter:1993tm,Morris:1995wd}. Moreover, their superpartner could carry fermionic currents through their zero modes \cite{Witten:1984eb,Jackiw:1981ee,Weinberg:1981eu,Davis:1995kk,Ringeval:2000kz,Peter:2000sw}. Furthermore, this complex microscopic structure could qualitatively modify the intercommutation process \cite{Shellard:1988ki,Laguna:1990it,Matzner:1988ky,Moriarty:1988qs,Moriarty:1988em,Shellard:1987bv} and thus the evolution of the cosmological string network, hence modifying its consequences on the CMB \cite{Kibble:1976sj,Kibble:1980mv,Copeland:1991kz,Austin:1993rg,Martins:1996jp,Martins:2000cs}. The cusps evaporation \cite{Srednicki:1986xg,Brandenberger:1986vj,Gill:1994ic,Olum:1998ag,Bhattacharjee:1989vu} also deserves more thoughts, since in our case, knowledge of the fields present in the string core implies knowledge of the relevant branching ratios into specific particles.



\subsubsection*{Acknowledgment}

I wish to thank Patrick Peter for many valuable discussions and suggestions, and also for a critical reading of the manuscript. I also thank R. Brandenberger and M. Sakellariadou for enlightening discussions and remarks.

\chapter[Bosonic structure of realistic SO(10) SUSY cosmic strings (article)]{Bosonic structure of realistic SO(10) SUSY cosmic strings}
\label{PartArticleCorde2}
%

\begin{figure}[h!]
\begin{center}
\includegraphics[scale=1.4]{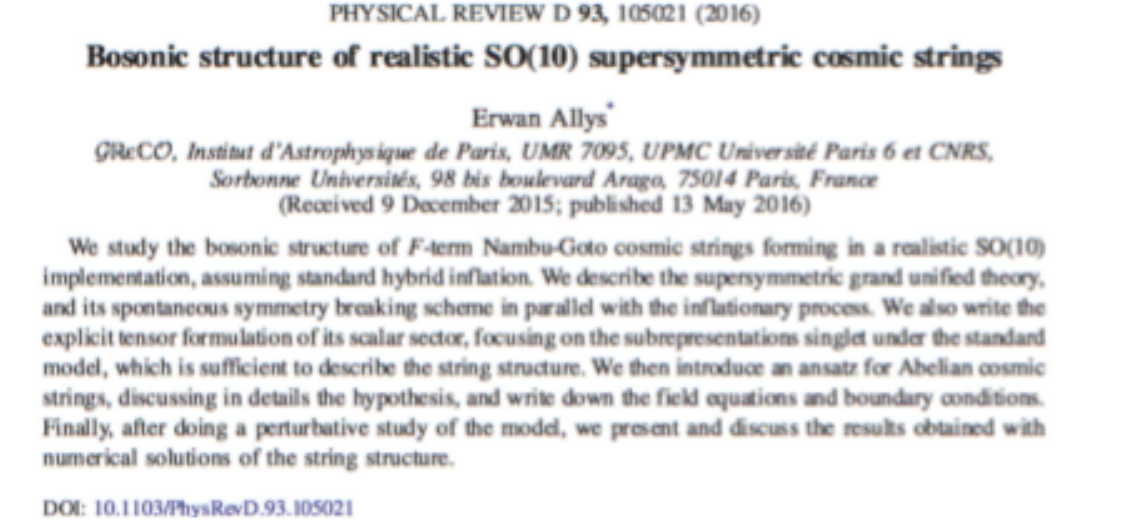}
\end{center}
\end{figure}

\section{Introduction}

The phenomenology of the early universe gives access to high energy physics models, such as Grand Unified Theories (GUTs). Indeed, we know that the Spontaneous Symmetry Breaking (SSB) down to the Standard Model (SM) of such theories around $10^{16}$GeV must produce topological defects, e.g. cosmic strings \cite{Kibble:1980mv,Hindmarsh:1994re}. The observation of such objects, for instance in the Cosmic Microwave Background (CMB) \cite{Bouchet:2000hd,Bevis:2007gh,Ringeval:2010ca,Ade:2013xla,Brandenberger:2013tr}, allows then to put constraints on the string energy per unit length, and thus on the GUT which led their formation. In the past, some work has already been done to study the structures and properties of such strings, see e.g. \cite{Kibble:1982ae,Ma:1992ky,Davis:1996sp,Davis:1997bs,Morris:1997ua,Davis:1997ny}.

In order to have a complete understanding of these objects, it is necessary to study them in a realistic GUT context, and not only through toy models which contain only the minimal field content necessary to describe this kind of defects. Such a work was done in Ref.~\cite{Allys:2015yda}, without using a specified GUT, and considering only the bosonic structure of the strings. We continue this study by considering a given SO(10) supersymmetric (SUSY) GUT, which has already been studied in a particle physics and a cosmological framework \cite{Aulakh:2002zr,Fukuyama:2004ps,Fukuyama:2004xs,Aulakh:2004hm,Bajc:2004xe,Aulakh:2005mw,Bajc:2005qe,Aulakh:2003kg,Cacciapaglia:2013tga,Fukuyama:2012rw,Garg:2015aga}. We also consider that the SSB scheme of the GUT takes place during a $F$-term hybrid inflation scenario~\cite{Copeland:1994vg,Linde:1993cn,Dvali:1994ms,Lyth:1998xn,Kyae:2005vg,Mazumdar:2010sa}, where the inflaton dynamics is driven by the field content of the GUT itself.

Having a complete description of these linear topological defects can give interesting results. On the one hand, their macroscopic properties could considerably change in comparison with the simpler models, which would modify the current constraints using cosmic strings observations. On the other hand, it gives access to the parameters of the GUT itself, and not only to the energy scale of formation of the strings. For example, having the energy per unit length of the string as function of the different parameters of the GUT permits to implement observational constraints on their ranges.

In a first part, we introduce the SUSY SO(10) GUT studied, and its SSB scheme in parallel with the inflationary process. We then focus in Sec.~\ref{CS2TensorFormulation} on the explicit tensor formulation of the theory. Special attention is payed to the formulation of the model as a function of the restricted representations which are singlet under the SM. Some of the calculations and results of this section are put in Appendix \ref{CS2AppendixA}. The cosmic strings are studied in Sec.~\ref{CS2PartCosmicStrings}, where we give an ansatz for their structure, and write the equations of motion as well as the
boundary conditions for all the fields. After writing the model with dimensionless variables, we perform a perturbative study of the string in Sec.~\ref{CS2PartScaling&Pert}. Finally, in Sec.~\ref{CS2PartNumericalSolution}, we present and discuss the numerical solutions of the strings, and their microscopic and macroscopic properties.

\section{SO(10) GUT, hybrid inflation, and SSB}
\subsection{GUT and field content}
\label{CS2PartIntroGUT}

We focus on a well-studied SO(10) SUSY GUT, which has already been considered in a particle physics \cite{Aulakh:2002zr,Fukuyama:2004ps,Fukuyama:2004xs,Aulakh:2004hm,Bajc:2004xe,Aulakh:2005mw,Bajc:2005qe,Aulakh:2003kg,Fukuyama:2012rw} and a cosmological \cite{Cacciapaglia:2013tga,Garg:2015aga} framework. The superpotential yields \cite{Bajc:2004xe,Martin:1997ns,Slansky:1981yr}
\begin{equation}
W= \frac{m}{2}\mathbf{\Phi}^2
+ m_{\Sigma}\mathbf{\Sigma}\overline{\mathbf{\Sigma}}
+ \frac{\lambda}{3}\mathbf{\Phi}^3
+ \eta\mathbf{\Phi}\mathbf{\Sigma}\overline{\mathbf{\Sigma}}
+ \kappa S (\mathbf{\Sigma}\overline{\mathbf{\Sigma}}-M^2),
\end{equation}
where $\mathbf{\Sigma}$ and $\mathbf{\overline{\Sigma}}$ are in the \textbf{126} and $\mathbf{\overline{126}}$ representations, $\mathbf{\Phi}$ is in the \textbf{210} representation, and the inflaton $S$ is a singlet of SO(10). It is the more general singlet term we can write, in addition with an $F$-term hybrid inflation coupling involving the inflaton and $\mathbf{\Sigma}$ and $\mathbf{\overline{\Sigma}}$, which are the only fields in complex conjugate representations. This last term is the simplest we can write which reproduces the inflation phenomenology \cite{Lyth:1998xn,Mazumdar:2010sa}. Additional terms in $S^2$, $S^3$ or $S \mathbf{\Phi}^2$ could be added, but they often generate mass or quartic terms for the inflaton at tree-level, and then spoil the inflation. A discussion of those terms can be found in Ref.~\cite{Cacciapaglia:2013tga}.

All the parameters introduced are complex, but we can use redefinitions of the phases of the superfields to set $m$, $m_{\Sigma}$, $\kappa$ and $M$ real, $\lambda$ and $\eta$ still being complex. The explicit component formulation will be given in Sec.~\ref{CS2TensorFormulation}. The reader should be reminded that as we work in a SUSY framework, all the components of the scalar fields are complex.

Note that the purpose of this article is to consider in details the complete GUT structure of the strings, rather than focus on the inflation, which is treated at a basic level. We could also consider an inflation led by fields out of the GUT sector, and recover the same kind of phenomenology for the strings. The advantage here is that, inflation being implemented by the GUT fields, we have a precise relation between the cosmological evolution and the breaking scheme of the GUT, see Sec.~\ref{CS2SSB&HI}. It is also possible to refine the model by including additional couplings to develop the inflationary phenomenology.

This superpotential is sufficient to describe the SSB of the GUT to the SM symmetry. In addition, a Higgs field in the \textbf{10} representation permits us to implement the electroweak symmetry breaking. Its characteristic scale is very different of that considered in this paper and we can omit it from now on. Another Higgs field in the \textbf{120} representation of SO(10) can be added in order to recover the complete SM fermion mass spectrum, but it can be omitted in a first approximation \cite{Aulakh:2005mw}.

%

From now on, we will focus on the bosonic sector of the theory, so we will not consider the superfields anymore, and focus instead on their bosonic part. Also, as we work with $F$-term hybrid inflation scenario, we assume that all the $D$-terms associated to the gauge generators take identically vanishing values, with no Fayet-Iliopoulos term \cite{Martin:1997ns}. This condition will permit us to impose some constraints on the fields thereafter.

\subsection{Lagrangian of the bosonic sector}

We take the signature of the metric to be $+2$. The Lagrangian of the bosonic sector gives
\begin{align}
\label{CS2LagFull}
\mathcal{L}=&-\frac{1}{4}\text{Tr}(F_{\mu \nu}F^{\mu \nu})
-(D_{\mu}\mathbf{\Phi})^\dagger(D^{\mu}\mathbf{\Phi})
-(D_{\mu}\mathbf{\Sigma})^\dagger(D^{\mu}\mathbf{\Sigma})\nonumber\\
& ~~ -(D_{\mu}\mathbf{\overline{\Sigma}})^\dagger(D^{\mu}\mathbf{\overline{\Sigma}})
-(\nabla_{\mu}S)^\ast(\nabla^{\mu}S)
-V(\mathbf{\Phi},\mathbf{\Sigma},\mathbf{\overline{\Sigma}},S).
\end{align}
The inflaton has no gauge covariant derivative since this is a singlet of the gauge group. We also take the following definitions
\begin{equation}
D_{\mu}X=(\nabla_{\mu}+ g A_{\mu})X,
\end{equation}
\begin{equation}
A_{\mu}=-iA_{\mu}^{a}\tau_X^a,
\end{equation}
\begin{equation}
F_{\mu\nu}=-F_{\mu\nu}^{a}\tau_X^a,
\end{equation}
\begin{equation}
F_{\mu\nu}=\nabla_\mu A_{\nu}-\nabla_\nu A_{\mu}+q [A_{\mu},A_{\nu}],
\end{equation}
where from now on $X$ is a general notation for all scalar fields, \emph{i.e.} $X\in \{\mathbf{\Sigma}, \overline{\mathbf{\Sigma}}, \mathbf{\Phi}, S \}$. We note $\tau_X^a$ the action of the generators of the gauge group in the representation of $X$, the index $a$ labeling the $45$ generators of SO(10). The \textbf{210} representation being real, we can use a basis where ${(\tau_{210}^a)}^\dagger=\tau_{210}^a$. It is not the case anymore for the \textbf{126} representation, which is complex: ${(\tau_{126}^a)}^\dagger$ is not anymore in the \textbf{126} but in the $\mathbf{\overline{126}}$ representation. However, we can choose a basis where $\tau_{\overline{126}}^a=-{(\tau_{ 126}^a)}^\dagger$.
Independently of the representation, we have 
\begin{equation}
[\tau_{a},\tau_{b}]=if_{ab}{}^c\tau_{c},
\end{equation}
with $f_{ab}{}^{c}$ the structure constants of SO$(10)$.

We first focus on the $F$-term part of the potential. They are defined by
\begin{equation}
\label{CS2DefFTerms}
F_X=\frac{\partial W}{\partial X},
\end{equation}
with $F_X$ in the conjugate representation of $X$.
It yields
\begin{equation}
\begin{array}{l}
F_S= \kappa  (\mathbf{\Sigma}\overline{\mathbf{\Sigma}}-M^2), \\
F_{\Phi}=m\mathbf{\Phi} + \lambda(\mathbf{\Phi} ^2)_{\mathbf{\Phi}} + \eta(\mathbf{\Sigma}\overline{\mathbf{\Sigma}})_{\mathbf{\Phi}},\\
F_{\Sigma}=m_{\Sigma}\overline{\mathbf{\Sigma}}+ \eta(\mathbf{\Phi}\overline{\mathbf{\Sigma}})_{\overline{\mathbf{\Sigma}}} + \kappa S \overline{\mathbf{\Sigma}}, \\
F_{\overline{\Sigma}}=m_{\Sigma}\mathbf{\Sigma}+ \eta(\mathbf{\Phi}\mathbf{\Sigma})_{\mathbf{\Sigma}} + \kappa S \mathbf{\Sigma},
\end{array}
\end{equation}
where we denote with $(X Y)_Z$ the term in the representation of $Z$ we can build from the product of the fields $X$ and $Y$. See Sec.~\ref{CS2AppendixDerivatives} for a more detailed explanation about how to obtain these results.

We finally obtain for the potential
\begin{equation}
V=\sum_X F_X^\dagger F_X\equiv\sum_X V_X,
\end{equation}
which is a sum of positive terms. 
It gives
\begin{equation}
\label{CS2PotentialS}
V_S= \kappa^2 (\mathbf{\Sigma}\overline{\mathbf{\Sigma}}-M^2)^2,
\end{equation}
\begin{align}
\label{CS2PotentialPhi}
V_\Phi=& m^2\mathbf{\Phi} \mathbf{\Phi}^\dagger 
+ |\eta|^2 (\mathbf{\Sigma}\overline{\mathbf{\Sigma}})_{\mathbf{\Phi}} (\mathbf{\Sigma}\overline{\mathbf{\Sigma}})_{\mathbf{\Phi}}^\dagger 
 + |\lambda|^2 (\mathbf{\Phi} ^2)_{\mathbf{\Phi}} (\mathbf{\Phi} ^2)_{\mathbf{\Phi}} ^\dagger 
 \nonumber\\
&~~ + \left[ \lambda \eta^* (\mathbf{\Phi} ^2)_{\mathbf{\Phi}}(\mathbf{\Sigma}\overline{\mathbf{\Sigma}})_{\mathbf{\Phi}}^\dagger 
+ m \lambda^* \mathbf{\Phi}(\mathbf{\Phi} ^2)_{\mathbf{\Phi}}^\dagger  
+ m \eta^* \mathbf{\Phi}(\mathbf{\Sigma}\overline{\mathbf{\Sigma}})_{\mathbf{\Phi}}^\dagger \right] + \text{h.c.},
\end{align}

\begin{align}
\label{CS2PotentialSigma}
V_{\overline{\Sigma}}= & m_{\Sigma}^2 \mathbf{\Sigma}\mathbf{\Sigma}^\dagger   + |\eta|^2 (\mathbf{\Phi}\mathbf{\Sigma})_{\mathbf{\Sigma}} (\mathbf{\Phi}\mathbf{\Sigma})_{\mathbf{\Sigma}}^\dagger
+ \kappa^2 SS^*\mathbf{\Sigma}\mathbf{\Sigma}^\dagger 
\nonumber\\
& ~~ +\left[ \eta \kappa S^{*} (\mathbf{\Phi}\mathbf{\Sigma})_{\mathbf{\Sigma}} \mathbf{\Sigma}^\dagger
+ m_{\Sigma} \kappa S^*  \mathbf{\Sigma}\mathbf{\Sigma}^\dagger
+  m_{\Sigma}  \eta^*\mathbf{\Sigma}(\mathbf{\Phi}\mathbf{\Sigma})_{\mathbf{\Sigma}}^{\dagger}\right] +\text{h.c.},
\end{align}
and
\begin{equation}
\label{CS2PotentialSigmaBar}
V_{\Sigma}=V_{\overline{\Sigma}} (\mathbf{\Sigma} \longleftrightarrow \mathbf{\overline{\Sigma}} ).
\end{equation}
Note that we did not include in what is called $V$ the $D$-term contribution to the potential. Indeed, this term having an identically vanishing value, it does not contribute to the dynamics of the fields. However, this condition will be imposed by the following, and gives some constraints on the fields.

\subsection{SSB scheme, hybrid inflation and topological defects}
\label{CS2SSB&HI}

We now turn to the cosmological evolution of the GUT, following Refs.~\cite{Cacciapaglia:2013tga,Allys:2015yda}, and also Ref.~\cite{Jeannerot:1995yn} for a model very close to the one we study. The SSB schemes take the following form
\begin{equation}
\text{SO}(10)\overset{\langle\Phi\rangle}{\relbar\joinrel\relbar\joinrel\longrightarrow}
G' \overset{\langle\Sigma\rangle}{\relbar\joinrel\relbar\joinrel\longrightarrow}
G_{\text{SM}} \times \mathbb{Z}_2,
\end{equation}
where $G_{\text{SM}}=3_C 2_L 1_Y$. We use from now on short notations for the gauge groups, $3_C 2_L 1_Y$ meaning SU(3)$_C$ $\times$SU(2)$_L \times$U(1)$_Y$, and so on. The $\mathbb{Z}_2$ factor appears in addition to the SM gauge group in order to suppress proton decay \cite{Martin:1992mq}. We assume that the first SSB step happens at $E_{\text{GUT}}\sim 10^{16}$GeV, and the second one slightly below, at $E\sim \left(10^{15}-10^{16}\right)$GeV.

At the onset of inflation, we can assume a very large value for the inflaton $S$ in comparison with all the other fields, as it is the case in chaotic inflation. To minimize the higher order terms containing it in Eqs.~(\ref{CS2PotentialSigma}) and~(\ref{CS2PotentialSigmaBar}), so that $V_{\bar{\Sigma}}\sim |\kappa S \mathbf{\Sigma}|^2$ and $V_\Sigma \sim |\kappa S \overline{\mathbf{\Sigma}}|^2$, the field $\mathbf{\Sigma}$ and $\mathbf{\overline{\Sigma}}$ must take a vanishing value. We thus have $V_{\Sigma}=0$, $V_{\overline{\Sigma}}=0$ and $V_S=\kappa^2 M^4$ at this step. In addition, we assume that the VEV taken by $\mathbf{\Phi}$ minimizes $V_\Phi$. Different VEVs for $\mathbf{\Phi}$ have this property, each one being associated with a different gauge group at this step $G'$.

After this first SSB, a false vacuum hybrid inflation begins. Indeed, the potential verifies $V=V_0 + \text{quant}.\text{corr}.$, with $V_0=\kappa^2 M^4$ and the inflaton slowly rolls along this flat direction at tree level until it reaches its critical value, thus ending inflation (see e.g. Ref.~\cite{Cacciapaglia:2013tga} for a detailed explanation about this phase). At this point, a new phase transition takes place, causing a SSB down to the SM gauge group. We assume the set of VEVs at this last step to be a global minimum for the potential, \emph{i.e.} implying $V=0$. As previously, this condition is fulfilled by several set of VEVs, each one defining an unbroken gauge symmetry.

A careful study of the possible SSB cascades has been done in \cite{Cacciapaglia:2013tga}, including stability of the inflationary valley. Only two of them are valid and permit us to recover the SM at low energy. Their respective intermediate symmetry group is $G'=3_C 2_L 2_R 1_{B-L}$ and $G'=3_C 2_L 1_R 1_{B-L}$. For both these intermediate symmetry groups, the topological defects produced are monopoles in the first SSB step, and cosmic strings in the second \cite{Jeannerot:2003qv}. The monopoles are washed out by the inflation, which takes place after their formation. In this paper, we study the cosmic strings which form in the SSB from $G'$ to $G_{\text{SM}}\times \mathbb{Z}_2$, at the end of inflation.

\subsection{Use of fields singlet under the SM}
\label{CS2PartSMSinglet}

The VEVs which have a non vanishing value at the last stage of SSB define the SM symmetry. It implies that they must be uncharged under $G_{\text{SM}}$, since they would otherwise break this symmetry group at this stage. These non vanishing VEVs thus are singlet under the SM. We will also assume there is no symmetry restoration, \emph{i.e.} all fields acquiring a non-zero VEV at a given step keep it non vanishing at later stages. It permits us to impose that all the non vanishing VEVs of the SSB cascade are also singlet under the SM.

We can consider all the restricted representations of the field content of the GUT we study, and look for sub-representations which are singlet under the SM. These are very few \cite{Bajc:2004xe,Slansky:1981yr}, and are listed below, giving their representations under the Pati-Salam group ($2_L 2_R 4_C$)
\begin{equation}
\label{CS2SMSinglet}
\begin{matrix}
   \mathbf{\Phi}_p=\mathbf{\Phi}(1,1,1), & \boldsymbol\sigma=\mathbf{\Sigma}(1,3,\overline{10}),  \\
   \mathbf{\Phi}_a=\mathbf{\Phi}(1,1,15), & \overline{\boldsymbol\sigma}=\overline{\mathbf{\Sigma}}(1,3,10), \\
   \mathbf{\Phi}_b=\mathbf{\Phi}(1,3,15), & S=S(1,1,1).
\end{matrix}
\end{equation}
Considering the \textbf{3} representation of $2_R$, it is sufficient to take the neutral component of $1_R \subset 2_R$, which is what we will do from now on. For the representations non singlet under $4_C$, their branching rules on $3_C 1_{B-L}$ are
\begin{equation}
\label{CS24RSingletDecomposition}
\begin{array}{l}
\mathbf{1}=\mathbf{1}(0),\\
\mathbf{10}=\mathbf{1}(2)+\mathbf{3}(2/3)+\mathbf{6}(-2/3),\\
\mathbf{15}=\mathbf{1}(0)+\mathbf{3}(-4/3)+\mathbf{\overline{3}}(4/3)+\mathbf{8}(0),
\end{array}
\end{equation}
where we denote with $\mathbf{n}(q)$ the representation of $3_C$ of dimensions $n$ which has a charge $q$ under $1_{B-L}$.
For these representations, we use their sub-representation singlet of $3_C$, which is unique. The explicit tensor formulation of these restricted representations will be given in Sec.~\ref{CS2PartSingletDecomposition}.

It permits us to describe in a short way the non vanishing VEVs appearing after the first SSB scheme and defining $G'$, for the two relevant schemes \cite{Bajc:2004xe,Cacciapaglia:2013tga}. For the first one, only $\langle \mathbf{\Phi}_a \rangle$ takes a non vanishing value, and the residual symmetry group is $G'_1=3_C 2_L 2_R 1_{B-L}$. For the second one, the three restricted representations singlet under the SM contained in $\mathbf{\Phi}$ take a non vanishing expectation value, and the residual symmetry group is $G'_2=3_C 2_L 1_R 1_{B-L}$.

Finally, and as discussed in Ref.~\cite{Allys:2015yda}, we can restrict the study of the microscopic structure of the string to a configuration where all the fields which are not singlet under the SM take an identically vanishing value. Indeed, the potential is at least quadratic in these fields, since it would otherwise be charged under the SM. So the solutions where all these fields take a vanishing value is a solution of their equations of motion. On the other hand, these fields must have a zero value at infinity since the vacuum is uncharged under the SM. This shows that the solution discussed previously is also compatible with the boundary conditions at infinity. We will assume this particular ansatz from now on. For a more detailed discussion about this assumption, see Ref.~\cite{Allys:2015yda}.

\section{Explicit tensor formulation}
\label{CS2TensorFormulation}
\subsection{Introduction and fields content}

In order to have a complete study of the model, including numerical solution, we need to write it in a component (here tensor) formulation, which is the purpose of this whole section. As explained in Sec \ref{CS2PartSMSinglet}, we will also restrict the study to only the fields which are singlet under the SM, which permits us to describe the problem in terms of only a few complex functions. In order to be as concise and clear as possible, a part of the calculations and results are given as an appendix, in Sec.~\ref{CS2AppendixA}.

The tensor formulation of the field content yields \cite{Bajc:2004xe,Aulakh:2000sn} : 
\begin{itemize}
\item $\mathbf{\Sigma}$ (\textbf{126}) is a fifth rank anti-symmetric tensor $\Sigma_{ijklm}$, self dual (in the sense of Hodge duality) :
\begin{equation}
\label{CS2self-dual}
\Sigma_{ijklm}=\frac{i}{5!}\epsilon_{ijklmabcde}\Sigma_{abcde},
\end{equation}
\item $\overline{\mathbf{\Sigma}}$ ($\mathbf{\overline{126}}$) is a fifth rank anti-symmetric tensor $\overline{\Sigma}_{ijklm}$, anti-self-dual :
\begin{equation}
\overline{\Sigma}_{ijklm}=-\frac{i}{5!}\epsilon_{ijklmabcde}\overline{\Sigma}_{abcde},
\end{equation}
\item $\mathbf{\Phi}$ (\textbf{210}) is a fourth rank anti-symmetric tensor $\Phi_{ijkl}$,
\item $S$ is a singlet
\end{itemize}
We remind the reader that in the tensor formulation of SO(10), all the indices go between $1$ and $10$.
 
\subsection{Superpotential, $F$-terms, and potential}
\label{CS2TensorW&F}

The superpotential, defined in Sec.~\ref{CS2PartIntroGUT}, yields
\begin{align}
W=& \frac{1}{2}m \Phi_{ijkl} \Phi_{ijkl}
+ m_{\Sigma} \Sigma_{ijklm} \overline{\Sigma}_{ijklm}
+ \frac{1}{3}\lambda\Phi_{ijkl} \Phi_{klmn}\Phi_{mnij} \nonumber \\
&~~ + \eta \Phi_{ijkl}\Sigma_{ijmno} \overline{\Sigma}_{klmno}
+ \kappa S (\Sigma_{ijklm} \overline{\Sigma}_{ijklm}-M^2)
\end{align}

Now, to compute the $F$-terms, we have to take the derivatives with respect to the different tensor components of the fields. However, we have to take into account the fact that they are not independent (due to the symmetry and duality properties of the tensors). The way to proceed is given in Sec.~\ref{CS2AppendixDerivativeComputation}, and the derivatives of the different terms and the associated notations are written in Sec.~\ref{CS2AppendixDerivatives}. Explicitly, we find
\begin{equation}
\displaystyle{F_S= \kappa  (\Sigma_{ijklm} \overline{\Sigma}_{ijklm}-M^2)},
\end{equation}
\begin{equation}
{\left(F_{\Phi}\right)}_{ijkl}= m \Phi_{ijkl} + \lambda \Phi_{[ij|ab}\Phi_{ab|kl]}
 + \eta \Sigma_{[ij|abc}\overline{\Sigma}_{|kl]abc},
\end{equation}
\begin{equation}
{\left(F_{\Sigma}\right)}_{ijklm}=m_{\Sigma} \overline{\Sigma}_{ijklm} + \frac{\eta }{2}\bigg(\Phi_{[ij|\alpha\beta}\overline{\Sigma}_{\alpha\beta |klm]} 
-\frac{i}{5!}\epsilon_{ijklmabcde}\Phi_{ab\alpha\beta}\overline{\Sigma}_{\alpha\beta cde}\bigg)  + \kappa S  \overline{\Sigma}_{ijklm},
\end{equation}
and
\begin{equation}
{\left(F_{\overline{\Sigma}}\right)}_{ijklm}=m_{\Sigma} \Sigma_{ijklm} +  \frac{\eta}{2}\bigg(\Phi_{[ij|\alpha\beta}\Sigma_{\alpha\beta |klm]}
+\frac{i}{5!}\epsilon_{ijklmabcde}\Phi_{ab\alpha\beta}\Sigma_{\alpha\beta cde}\bigg) + \kappa S  \Sigma_{ijklm}.
\end{equation}
Note that in these $F$-terms, we cannot obtain $F_{\overline{\Sigma}}$ from $F_{\Sigma}$ by only changing $\Sigma_{abcde}$ in $\overline{\Sigma}_{abcde}$.

We do not write the full tensorial expression of the potential at this step, which is obtained by injecting the results of Sec.~\ref{CS2AppendixDerivatives} in Eq.~(\ref{CS2PotentialS}) to (\ref{CS2PotentialSigmaBar}).
 
\subsection{Singlet decomposition and $D$-terms}
\label{CS2PartSingletDecomposition}

Let us consider now the restricted representations which are singlet under the SM. In addition, they are uncharged under any continuous non abelian symmetry which commutes with the SM symmetry. It implies that we can describe their dynamics by a single complex function. It yields, e.g. 
\begin{equation}
\mathbf{\Phi}_{a}  = a(x^\mu) {\langle \mathbf{\Phi}_{a}\rangle}_0,
\end{equation}
where $a(x^\mu)$ is a complex function of the space-time, and ${\langle \mathbf{\Phi}_{a}\rangle}_0$ is a constant vector in the representation space. Following the conventions of \cite{Allys:2015yda}, we choose to work with normalized constant vectors, \emph{i.e.} with ${\langle \mathbf{\Phi}_{a}\rangle}_0 {\langle \mathbf{\Phi}_{a}\rangle}_0^\dagger=1$. 

We can now write these singlets under the SM in a tensor formulation \cite{Bajc:2004xe}, following the notations introduced in Eq.~(\ref{CS2SMSinglet}),
\begin{equation}
\label{CS2SingletDefinition}
\left\{
\begin{array}{l}
\displaystyle{ \frac{p}{\sqrt{4!}}= \Phi_{1234} },\\
\displaystyle{\frac{a}{\sqrt{4!3}}= \Phi_{5678} = \Phi_{5690} 
= \Phi_{7890} },\\
\displaystyle{\frac{b}{\sqrt{4!6}}= \Phi_{1256} = \Phi_{1278} 
= \Phi_{1290} }\\
\displaystyle{~~~~~~~ = \Phi_{3456} 
= \Phi_{3478} = \Phi_{3490} },\\
\displaystyle{\frac{1}{\sqrt{5!2^5}}(i)^{(-a-b+c+d+e)}\sigma}=\Sigma_{a+1,b+3,c+5,d+7,e+9}, \\
\displaystyle{\frac{1}{\sqrt{5!2^5}}(-i)^{(-a-b+c+d+e)}\overline{\sigma}} = \overline{\Sigma}_{a+1,b+3,c+5,d+7,e+9},
\end{array}
\right.
\end{equation}
where the complex functions are $p$, $a$, $b$, $\sigma$ and $\overline{\sigma}$. In the last two equations, the indices $a$, $b$, $c$, $d$ and $e$ are either $0$ or $1$.

The $D$-term condition permits us to impose additional constraints  on the fields. The general expression is
\begin{equation}
\label{CS2D-term}
D^a=-g\sum_X (X^\dagger \tau^a_X X),
\end{equation}
for the $D$-term associated to the generator $\tau^a$ (in the case of a vanishing Fayet-Iliopoulos term) \cite{Martin:1997ns}. The associated potential in the Lagrangian is
\begin{equation}
V_{D} = \frac{1}{2} \sum_a D^a D^a
\end{equation}
As we work in the framework of an $F$-term theory, all the $D$-terms identically vanish. The $D$-term condition associated to the generator of $1_{B-L}$ simplifies a lot since only the SM singlets associated to $\mathbf{\Sigma}$ and $\mathbf{\overline{\Sigma}}$ are charged under this group [see Eqs.~\ref{CS24RSingletDecomposition}], yielding \cite{Slansky:1981yr,Harada:2003sb}
\begin{equation}
D_{1_{B-L}} = \sqrt{\frac{3}{8}}\left( 2{ \boldsymbol\sigma }^\dagger \boldsymbol\sigma  -2 { \overline{\boldsymbol\sigma} }^\dagger \overline{\boldsymbol\sigma}\right)  = 0.
\end{equation}
In addition, as the phase of the inflaton $S$ have been rephased in order to make $M$ real, it ensures that the global minimum of the potential is reached when $\boldsymbol\sigma \overline{\boldsymbol\sigma}=M^2 \in \mathbb{R}$. Both these results impose that $\overline{\boldsymbol\sigma}=\boldsymbol\sigma^\dagger$, which finally gives $\overline{\sigma}=\sigma^*$.

To summarize, we can write the fields in the VEV directions defined in Eq.~(\ref{CS2SingletDefinition}) 
\begin{equation}
\left\{
\begin{array}{l}
\displaystyle{ \langle \boldsymbol\Sigma \rangle ={\langle \overline{\boldsymbol\Sigma} \rangle }^\dagger =  \sigma {\langle \boldsymbol\sigma \rangle }_0},\\
\displaystyle{\langle \mathbf{\Phi} \rangle = a {\langle \mathbf{\Phi}_a \rangle}_0 + b {\langle \mathbf{\Phi}_b \rangle}_0 + p {\langle \mathbf{\Phi}_p \rangle}_0 },
\end{array}
\right.
\end{equation}
with the normalization conditions giving
\begin{equation}
\label{CS2Orthonormalization}
\begin{array}{l}
{\langle \mathbf{\Phi}_p\rangle}_0 {\langle \mathbf{\Phi}_p \rangle}_0^\dagger =1,\\
{\langle \mathbf{\Phi}_a \rangle}_0 {\langle \mathbf{\Phi}_a \rangle}_0^\dagger =1,\\
{\langle \mathbf{\Phi}_b \rangle}_0 {\langle \mathbf{\Phi}_b \rangle}_0^\dagger =1,\\
\displaystyle{ {\langle \boldsymbol\sigma \rangle }_0 {\langle \boldsymbol\sigma \rangle }_0^\dagger = {\langle \boldsymbol\sigma \rangle }_0{\langle \overline{\boldsymbol\sigma} \rangle }_0 =1.}
\end{array}
\end{equation}
The other scalar products vanish, since we cannot construct scalar terms with two fields which are not in conjugate representations.

Finally, the different contractions between the fields written in terms of the few complex functions introduced previously can be found in Appendix~\ref{CS2AppendixSelectionRules}.

\subsection{Superpotential and potential in singlet form}
\label{CS2PartW&VSinglet}

We can now write the superpotential and the $F$-term scalar potential terms in term of the few complex functions introduced in the previous section. The superpotential gives 
\begin{align}
W=&\frac{m}{2}\left(p^2+a^2+b^2 \right)+m_{\Sigma}\sigma\sigma^*+\frac{\lambda}{3}\left(\frac{a^3}{9\sqrt{2}}
+\frac{ab^2}{3\sqrt{2}} +\frac{pb^2}{2\sqrt{6}}\right) 
\nonumber\\
&~~ +\eta \sigma \sigma^*\left(\frac{p}{10\sqrt{6}}+\frac{a}{10\sqrt{2}}-\frac{b}{10} \right)
+\kappa S \left( \sigma \sigma^* -M^2\right).
\end{align}
In a similar way, the potential can be written by using the expressions given in Sec.~\ref{CS2AppendixSelectionRules} in the Eq.~(\ref{CS2PotentialS}) to (\ref{CS2PotentialSigmaBar}). However, it is possible to write it in a more convenient form.

For this purpose, we can introduce the $F$-terms associated with the restricted representations singlet under the SM. Indeed, the only non vanishing terms in the $F$-term associated to $\mathbf{\Phi}$ are
\begin{equation}
\label{CS2EqFp}
\frac{F_p}{2 \sqrt{6}}={\left(F_{\Phi}\right)}_{1,2,3,4}=\frac{m p}{2 \sqrt{6}} + \frac{\lambda b^2}{72} +\frac{ \eta\sigma\sigma^*}{120},
\end{equation}
\begin{equation}
\label{CS2EqFa}
\frac{F_a}{6 \sqrt{2}}={\left(F_{\Phi}\right)}_{5,6,7,8}={\left(F_{\Phi}\right)}_{5,6,9,10}={\left(F_{\Phi}\right)}_{7,8,9,10}
=\frac{m a}{6\sqrt{2}}+\frac{\lambda}{3}\left( \frac{a^2}{36} + \frac{b^2}{36} \right) + \frac{\eta \sigma \sigma^*}{120},
\end{equation}
and
\begin{align}
\label{CS2EqFb}
\frac{F_b}{12}=& {\left(F_{\Phi}\right)}_{1,2,5,6}={\left(F_{\Phi}\right)}_{1,2,7,8}={\left(F_{\Phi}\right)}_{1,2,9,10} \nonumber\\
=&{\left(F_{\Phi}\right)}_{3,4,5,6}={\left(F_{\Phi}\right)}_{3,4,7,8}={\left(F_{\Phi}\right)}_{3,4,9,10}\\
=& \frac{m b}{12}+\frac{\lambda}{3}\left( \frac{ab}{18\sqrt{2}} + \frac{bp}{12\sqrt{6}} \right) - \frac{\eta \sigma \sigma^*}{120}.\nonumber
\end{align}
$F_{\Phi}$ being in the same representation as $\mathbf{\Phi}$, they can be identified as the terms in the representations of $\mathbf{\Phi}_p$, $\mathbf{\Phi}_a$ and $\mathbf{\Phi}_b$ appearing in its branching rules. We also introduce, without specifying anymore all the sets of indices obtained by considering the antisymmetric and self-dual configurations [see Eq.~(\ref{CS2SingletDefinition})],
\begin{equation}
\frac{F_\sigma}{16\sqrt{15}}={\left(F_{\overline{\Sigma}}\right)}_{1,3,5,7,9}= \frac{m_\Sigma \sigma}{16\sqrt{15}} +\frac{\eta  \sigma}{960\sqrt{5}}\left( \sqrt{6}a 
-2\sqrt{3}b+\sqrt{2}p \right)+  \frac{\kappa S\sigma}{16\sqrt{15}}.
\end{equation}
Finally, we have
\begin{equation}
F_S = \kappa (\sigma\sigma^* - M^2 ).
\end{equation}

These definitions permit us to write the $F$-term scalar potential in a simpler form,
\begin{equation}
\label{CS2EqPotWithFterms}
\begin{split}
V & = V_\Phi + V_\Sigma + V_{\overline{\Sigma}} + V_S \\
&= {F_{\Phi}}{F_{\Phi}}^\dagger + {F_{\Sigma}}{F_{\Sigma}}^\dagger + {F_{\overline{\Sigma}}}{F_{\overline{\Sigma}}}^\dagger + F_S F_S^*\\
&= \left| F_p\right|^2+\left| F_a\right|^2+\left| F_b\right|^2 + 2 \left| F_\sigma \right|^2 + |F_S|^2.
\end{split}
\end{equation}
This formulation of the potential is useful since it shows explicitly the sum of positive terms. So, when doing only a static study of the problem as done in Sec.~\ref{CS2SSB&HI}, \emph{i.e.} when not comparing the different terms, it is possible to work with these few simplified $F$-terms only, as often done in the literature. Note that the $F$-terms can indeed be recovered from the usual definition 
\begin{equation}
F_a = \frac{\partial W}{\partial a},
\end{equation}
and so on. However, the simple form of the potential given in Eq.(\ref{CS2EqPotWithFterms}) is permitted only because we chose normalized fields. 

Before going on, let us mention that other papers use different conventions when defining the superpotential and the kinetic part of the Lagrangian, as well as the singlet configurations in tensor formulation (which can be not normalized). The normalization choice we use is useful to properly recover the abelian Higgs model in a given limit discussed in Sec.~\ref{CS2PartOdGToyModel}. We give in Sec.~\ref{CS2AppendixAlternativeFormulation} the link between the expressions of the present paper, and the expressions found in \cite{Aulakh:2003kg,Bajc:2004xe,Cacciapaglia:2013tga}. All the results obtained are indeed compatible with the previous works on the subject.

\section{Abelian cosmic strings}
\label{CS2PartCosmicStrings}
\subsection{Introduction, cosmic strings studied}
\label{CS2PartIntroCosmicStrings}

We now turn to the study of the cosmic strings which form at the second step of SSB which ends hybrid inflation, at $E\sim \left(10^{15}-10^{16}\right)$GeV (see Sec.~\ref{CS2SSB&HI}). As we saw from a cosmological study, the two possible SSBs at this step are from  $G'_1=3_C 2_L 2_R 1_{B-L}$ or $G'_2=3_C 2_L 1_R 1_{B-L}$ to $G_{\text{SM}}\times\mathbb{Z}_2$. In both cases, only cosmic strings form at this stage.
As detailed in \cite{Allys:2015yda}, these strings cannot connect with monopoles. From now on, we use a set of cylindrical coordinates $(r, \theta, z, t)$ based on the location of the string, and taken to be locally aligned along the $z$-axis ar $r=0$. We also focus on strings with fields which are functions only of $r$ and $\theta$, \emph{i.e.} Nambu-Goto strings. 

In the case of $G'_2 \rightarrow G_{\text{SM}}\times\mathbb{Z}_2$, only abelian strings associated with the generator of $1_{B-L}$ can form. But in the other case, $G'_1 \rightarrow G_{\text{SM}}\times\mathbb{Z}_2$, other non abelian-strings could also form, see e.g. Refs.~\cite{Aryal:1987sn,Ma:1992ky,Davis:1996sp}. We will focus in both cases on the abelian strings which could form, associated with the generator of $1_{B-L}$. As explained in Sec.~\ref{CS2PartSingletDecomposition}, only the non-zero VEVs associated to $\mathbf{\Sigma}$ and $\mathbf{\overline{\Sigma}}$ are charged under this abelian group. As these fields are also the fields which are in conjugate representations and coupled with the inflaton in the superpotential, these strings are called single field strings following \cite{Allys:2015yda}.


\subsection{Ansatz and equation of motion}
\label{CS2Ansatz}

In order to have unified notation with \cite{Allys:2015yda}, we call U$(1)_{\text{str}}=1_{\text{str}}$ the abelian symmetry related with the cosmic string ($1_{\scriptscriptstyle{(B-L)}}$ here), and $\tau_{\text{str}}$ the associated generator. As a first part of the ansatz, we assume that all the fields which are not singlet under the SM take an identically vanishing value, since it verifies their equations of motion, as discussed in Sec.~\ref{CS2PartSMSinglet}. Then, we also assume that the only gauge field which does not vanish is the one associated with this generator $\tau_{\text{str}}$, which forms the string \cite{Aryal:1987sn,Ma:1992ky,Peter:1992dw,Hindmarsh:1994re,Allys:2015yda}. In order to simplify the notation, we normalize the charges associated to $1_{\text{str}}=1_{B-L}$ to have $q_\Sigma=1$ and $q_{\overline{\Sigma}}=-1$. Thus, the kinetic term, yields [using Eq.~(\ref{CS2Orthonormalization})]
\begin{equation}
K= - 2\left|\left(\nabla_\mu-ig A_\mu^{\text{str}}\right)\sigma\right|^{2}
 - \left|\left(\nabla_\mu p\right)\right|^{2}  - \left|\left(\nabla_\mu a\right)\right|^{2} 
 - \left|\left(\nabla_\mu b\right)\right|^{2} - \left|\left(\nabla_\mu S\right)\right|^{2} 
  - \frac{\text{Tr} \left({\tau_{\text{str}}}^2 \right) }{4}F_{\mu \nu}^{\text{str}} F^{\mu \nu \,\text{str}}.
\end{equation}
The potential written in terms of the complex functions describing the dynamic of the singlets of the SM can be found in Sec.~\ref{CS2PartW&VSinglet}.

The complete form of the ansatz is \cite{Aryal:1987sn,Ma:1992ky,Peter:1992dw,Hindmarsh:1994re,Allys:2015yda}
\begin{equation}
\left\{
\begin{array}{l}
p=p(r),\\
a=a(r),\\
b=b(r),\\
\sigma=f(r)e^{i n \theta},\\
\displaystyle{A_\mu=A_{\theta}^{\text{str}}(r)\tau^{\text{str}}\delta_\mu^\theta },\\
S=S(r),
\end{array}
\right.
\end{equation}
where the integer $n$ is the winding number. In this ansatz, $f$ and $Q$ are real fields, while $a$, $b$, $p$ and $S$ are complex. This ansatz gives the following equations of motion
\begin{equation}
\begin{array}{l}
\displaystyle{2 \left( f''+\frac{f'}{r}\right) =\frac{fQ^2}{r^2}+\frac{1}{2}\frac{\partial V}{\partial f}},\\ 
\displaystyle{p''+\frac{p'}{r}=\frac{\partial V}{\partial p^*}},\\[6pt] 
\displaystyle{a''+\frac{a'}{r}=\frac{\partial V}{\partial a^*}},\\ [6pt] 
\displaystyle{b''+\frac{b'}{r}=\frac{\partial V}{\partial b^*}},\\ [6pt] 
 \displaystyle{S''+\frac{S'}{r}=\frac{\partial V}{\partial S^*}},\\
\displaystyle{\text{Tr} \left({\tau_{\text{str}}}^2 \right) \left(Q''-\frac{Q'}{r}\right)=2 g^2 f^2Q},
\end{array}
\end{equation}
where a prime means a derivative with respect to the radial
coordinate $'\equiv \text{d}/\text{d}r$, and where we introduced the field
\begin{equation}
\label{CS2defQ}
Q(r)=n-gA_\theta^{\text{str}}(r),
\end{equation}
which is a real field function of $r$ only. Note that this whole ansatz is the minimal structure one, which is developed in \cite{Allys:2015yda}.

Finally, we can reduce the model to the following effective Lagrangian
\begin{equation}
\label{CS2LagEffDim}
\mathcal{L}_{eff}= - 2{f'}^2 - \frac{\text{Tr} \left({\tau_{\text{str}}}^2 \right) }{g^2}\frac{Q'^2}{2 r^2}- s'^\ast s - p'^\ast p'
  -  a'^\ast a' - b'^\ast b' -\frac{f^2Q^2}{r^2}-V(\sigma,a,b,p,s).
\end{equation}
 
\subsection{Boundary conditions}
\label{CS2PartBoundaryConditions}

Let consider first the boundary conditions at infinity. For the scalar fields, they take the non vanishing VEVs discussed in Sec.~\ref{CS2SSB&HI}, which are a global minimum of the potential and define the SM gauge symmetry. Their values will be given in the Sec.~\ref{CS2PartScaling} in a dimensionless form. For the gauge field, we have
\begin{equation}
\lim_{r \to \infty} A_{\theta}^{\text{str}}(r)=\frac{n}{g},
\end{equation}
in order to properly cancel $D_\mu \sigma$ at infinity, which gives in term of the field $Q$
\begin{equation}
\lim_{r \to \infty} Q(r)=0.
\end{equation}

Concerning the values of the fields at the center of the strings, topological arguments and symmetry considerations give  \cite{Hindmarsh:1994re,Kibble:1976sj} 
\begin{equation}
\begin{array}{l}
f(0)=0,\\
Q(0)=n.
\end{array}
\end{equation}
For the other fields, the cylindrical symmetry around the string gives, assuming that they have a non vanishing value at the center of the string,
\begin{equation}
\frac{\text{d} x}{\text{d} r}(0)=0,
\end{equation}
for $x=p,a,b$, and $S$.

At this point, we have the equations of motion and the boundary conditions for the whole field content of the model. So it is completely defined in a mathematical point of view. 

\subsection{Equation of state of the cosmic string}

Without solving the equations of motion, it is possible to obtain the equation of state of the cosmic string. Indeed, we chose an ansatz for the fields where they only depend on $r$ and $\theta$. So, nothing in the configuration we are interested in can depend on internal string world sheet coordinates, here locally $z$ and $t$. The equation of state then gives \cite{Allys:2015yda,Hindmarsh:1994re}
\begin{equation}
U=T,
\end{equation}
where $U$ is the energy per unit length and $T$ is the tension of the string. This equation is the Nambu-Goto equation of state, which is Lorentz-invariant along the world sheet.

Thus, the only macroscopic parameter we will consider in the following is the energy per unit length, defined by
\begin{equation}
\label{CS2DefU}
U=2\pi \int r \text{d}r \mathcal{L}.
\end{equation}

\section{Dimensionless model, perturbative study}
\label{CS2PartScaling&Pert}

\subsection{Dimensionless model}
\label{CS2PartScaling}

To work with a dimensionless model, we introduce the following new variables (denoted by a tilde)
\begin{equation}
\begin{matrix}
\displaystyle{ r = \frac{\tilde{r}}{\kappa M},} & \displaystyle{f = M \tilde{f},} & \displaystyle{S = \frac{m_{\mathbf{\Sigma}}}{\kappa} \tilde{S},}  \\[7pt]
  \displaystyle{a = \frac{m}{\lambda} \tilde{a},} & \displaystyle{b = \frac{m}{\lambda} \tilde{b},} & \displaystyle{p = \frac{m}{\lambda} \tilde{p},}\\[7pt]
  \displaystyle{F_X = \kappa M^2 \tilde{F_X},} & \displaystyle{V = \kappa^2 M^4 \tilde{V},} & \displaystyle{g^2 = \kappa^2 \tilde{g}^2}.
\end{matrix}
\end{equation}
Also, we introduce a set of dimensionless parameters, 
\begin{equation}
\begin{matrix}
\displaystyle{\alpha_1 = \frac{m}{\lambda M},} & \displaystyle{\alpha_2 = \frac{m_{\mathbf{\Sigma}}}{\kappa M},} \\[7pt]
\displaystyle{\alpha_3 = \frac{\eta}{\lambda},} & \displaystyle{\alpha_4 = \frac{\eta}{\kappa }},
\end{matrix}
\end{equation}
which, in addition to $\tilde{g}$, are the free parameters of the theory. As $\alpha_2$ and $g$ are real, while the other $\alpha_i$ are complex, the total parameter space of the model is of dimension 7 (the phases of $\alpha_1$, $\alpha_3$ and $\alpha_4$ are not independent).

Finally, the integrated Lagrangian over the radial coordinates $(r,\theta)$ gives
\begin{align}
\label{CS2LagDensFull}
-\frac{L}{M^{2}}=& 2\pi \int \tilde{r} \text{d}\tilde{r} \left[  2\left(\tilde{f}^\prime\right)^2 +\frac{\text{Tr} \left({\tau_{\text{str}}}^2 \right) }{\tilde{g}^2}\frac{Q'^2}{2 \tilde{r}^2}
+  |\alpha_2|^2 \tilde{S}'^\ast \tilde{S}'
  \right. \nonumber \\ 
& \left. ~~
 + |\alpha_1|^2 \tilde{p}'^\ast \tilde{p}' 
 + |\alpha_1|^2 \tilde{a}'^\ast \tilde{a}' 
 +  |\alpha_1|^2 \tilde{b}'^\ast \tilde{b}'
+\frac{\tilde{f}^2Q^2}{\tilde{r}^2}+\tilde{V}(\tilde{\sigma},\tilde{a},\tilde{b},\tilde{p},\tilde{S}) \right],
\end{align}
with 
\begin{align}
\label{CS2PotentialFull}
\tilde{V}=&{\left|\frac{\alpha_1^2\alpha_4}{\alpha_3}\right|}^2 \Bigg( 
 {\left|\tilde{p}+\frac{\tilde{b}^2}{6\sqrt{6}}+\frac{\alpha_3}{\alpha_1^2}\frac{\tilde{f}^2}{10\sqrt{6}} \right|}^2
+  {\left|\tilde{a}+\frac{\tilde{a}^2}{9\sqrt{2}}+\frac{\tilde{b}^2}{9\sqrt{2}}+\frac{\alpha_3}{\alpha_1^2}\frac{\tilde{f}^2}{10\sqrt{2}} \right|}^2
\nonumber 
+ \left|\tilde{b} 
+\frac{\sqrt{2}\tilde{a}\tilde{b}}{9}+\frac{\tilde{b}\tilde{p}}{3\sqrt{6}}
-\frac{\alpha_3}{\alpha_1^2}\frac{\tilde{f}^2}{10} \right|^2 \Bigg)
\\
& ~~
 + 2 {\left| \alpha_2 \right|}^2
\bigg| \tilde{f}
 +\frac{\alpha_1\alpha_4}{\alpha_2}\frac{\tilde{f}}{10}\left(\frac{\tilde{a}}{\sqrt{2}}-\tilde{b} + \frac{\tilde{p}}{\sqrt{10}} \right) +\tilde{S}\tilde{f}\bigg|^2 + {\left| \tilde{f}^2-1 \right|}^2.
\end{align}

We can also write in a dimensionless form the sets of VEVs before and after the end of inflation, obtained as explained in Sec.~\ref{CS2SSB&HI}. The analytical solutions being cumbersome, we give here only a Taylor expansion. The expansion parameter we consider is
\begin{equation}
\label{CS2DefX}
x=\frac{\alpha_3}{\alpha_1^2}=\frac{\eta\lambda M^2}{m^2},
\end{equation}
which is often small in comparison with unity in such GUT models. Indeed, we expect the second step of symmetry breaking to appear at lower energy than the first step, and we mainly stay in a regime where the coupling constant are at most of order $1$. Note however that even if these expansions are convenient to describe the VEVs when $x\ll 1$, it is necessary to use the complete analytical solutions when it is not the case anymore.

For the first SSB scheme, going through $G'_2=3_C 2_L 2_R 1_{B-L}$, we have
\begin{equation}
\label{CS2VEV0}
\left\{
\begin{array}{l}
\tilde{\sigma}_0=0,\\
\tilde{a}_0=-9 \sqrt{2},\\
\tilde{b}_0=0,\\
\tilde{p}_0=0,\\
\end{array}
\right.
\end{equation}
before the end of inflation, and
\begin{equation}
\label{CS2VEV1}
\left\{
\begin{array}{l}
|\tilde{\sigma}_1|=1\\
\displaystyle{\tilde{a}_1=-9\sqrt{2}+\frac{x}{10\sqrt{2}}}\\
\displaystyle{\tilde{b}_1=-\frac{x}{10}}\\
\displaystyle{\tilde{p}_1=-\frac{x}{10\sqrt{6}}}\\
\displaystyle{\tilde{S}_1=-1+\frac{\alpha_1 \alpha_4}{\alpha_2}\left(\frac{9}{10} -\frac{x}{75} \right)}\\
\end{array}
\right.
\end{equation}
after the end of inflation. We recall that this previous set of VEVs defines the boundary conditions for the fields at infinity, as explained in Sec.~\ref{CS2PartBoundaryConditions}. The set of VEV for the scheme going through $G'_1=3_C 2_L 1_R 1_{B-L}$ is given in the Appendix~\ref{CS2AppendixVEVG2}. 

\subsection{Toy-model limit}
\label{CS2PartOdGToyModel}

We will compare the results obtained to the abelian Higgs model. This toy model contains two scalar fields of opposite charges under a local U(1) gauge symmetry, $\Sigma$ and $\overline{\Sigma}$, and has for Lagrangian\footnote{
Note that if one wants to use a model where the kinetic part of the Lagrangian only contains a term in 
\begin{equation}
K_\Sigma = -(D_{\mu}\Sigma)(D^{\mu}\overline{\Sigma}),
\end{equation}
it it possible to make a link between both these toy-model considering $\tilde{f}=\sqrt{2} f$, $\tilde{\kappa}=\kappa/2$ and $\tilde{M} = \sqrt{2}M$, labeling with a tilde the expressions to use in the model with one single kinetic term. Indeed, the factor $2$ in the kinetic term will vanish, and the superpotential will stay identical.} \cite{Davis:1997bs,Aryal:1987sn,Peter:1992dw,Hindmarsh:1994re}
\begin{equation}
\label{CS2LagToyModel}
\mathcal{L}=-(D_{\mu}\Sigma)^\dagger(D^{\mu}\Sigma)
-(D_{\mu}\overline{\Sigma})^\dagger(D^{\mu}\overline{\Sigma})
-\frac{1}{4} F_{\mu\nu}F^{\mu\nu}-\kappa^2 \left| \Sigma \overline{\Sigma}-M^2 \right| ^2.
\end{equation}
In order to describe it in a SUSY formalism, we have to introduce another field $S$, uncharged under the local symmetry, and use for the superpotential 
\begin{equation}
W=\kappa S \left( \overline{\Sigma} \Sigma -M^2\right).
\end{equation}
This yields the Lagrangian of Eq.~(\ref{CS2LagToyModel}) when taking an ansatz where $S$ identically vanishes. In this toy model, the characteristic radii are $(\kappa M)^{-1}$ for $\Phi$ and $M^{-1}$ for the gauge field (see Appendix \ref{CS2AppendixRCharac}).

This toy model can also be recovered from our realistic model, taking the limit $\eta \rightarrow 0$, and an ansatz where $\tilde{S}=-1$, and where the fields $a$, $b$ and $p$ identically take the value they have at infinity. Indeed, the fields associated to $\mathbf{\Phi}$ fully decouple to the string, $V_\Sigma$ and $V_{\overline{\Sigma}}$ vanish due to value of $\tilde{S}$ (see e.g. Eq.~(\ref{CS2PotentialFull}), with $\alpha_4=0$), and $V_S$ reduces to the potential term of Eq.~(\ref{CS2LagToyModel}). Note that we properly recover the toy-model case in this limit due to the normalization choice of Eq.~(\ref{CS2SingletDefinition}). 

\subsection{Perturbative study}
\label{CS2PartPerturbativeStudy}

We now perform a perturbative study of the condensation of the field $\mathbf{\Phi}$ in the string, in a certain range of parameter, following Ref.~\cite{Allys:2015yda}. For this purpose, considering the modifications of the fields $\tilde{a}$, $\tilde{b}$ and $\tilde{p}$ in a perturbative way when $\tilde{f}$ goes from $0$ to $1$, we obtain the characteristic scale of variation, e.g. for $\tilde{a}$, of [see Eq.~(\ref{CS2PotentialFull})]\footnote{The small parameter we introduced in Eq.~(\ref{CS2DefX}) in order to describe with a Taylor expansion the set of VEV at infinity appears in this characteristic scale of perturbation. This result is somehow natural, since we consider in both cases the static configurations which minimize the potential taking into account the constraint $\tilde{f}=0$ or $\tilde{f}=1$.}
\begin{equation}
\label{CS2DefXPert}
\delta \tilde{a}_0 =\frac{\lambda \eta M^2}{ 10 \sqrt{2} m^2}=\frac{x}{10\sqrt{2}}.
\end{equation}
As we consider models where the end of inflation appears at a lower scale than the GUT scale, and with coupling constant often smaller than unity, this characteristic scale of variations is in most cases smaller than the scale of variation of $\tilde{f}$, which is $1$.  Finally, the ratio between the characteristic variation of $\tilde{a}$ and its dominant contribution at infinity $\tilde{a_0}$ gives [see Eqs.~(\ref{CS2VEV0}) and~(\ref{CS2VEV1})]
\begin{equation}
\frac{\delta \tilde{a}_0}{\tilde{a}_0} = \frac{\lambda \eta M^2}{ 180 m^2}=\frac{x}{180},
\end{equation}
which also legitimates the perturbation study.
This result is in accord with Ref.~\cite{Allys:2015yda}, which estimated this term to be of order $x/N$, with $N$ the characteristic dimension of the representations used. 

However, we could introduce a more precise estimate for the condensation of the fields coming from $\mathbf{\Phi}$ into the core of the string. Indeed, the value $\delta \tilde{a}_0$ we computed describes the differences between the configurations which minimize the potential in the center of the string and at infinity. But the actual value of this field $\tilde{a}$ results in a competition between the kinetic and the potential terms. In order to estimate this value, we can approximate at a linear order the contribution of this field to the Lagrangian close to the center of the string to 
\begin{equation}
\label{CS2LagPert}
|\alpha_1|^2 \mathcal{L}_{\text{pert}} \simeq \left( \frac{\text{d} \delta\tilde{a}}{\text{d} \tilde{r}}\right)^2 + \left( \frac{m}{\kappa M } \right)^2 \left(\delta \tilde{a} -\delta\tilde{a}_0 \right)^2.
\end{equation}
Then, when $m/(\kappa M) \gg 1$, we obtain that $\delta \tilde{a} \simeq \delta \tilde{a}_0$. On the other hand, when $m/(\kappa M) \ll 1$, we obtain at dominant order that 
\begin{equation}
\delta \tilde{a}\simeq  \left( \frac{m}{\kappa M } \right) \delta\tilde{a}_0.
\end{equation}
To compute the kinetic term, we assume that the characteristic radius of the fields $a$, $b$ and $p$ is the same as the characteristic radius of $f$, \emph{i.e.} $(\kappa M)^{-1}$, since these fields have a direct coupling in the Lagrangian with $f$ only, and not $Q$.

When the perturbative study is possible, we can now estimate the variation of the energy per unit length of the string due to the condensation of the additional fields into the string. For this purpose, it is sufficient to consider the kinetic contribution of the fields condensing in the core of the string, \emph{i.e.} $\tilde{a}$, $\tilde{b}$ and $\tilde{p}$. Indeed, without this term, the potential would only play the role of a Lagrange multiplier for these fields, and it would not add any contribution to the energy of the string. An additional way to check this assumption is to verify that the kinetic and potential contributions of these fields to the Lagrangian density are similar.

These assumptions finally give a characteristic modification to the Lagrangian density of order 
\begin{equation}
\label{CS2EqLArrach}
\delta \mathcal{L} \simeq |\alpha_1^{2}| \left( | \delta \tilde{a}|^2 + |\delta \tilde{b} |^2 +|\delta \tilde{p} |^2  \right),
\end{equation}
since in the Eq.~(\ref{CS2LagDensFull}), the dimensionless radius used is $\tilde{r}$, in units of $(\kappa M)^{-1}$.
It yields, considering $U_0 \simeq M^2$ the energy per unit length of the toy-model string, and $\delta U$ the modification of the energy per unit length of the string due to the condensation of $\mathbf{\Phi}$ in the core of the string, 
\begin{equation}
\frac{\delta U}{U_0} \simeq \int \tilde{r} \text{d}\tilde{r} ~ \delta \mathcal{L} \simeq \delta \mathcal{L},
\end{equation}
which gives 
\begin{equation}
\label{CS2PredictionUKappaPetit}
\frac{\delta U}{U_0} \simeq \frac{\eta^2 M^2}{60 m^2},
\end{equation}
when $m/(\kappa M) \gg 1$, and 
\begin{equation}
\label{CS2PredictionUKappaGrand}
\frac{\delta U}{U_0} \simeq \frac{\eta^2 }{60 \kappa^2},
\end{equation}
when $m/(\kappa M) \ll 1$.
These evaluations of the modification of the energy per unit length of the string due to its realistic structure are relevant only when the perturbative approach is verified, \emph{i.e.} when $x \ll 1$, and also when $\delta U / U_0$ does not approach unity. 

In Ref.~\cite{Allys:2015yda}, an estimate of the maximal modification of the energy per unit length from standard toy models was computed, considering the contribution from the scalar potential to the energy per unit length due to the condensation of the additional field in the core. This maximal estimate of $\delta U / U_0 \simeq \eta^2 /(N \kappa^2)$, with $N$ the characteristic dimension of the representations, is compatible with the results of Eqs.~(\ref{CS2PredictionUKappaPetit}) and~(\ref{CS2PredictionUKappaGrand}). The two results are very close in the second case, since as the additional fields barely condensate in the core of the string, the entire potential contribution used in Ref.~\cite{Allys:2015yda} has to be taken into account.

Note that we left aside at this point the contribution of the inflaton field $S$. It is possible to check that this contribution is of same order or lower than the contributions of the fields $a$, $b$ and $p$. On the other hand, it can be understood by the fact that $S$ has no characteristic scale, which is necessary for it in order to play the role of the inflaton.

\section{Numerical solution}
\label{CS2PartNumericalSolution}

\subsection{Implementation in a real case}
\label{CS2PartNumericalImplementation}

In order to simplify the model, we assume in the following that all the parameters, including $\lambda$ and $\eta$, are real. The numerical solution then reduces to a parameter space of $5$ dimensions, described e.g. by the $\alpha_i$ and $g$  which are real. Given this assumption, the boundary conditions for the fields at infinity given in Eq.~(\ref{CS2VEV1}) are also real, since $x$ is real\footnote{When considering the analytical solutions, this result is valid only until $x\sim 22$.}.

With these real parameters, and since the potential is real, all the imaginary parts of the fields must appear at least in a quadratic form. It implies that the configuration where all the imaginary parts of the fields take an identically vanishing value is solution of the equations of motion. This solution is also compatible with the boundary conditions. We thus consider this ansatz from now on, and simplify the fields to real ones.

Moreover, the study of the minima of the potential with the constraint $\sigma=0$ has already been done since it is the configurations for the non vanishing VEVs before the end of inflation. Now, as this configuration is also real, it is compatible with the previous reality assumption done. Indeed, as discussed in \cite{Witten:1984eb}, the solutions for the fields in the core of the string result in a competition between the kinetic and the potential terms, and we expect fields to take values between the boundary conditions at infinity and the configuration which minimizes the potential in the center of the string.

\begin{figure}[t]
\begin{center}
\includegraphics[scale=0.4]{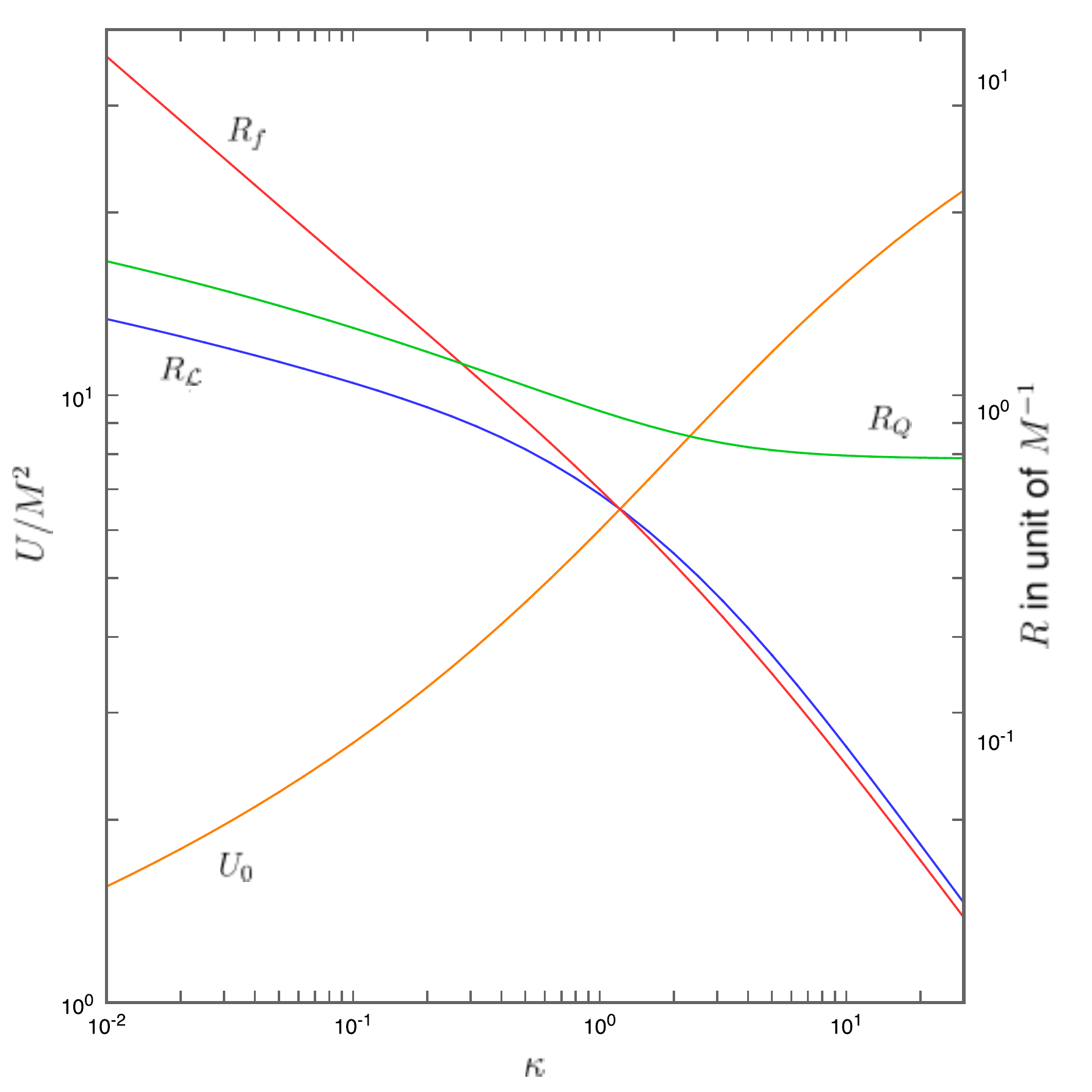}
\end{center}
 \caption{Energy per unit length $U$ and characteristic radius of the string ($R_{\mathcal{L}}$) and of the fields $f$ and $Q$ (resp. $R_f$ and $R_Q$) for the toy model limit ($\eta=0$). $U$ is in unit of $M^2$ and the radiuses in units of $M^{-1}$. }
 \label{CS2Graph_eta=0}
\end{figure}

We suppose that this assumption will not change the results more than a small numerical factor. On the one hand, if the coupling constants $\lambda$ and $\eta$ were complex, there is no reason for their real and imaginary parts to be slightly different. Then, even if we were taking into account the imaginary part of the fields (using complex coupling constants), they would have similar equations of motion than their associated real part, and so would have comparable contributions.

In order to compute numerical solutions, we use a successively over-relaxed method to solve the equations of motion, after writing the whole model on a finite lattice (see e.g. Ref.~\cite{Adler:1983zh}). For this purpose, we convert the integral to a finite range one, introducing a variable $\rho = \tan r$. The numerical solution solves the equation of motion by minimizing the Lagrangian, which reduces to an algebraic function of the fields after being written on the lattice. For this purpose, we use successive Newton iterations, introducing an over-relaxation parameter $\omega$. For example, computing the root of an equation $f(x)=0$, the value of $x$ at the iteration $n+1$ thus gives
\begin{equation}
x^{n+1} = x^n - \omega \frac{f(x^n)}{f^\prime (x^n)}.
\end{equation}
This whole method is called the successively over-relaxed Gauss-Seidel iteration. For this kind of problem, keeping $0<\omega<2$ provides that the Lagrangian monotonously decreases at each step, the convergence being exponential for values of $\omega$ close to $2$. The precision of the result can be evaluated and is of the order of the square root of the modification of the Lagrangian in one step of over-relaxation at the end of the implementation, around $10^{-8}$ in our case. Also, we obtain the same numerical values with an accuracy better than $10^{-8}$ when either imposing the values of the fields at infinity or only a vanishing condition for the derivative of the fields. The lattices used contain $2000$ nodes. A scale factor on the radius is used in order to adapt the characteristic size of this lattice to the one of the string.

\begin{figure}[t]
\begin{center}
\includegraphics[scale=1.3]{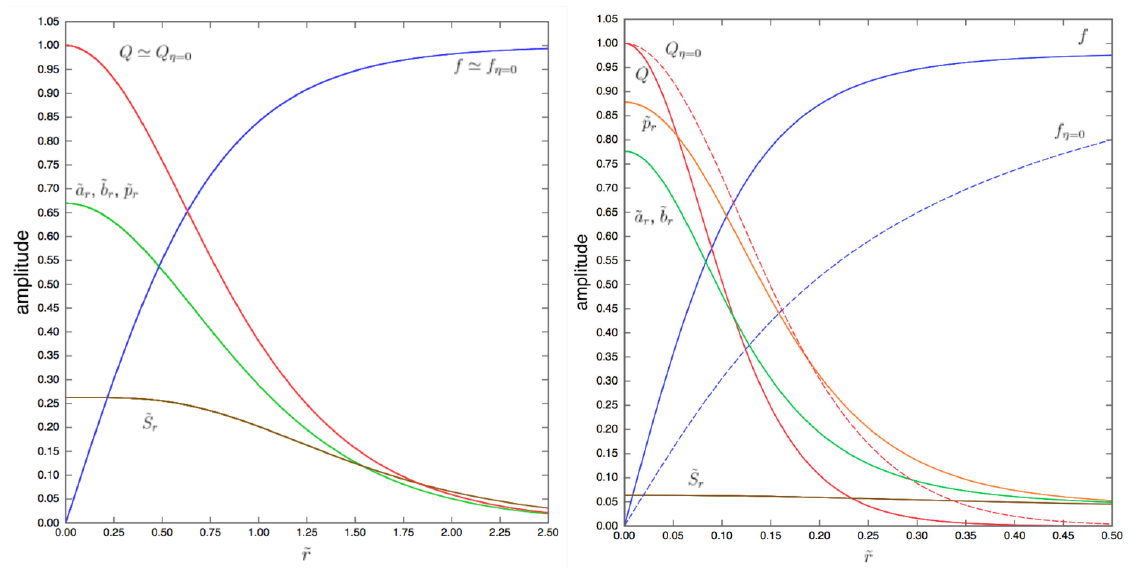}
\end{center}
 \caption{Left: Structure of the cosmic string for $m/M=2$ and $\kappa=\lambda=\eta=1$. The values of $f$ and $Q$ cannot be distinguished between the case $\eta=1$ and $\eta=0$. The curves of $\tilde{a}_r$, $\tilde{b}_r$ and $\tilde{p}_r$ cannot either be distinguished. Right: Structure of the cosmic string for $m/M=2$, $\kappa=0.1$ $\lambda=1$ and $\eta=10$. In dashed lines are the values of $f$ and $Q$ in the toy model limit $\eta=0$. The curves of $\tilde{a}_r$ and $\tilde{b}_r$ cannot be distinguished.}
 \label{CS2Graph_Structure1}
 \label{CS2Graph_Structure2}
\end{figure}

\subsection{Range of parameters, high-coupling limit}
\label{CS2PartRangeParameters}

Let us now discuss the range of parameter chosen for the numerical solution. This part of the investigation permits to test the range of parameters for which the perturbative expansion is not valid, \emph{i.e.} a high coupling limit for the additional fields condensing in the string. This limit is mainly achieved for large $M/m$ and $\eta$ and small $\kappa$, which we will consider.

The two masses $m$ and $m_{\Sigma}$ are set to be equal, presumably around $E_{\text{GUT}}\sim 10^{16}$GeV, and the energy scale $M$, characteristic of the end of inflation, takes values between $m$ and $m/20$. It is not possible to go to values of M/m larger than 1, since it is not compatible with the cosmological evolution we assumed (\emph{i.e.} the order of the phase transition). We consider values of $\eta$ up to 10, which is already a high coupling in what concerns the GUT sector. Following the discussions of Ref.~\cite{Jeannerot:1995yn}, we take for $\kappa$ values between $0.01$ and $30$. The upper limit taken for $\lambda$ is one, but as it has a very small impact on the string, we leave it aside in most of the results presented here. The limit where the coupling constants and the mass ratio go to zero are well defined and described most in the case by the perturbative expansion, as discussed below. We considered values of coupling constants $\lambda$ and $\eta$ down to $10^{-2}$. For all the solutions, we take $g=1$, and a winding number unity. We also take $\text{Tr}( {\tau_{\text{str}}}^2)=2/5$ \cite{Ma:1992ky}.

Around 2000 different sets of parameters have been examined in the whole range discussed above.

\subsection{Toy model limit}
\label{CS2PartDefRadius}

To describe the structure of the string, we define different characteristic radiuses, related to the string itself, or to a field. When normalizing a field in order for it to have the value of $1$ in the center of the string, and $0$ at infinity, the characteristic radius of this field verifies $\phi(r_\phi)=0.40$. The characteristic radius of the string verifies the same property for the Lagrangian density $\mathcal{L}$.

The microscopic structure of the toy model string, \emph{i.e.} the fields as functions of the radial coordinate $\tilde{r}$, can be found in Fig. \ref{CS2Graph_Structure1}. Still for the toy-model case, we give in Fig.~\ref{CS2Graph_eta=0} the value of the energy per unit length in unit of $M^2$, as well as the characteristic radius in units of $M^{-1}$ of $f$, $Q$ and of the Lagrangian density of the string, both as functions of $\kappa$. We properly recover the results of Sec.~\ref{CS2AppendixRCharac}. The energy per unit length of this toy-model string verifies $U \simeq (1-20)M^2$, a common result found in the literature \cite{Kibble:1982ae,Ma:1992ky,Davis:1996sp,Davis:1997bs,Ferreira:2002mg,Morris:1997ua,Davis:1997ny,Adler:1983zh,Peter:1992dw}. From now on, we will compare the results obtained in the realistic implementation of the cosmic string to these particular numerical values, taking the associated toy-model for which all the parameters are the same but $\eta$ goes to zero.

\subsection{Microscopic structure of the realistic string}

\begin{figure}[t]
\begin{center}
\includegraphics[scale=1.3]{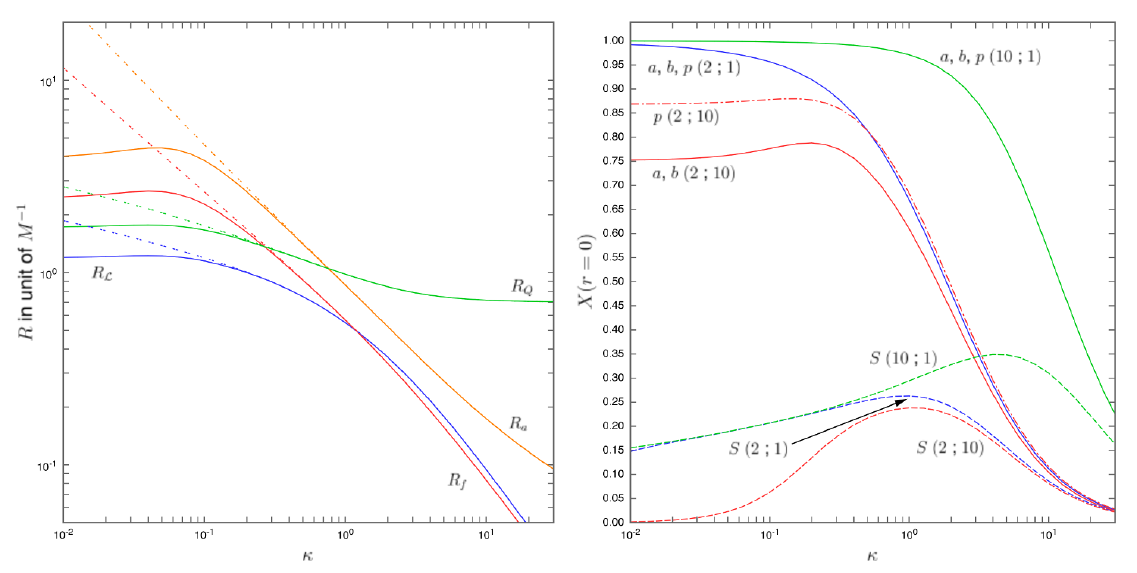}
\end{center}
 \caption{Left: Characteristic radius of the string $R_{\mathcal{L}}$, and of the fields $f$, $Q$ and $a$ ($R_f$, $R_Q$ and $R_a$) as functions of $\kappa$, obtained for $\eta=10$ in solid lines, and for $\eta=0.1$ in dashed lines. In both case, $m/M=10$ and $\lambda=1$. Right: Value at $\tilde{r}=0$ of $\tilde{a}_r$, $\tilde{b}_r$, $\tilde{p}_r$ and $\tilde{S}_r$, as functions of $\kappa$. The different sets of parameters are labeled by $(m/M$ ; $\eta)$, with $m/M=2$ and $\eta=1$ in blue, $m/M=2$ and $\eta=10$ in red, and $m/M=10$ and $\eta=1$ in green, while $\lambda=1$ in all the configurations. For the blue and green curves, $a$, $b$ and $p$ cannot be distinguished and are in solid lines. For the red curves, $a$ and $b$ cannot be distinguished and are in solid lines, while $p$ is in dashed-pointed line. In all cases, $S$ is plotted in dashed lines. }
 \label{CS2Graph_Rayons_EtaFixe}
 \label{CS2Graph_X_0}
\end{figure}

In order to show a graphic representation of the fields $\tilde{a}$, $\tilde{b}$, $\tilde{p}$ and $\tilde{S}$, we normalize them between $0$ and $1$, taking $0$ for their values at infinity, and $1$ for the static configuration which minimizes the potential at the center of the string, \emph{i.e.} for $f=0$. For the inflaton field, we choose the configuration which minimizes the potential in the center of the string for small but not vanishing values of $\tilde{f}$. Indeed, when $\tilde{f}=0$, the inflaton field has a flat potential at tree level (see discussion of Sec.~\ref{CS2SSB&HI}). We denote these normalized fields $\tilde{a}_r$, $\tilde{b}_r$, $\tilde{p}_r$ and $\tilde{S}_r$. This additional normalization for these fields uses the characteristic scale of variation of the problem: the value of these fields in the center of the string tends to $1$ if the potential term is dominant, and to zero if the kinetic term is dominant.


Two microscopic structures of cosmic strings are given in Fig.~\ref{CS2Graph_Structure1}. Note that we plot in dashed lines the values of the fields $f$ and $Q$ in the toy-model limit of these strings. In the first configuration plotted, $f$ and $Q$ are very close to their toy-model values, and cannot be distinguished from them in the graphic. The values of the fields $\tilde{a}_r$, $\tilde{b}_r$ and $\tilde{p}_r$ are also very close, and the associated curves merge together. In the second configuration, the realistic structure of the string causes its radius to lower. In this configuration, the fields $\tilde{a}_r$ and $\tilde{b}_r$ are still very close in values, and cannot be distinguished in the graph. Note that in the second figure, the fields $\tilde{f}$, $\tilde{a}_r$, $\tilde{b}_r$, $\tilde{p}_r$ and $\tilde{S}$ properly converge at high radii. In this range, they recover the behavior they have for the same set of parameters but $\eta=1$.

In both configurations, the value of the perturbation parameter of order $x/10$ [see Eq.~(\ref{CS2DefXPert}) and the subsequent discussion] is respectively $1/40$ and $1/4$, which is in agreement with the results observed, \emph{i.e.} minor modifications from the toy model in Fig.~\ref{CS2Graph_Structure1} (left), and sizable modifications in Fig.~\ref{CS2Graph_Structure2} (right).

The similar behavior of the fields $\tilde{a}_r$, $\tilde{b}_r$ and $\tilde{p}_r$ can also be understood with the perturbative approach. In this limit, only the linear terms in these fields can be considered in $F_a$, $F_b$ and $F_p$, see Eqs.~(\ref{CS2EqFp}-\ref{CS2EqFb}) or (\ref{CS2LagDensFull}), and they all have the same coupling with $\tilde{f}$ only. Thus, these fields have very close values after normalization. In Fig.~\ref{CS2Graph_Structure2}, one can note that the fields $\tilde{a}_r$ and $\tilde{b}_r$ still have similar values, which is not the case anymore for the field $\tilde{p}_r$. This difference can be explained by the fact that $\tilde{a}_r$ and $\tilde{b}_r$ have quadratic terms in the $F$-terms, whereas $\tilde{p}_r$ only has linear terms. Note that in all these cases, one can verify that the kinetic and potential contributions to the Lagrangian density are of the same order.

\begin{figure}[t]
\begin{center}
\includegraphics[scale=1.3]{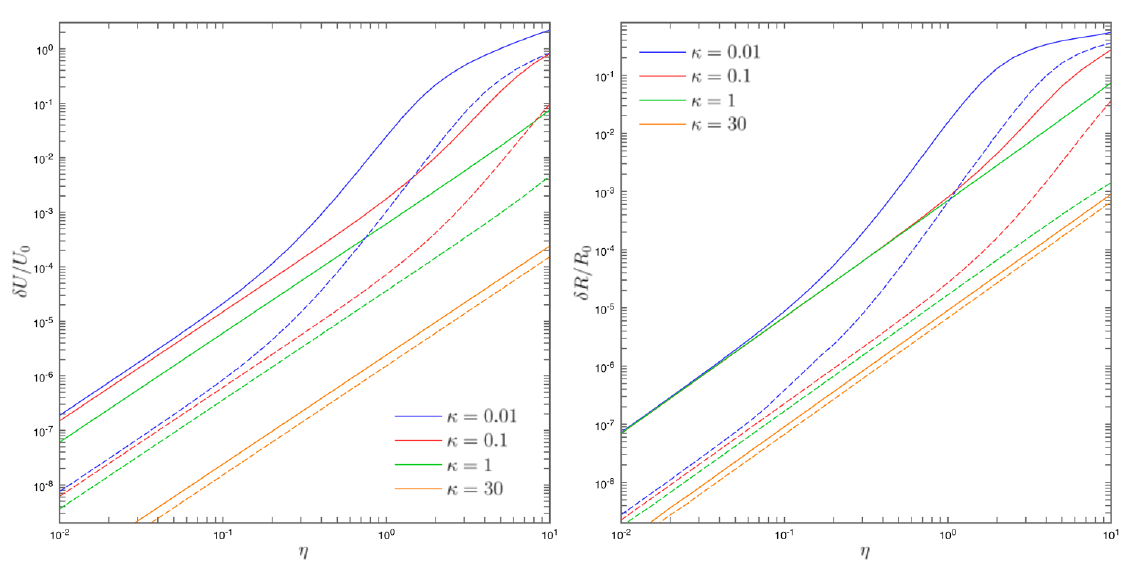}
\end{center}
 \caption{Left: Normalized modifications of the energy per unit length from the toy model values, $\delta U / U_0$ as functions of $\eta$, for $\lambda=1$, and for various values of $\kappa$ and $m/M$. The constant lines are for $m/M=2$, and the dashed lines for $m/M=10$. Right: Normalized modifications of the radius of the string from the toy model values, $\delta R_{\mathcal{L}} / R_{\mathcal{L},0}$ as functions of $\eta$, for $\lambda=1$, and for various values of $\kappa$ and $m/M$. The constant lines are for $m/M=2$, and the dashed lines for $m/M=10$.}
 \label{CS2Graph_GlobU_KappaFixe}
 \label{CS2Graph_GlobR_KappaFixe}
\end{figure}

In order to study the general shape of the microscopic structure, the radii of the string and of different fields are given in Fig.~\ref{CS2Graph_Rayons_EtaFixe}. As before, two different behaviors appear in this figure. For $\eta=0.1$, the modification from the toy-model due to the realistic structure of the string is negligible (see Fig.~\ref{CS2Graph_eta=0}). For $\eta=10$, important modifications appear for small values of $\kappa$. In this last limit, the perturbative approach considered before is not valid anymore. We also see that the coupling of the string-forming Higgs fields with other fields of higher energy tightens the radius of the string. 

To describe the condensation of the fields $\tilde{a}$, $\tilde{b}$, $\tilde{p}$ and $\tilde{S}$ in the string, we give in Fig.~\ref{CS2Graph_X_0} the values of $\tilde{a}_r$, $\tilde{b}_r$, $\tilde{p}_r$ and $\tilde{S}_r$ in the center of the string, \emph{i.e.} at $\tilde{r}=0$, for different configurations. When the perturbative study is possible, we recover the behavior discussed in Sec.~\ref{CS2PartPerturbativeStudy} for the fields  $\tilde{a}$, $\tilde{b}$ and $\tilde{p}$, with a constant value around $1$ for $m/(\kappa M) \gg 1$, and a limit in $m/(\kappa M)$ for $m/(\kappa M) \ll1$. Note that a numerical coefficient appears in the asymptotic behavior, coming from the rough estimate done in Eq.~(\ref{CS2EqLArrach}), and of the form 
\begin{equation}
\label{CS2DefBeta}
{a}_r (0) \sim \frac{\beta m}{\kappa M},
\end{equation}
with $\beta$ of order $\simeq 0.5$.

Such a reasoning for the inflaton field $S$ is not as simple, since its kinetic and potential terms in dimensionless forms have similar prefactors in the Lagrangian, see Eqs.~(\ref{CS2LagDensFull}) and~(\ref{CS2PotentialFull}). It could explain nevertheless why the inflaton field never fully condensates in the string, \emph{i.e.} with values of $\tilde{S}_r$ close to unity. The inflaton also has an asymptotic behavior in $m/(\kappa M)$ for $m/(\kappa M) \ll1$.

In the previous results, we left aside the study of the microscopic structure as a function of the parameter $\lambda$. Indeed, varying this parameter from $0.01$ to $1$ in the previous configurations only has a minor impact on the model, and barely modifies the graphic results obtained.

\begin{figure}[t]
\begin{center}
\includegraphics[scale=1.3]{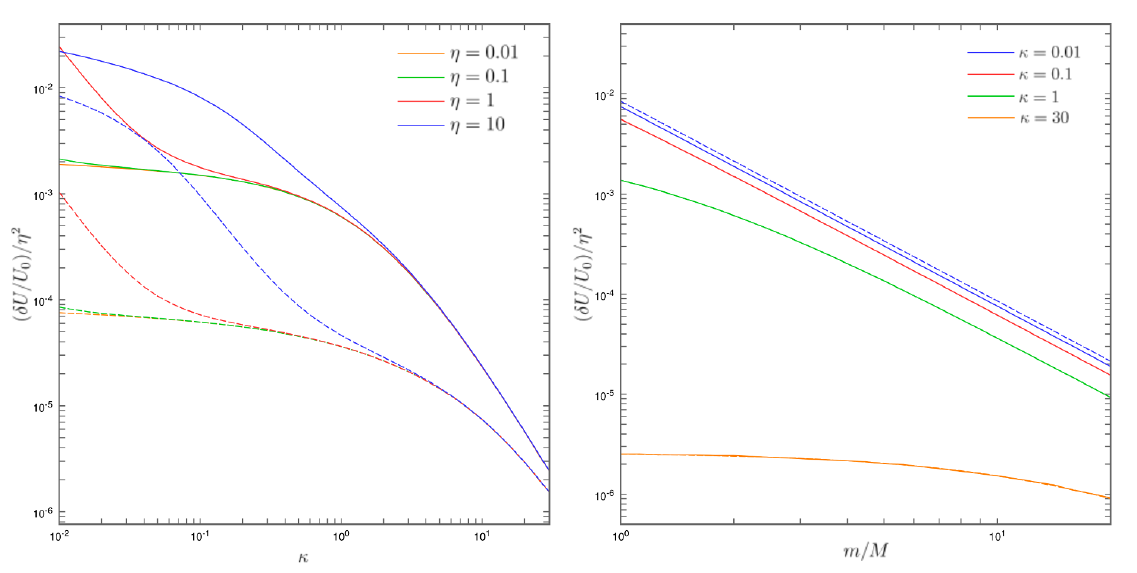}
\end{center}
 \caption{Left: Normalized modification of the energy per unit length from the toy-model values $\delta U / U_0$, divided by $\eta^2$, as functions of $\kappa$, for various values of $\eta$ and $m/M$, and for $\lambda=1$. The constant lines are for $m/M=2$, and the dashed lines for $m/M=10$. Right: Normalized modifications of the energy per unit length from the toy-model values, $\delta U / U_0$, divided by $\eta^2$, as functions of $m/M$, for various values of $\eta$ and $\kappa$, and for $\lambda=1$. The constant lines are for $\eta=0.01$, and the dashed lines for $\eta=0.1$. These curves can be distinguished for $\kappa=0.01$ only.}
 \label{CS2Graph_GlobU_EtaFixe_Eta2}
 \label{CS2Graph_GlobU_xVariable_Eta2}
\end{figure}

\subsection{Macroscopic properties of the realistic string}

Let us turn now to the modifications of the energy per unit length and the radius of the strings, as functions of the different GUT parameters. Fig.~\ref{CS2Graph_GlobU_KappaFixe} shows the modifications of the energy per unit length from the toy-model limit, \emph{i.e.} $(U-U_0)/U_0$ as a function of $\eta$ where $U_0$ is the energy of the toy-model associated to the same GUT parameters (described in Sec.~\ref{CS2PartOdGToyModel}) which can be found in Sec.~\ref{CS2PartDefRadius}. Similarly, Fig.~\ref{CS2Graph_GlobR_KappaFixe} shows the evolution of $(R_{\mathcal{L}}-R_{\mathcal{L},0})/R_{\mathcal{L},0}$, as a function of $\eta$, where $R_{\mathcal{L},0}$ is defined similarly to $U_0$ and can be found in Sec.~\ref{CS2PartDefRadius}.
The results obtained in Sec.~\ref{CS2PartPerturbativeStudy} are verified, with a behavior proportional to $\eta^2$ when the perturbative study is valid. 

In addition, we see that the modifications of the radius of the string are close to the modifications of the energy per unit length. This result is verified for the whole range of parameters studied here. It is not particularly relevant to discuss the small differences between these both curves, since there is some arbitrariness in the definition of the radii. Also, an inflection point appears in both these figures when the radius of the string is modified more than a few percent. At this point, the decrease of the radius of the string seems to partially balance the augmentation of the energy due to the condensation of the additional fields in the core of the string. It can also be explained by the fact that the parameter $m$ becomes important in the description of the string, parameter which is associated with a typical length $\sim m^{-1}$ smaller than $M^{-1}$ the typical length of the toy-model string.

The dependencies in $\kappa$ and $m/M$ of the modification of the energy per unit length are plotted in~\ref{CS2Graph_GlobU_EtaFixe_Eta2} and \ref{CS2Graph_GlobU_xVariable_Eta2} respectively. In both of these figures, the modification of energy per unit length is divided by $\eta^2$. We recover the results obtained in the perturbative study, see Eqs.~(\ref{CS2PredictionUKappaPetit}) and~(\ref{CS2PredictionUKappaGrand}), \emph{i.e.} $\delta U/U_0 \simeq \eta^2 M^2/(60 m^2)$ for large $m/(\kappa M)$, and $\delta U/U_0 \simeq \eta^2/(60 \kappa^2)$ for small $m/(\kappa M)$. Non perturbative sets of parameters are also presented in Fig.~\ref{CS2Graph_GlobU_EtaFixe_Eta2}. 

Finally, it is possible to evaluate the numerical parameter which appears in Eqs.~(\ref{CS2PredictionUKappaPetit}) and~(\ref{CS2PredictionUKappaGrand}), and was approximate to $1/60$. For the behavior at small $\kappa$, we obtain values close to $1/120$ with is in good agreement with the perturbative result. In the case of large $\kappa$, we obtain results around $1/500$. However, it is also in good agreement with the perturbative results after taking into account the additional parameter $\beta$ defined in Eq.~(\ref{CS2DefBeta}).

In what concerns the macroscopic properties of the string, we also left aside the variations with respect to the parameter $\lambda$. Indeed, varying this parameter from $0.01$ to $1$ also barely modifies the graphic results given here.

\section{Conclusion and discussion}

In this paper, we performed a complete study of the realistic structure of cosmic strings forming in a given SO(10) SUSY GUT. Writing this GUT with tensorial representations, we showed that it was possible to simplify this study to a few complex functions describing the dynamics of the sub-representations which are singlet under the SM symmetry. We gave a full ansatz for a string in this context, and performed a perturbative study of the model obtained. We then presented numerical solutions of the string structure, and discussed their microscopic and macroscopic properties, which involve a rich phenomenology.

The numerical results showed that the modification of the energy per unit length from standard toy model strings is modified by a factor slightly higher than unity in the high coupling limit, which is already important with regards to CMB constraints. Note that we tried to investigate the largest possible available range of parameters for which the modifications of the macroscopic properties of the strings is sizable, see discussion of Sec.~\ref{CS2PartRangeParameters}. Getting stronger modifications would require more extreme values of the parameters, values which would then be questionable for the reasons discussed above. Whatever, it shows that in this high-coupling limit, the simplest toy-models seem not to be appropriate models to describe the macroscopic properties of the strings, and corrections due to their realistic structure should be taken into account. In addition, the contribution of the inflaton field to the macroscopic characteristics of the strings appears to be negligible. It is an additional indication that including or not the inflaton in the GUT field content has no major impact on the cosmic strings properties.

The perturbative expansion is in very good agreement with numerical solutions in the wide range of parameters where this approach is possible. The precise microscopic structure gives several different criteria to test the perturbative results, and strengthens their relevance. These results show that a perturbative expansion of the realistic structure of cosmic strings around the toy-mode one is a reliable method to study them. It is particularly useful since such studies are often permitted, especially with the wide numerical factors coming from the large dimensional representations used in GUT. It enhances the result that the modifications of the macroscopic properties of so-called single-field strings, \emph{i.e.} with no coupling of the form $\beta \mathbf{\Phi}\mathbf{\Sigma}\overline{\mathbf{\Sigma}}$ between the singlets of the SM in the superpotential, become sizable in a very high coupling limit~\cite{Allys:2015yda}, as it is the case here. Also, the present work shows that when a perturbative study is not possible anymore, the modifications of the structure and properties of the strings can become important. It strengthens the idea that in the case of the so-called many-fields strings, \emph{i.e.} with a coupling of the form $\beta \mathbf{\Phi}\mathbf{\Sigma}\overline{\mathbf{\Sigma}}$~\cite{Allys:2015yda}, a complete study is necessary and should be done in the future, since no perturbative discussion is possible in most of the range of parameters.

In this work, we left aside precise considerations about the stability of the ansatz we used. We should consider this in more details in the future. Other properties of the strings should also be studied in the realistic GUT context considered here, such as the existence of bosonic currents in the core of the string \cite{Witten:1984eb,Peter:1992dw,Peter:1992nz,Peter:1993tm,Morris:1995wd}. Moreover, and as we work in a supersymmetric framework, their superpartner could carry fermionic currents through their zero modes \cite{Witten:1984eb,Jackiw:1981ee,Weinberg:1981eu,Davis:1995kk,Ringeval:2000kz,Peter:2000sw}. The effect of this complete structure to the processes of intercommutation \cite{Shellard:1988ki,Laguna:1990it,Matzner:1988ky,Moriarty:1988qs,Moriarty:1988em,Shellard:1987bv} should also be investigated, keeping in mind that modifications of such properties of the cosmic strings could affect the network evolution, and thus the observational consequences on the CMB \cite{Kibble:1976sj,Kibble:1980mv,Copeland:1991kz,Austin:1993rg,Martins:1996jp,Martins:2000cs}. Finally, the possibility of formation of non-abelian strings could also be studied in this context, and gives some interesting phenomenology\cite{Aryal:1987sn,Ma:1992ky,Davis:1996sp}.


\subsubsection*{Acknowledgment}

I wish to thank P. Peter and M. Sakellariadou for many valuable discussions and suggestions, and also for a critical reading of the manuscript, which permitted important improvements on the present paper. I also thank J. Allys and J.-B. Fouvry for useful advice on numerical solutions and data processing.


\section{Appendix: Intermediate calculations}
\label{CS2AppendixA}
\subsection{Computation of the derivatives}
\label{CS2AppendixDerivativeComputation}

We present here how to take derivative with respect to the fields of the GUT in a tensor formulation. We take as an example the computation of $(\partial \{\mathbf{\Phi}\mathbf{\Sigma}\overline{\mathbf{\Sigma}}\} )/ (\partial \mathbf{\Sigma})$. Computing this derivative, we have to take into account that the different components of the multiplet $\mathbf{\Sigma}$ are not independent. 

Starting from
\begin{equation}
\mathbf{\Phi}\mathbf{\Sigma}\overline{\mathbf{\Sigma}} = \Sigma_{ijklm}\Phi_{ij\alpha\beta}\overline{\Sigma}_{\alpha\beta klm},
\end{equation}
we can use the fact that $\mathbf{\Sigma}$ is self-dual, see Eq. (\ref{CS2self-dual}), to write 
\begin{equation}
\mathbf{\Phi}\mathbf{\Sigma}\overline{\mathbf{\Sigma}} = \frac{1}{2}\left(\Sigma_{ijklm}+\frac{i}{5!}\epsilon_{ijklmabcde}\Sigma_{abcde}\right)\Phi_{ij\alpha\beta}\overline{\Sigma}_{\alpha\beta klm}.
\end{equation}
Now, as $\mathbf{\Sigma}$ is totally antisymmetric, we can write
\begin{equation}
\mathbf{\Phi}\mathbf{\Sigma}\overline{\mathbf{\Sigma}} = \frac{1}{2}\left(\Sigma_{[ijklm]}+\frac{i}{5!}\epsilon_{ijklmabcde}\Sigma_{abcde}\right)\Phi_{ij\alpha\beta}\overline{\Sigma}_{\alpha\beta klm},
\end{equation}
where the antisymmetrization is on all the indices of $\mathbf{\Sigma}$. There is no need to make it appear explicitly in the Levi-Civita symbol since it is already totally antisymmetric. Then, we wrote all the component of $\mathbf{\Sigma}$ which are not independent. It gives after a relabeling of the indices :
\begin{equation}
\mathbf{\Phi}\mathbf{\Sigma}\overline{\mathbf{\Sigma}} = \Sigma_{ijklm}\frac{1}{2}\bigg(\Phi_{[ij|\alpha\beta}\overline{\Sigma}_{\alpha\beta |klm]}
-\frac{i}{5!}\epsilon_{ijklmabcde}\Phi_{ab\alpha\beta}\overline{\Sigma}_{\alpha\beta cde}\bigg).
\end{equation}
The antisymmetrization of the first product in the parentheses is on the indices $(i,j,k,l,m)$, and it is defined by
\begin{multline}
\label{CS2antisym1}
\Phi_{[ij|\alpha\beta}\overline{\Sigma}_{\alpha\beta |klm]}=\frac{1}{10}\big( 
\Phi_{ij\alpha\beta}\overline{\Sigma}_{\alpha\beta klm} 
-\Phi_{ik\alpha\beta}\overline{\Sigma}_{\alpha\beta jlm}
-\Phi_{il\alpha\beta}\overline{\Sigma}_{\alpha\beta kjm}
-\Phi_{im\alpha\beta}\overline{\Sigma}_{\alpha\beta klj}
+\Phi_{jk\alpha\beta}\overline{\Sigma}_{\alpha\beta ilm}\\
+\Phi_{jl\alpha\beta}\overline{\Sigma}_{\alpha\beta kim}
+\Phi_{jm\alpha\beta}\overline{\Sigma}_{\alpha\beta kli}
-\Phi_{kl\alpha\beta}\overline{\Sigma}_{\alpha\beta jim}
-\Phi_{km\alpha\beta}\overline{\Sigma}_{\alpha\beta jli}
+\Phi_{lm\alpha\beta}\overline{\Sigma}_{\alpha\beta jki}
\big).
\end{multline}
Finally, we have 
\begin{equation}
\label{CS2FinalDerivative}
\frac{\partial (\mathbf{\Phi}\mathbf{\Sigma}\overline{\mathbf{\Sigma}})}{\partial \mathbf{\Sigma}}=\frac{1}{2}\bigg(\Phi_{[ij|\alpha\beta}\overline{\Sigma}_{\alpha\beta |klm]}
-\frac{i}{5!}\epsilon_{ijklmabcde}\Phi_{ab\alpha\beta}\overline{\Sigma}_{\alpha\beta cde}\bigg).
\end{equation}
And we can note that this is the term in the $\mathbf{\overline{126}}$ representation in the contraction between the $\mathbf{210}$ and $\mathbf{\overline{126}}$ representation, since this is totally antisymmetric in its five indices, and anti-self-dual (it can be checked explicitly using\footnote{The general expression is 
\begin{equation}
\epsilon_{i_1 \dots i_k~i_{k+1}\dots i_n} \epsilon^{i_1 \dots i_k~j_{k+1}\dots j_n} = k!(n-k)!~\delta_{[ i_{k+1}}{}^{j_{k+1}} \dots \delta_{i_n ]}{}^{j_n}.
\end{equation}
} $\epsilon_{abcdeijklm}\epsilon^{abcdepqrst}=(5!)^2 \delta_{[a}^p  \delta_b^q  \delta_c^r  \delta_d^s  \delta_{e]}^t$).
So, this can be written
\begin{equation}
\frac{\partial (\mathbf{\Phi}\mathbf{\Sigma}\overline{\mathbf{\Sigma}})}{\partial \mathbf{\Sigma}}=(\mathbf{\Phi}\mathbf{\overline{\Sigma}})_{\overline{\Sigma}}.
\end{equation}

In fact, this property can also permit us to calculate this derivative in another manner, by group considerations. Indeed, we considered the singlet term built from the product $\mathbf{\Phi}\mathbf{\Sigma}\overline{\mathbf{\Sigma}}$, so from $\mathbf{210}\times\mathbf{126}\times \mathbf{\overline{126}}$, which can be written $\mathbf{126}\times (\mathbf{210}\times \mathbf{\overline{126}})$.Then, the branching rules for the second term give \cite{Slansky:1981yr}
\begin{equation}
\label{CS2BranchingRules210-126}
\mathbf{210}\times \mathbf{\overline{126}} = \mathbf{10} + \mathbf{120} + \mathbf{\overline{126}} + \mathbf{320} + \cdots.
\end{equation}
But now, the only possibility to have a singlet term from the contraction of Eq.~(\ref{CS2BranchingRules210-126}) with the \textbf{126} representation comes with the product $\mathbf{126}\times\mathbf{\overline{126}}$, the other terms giving a vanishing value.
So it is possible to write this singlet term as
\begin{equation}
{\mathbf{1}}_{\mathbf{210}\times\mathbf{126}\times \mathbf{\overline{126}}} \ni \mathbf{126} \times (\mathbf{210}\times \mathbf{\overline{126}})_{\mathbf{\overline{126}}}
\end{equation}
Now, it is straightforward to take the derivative with respect to the $\mathbf{126}$, since we already simplified the term we have to consider, and it gives $(\mathbf{\Phi}\mathbf{\overline{\Sigma}})_{\overline{\Sigma}}$. We see that we could in fact compute this derivative only by considering the appropriate part in the product $\mathbf{\Phi}\mathbf{\overline{\Sigma}}$, which here have to be in the $\mathbf{\overline{126}}$ representation. So, as this representation is totally antisymmetric and anti-self-dual, it was sufficient to take the antisymmetric anti-self-dual part of this product\footnote{In a similar way we decompose a rank two tensor in its symmetric and anti-symmetric part, we can decompose a tensor in its self-dual and anti-self-dual part.}. Indeed, a tensor which is anti-self-dual or totally symmetric gives a vanishing expression when contracted with a totally antisymmetric or anti-self-dual tensor. This proves in another method the result of Eq.~(\ref{CS2FinalDerivative}).

Thus, as the singlet term coming from such a product can be written in different manner 
\begin{equation}
\mathbf{1}_{\mathbf{210}\times\mathbf{126}\times \mathbf{\overline{126}}}
=\mathbf{210}\times (\mathbf{126}\times \mathbf{\overline{126}})_{210}
=\mathbf{\overline{126}} 
\times (\mathbf{210}\times \mathbf{126})_{126}
=\mathbf{126} \times (\mathbf{210}\times \mathbf{\overline{126}})_{\overline{126}},
\end{equation}
we can use this method to compute all the derivative we want in a direct way.

\subsection{Derivative terms and associated notations}
\label{CS2AppendixDerivatives}
 
Using the methods explained in the former section, we obtain, in addition to Eq.~(\ref{CS2FinalDerivative})
\begin{equation}
\frac{\partial (\mathbf{\Phi}\mathbf{\Sigma}\mathbf{\overline{\Sigma}})}{\partial \mathbf{\overline{\Sigma}}}=
(\mathbf{\Phi}\mathbf{\Sigma})_{\Sigma}=
\frac{1}{2}\bigg(\Phi_{[ij|\alpha\beta}\Sigma_{\alpha\beta |klm]}
+\frac{i}{5!}\epsilon_{ijklmabcde}\Phi_{ab\alpha\beta}\Sigma_{\alpha\beta cde}\bigg),
\end{equation}
\begin{equation}
\frac{\partial (\mathbf{\Phi}\mathbf{\Sigma}\overline{\mathbf{\Sigma}})}{\partial \mathbf{\Phi}}=
(\mathbf{\Sigma}\mathbf{\overline{\Sigma}})_\Phi=
\Sigma_{[ij|abc}\overline{\Sigma}_{|kl]abc},
\end{equation}
and
\begin{equation}
\frac{\partial (\mathbf{\Phi}\mathbf{\Phi}\mathbf{\Phi})}{\partial \mathbf{\Phi}}=(\mathbf{\Phi}\mathbf{\Phi})_{\Phi}=3 \Phi_{[ij|ab}\Phi_{ab|kl]},
\end{equation}
where the antisymmetrization of the different products are defined as in Eq. (\ref{CS2antisym1}), by 
\begin{align}
\Sigma_{[ij|abc}\overline{\Sigma}_{|kl]abc}=&\frac{1}{6} \big(
\Sigma_{ijabc}\overline{\Sigma}_{klabc}
-\Sigma_{ikabc}\overline{\Sigma}_{jlabc}
-\Sigma_{ilabc}\overline{\Sigma}_{kjabc}\nonumber  \\
& ~
+\Sigma_{jkabc}\overline{\Sigma}_{ilabc}
+\Sigma_{jlabc}\overline{\Sigma}_{kiabc}
-\Sigma_{klabc}\overline{\Sigma}_{jiabc}\big),
\end{align}
and
\begin{equation}
\Phi_{[ij|ab}\Phi_{ab|kl]}=\frac{1}{3}\big(
\Phi_{ijab}\Phi_{abkl}-\Phi_{ikab}\Phi_{abjl}
+\Phi_{ilab}\Phi_{abjk}\big).
\end{equation}


\subsection{Selection rules}
\label{CS2AppendixSelectionRules}

For the quadratic contractions, we have
\begin{equation}
\langle \mathbf{\Phi} \rangle \langle \mathbf{\Phi} \rangle^\dagger =pp^*+aa^*+bb^*,
\end{equation}
and
\begin{equation}
\langle \mathbf{\Sigma} \rangle \langle \mathbf{\Sigma} \rangle ^\dagger=\langle \mathbf{\Sigma} \rangle \langle \overline{\mathbf{\Sigma}} \rangle   = \sigma \sigma^*,
\end{equation}
which comes from the normalization of the vectors. 

The cubic contractions give
\begin{equation}
\langle \mathbf{\Phi} \mathbf{\Phi} \mathbf{\Phi} \rangle _{\mathbf{1}}=\langle \mathbf{\Phi} \mathbf{\Phi} \rangle _{\Phi} \langle \mathbf{\Phi} \rangle ^{\text{T}} = \frac{1}{9\sqrt{2}}a^3
 + \frac{ab^2}{3\sqrt{2}} + \frac{pb^2}{2\sqrt{6}},
\end{equation}
\begin{equation}
\langle \mathbf{\Phi} \mathbf{\Phi} \rangle _{\Phi} \langle \mathbf{\Phi} \rangle ^\dagger 
= \frac{1}{9\sqrt{2}}a^2a^* + \left[ \frac{1}{6\sqrt{6}}\left(2b^*bp + b^2p^*\right) 
+ \frac{1}{9\sqrt{2}}\left(2b^*ba + b^2a^*\right)\right],
\end{equation}
and
\begin{equation}
\langle \mathbf{\Sigma} \overline{\mathbf{\Sigma}} \mathbf{\Phi} \rangle_{\mathbf{1}}=\langle \mathbf{\Phi} \mathbf{\Sigma} \rangle _{\Sigma} \langle \mathbf{\Sigma} \rangle^\dagger=\langle \mathbf{\Phi} \overline{\mathbf{\Sigma}} \rangle _{\overline{\Sigma}} \langle \overline{\mathbf{\Sigma}} \rangle^\dagger 
=\langle \mathbf{\Phi} \rangle  \langle \mathbf{\Sigma} \overline{\mathbf{\Sigma}} \rangle _\Phi ^\dagger 
= \sigma \sigma^* \left( \frac{1}{10\sqrt{6}} p + \frac{1}{10\sqrt{2}} a - \frac{1}{10} b \right).
\end{equation}

Finally, the quartic contractions give
\begin{equation}
\langle \mathbf{\Phi} \mathbf{\Phi} \rangle_\Phi \langle \mathbf{\Phi} \mathbf{\Phi} \rangle_\Phi^{\text{T}}=
\frac{a^4}{164}+\frac{a^2b^2}{27} + \frac{7b^4}{648}+\frac{2ab^2p}{27\sqrt{3}}+\frac{b^2p^2}{54},
\end{equation}
\begin{align}
\langle \mathbf{\Phi} \mathbf{\Phi} \rangle_\Phi \langle \mathbf{\Phi} \mathbf{\Phi} \rangle _\Phi ^\dagger = &
\frac{(aa^*)^2}{164}  +\frac{7(bb^*)^2 }{648}
+ \frac{1}{162} \left({a^*}^2b^2 + a^2 {b^*}^2+ 4 aa^*bb^*\right) \nonumber \\
&~~ 
+\frac{bb^*pp^*}{54}
+\frac{1}{27\sqrt{3}}\left( abb^*p^*+a^*b^*bp \right),
\end{align}
\begin{align}
\langle \mathbf{\Phi} \overline{\mathbf{\Sigma}} \rangle _{\overline{\Sigma}} \langle \mathbf{\Phi} \overline{\mathbf{\Sigma}} \rangle _{\overline{\Sigma}} ^\dagger
= \langle \mathbf{\Phi} \mathbf{\Sigma} \rangle_\Sigma \langle \mathbf{\Phi} \mathbf{\Sigma} \rangle _\Sigma ^\dagger
= & \frac{1}{600}\sigma\sigma^* \left(\sqrt{3}a-\sqrt{6}b+p\right)\left(\sqrt{3}a-\sqrt{6}b+p\right)^* \nonumber \\
&  =\frac{1}{600}\sigma\sigma^* \left|\sqrt{3}a-\sqrt{6}b+p\right|^2 ,
\end{align}
\begin{equation}
\langle \mathbf{\Phi} \mathbf{\Phi} \rangle_\Phi \langle \mathbf{\Sigma} \overline{\mathbf{\Sigma}} \rangle _\Phi^\dagger  
= \frac{a^2\sigma\sigma^*}{180}+\frac{b^2\sigma\sigma^*}{120}
-\frac{ab\sigma\sigma^*}{45\sqrt{2}}-\frac{bp\sigma\sigma^*}{30\sqrt{6}},
\end{equation}
and
\begin{equation}
\langle \mathbf{\Sigma} \overline{\mathbf{\Sigma}} \rangle _\Phi \langle \mathbf{\Sigma} \overline{\mathbf{\Sigma}} \rangle _\Phi ^\dagger
=\frac{1}{60} (\sigma\sigma^*)^2.
\end{equation}

\subsection{Alternative formulation}     
\label{CS2AppendixAlternativeFormulation}

We do here the correspondence with \cite{Aulakh:2003kg,Bajc:2004xe,Cacciapaglia:2013tga}, with a tilde to denote the alternative notations. These papers take the following definitions :
\begin{equation}
W= \frac{\tilde{m}}{4!}\mathbf{\Phi}^2
+ \frac{\tilde{m_{\mathbf{\Sigma}}}}{5!}\mathbf{\Sigma}\overline{\mathbf{\Sigma}}
+ \frac{\tilde{\lambda}}{4!}\mathbf{\Phi}^3 
+ \frac{\tilde{\eta}}{4!}\mathbf{\Phi}\mathbf{\Sigma}\overline{\mathbf{\Sigma}}
+ \kappa S (\frac{\mathbf{\Sigma}\overline{\mathbf{\Sigma}}}{5!}-M^2),
\end{equation}
and
\begin{equation}
\left\{
\begin{array}{l}
 \tilde{p}= \Phi_{1234} ,\\
\tilde{a}= \Phi_{5678} = \Phi_{5690} = \Phi_{7890} ,\\
\tilde{b}= \Phi_{1256} = \Phi_{1278} = \Phi_{1290} \\
~~ = \Phi_{3456} = \Phi_{3478} = \Phi_{3490} ,\\
\frac{1}{\sqrt{2^5}}\tilde{\sigma} =\Sigma_{1,3,5,7,9}, \\
\frac{1}{\sqrt{2^5}} \tilde{\overline{\sigma}}= \overline{\Sigma}_{1,3,5,7,9},
\end{array}
\right.
\end{equation}
where we did not note all the possible configuration for $\mathbf{\Sigma}$ and $\mathbf{\overline{\Sigma}}$.
The link between the different definitions is
\begin{equation}
\left\{
\begin{array}{l}
\displaystyle{\frac{\tilde{m}}{4!}=\frac{m}{2}},\\[6pt]
\displaystyle{\frac{\tilde{m}_{\Sigma}}{5!}=m_{\Sigma}},\\[6pt]
\displaystyle{\frac{\tilde{\lambda}}{4!}=\frac{\lambda}{3}},
\end{array}
\right.
~~~~~~~~~
\left\{
\begin{array}{l}
\displaystyle{\frac{\tilde{\eta}}{4!}=\eta,}\\[6pt]
\displaystyle{\frac{\tilde{\kappa}}{5!}=\kappa,}\\[6pt]
\sqrt{5!}\tilde{M}=M,
\end{array}
\right.
\end{equation}
and
\begin{equation}
\left\{
\begin{array}{l}
\displaystyle{\tilde{p}=\frac{p}{\sqrt{4!}},}\\[6pt]
\displaystyle{\tilde{a}=\frac{a}{\sqrt{4!3}},}\\[6pt]
\displaystyle{\tilde{b}=\frac{b}{\sqrt{4!6}},}
\end{array}
\right.
~~~~~~~~~
\left\{
\begin{array}{l}
\displaystyle{\tilde{\sigma}=\frac{\sigma}{\sqrt{5!}}},\\[6pt]
\displaystyle{\tilde{\overline{\sigma}}=\frac{\overline{\sigma}}{\sqrt{5!}}.}
\end{array}
\right.
\end{equation}

Thus, expressing them with these particular conventions, we obtain 
\begin{equation}
W=\tilde{m}(\tilde{p}^2+3\tilde{a}^2+6\tilde{b}^2)+2\tilde{\lambda}(\tilde{a}^3+3\tilde{p}\tilde{b}^2+6\tilde{a}\tilde{b}^2)
+\tilde{m}_{\Sigma}\tilde{\sigma}\tilde{\overline{\sigma}}
+ \tilde{\eta}\tilde{\sigma}\overline{\sigma}(\tilde{p}+3\tilde{a}-6\tilde{b}) 
+\tilde{\kappa} s (\tilde{\sigma}\tilde{\overline{\sigma}}- \tilde{M}^2),
\end{equation}
for the superpotential, and
\begin{equation}
\left\{
\begin{array}{l}
\displaystyle{ F_{p}=\frac{1}{2\sqrt{6}} \left( 2\tilde{m}\tilde{p}+6\tilde{\lambda} \tilde{b}^2+\tilde{\eta} \tilde{\sigma} \tilde{\overline{\sigma}}\right)},\\
\displaystyle{ F_{a}=\frac{1}{6\sqrt{2}} \left[ 3\left(2\tilde{m}\tilde{a}+2\tilde{\lambda} (2\tilde{b}^2+\tilde{a}^2)+\tilde{\eta} \tilde{\sigma}\tilde{\overline{\sigma}}\right) \right]},\\
\displaystyle{ F_{b}=\frac{1}{12} \left[ 6\left(2\tilde{m}\tilde{b}+2\tilde{\lambda} \tilde{b}(2\tilde{a}+\tilde{p})-\tilde{\eta} \tilde{\sigma}\tilde{\overline{\sigma}}\right) \right]},\\
\displaystyle{ F_{\sigma}=\frac{1}{2\sqrt{30}}\left[ \tilde{\overline{\sigma}}\left(\tilde{m}_{\Sigma}+\tilde{\eta} (\tilde{p}+3\tilde{a}-6\tilde{b})+\tilde{\kappa} s\right)\right]},\\
\displaystyle{ F_{\overline{\sigma}}=\frac{1}{2\sqrt{30}}\left[\tilde{\sigma}\left(\tilde{m}_{\Sigma}+\tilde{\eta} (\tilde{p}+3\tilde{a}-6\tilde{b})+\tilde{\kappa} s\right)\right]},\\
\displaystyle{ F_{S}=\tilde{\kappa} (\tilde{\sigma} \tilde{\overline{\sigma}}-\tilde{M}^2)},
\end{array}
\right.
\end{equation}
for the $F$-terms. Ref.~\cite{Bajc:2004xe} and \cite{Cacciapaglia:2013tga} also introduce different $F$-terms, which are defined by, e.g. 
\begin{equation}
\tilde{F}_{\tilde{a}}=\frac{\partial{W}}{\partial\tilde{a}}.
\end{equation}
They are related to the $F$-terms by
\begin{equation}
\left\{
\begin{array}{l}
\displaystyle{ F_{p}=\frac{1}{2\sqrt{6}}  \tilde{F}_{\tilde{p}}, }\\[6pt]
\displaystyle{F_{a}=\frac{1}{6\sqrt{2}}\tilde{F}_{\tilde{a}},}\\[6pt]
\displaystyle{F_{b}=\frac{1}{12}\tilde{F}_{\tilde{b}},}
\end{array}
\right.
~~~~~~~~~
\left\{
\begin{array}{l}
\displaystyle{F_{\sigma}=\frac{1}{2\sqrt{30}}\tilde{F}_{\tilde{\sigma}},}\\[6pt]
\displaystyle{F_{\overline{\sigma}}=\frac{1}{2\sqrt{30}}\tilde{F}_{\tilde{\overline{\sigma}}},}\\[6pt]
\displaystyle{F_{S}=\tilde{F}_{\tilde{S}}.}
\end{array}
\right.
\end{equation}
Finally, as functions of these $\tilde{F}$-terms, the potential terms are
\begin{equation}
V_{\Phi}={F_{\Phi}}{F_{\Phi}}^\dagger=\frac{1}{24}\tilde{F}_{\tilde{p}}\tilde{F}_{\tilde{p}}^*+\frac{1}{72}\tilde{F}_{\tilde{a}}\tilde{F}_{\tilde{a}}^*+\frac{1}{144}\tilde{F}_{\tilde{b}}\tilde{F}_{\tilde{b}}^*,
\end{equation}
and 
\begin{equation}
V_{\Sigma}=V_{\overline{\Sigma}}={F_{\Sigma}}{F_{\Sigma}}^\dagger = \frac{1}{120} \tilde{F}_{\tilde{\sigma}}\tilde{F}_{\tilde{\sigma}}^*.
\end{equation}
At this point, we can note that the potential cannot be obtained simply by summing the square of the norms of the $\tilde{F}$ terms, but that some numerical coefficients appear. It is particularly important when comparing the values of these different potential terms, since these coefficients can considerably modify the results obtained.

\subsection{Set of VEV for the scheme of $G'_2$}
\label{CS2AppendixVEVG2}

The values of the non-vanishing VEV for the SSB scheme going through $G'_2=3_C 2_L 1_R 1_{B-L}$ are
\begin{equation}
\left\{
\begin{array}{l}
\tilde{\sigma}_0=0,\\
\tilde{a}_0=-18 \sqrt{2},\\
\tilde{b}_0=\pm 18 \text{i}, \\
\tilde{p}_0= 9 \sqrt{6},\\
\end{array}
\right.
\end{equation}
before the end of inflation, and
\begin{equation}
\left\{
\begin{array}{l}
\tilde{\sigma}_1=|1|,\\
\displaystyle{\tilde{a}_1=-18 \sqrt{2}+\frac{\left(\frac{1}{50} \pm \frac{i}{25}\right) x}{\sqrt{2}}},\\[7pt]
\displaystyle{\tilde{b}_1=\pm 18 i+\left(\frac{3}{100}\pm\frac{i}{100}\right) x},\\[7pt]
\displaystyle{\tilde{p}_1=9 \sqrt{6}-\frac{\left(\frac{1}{25}\pm\frac{9 i}{50}\right) x}{\sqrt{6}}},\\[7pt]
\displaystyle{\tilde{S}_1=-1+\frac{\alpha_1 \alpha_3}{\alpha_2}\left(\frac{4\pm 3 i}{1500}\right) \left[x+(540\pm 270 i)\right]},\\
\end{array}
\right.
\end{equation}
after the end of inflation.

\subsection{Characteristic radii for the toy-model string}
\label{CS2AppendixRCharac}

Using the Lagrangian of Eq. (\ref{CS2LagToyModel}), two characteristic radius appear for this toy-model limit. Indeed, writing the equations of motion for $f$ and $Q$ in a dimensionless form, we obtain
\begin{equation}
\displaystyle{ 2 \left( \frac{ \text{d}^2 f}{ \text{d} \tilde{r}^2} + \frac{1}{\tilde{r}} \frac{ \text{d} f} { \text{d} \tilde{r}}\right) = \frac{f
Q^2}{\tilde{r}^2} + 2 f \left( f^2-1\right) },
\end{equation}
and 
\begin{equation}
\displaystyle{\text{Tr} \left({\tau_{\text{str}}}^2 \right) \left(\frac{ \text{d}^2 Q}{ \text{d} \tilde{r}^2} - \frac{1}{\tilde{r}} \frac{ \text{d} Q}{ \text{d} \tilde{r}}\right) = \frac{2 g^2}{\kappa^2} f^2 Q}.
\end{equation}
The first equation is properly written in a dimensionless form, and as $\tilde{r} = r \kappa M$, we identify the characteristic radius of the field $f$ to be $r_f \sim {\kappa M}^{-1}$. In regard to the field $Q$, we obtain a properly dimensionless equation of motion after introducing $\rho = \tilde{r}/\kappa$, yielding 
\begin{equation}
\displaystyle{\text{Tr} \left({\tau_{\text{str}}}^2 \right)  \left(\frac{ \text{d}^2 Q}{ \text{d} \rho^2} - \frac{1}{\rho} \frac{ \text{d} Q}{ \text{d} \rho}\right) = 2 g^2 f^2 Q}.
\end{equation}
It gives the characteristic radius for $Q$ to be $r_Q \sim M^{-1}$, since $\rho = r M$.

Taking into account the previous results, we can broadly evaluate their contribution to the Lagrangian density, from Eq.~(\ref{CS2LagEffDim}). The contribution of the field $f$ is of order $\kappa^2 M^4$, and the one of the field $Q$ of order $M^4$ (we estimate $g$ and $\text{Tr} \left({\tau_{\text{str}}}^2 \right) $ to be of order unity). So, in the limit $\kappa \geq 1$, the main contribution of the Lagrangian density comes from the field $f$, and for $\kappa\leq1$, the main contribution comes from the field $Q$. It means that in both cases, the characteristic radius of the string will be either $r_f\sim (\kappa M)^{-1}$, or $r_Q \sim M^{-1}$, see Sec.~\ref{CS2PartDefRadius}. Note that whatever the limit we consider, the characteristic energy per unit length of the string is always of order $M^2$.




\part{Modèles de Galiléons scalaires}
\label{ChapterGalileons}

%
%
%

%

\chapter{Gravité modifiée et modèles de Galiléons scalaires}
\section{Introduction}

\noindent
Rares sont les théories dont la validité n'a jamais été remise directement en question après plus d'un siècle de tests expérimentaux et observationnels, comme l'est la relativité générale. Cette propriété donne de fait une place unique à la relativité générale en tant que description d'une interaction fondamentale. Malgré cette absence de remise en question directe, la modification de la gravité à grande distance permettrait cependant d'expliquer naturellement certains problèmes de la physique moderne, comme la nature de l'énergie noire, voire de la matière noire, justifiant la recherche de théories dites de gravité modifiée. Cette recherche est cependant complexifiée par le succès à la fois théorique et expérimental de la relativité générale, rendant difficile de modifier celle-ci tout en gardant une théorie viable ayant les propriétés attendues d'une théorie de la gravitation, et imposant d'introduire un mécanisme d'écrantage pour obtenir des résultats compatibles avec les vérifications actuelles de la relativité générale. 

Après avoir décrit certains aspects de la relativité générale, nous discutons les principales motivations et difficultés liées à la construction de théories de gravité modifiée. Les modèles de Galiléons\footnote{Dans ce document et sauf mention contraire, les théories de Galiléons scalaires désignent les théories les plus générales avec des équations du mouvement d'ordre 2 au plus en espace courbe. Historiquement, les Galiléons désignaient uniquement les théories en espace plat avec des équations du mouvement d'ordre 2 seulement, et les Galiléons généralisés désignaient les théories en espace plat avec des équations du mouvement d'ordre 2 au plus. Le terme de \og Galiléon\fg{} tel qu'il est utilisé ici désigne donc à strictement parler les théories covariantes de Galiléons généralisés. Cette appellation fait cependant plus de sens avec la recherche actuelle de théories de Galiléons généralisées à d'autres champs qu'un seul champ scalaire.} scalaires sont ensuite introduits de façon autonome en tant que théorie de gravité modifiée. Leur rôle particulier en tant que théorie la plus générale décrivant des couplages entre la métrique et un scalaire et ayant des équations du mouvement d'ordre 2 au plus est discuté dans un deuxième temps.

\section{La relativité générale, succès et défauts}
\label{PartIntroRelativiteGenerale}

La relativité générale, développée par Einstein et finalisée en 1915, est le modèle standard actuel de la gravitation. C'est une théorie relativiste qui traite l'espace-temps comme une variété Riemannienne, de métrique $g_{\mu\nu}$. Les forces de gravitation sont traitées comme une manifestation de la courbure de l'espace-temps, dans lequel les points matériels suivent des trajectoires correspondant à des géodésiques. Plus précisément, le couplage entre la métrique et les champs de matière -- à savoir les champs du Modèle Standard de la physique des particules -- est universel, et se fait de façon minimale en remplaçant dans la formulation des lois décrivant la matière dans le cadre de la relativité restreinte la métrique de Minkowski $\eta_{\mu\nu}$ par $g_{\mu\nu}$ et les dérivées partielles $\partial_\mu$ par des dérivées covariantes $\nabla_\mu$\footnote{Cette contribution s'additionne si nécessaire aux champs de jauge nécessaires pour construire une dérivée covariante de jauge.}. C'est ce couplage minimal qui permet à la relativité générale de vérifier le principe d'équivalence d'Einstein, à savoir l'universalité de la chute libre, ainsi que le fait de retrouver localement les lois régissant la matière en l'absence de gravité dans les référentiels en chute libre, indépendamment des positions et vitesses initiales et pour tous les systèmes dont l'énergie de liaison gravitationnelle est négligeable~\cite{Will:1993ns,Clifton:2011jh}.

Les équations de la relativité générale sont régies par l'action d'Einstein-Hilbert, 
\begin{equation}
\label{EqEinsteinHilbert}
S_{EH} = \frac{1}{16\pi G_{\text{N}}} \int \sqrt{-g}~ \text{d}^4 x \left[R(g_{\mu\nu}) - 2\Lambda \right] + \int \text{d}^4 x ~ \mathcal{L}_{\text{m}} \left(g_{\mu\nu},\psi\right),
\end{equation}
où $g$ est le déterminant de la métrique, $\mathcal{L}_{\text{m}}$ le Lagrangien du Modèle Standard dont les champs ont été rassemblés dans la notation $\psi$, $\Lambda$ la constante cosmologique, et $R$ le scalaire de Ricci. Ce scalaire s'obtient à partir de la courbure de Riemann
\begin{equation}
\label{EqTenseurRiemann}
R^{\alpha}{}_{\mu\beta\nu} = \partial_\beta \Gamma^{\alpha}{}_{\mu\nu} - \partial_\nu \Gamma^{\alpha}{}_{\mu\beta} + \Gamma^{\alpha}{}_{\sigma\beta} \Gamma^{\sigma}{}_{\mu\nu} - \Gamma^{\alpha}{}_{\sigma\nu}\Gamma^{\sigma}{}_{\mu\beta},
\end{equation}
où les $\Gamma^{\lambda}{}_{\mu\nu}$ sont les symboles de Christoffel définis par
\begin{equation}
\label{EqChristoffel}
\Gamma^{\lambda}{}_{\mu\nu} = \frac12 g^{\lambda\sigma}\left(\partial_\mu g_{\nu\sigma}+\partial_\nu g_{\mu\sigma} - \partial_\sigma g_{\mu\nu} \right).
\end{equation}
Les symboles de Christoffel sont symétriques sur leurs indices inférieurs, et ne sont pas des tenseurs.
Le tenseur $R^{\alpha}{}_{\mu\beta\nu}$ permet de calculer le tenseur de Ricci 
\begin{equation}
\label{EqTenseurRicci}
R_{\mu\nu} \equiv R^{\alpha}{}_{\mu\alpha\nu},
\end{equation}
qui en est sa seule trace indépendante, ainsi que le scalaire de Ricci  
\begin{equation}
\label{EqScalaireRicci}
R \equiv g^{\mu\nu} R_{\mu\nu}.
\end{equation}
L'action du champ de matière fait intervenir indirectement la métrique dans les dérivées covariantes $\nabla_\mu$. Ces dérivées covariantes font apparaitre les symboles de Christoffel définis dans l'équation~\eqref{EqChristoffel}, et donnent par exemple pour des vecteurs
\begin{equation}
\label{EqDevCovRG}
\nabla_\mu A^\nu = \partial_\mu A^\nu + \Gamma^\nu{}_{\mu\alpha} A^\alpha, ~~~~\text{et} ~~~~ \nabla_\mu A_\nu = \partial_\mu A_\nu - \Gamma^\alpha{}_{\mu\nu} A_\alpha,
\end{equation}
ces dérivées commutant avec la métrique~:
\begin{equation}
\nabla_\lambda g_{\mu\nu} = 0.
\end{equation}

Les équations dérivant de l'action d'Einstein-Hilbert sont les équations d'Einstein,
\begin{equation}
\label{EqEqEinstein}
G_{\mu\nu} = 8\pi G T_{\mu\nu} - \Lambda g_{\mu\nu},
\end{equation}
où l'on a introduit le tenseur d'Einstein
\begin{equation}
G_{\mu\nu} \equiv R_{\mu\nu} - \frac12 R g_{\mu\nu},
\end{equation}
et le tenseur d'énergie-impulsion
\begin{equation}
T_{\mu\nu} \equiv \frac{2}{\sqrt{-g}} \frac{\delta \mathcal{L}_m}{\delta g_{\mu\nu}}.
\end{equation}
Ces deux tenseurs sont symétriques. Ils sont aussi conservés, à savoir qu'ils vérifient
\begin{equation}
\nabla_{\mu} G^{\mu\nu} = 0, ~~~~ \text{et}~~~~  \nabla_{\mu} T^{\mu\nu}=0,
\end{equation}
et les équations d'Einstein sont donc de divergence nulle. Dans l'action d'Einstein-Hilbert donnée par l'équation~\eqref{EqEinsteinHilbert}, les deux quantités lagrangiennes, à savoir $\mathcal{L}_{\text{m}}$ et $\mathcal{L}_{\text{g}}=\sqrt{-g}\left(R-2\Lambda\right)/16\pi G$ se transforment comme des quantités scalaires, et vérifient donc
\begin{equation}
\bar{\mathcal{L}} = \text{det}\left(\frac{\partial x^\mu}{\partial \bar{x}^\nu} \right) \mathcal{L},
\end{equation}
sous un changement de coordonnées $\bar{x}^\mu = \bar{x}^\mu (x^\nu)$. Les changements de coordonnées dans l'intégrant faisant aussi apparaitre le Jacobien de la transformation,
\begin{equation}
\text{d}^4 \bar{x} = \text{det}\left(\frac{\partial \bar{x}^\nu}{\partial x^\mu} \right) \text{d}^4 x,
\end{equation}
l'action d'Einstein-Hilbert est invariante sous les transformations quelconques de coordonnées, aussi appelées reparamétrisations. Cette symétrie implique notamment que les équations résultantes de la variation de cette action par rapport à la métrique sont de divergence nulle, comme discuté plus haut~\cite{Clifton:2011jh}. 

Jusqu'à présent, toutes les expériences testant directement la relativité générale sont compatibles avec ses prédictions~\cite{Will:2014kxa,PDG2016}. L'universalité du couplage entre la matière et la gravité est testé jusqu'à de l'ordre de $10^{-13}$. Les prédictions en champ faible, testées principalement sur le système solaire, sont vérifiées jusqu'à de l'ordre de $10^{-5}$. Les prédictions en champ fort, testées principalement sur les pulsars binaires, sont quant à elles vérifiées jusqu'à de l'ordre de $10^{-3}$ au minimum. La détection interférométrique d'ondes gravitationnelles, des oscillations de la métrique pouvant se propager, a également été effectuée en 2015~\cite{PhysRevLett.116.061102}. Ces résultats confortent le statut qu'a la relativité générale de modèle standard pour l'interaction gravitationnelle, étant donné que cette théorie a passé avec succès l'intégralité des tests expérimentaux et observationnels effectués depuis sa découverte il y a plus d'un siècle.

Dans un cadre cosmologique, certaines observations questionnent cependant la validité de la relativité générale. Le modèle cosmologique standard~\cite{PeterUzan} nécessite notamment l'introduction de matière non relativiste n'interagissant pas avec le Modèle Standard de la physique des particules. Cette matière, appelée matière noire ou matière noire froide (CDM pour \og Cold Dark Matter \fg{}), est notamment nécessaire pour expliquer les courbes de rotation des galaxies ou leur potentiel gravitationnel~\cite{1999coph.book.....P}. D'autre part, l'observation de l'accélération de l'expansion de l'univers comme les différentes observations des grandes structures de l'univers nécessite une constante cosmologique $\Lambda$ non nulle~\cite{PeterUzan,Clifton:2011jh}. Bien que cette constante cosmologique puisse correspondre à une contribution énergétique du vide, sa valeur n'est cependant pas du tout compatible avec les estimations du Modèle Standard~\cite{Martin:2012bt,Burgess:2013ara}. Il n'est pas clair à l'heure actuelle que ces phénoménologies soient liées à une modification à longue distance des interactions gravitationnelles par rapport à la relativité générale, ou bien à une mauvaise description de la matière/énergie qui gravite. Ces observations représentent cependant un domaine où tester de potentielles nouvelles théories de la gravité. 

Un autre problème inhérent à la relativité générale est lié à sa formulation en tant que théorie quantique. Le Lagrangien décrivant la dynamique du champ gravitationnel dans l'équation~\eqref{EqEinsteinHilbert} a en effet un couplage dimensionné. Si le champ gravitationnel est traité comme une perturbation et quantifié suivant les méthodes usuelles de théorie quantique des champs, la constante de couplage qui lui est associé est alors de dimension $-2$ (en masse), avec comme énergie caractéristique l'énergie de Planck $E_P \simeq 1.2 \times 10^{19}$GeV. Suivant les considérations discutées dans la section~\ref{TheorieFonda&Effect}, cela signifie notamment que les effets liés à la quantification de la relativité générale sont complètement négligeables à des échelles d'énergie petites devant l'échelle de Planck, ce qui légitime la description quantique de la physique des particules indépendamment de l'interaction gravitationnelle. Cela signifie également que la relativité générale est non-renormalisable, et qu'il n'est pas possible d'extraire des résultats des calculs liés à sa version quantique aux énergies proches et au-delà de l'énergie de Planck. Partant du principe qu'il est nécessaire de décrire la gravité dans un cadre quantique aux échelles de Planck, la relativité générale apparait alors comme une théorie effective classique, résultant d'une théorie quantique cohérente dont les déviations phénoménologiques par rapport à la relativité générale doivent apparaitre au plus tard aux échelles de Planck. Une telle phénoménologie de gravité quantique non-perturbative ne semble à l'heure actuelle atteignable qu'au centre des trous noirs ou au niveau du big-bang.

\section{Différentes approches de la relativité générale}
\label{PartDifferentesApprochesRG}

\noindent
Avant de discuter des possibles théories de gravité modifiée, il est intéressant d'étudier certaines approches de la relativité générale complémentaires à la description donnée dans la section~\ref{PartIntroRelativiteGenerale}.  Celles-ci donnent en pratique des résultats similaires -- et se ramènent plus ou moins aux mêmes propriétés mathématiques --, mais à partir de motivations physiques \emph{a priori} très différentes. Ces approches mettent notamment en perspective le lien entre les différents objets de la relativité générale, et leurs propriétés. Elles permettent également de montrer en quoi la relativité générale est une théorie très particulière, qu'il n'est pas aisé de modifier sans perdre certaines des propriétés attendues d'une théorie de la gravitation.

Dans une approche liée à la théorie des groupes similaire à celle détaillée dans le chapitre~\ref{ChapterModelBuilding}, la relativité générale peut être introduite comme la théorie décrivant des particules sans masse d'hélicité 2 dans l'espace de Minkowski. La façon la plus simple de décrire de telles particules de façon covariante nécessite l'introduction d'un tenseur de rang 2 symétrique, $g_{\mu\nu}$, décrivant les deux états d'hélicité $\pm2$, et possédant une symétrie de jauge infinitésimale $\delta g_{\mu\nu} = - \partial_\mu \xi_\nu - \partial_\nu \xi_\mu$ avec $\xi^\mu(x^\alpha)$ un champ arbitraire (correspondant à une transformation infinitésimale des coordonnées $\delta x^\mu = \xi^\mu$ lorsque $g_{\mu\nu}$ est la métrique). On montre alors que les théories décrivant l'auto-interaction d'un tel champ se réduisent à la relativité générale~\cite{PhysRev.98.1118,PhysRev.138.B988,Deser:1969wk}, dont les symétries contiennent les reparamétrisations quelconques. D'autre part, la nécessité d'écrire des couplages invariants de Lorentz entre un tel champ et les champs de matière implique un couplage identique pour tous les champs, et donc le principe d'équivalence~\cite{Weinberg:1995mt}. Dans ce cadre, le principe d'équivalence et l'invariance sous les reparamétrisations apparaissent alors comme une conséquence de l'étude de particules sans masse d'hélicité~ 2.

Du point de vue géométrique, la relativité générale traite l'espace-temps comme une variété Riemannienne. Questionnant cette hypothèse, on peut considérer le cas d'une variété non Riemannienne. La construction de dérivées covariantes comme dans l'équation~\eqref{EqDevCovRG} repose alors sur l'introduction d'une connexion linéaire, $C^\alpha{}_{\mu\nu}$, généralisant le rôle du symbole de Christoffel. Cette connexion a une contribution symétrique, qui correspond au symbole de Christoffel en relativité générale, et une contribution antisymétrique, la torsion. 
Contrairement à la contribution symétrique de la connexion, la torsion correspond à une quantité tensorielle. Le principe d'équivalence d'Einstein implique alors que la torsion est nulle dans le vide, puisqu'elle peut être annulée en tout point pour un référentiel donné en supposant que le vide est décrit localement par l'espace de Minkowski. Supposant alors que la variété est métrique\footnote{La non-métricité d'une variété est également quantifiée par un tenseur, le tenseur de non-métricité $Q_{\alpha\mu\nu}=\nabla_\alpha g_{\mu\nu}$. Retrouver localement des dérivées partielles pour un système en chute libre grâce au principe d'équivalence impose donc également que ce tenseur soit nul.},
comme c'est le cas en relativité générale, l'annulation de la torsion implique alors que l'espace-temps est une variété Riemannienne. La connexion est alors dite de Levi-Civita, et correspond aux symboles de Christoffel qui se calculent à partir de la métrique comme dans l'équation~\eqref{EqChristoffel}. Dans une variété Riemannienne quadri-dimensionnelle, le seul 2-tenseur symétrique conservé écrit à partir de la métrique et de ses dérivées est alors le tenseur d'Einstein $G_{\mu\nu}$~\cite{Aldrovandi:1996ke}, ce qui identifie de manière unique les équations de la dynamique du champ de gravitation dans le vide pour une théorie invariante par reparamétrisation (puisqu'on cette symétrie implique des équations du mouvement conservées, et que la variation d'un scalaire par rapport à $g_{\mu\nu}$ donne un 2-tenseur symétrique conservé). ll y a donc un lien très fort entre le principe d'équivalence, le fait de décrire l'espace-temps par une variété Riemannienne, et la forme des équations d'Einstein.

Le résultat que l'on vient d'énoncer sur le rôle particulier de $G_{\mu\nu}$ dans une variété Riemannienne correspond au théorème de Lovelock~\cite{Lovelock:1971yv,lovelock1972four}, qui généralise un théorème antérieur de Cartan~\cite{E1922}. Énoncé dans un cadre de théorie des champs, ce théorème stipule qu'à partir d'une action fonction d'un tenseur métrique seulement (et donc associé à la métrique d'une variété Riemannienne comme discuté plus haut),
\begin{equation}
S = \int \text{d}^4 x ~\mathcal{L}\left(g_{\mu\nu}\right),
\end{equation}
les seules équations d'Euler-Lagrange du second ordre qu'il est possible d'obtenir dans un espace à quatre dimensions sont de la forme
\begin{equation}
\alpha\sqrt{-g} \left(R^{\mu\nu} - \frac12 g^{\mu\nu} R \right) + \lambda \sqrt{-g} g^{\mu\nu}=0,
\end{equation}
où $\alpha$ et $\lambda$ sont des constantes arbitraires, et où $R^{\mu\nu}$ et $R$ sont les tenseurs et scalaires de Ricci définis à partir de la métrique comme dans les équations~\eqref{EqTenseurRiemann} à~\eqref{EqScalaireRicci}. Une action donnant de telles équations n'est pas définie de manière unique, mais correspond à l'action d'Einstein-Hilbert (introduisant les nécessaires pré-facteurs en $\alpha$ et $\lambda$) additionnée à deux contributions en dérivées totales ne contribuant pas aux équations du mouvement, nommées invariants de Lovelock. Ces contributions ne contribuant pas aux équations du mouvement sont de la forme
\begin{equation}
\label{EqInvLovelock}
\mathcal{S}_{\text{Lovelock}}  = \int \text{d}^4 x ~ \left[\beta \epsilon^{\mu\nu\rho\lambda}R^{\alpha\beta}{}_{\mu\nu} R_{\alpha\beta\rho\lambda} + \gamma\sqrt{-g}\left( R^2 - 4 R^\mu{}_\nu R^\nu{}_\mu + R^{\mu\nu}{}_{\rho\lambda} R^{\rho\lambda}_{\mu\nu}\right) \right].
\end{equation}
La contribution de préfacteur $\beta$, nommé scalaire de Pontryagin, est une dérivée totale quelque soit le nombre de dimensions. Ce n'est pas le cas de la contribution en $\gamma$, nommé invariant de Gauss-Bonnet ou $R^2_{GB}$, qui n'est plus une dérivée totale lorsque l'espace-temps a plus de quatre dimensions.

Pour conclure cette section, il est également intéressant de questionner la possibilité d'obtenir la relativité générale en rendant locales les symétries continues du groupe de Poincaré définissant la relativité restreinte. La théorie ainsi obtenue est une théorie de jauge de la gravité, appelée théorie d'Einstein-Cartan (ou d'Einstein–Cartan–Sciama–Kibble)~\cite{cartan1922generalisation,doi:10.1063/1.1703702,RevModPhys.36.463}. Cette théorie correspond à une modification de la relativité générale, où l'énergie-impulsion source toujours la courbure, mais où la torsion est sourcée par le spin des particules~\cite{RevModPhys.48.393,Ramond:1981pw} (et est donc nulle dans le vide). Une telle torsion mène à l'apparition d'un couplage entre spins, qui ne se propage cependant pas. De telles interactions ne pouvant être testées que dans les états extrêmes de la matière, liés à des trous noirs ou au big-bang, la théorie d'Einstein-Cartan est à l'heure actuelle aussi compatible avec les observations que la relativité générale~\cite{Blagojevic:2013xpa}. Par l'écriture de dérivées covariantes, cette théorie de jauge de la gravitation fait également apparaitre le lien entre la connexion de l'espace et la structure de groupe de jauge, et donc entre les approches géométriques et algébriques de la relativité générale.

\section{La gravité modifiée, motivations et tests expérimentaux}
\label{PartGraviteModifieeGenerale}

\noindent
Les problèmes inhérents à la relativité générale comme théorie de la gravitation, principalement l'observation du secteur sombre de l'univers et la difficulté à en effectuer une quantification cohérente, comme discuté dans la section~\ref{PartIntroRelativiteGenerale}, mène à rechercher des théories de gravité modifiée. De telles théories modifient dans un cadre non-quantique et à grande distance cette interaction, afin d'expliquer la phénoménologie de l'énergie noire (voire de la matière noire). Décrivant des théories différentes de la relativité générale, ces modèles de gravité modifiée pourraient alors être quantifiables à haute énergie, fournissant par là même une théorie quantique de la gravitation. Un obstacle important à la formulation de telles théories est qu'elles doivent être compatibles avec les prédictions de la relativité générale aux échelles intermédiaires, ce qui nécessite généralement l'introduction d'un mécanisme dit d'écrantage\footnote{Le mécanisme de Vainshtein~\cite{VAINSHTEIN1972393,Deffayet:2001uk,Babichev:2009jt,Babichev:2009us,Babichev:2010jd,Babichev:2013usa} fourni un exemple de mécanisme d'écrantage, apparaissant dans les théories de Galiléons discutées dans la section~\ref{PartIntroGalileonScalaire} comme pour la gravité massive~\cite{ArkaniHamed:2002sp,deRham:2010ik,deRham:2010kj,Hinterbichler:2011tt,deRham:2014zqa}. Dans ce mécanisme, des effets non-linéaires dus à des contributions en dérivées secondes deviennent importants et gèlent certains degrés de liberté, qui ne se propagent plus. Cet effet devient important dans les zones où la courbure de l'espace dépasse certaines valeurs, et donc notamment près des corps massifs.}. Elles donnent aussi généralement des prédictions différentes de la relativité générale dans le régime de couplage fort, où les non-linéarités des théories doivent être prises en compte.

Une difficulté importante de la construction de théories de gravité modifiée vient des propriétés très particulières de la relativité générale discutées dans la section~\ref{PartDifferentesApprochesRG}. Dans un espace à 4 dimensions, construire une théorie métrique vérifiant les principes d'équivalence et l'invariance sous les difféomorphismes mène en effet uniquement à la relativité générale, et pas à une classe de théories comprenant la relativité générale. Rajouter naïvement des termes impliquant une métrique au Lagrangien de Einstein-Hilbert fait donc perdre \emph{a priori} la plupart des propriétés recherchées pour une théorie de la gravitation. De manière générale, et étant donné qu'une théorie décrivant une particule sans masse d'hélicité 2 redonne de manière unique la relativité générale, modifier la gravité nécessite donc de changer ses degrés de liberté. La discussion de la section~\ref{PartDifferentesApprochesRG} donne également qu'une théorie de la gravitation ne pourra pas être décrite uniquement par une métrique d'une variété Riemannienne, mais devra prendre en compte par exemple un couplage entre cette métrique et d'autres champs, changeant la dynamique de la métrique dans un espace vide de matière.

Construire une théorie de gravité modifiée repose alors sur l'introduction de champs en plus du tenseur métrique (ou le remplaçant), sur l'utilisation de dérivées d'ordre supérieur à 2 dans les équations du mouvement, sur l'introduction d'un espace-temps à plus de 4 dimensions, ou sur l'abandon de certaines autres hypothèses de la relativité générale (comme le fait de chercher une équation tensorielle à deux indices ou de considérer une théorie locale)~\cite{Clifton:2011jh}. Il est alors nécessaire de vérifier que les théories résultantes ne contiennent pas certaines pathologies des théories effectives, comme des fantômes, liés généralement à des termes cinétiques dans le Lagrangien de signes opposés à ceux des termes conventionnels, et qui étant associés à des états d'énergie négative déstabilisent le vide par la création simultanée de particules d'énergie négative et positive\footnote{Une telle déstabilisation nécessite un couplage entre ces fantômes et les champs ordinaires. 
Dans le cadre de théories décrivant la gravitation, un tel couplage est attendu au minimum avec la métrique.}~\cite{Cline:2003gs,Carroll:2003st}. L'apparition de tels fantômes est une conséquence générique, bien que non systématique, de la présence de dérivées d'ordre supérieur à 3 dans les équations du mouvement. De telles instabilités sont décrites par le théorème d'Ostrogradsky~\cite{Woodard:2006nt,deUrries:1995ty,deUrries:1998obu}. D'autres types de pathologies sont aussi à vérifier, comme par exemple la présence d'instabilités de gradient, d'instabilités tachyoniques, ou l'apparition de modes superluminiques ou de phénomènes non locaux~\cite{Joyce:2014kja}.

Plusieurs classifications des théories de gravité modifiée ont été proposées~\cite{Clifton:2011jh,Berti:2015itd}, basées sur les motivations initiales de ces théories, comme ajouter un champ particulier se couplant à la métrique, violer l'invariance de Lorentz, décrire une théorie de gravité massive, introduire des dérivées d'ordre supérieur à deux, etc. Comme on l'a vu précédemment, toutes ces théories demandent néanmoins de modifier les degrés de liberté de la relativité générale, et donc généralement à en rajouter. Il est alors souvent possible d'écrire ces théories via un couplage entre une métrique et un autre champ. Dans ce cadre, il est courant que des théories de gravité modifiée deviennent équivalentes dans une certaine limite\footnote{Une méthode de classification consisterait peut-être à cataloguer les théories en fonction des symétries qu'elles contiennent ainsi que des degrés de liberté qu'elle décrivent (en lien avec ces symétries). Cela permettrait notamment plus facilement d'identifier dans quelles limites (comme le fait de négliger des degrés de liberté ou des symétries) une théorie donnée s'identifie à une autre.}. On peut trouver des revues de ces théories dans les références~\cite{Clifton:2011jh,deRham:2014zqa,Joyce:2014kja,Berti:2015itd,Koyama:2015vza}.

Les tests observationnels de la gravité modifiée sont multiples. Il est d'une part nécessaire que ces théories reproduisent les résultats cosmologiques sans introduire une constante cosmologique, puisque ces théories ont pour motivation d'expliquer la phénoménologie liée à l'énergie noir. Il est d'autre part nécessaire que les prédictions des théories de gravité modifiée se ramènent à celles de la relativité générale dans les régimes où celle-ci a été testée avec précision. Ces deux critères peuvent être vérifiés de manière systématique avec l'introduction de paramètres post-Friedmanniens pour les observations cosmologiques~\cite{Hu:2007pj,Caldwell:2007cw,Daniel:2010ky,Zhao:2010dz,Clifton:2011jh} ou post-Newtoniens pour les tests directs de la relativité générale~\cite{Thorne:1970wv,Will:1993ns,Will:2014kxa}. Bien que contraignant fortement les théories de gravité modifiée, ces observations ne permettent cependant pas de les discriminer vis-à-vis de la relativité générale et du modèle standard de la cosmologie. Une observation discriminante (un \og smoking gun \fg{}) pourrait être apportée par les observations futures de dynamique gravitationnelle en champ fort, comme celle des systèmes binaires d'objets compacts tels que des trous noirs ou des étoiles à neutron~\cite{Berti:2015itd}, et qui sont permises par le développement actuel de l'astronomie gravitationnelle~\cite{TheLIGOScientific:2014jea,TheVirgo:2014hva,AmaroSeoane:2012je,Moore:2014lga}.

\section{Modèles de Galiléons scalaires}
\label{PartIntroGalileonScalaire}

\noindent
Les modèles de Galiléons scalaires -- ou plus simplement de Galiléons -- sont des théories de gravité modifiée couplant la métrique $g_{\mu\nu}$ à un champ scalaire $\pi$, le Galiléon. Ces modèles ont été développés dès 2009~\cite{Nicolis:2008in,Deffayet:2009wt,Deffayet:2009mn,Deffayet:2011gz,Deffayet:2013lga}, mais ils avaient été obtenus par d'autres approches précédemment. Ils représentent une formulation alternative des théories de Horndeski~\cite{Horndeski:1974wa}, décrivant toutes les théories tenseurs-scalaires avec des équations au maximum du second-ordre, et avaient également été obtenus dans le cadre d'un espace plat~\cite{Fairlie:1991qe,Fairlie:1992he,Fairlie:1992nb}. Ces théories sont motivées et présentées ici de manière auto-cohérente en tant que théorie de gravité modifiée, en montrant quelles hypothèses y mènent (des présentations différentes sont évidemment possibles, voir par exemple~\cite{Deffayet:2013lga,Clifton:2011jh,Curtright:2012gx}). 

Les théories de Galiléons sont obtenues en essayant de coupler un champ scalaire $\pi$ avec une métrique $g_{\mu\nu}$ de telle façon que la relativité générale soit retrouvée lorsque le champ scalaire s'annule identiquement. L'action la plus simple décrivant de telles théories est de la forme
\begin{equation}
\mathcal{S} = \int \sqrt{-g} ~ \text{d}^4x \left[\frac{1}{16\pi G_{\text{N}}}R - \nabla_\mu \pi \nabla^\mu \pi - V(\pi)\right].
\end{equation}
Elle contient le Lagrangien d'Einstein-Hilbert (sans constante cosmologique) et le Lagrangien standard d'un champ scalaire. Une telle action ne conduit qu'aux équations d'Einstein données par l'équation~\eqref{EqEqEinstein}, et toute la liberté permise par un potentiel $V(\pi)$ arbitraire ne peut que modifier la contribution de $\pi$ au tenseur d'énergie-impulsion, ne modifiant donc pas la dynamique de la métrique elle-même mais seulement son terme de source. Une façon de faire apparaitre un couplage entre le champ scalaire et des termes de courbure consiste alors à introduire des dérivées d'ordre~2  du champ scalaire dans le Lagrangien. Les équations du mouvement contiendront en effet des dérivées d'ordres 3 et 4, qui peuvent commuter pour donner des termes de courbure, comme des termes en la dérivée première de la courbure. Ce type de construction mène aux théories de Galiléons.

Le problème d'une telle construction réside dans le fait que des équations du mouvement contenant des dérivées troisièmes et quatrièmes en le champ scalaire conduisent généralement à des instabilités, comme discuté dans la section~\ref{PartGraviteModifieeGenerale}. Une manière d'éviter ces instabilités consiste à ne garder que des Lagrangiens contenant des dérivées secondes du champ scalaire mais donnant des équations du mouvement au maximum du second ordre en ce champ\footnote{Une telle hypothèse revient à chercher des théories du type Horndeski. Une possibilité alternative consiste à garder des équations du mouvement d'ordre supérieur à~2 tout en assurant qu'elles n'impliquent pas de fantômes. Ces théories sont dites au-delà de Horndeski~\cite{Gleyzes:2014dya,Langlois:2015cwa,Langlois:2015skt}.}. Les théories de Galiléons sont obtenues en construisant en espace plat les Lagrangiens $\mathcal{L}(\pi,\partial_\mu \pi, \partial_\mu\partial_\nu \pi)$ donnant des équations du mouvement au maximum du second ordre, puis en les rendant covariants pour obtenir les théories de gravité modifiée qui leur sont associées.

Il est commode de commencer par étudier les Lagrangiens donnant des équations du mouvement d'ordre~2 seulement en espace plat~\footnote{Ces Lagrangiens, dont l'action est invariante sous les transformations
\begin{equation}
\label{EqTransfoGalilee}
\pi\rightarrow \pi + b_\mu x^\mu + c,
\end{equation}
qui sont similaires aux transformations de Galilée, ont donné leur nom aux théories de Galiléons.}. Ils s'écrivent à partir d'un tenseur invariant formé à partir des tenseurs de Levi-Civita, 
\begin{equation}
\displaystyle{\delta^{j_1 \cdots j_n}_{i_1 \cdots
i_n} =  n! \delta^{j_1
\cdots j_n}_{[i_1 \cdots i_n]} = \frac{1}{(D-n)!}\epsilon^{i_1\cdots i_n \sigma_1 \cdots \sigma_{D-n}}
\epsilon_{j_1\cdots j_n \sigma_1 \cdots \sigma_{D-n}},}
\label{EqDeltaMult}
\end{equation}
avec $n$ prenant des valeurs entre $1$ et $D$ la dimension de l'espace-temps. Comme ce tenseur est complètement antisymétrique sur ses indices $\mu_i$ et $\nu_i$ respectivement, les Lagrangiens de la forme
\begin{equation}
\displaystyle{\mathcal{L}^{\text{Gal.}}_{m+1} =  \pi  \delta^{\mu_1 \cdots \mu_m}_{\nu_1\cdots\nu_m} \partial_{\mu_1} \partial^{\nu_1} \pi \partial_{\mu_2}\partial^{\nu_2}  \pi \cdots \partial_{\mu_m}\partial^{\nu_m} \pi,}
\end{equation}
donnent bien des équations du second ordre pour $m=1,\cdots,4$ en dimension 4 (ces Lagrangiens sont indicés conventionnellement par $m+1$, car ils contiennent $m+1$ champs scalaires). Calculant les équations du mouvement de ces Lagrangiens, il n'est en effet pas possible d'écrire un champ scalaire avec deux dérivées en $\mu_i$ ou en $\nu_i$, puisque le terme résultant s'annulerait identiquement après contraction avec $\delta^{j_1 \cdots j_n}_{i_1 \cdots
i_n} $, et les termes obtenus sont tous du second ordre. Les équations du mouvement de ces termes sont simplement~:
\begin{equation}
\displaystyle{EOM^{\text{Gal.}}_{m+1} =  \delta^{\mu_1 \cdots \mu_m}_{\nu_1\cdots\nu_m} \partial_{\mu_1} \partial^{\nu_1} \pi \partial_{\mu_2}\partial^{\nu_2}  \pi \cdots \partial_{\mu_m}\partial^{\nu_m} \pi=0,}
\end{equation}
qui ne contiennent que des dérivées du second ordre. 

Les Lagrangiens donnant des équations du mouvement d'ordre~2 seulement peuvent être généralisés en Lagrangiens donnant des équations du mouvement d'ordre~2 au plus, de la forme 
\begin{equation}
\label{EqLagGenGal}
\displaystyle{\mathcal{L}^{\text{Gen.Gal.}}_{m+1} = f_{m+1}\left(\pi,X\right) \delta^{\mu_1 \cdots \mu_m}_{\nu_1\cdots\nu_m} \partial_{\mu_1} 
\partial^{\nu_1} \pi \cdots \partial_{\mu_m}\partial^{\nu_m} \pi},
\end{equation}
toujours pour $m=1,\cdots,4$, et où les $f_i$ sont des fonctions arbitraire de $\pi$ et de $X=\partial_\alpha \pi \partial^\alpha \pi$. Ces Lagrangiens, en plus d'une fonction arbitraire $f_1\left(\pi,X\right)$, généralisent les modèles de Galiléons avec des équations du mouvement d'ordre 2 seulement\footnote{Ils ont été initialement appelés modèles de Galiléons généralisés, car leurs équations du mouvement ne contiennent pas que des dérivées secondes, et ne sont donc pas invariantes sous les transformations de Galilée définies équation~\eqref{EqTransfoGalilee}.}. Une observation de ces Lagrangiens montre que des dérivées troisièmes dans les équations du mouvement peuvent apparaitre après différentiation d'un terme en $\partial_\alpha \pi \partial^\alpha\pi $ contenu dans une fonction arbitraire $f_i$ -- une dérivée $\partial_\alpha$ pouvant affecter un terme en $\partial_{\mu_i}\partial^{\nu_i} \pi$ --, et de même lorsqu'on différencie le Lagrangien par rapport à un terme $\partial_{\mu_i}\partial^{\nu_i} \pi$ -- ce qui peut faire agir une dérivée $\partial_{\mu_i}\partial^{\nu_i} $ sur un terme $\partial_\alpha \pi \partial^\alpha\pi $ contenu dans une fonction $f_i$. Cependant, les termes apparaissant dans ces deux types de différentiations s'identifient par paires, et s'annulent deux à deux dans les équations du mouvement et en espace plat, car ils sont obtenus par une et deux intégrations par partie, respectivement.

Les raisonnements précédents ont été faits en espace plat, où il est possible de permuter les dérivées partielles. Après avoir rendu covariante la théorie, la commutation de dérivées de scalaires d'ordre jusqu'à~4 va faire apparaitre des tenseurs de courbure et leur dérivées premières, comme attendu. Cependant, la présence de dérivées premières de la courbure fait apparaitre des dérivées troisièmes de la métrique, ce qui induit des instabilités. Pour éviter ces dérivées troisièmes, il est nécessaire d'introduire des contre-termes, couplant directement dans le Lagrangien des termes de courbure et le champ scalaire, et impliquant des équations complètes du second ordre pour la métrique. Les Lagrangiens de l'équation~\eqref{EqLagGenGal} une fois rendu covariants, ainsi que ces nécessaires contre-termes décrivent alors les théories de gravité modifiée de Galiléons~\cite{Deffayet:2009wt}.

La forme finale des Lagrangiens des modèles de Galiléons est alors\footnote{On remarquera que l'ordre des fonctions $\delta^{j_1 \cdots j_n}_{i_1 \cdots i_n}$ a diminué d'une unité par rapport aux expressions données dans l'équation~\eqref{EqLagGenGal}. L'écriture compacte de l'équation~\eqref{EqCovGal1} n'est cependant qu'une formulation simplifiée permise par l'ajout de dérivées totales~\cite{Deffayet:2011gz}.}
\begin{equation}
\label{EqCovGal1}
\begin{array}{l}
\displaystyle{\mathcal{L}_1 = f_1\left(\pi\right),}\vspace{0.25cm}\\
\displaystyle{\mathcal{L}_2 = f_2\left(\pi,Y\right),}\vspace{0.25cm}\\
\displaystyle{\mathcal{L}_3 = f_3\left(\pi,Y\right) \delta^{\mu_1}_{\nu_1}\nabla_{\mu_1} \nabla^{\nu_1}\pi,}\vspace{0.25cm}\\
\displaystyle{\mathcal{L}_4 = f_4\left(\pi,Y\right) R +    f_{4,Y}\left(\pi,Y\right)\delta^{\mu_1\mu_2}_{\nu_1\nu_2}\nabla_{\mu_1} \nabla^{\nu_1}\pi \nabla_{\mu_2}\nabla^{\nu_2} \pi,}\vspace{0.2cm}\\
\displaystyle{\mathcal{L}_5 = f_5\left(\pi,Y\right) G_{\mu\nu}\nabla^\mu \nabla^\nu \pi - \frac16 f_{5,Y}\left(\pi,Y\right)\delta^{\mu_1\mu_2\mu_3}_{\nu_1\nu_2\nu_3}\nabla_{\mu_1} \nabla^{\nu_1}\pi \nabla_{\mu_2}\nabla^{\nu_2} \pi\nabla_{\mu_3}\nabla^{\nu_3} \pi,}
\end{array}
\end{equation}
où l'on a utilisé la notation $Y=-\frac12 \partial_\alpha \pi \partial^\alpha\pi$ pour correspondre aux notations usuelles de la littérature, et où $f_{i,Y}$ est la dérivée par rapport à $Y$ de la fonction arbitraire $f_i$. Le terme $\mathcal{L}_1$, dit de "tadpole", est généralement omis. La forme développée de ces Lagrangiens est
\begin{equation}
\label{EqCovGal2}
\begin{array}{l}
\displaystyle{\mathcal{L}_3 = f_3\left(\pi,Y\right) \square \pi,}\vspace{0.25cm}\\
\displaystyle{\mathcal{L}_4 = f_4\left(\pi,Y\right) R +    f_{4,Y}\left(\pi,Y\right) \left[\left(\square\pi\right)^2 - \left(\nabla_\mu\nabla_\nu \pi \right)\left(\nabla^\mu\nabla^\nu\pi \right) \right],}\vspace{0.2cm}\\
\displaystyle{\mathcal{L}_5 = f_5\left(\pi,Y\right) G_{\mu\nu}\nabla^\mu \nabla^\nu \pi - \frac16 f_{5,Y}\left(\pi,Y\right)\left[\left(\square\pi\right)^3 \right.}\vspace{0.25cm}\\
\displaystyle{~~~~ ~~~~ - 3\square\pi\left(\nabla_\mu\nabla_\nu \pi \right)\left(\nabla^\mu\nabla^\nu\pi \right) + 2 \left(\nabla_\mu\nabla_\nu \pi \right)\left(\nabla^\mu\nabla_\sigma \pi \right)\left(\nabla^\nu\nabla^\sigma \pi \right)\Big],}
\end{array}
\end{equation}
où l'on a utilisé la notation $\square \pi = \nabla_\alpha \nabla^\alpha \pi$. 

\section{Gravité modifiée et Galiléons scalaires}

\noindent
Les théories de Galiléons scalaires ont été introduites de façon autonome dans la section précédente. Elles représentent cependant une classe très particulière de théories, qui englobe toutes les dynamiques possibles qui peuvent être écrites pour une théorie tenseur-scalaire ayant des équations du mouvement au plus du second ordre~\cite{Horndeski:1974wa,Deffayet:2011gz,Kobayashi:2011nu}.  Cette propriété est synthétisée par le théorème suivant~:
\vspace{-0.3cm}
\paragraph{Théorème (Horndeski/Galiléons)~:}\textit{ Toute théorie couplant un scalaire $\pi$ et une métrique $g_{\mu\nu}$ dans un espace de dimension 4 et ayant des équations du mouvement au plus du second ordre pour ces deux champs peut être décrite par les théories de Galiléons scalaires, et plus exactement par les théories covariantes de Galiléons généralisés. Ces théories ont comme action}
\begin{equation}
S_{\text{Gal.}} = \frac{1}{16\pi G_{\text{N}}} \int \sqrt{-g}~\text{d}^4 x ~ R + \int \sqrt{-g} ~ \text{d}^4 x \sum_{i=1}^5 \mathcal{L}_{i}(\pi,g_{\mu\nu}),
\end{equation}
\textit{où les Lagrangiens $\mathcal{L}_{i}$ sont ceux définis dans les équations~\eqref{EqCovGal1} et~\eqref{EqCovGal2}.} 

\vspace{0.2cm}
La généralité de cette classe de théories a initialement été obtenue en 1974 par Horndeski~\cite{Horndeski:1974wa}, mais l'expression de ces théories sous forme d'action ne correspondait pas à la formulation des Galiléons. Les théories de Galiléons ont ensuite été obtenues, et il a été montré qu'elles décrivaient exactement toutes les dynamiques étudiées par Horndeski~\cite{Deffayet:2011gz,Kobayashi:2011nu} (sous une forme cependant plus commode que celle donnée initialement par Horndeski). 

Toutes les théories de gravité modifiée couplant un scalaire avec la métrique et se ramenant aux hypothèses précédentes peuvent être décrite par les Galiléons. On peut citer par exemple les théories Einstein-dilaton-Gauss-Bonnet~\cite{Kanti:1995vq}, décrivant un couplage entre une métrique $g_{\mu\nu}$ et un champ scalaire $\pi$ via une action
\begin{equation}
S_{EdGB} = \frac{1}{16\pi G_{\text{N}}} \int \sqrt{-g}~\text{d}^4 x \left[R - 2\nabla_\mu \pi \nabla^\mu \pi -V(\pi) + f(\pi)R^2_{GB}\right],
\end{equation}
où $f$ est une fonction arbitraire et où $R^2_{GB}$ est l'invariant de Gauss-Bonnet défini dans l'équation~\eqref{EqInvLovelock}. Ces théories se réduisent à un cas particulier des théories de Galiléons~\cite{Kobayashi:2011nu,Berti:2015itd}. En prenant par exemple le cas $f(\phi)=\alpha \phi$, cette théorie est équivalente à une théorie de Galiléons en prenant comme fonctions arbitraires $f_2=X/2$, $f_3=0$, $f_4=1/2$ et $f_5=-2\alpha \text{ln}\left|X\right|$, où l'on utilise les notations de l'équation~\eqref{EqCovGal1}. C'est aussi le cas des théories dites de $k$-essence~\cite{ArmendarizPicon:2000ah} et de $k$-mouflage~\cite{Babichev:2009ee}, qui peuvent être décrites par des modèles de Galiléons~\cite{Deffayet:2011gz}. 

Les théories de Galiléons décrivent aussi la limite de découplage de certaines théories de gravité modifiée décrivant plus d'un degré de liberté additionnel à la métrique, lorsque tous les degrés de liberté sauf un se découplent, ce dernier pouvant être traité comme un champ scalaire. C'est par exemple le cas des théories de gravité massive~\cite{ArkaniHamed:2002sp,deRham:2010ik,deRham:2010kj,Hinterbichler:2011tt,deRham:2014zqa}, où des Lagrangiens de Galiléons de type $\mathcal{L}_3$ à $\mathcal{L}_5$ sont obtenus dans une limite de découplage~\footnote{Dans la limite de découplage, il est possible de décomposer les modes de la gravité massive en deux modes tensoriels d'hélicité~$\pm2$, deux modes vectoriels d'hélicité~$\pm1$ et un mode scalaire d'hélicité nulle. Ce sont ces modes qu'on discute ici. Voir aussi la discussion de la section~\ref{PartRepCovariante} sur les représentations unitaires du groupe de Lorentz inhomogène contenues dans les représentations covariantes.} où les modes d'hélicité 1 ne sont pas sourcés par la matière et où reste seulement le mode d'hélicité 0 qui peut être décrit par un champ scalaire se couplant à la métrique. Certaines termes des théories de Galiléons peuvent aussi être identifiés dans des constructions de branes~\cite{deRham:2010eu,Hinterbichler:2010xn,Goon:2011qf,Trodden:2011xh,Burrage:2011bt}. Ces théories décrivent par exemple la gravité dans un espace-temps à 5 dimensions, où l'espace-temps quadri-dimensionnel correspond à une brane. Il est alors possible de se placer dans un choix de coordonnées où la coordonnée repérant la position de la brane quadri-dimensionnelle dans le volume à 5 dimensions est décrite par une champ scalaire, s'identifiant au Galiléon. Les théories tenseurs-scalaires ainsi obtenues sont contenues dans les modèles de Galiléons, et plus particuliérement les DBI-Galiléons~\cite{Clifton:2011jh}. 

Le théorème cité précédemment donne une place particulière aux modèles de Galiléons, puisque ceux-ci décrivent l'intégralité des théories possibles modifiant la gravité par l'ajout d'un degré de liberté scalaire et avec des équations du mouvement d'ordre~2. Il est donc suffisant d'étudier les modèles de Galiléons pour tester les phénoménologies associées à toutes ces théories, permettant une investigation systématiques et exhaustive des extensions les plus simples de la relativité générale -- puisque ne lui ajoutant qu'un degré de liberté scalaire -- et viables mathématiquement\footnote{Le reproche principal que l'on peut faire à une telle approche est qu'elle est basée sur des considérations mathématiques plutôt que physiques.}. À l'heure actuelle, certaines théories de Galiléons sont aussi compatibles avec les observations cosmologiques que le modèle $\Lambda$CDM~\cite{Neveu:2016gxp}. 

Pour les Galiléons scalaires en 4 dimensions, la théorie la plus générale en espace courbe a été obtenue en rendant covariante la théorie la plus générale en espace plat tout en continuant à demander des équations du mouvement du second ordre. Pour investiguer les théories de gravité modifiées couplant d'autres types de champs à la métrique, comme des modèles impliquant par exemple plusieurs champs scalaires ou des champs vectoriels, Il semble donc efficace de rechercher les théories associées les plus générales en espace plat, puis de les rendre covariantes en rajoutant les contre-termes nécessaires pour garder des équations du mouvement d'ordre 2 ; et la recherche de théories les plus générales en espace plat est donc d'un grand intérêt pour construire des théories de gravité modifiée. Bien que cela soit le cas pour les Galiléons scalaire dans un espace à 4 dimension, il n'est cependant pas prouvé que cette procédure donne effectivement la théorie la plus générale après couplage avec la métrique, et il possible\footnote{Si c'est le cas, il devrait être possible d'identifier des contributions en espace courbe qui s'annulent identiquement en espace plat.} que certains termes en espace courbe ne soient pas obtenues ainsi~\cite{Deffayet:2013lga}. 

\section{Conclusion}

\noindent
Après avoir décrit certaines approches théoriques de la relativité générale, on a discuté les points qui motivent la recherche de théories de gravité modifiée au-delà de la relativité générale. Les théories de Galiléons scalaires ont notamment été introduites, et leur place en tant que théorie la plus générale couplant un scalaire avec la métrique a été discutée, sous des hypothèses d'équations du mouvement du second ordre. Dans l'idée d'investiguer les possibles théories de gravité modifiée de façon systématique, une approche consiste à reproduire les résultats obtenus pour les Galiléons scalaires pour des théories couplant d'autres types de champs à la métrique, comme plusieurs scalaires ou des champs vectoriels. Je présente dans la suite de ce document les articles publiés suite à ces travaux.

Les théories de multi-Galiléons scalaires représentent une première extension possible des théories de  Galiléons. Ces théories décrivent un couplage entre une métrique $g_{\mu\nu}$ et un ensemble de champs scalaires $\pi^{\a}$ en auto-interaction (les indices distinguant les différents champs scalaires sont indiqués en rouge afin de les distinguer facilement des indices d'espace-temps). De telles théories ont été initialement étudiées en espace plat avant d'être rendues covariantes, comme cela a été fait pour les Galiléons scalaires~\cite{Padilla:2010de,Padilla:2010tj,Padilla:2010ir,Hinterbichler:2010xn,Padilla:2012dx}, et la théorie la plus générale en espace plat a été proposée~\cite{Sivanesan:2013tba}. Des termes supplémentaires à cette théorie ont cependant été introduits de diverses façons dans~\cite{Deffayet:2010zh,Kobayashi:2013ina,Ohashi:2015fma}. Dans l'article~\cite{Allys:2016hfl} reproduit dans le chapitre~\ref{PartArticleMultigal}, j'ai proposé une nouvelle classe de Lagrangiens de multi-Galiléons scalaires en espace plat, qui n'est pas inclue dans la théorie la plus générale de~\cite{Sivanesan:2013tba}, et qui contient notamment les termes qui avaient été obtenus dans~\cite{Deffayet:2010zh,Kobayashi:2013ina,Ohashi:2015fma}. J'ai étudié en détails les propriétés de cette classe de modèles, notamment leur propriétés de symétrie, et montré que seul un petit nombre de dynamiques sont possibles malgré le fait que beaucoup de Lagrangiens puisse être écrits. J'applique notamment ces propriétés à l'investigation des multi-Galiléons dans les représentations fondamentales et adjointes des groupes SO($N$) et SU($N$). La version covariante de cette classe de modèle a été étudiée ultérieurement dans~\cite{Akama:2017jsa}.

\newpage
\chapter[New termes for scalar multi-Galileon models (article)]{New terms for scalar multi-Galileon models}
\label{PartArticleMultigal}

\begin{figure}[h!]
\begin{center}
\includegraphics[scale=1.4]{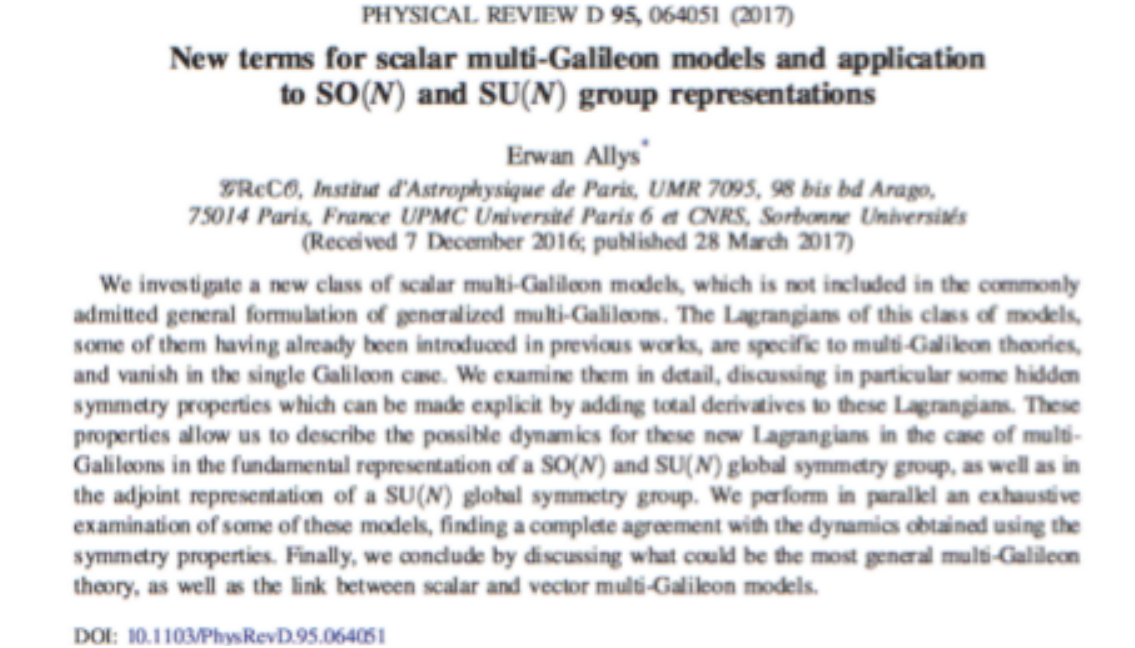}
\end{center}
\end{figure}

\section{Introduction}

In recent years, the attempts to investigate in a systematic way the different classes of modified gravity theories have been very successful. Galileon theories, describing models involving one scalar field coupling to general relativity, have been extensively studied \cite{Nicolis:2008in,Deffayet:2009wt,Deffayet:2009mn,Deffayet:2011gz,deRham:2011by,Joyce:2014kja}. Their most general extension has been especially proven in Ref.~\cite{Deffayet:2011gz}. This approach to modify gravity has found multiple applications in cosmology, for example on the subject of dark energy \cite{Chow:2009fm,Silva:2009km,Deffayet:2010qz,Kobayashi:2010wa,Gannouji:2010au,Tsujikawa:2010zza,DeFelice:2010pv,Ali:2010gr,DeFelice:2010nf,Mota:2010bs,Nesseris:2010pc,Easson:2011zy,Gleyzes:2013ooa,Gabadadze:2016llq,Salvatelli:2016mgy,Shahalam:2016kkg,Minamitsuji:2016qyc,Saridakis:2016ahq,Biswas:2016bwq} or inflation \cite{Creminelli:2010ba,Kobayashi:2010cm,Mizuno:2010ag,Burrage:2010cu,Creminelli:2010qf,Kamada:2010qe,Libanov:2016kfc,Banerjee:2016hom,Hirano:2016gmv,Brandenberger:2016vhg,Nishi:2016wty}. It has even been shown that some Galileon models are as compatible with cosmological data as the $\Lambda$CDM model \cite{Neveu:2016gxp}.

Several attempts have been made to investigate theories going beyond the standard single scalar Galileon theory. For example, the possibility to build vector Galileon models with vector fields propagating three degrees of freedom have been investigated \cite{Deffayet:2010zh,Heisenberg:2014rta,Tasinato:2014eka,Allys:2015sht,Jimenez:2016isa,Allys:2016jaq}, as well as their first cosmological applications \cite{Tasinato:2014eka,Tasinato:2014mia,Hull:2014bga,DeFelice:2016cri,DeFelice:2016yws,DeFelice:2016uil,Heisenberg:2016wtr}. The possibility to have several scalar fields has also been investigated \cite{Padilla:2010de,Padilla:2010tj,Padilla:2010ir,Zhou:2010di,Hinterbichler:2010xn,Andrews:2010km,Padilla:2012dx,Sivanesan:2013tba,Garcia-Saenz:2013gya,Charmousis:2014zaa,Saridakis:2016mjd}. Such models are called multi-Galileon ones, and can be considered for arbitrary internal indices, as well as for multi-Galileons in given group representations. A formulation of what would be the most general multi-Galileon theory has been especially discussed in Refs.~\cite{Padilla:2012dx,Sivanesan:2013tba}. However, it has been shown that some models are not included in this previously derived general action, for example the multifield Dirac-Born-Infeld Galileons~\cite{Kobayashi:2013ina}. Additional terms have also been derived e.g. in Ref.~\cite{Deffayet:2010zh} for arbitrary $p$-forms or Ref.~\cite{Ohashi:2015fma} for bi-Galileon theory that do not enter this general class of models. All these extra terms are included in the present construction.

In this paper, we discuss a class of terms satisfying the standard multi-Galileon hypotheses, but which are not included in the general action of Refs.~\cite{Padilla:2012dx,Sivanesan:2013tba}. These Lagrangians, which we call extended multi-Galileon ones, are specific to multi-Galileon models, and identically vanish in the single Galileon case. In the first part, we introduce these new Lagrangians, and examine their different properties, including hidden symmetry properties, \emph{i.e.} which can be made explicit by adding conserved currents to the Lagrangians. Then, in Secs.~\ref{MGPartFundRep} and~\ref{MGPartAdjRep}, we use the previous properties to investigate all the possible dynamics for extended multi-Galileon Lagrangians in the fundamental representation of a SO($N$) or SU($N$) global symmetry group, and in the adjoint representation of a SU($N$) global symmetry group. A similar work has previously been done in~\cite{Padilla:2010ir} for multi-Galileon Lagrangians with equations of motion of order two only. For some of these models, we also perform in parallel an exhaustive examination of all the possible Lagrangians. The results of these systematic investigations, mostly given in Appendices~\ref{MGAppEpsSON} and~\ref{MGAppPartSUN}, are in complete agreement with the dynamics obtained using the complete symmetry properties of the Lagrangians, which strengthens our examination. This investigation also allows us to examine the internal properties of the model, e.g. the link between the possible alternative Lagrangian formulations. We conclude the paper in Sec.~\ref{MGPartConclusion}, in particular by discussing what could be the most general multi-Galileon theory, as well as the link between scalar and vector multi-Galileon models.

\section{Extended multi-Galileon Theory}
\subsection{Presentation}
\label{MGIntroMultiGal}

We discuss in this paper possible Lagrangian terms for multi-Galileon theories in flat spacetime (which we assume is four-dimensional). These terms are built from multi-Galileon fields only (which we also call simply multi-Galileons), \emph{i.e.} scalar fields with internal indices $\pi^\a$. These fields can lie in given group representations, or only be parts of generic nonlinear sigma models. The multi-Galileon theories satisfy the following conditions
\begin{itemize}
\item[i)] \textsl{The Lagrangians contain up to second-order derivatives of the multi-Galileons.}
\item[ii)] \textsl{The Lagrangians are polynomial in the second-order derivatives of the multi-Galileons.}
\item[iii)] \textsl{The field equations contain up to second-order derivatives of the multi-Galileon fields.}
\end{itemize}
The third condition is necessary in order for the theory not to include the Ostrogradski instability~\cite{Woodard:2006nt,Woodard:2015zca}. See also~\cite{Motohashi:2014opa} for a discussion of the instability coming from the third-order derivatives in the equations of motion, especially in the case of multifield theories. It has been proven in Ref.~\cite{Sivanesan:2013tba} that the conditions \textbf{i} and \textbf{iii} imply the condition \textbf{ii}. We however leave it as a hypothesis here.

The starting point for the study of such theories has been the single Galileon case, which has already been examined in detail~\cite{Nicolis:2008in,Deffayet:2009wt,Deffayet:2009mn,Deffayet:2011gz,deRham:2011by}. The generalized Galileon theory has been proven to be the most general one with the hypotheses given below, when considering the case of a unique scalar field~\cite{Deffayet:2011gz}. Its construction begins from the most general theory giving  only second-order equations of motion, \emph{i.e.} Lagrangians of the form
\begin{equation}
\displaystyle{\mathcal{L}^{\text{Gal}}_0 = \delta^{\mu_1 \cdots \mu_m}_{\nu_1\cdots\nu_m} \pi
\partial_{\mu_1}\partial^{\nu_1} \pi \cdots \partial_{\mu_m}\partial^{\nu_m} \pi,}
\label{MGEqLagGalIni0}
\end{equation}
or equivalently, up to a total derivative,
\begin{equation}
\displaystyle{\mathcal{L}^{\text{Gal}}_1 = \delta^{\mu_1 \cdots \mu_m}_{\nu_1\cdots\nu_m} \partial_{\mu_1} \pi
\partial^{\nu_1} \pi \partial_{\mu_2}\partial^{\nu_2}  \pi \cdots \partial_{\mu_m}\partial^{\nu_m} \pi,}
\label{MGEqLagGalIni1}
\end{equation}
with $m$ taking values between $1$ and $4$, and where 
\begin{equation}
\displaystyle{\delta^{j_1 \cdots j_n}_{i_1 \cdots
i_n} =  n! \delta^{j_1
\cdots j_n}_{[i_1 \cdots i_n]} = \frac{1}{(D-n)!}\epsilon^{i_1\cdots i_n \sigma_1 \cdots \sigma_{D-n}}
\epsilon_{j_1\cdots j_n \sigma_1 \cdots \sigma_{D-n}} =   \delta^{j_1}_{i_1} \cdots \delta^{j_n}_{i_n} \pm \cdots,}
\label{MGEqDeltaMult}
\end{equation}
for $n$ running from 1 to 4 (in a four-dimensional spacetime). Then, multiplying those terms by an arbitrary function of $\pi$ and its first derivative still gives second-order equations of motion. This is due to the fact that all the third-order derivatives in the equations of motion coming from the variation of the first-order derivatives in the arbitrary function and the second-order derivatives in the initial Galileon Lagrangian will cancel each other out. Indeed, both terms produce the same third-order derivative contribution, but from a different number of integrations by parts, one and two respectively, granting them an opposite sign (see Ref.~\cite{Deffayet:2011gz} for a detailed discussion of this property).

Multi-Galileon Lagrangians have been obtained in the same way. The Lagrangians giving only second-order equations of motion were first examined in Refs.~\cite{Deffayet:2010zh,Padilla:2010de,Padilla:2010tj}, and consist in adding internal indices to the Galileon Lagrangians of Eqs.~\eqref{MGEqLagGalIni0} or \eqref{MGEqLagGalIni1}. One could then consider the possibility to multiply these Lagrangians by an arbitrary function of the multi-Galileons and their first-order derivatives. However, the property of cancellation of third-order derivative terms discussed in the single Galileon case is not valid anymore, since the internal indices can break the symmetry between the contributions to the equations of motion coming from one or two integrations per part, respectively. Thus, the way to overcome this difficulty is to ensure the necessary symmetry between these pairs of terms~\cite{Sivanesan:2013tba,Padilla:2012dx}. Following this procedure, the related Lagrangians are of the form 
\begin{equation}
\displaystyle{\mathcal{L}^{\text{multiGal}} = A^{\au\cdots \am}\left(X_{\a\b},\pi_\EAc \right) 
\delta^{\mu_1 \cdots \mu_m}_{\nu_1\cdots\nu_m} 
\partial_{\mu_1}\partial^{\nu_1} \pi_{\au} \cdots \partial_{\mu_m}\partial^{\nu_m} \pi_{\am},}
\label{MGEqMultiGalVish}
\end{equation}
with $X_{\a\b}= (1/2) \partial_\rho \pi_\a \partial^\rho \pi_\b$, and with the property that $\partial A^{\au\cdots \am} /\partial X_{\a\b}$ is symmetric in all of its indices $(\au, \cdots,\am,\a,\b)$. The theory spanned by these Lagrangian was determined to be the most general one in Refs.~\cite{Sivanesan:2013tba,Padilla:2012dx}.

However, some authors found that other terms verifying the same hypotheses \textbf{i}-\textbf{iii} are not included in these Lagrangians~\cite{Kobayashi:2013ina,Ohashi:2015fma}. One could then ask if the construction done by multiplying the Lagrangians giving second-order-only equations of motion by such an arbitrary function, while still verifying the above conditions \textbf{i}-\textbf{iii} produce all the possible terms. Indeed, including this arbitrary function which satisfies the symmetry condition introduced in Refs.~\cite{Sivanesan:2013tba,Padilla:2012dx} is a sufficient condition to include extra first-order derivatives in the action, but not a necessary one. Another way to incorporate additional first-order derivatives into such a Lagrangian while keeping second-order equations of motion, if the Lagrangian does not already contain more than two second-order derivatives, is to use the antisymmetry of the term given in Eq.~\eqref{MGEqDeltaMult}. We can indeed write the following Lagrangians, introduced especially in Ref.~\cite{Deffayet:2010zh}:
\begin{equation}
\displaystyle{\begin{array}{l}
\mathcal{L}^{\text{ext}}_{_I} =  \delta^{\mu_1 \mu_2 \mu_3}_{\nu_1\nu_2\nu_3} \partial_{\mu_1}\pi^\a  
\partial_{\mu_2}\pi^\b\partial^{\nu_1}\pi^\EAc\partial^{\nu_2}\pi^\d \partial_{\mu_3}\partial^{\nu_3}\pi^\e,\\
\mathcal{L}^{\text{ext}}_{_{II}} =  \delta^{\mu_1 \mu_2 \mu_3 \mu_4}_{\nu_1\nu_2\nu_3\nu_4} \partial_{\mu_1}\pi^\a  \partial_{\mu_2}\pi^\b\partial^{\nu_1}\pi^\EAc\partial^{\nu_2}\pi^\d \partial_{\mu_3}\partial^{\nu_3}\pi^\e\partial_{\mu_4}\partial^{\nu_4}\pi^\f,\\
\mathcal{L}^{\text{ext}}_{_{III}} =  \delta^{\mu_1 \mu_2 \mu_3 \mu_4}_{\nu_1\nu_2\nu_3\nu_4} \partial_{\mu_1}\pi^\a  \partial_{\mu_2}\pi^\b\partial_{\mu_3}\pi^\EAc\partial^{\nu_1}\pi^\d\partial^{\nu_2}\pi^\e\partial^{\nu_3}\pi^\f \partial_{\mu_4}\partial^{\nu_4}\pi^\g,
\end{array}}
\label{MGEqExtGal}
\end{equation}
where we leave the internal indices totally free for the moment. We call these particular Lagrangians extended multi-Galileon ones. 

Due to the contraction of all derivative terms with a prefactor of the form $\delta^{j_1 \cdots j_n}_{i_1 \cdots i_n} $, it is straightforward to see that these Lagrangians give second-order equations of motion. In addition, their particular structure, implying antisymmetric properties between the internal indices, makes them different from the Lagrangians of Eq.~\eqref{MGEqMultiGalVish}. To see this, one can consider the case of $\mathcal{L}^{\text{ext}}_{_I}$. Terms of the same order from Lagrangians of the form of Eq.~\eqref{MGEqMultiGalVish} are $\mathcal{L}_1 = \partial_\mu\pi^\a \partial^\mu\pi^\b \partial_\nu \pi^\EAc \partial^\nu \pi^\d \partial_\rho\partial^\rho \pi^\e$ and $\mathcal{L}_2 = \pi ^\a\partial_\mu\pi^\b \partial^\mu\pi^\EAc   \delta^{\mu_1 \mu_2 }_{\nu_1\nu_2} \partial_{\mu_3}\partial^{\nu_3}\pi^\d\partial_{\mu_4}\partial^{\nu_4}\pi^\e$. They have to be symmetric in all the internal indices for the first one, and all the internal indices but $\a$ in the second, which means that an antisymmetry under the exchanges of two pairs of indices is not possible. In addition, they do not vanish in the single Galileon limit, whereas $\mathcal{L}^{\text{ext}}_{_I}$ vanishes in this limit. It means that they can be obtained by adding internal indices to single Galileon models, which is not the case for $\mathcal{L}^{\text{ext}}_{_I}$, making this last Lagrangian a purely multi-Galileon one.

These extended Lagrangians can be considered as a basis, in addition to the Lagrangians giving second-order-only equations of motion, to build the generalized multi-Galileon models. Indeed, it is still possible to include an arbitrary function of $\pi^\a$ and its first derivative in front of them, as long as one ensures that the third-orders derivatives appearing in the equations of motion cancel each other out thanks to additional symmetry properties of the internal indices. Note that this kind of generalized multi-Galileon Lagrangians was already partially written in the literature. In the following sections, we will focus on the extended multi-Galileon Lagrangians only.

\subsection{General properties}
\label{MGPartPropGen}

Let us first consider the symmetry properties of these extended Lagrangians. The properties of $\delta^{j_1 \cdots j_n}_{i_1 \cdots i_n} $ impose that the Lagrangians are completely antisymmetric by exchange of the fields with $\partial_{\mu_i}$ derivatives only, as well as by exchange of the fields with $\partial^{\nu_i}$ derivatives only. They are also symmetric by exchange of the groups of fields with $\partial_{\mu_i}$ derivatives only and $\partial^{\nu_i}$ derivatives only, since the $\mu_i$ and $\nu_i$ can be exchanged without modifying the Lagrangian. Finally, $\mathcal{L}^{\text{ext}}_{_{II}}$ is symmetric by exchange of the fields with second-order derivatives. To summarize, and taking the example of 
\begin{equation}
\displaystyle{\mathcal{L}^{\text{ext}}_{_{II}} =  \delta^{\mu_1 \mu_2 \mu_3 \mu_4}_{\nu_1\nu_2\nu_3\nu_4} \partial_{\mu_1}\pi^\a  \partial_{\mu_2}\pi^\b\partial^{\nu_1}\pi^\EAc\partial^{\nu_2}\pi^\d \partial_{\mu_3}\partial^{\nu_3}\pi^\e\partial_{\mu_4}\partial^{\nu_4}\pi^\f,}
\end{equation}
this Lagrangian is 
\begin{itemize}
\item[i)] \textsl{Antisymmetric under the exchanges $\a\leftrightarrow \b$ or $\EAc\leftrightarrow\d$},
\item[ii)] \textsl{Symmetric under the exchange $\e\leftrightarrow\f$},
\item[iii)] \textsl{Symmetric under exchange $(\a,\b)\leftrightarrow (\EAc,\d)$}.
\end{itemize}
These symmetries are very restrictive, particularly for models where there are only a few internal states. For example, in the case of a bi-Galileon model, and for $\mathcal{L}^{\text{ext}}_{_I}$, only two independent terms are possible out of the 32 initial configurations, \emph{i.e.} $\mathcal{L}_1=  \delta^{\mu_1 \mu_2 \mu_3}_{\nu_1\nu_2\nu_3} \partial_{\mu_1}\pi^{\color{red}1\color{black}}
\partial_{\mu_2}\pi^{\color{red}2\color{black}}\partial^{\nu_1}\pi^{\color{red}1\color{black}}\partial^{\nu_2}\pi^{\color{red}2\color{black}} \partial_{\mu_3}\partial^{\nu_3}\pi^{\color{red}1\color{black}}$ and $\mathcal{L}_2=  \delta^{\mu_1 \mu_2 \mu_3}_{\nu_1\nu_2\nu_3} \partial_{\mu_1}\pi^{\color{red}1\color{black}} \partial_{\mu_2} \pi^{
\color{red}2\color{black}} \partial^{\nu_1}\pi^{\color{red}1\color{black}}$ $\times \partial^{\nu_2}\pi^{\color{red}2\color{black}} \partial_{\mu_3}\partial^{\nu_3}\pi^{\color{red}2\color{black}}$.

Another symmetry of this Lagrangian can be seen indirectly, with the following property:
\begin{itemize}
\item[\textbf{a)}] \textsl{If there is an antisymmetry between the internal indices of fields with first and second-order derivatives in a Lagrangian $\mathcal{L}^{\text{ext}}_n$, this Lagrangian is a total derivative. In the case of $\mathcal{L}^{\text{ext}}_{_I}$, it would be for example an antisymmetry between fields with indices $\a$ and $\e$, or $\EAc$ and $\e$.}
\end{itemize}
Let us prove it in the case of  $\mathcal{L}^{\text{ext}}_{_I}$. For this purpose, we can consider the contraction of the internal indices with a function which contains no fields. This function can be arbitrary in a general sigma model, and has to be built from primitive invariant tensors in the case of multi-Galileons in a given group representation. This function has to verify the symmetry properties detailed previously. So, we can write 
\begin{equation}
\label{MGEqLExt1Cont}
\displaystyle{\mathcal{L}^{\text{ext}}_{I,A} = \delta^{\mu_1 \mu_2 \mu_3}_{\nu_1\nu_2\nu_3} \partial_{\mu_1}\pi^\a  
\partial_{\mu_2}\pi^\b\partial^{\nu_1}\pi^\EAc\partial^{\nu_2}\pi^\d \partial_{\mu_3}\partial^{\nu_3}\pi^\e A_{[\a\b][\EAc\d]\e},}
\end{equation}
where the square brackets mean antisymmetrization. Suppose that in addition to these symmetry properties, there is an antisymmetry of $A$ under the exchange $\EAc\leftrightarrow\e$, without loss of generality. In this case, $A$ is antisymmetric on the group of indices $(\EAc,\d,\e)$. Indeed, the antisymmetry under the exchange $\b\leftrightarrow\EAc$ is forced, since the symmetric configuration always vanishes. Then, the only possibility for a tensor that is antisymmetric under the exchanges $\EAc\leftrightarrow\d$ and $\EAc\leftrightarrow\e$ is to be completely antisymmetric on these three indices. It is then possible to build the following current
\begin{equation}
\displaystyle{J^{\mu_1} = \delta^{\mu_1 \mu_2 \mu_3}_{\nu_1\nu_2\nu_3} \pi^\a  
\partial_{\mu_2}\pi^\b\partial^{\nu_1}\pi^\EAc\partial^{\nu_2}\pi^\d \partial_{\mu_3}\partial^{\nu_3}\pi^\e A_{[\a\b][\EAc\d]\e}},
\end{equation}
which is not vanishing if the Lagrangian is not vanishing. 
Using the symmetry properties of $A$, one can compute that
\begin{equation}
\displaystyle{\partial_{\mu_1} J^{\mu_1} = \mathcal{L}^{\text{ext}}_{1,a} +2 \delta^{\mu_1 \mu_2 \mu_3}_{\nu_1\nu_2\nu_3}  \pi^\a  \partial_{\mu_2}\pi^\b\partial_{\mu_1}\partial^{\nu_1}\pi^\EAc\partial^{\nu_2}\pi^\d \partial_{\mu_3}\partial^{\nu_3}\pi^\e A_{[\a\b][\EAc\d]\e}.}
\end{equation}
However, the second term vanishes, since $A$ is antisymmetric by exchange of the indices $\EAc\leftrightarrow\e$, which proves that the $\mathcal{L}^{\text{ext}}_{1,a}$ is a total derivative. Another way to prove this property is by introducing the current 
\begin{equation}
\displaystyle{J^{\mu_3} = \delta^{\mu_1 \mu_2 \mu_3}_{\nu_1\nu_2\nu_3} \partial_{\mu_1} \pi^\a  
\partial_{\mu_2}\pi^\b\partial^{\nu_1}\pi^\EAc\partial^{\nu_2}\pi^\d \partial^{\nu_3}\pi^\e A_{[\a\b][\EAc\d]\e}},
\end{equation}
whose divergence gives 3 times the expected Lagrangian. This current clearly shows that it is the contribution totally antisymmetric here in $(\EAc,\d,\e)$ which is a total derivative, since this is the only one which gives a nonvanishing $J^{\mu_3}$. A similar calculation can be done for the other extended multi-Galileon Lagrangians.

This allows us to prove another symmetry property, which will turn out to be very useful in the following:
\begin{itemize}
\item[\textbf{b)}] \textsl{The extended multi-Galileon Lagrangians with a nontrivial dynamics can be written up to a total derivative in a form that has a complete symmetry under the exchange of the internal indices of one field with a single $\partial_{\mu_j}$ derivative, one field with a single $\partial^{\nu_j}$ derivative, and all the fields with a $\partial_{\mu_k} \partial^{\nu_k}$ derivative, and no other fields. Without loss of generality, we can assume this symmetry between the fields with $\partial_{\mu_1}$, $\partial^{\nu_1}$, $\partial_{\mu_3} \partial^{\nu_3}$ and $\partial_{\mu_4} \partial^{\nu_4}$ (if it exists) derivatives in the Lagrangians of Eq.~\eqref{MGEqExtGal}. In this form, the other indices of fields with one derivatives can be symmetrized by pairs, e.g. the fields with $\partial_{\mu_2}$ and $\partial^{\nu_2}$ derivatives, and so on.}
\end{itemize}
Two independent proofs of this property are given in Appendix~\ref{MGAppProofB}.

An important remark is that property~\textbf{b} does not claim that the Lagrangians which have a nontrivial dynamics possess these particular symmetry properties, but that they can be written as Lagrangians with these symmetry properties plus antisymmetric total derivatives. In other words, the complete symmetry of a Lagrangian can be hidden by total derivatives. It is linked to the fact that starting from such a symmetric Lagrangian, it is always possible to add conserved currents which do not possess these symmetries. On the other hand, this property shows that it is sufficient to look for Lagrangians which have these complete symmetry properties.

Finally, a last property of these extended multi-Galileon Lagrangians can be proven:
\begin{itemize}
\item[\textbf{c)}] \textsl{The equations of motion for the extended multi-Galileon Lagrangians are total derivatives.}
\end{itemize}
It can be shown from the fact that these Lagrangians contain only derivatives of the scalar fields. The equations of motion have thus the following form,
\begin{equation}
\displaystyle{0= \left[\partial_\mu \partial_\nu \left(\frac{\partial}{\partial (\partial_\nu \partial_\mu \pi_\al)} \right)- \partial_\mu \left(\frac{\partial}{\partial (\partial_\mu \pi_\al)}\right) \right] \mathcal{L}^{\text{ext}}_n=
\partial_\mu J^{\mu\al}}
\end{equation}
This last property is useful to verify the consistency of calculations. Indeed, the generalized Galileon actions obtained by multiplying the extended multi-Galileon Lagrangians by functions of $\pi^\a$ and its first derivative do not verify this property anymore.

\subsection{Extended multi-Galileon in group representations}
\label{MGPartPropGroup}

In the following, to reduce the possible contractions between the internal indices, we will work with multi-Galileons lying in a given group representation. We suppose that this group transformation describes a global symmetry of the model, from an effective field theory point of view. Thus, we assume that the Lagrangians behave as singlets under the action of this symmetry group. The representations we will take as examples will be the fundamental representation of a SO($N$) or SU($N$) symmetry group, and the adjoint representation of a SU($N$) symmetry group. A similar work has previously been done in Ref.~\cite{Padilla:2010ir} for multi-Galileon Lagrangians with equations of motion of order two only.

In these representations, it is possible to write the Lagrangians as a contraction between the terms given in Eq.~\eqref{MGEqExtGal} and a prefactor term which is built only from primitive invariant tensors of the multi-Galileon representation, simply called primitive invariants in the following. The primitive invariants of a given group representation are the set of invariant tensors (under the action of the group) with the minimal possible numbers of indices and from which all invariant tensors can be built, using sums, products, and contractions. Primitive invariants of more than two indices are traceless, since another invariant with less indices could otherwise be formed by contracting two indices of the first invariant. For simple groups, primitive invariants can be written in a form where they are either totally symmetric or totally antisymmetric~\cite{Fuchs:1997jv}. Coming back to the extended multi-Galileon Lagrangians, the first of them can be written as
\begin{equation}
\displaystyle{\mathcal{L}^{\text{ext}}_{_I} =  A_{\a\b\EAc\d\e}\delta^{\mu_1 \mu_2 \mu_3}_{\nu_1\nu_2\nu_3} \partial_{\mu_1}\pi^\a  
\partial_{\mu_2}\pi^\b\partial^{\nu_1}\pi^\EAc\partial^{\nu_2}\pi^\d \partial_{\mu_3}\partial^{\nu_3}\pi^\e,}
\end{equation}
with $A_{\a\b\EAc\d\e}$ built from primitive invariants. For example, in the fundamental representation of a SO($N$) symmetry group in a vector representation, the primitive invariants are only $\delta_{\a\b}$ and $\epsilon_{\au\ldots \aN}$. In the adjoint representation of a SU($N$) symmetry group described with one-tensors $\pi^\a$ for $\a \in \{1,\cdots,N^2-1\}$, the primitive invariants are $\delta_{\a\b}$, $f_{\a\b\EAc}$ the structure constants tensor, and $d_{\a\b\EAc}$ the completely symmetric invariant (for $N\geq 3$) \cite{Slansky:1981yr,Fuchs:1997jv,Ramond:2010zz}. These results will be discussed in detail in Secs.~\ref{MGPartFundRep} and~\ref{MGPartAdjRep}. In the following discussion, we will label the representations with only one internal index, to simplify the notation.

Note that when computing the equations of motion, there is a derivation with respect to a field. It means that the equations of motion will be in the conjugate representation of this field\footnote{We remind that the fundamental representation of the orthogonal groups and the adjoint representation of all simple groups are real and self-conjugate, while the fundamental representations of SU($N$) are complex-conjugate and in pair (for $N\geq3$) \cite{Slansky:1981yr}.}. So, the equations of motion will be linear combinations of terms with one free index in the primitive invariant part, that we will label $\al$, yielding
\begin{equation}
\displaystyle{\begin{array}{l}
EOM^{\text{ext}}_{_I}=  A^{\cdots\al\cdots} \delta^{\mu_1 \mu_2 \mu_3}_{\nu_1\nu_2\nu_3} \partial_{\mu_1}\pi \partial^{\nu_1}\pi\partial_{\mu_2}\partial^{\nu_2}\pi \partial_{\mu_3}\partial^{\nu_3}\pi\\
EOM^{\text{ext}}_{_{II}} =   A^{\cdots\al\cdots} \delta^{\mu_1 \mu_2 \mu_3 \mu_4}_{\nu_1\nu_2\nu_3\nu_4} \partial_{\mu_1}\pi \partial^{\nu_1}\pi\partial_{\mu_2}\partial^{\nu_2} \pi \partial_{\mu_3}\partial^{\nu_3}\pi\partial_{\mu_4}\partial^{\nu_4}\pi,\\
EOM^{\text{ext}}_{_{III}} =  A^{\cdots\al\cdots} \delta^{\mu_1 \mu_2 \mu_3 \mu_4}_{\nu_1\nu_2\nu_3\nu_4} \partial_{\mu_1}\pi  \partial_{\mu_2}\pi\partial^{\nu_1}\pi\partial^{\nu_2}\pi\partial_{\mu_3}\partial^{\nu_3}\pi\partial_{\mu_4}\partial^{\nu_4}\pi,
\end{array}}
\end{equation}
where we did not give explicitly the other contractions between group indices, and where in the whole paper the equations of motion read $EOM=0$. Note that the equations of motion for $\mathcal{L}^{\text{ext}}_{_I}$ and $\mathcal{L}^{\text{ext}}_{_{II}}$ do not reduce to multi-Galileon Lagrangians, since they are not in a singlet representation of the symmetry group.

To strengthen the examination of possible Lagrangians, we will also proceed to a systematic search of possible terms without using property~\textbf{b}. To study the possible Lagrangians that can be written, one has to consider the possible contractions between the independent primitive invariant prefactors and the Lagrangians given in Eq.~\eqref{MGEqExtGal}. Thanks to the symmetries of these Lagrangians and to property~\textbf{a}, only few possibilities will remain in most cases. This leads us to the following useful property:
\begin{itemize}
\item[\textbf{d)}] \textsl{If for a given primitive invariant prefactor, it is possible to write only one non-vanishing Lagrangian $\mathcal{L}_1$ which is not a total derivative by contracting this prefactor to a $\mathcal{L}^{\text{ext}}_i$, and if it is possible to write a nonvanishing current by taking off one of the derivatives of a second-order derivative contribution in this Lagrangian, then this Lagrangian is a total derivative.}
\end{itemize}
To prove this result, it is sufficient to consider that the total derivative obtained when taking the divergence of the previously formed current only reduces to contractions of the given primitive invariant prefactor with $\mathcal{L}^{\text{ext}}_i$. Thus, the divergence of this conserved current will give $\mathcal{L}_1$ and Lagrangians which are either vanishing or total derivatives. It shows that $\mathcal{L}_1$ is also a total derivative. Using the same method, one can show the following property:{Let us call $J^\mu_{1}$ one of these currents associated to $\mathcal{L}_{1}$. Its divergence gives $\mathcal{L}_{1}$ and additional Lagrangians. Either all these Lagrangians vanish or are total derivatives, or they contain $\mathcal{L}_{2}$. In the first case, $\mathcal{L}_{1}$ is a total derivative, in the second case, $\mathcal{L}_{1}$ implies the same dynamics as $\mathcal{L}_{2}$.}
\begin{itemize}
\item[\textbf{e)}] \textsl{If for a given primitive invariant prefactor, it is possible to write only two non-vanishing Lagrangians $\mathcal{L}_j$ which are not total derivatives by contracting this prefactor with a $\mathcal{L}^{\text{ext}}_i$, and if it is possible to write a nonvanishing current by taking off one of the derivatives of a second-order derivative term of one of these Lagrangians, then they imply at most one nontrivial dynamics.}
\end{itemize}
Both these properties will be useful in the following.

\section{Extended Multi-Galileons in SO($N$) and SU($N$) fundamental representations}
\label{MGPartFundRep}

\subsection{SO($N$) fundamental representation}
\label{MGPartIntroSON}

In this section, we consider real multi-Galileon fields which transform in the fundamental representation of a SO($N$) global symmetry group. They are labeled as vectors with group indices going from $1$ to $N$. The action of the group on them corresponds to the defining rotation matrices of SO($N$):
\begin{equation}
\pi^\a \rightarrow O^{\a}{}_{\b} \pi^{\b}.
\end{equation}
The representation being self-conjugate and real, the upper and lower indices are equivalent, and can be exchanged with the flat metric $\delta_{\a\b}$. The only primitive invariants of the orthogonal group in this representation are the Kronecker delta $\delta_{\a\b}$ and the Levi-Civita tensor $\epsilon_{\au\cdots\aN}$. 

Following property~\textbf{b} and keeping in mind the Lagrangians of Eq.~\eqref{MGEqExtGal}, it is necessary to build from the primitive invariants totally symmetric terms of rank 3 for $\mathcal{L}^{\text{ext}}_{_I}$ and $\mathcal{L}^{\text{ext}}_{_{III}}$, and of rank 4 for $\mathcal{L}^{\text{ext}}_{_{II}}$. It is only possible for rank 4, yielding $\delta_{(\a\b}\delta_{\EAc\d)}$, where the parentheses denote a symmetrization with the following normalization
\begin{equation}
\label{MGEqNormDeltaSON}
\delta_{(\a\b}\delta_{\EAc\d)} \equiv \delta_{\a\b}\delta_{\EAc\d} + \delta_{\a\EAc}\delta_{\b\d}+ \delta_{\a\d}\delta_{\b\EAc}.
\end{equation}
Thus, the only viable dynamics in this representation is described by the Lagrangian
\begin{equation}
\mathcal{L}^{\text{SO}(N)} = \delta^{\mu_1 \mu_2 \mu_3 \mu_4}_{\nu_1\nu_2\nu_3\nu_4} \partial_{\mu_1}\pi^\a  \partial_{\mu_2}\pi^\b\partial^{\nu_1}\pi^\EAc\partial^{\nu_2}\pi^\d \partial_{\mu_3}\partial^{\nu_3}\pi^\e\partial_{\mu_4}\partial^{\nu_4}\pi^\f \delta_{(\a\EAc}\delta_{\e\f)}\delta_{\b\d}.
\label{MGEqLagUniSON}
\end{equation}
Note that as expected, the invariant prefactor is also symmetric under the exchange $\b\leftrightarrow\d$. Its equation of motion yields
\begin{multline}
\displaystyle{EOM^{\text{SO}(N)} = 8 \delta^{\mu_1 \mu_2 \mu_3 \mu_4}_{\nu_1\nu_2\nu_3\nu_4}
\left[ \partial_{\mu_1}\pi^\b \partial^{\nu_1}\pi^\al\partial_{\mu_2}\partial^{\nu_2}\pi_\b \partial_{\mu_3}\partial^{\nu_3}\pi^\EAc\partial_{\mu_4}\partial^{\nu_4}\pi_\EAc 
\right. }\\\displaystyle{ \left. 
- 2  \partial_{\mu_1}\pi^\b \partial^{\nu_1}\pi_\b \partial_{\mu_2}\partial^{\nu_2}\pi^\al \partial_{\mu_3}\partial^{\nu_3}\pi^\EAc\partial_{\mu_4}\partial^{\nu_4}\pi_\EAc
+ \partial_{\mu_1}\pi^\b \partial^{\nu_1}\pi^\EAc\partial_{\mu_2}\partial^{\nu_2}\pi^\al \partial_{\mu_3}\partial^{\nu_3}\pi_\b\partial_{\mu_4}\partial^{\nu_4}\pi_\EAc\right],}
\label{MGEqEOMSON}
\end{multline}
in its simpler form, where $\al$ is the free index denoting the fact that this equation is in the fundamental representation of the SO($N$) symmetry group. Finally, one can verify that 
\begin{equation}
\displaystyle{EOM^{\text{SO}(N)}  = 8 ~ \partial_{\mu_2} \left[J_1^{\mu_2} - J_2^{\mu_2} \right],}
\end{equation}
with 
\begin{equation}
\displaystyle{
\begin{array}{l}
J_1^{\mu_2} = \delta^{\mu_1 \mu_2 \mu_3 \mu_4}_{\nu_1\nu_2\nu_3\nu_4} \partial_{\mu_1}\pi^\b \partial^{\nu_1}\pi^\al\partial^{\nu_2}\pi_\b \partial_{\mu_3}\partial^{\nu_3}\pi^\EAc\partial_{\mu_4}\partial^{\nu_4}\pi_\EAc,\\
J_2^{\mu_2} = \delta^{\mu_1 \mu_2 \mu_3 \mu_4}_{\nu_1\nu_2\nu_3\nu_4} \partial_{\mu_1}\pi^\b \partial^{\nu_1}\pi_\b \partial^{\nu_2}\pi^\EAc \partial_{\mu_3}\partial^{\nu_3}\pi^\al\partial_{\mu_4}\partial^{\nu_4}\pi_\EAc,
\end{array}}
\end{equation}
as expected from property~\textbf{c}.

\subsection{Exhaustive examination and alternative formulations}

In this section, we make an exhaustive investigation of all the possible Lagrangian terms for multi-Galileons in the fundamental representation of a SO($N$) symmetry group, \emph{i.e.} without using property~\textbf{b}. For that purpose, we consider the prefactors built from the primitive invariants of this representation that give nonvanishing contractions with the $\mathcal{L}^{\text{ext}}_j$ Lagrangians of Eq.~\eqref{MGEqExtGal}. It allows us to verify that the Lagrangian introduced in Sec.~\ref{MGPartIntroSON} describes the only possible dynamic for extended multi-Galileons in this representation. It also allows us to describe in more detail the alternative formulations of the model and their properties. 

Let us first consider the contractions with $\delta_{\a\b}$ only. These contractions are only possible with $\mathcal{L}^{\text{ext}}_{_{II}}$, since it is the only Lagrangian with an even number of fields. Then, taking into account the symmetries of this Lagrangian, the only two independent non-vanishing contractions are 
\begin{equation}
\displaystyle{
\begin{array}{l}
\mathcal{L}^{\text{SO}(N)}_1 = \delta^{\mu_1 \mu_2 \mu_3 \mu_4}_{\nu_1\nu_2\nu_3\nu_4} \partial_{\mu_1}\pi^\a  \partial_{\mu_2}\pi^\b\partial^{\nu_1}\pi_\a\partial^{\nu_2}\pi_\b \partial_{\mu_3}\partial^{\nu_3}\pi^\EAc\partial_{\mu_4}\partial^{\nu_4}\pi_\EAc,\\
\mathcal{L}^{\text{SO}(N)}_2 = \delta^{\mu_1 \mu_2 \mu_3 \mu_4}_{\nu_1\nu_2\nu_3\nu_4} \partial_{\mu_1}\pi^\a  \partial_{\mu_2}\pi^\b\partial^{\nu_1}\pi^\EAc\partial^{\nu_2}\pi_\b \partial_{\mu_3}\partial^{\nu_3}\pi_\a\partial_{\mu_4}\partial^{\nu_4}\pi_\EAc.\\
\end{array}}
\label{MGEqLagEqSON}
\end{equation}
Following property~\textbf{e}, it is straightforward to show that they are related by a total derivative. Indeed, taking 
\begin{equation}
\displaystyle{J^{\mu_4}_{1\leftrightarrow2} = \delta^{\mu_1 \mu_2 \mu_3 \mu_4}_{\nu_1\nu_2\nu_3\nu_4} \partial_{\mu_1}\pi^\a  \partial_{\mu_2}\pi^\b\partial^{\nu_1}\pi_\a\partial^{\nu_2}\pi_\b \partial_{\mu_3}\partial^{\nu_3}\pi^\EAc\partial^{\nu_4}\pi_\EAc,}
\end{equation}
we get
\begin{equation}
\label{MGEqSONCurrent}
\displaystyle{\partial_{\mu_4} J^{\mu_4}_{1\leftrightarrow2} = \mathcal{L}^{\text{SO}(N)}_1 - 2 \mathcal{L}^{\text{SO}(N)}_2.}
\end{equation}
In addition, these two Lagrangians can be related to $\mathcal{L}^{\text{SO}(N)}$, yielding
\begin{equation}
\mathcal{L}^{\text{SO}(N)} = \mathcal{L}^{\text{SO}(N)}_1 + 2 \mathcal{L}^{\text{SO}(N)}_2.
\label{MGEqSONSum}
\end{equation}
Using Eqs.~\eqref{MGEqSONCurrent} and~\eqref{MGEqSONSum}, we see that all three Lagrangians imply the same dynamics. These alternative Lagrangians illustrate the discussion following property~\textbf{b}, since they have a nontrivial dynamic and do not explicitly have the set of four indices which are totally symmetric under the exchange described in property~\textbf{b}. However, this symmetry can be recovered as expected by adding the total derivative of a current with a set of three indices that are totally symmetric under exchange, \emph{i.e.} the lower indices $\a$, $\b$ and $\EAc$ of the current $J^{\mu_4}_{1\leftrightarrow2}$. One can note that the most symmetric version of the Lagrangians is not necessarily the simplest one. For example, the equation of motion given in Eq.~\eqref{MGEqEOMSON} is easier to obtain from the alternative formulations given in Eq.~\eqref{MGEqLagEqSON}.

The Lagrangians that can be constructed with primitive invariant prefactors containing antisymmetric Levi-Civita tensors are examined in Appendix.~\ref{MGAppEpsSON}. None of them give a nontrivial dynamics, as expected. This concludes the complete investigation of possible Lagrangian terms for extended multi-Galileons in the fundamental representation of a SO($N$) global symmetry group, showing that only one dynamics is possible, described by the Lagrangian of Eq.~\eqref{MGEqLagUniSON}. 

\subsection{SU($N$) fundamental representation}

We discuss here the case of complex multi-Galileon fields which transform in the fundamental representation of a SU($N$) global symmetry group. In this case, we have to consider also its complex-conjugate representation. For SU(2), both representations are equivalent, but the fundamental representation is pseudoreal. Then, there exists no basis in which the action of the group elements on the fundamental representation and its complex-conjugate representation are equal, and it is thus better to distinguish them. For SU($N$) with $N\geq3$, which we focus on from now on, the complex-conjugate representations are inequivalent, and have to be distinguished.

We use vector notations with upper indices for the fundamental representation, labeling the multi-Galileon fields by $\pi^\a$ with $\a$ from $1$ to $N$. The complex-conjugates of these fields are labeled with lower indices, \emph{i.e.} denoting  $\pi_\a = \left(\pi^\a\right)^*$. With these notations, they transform under the action of the defining matrices of SU($N$) as 
\begin{equation}
\pi^\a \rightarrow U^{\a}{}_{\b} \pi^\b, ~~~ ~~~ ~~~ \pi_\a \rightarrow U^{\b}{}_{\a} \pi_\b.
\end{equation}
It is important to keep in mind that lower and upper indices label two different representations, and cannot be exchanged by the application of a group metric, as in the SO($N$) case. There are three primitive invariants in this representation, the Kronecker delta $\delta^{\a}{}_{\b}$ and two Levi-Civita tensors $\epsilon_{\au\cdots\aN}$ and $\epsilon^{\au\cdots\aN}$.

To obtain the possible nontrivial Lagrangians satisfying property~\textbf{b} in this case, one has to consider the possible totally symmetric terms built from primitive invariants of ranks 3 and 4. As in the SO($N$) case, it is only possible to build such a term for rank 4, by symmetrizing two Kronecker deltas. This symmetrization has to be done independently of the representations of the fields. The symmetric tensor thus yields
\begin{equation}
\displaystyle{
\delta^{(\a}{}_{\b} \delta^{\EAc}{}_{\d)} \equiv \delta^{\a}{}_{\b} \delta^{\EAc}{}_{\d} + \delta^{\a}{}_{\b} \delta^{\d}{}_{\EAc}  + \delta^{\a}{}_{\EAc} \delta^{\b}{}_{\d} + \delta^{\a}{}_{\EAc} \delta^{\d}{}_{\b} + \delta^{\a}{}_{\d} \delta^{\b}{}_{\EAc} + \delta^{\a}{}_{\d} \delta^{\EAc}{}_{\b}.
}
\end{equation}
In this equation, one has to pay attention to the fact that the lower indices have to be contracted with $\pi^\a$ fields, while the upper indices have to be contracted with complex-conjugate $\pi_\a$ fields. Here, the alphabetical order merely is an indication of the position of the field in the Lagrangian. Then, the only possible Lagrangian is 
\begin{equation}
\displaystyle{
\mathcal{L}^{\text{SU}(N)} = \delta^{\mu_1 \mu_2 \mu_3 \mu_4}_{\nu_1\nu_2\nu_3\nu_4} \partial_{\mu_1}\pi_\a  \partial_{\mu_2}\pi_\b\partial^{\nu_1}\pi^\EAc\partial^{\nu_2}\pi^\d \partial_{\mu_3}\partial^{\nu_3}\pi_\e\partial_{\mu_4}\partial^{\nu_4}\pi^\f \delta^{(\a}{}_{\EAc} \delta^{\e}{}_{\f)} \delta^{\b}{}_{\d} +  \text{h.c.},
}
\label{MGEqLagSUNFund}
\end{equation}
whose equations of motion are similar but more involved than the equations of motion of the real case.

Looking for alternative formulations, one has to pay attention that it is \emph{a priori} possible to build more Lagrangians in the SU($N$) case than in the SO($N$) case. First, the necessity to distinguish fields with upper and lower indices allows us to write several Lagrangians which reduce to the ones of the SO($N$) case in the limit where the fields become real (\emph{i.e.} with a real representation in which complex-conjugate fields are equivalent). For example, the two Lagrangians\footnote{
Note that here, $\mathcal{L}^{\text{SU}(N)}_1$ to $\mathcal{L}^{\text{SU}(N)}_3$ are written with expressions which are already real, thanks to their symmetry properties. We however prefer to write them explicitly with the h.c. contribution, to remind the reader that they are built from complex-conjugate fields. In addition, it simplifies the calculations when dealing with other Lagrangians which do not have this property, as it is the case e.g. for
\begin{equation}
\displaystyle{
\mathcal{L}^{\text{SU}(N)}_4 =  \delta^{\mu_1 \mu_2 \mu_3 \mu_4}_{\nu_1\nu_2\nu_3\nu_4} \partial_{\mu_1}\pi^\a  \partial_{\mu_2}\pi^\b\partial^{\nu_1}\pi_\b\partial^{\nu_2}\pi^\EAc \partial_{\mu_3}\partial^{\nu_3}\pi_\a\partial_{\mu_4}\partial^{\nu_4}\pi_\EAc + \text{h.c.}.
}
\end{equation}
}
\begin{equation}
\displaystyle{
\begin{array}{l}
\mathcal{L}^{\text{SU}(N)}_1 =  \delta^{\mu_1 \mu_2 \mu_3 \mu_4}_{\nu_1\nu_2\nu_3\nu_4} \partial_{\mu_1}\pi^\a  \partial_{\mu_2}\pi^\b\partial^{\nu_1}\pi_\a\partial^{\nu_2}\pi_\b \partial_{\mu_3}\partial^{\nu_3}\pi^\EAc\partial_{\mu_4}\partial^{\nu_4}\pi_\EAc + \text{h.c.},\\
\mathcal{L}^{\text{SU}(N)}_2 =  \delta^{\mu_1 \mu_2 \mu_3 \mu_4}_{\nu_1\nu_2\nu_3\nu_4} \partial_{\mu_1}\pi^\a \partial_{\mu_2}\pi_\b\partial^{\nu_1}\pi_\a\partial^{\nu_2}\pi^\b \partial_{\mu_3}\partial^{\nu_3}\pi^\EAc\partial_{\mu_4}\partial^{\nu_4}\pi_\EAc + \text{h.c.},
\end{array}
}
\end{equation}
reduce to $\mathcal{L}^{\text{SO}(N)}_1$ of Eq.~\eqref{MGEqLagEqSON} in the real case. Similarly, three different Lagrangians can be written in the SU($N$) case which reduce to $\mathcal{L}^{\text{SO}(N)}_2$ of Eq.~\eqref{MGEqLagEqSON}. Then, dealing with two representations which are not equivalent, it is also possible to write down Lagrangians which would vanish due to symmetry considerations in the SO($N$) limit. One can for example consider
\begin{equation}
\displaystyle{
\mathcal{L}^{\text{SU}(N)}_3 =  \delta^{\mu_1 \mu_2 \mu_3 \mu_4}_{\nu_1\nu_2\nu_3\nu_4} \partial_{\mu_1}\pi^\a  \partial_{\mu_2}\pi_\a\partial^{\nu_1}\pi_\b\partial^{\nu_2}\pi^\EAc \partial_{\mu_3}\partial^{\nu_3}\pi^\b\partial_{\mu_4}\partial^{\nu_4}\pi_\EAc + \text{h.c.},
}
\end{equation}
which identically vanishes in the real limit. This indicates that the alternative formulations of $\mathcal{L}^{\text{SU}(N)}$ are \emph{a priori} more numerous than in the real case.

\subsection{Summary}

For the extended multi-Galileon models in the fundamental representation of a global SO($N$) symmetry group, the only nontrivial dynamics is given by 
\begin{equation}
\label{MGEqLagFinSON}
\mathcal{L}^{\text{SO}(N)} = \delta^{\mu_1 \mu_2 \mu_3 \mu_4}_{\nu_1\nu_2\nu_3\nu_4} \partial_{\mu_1}\pi^\a  \partial_{\mu_2}\pi^\b\partial^{\nu_1}\pi^\EAc\partial^{\nu_2}\pi^\d \partial_{\mu_3}\partial^{\nu_3}\pi^\e\partial_{\mu_4}\partial^{\nu_4}\pi^\f \delta_{(\a\EAc}\delta_{\e\f)}\delta_{\b\d}.
\end{equation}
Equivalent Lagrangians are given in Eq.~\eqref{MGEqLagEqSON}, and its equation of motion can be found in Eq.~\eqref{MGEqEOMSON}. There is also one possible nontrivial dynamics in the case of the fundamental representation of a global SU($N$) symmetry group, where the only possible dynamics is described by the Lagrangian given in Eq.~\eqref{MGEqLagSUNFund}. However, more equivalent Lagrangians can \emph{a priori} be written in this case.

This whole section showed that as discussed previously, the properties of the extended multi-Galileon models are very constraining and allow us to build only a few independent nontrivial dynamics, even if it is possible to write a lot of nonvanishing Lagrangians. Properly used, these properties drastically simplify the research of possible nontrivial Lagrangian terms. 

\section{Extended multi-Galileons in the SU($N$) adjoint representation}
\label{MGPartAdjRep}
\subsection{SU($N$) adjoint representation}
\label{MGPartIntroSUN}

In this section, we consider multi-Galileon fields in the adjoint representation of a global SU($N$) symmetry group. We will denote these fields with one single index going from 1 to $N^2-1$ as the dimension of the group (\emph{i.e.} using the adjoint module of the group). The action of the group elements labeled with $\a$ on a multi-Galileon $\pi_\b$ yields
\begin{equation}
\pi_\b ~~~ \rightarrow ~~~ (T_\a)_\b{}^\EAc \pi_\EAc = f_{\a\b}{}^\EAc \pi_\b,
\end{equation}
where $f_{\a\b}{}^\EAc$ the structure constants of SU($N$). We chose here to focus on the adjoint representation of a SU($N$) rather than a SO($N$) symmetry group, due to the fact that the low-rank SO($N$) groups can be related to other symmetry groups thanks to local isomorphisms. This is for example the case for SO(3) and SU(2), SO(4) and SU(2)$\times$SU(2), SO(5) and Sp(4), and SO(6) and SU(4).

The metric on this representation is $g_{\a\b}=f_{\a\al}{}^\be f_{\b\be}{}^\al$, and can be used to raise and lower the indices. It allows us to work with the completely antisymmetric forms of the structure constant, $f_{\a\b\EAc}$, as well as $f^{\a\b\EAc}$. The primitive invariants in this representation are $g_{\a\b}$, $f_{\a\b\EAc}$, and $d_{\a\b\EAc}$ which is the symmetric primitive invariant for $N\geq3$ \cite{Fuchs:1997jv,deAzcarraga:1997ya,Ramond:2010zz}. In the SU(3) case, the link between this basis and the one of the defining matrices of SU($N$) was discussed e.g. in Ref.~\cite{Dittner:1972hm}. We choose a basis where the group metric is proportional to the Kronecker delta $\delta_{\a\b}$. In addition, we can impose the following normalizations \cite{Fuchs:1997jv}
\begin{equation}
f^{\a\b\EAc}f^{\a\b\d}= N \delta^{\EAc\d}, ~~~ ~~~ d^{\a\b\EAc}d^{\a\b\d} = \left(N - \frac4N \right) \delta^{\EAc\d}.
\end{equation}
The values of $f_{\a\b\EAc}$ and $d_{\a\b\EAc}$ in the SU(3) case can be found e.g. in Ref.~\cite{deAzcarraga:1997ya}. Concerning the contractions of these primitive invariants, some properties are very useful. First, the contractions of primitive invariants with the Kronecker delta tensor can be omitted since this tensor only raises or lowers the group indices. Then, the contractions between $f$ and $d$ primitive invariants of ranks zero to three, \emph{i.e.} with zero to three indices not contracted together, are also known~\cite{Metha:1983mng,Fuchs:1997jv}; they vanish for ranks zero and one, are proportional to the Kronecker delta for rank two, and proportional to $f$ or $d$ for rank three\footnote{The result is proportional to $f$ if there is an odd number of $f$ and an even number of $d$, and to $d$ if there is an even number of $f$ and an odd number of $d$. The case where the numbers of $f$ and $d$ have the same parity is not possible since it would not be possible to have three indices not contracted.}. 

Following property~\textbf{b}, we will discuss the possible independent dynamics by considering the totally symmetric terms of ranks 3 and 4 that can be built from primitive invariants. It will allow us to build only a few independent Lagrangians, that we will examine in Sec.~\ref{MGPartSUNLag}. In Appendix~\ref{MGAppPartSUN}, we also make an exhaustive examination of all the possible Lagrangians at the order of $\mathcal{L}^{\text{ext}}_{_I}$. Finding similar Lagrangians by both methods corroborates our approach, and shows that we effectively describe all the possible dynamics for multi-Galileon fields in the adjoint representation of a SU($N$) symmetry group. The cases of SU(2) and SU(3) symmetry groups  are then discussed in Sec.~\ref{MGPartSU2&3}. 

\subsection{Multi-Galileon Lagrangians}
\label{MGPartSUNLag}

To obtain the Lagrangians implying the different nontrivial dynamics of this model, we need to discuss the possible totally symmetric invariants of ranks 3 and 4. For rank 3, only one such invariant is possible, $d_{\a\b\EAc}$, which is nonzero only for $N\geq3$. For rank 4, two such invariants are possible. The first one can be built from the Kronecker delta, and gives $\delta_{(\a\b}\delta_{\EAc\d)}$, where we take the normalization of Eq.~\eqref{MGEqNormDeltaSON}. The second one is the Casimir invariant at order $4$, for $N\geq 4$ (see e.g. Refs.~\cite{Fuchs:1997jv,Ramond:2010zz}), and can be written as
\begin{equation}
\label{MGPrimInvFour}
\displaystyle{d_{\be(\a\b} d^{\be}{}_{\EAc\d)} \equiv d_{\be\a\b} d^{\be}{}_{\EAc\d} + d_{\be\a\EAc} d^{\be}{}_{\b\d} +d_{\be\a\d} d^{\be}{}_{\b\EAc}.}
\end{equation}
With the knowledge of these totally symmetric invariants, one can thus investigate the possible Lagrangians with a nontrivial dynamics, using property~\textbf{b} as well as the form of the Lagrangians given in Eq.~\eqref{MGEqExtGal}. 

At the order of $\mathcal{L}^{\text{ext}}_{_I}$, only one Lagrangian is possible, yielding
\begin{equation}
\label{MGLagFinSUN1}
\displaystyle{\mathcal{L}^{\text{Adj},1}_{_I} =  \delta^{\mu_1 \mu_2 \mu_3}_{\nu_1\nu_2\nu_3} \partial_{\mu_1}\pi^\a  
\partial_{\mu_2}\pi^\b\partial^{\nu_1}\pi^\EAc\partial^{\nu_2}\pi^\d \partial_{\mu_3}\partial^{\nu_3}\pi^\e d_{\a\EAc\e}\delta_{\b\d}.}
\end{equation}
Note that a systematic examination of all the possible terms at this order is presented in Appendix~\ref{MGAppPartSUN}, showing that the dynamics implied by this Lagrangian is the only nontrivial one. The equations of motion of this Lagrangian are given in Eq.~\eqref{MGAppEqEOMSUN}.
At the order of $\mathcal{L}^{\text{ext}}_{_{II}}$, the two possible fourth-rank totally symmetric invariants can be used. It yields the two following Lagrangians
\begin{equation}
\label{MGLagFinSUN2}
\displaystyle{
\begin{array}{l}
\mathcal{L}^{\text{Adj},1}_{_{II}} =  \delta^{\mu_1 \mu_2 \mu_3 \mu_4}_{\nu_1\nu_2\nu_3\nu_4} \partial_{\mu_1}\pi^\a  \partial_{\mu_2}\pi^\b\partial^{\nu_1}\pi^\EAc\partial^{\nu_2}\pi^\d \partial_{\mu_3}\partial^{\nu_3}\pi^\e\partial_{\mu_4}\partial^{\nu_4}\pi^\f \delta_{(\a\EAc}\delta_{\e\f)}\delta_{\b\d},\\
\mathcal{L}^{\text{Adj},2}_{_{II}} =  \delta^{\mu_1 \mu_2 \mu_3 \mu_4}_{\nu_1\nu_2\nu_3\nu_4} \partial_{\mu_1}\pi^\a  \partial_{\mu_2}\pi^\b\partial^{\nu_1}\pi^\EAc\partial^{\nu_2}\pi^\d \partial_{\mu_3}\partial^{\nu_3}\pi^\e\partial_{\mu_4}\partial^{\nu_4}\pi^\f d_{\be(\a\EAc} d^{\be}{}_{\e\f)}\delta_{\b\d}.
\end{array}
}
\end{equation}
At the order of $\mathcal{L}^{\text{ext}}_{_{III}}$, one needs a rank-three symmetric tensor, which can only be $d_{\a\b\EAc}$. Then, four indices still have to be contracted, yielding the following possibilities:
\begin{equation}
\label{MGLagFinSUN3}
\displaystyle{
\begin{array}{l}

\mathcal{L}^{\text{Adj},1}_{_{III}} =  \delta^{\mu_1 \mu_2 \mu_3 \mu_4}_{\nu_1\nu_2\nu_3\nu_4} \partial_{\mu_1}\pi^\a  \partial_{\mu_2}\pi^\b\partial_{\mu_3}\pi^\EAc\partial^{\nu_1}\pi^\d\partial^{\nu_2}\pi^\e\partial^{\nu_3}\pi^\f \partial_{\mu_4}\partial^{\nu_4}\pi^\g d_{\a\d\g} \delta_{\b\e}\delta_{\EAc\f}.\\
\mathcal{L}^{\text{Adj},2}_{_{III}} =  \delta^{\mu_1 \mu_2 \mu_3 \mu_4}_{\nu_1\nu_2\nu_3\nu_4} \partial_{\mu_1}\pi^\a  \partial_{\mu_2}\pi^\b\partial_{\mu_3}\pi^\EAc\partial^{\nu_1}\pi^\d\partial^{\nu_2}\pi^\e\partial^{\nu_3}\pi^\f \partial_{\mu_4}\partial^{\nu_4}\pi^\g d_{\a\d\g} d^{\be}{}_{\b\e} d_{\be\d\f},
\end{array}
}
\end{equation}
There are possible contractions with terms in $ d_{\a\d\g}f^{\be}{}_{\b\EAc} f_{\be\e\f}$ and $d_{\a\d\g}f^{\be}{}_{\b\e} f_{\be\EAc\f}$, which are related by the Jacobi identity. But these terms would in fact identically vanish, since they imply an antisymmetrical contraction between $\a$ and $\d$. For example, exchanging the positions of $\b$ and $\e$ thanks to a $f^{\be}{}_{\b\e}$ terms shows the antisymmetry between $\a$ and $\e$. Then, and starting from the initial configuration, exchanging the positions of $\a$ and $\e$ thanks to the previous antisymmetry puts the Lagrangian in a form which involves a $\partial^{\nu_1} \pi^{\d} \partial^{\nu_2} \pi^{\a}$ part which identically vanishes due to the contractions with $\delta^{\mu_1 \mu_2 \mu_3 \mu_4}_{\nu_1\nu_2\nu_3\nu_4}$ and $d_{\a\d\g}$. 

Finally, we exactly recover the symmetries of property~\textbf{b}. Indeed, the Lagrangians built from $\mathcal{L}^{\text{ext}}_{_I}$ and $\mathcal{L}^{\text{ext}}_{_{II}}$ are symmetric under the exchange $\b\leftrightarrow\d$, and the Lagrangians built from $\mathcal{L}^{\text{ext}}_{_{III}}$ are symmetric by pairs for the set of indices $(\b,\EAc,\e,\f)$. These Lagrangians describe the only possible dynamics of extended multi-Galileon models with fields in the adjoint representation of a SU($N$) symmetry group. The model is simplified in the case of $N=2$ or $N=3$, as discussed in Sec.~\ref{MGPartSU2&3}. Note also that the properties we introduced drastically simplify the investigation of such a model. For example, investigating $\mathcal{L}^{\text{ext}}_{_{III}}$ would otherwise imply considering the 1557 singlet configurations built from seven adjoint fields in the case of SU(4) ~\cite{Slansky:1981yr,Feger:2012bs}.

\subsection{SU(2) and SU(3) cases}
\label{MGPartSU2&3}

In the previous section, we investigated the case of a general SU($N$) symmetry group. However, this study can be simplified in the case of a SU(3) or SU(2) symmetry group. In the case of SU(3), there is only one fourth-rank Casimir symmetric invariant. Indeed, one has the following relations between the primitive invariants~\cite{Fuchs:1997jv}:
\begin{equation}
 d_{\a\b\be}d_{\EAc\d\be} = \frac12\left(\delta_{\a\EAc}\delta_{\b\d} + \delta_{\b\EAc}\delta_{\a\d} - \delta_{\a\b}\delta_{\EAc\d} \right) +  f_{\a\EAc\be}f_{\b\d\be} +  f_{\a\d\be}f_{\b\EAc\be} ,
\end{equation}
allowing us to write as expected $d_{\be(\a\b} d^{\be}{}_{\EAc\d)}$ as a function of the Kronecker deltas and terms implying the structure constant tensors which are total derivative contributions thanks to property \textbf{a}. The same relation can be used to write the Lagrangian $\mathcal{L}^{\text{Adj},2}_{_{III}}$ as a linear combination of $\mathcal{L}^{\text{Adj},1}_{_{III}}$ plus some terms which give identically vanishing Lagrangians as explained in the previous section. Thus, the Lagrangians which describe the possible independent dynamics for extended multi-Galileon models in the adjoint representation of a SU(3) symmetry group are $\mathcal{L}^{\text{Adj},1}_{_I}$, $\mathcal{L}^{\text{Adj},1}_{_{II}}$ and $\mathcal{L}^{\text{Adj},1}_{_{III}}$.

The case of SU(2) is even simpler. In this symmetry group, there is no $d_{\a\b\EAc}$ primitive invariant. Therefore, most of the Lagrangians of the general SU($N$) case vanish. The only possible dynamics is thus described by the $\mathcal{L}^{\text{Adj},1}_{_{II}}$ Lagrangian. One can note that as SU(2) is locally isomorphic to SO(3), one should recover the same results for the three-dimensional representations of both groups. It is indeed the case, since the result for the adjoint representation of SU(2) is exactly the one we found for the fundamental representation of SO(3) in Sec.~\ref{MGPartFundRep}.

\section{Conclusion and discussions}
\label{MGPartConclusion}

In this paper, we investigated possible terms included in the generalized multi-Galileon theories, which we call extended multi-Galileon Lagrangians. Some of these terms were already introduced in the literature: the possibility for more than two fields with first-order derivative only to be contracted with the same $\delta^{\mu_1\cdots\mu_m}_{\nu_1\cdots\nu_m}$ was discussed in Ref.~\cite{Deffayet:2010zh} for a theory containing several $p$-forms, and the Lagrangian $\mathcal{L}^{\text{ext}}_{_I}$ for biscalar theories was given in Ref.~\cite{Ohashi:2015fma}. We performed here a systematic examination of those Lagrangians. We first discussed the general properties of these Lagrangians, showing that they are strongly constrained by symmetry relations, which can be explicit or hidden up to a total derivative. Then, using these symmetry properties, we examined the possible dynamics for multi-Galileons in the fundamental representation of a SO($N$) or SU($N$) symmetry group, and for the adjoint representation of a SU($N$) symmetry group. In the case of the fundamental representation of SO($N$) and of certain terms of the adjoint of SU($N$), we also performed in parallel a complete investigation of the possible nontrivial dynamics. The results of these investigations are in complete agreement with the Lagrangians built from symmetry considerations, and also allowed us to discuss internal properties of the model.

A next step would be to investigate what would be the most general theory for multi-Galileon fields. Following the works of Refs.~\cite{Padilla:2010de,Padilla:2010tj,Sivanesan:2013tba,Padilla:2012dx}, we would expect Lagrangians of the form 
\begin{multline}
\displaystyle{
\mathcal{L}= A^{\color{red}a_1 b_1  \color{black}\hspace{-0.05cm}\cdots  \color{red}a_n b_n c_1 \color{black}\hspace{-0.05cm}\cdots \color{red}c_{(m-n)}\color{black}}\left(X_{\a\b},\pi_\EAc \right) 
\delta^{\mu_1 \cdots \mu_m}_{\nu_1\cdots\nu_m} 
\partial_{\mu_1} \pi_{\color{red}a_1\color{black}} \partial^{\nu_1} \pi_{\color{red}b_1\color{black}} \hspace{-0.15cm}\cdots \partial_{\mu_n} \pi_{\color{red}a_n\color{black}} \partial^{\nu_n} \pi_{\color{red}b_n\color{black}}
}\\\displaystyle{
\partial_{\mu_{(n+1)}}\partial^{\nu_{(n+1)}} \pi_{\color{red}c_1\color{black}}
\hspace{-0.15cm}\cdots
\partial_{\mu_{m}}\partial^{\nu_{m}} \pi_{\color{red}c_{(m-n)}\color{black}}
 }
\end{multline}
where $X_{\a\b}= (1/2) \partial_\rho \pi_\a \partial^\rho \pi_\b$, $m$ goes from 1 to 4, and $n$ from 0 to $m-1$, and where we also expect $\partial A^{\color{red}a_1 b_1  \color{black}\hspace{-0.05cm}\cdots  \color{red}a_n b_b c_1 \color{black}\hspace{-0.05cm}\cdots \color{red}c_{(m-n)}\color{black}} /\partial X_{\a\b}$ to be symmetric in its indices $(\color{red}c_1\color{black},\cdots,\color{red}c_{(m-n)}\color{black},\a,\b)$ in order to have second-order equations of motion. An investigation of these Lagrangians for multi-Galileons in group representations of a global symmetry group would also allow us to have a better understanding of their properties. It would then be interesting to investigate this kind of models while allowing the theory to be degenerate, with Lagrangians potentially involving third-order-derivatives equations of motion, \emph{i.e.} in a beyond-Horndeski context~\cite{Gleyzes:2014dya,Langlois:2015cwa,Langlois:2015skt,Motohashi:2016ftl,Crisostomi:2016czh}.

The link with vector multi-Galileon models can also be very fruitful. Vector Galileon models have been examined in Refs.~\cite{Heisenberg:2014rta,Tasinato:2014eka,Allys:2015sht,Jimenez:2016isa,Allys:2016jaq}, and their cosmological applications have been explored e.g. in Refs.~\cite{Tasinato:2014eka,Tasinato:2014mia,Hull:2014bga,DeFelice:2016cri,DeFelice:2016yws,DeFelice:2016uil,Heisenberg:2016wtr}. Those models are built from the same requirements as those of scalar Galileons, in addition to the requirement that the vector field propagate at most three degrees of freedom. Its longitudinal component described by considering the pure scalar part of the vector only, \emph{i.e.} considering the $\partial_\mu \pi$ contribution to $A_\mu$ in its scalar-vector decomposition, should also have second-order equations of motion.

Recently, models of vector multi-Galileons have been developed \cite{Allys:2016kbq,Jimenez:2016upj}. The case of vector Galileons lying in the adjoint representation of a SU(2) global symmetry is particularly interesting since it can source or contribute to the inflation while keeping isotropy \cite{Golovnev:2008cf,ArmendarizPicon:2004pm}. In those models, vector multi-Galileons are denoted $A_\mu^\a$, and transform for the group indices similarly to the scalar multi-Galileons, see Sec.~\ref{MGPartIntroSUN}. Starting from a vector Lagrangian, and considering its pure longitudinal contribution, \emph{i.e.} doing the replacement $A_\mu^\a \rightarrow \partial_\mu \pi^\a$, one recovers a scalar multi-Galileon model. In fact, this link is possible for the vector Lagrangians whose derivative parts can be written as a function of its symmetric form $S_{\mu\nu}^\a = \partial_\mu A_\nu^\a + \partial_\nu A_\mu^\a$ only, the antisymmetric one vanishing in the pure scalar sector.

Thus, it is also possible to obtain vector Lagrangians when starting from the scalar ones. Actually, at least all the terms that are functions of $A_\mu^\a$ and $S_{\mu\nu}^\a$ can be obtained from the scalar sector, which makes a very strong link between both theories. For that purpose, it is sufficient to consider the scalar multi-Galileon Lagrangian formulations given e.g. in Eqs.~\eqref{MGEqMultiGalVish} and~\eqref{MGEqExtGal}, and to do the replacement\footnote{Assuming that we consider Lagrangian forms which contain only first- and second-order derivatives of the scalar field, and no fields without any derivatives.} $\partial_\mu \pi^\a \rightarrow A_\mu^\a$. The second-order derivative of scalars can be promoted to $S_{\mu_i}{}^{\nu_i\a}$ or $G_{\mu_i}{}^{\nu_i\a}=\partial_{\mu_i} A^{\nu_i\a} - \partial^{\nu_i }A_{\mu_i}^\a$ terms, giving different kinds of terms. This procedure produces viable Lagrangians in the vector sector\footnote{It was shown in Ref.~\cite{Allys:2016jaq} that all Lagrangians built from contractions of one or two Levi-Civita tensors with vector Galileons and its first derivative propagate at most three degrees of freedom, as desired. This result, discussed in the single Galileon case, can be immediately extended to the multi-Galileon case.}.

The important point is that Lagrangians which are equivalent up to a total derivative in the scalar sector can not be related anymore when doing the replacement $\partial_\mu \pi^\a \rightarrow A_\mu^\a$. Indeed, the divergence of the associated currents could produce terms in $\partial_{\mu_i}\partial^{\nu_j}A_{\mu_k}^\a = (1/2)\partial^{\nu_j}G_{\mu_j\mu_k}^\a$, which vanish in the scalar sector. Several Lagrangians in the vector sector can then be obtained when starting from alternative equivalent formulations of a given dynamic in the scalar sector before promoting scalars to vectors. Such an examination was performed in Ref.~\cite{Allys:2016kbq}, where all the equivalent formulations of the pure scalar sector were detailed at a given order, and a comparison with the vector sector was done. Then, any examination of a vector multi-Galileon model should be performed in parallel to an examination of the associated scalar multi-Galileon model. It would be interesting to investigate this link in more detail in future works.


\subsection*{Acknowledgments} 

I wish to thank P.~Peter and Y.~Rodriguez for their support and advice, especially at the early stage of the project. I thank G.~Esposito-Farese for many valuable discussions and suggestions, and for a critical reading of the manuscript. I also thank C.~Deffayet, P.~Saffin and V.~Sivanesan for enlightening discussions, and T.~Marchand and J.~Van Dijk for their comments on my first draft.

\section{Appendix}
\subsection{Proof of property~\textbf{b}}
\label{MGAppProofB}

Before giving the proof of the property~\textbf{b}, let us recall some properties about the symmetry properties of a group of indices. To consider the symmetry properties under the exchange of an ensemble of variables, we will refer to the states with symmetry properties associated to Young diagrams, describing the different representations of the permutation group~\cite{Landau:1991wop}. These symmetrized state are such that the symmetry cannot be higher, which means that applying a symmetrization or anti-symmetrization on them gives either zero or a linear combination of states whose symmetry properties are related to other Young diagrams. One has to pay attention to the fact that symmetrizing or antisymmetrizing on two indices can change the symmetry properties of both these indices. For example, starting from a tensor that is symmetric under the exchange $\a\leftrightarrow\b$, and symmetrizing the indices $\b$ and $\EAc$, the result can be not symmetric anymore under the exchange $\a\leftrightarrow\b$. On the other hand, a (anti)symmetry on more than two indices implies a complete (anti)symmetry under exchange of the group of indices. For example, if there is a symmetry under the exchanges $\a\leftrightarrow\b$ and $\b\leftrightarrow\EAc$, then there is a complete symmetry for the $(\a,\b,\EAc)$ group of indices.

Let us first prove the property~\textbf{b} in the case of the $\mathcal{L}^{\text{ext}}_{_I}$, still using the notation of Eq.~\eqref{MGEqLExt1Cont}, \emph{i.e.} considering a Lagrangian with an arbitrary $A^0_{[\a\b][\EAc\d]\e}$ prefactor. We will proceed in two steps, introducing at each step a current allowing us to improve the symmetry properties of the Lagrangian we started with. In a first step, we remove the contribution to $A^0_{[\a\b][\EAc\d]\e}$ that is totally antisymmetric for the group of indices $(\a,\b,\EAc,\d)$, which is a total derivative. To see it, one can take into account that the equations of motion will be linear combination of terms of the form $\delta^{\mu_1 \mu_2 \mu_3}_{\nu_1\nu_2\nu_3} \partial_{\mu_1}\pi^\al \partial^{\nu_1}\pi^\be\partial_{\mu_2}\partial^{\nu_2} \pi^{\color{red}\gamma\color{black}}\partial_{\mu_3}\partial^{\nu_3}\pi^{\color{red}\delta\color{black}}$ with internal indices contracted with four of the indices of $A_{[\a\b\EAc\d]\e}$. However, due to the fact that each term in the equations of motion is symmetric under the exchanges $\al\leftrightarrow\be$ and $\color{red}\gamma\color{black}\leftrightarrow\color{red}\delta\color{black}$, no non-vanishing contractions can be done with such an antisymmetric prefactor, and the equations of motion identically vanish. We call the new prefactor obtained this way $A^1_{[\a\b][\EAc\d]\e}$.

In fact, we removed in this step all the components of $A^0_{[\a\b][\EAc\d]\e}$ which are antisymmetric under the exchange of one index of $(\a,\b)$ and one index of $(\EAc,\d)$. To see it, let us assume for example that there is an antisymmetry on $\a\leftrightarrow\EAc$. Then, as there is a symmetry under the exchange $(\a,\b)\leftrightarrow(\EAc,\d)$, there is also an antisymmetry on $\b\leftrightarrow\d$. In addition, permuting e.g. $\a$ and $\EAc$ in the Lagrangian, and using the symmetry properties of $\delta^{\mu_1 \mu_2 \mu_3}_{\nu_1\nu_2\nu_3}$, one can show that the symmetric configurations in $\EAc\leftrightarrow\b$ and $\a\leftrightarrow\d$ identically vanish, which finally implies that there is a complete antisymmetry for the $(\a,\b,\EAc,\d)$ group of indices. This result is due to the strong symmetry conditions imposed by the structure of the extended multiGalileon Lagrangians, and will be useful in the following. It is then possible to symmetrize two pairs of indices between the $(\a,\b)$ and $(\EAc,\d)$ groups. Without loss of generality, we can consider that the pairs are $\a\leftrightarrow\EAc$ and $\b\leftrightarrow\d$, absorbing some minus signs in $A^1_{[\a\b][\EAc\d]\e}$ if necessary.

Let us now focus on the $(\a,\b,\e)$ group of indices. Using the property~\textbf{a}, it is possible to remove the contribution totally antisymmetric in $\a$, $\b$ and $\e$, which is a total derivative. The associated conserved current is obtained by removing $\partial^{\nu_3}$ from $\mathcal{L}^{\text{ext}}_{_I}$, and reads
\begin{equation}
\displaystyle{
J_{\nu_3}^{1} =\frac13 \delta^{\mu_1 \mu_2 \mu_3}_{\nu_1\nu_2\nu_3} \partial_{\mu_1}\pi^\a  
\partial_{\mu_2}\pi^\b\partial^{\nu_1}\pi^\EAc\partial^{\nu_2}\pi^\d \partial_{\mu_3}\pi^\e A^1_{[\a\b][\EAc\d]\e} = \frac13 \delta^{\mu_1 \mu_2 \mu_3}_{\nu_1\nu_2\nu_3} \partial_{\mu_1}\pi^{[\a}  
\partial_{\mu_2}\pi^\b\partial_{\mu_3}\pi^{\e]}\partial^{\nu_1}\pi^\EAc\partial^{\nu_2}\pi^\d  A^1_{[\a\b][\EAc\d]\e}.}
\end{equation}
The initial symmetry of the group of indices $(\a,\b,\e)$ being described by a linear combination of Young diagrams, and the totally antisymmetric one having been removed, each other terms are symmetric on at least two indices. This pair of indices cannot be $\a$ and $\b$, and we can always consider it as $\a$ and $\e$ (permuting $\partial_{\mu_1}$ and $\partial_{\mu_2}$ when necessary). In addition, the current $J_{\nu_1}^{1} $ does not contain a part which is antisymmetric on the group of indices $(\a,\b,\EAc,\d)$. Indeed, such a part would be in fact antisymmetric for the group of indices $(\a,\b,\EAc,\d,\e)$, and the only way to obtain it by antisymmetrizing on $\a\leftrightarrow\e$ (which entirely described the antisymmetrization done here due to the forced antisymmetry in $\a\leftrightarrow\b$) would be to start from a configuration already antisymmetric in $(\a,\b,\EAc,\d)$, which is excluded. In other words, and starting from a configuration which contains two pairs of indices that are symmetric by exchange, e.g. $\a\leftrightarrow\EAc$ and $\b\leftrightarrow\d$, it is not possible to obtain a completely antisymmetric configuration by antisymmetrizing some indices. 

The new prefactor obtained at this step is called $A^2_{[\a\b][\EAc\d]\e}$. This prefactor is symmetric by exchange under $\a\leftrightarrow\e$. As $A^2_{[\a\b][\EAc\d]\e}$ is a sum of two terms which do not contain any antisymmetric part in $(\a,\b,\EAc,\d)$, it can also be put in a form that is symmetric by exchange under $\a\leftrightarrow\EAc$ and $\b\leftrightarrow\d$ as explained previously. The only possibility for a term that is symmetric under the exchanges $\a\leftrightarrow\EAc$ and $\a\leftrightarrow\e$ is finally to be completely symmetric under exchange of $(\a,\EAc,\e)$ indices, which proves property \textbf{b}. It is indeed not possible to symmetrize on more indices, since this antisymmetrization would involve indices which are already antisymmetric by exchange. The demonstration of property~\textbf{b} is similar for the other Lagrangian terms. Note that this proof also allows us to obtain the additional symmetries of the Lagrangian. For $\mathcal{L}^{\text{ext}}_{_I}$ and $\mathcal{L}^{\text{ext}}_{_{II}}$, we can impose a symmetry under the exchange $\b\leftrightarrow\d$, and for $\mathcal{L}^{\text{ext}}_{_{III}}$ a symmetry under the exchanges $\b\leftrightarrow\e$ and $\EAc\leftrightarrow\f$ [using the notations of Eq.~\eqref{MGEqExtGal}].

Another proof of property~\textbf{b} can be done using Young diagrams to describe the symmetry properties under exchange of internal indices. Indeed, the symmetry under permutations of indices of the Lagrangians with a nontrivial dynamics can be associated to a Young diagram, or at least a linear combination of them. However, some symmetries are "forced", due to the presence of the $\delta^{\mu_1\cdots}_{\nu_1\cdots}$ term. These "forced" symmetries impose some subblocks of the Young diagrams describing the complete symmetry properties of the Lagrangians. Then, considering the possible complete diagrams formed from these subblocks, and taking into account the property~\textbf{a} as well as the form of the equations of motion, one recovers the result of property~\textbf{b}.

Let us apply it to $\mathcal{L}^{\text{ext}}_{_I}$, using the notation of Eq.~\eqref{MGEqLExt1Cont}, \emph{i.e.} studying the symmetry properties of the $A_{[\a\b][\EAc\d]\e}$ prefactor. Due to the symmetrization of this tensor under the exchanges $\a\leftrightarrow\b$ and $\EAc\leftrightarrow\d$, the possible Young diagrams describing $A_{[\a\b][\EAc\d]\e}$ contain the following blocks:

\vspace{0.1cm}
\hfill
\begin{tabular}{|c|}
\hline
$\a$ \\
\hline
$\b$ \\
\hline
\end{tabular}
\hspace{2cm}
\begin{tabular}{|c|}
\hline
$\EAc$ \\
\hline
$\d$ \\
\hline
\end{tabular}
\hspace{2cm}
\begin{tabular}{|c|}
\hline
$\e$ \\
\hline
\end{tabular}
\hfill ~
\newline

Taking into account property~\textbf{a}, the Young diagrams describing $A_{[\a\b][\EAc\d]\e}$ cannot have the index $\e$ in the same column as any other indices. Thus, the only possible Young diagrams are 

\vspace{0.1cm}
\hfill
\begin{tabular}{c}
\begin{Young}
     $\a$  &  $\EAc$ &  $\e$ \cr
     $\b$  &  $\d$  \cr
\end{Young}
\end{tabular}
\hspace{2cm}
\begin{tabular}{c}
\begin{Young}
     $\a$  &  $\e$ \cr
     $\b$ \cr
     $\EAc$ \cr
     $\d$ \cr
\end{Young}
\end{tabular}
\hfill ~
\newline

The first configuration, which can be symmetrized on $(\a,\EAc,\e)$, is exactly the configuration described by property~\textbf{b}. Note that this configuration can also be symmetrized on $(\b,\d)$, as discussed previously. The second configuration is antisymmetric on $(\a,\b,\EAc,\d)$, and gives identically vanishing equations of motions, as explained before: it is a total derivative and can be omitted. This proves property \textbf{b}. 

Similar reasoning can be applied for $\mathcal{L}^{\text{ext}}_{_{II}}$ and $\mathcal{L}^{\text{ext}}_{_{III}}$, yielding respectively the following diagrams:

\vspace{0.1cm}
\hfill
\begin{tabular}{c}
\begin{Young}
     $\a$  &  $\EAc$ &  $\e$ & $\f$ \cr
     $\b$  &  $\d$  \cr
\end{Young}
\end{tabular}
\hspace{2cm}
\begin{tabular}{c}
\begin{Young}
     $\a$  &  $\d$ &  $\g$ \cr
     $\b$  &  $\e$  \cr
     $\EAc$  &  $\f$  \cr
\end{Young}
\end{tabular}
\hfill ~
\newline

These Young diagrams also show the additional symmetry properties discussed before.

\subsection{Terms with Levi-Civita tensors in the SO($N$) fundamental representation}
\label{MGAppEpsSON}

We consider in this section the possibility to build Lagrangians in the fundamental representation of a SO($N$) symmetry group with a prefactor containing Levi-Civita tensors. 
Note that property~\textbf{a}, in addition to the symmetry properties of the $\mathcal{L}^{\text{ext}}_j$ Lagrangians,
implies that no such Levi-Civita tensors can be contracted with $\partial_{\mu_i}\partial^{\nu_i} \pi^\a$ terms. Indeed, the other indices could be contracted either with a similar second-order derivative term, and the Lagrangian would identically vanish due to symmetry considerations, either with a first-order derivative term, and the Lagrangian would be a total derivative thanks to property~\textbf{a}. In addition, it is not necessary to consider contractions of Levi-Civita tensors together, since these contractions could be written with Kronecker delta only.

Let us begin with the case of SO(3), with a three-index Levi-Civita tensor $\epsilon_{\a\b\EAc}$. The only Lagrangian that can be written from $\mathcal{L}^{\text{ext}}_{_I}$ is
\begin{equation}
\mathcal{L}^{\text{SO}(3)}_4 = \displaystyle{ \delta^{\mu_1 \mu_2 \mu_3}_{\nu_1\nu_2\nu_3} \partial_{\mu_1}\pi^\a  
\partial_{\mu_2}\pi^\b\partial^{\nu_1}\pi^\EAc\partial^{\nu_2}\pi^\d \partial_{\mu_3}\partial^{\nu_3}\pi_\d \epsilon_{\a\b\EAc}}.
\end{equation}
Following property~\textbf{d}, we expect it to be a total derivative. It can be shown using the current
\begin{equation}
J^{3}_{\nu_3} =  \delta^{\mu_1 \mu_2 \mu_3}_{\nu_1\nu_2\nu_3} \partial_{\mu_1}\pi^\a  
\partial_{\mu_2}\pi^\b\partial^{\nu_1}\pi^\EAc\partial^{\nu_2}\pi^\d \partial_{\mu_3}\pi_\d \epsilon_{\a\b\EAc},
\end{equation}
in addition to property~\textbf{a}. No Lagrangian can be built from $\mathcal{L}^{\text{ext}}_{_{II}}$, since two Levi-Civita tensors would be involved, and thus there would be contractions between these tensors and $\partial_{\mu_i}\partial^{\nu_i} \pi^\a$ terms. With $\mathcal{L}^{\text{ext}}_{_{III}}$, only one term can be built,
\begin{equation}
\displaystyle{
\mathcal{L}^{\text{SO}(3)}_4 = \delta^{\mu_1 \mu_2 \mu_3 \mu_4}_{\nu_1\nu_2\nu_3\nu_4} \partial_{\mu_1}\pi^\a  \partial_{\mu_2}\pi^\b\partial_{\mu_3}\pi^\d\partial^{\nu_1}\pi^\EAc\partial^{\nu_2}\pi_\d\partial^{\nu_3}\pi^\e \partial_{\mu_4}\partial^{\nu_4}\pi_\e\epsilon_{\a\b\EAc}.
}
\end{equation}
However, we can show following property~\textbf{d} that it is a total derivative. The associated current is
\begin{equation}
\displaystyle{J^{4}_{\nu_4} = \delta^{\mu_1 \mu_2 \mu_3 \mu_4}_{\nu_1\nu_2\nu_3\nu_4} \partial_{\mu_1}\pi^\a  \partial_{\mu_2}\pi^\b\partial_{\mu_3}\pi^\d\partial^{\nu_1}\pi^\EAc\partial^{\nu_2}\pi_\d\partial^{\nu_3}\pi^\e \partial_{\mu_4}\pi_\e\epsilon_{\a\b\EAc},}
\end{equation}
which gives $\partial^{\nu_4} J^{4}_{\nu_4} = 2\mathcal{L}^{\text{SO}(N)}_4$.

Considering the case of SO(4), the only additional possible contraction is with $\mathcal{L}^{\text{ext}}_{_{II}}$, giving
\begin{equation}
\displaystyle{\mathcal{L}^{\text{SO}(4)}_5 =  \delta^{\mu_1 \mu_2 \mu_3 \mu_4}_{\nu_1\nu_2\nu_3\nu_4} \partial_{\mu_1}\pi^\a  \partial_{\mu_2}\pi^\b\partial^{\nu_1}\pi^\EAc\partial^{\nu_2}\pi^\d \partial_{\mu_3}\partial^{\nu_3}\pi^\e\partial_{\mu_4}\partial^{\nu_4}\pi_\e \epsilon_{\a\b\EAc\d}.}
\end{equation}
Following property~\textbf{d}, we can show that it is a total derivative, using the current
\begin{equation}
\displaystyle{J_{5}^{\mu_3} =  \delta^{\mu_1 \mu_2 \mu_3}_{\nu_1\nu_2\nu_3} \partial_{\mu_1}\pi^\a  
\partial_{\mu_2}\pi^\b\partial^{\nu_1}\pi^\EAc\partial^{\nu_2}\pi^\d  \partial^{\nu_3}\pi^\e\partial_{\mu_4}\partial^{\nu_4}\pi_\e \epsilon_{\a\b\EAc\d},}
\end{equation}
in addition to property~\textbf{a}.

For SO(5), the only possible contraction is with $\mathcal{L}^{\text{ext}}_{_{III}}$, yielding 
\begin{equation}
\mathcal{L}^{\text{SO}(5)}_6 = \delta^{\mu_1 \mu_2 \mu_3 \mu_4}_{\nu_1\nu_2\nu_3\nu_4} \partial_{\mu_1}\pi^\a  \partial_{\mu_2}\pi^\b\partial_{\mu_3}\pi^\EAc\partial^{\nu_1}\pi^\d\partial^{\nu_2}\pi^\e\partial^{\nu_3}\pi^\f \partial_{\mu_4}\partial^{\nu_4}\pi_\f \epsilon_{\a\b\EAc\d\e}.
\end{equation}
One more time, following property~\textbf{d}, we expect this Lagrangian to be a total derivative. It is shown by introducing the current 
\begin{equation}
\displaystyle{J_{6,\nu_4}=  \delta^{\mu_1 \mu_2 \mu_3 \mu_4}_{\nu_1\nu_2\nu_3\nu_4} \partial_{\mu_1}\pi^\a  \partial_{\mu_2}\pi^\b\partial_{\mu_3}\pi^\EAc\partial^{\nu_1}\pi^\d\partial^{\nu_2}\pi^\e\partial^{\nu_3}\pi^\f \partial_{\mu_4} \pi_\f \epsilon_{\a\b\EAc\d\e},}
\end{equation}
in addition to property~\textbf{a}.

\subsection{Exhaustive examination of $\mathcal{L}^{\text{ext}}_{_I}$ in the SU($N$) adjoint representation}
\label{MGAppPartSUN}
\paragraph{Introduction}

We consider in this section all the possible independent Lagrangians which can be built from $\mathcal{L}^{\text{ext}}_{_I}$ in the case of multi-Galileons in the adjoint representation of a SU($N$) symmetry group, without using property~\textbf{b}. For this purpose, it is necessary to produce a basis of independent prefactors built from contractions of the primitive invariants only. The properties of the primitive invariants given in Sec.~\ref{MGPartIntroSUN} are thus very useful. One can also note that it is not necessary to consider too many contractions of indices. For example, the authors of Ref.~\cite{Dittner:1972hm} showed that the rank-seven or higher contractions of primitive invariants can be described with the contractions of lower ranks in the SU(3) case.  

These prefactors can be obtained from an explicit construction of the product representations. We give here an example for the singlet built from four adjoint representations of a SU(3) symmetry group. It is possible to build eight such singlets \cite{Slansky:1981yr,Feger:2012bs}. They can be identified through the product 
\begin{equation}
\boldsymbol{8} \times \boldsymbol{8}  = \boldsymbol{1} + \boldsymbol{8_s}+\boldsymbol{8_a} + \boldsymbol{10_a}+\boldsymbol{\overline{10}_a} + \boldsymbol{27_s},
\end{equation}
where the subscript $a$ or $s$ mean that the representations are symmetric or antisymmetric under the exchange of the two initial adjoint representations. Denoting for example the two initial adjoint representations as $\phi^\a$ and $\psi^\b$, the $\boldsymbol{8_s}$ representation is described by $d^{\a\b\EAc} \phi_\b \psi_\EAc$, and the $\boldsymbol{27_s}$ representation by $S^{\a\b}=\phi^{(\a}\psi^{\b)} + c.t.$, with the last term denoting counterterms such that $S_\a^\a=0$ and $d^{\a\b\EAc}S_{\b\EAc}=0$. This product allows us to identify the singlets from the product of four adjoint representations as those which appear in the product $\left(\boldsymbol{8} \times \boldsymbol{8}\right) \times \left(\boldsymbol{8} \times \boldsymbol{8}\right)$ as products of conjugate representations, since it is the only way to build a singlet from a product of two representations\footnote{We remind that the order of appearance of the different fields in this construction is not important. It is due to the fact that (denoting with $\boldsymbol{R_j}$ a given $j$ representation) if $\boldsymbol{R_1}\times\boldsymbol{R_2}$ contains $\boldsymbol{R_i}$, then $\boldsymbol{R_1}\times\boldsymbol{\bar R_i}$ contains $\boldsymbol{\bar R_2}$, etc.~\cite{Slansky:1981yr}}.

This method, even if exhaustive, becomes quite involved when high-dimensional representations appear in the products. In addition, it is not necessary to express the Lagrangians in terms of the irreducible representations which appear in the intermediate products of fields. Another equivalent method consists in listing all the independent prefactors built from primitive invariants at each order. For example, the only possibility with two fields is $\delta_{\a\b}$, and the two possibilities with three fields are $f_{\a\b\EAc}$ and $d_{\a\b\EAc}$. This number grows rapidly with the number of fields. The number of independent prefactors at each rank can be easily computed from group-theoretical calculations~\cite{Slansky:1981yr,Feger:2012bs}. The number of such combinations at different orders and for different SU($N$) groups are given in the following table.

\begin{center}
\begin{tabular}{|l|p{0.75cm}|p{0.75cm}|p{0.75cm}|p{0.75cm}|p{0.75cm}|p{0.75cm}|p{0.75cm}|}
\hline
$\#$ adjoint rep. & 1 & 2 & 3 & 4 & 5 & 6 & 7\\
\hline
$\#$ SU($2$) singlets & 0  & 1 & 1 & 3 & 6 & 15 & 36 \\
\hline
$\#$ SU($3$) singlets & 0  & 1 & 2 & 8 & 32 & 145 & 702 \\
\hline
$\#$ SU($4$) singlets & 0  & 1 & 2 & 9 & 43 & 245 & 1557 \\
\hline
$\#$ SU($5$) singlets & 0  & 1 & 2 & 9 & 44 & 264 & 1824 \\
\hline
\end{tabular}
\end{center}

We will use a third method in this paper. We saw in Sec.~\ref{MGPartFundRep} that the properties of the extended multi-Galileon Lagrangians are quite restrictive, particularly the symmetry properties. Thus, contractions with only a small number of primitive invariant prefactors will allow nontrivial dynamics at first glance. We will then proceed by considering the possible contractions for each kind of primitive invariant prefactor. After having obtained the nontrivial Lagrangians, we will finally verify their independence taking into account the relations between the primitive invariants contractions.

\paragraph{Terms without primitive invariants contracted together}

The first possibility is to contract $\mathcal{L}^{\text{ext}}_{_I}$ with one Kronecker delta and one structure constant. However, this term has already been investigated in the case of the fundamental 
representation of a SO($3$) symmetry group, since the only property of antisymmetry of $\epsilon_{\a\b\EAc}$ has been used. One nontrivial dynamic is possible, described e.g. by the Lagrangian given in Eq.~\eqref{MGEqLagFinSON} whose equation of motion is given in Eq.~\eqref{MGEqEOMSON}.

The other possibility is to contract $\mathcal{L}^{\text{ext}}_{_I}$ with one Kronecker delta and one symmetric $d_{\a\b\EAc}$ invariant. Only one such Lagrangian is possible, \emph{i.e.} 
\begin{equation}
\displaystyle{\mathcal{L}^{\text{SU}(N)}_1 =   \delta^{\mu_1 \mu_2 \mu_3}_{\nu_1\nu_2\nu_3} \partial_{\mu_1}\pi^\a \partial_{\mu_2}\pi^\d\partial^{\nu_1}\pi^\b\partial^{\nu_2}\pi_\d \partial_{\mu_3}\partial^{\nu_3}\pi^\EAc d_{\a\b\EAc}.}
\end{equation}
Property~\textbf{d} cannot be applied here, and this Lagrangian is not a total derivative. Indeed, the currents that can be formed by removing $\partial_{\mu_3}$ or $\partial^{\nu_3}$ from this Lagrangian trivially vanish. The equation of motion thus gives
\begin{multline}
\displaystyle{EOM^{\text{SU}(N)}_1 = 
\delta^{\mu_1 \mu_2 \mu_3}_{\nu_1\nu_2\nu_3} \left[
4 \partial_{\mu_1}\pi^\b \partial^{\nu_1}\pi^\d \partial_{\mu_2}\partial^{\nu_2}\pi^\EAc \partial_{\mu_3}\partial^{\nu_3}\pi_\d d^{\al}{}_{\b\EAc}
-3 \partial_{\mu_1}\pi^\d \partial^{\nu_1}\pi_\d \partial_{\mu_2}\partial^{\nu_2}\pi^\b \partial_{\mu_3}\partial^{\nu_3}\pi^\EAc d^{\al}{}_{\b\EAc}
\right.} \\ \displaystyle{\left.
- \partial_{\mu_1}\pi^\b \partial^{\nu_1}\pi^\EAc\partial_{\mu_2}\partial^{\nu_2}\pi^\d \partial_{\mu_3}\partial^{\nu_3}\pi_\d d^{\al}{}_{\b\EAc}
+2 \partial_{\mu_1}\pi^\a \partial^{\nu_1}\pi^\al\partial_{\mu_2}\partial^{\nu_2}\pi^\b \partial_{\mu_3}\partial^{\nu_3}\pi^\EAc d_{\a\b\EAc}
\right.} \\ \displaystyle{\left.
-2 \partial_{\mu_1}\pi^\a \partial^{\nu_1}\pi^\b\partial_{\mu_2}\partial^{\nu_2}\pi^\al \partial_{\mu_3}\partial^{\nu_3}\pi^\EAc d_{\a\b\EAc},
\right]}
\label{MGAppEqEOMSUN}
\end{multline}
where $\al$ is the free group index of the equation of motion, since this equation of motion is in the adjoint representation of the SU($N$) symmetry group.
This equation of motion is a total derivative, as expected. Indeed, introducing the currents
\begin{equation}
\displaystyle{\left\{
\begin{array}{l}
J_1^{\mu_2} = \delta^{\mu_1 \mu_2 \mu_3}_{\nu_1\nu_2\nu_3} \partial_{\mu_1}\pi^\a \partial^{\nu_1}\pi^\al\partial^{\nu_2}\pi^\b \partial_{\mu_3}\partial^{\nu_3}\pi^\EAc d_{\a\b\EAc},\\
J_2^{\mu_2} = \delta^{\mu_1 \mu_2 \mu_3}_{\nu_1\nu_2\nu_3} \partial_{\mu_1}\pi^\b \partial^{\nu_1}\pi^\d\partial^{\nu_2}\pi^\EAc \partial_{\mu_3}\partial^{\nu_3}\pi_\d d^{\al}{}_{\b\EAc},\\
J_3^{\mu_2} = \delta^{\mu_1 \mu_2 \mu_3}_{\nu_1\nu_2\nu_3}\partial_{\mu_1}\pi_\d \partial^{\nu_1}\pi^\b\partial^{\nu_2}\pi^\d \partial_{\mu_3}\partial^{\nu_3}\pi^\EAc d^{\al}{}_{\b\EAc},
\end{array}
\right.}
\end{equation}
one can verify that 
\begin{equation}
EOM^{\text{SU}(N)}_1 = \partial_{\mu_2} \left[2 J_1^{\mu_2} + J_2^{\mu_2} + 3 J_3^{\mu_2} \right].
\end{equation}

\paragraph{Terms with primitive invariants contracted together}

As discussed before, it is not necessary to consider the contractions between $\delta_{\a\b}$ and the other primitive invariants, since they only raise or lower the indices. We thus focus on the contractions of $f_{\a\b\EAc}$ and $d_{\a\b\EAc}$. As $\mathcal{L}^{\text{ext}}_{_I}$ contains five fields, it is sufficient to consider only the rank-five contractions, \emph{i.e.} with five indices not contracted together. Indeed, considering the rank-four contractions, it would then be necessary to form a singlet from the single remaining field, which is not possible. To describe the rank-five contractions of the primitive invariants, it is sufficient to consider the contractions of only three $f$ or $d$. We will then consider them successively. The contractions of more primitive invariants will only reduce to those already considered thanks to the different structure properties of the group.

Concerning the contractions with a prefactor of the form $f_{\be \a\b}f_{\ga \EAc\d}f^{\be\ga}{}_{\e}$, no contractions are possible due to symmetry considerations. Concerning the prefactor of the form $f_{\be \a\b}d_{\ga \EAc\d}d^{\be\ga}{}_{\e}$, two Lagrangians can be written, yielding
\begin{equation}
\displaystyle{\begin{array}{l}
\mathcal{L}^{\text{SU}(N)}_2 =   \delta^{\mu_1 \mu_2 \mu_3}_{\nu_1\nu_2\nu_3} \partial_{\mu_1}\pi^\a \partial_{\mu_2}\pi^\b\partial^{\nu_1}\pi^\EAc\partial^{\nu_2}\pi^\e \partial_{\mu_3}\partial^{\nu_3}\pi^\d f_{\be \a\b}d_{\ga \EAc\d}d^{\be\ga}{}_{\e},\\
\mathcal{L}^{\text{SU}(N)}_3 =   \delta^{\mu_1 \mu_2 \mu_3}_{\nu_1\nu_2\nu_3} \partial_{\mu_1}\pi^\a \partial_{\mu_2}\pi^\EAc\partial^{\nu_1}\pi^\b\partial^{\nu_2}\pi^\e \partial_{\mu_3}\partial^{\nu_3}\pi^\d f_{\be \a\b}d_{\ga \EAc\d}d^{\be\ga}{}_{\e}.
\end{array}}
\end{equation}
However, building currents by removing $\partial^{\nu_3}$ from $\mathcal{L}^{\text{SU}(N)}_2$ and $\partial_{\mu_3}$ from $\mathcal{L}^{\text{SU}(N)}_3$, and using property~\textbf{a} as well as symmetry properties, one can show that both Lagrangians are total derivatives.
The two Lagrangians built from the contractions with $f_{\be \a\b}f_{\ga \EAc\d}d^{\be\ga}{}_{\e}$, as well as both Lagrangians built with $f_{\be \a\b}d_{\ga \EAc\d}d^{\be\ga}{}_{\e}$, are also total derivatives for similar reasons. 

Let us turn to the contractions with a prefactor in $d_{\be \a\b}d_{\ga \EAc\d}f^{\be\ga}{}_{\e}$. This prefactor allows us to build only one Lagrangian, 
\begin{equation}
\displaystyle{
\mathcal{L}^{\text{SU}(N)}_4 =   \delta^{\mu_1 \mu_2 \mu_3}_{\nu_1\nu_2\nu_3} \partial_{\mu_1}\pi^\a \partial_{\mu_2}\pi^\EAc\partial^{\nu_1}\pi^\b\partial^{\nu_2}\pi^\e \partial_{\mu_3}\partial^{\nu_3}\pi^\d d_{\be \a\b}d_{\ga \EAc\d}f^{\be\ga}{}_{\e}.
}
\end{equation}
This Lagrangian cannot be written as a total derivative, since the currents obtained by removing $\partial_{\mu_3}$ or $\partial^{\nu_3}$ vanish thanks to symmetry considerations. However, it is possible to link this Lagrangian with other ones thanks to the structure properties of SU($N$). The primitive invariants verify the following relations (see e.g. Ref.~\cite{Fuchs:1997jv})
\begin{equation}
\label{MGEqPropStructSUN}
\displaystyle{
f_{\a\d\be}d^{\be}{}_{\b\EAc} + f_{\b\d\be}d^{\be}{}_{\EAc\a} +f_{\EAc\d\be}d^{\be}{}_{\a\b} =0.
}
\end{equation}
This relation implies that 
\begin{equation}
\mathcal{L}^{\text{SU}(N)}_4 = \mathcal{L}^{\text{SU}(N)}_2 + \mathcal{L}^{\text{SU}(N)}_3,
\end{equation}
and thus that $\mathcal{L}^{\text{SU}(N)}_4$ is a total derivative. Note that this result cannot be seen directly from the equation of motion of $\mathcal{L}^{\text{SU}(N)}_4$ without using Eq.~\eqref{MGEqPropStructSUN}. It is due to the fact that we used a basis of primitive invariant prefactors which is convenient, but with terms which could be not linearly independent.

Finally, one can consider the possible contraction with a $d_{\be \a\b} d_{\ga \EAc\d} d^{\be\ga}{}_{\e}$ prefactor. It is possible to build two such Lagrangians, \emph{i.e.} 
\begin{equation}
\displaystyle{\begin{array}{l}
\mathcal{L}^{\text{SU}(N)}_5 =   \delta^{\mu_1 \mu_2 \mu_3}_{\nu_1\nu_2\nu_3} \partial_{\mu_1}\pi^\a \partial_{\mu_2}\pi^\EAc\partial^{\nu_1}\pi^\b\partial^{\nu_2}\pi^\d \partial_{\mu_3}\partial^{\nu_3}\pi^\e d_{\be \a\b} d_{\ga \EAc\d} d^{\be\ga}{}_{\e},\\
\mathcal{L}^{\text{SU}(N)}_6 =   \delta^{\mu_1 \mu_2 \mu_3}_{\nu_1\nu_2\nu_3} \partial_{\mu_1}\pi^\a \partial_{\mu_2}\pi^\EAc\partial^{\nu_1}\pi^\b\partial^{\nu_2}\pi^\e \partial_{\mu_3}\partial^{\nu_3}\pi^\d d_{\be \a\b} d_{\ga \EAc\d} d^{\be\ga}{}_{\e}.
\end{array}}
\end{equation}
As expected from property~\textbf{e}, they are related by a total derivative. Indeed, the current:
\begin{equation}
\displaystyle{
J^{\mu_3} = \delta^{\mu_1 \mu_2 \mu_3}_{\nu_1\nu_2\nu_3}  \partial_{\mu_1}\pi^\a \partial_{\mu_2}\pi^\EAc\partial^{\nu_1}\pi^\b\partial^{\nu_2}\pi^\d \partial^{\nu_3}\pi^\e d_{\be \a\b} d_{\ga \EAc\d} d^{\be\ga}{}_{\e},
}
\end{equation}
gives
\begin{equation}
\displaystyle{
\partial_{\mu_3}J^{\mu_3} = \mathcal{L}^{\text{SU}(N)}_5 - 2 \mathcal{L}^{\text{SU}(N)}_6.
}
\end{equation}
Then, it is also possible to use relations between the primitive invariants of SU($N$), especially \cite{Fuchs:1997jv}
\begin{equation}
\label{MGEqPropStructSUN2}
\displaystyle{
f_{\a\b\be}f_{\EAc\d}{}^{\be}= \frac2N\left(\delta_{\a\EAc}\delta_{\b\d} - \delta_{\b\EAc}\delta_{\a\d} \right) + d_{\a\EAc\be}d_{\b\d}{}^{\be} - d_{\a\d\be}d_{\b\EAc}{}^{\be},
}
\end{equation}
which allows us to write down 
\begin{equation}
\displaystyle{
\mathcal{L}^{\text{SU}(N)}_6 - \mathcal{L}^{\text{SU}(N)}_5 = \frac2N\mathcal{L}^{\text{SU}(N)}_1 + \text{total derivative},
}
\end{equation}
with the total derivative being obtained thanks to property~\textbf{a}. It implies that both $\mathcal{L}^{\text{SU}(N)}_5$ and $\mathcal{L}^{\text{SU}(N)}_6$ involve the same dynamics as $\mathcal{L}^{\text{SU}(N)}_1$. 

This finally shows that only one nontrivial dynamics allowed at the order of $\mathcal{L}^{\text{ext}}_{_I}$ is described by the Lagrangian $\mathcal{L}^{\text{SU}(N)}_1$. It is exactly what is predicted from the construction of Sec.~\ref{MGPartSUNLag} with a third-rank symmetric tensor as implied by property~\textbf{b}.



%
%


\part{Modèles de Galiléons vectoriels}
\label{ChapterGalVec}
\chapter{Introduction aux articles}
\label{PartIntroProca}

\noindent
Les modèles de Galiléons étudiés jusqu'à présent couplaient des champs scalaires avec la métrique. 
Une autre extension possible des théories de Galiléons consiste à coupler des champs vectoriels avec la métrique. La théorie la plus générale en espace plat pour des champs vectoriels $A^\mu$ et invariante sous les transformations de jauge $A_\mu \rightarrow A_\mu + \partial_\mu \phi$ a été étudiée dans~\cite{Deffayet:2013tca}, où il a été montré que les équations du mouvement de tels modèles en 4 dimensions étaient au mieux linéaires en les dérivées secondes de $A_\mu$. Ce résultat a mené à la construction de modèles de Galiléons décrivant des champs vectoriels sans symétries de jauge, et propageant donc trois degrés de liberté. Ces théories incluent notamment un terme de masse $m^2 A^\mu A_\mu$ pour le champ vectoriel, et sont donc des généralisation des théories de Proca\footnote{Les théories de Proca sont usuellement distinguées en théories abéliennes ou non-abéliennes selon qu'elle contiennent un ou plusieurs champs vectoriels massifs. Cette dénomination vient de l'écriture de la théorie de Proca comme l'extension massive d'une théorie de jauge, l'extension d'une théorie de jauge abélienne n'ayant en effet qu'un seul champ, au contraire de celles des théories de jauge non-abéliennes.}. Ces modèles ont été introduits dans les articles~\cite{Heisenberg:2014rta,Tasinato:2014eka}, et ont été développés ultérieurement dans~\cite{Allys:2015sht,Allys:2016jaq}, qui seront présentés dans les chapitres~\ref{PartArticleProca1} et~\ref{PartArticleProca2}, ainsi que dans~\cite{Jimenez:2016isa}. Je récapitule ici leurs principales propriétés, avant d'introduire les travaux présentés dans les chapitres~\ref{PartArticleProca1} à~\ref{PartArticleProcaSU2}.

Pour décrire ces théories, il est commode d'introduire les contributions symétriques et antisymétriques des dérivées premières du champ vectoriel $A_\mu$, à savoir
\begin{equation}
S_{\mu\nu} = \partial_\mu A_\nu + \partial_\nu A_\mu,
\end{equation}
et 
\begin{equation}
F_{\mu\nu} = \partial_\mu A_\nu - \partial_\nu A_\mu,
\end{equation}
qui s'identifie au tenseur de Faraday des théories de jauge abélienne. La conclusion des travaux~\cite{Heisenberg:2014rta,Tasinato:2014eka,Allys:2015sht,Allys:2016jaq,Jimenez:2016isa} est que la théorie la plus générale pour des Galiléons vectoriels en espace plat est
\begin{equation}
\label{EqActionVectorGalileon}
\displaystyle{ 
\mathcal{S} = \int\text{d}^4 x ~ \left( -\frac{1}{4} F_{\mu\nu} F^{\mu\nu} + \sum_{i=2}^6 \mathcal{L}_i  \right), 
}
\end{equation}
avec 
\begin{equation}
\label{EqVectorGal1}
\displaystyle{
\begin{split}
\mathcal{L}_2 &= f_2\left( A_\mu, F_{\mu\nu}, \tilde{F}_{\mu\nu} \right), \\
\mathcal{L}_3 &= f_3\left( X\right)  \delta_{\mu_1}^{\nu_1} S_{\nu_1}{}^{\mu_1},\\
\mathcal{L}_4 &=  f_4 \left( X\right) \delta_{\mu_1 \mu_2}^{\nu_1\nu_2}S_{\nu_1}{}^{\mu_1}S_{\nu_2}{}^{\mu_2} , \\
\mathcal{L}_5 &=  f_5\left( X\right)\delta_{\mu_1 \mu_2 \mu_3}^{\nu_1\nu_2 \nu_3} S_{\nu_1}{}^{\mu_1}S_{\nu_2}{}^{\mu_2}S_{\nu_3}{}^{\mu_3}  +  g_5\left( X\right) \delta_{\mu_1 \mu_2 \mu_3}^{\nu_1\nu_2 \nu_3}S_{\nu_1}{}^{\mu_1} F_{\nu_2\nu_3}F^{\mu_2\mu_3}, \\
\mathcal{L}_6 &= g_6 \left( X\right) \delta_{\mu_1 \mu_2 \mu_3\mu_4}^{\nu_1\nu_2 \nu_3\nu_4}S_{\nu_1}{}^{\mu_1}S_{\nu_2}{}^{\mu_2}  F_{\nu_3\nu_4}F^{\mu_3\mu_4},
\end{split}
}
\end{equation}
où $f_i$ et $g_i$ sont des fonctions arbitraires de $X=A_\alpha A^\alpha$, $f_2$ pouvant également dépendre du tenseur de Faraday $F_{\mu\nu}$ et de son dual de Hodge
\begin{equation}
\tilde{F}_{\mu\nu} = \frac12 \epsilon_{\mu\nu\rho\sigma}F^{\rho\sigma}.
\end{equation} 
Le terme cinétique de Maxwell isolé dans l'action complète donnée par l'équation~\eqref{EqActionVectorGalileon} permet de fixer la normalisation du champ vectoriel. Ces Lagrangiens décrivent les termes qui conservent la parité comme les termes qui la violent, même si des termes de violation de parité ne sont contenus que dans $\mathcal{L}_2$~\cite{Allys:2016jaq}.
Une expression développée des Lagrangiens $\mathcal{L}_3$ à $\mathcal{L}_6$ donne
\begin{equation}
\displaystyle{
\begin{split}
\mathcal{L}_3 &= f_3\left( X\right) S_{\mu}{}^{\mu}, \\
\mathcal{L}_4 &=  f_4 \left( X\right) \left[ (S_{\mu}{}^{\mu})^2-S_{\rho}{}^{\sigma} S_{\sigma}{}^{\rho} \right], \\
\mathcal{L}_5 &=  f_5\left( X\right)\left[ (S_{\mu}{}^{\mu})^3-3 (S_{\mu}{}^{\mu}) 
S_{\rho}{}^{\sigma}S_{\sigma}{}^{\rho}  + 2 S_{\rho}{}^{\sigma}S_{\sigma}{}^{\gamma}
S_{\gamma}{}^{\rho}  \right] +  g_5\left( X\right) 
\tilde{F}^{\alpha\mu} \tilde{F}^\beta{}_\mu S_{\alpha \beta },\\
\mathcal{L}_6 &= g_6 \left( X\right) \tilde{F}^{\alpha\beta} \tilde{F}^{\mu\nu}
 S_{\alpha\mu} S_{\beta\nu}.
 \end{split}
}
\end{equation}
En rendant ces théories covariantes, il est nécessaire comme dans le cas des Galiléons scalaires d'introduire des contre-termes dépendant de la courbure. De telles contributions sont nécessaires pour les termes faisant apparaître les fonctions arbitraires $f_4$, $f_5$ et $g_6$, et s'écrivent 
\begin{equation}
\label{EqVectorGal2}
\displaystyle{
\begin{split}
\mathcal{L}_4^{\text{Cov.}} &= f_4\left( Y \right) R + f_{4,Y} \left( Y\right) \delta_{\mu_1 \mu_2}^{\nu_1\nu_2} S_{\nu_1}{}^{\mu_1}S_{\nu_2}{}^{\mu_2},\\
\mathcal{L}_5^{\text{Cov.}}  &= f_5\left( Y\right) G^{\mu\nu} S_{\mu\nu}- \frac16 f_{5,Y}\left( Y\right)\delta_{\mu_1 \mu_2 \mu_3}^{\nu_1\nu_2 \nu_3}S_{\nu_1}{}^{\mu_1}S_{\nu_2}{}^{\mu_2}S_{\nu_3}{}^{\mu_3},\\
\mathcal{L}_6^{\text{Cov.}}  &= g_6 \left( Y \right) L^{\mu\nu\rho\sigma} F_{\mu\nu}F_{\rho\sigma} +  \frac{1}{2} g_{6,Y} \left( Y\right) \tilde{F}^{\alpha\beta} \tilde{F}^{\mu\nu}
 S_{\alpha\mu} S_{\beta\nu} ,
\end{split}
}
\end{equation}
où l'on a introduit $Y=-\frac12 A_\alpha A^\alpha$ et la notation $f_{i,Y}=\text{d}f_i/\text{d}Y$. Le tenseur $L^{\mu\nu\rho\sigma}$ est le dual de Hodge du tenseur de Riemann~:
\begin{equation}
L^{\mu\nu\rho\sigma} = \frac14 \epsilon^{\mu\nu\alpha\beta}\epsilon^{\rho\sigma\gamma\delta} R_{\alpha\beta\gamma\delta}.
\end{equation}
Bien que ce ne soit pas visible à première vue, cette action covariante des Galiléons vectoriels décrit également des contributions comme $G^{\mu\nu} A_\mu A_\nu$, qui peut être écrit à l'aide d'intégrations par partie comme la somme d'un terme cinétique usuel $F_{\mu\nu} F^{\mu\nu}$ et d'un Lagrangien de type $\mathcal{L}_4^{\text{Cov.}}$ avec $f_{4} = A_\mu A^\mu$~\cite{Jimenez:2016isa}.

Une point important dans l'étude des théories de Galiléons vectoriels est de vérifier que les champs $A_\mu$ ne propagent que trois degrés de liberté, étant entendu que le degré de liberté associé à $A_0$ correspond à un fantôme qui ne doit donc pas se propager~\cite{Heisenberg:2014rta}. Il est nécessaire pour cela d'identifier une contrainte primaire et une contrainte secondaire permettant d'éliminer totalement le degré de liberté lié à la composante temporelle du champ vectoriel. Lors de l'étude d'un Lagrangien donné $\mathcal{L}$, une manière d'assurer la présence d'une contrainte primaire associée à $A_0$ est d'avoir une valeur propre nulle vis-à-vis de ce champ pour le Hessien
\begin{equation}
\mathcal{H}^{\mu\nu} = \frac{\partial^2 \mathcal{L}}{\partial(\partial_0 A_\mu)\partial(\partial_0 A_\nu)},
\end{equation}
ce qui est vérifié en imposant\footnote{Cette condition peut sembler un peu forte, mais dans le cas où le Hessien a une valeur propre nulle associée à une combinaison linéaire à coefficients constants des différentes composantes de $A_\mu$ -- les trois autres valeurs propres étant positives --, il est possible de faire une transformation de Lorentz pour se ramener à la condition de l'équation~\eqref{ConditionHessienne}.}
\begin{equation}
\label{ConditionHessienne}
\mathcal{H}^{0i}=0~~~~ \text{et} ~~~~ \mathcal{H}^{00}=0,
\end{equation} 
pour $i=1,\cdots,3$. Cette condition impose en effet l'absence de dérivées temporelles de $A_0$ dans le Lagrangien, et donc une contrainte primaire liée à la nullité de son moment conjugué
\begin{equation}
\Pi^0 = \frac{\partial \mathcal{L}}{\partial(\partial_0 A_0)} = 0.
\end{equation}
Il est ensuite nécessaire de vérifier que cette contrainte primaire implique bien une contrainte secondaire viable après dérivation par rapport au temps~\cite{Heisenberg:2014rta}.

Afin d'identifier les différents modes se propageant dans les théories de Galiléons vectoriels, il est possible d'utiliser la décomposition de Stückelberg (aussi appelée astuce de Stückelberg, ou \og Stückelberg trick \fg{}). Celle-ci décompose les trois degrés de liberté du champ vecteur $A_\mu$ en deux degrés de liberté associés à un champ vectoriel $\bar A_\mu$ possédant une symétrie de jauge, et un degré de liberté décrit par un champ scalaire $\pi$ et associé au mode de propagation longitudinal du vecteur massif. Cette décomposition se fait en écrivant
\begin{equation}
A_\mu = \bar A_\mu + \frac{1}{m}\partial_\mu \pi,
\end{equation} 
avec $\bar A_\mu$ qui vérifie
\begin{equation}
\partial^\mu \bar A_\mu = 0.
\end{equation}
Lors de l'étude des modèles de Galiléons vectoriels, il est demandé comme hypothèse que le mode longitudinal $\pi$ implique des équations du second ordre. Pour les Galiléons vectoriels, la limite de découplage $m\rightarrow 0$ est bien définie\footnote{Il n'y a donc pas de discontinuité de vDVZ comme dans le cas de la gravité massive~\cite{deRham:2014zqa}.}, et résulte en une théorie décrivant un champ de jauge et un champ scalaire sans masse~\cite{Heisenberg:2014rta}.

La décomposition de Stückelberg permet d'identifier trois types de contribution dans les Lagrangiens de Galiléons vectoriels~: les contributions purement scalaires qui ne dépendent que du mode longitudinal $\pi$, les contributions purement vectorielles qui ne dépendent que de $\bar A_\mu$, et les contributions mixtes qui font apparaître $\pi$ et $\bar A_\mu$. La décomposition des dérivées du champ $A_\mu$ en ses composantes symétrique $S_{\mu\nu}$ et antisymétrique $F_{\mu\nu}$ permet alors d'identifier aisément les composantes purement scalaires, car $F_{\mu\nu}$ s'annule identiquement pour le mode longitudinal, alors que $S_{\mu\nu}$ se ramène à $\partial_\mu \partial_\nu\pi $. Les contributions purement scalaires se ramenant exactement aux hypothèses des modèles de Galiléons scalaires, elles doivent être inclues dans cette dernière classe de modèles. Cette identification se fait aisément en effectuant dans l'équation~\eqref{EqVectorGal1} les remplacements de $F_{\mu\nu}$ et de $S_{\mu\nu}$ décrits précédemment, les termes obtenus s'identifiant identiquement aux Lagrangiens de Galiléons scalaires généralisés donnés dans l'équation~\eqref{EqLagGenGal} (la seule différence étant que les fonctions arbitraires ne peuvent pas dépendre de $\pi$ mais seulement de $\partial_\alpha \pi \partial^\alpha\pi$).

Les premiers développements des Galiléons vectoriels ont été faits dans les articles~\cite{Heisenberg:2014rta,Tasinato:2014eka}, qui ont obtenu les termes $\mathcal{L}_2$ à $\mathcal{L}_5$ de l'équation~\eqref{EqVectorGal1}. Ces investigations ne s'intéressaient qu'aux termes conservant la parité, et les fonctions arbitraires $f_5$ et $g_5$ apparaissant dans ces Lagrangiens n'étaient pas identifiées comme des fonctions indépendantes. Estimant que ces investigations n'avaient pas obtenu tous les termes possibles, nous avons effectué une recherche systématique de termes dans l'article~\cite{Allys:2015sht}, reproduit dans le chapitre~\ref{PartArticleProca1}. Cette recherche systématique consiste à écrire tous les termes Lagrangiens possibles à chaque ordre, puis à calculer quelles combinaisons linéaires de ces termes vérifient les conditions Hessiennes de l'équation~\eqref{ConditionHessienne}. Cette investigation a permis d'obtenir le terme $\mathcal{L}_6$ de l'équation~\eqref{EqVectorGal1}, et a aussi montré que les fonctions $f_5$ et $g_5$ qui apparaissent dans $\mathcal{L}_5$ sont indépendantes.

L'investigation systématique de l'article~\cite{Allys:2015sht} a été poussée jusqu'à l'ordre de $\mathcal{L}_7$ -- c'est à dire pour des Lagrangiens contenant cinq dérivées premières de $A_\mu$ -- pour le secteur conservant la parité, et nous avons obtenu deux termes supplémentaires à ceux mentionnés précédemment, notés $\mathcal{L}_7^{\text{Perm,1}}$ et $\mathcal{L}_7^{\text{Perm,2}}$. Dans le secteur de violation de parité, les investigations ont été poussées jusqu'à l'ordre de $\mathcal{L}_6$, et ont permis d'obtenir deux termes supplémentaires à ceux mentionnés précédemment, notés $\mathcal{L}^\epsilon_5$ et $\mathcal{L}^\epsilon_6$. Comme discuté ensuite, ces termes Lagrangiens ne sont pas pertinents~: ils sont nuls, correspondent à des dérivées totales, ou bien sont déjà compris dans $\mathcal{L}_2$. Leur apparition est en fait inhérente à la méthode d'investigation de termes utilisée, qui cherche de façon systématique des termes acceptables au niveau du Lagrangien, et qui demande de vérifier après-coup que les Lagrangiens obtenus ont bien des dynamiques non-triviales et indépendantes. Une mauvaise compréhension de ce défaut de la méthode d'investigation utilisée a mené à postuler un nombre infini de termes pour les Galiléons vecteurs, un résultat que nous avons révisé par la suite.

Le secteur de parité conservée a été corrigé subséquemment dans l'article~\cite{Jimenez:2016isa}, qui a montré que $\mathcal{L}_7^{\text{Perm,1}}$ et $\mathcal{L}_7^{\text{Perm,2}}$ étaient trivialement nuls en vertu du théorème de Cayley-Hamilton. Nous avons corrigé le secteur de violation de parité avec des collaborateurs dans l'article~\cite{Allys:2016jaq}, reproduit dans le chapitre~\ref{PartArticleProca2}, en montrant à l'aide d'une identité sur les tenseurs antisymétriques que les deux termes $\mathcal{L}^\epsilon_5$ et $\mathcal{L}^\epsilon_6$ sont respectivement une dérivée totale et inclus dans $\mathcal{L}_2$. Dans cet article, nous avons aussi discuté la construction systématique à partir de tenseurs de Levi-Civita de Lagrangiens vérifiant la condition Hessienne de l'équation~\eqref{ConditionHessienne}. Cette construction mène explicitement et uniquement à tous les termes obtenus précédemment\footnote{Un terme additionnel, nommé $\mathcal{L}_4^{\text{bis}}$ a également été introduit dans cet article. On peut cependant montrer qu'il se ramène à $\mathcal{L}_2$ par l'introduction d'une dérivée totale non-triviale, voire~\cite{Rodriguez:2017ckc}.} et donnés dans l'équation~\eqref{EqVectorGal1}. Ce dernier résultat, en conjonction de la recherche systématique de l'article~\cite{Allys:2015sht} du chapitre~\ref{PartArticleProca1} (après correction), donne finalement une indication très forte sur le fait que la théorie décrite par l'action~\eqref{EqActionVectorGalileon} décrit la dynamique la plus générale qu'on puisse écrire pour des Galiléons vectoriels dans un espace à 4 dimensions.

Les modèles de Galiléons impliquant des champs vectoriels dans la représentation adjointe d'un groupe de symétrie global SU(2) ont finalement été investigués dans l'article~\cite{Allys:2016kbq}, reproduit dans le chapitre~\ref{PartArticleProcaSU2}. La méthode d'investigation introduite dans l'article du chapitre~\ref{PartArticleProca1} y est améliorée, en vérifiant explicitement par des considérations de théories des groupes que tous les termes possibles sont écrits à chaque ordre, et en prenant soin de ne garder que des Lagrangiens produisant des dynamiques non-triviales et indépendantes. L'application de cette méthode d'investigation est utilisée pour construire explicitement tous les Lagrangiens possibles ayant jusqu'à six indices de Lorentz contractés. Un soin particulier est aussi porté à l'étude du lien entre les modèles de multi-Galiléons scalaires et vectoriels.

\chapter[Generalized Proca action for an Abelian vector field (article)]{Generalized Proca action for an Abelian vector field}
\label{PartArticleProca1}

\begin{figure}[h!]
\begin{center}
\includegraphics[scale=1.0]{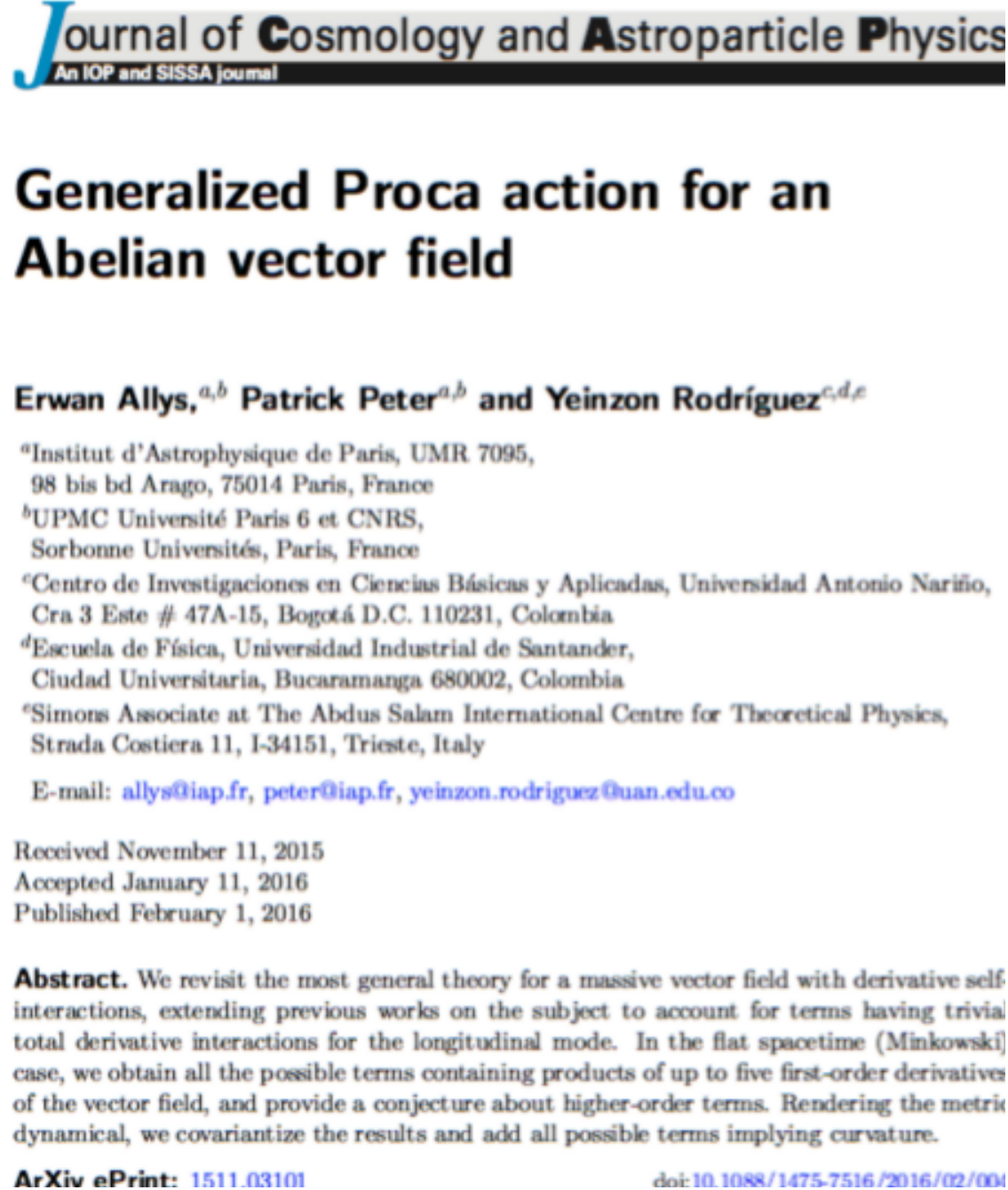}
\end{center}
\end{figure}

\section{Introduction}

Modifying general relativity in order to account for otherwise unresolved issues like,
e.g., the cosmological constant or even dark matter, has become common
practice lately. One way of doing so, following the scalar-tensor
proposal of the \oldstylenums{1930}s, is the so-called galileon method,
which provides a means to write down the most general theory allowing
for Ostrogradski-instability-free \cite{ostro,Woodard:2006nt}
second-order equations of motion
\cite{Nicolis:2008in,Deffayet:2009wt,Deffayet:2009mn,Deffayet:2011gz,Deffayet:2013lga},
 which happens to be equivalent \cite{Kobayashi:2011nu}, in the
 single-field case\footnote{It is a subset in the multi-field case.} and
 in four dimensions, to that proposed much earlier by Horndeski
 \cite{Horndeski:1974wa}.  The latter author, indeed, had generalized
 this idea to what, today, should appropriately be named ``vector
 Galileon'', an Abelian vector field, with an action including sources,
 and with the assumption of recovering the Maxwell equations in flat
 spacetime \cite{Horndeski:1976gi}.  Related works were developed
 recently that extended the applicability of the word ``vector
 Galileon'':  not invoking gauge invariances but having several vector
 fields and sticking to purely second-order field equations
 \cite{Deffayet:2010zh}, coupling an Abelian vector field with a scalar
 field in the framework of Einstein gravity \cite{Fleury:2014qfa}, or
 invoking an Abelian gauge invariance for just one vector field in flat
 spacetime and sticking to purely second-order field equations
 \cite{Deffayet:2013tca}.  The latter work became a no-go theorem that,
 however, is not
the end of the vector Galileon theories:
it suffices to drop the U(1) invariance hypothesis to obtain more terms
in a non-trivial and viable theory. This procedure thus naturally
generalizes the Proca theory for a massive vector field in the sense
that it describes a vector field with second-order equations of motion
for three propagating degrees of freedom, as required for the finite-dimensional representation of the Lorentz group leading to a massive
spin 1 field.

In previous works \cite{Heisenberg:2014rta,Tasinato:2014eka}, such a
theory was elaborated, introducing some fruitful ideas. A thorough
examination of these works revealed that some terms that have been
proposed are somehow redundant, while others appear to be missing. The
purpose of the present article is thus to complete these pioneering
works, investigating in a hopefully exhaustive way all the terms which
can be included in such a generalized Proca theory under our
assumptions\footnote{Some general works on general ghost-free Lagrangians 
published recently, see Refs. \cite{Li:2015vwa,Li:2015fxa}, can also be applied to the generalization of the Proca theory.}. The most crucial constraint in the building up of the
theory is the demand that only three polarizations should be able to
propagate, namely two transverse and one longitudinal, the latter
being reducible to the scalar formulation of the Galileon, thus providing
the required link between the vector and scalar theories.
On the other hand, considering the transverse modes, we found
numerous differences between the two theories, as in particular we
obtained many more terms in the vector case. Indeed, whereas the scalar
Lagrangian comprises a finite number of possibilities, we conjecture
that the vector one is made up of an infinite tower of such terms.

The structure of the rest of the article goes as follows. First, we recall
previous works by introducing the generalized Abelian Proca theory,
and explicit the assumptions under which our Lagrangian is constructed.
Recalling as a starting point the terms already obtained in Refs. \cite{Heisenberg:2014rta}
and \cite{Tasinato:2014eka}, we use their results to understand
what kind of other terms could also be compatible with our assumptions.
We then move on to these extra terms in Sec. \ref{PA1AdditionalTerms}, first
introducing a general method of investigation of these new terms, then
applying the latter to terms containing products of up to five first-order
derivatives of the vector field. We conclude this part by discussing what 
the complete generalized Proca theory could be. Finally, in
Sec. \ref{PA1Covariantization}, we consider the extension of the Proca
theory to a curved spacetime. We need to add yet more terms necessary
to render the Lagrangian healthy (in the sense defined above), and this
also leads to new interactions with gravity.

\section{Generalized Abelian Proca theory}

The Proca theory describes the dynamics of a vector field with a mass
term in its Lagrangian, thus explicitly breaking the U(1) gauge
invariance usually associated with the Maxwell field $A_\mu$. The Proca
action therefore reads
\begin{equation}
\mathcal{S}_\mathrm{Proca}  = \int \mathcal{L}_\mathrm{Proca}\,\text{d}^4x 
= \int \left( -\frac14 F_{\mu\nu} F^{\mu\nu} + \frac12 m_A^2 X \right)\,\text{d}^4x,
\label{PA1Proca}
\end{equation}
where the antisymmetric Faraday tensor is $F_{\mu\nu} \equiv
\partial_\mu A_\nu -\partial_\nu A_\mu$, and we have defined, for further
convenience, the shorthand notation $X\equiv A_\mu A^\mu$. Following and
extending Ref.~\cite{Heisenberg:2014rta}, we want to provide the
broadest possible generalization of this action under a set of
specific assumptions to which we now turn.

Before we move on to curved spacetime in Sec.~\ref{PA1Covariantization},
we shall in what follows consider that our field lives in a non-dynamical
Minkowski metric $g^{\mu \nu}=\eta^{\mu \nu}=\rm{diag}(-1,+1,+1,+1)$. We shall
also make use of the notation $(\partial \cdot A)=\partial_\mu A^\mu$
for simplification and notational convenience.

\subsection{Theoretical assumptions}
\label{PA1conditions}

The idea is to add to the minimal Proca action in Eq. (\ref{PA1Proca}) all
acceptable terms containing not only functions of $X$ but also
derivative self-interactions under a set of suitable conditions.
In order to explicit those conditions, let us split the vector into
a scalar/vector decomposition
\begin{equation}
A_\mu=\partial_\mu \pi + \bar{A}_\mu \ \ \ \hbox{or} \ \ \ A
= \text{d} \pi + \bar{A},
\end{equation}
 where $\pi$ is a scalar field, called the St\"uckelberg field, and
 $\bar{A}_\mu$ is a divergence-free vector ($\partial_\mu \bar
 A^\mu=0$), containing the curl part of the field, \emph{i.e.} that for which
 the Faraday tensor is non vanishing (in differential form terms, it is
 the non-exact part of the form, with the Faraday form now equal to
 $F=\text{d} A = \text{d}\bar{A}$). The conditions we want to impose on the theory
 in order that it makes (classical) sense are reminiscent of the
 galileon conditions, namely we demand
\begin{itemize}
\item[a)] at most second-order equations of motion for all physical
degrees of freedom, \emph{i.e.}, for both $A_\mu$ and $\pi$,
\item[b)] at most second-order derivative terms in $\pi$ in the action,
and first-order derivative terms for $A_\mu$,
\item[c)] only three propagating degrees of freedom for the vector field;
 in other words, there should be no propagation of its zeroth component.
\end{itemize} 
The first condition ensures stability, as discussed e.g. in
Ref.~\cite{Woodard:2006nt}\footnote{This result is indirect: accepting a
third-order derivative in the equations of motion would imply at least a
second-order time derivative in the Lagrangian, which is not degenerate
but yields a Hamiltonian which is unbounded from below.}. Because we
demand first-order derivatives in $A_\mu$, the second condition implies
that the first is automatically satisfied for $A_\mu$; the first
condition is therefore necessary to implement only for the scalar part
$\pi$ of the vector field. As for the third condition, it stems from the
definition of a vector as a unit spin Lorentz-group representation, a spin
$s$ object having $(2s+1)$ propagating degrees of freedom. In Ref.
\cite{Heisenberg:2014rta}, an extra condition was implicit, namely that the
longitudinal mode should not have trivial total derivative interactions; the
terms we obtain below and that were not written down in this reference
stem precisely from our relaxing of this condition.

For a given Lagrangian function $\mathcal{L}(A_\mu)$, the number of
actually propagating degrees of freedom must be limited to three. To enforce
this requirement, we compute the Hessian matrix $\mathcal{H}^{\mu\nu}$
associated with the Lagrangian term considered through
\begin{equation}
\mathcal{H}^{\mu\nu}=\frac{\partial^2 \mathcal{L}}{\partial(\partial_0
A_\mu)\partial(\partial_0A_\nu)},
\label{PA1Hess}
\end{equation}
and demand that it should have a vanishing eigenvalue associated to the
absence of propagation of the time component of $A_\mu$. A sufficient
condition to achieve this goal is to ask that $\mathcal{H}^{00}=0$ and
$\mathcal{H}^{0i}=0$. One can show that this condition is the unique one
permitting a vanishing eigenvalue: since different derivatives of
the vector field are independent, the only way to diagonalize the
Hessian without any vanishing eigenvalue is to cancel its symmetric terms
with one another, and this, in turn, is not possible by means of a Lorentz
transformation.

\subsection{The Heisenberg action}

Let us summarize Ref.~\cite{Heisenberg:2014rta}, switching to notations
agreeing with those we use here. The Lagrangian terms which appear in
addition to the standard kinetic term will be called $\mathcal{L}_n$, a
notation coming from that which is usual in the Galileon theory
\cite{Deffayet:2013lga}. There, $N = n + 2$ is the number of scalar
field factors appearing in each Lagrangian, and $M=n-2$ is the number of
second derivatives of the Galileon field (we do not count the
arbitrary functions of $X$ in this power counting). Keeping the same
notation,
now applied to the scalar part $\pi$ of the vector field, one can write
the general Lagrangian as the usual Maxwell Lagrangian with a
generalized mass term, \emph{i.e.}, one
replaces the Lagrangian in Eq.~(\ref{PA1Proca}) by
\begin{equation}
\mathcal{L}_\mathrm{Proca}^\mathrm{gen.}=
-\frac{1}{4}F_{\mu\nu}F^{\mu\nu}+\sum_{n=2}^5
\mathcal{L}_n,
\end{equation}
with
\begin{equation}
\begin{split}
\mathcal{L}_2 = & f_2(A_\mu, F_{\mu\nu},\tilde{F}_{\mu\nu}),\cr
\mathcal{L}_3 = & f_3^{\mathrm{Gal}}(X) (\partial \cdot A),\cr
\mathcal{L}_4 = & f_4^{\mathrm{Gal}}(X) \left[ \left( \partial_\mu A_\nu
                                 \partial^\nu A^\mu \right) -
                                 \left( \partial\cdot A \right) \left( \partial\cdot A \right) \right],\cr                                 
\mathcal{L}_5 = & f_5^{\mathrm{Gal}}(X) \left[\left(\partial\cdot A\right)^3 - 3 \left(\partial\cdot A\right)
  \left(\partial_ {\nu}A_ {\rho} \partial^ {\rho}A^ {\nu}\right) + 2
  \left(\partial_{\mu}A^ {\nu} \partial_ {\nu}A^ {\rho} \partial_
       {\rho}A_ {\mu}\right) \right] \cr  &+ f_5^{\text{Perm}}(X)
\left[ \left(\partial\cdot A\right) \left[ \left(\partial_ {\rho}A_
    {\nu} \partial^ {\rho}A^ {\nu}\right) - \left(\partial_ {\nu}A_
    {\rho} \partial^ {\rho}A^ {\nu}\right) \right]+ \left(\partial_
  {\mu}A_ {\rho} \partial^ {\nu}A^ {\mu} \partial^ {\rho}A_ {\nu}
  \right) - \left(\partial^ {\nu}A^ {\mu} \partial_ {\rho}A_ {\nu}
  \partial^ {\rho}A_ {\mu}\right)\right],
\end{split}
\label{PA1LagHeisenberg}
\end{equation}
where $f_2(A_\mu, F_{\mu\nu},\tilde{F}_{\mu\nu})$ is an arbitrary
function of all scalars which can be constructed from $A_\mu$,
$F_{\mu\nu}$  and $\tilde{F}_{\mu\nu}\equiv
\frac{1}{2}\epsilon_{\mu\nu\alpha\beta}F^{\alpha\beta}$, the latter
being the Hodge dual of the Faraday tensor\footnote{Here and in
what follows, we denote by $\epsilon^{\sigma_1\sigma_2\dots \sigma_D}$
the totally antisymmetric Levi-Civita tensor in $D$ dimensions (in the
case $D=4$ we are concerned with in the rest of this paper, we write
$\epsilon^{\mu\nu\alpha\beta}$).}. All other functions $f$ are
independent arbitrary functions of $X$ only.

These expressions actually differ from those given in
Ref.~\cite{Heisenberg:2014rta}. First, we did not add a term of the form
$f(X) \left[ \left(\partial_ {\nu}A_ {\mu} \partial^ {\nu}A^ {\mu}\right) -
\left(\partial_ {\mu}A_ {\nu} \partial^ {\nu}A^ {\mu}\right) \right]$ in
$\mathcal{L}_4$, as such a term is merely equal to
$\frac12 f(X) F_{\mu\nu}F^{\mu\nu}$ and is therefore
already contained in $f_2$. Moreover, both terms in $\mathcal{L}_5$
have different arbitrary functions as prefactors. In fact, this general
form is sufficient to verify the conditions given in Sec.
\ref{PA1conditions}; this will be explained in Sec. \ref{PA1L3->5}.

The Lagrangians given by Eqs. (\ref{PA1LagHeisenberg}) contain different
kinds of terms with various origins. The first contributors, with
prefactors given by the arbitrary functions $f_n^{\text{Gal}}$,
originate from the scalar part of the vector field:  setting $\bar{A}\to
0$, \emph{i.e.} $A_\mu\to \partial_\mu \pi$, they are indeed nonvanishing. As,
for consistency, they must verify the hypothesis of Galileon theory,
they should thus come straightforwardly from the Galileon theory and
therefore ought to be equivalent to the only Lagrangians satisfying such
properties as they have been classified in Ref.~\cite{Deffayet:2011gz}.
One way of defining them explicitly consists, for instance, in taking the
specific form chosen e.g. in Ref.~\cite{Deffayet:2011gz,Deffayet:2013lga} (where it was
called $\mathcal{L}^3_N$)
\begin{equation}
\label{PA1LagGalileon}
\mathcal{L}^\text{Gal}_{N} =\left[ A_{(2n)}^{\mu_1\mu_2\dots
  \mu_n\nu_1\nu_2\dots \nu_n} \pi_{\lambda} \pi^{\lambda}
\right] \pi_{\mu_1 \nu_1} \dots \pi_{\mu_n \nu_n},
\end{equation}
with 
\begin{equation}
A_{(2m)}^{\mu_1\mu_2\dots \mu_m\nu_1\nu_2\dots \nu_m} =
\frac{1}{(D-m)!} \epsilon^{\mu_1\mu_2\dots \mu_m\sigma_1\sigma_2\dots
  \sigma_{D-m}} \epsilon^{\nu_1\nu_2\dots \nu_m}{}_{ \sigma_1\dots
  \sigma_{D-m}},
\end{equation}
where $\pi_{\mu\nu\cdots}\equiv\partial_\mu\partial_\nu\cdots\pi$ is
a convenient shorthand notation for the derivatives. 
The definition which appears to be unambiguous then arises from the
following procedure: we substitute, in Eq.~\eqref{PA1LagGalileon}, the
first derivative of the scalar field by a component of the vector field
($\pi_{\mu\nu\cdots}\to A_{\mu,\nu\cdots}$), keeping all indices the way
they appear in the original form. From the point of view of the scalar
part of the Lagrangian, this leads to the same Galileon action, and it
would also if we were to permute derivatives (indices in this case). For
example, we could take in $\mathcal{L}_5$, a term having the special
form
\begin{equation}
f_5^{\text{Gal,alt}}(X) \left[\left(\partial\cdot A\right)^3 -
  \left(\partial\cdot A\right) \left[ \left(\partial_ {\nu}A_ {\rho}
    \partial^ {\rho}A^ {\nu}\right) + 2 \left(\partial_ {\rho}A_ {\nu}
    \partial^ {\rho}A^ {\nu}\right)\right] + 2 \left(\partial^ {\nu}A^
       {\mu} \partial_ {\rho}A_ {\nu} \partial^ {\rho}A_ {\mu}\right)
       \right] ,
\end{equation}
which would also reduce to Eq.~(\ref{PA1LagGalileon}) in the scalar case.
This term is also obtained by taking
$f_5^{\text{Perm}}=-2f_5^{\text{Gal}}$. Both terms contained in
$\mathcal{L}_5$ could thus be obtained by using the Galileon Lagrangian
and shuffling around the second-order derivatives before doing the
replacement by the vector field. The choice proposed above however provides a
uniquely defined action when it comes to the vector field case.

The second category of terms, including $f_2$ (apart from terms containing
only $X$) and $f_5^{\text{Perm}}$, gives a vanishing
contribution when going to the scalar sector, due to the fact that $
\partial_\mu \partial_\nu \pi = \partial_\nu \partial_\mu \pi$, which is
not verified by $\partial_\mu \bar{A}_\nu$ by definition. This is
trivial for the terms containing the Faraday tensor, since $A_\mu\to
\partial_\mu\pi$ implies $F_{\mu\nu}\to 0$ identically, and can be
verified explicitly for the term including $f_5^{\text{Perm}}$. These
terms can also be seen to lead to only three propagating degrees of
freedom for the vector field. Given this fact, one is naturally led to
ask whether similar kind of terms, including more derivatives, could be
possible. They are discussed in the following section.

\section{Additional terms}
\label{PA1AdditionalTerms}
\subsection{Procedure of investigation}
\label{PA1procedure}

In order to find all the possible terms satisfying the conditions of
Sec. \ref{PA1conditions}, we developed a systematic procedure, which we
explain below by considering the term $\mathcal{L}_n$ of the
Lagrangian, containing $(n-2)$ first-order derivatives of the vector
field\footnote{The index $n$ appearing in our Lagrangians is inconvenient
but we keep it in order to follow previous conventions developed historically
from the Galileon action.}.
The final action will then consist of a sum over all such possible terms,
weighted by arbitrary functions of $X$, as this can in no way change
anything in the discussion of the Hessian.

We first list all the possible terms which can be written as
contractions of $(n-2)$ first-order derivatives of $A_\mu$. We call them
$\mathcal{L}^{\text{test}}_{n,i}$, where $i$ labels the different terms
which can be written for a given $n$. These test Lagrangians are then
linearly combined to provide the most general term at a given order $n$,
whose Hessian, in Eq. \eqref{PA1Hess}, is computed. We apply our requirements
($\mathcal{H}^{0\mu}=0$ for all $\mu=0,\cdots,3$) to derive relations
among the coefficients of the linear combination, and to finally obtain
the relevant terms giving only three propagating degrees of freedom.
Those fall into two distinct categories, namely those whose scalar
sector vanishes or not. The latter necessarily reduce to the
Galileon action, while the former are new, purely vectorial, terms
which, in turn, can either be constructed from $F_{\mu\nu}$ only, or be
the terms we are interested in. Any term leading to a non-trivial
dynamics for the scalar part that would be nonvanishing should be then
set to zero in order to comply with the requirement that the scalar
action is that provided by the Galileon.

We are investigating terms that contain only derivatives  and not the
vector field itself in the contractions over Lorentz indices, as all
possible such terms can always be reduced to those studied below and a
total derivative. The typical example we have in mind is a term of the
form
$$
A_\alpha A^\mu\partial_\mu A^\alpha = \frac12 \left[ \partial_\mu
\left( X A^\mu\right) - X \left( \partial \cdot A\right) \right],
$$
which is then equivalent to those we discuss below, up to a total
derivative. Note that a similar term including an arbitrary function of
$X$ in front would lead to second-order derivatives in $A$, and is thus
excluded by construction.

Finally, we would like to mention at this point that we are here only
considering terms involving the metric, \emph{i.e.}, we build scalars for the
Lagrangian by contracting the indices of quantities such as
$\partial_\mu A_\nu$ using the metric $g^{\mu\nu}$. Below, in
Sec.~\ref{PA1sec:eps}, we will consider another possible case, that for
which scalars are also built through contractions with the completely
antisymmetric tensor $\epsilon^{\alpha\beta\mu\nu}$; these extra
terms and those first considered produce independent Hessian variations
and can thus be studied independently.

\subsection{From $\mathcal{L}_3$ to $\mathcal{L}_5$}
\label{PA1L3->5}

The first term appearing in our expansion is $\mathcal{L}_3$, whose only
test Lagrangian can be written as $(\partial \cdot A)$, which has a
vanishing Hessian, and is nothing but the Galileon vector term
\begin{equation}
\mathcal{L}_{3}^{\text{Gal}} =  \left(\partial \cdot A\right).
\end{equation}

The case of $\mathcal{L}_4$ is slightly more involved, the test Lagrangians
reading
\begin{equation}
\mathcal{L}^{\text{test}}_{4,1} = \left(\partial\cdot A\right)^2,\ \ \ \ 
\mathcal{L}^{\text{test}}_{4,2} = \left(\partial_\nu A_\mu \partial^\nu A^\mu \right),\ \ \ \ \hbox{and}
\ \ \ \ \mathcal{L}^{\text{test}}_{4,3} = \left(\partial_\mu A_\nu \partial^\nu A^\mu \right).
\end{equation}
Setting the full test Lagrangian at this order to be
\begin{equation}
\mathcal{L}_{4}^{\text{test}} = x_1 \mathcal{L}^{\text{test}}_{4,1} + x_2
\mathcal{L}^{\text{test}}_{4,2} + x_3 \mathcal{L}^{\text{test}}_{4,3},
\end{equation}
we find the following relevant Hessian terms,
\begin{equation}
\mathcal{H}_{4}^{00} = 2 \left(x_{1} + x_{2} + x_{3}\right) \ \ \ \ \hbox{and}
\ \ \ \ \mathcal{H}_{4}^{0i} = 0,
\end{equation}
so that the solutions ensuring only three propagating degrees of freedom are
\begin{equation}
\mathcal{L}_{4}^{\text{Gal}} = \left(\partial_ {\mu}A_ {\nu} \partial^
        {\nu}A^ {\mu}\right) - \left(\partial\cdot A\right)
        \left(\partial\cdot A\right) ,
\end{equation}
and
\begin{equation}
\mathcal{L}_{FF} = F_{\mu\nu}F^{\mu\nu}=\left(\partial_ {\nu}A_ {\mu}
\partial^ {\nu}A^ {\mu}\right) - \left(\partial_ {\mu}A_ {\nu}
\partial^ {\nu}A^ {\mu}\right).
\end{equation}
We identify the first term with the Galileon vector term, and the second
to the only term at this order which can be built from the field tensor.
This term was expected, and we will
not consider it any further since it is already included in
$\mathcal{L}_{2}$. We also check that, as these terms independently
verify the required conditions, they are effectively independent.

Let us move on to the following order, namely $\mathcal{L}_5$. The test
Lagrangians are
\begin{equation}
\begin{split}
\mathcal{L}^{\text{test}}_{5,1} & = \left(\partial\cdot A\right)^3,\\
\mathcal{L}^{\text{test}}_{5,2} & = \left(\partial\cdot A\right) \left(\partial_
       {\rho}A_ {\nu} \partial^ {\rho}A^\nu\right),\\
\mathcal{L}^{\text{test}}_{5,3} & = \left(\partial\cdot
        A\right) \left(\partial_ {\nu}A_ {\rho} \partial^ {\rho}A^\nu\right),\\
\mathcal{L}^{\text{test}}_{5,4} & = \left(\partial_ {\mu}A_
        {\rho} \partial^ {\nu}A^ {\mu} \partial^ {\rho}A_\nu\right),\\
\mathcal{L}^{\text{test}}_{5,5} & = \left(\partial^ {\nu}A^
        {\mu} \partial_ {\rho}A_ {\nu} \partial^ {\rho}A_\mu\right).
\end{split}
\end{equation}
Using 
\begin{equation}
\mathcal{L}_{5}^{\text{test}} = x_{1} \mathcal{L}^{\text{test}}_{5,1} +
x_{2} \mathcal{L}^{\text{test}}_{5,2} + x_{3}
\mathcal{L}^{\text{test}}_{5,3} + x_{4} \mathcal{L}^{\text{test}}_{5,4}
+x_{5} \mathcal{L}^{\text{test}}_{5,5},
\end{equation}
we obtain the relevant Hessian terms\footnote{Although the Hessian looks
different from that of Ref.~\cite{Heisenberg:2014rta}, it actually is
equivalent, the difference stemming from the fact that in the latter work,
$\mathcal{H}_{5}^{00}$ was decomposed on $\partial_i A^i$ and
$\partial_0 A^0$, while we chose to decompose it on $\partial_\mu A^\mu$
and $\partial_0 A^0$.}
\begin{equation}
\begin{split}
\mathcal{H}_{5}^{00} & = -2 \left(-3 x_{1} - x_{2} - x_{3}\right)
\left(\partial \cdot A\right) -2 \bigl[2 x_{2} + 2 x_{3} + 3
  \left(x_{4} + x_{5}\right) \bigr]
\left(\partial^{0}A^{0}\right),\\
\mathcal{H}_{5}^{0i} & = -2
\left(x_{2} + x_{5}\right) \left(\partial^{0}A^{i}\right) - \left(2
x_{3} + 3 x_{4} + x_{5}\right) \left(\partial^{i}A^{0} \right),
\end{split}
\end{equation}
which, when vanishing, provide the following solutions:
\begin{equation}
\mathcal{L}_{5}^{\text{Gal}} = \left(\partial\cdot A\right)^3 - 3
\left(\partial\cdot A\right) \left(\partial_ {\nu}A_ {\rho} \partial^
     {\rho}A^ {\nu}\right) + 2 \left(\partial_{\mu}A^ {\nu} \partial_
     {\nu}A^ {\rho} \partial_ {\rho}A_ {\mu}\right),
     \label{PA1Gal5}
\end{equation}
and
\begin{equation}
\mathcal{L}_{5}^{\text{Perm}} = \frac{1}{2}\left(\partial\cdot A\right) F_{\mu\nu}
F^{\mu\nu} + \partial_\rho A_\nu \partial^\nu A_\mu F^{\mu\rho}.
\label{PA1Perm5}
\end{equation}
The first, Eq. \eqref{PA1Gal5}, is the Galileon scalar Lagrangian, verifying
all the imposed conditions, leading to second-order equations of motion,
even multiplied by any function of $X=\partial_\mu \pi \partial^\mu
\pi$; this is shown in Ref. \cite{Deffayet:2011gz}. The second one,
Eq. \eqref{PA1Perm5}, gives no dynamics for the Galileon scalar, since it
vanishes in the scalar sector, as can be seen from the fact that it can
be factorized in terms of the strength tensor $F_{\mu\nu}$ (but
not exclusively, since it would otherwise be gauge invariant and thus
reducible to functions of $F_{\mu\nu}F^{\mu\nu}$ and $F_{\mu\nu}\tilde
F^{\mu\nu}$). It does however also satisfy our conditions, even
multiplied by any function of $X$.

This complete the proof that Eq.~(\ref{PA1LagHeisenberg}), as originally
obtained in Ref.~\cite{Heisenberg:2014rta}, fulfills the required constraints
for a generic classical second-order action for a vector field. We shall now
discuss the possibility of extra terms, not present in previous works.

\subsection{Higher-order actions}

Since the scalar Galileon action stops at the level discussed above, it sounds
reasonable to assume the same to apply for the vector field, in particular in
view of the fact that this field contains a scalar part, and to consider Eq. (\ref{PA1LagHeisenberg})
to provide the most general second-order classical vector theory. This is not
what happens, in practice, as we show below, as one can indeed find extra
terms, which do vanish in the limit $A_\mu\to\partial_\mu\pi$, leaving a
non-trivial dynamics for the divergence-free part $\bar A_\mu$.

\paragraph{Fourth power derivatives: $\mathcal{L}_6$.}
\label{PA1L6}

We begin our examination of the higher powers in derivatives by
concentrating on $\mathcal{L}_{6}$, which involves therefore four powers
of the field gradient. Since the actual calculations imply very large
and cumbersome terms, and because those are not so important for the
understanding of this work, we have regrouped this calculation in
Appendix \ref{PA1appL6}.

We find that there are four Lagrangians verifying the Hessian condition.
They are
\begin{equation}
\begin{split}
\mathcal{L}_{6}^{\mathrm{Gal}} =& \left(\partial\cdot A\right)^4 - 2
\left(\partial\cdot A\right) ^2 \left[\left(\partial_ {\rho}A_
  {\sigma} \partial^ {\sigma}A^ {\rho}\right) +2 \left(\partial_
  {\sigma}A_ {\rho} \partial^ {\sigma}A^ {\rho}\right) \right] + 8
\left(\partial\cdot A\right) \left(\partial^ {\rho}A^ {\nu} \partial_
     {\sigma}A_ {\rho} \partial^ {\sigma}A_ {\nu}\right) -
     \left(\partial_ {\mu}A_ {\nu} \partial^ {\nu}A^ {\mu}\right)^2 \\ &+
     4 \left(\partial_ {\nu}A_ {\mu} \partial^ {\nu}A^ {\mu}\right)
     \left(\partial_ {\rho}A_ {\sigma} \partial^ {\sigma}A^
          {\rho}\right) -2 \left(\partial_ {\nu}A^ {\sigma} \partial^
          {\nu}A^ {\mu} \partial_ {\rho}A_ {\sigma} \partial^ {\rho}A_
          {\mu}\right) - 4 \left(\partial^ {\nu}A^ {\mu} \partial^
          {\rho}A_ {\mu} \partial_ {\sigma}A_ {\rho} \partial^
          {\sigma}A_ {\nu}\right),
\end{split}
\label{PA1Gal6}
\end{equation}
which should \emph{a priori} give a dynamics for the scalar part, and
\begin{equation}
\begin{split}
\mathcal{L}_{6}^{\text{Perm}} =&  \left(\partial\cdot A\right)^2
F^{\mu\nu} F_{\mu\nu}  - \left(\partial_ {\rho}A_
     {\sigma} \partial^ {\sigma}A^ {\rho}\right)  F^{\mu\nu}F_{\mu\nu}
        + 4 \left(\partial\cdot A\right)
     \partial^\rho A^\nu \partial^\sigma A_\rho F_{\nu\sigma} \\
          & + \partial^\mu A_\nu F^\nu{}_\rho F^\rho{}_\sigma F^\sigma{}_\mu
          - 4 \,\partial^\mu A _\nu \partial^\nu A_\rho \partial^\rho A_\sigma F^\sigma{}_\mu.
\end{split}
\label{PA1Perm6}
\end{equation}

The first term, Eq. \eqref{PA1Gal6}, contains four second-order derivatives of
the scalar part and, thus, because it is of order higher than 3, cannot
verify the conditions of
Sec.~\ref{PA1conditions}~\cite{Nicolis:2008in,Deffayet:2011gz}. We shall
consequently discard it.

There are two more terms which satisfy the required conditions, namely
$\mathcal{L}_{FF\cdot FF} =(F_{\mu\nu}F^{\mu\nu})^2$ and
$\mathcal{L}_{FFFF} = F^\mu{}_\nu F^\nu{}_\rho F^\rho{}_\sigma
F^\sigma{}_\mu$.  Since these terms are built out only of the
field strength tensor, they do give a dynamics to the vector field,
but are already included in $\mathcal{L}_{2}$, and we therefore will not
consider them any further.

The terms in Eq.~\eqref{PA1Perm6} exhibit explicit powers of
($\partial \cdot A$), and therefore cannot be built
out of simple products of the field tensor. The Lagrangian
$\mathcal{L}_{6}^{\text{Perm}}$ vanishes in the scalar sector and verifies all the
conditions we want to impose, even multiplied by an arbitrary function
of $X$. We consider the general theory of a vector field, and we thus
must add to Eq. \eqref{PA1LagHeisenberg} the term
\begin{equation}
\mathcal{L}_{6}=f_6^{\text{Perm}}(X)\mathcal{L}_{6}^{\text{Perm}},
\end{equation}
with $f_6^{\text{Perm}}$ being an arbitrary function of $X$.

\paragraph{Fifth power derivatives: $\mathcal{L}_7$.}
\label{PA1L7}

Having found an extra non-trivial dynamical term, we now move on to
including yet one more power in the derivatives. Even more cumbersome
than those leading to $\mathcal{L}_6$, the calculations yielding
$\mathcal{L}_7$ are reproduced in Appendix \ref{PA1appL7}. We obtain

\begin{equation}
\begin{split}
\mathcal{L}_{7}^{\text{Gal}} =& 3 \left(\partial \cdot A\right)^5-
\left(\partial \cdot A\right)^3 10\left[ \left(\partial_ {\sigma}A_
  {\gamma} \partial^ {\gamma}A^ {\sigma}\right) - 2\left(\partial_
  {\gamma}A_ {\sigma} \partial^ {\gamma}A^ {\sigma} \right)\right]+ 60
\left(\partial \cdot A\right)^2 \left(\partial^ {\sigma}A^ {\rho}
\partial_ {\gamma}A_ {\sigma} \partial^ {\gamma}A_ {\rho}\right) \\ 
& + 15 \left(\partial \cdot A\right)\left[ \left(\partial_ {\nu}A_ {\rho}
  \partial^ {\rho}A^ {\nu} \right)\left(\partial_ {\sigma}A_ {\gamma}
  \partial^ {\gamma}A^ {\sigma}\right) +2 \left(\partial_ {\rho}A_
          {\nu} \partial^ {\rho}A^ {\nu}\right)\left( \partial_
          {\sigma}A_ {\gamma} \partial^ {\sigma}A^ {\gamma}\right) -
          4\left(\partial_ {\rho}A^ {\gamma} \partial^ {\rho}A^ {\nu}
          \partial_ {\sigma}A_ {\gamma} \partial^ {\sigma}A_
                  {\nu}\right) \right. \\
& \left.  - 4\left(\partial^
                  {\rho}A^ {\nu} \partial^ {\sigma}A_ {\nu} \partial_
                  {\gamma}A_ {\sigma} \partial^ {\gamma}A_
                  {\rho}\right) \right] + 20 \left( \partial_ {\rho}A_
{\gamma} \partial^ {\sigma}A^ {\rho} \partial^ {\gamma}A_
{\sigma}\right)\left[ \left(\partial_ {\nu}A_ {\mu} \partial^ {\nu}A^
  {\mu}\right) - \left(\partial_ {\mu}A_ {\nu} \partial^ {\nu}A^
  {\mu}\right) \right] \\
& - 60 \left(\partial_ {\nu}A_ {\mu} \partial^
{\nu}A^ {\mu} \right)\left(\partial^ {\sigma}A^ {\rho} \partial_
{\gamma}A_ {\sigma} \partial^ {\gamma}A_ {\rho} \right) + 60
\left(\partial^ {\nu}A^ {\mu} \partial_ {\rho}A^ {\gamma} \partial^
     {\rho}A_ {\mu} \partial_ {\sigma}A_ {\gamma} \partial^ {\sigma}A_
     {\nu}\right)\\
& + 60 \left(\partial^ {\nu}A^ {\mu} \partial^
     {\rho}A_ {\mu} \partial^ {\sigma}A_ {\nu} \partial_ {\gamma}A_
     {\sigma} \partial^ {\gamma}A_ {\rho}\right) - 60 \left(\partial^ {\nu}
     A^{\mu} \partial_ {\rho}A_ {\gamma}
     \partial^ {\rho}A_ {\mu} \partial^ {\sigma}A_ {\nu} \partial^
             {\gamma}A_ {\sigma}\right)\\
& + 12 \left(\partial_ {\mu}A_
             {\gamma} \partial^ {\nu}A^ {\mu} \partial^ {\rho}A_ {\nu}
             \partial^ {\sigma}A_ {\rho} \partial^ {\gamma}A_
                     {\sigma}\right) ,
\end{split}
\label{PA1Gal7}
\end{equation}
which, similarly to Eq. \eqref{PA1Gal6}, must be discarded as it leads to higher-order
equations of motion for the scalar part, and two extra terms, namely
\begin{equation}
\begin{split}
\mathcal{L}_{7}^{\text{Perm,1}} = &  \left(\partial \cdot
A\right)^3F^{\mu\nu} F_{\mu\nu} + 6 \left(\partial \cdot A\right) ^2 
\partial^\mu A^\nu \partial^\rho A_\mu F_{\nu \rho} + 3 \left(\partial
\cdot A\right) \left[ \left(\partial_  {\nu}A_ {\rho} \partial^ {\rho}A^
{\nu} \right)^2 - \left( \partial_ {\rho}A_ {\nu} \partial^ {\rho}A^
{\nu} \right)^2 \right] \\
& + 3 \left(\partial \cdot A\right)
\left(\partial^\mu A_\nu F^\nu{}_\rho F^\rho{}_\sigma F^\sigma{}_\mu - 4
\partial^\mu A _\nu \partial^\nu A_\rho \partial^\rho A_\sigma
F^\sigma{}_\mu\right) + 4 \left(\partial_\mu A_\nu \partial^\mu A^\nu
\right)\left(\partial^\rho A^\sigma \partial_\gamma A_\rho \partial
^\gamma A_\sigma \right)\\
&  - 4 \left(\partial_\mu A_\nu \partial^\nu
A^\mu \right)\left(\partial^\rho A^\sigma \partial_\gamma A_\rho
\partial _\sigma A^\gamma \right)  + 2 \left(\partial_\mu A_\nu
\partial^\mu A^\nu \right) \left(\partial^\rho A_\sigma \partial_\gamma
A_\rho F^{\gamma \sigma} \right) \\
& - 6  \partial^\mu A_\nu
F^\nu{}_\rho F^\rho{}_\sigma F^\sigma{}_\gamma F^\gamma{}_\mu 
 + 12 \partial^\mu A_\nu \partial^\nu A_\rho \partial^\rho A_\sigma
\partial^\sigma A_\gamma F^\gamma{}_\mu ,
\end{split}
\label{PA11Perm7}
\end{equation}
and
\begin{equation}
\begin{split}
\mathcal{L}_{7}^{\text{Perm,2}} = & \frac{1}{4}\left(\partial \cdot
A\right)\big[\left(F_{\mu\nu} F^{\mu\nu}\right)^2 - 4\partial^\mu A_\nu
F^\nu{}_\rho F^\rho{}_{\sigma} F^\sigma{}_\mu \big] + \left(
F^{\mu\nu}F_{\mu \nu}\right) \partial^\sigma A^\rho \partial^\gamma
A_\sigma F_{\rho \gamma}  \\
&  +2 \left( \partial^\mu A_\nu
F^\nu{}_\rho F^\rho{}_\sigma F^\sigma{}_\gamma F^\gamma{}_\mu \right),
\end{split} \label{PA12Perm7}
\end{equation}
both of which vanish in the limit $A_\mu\to\partial_\mu\pi$. It can
be checked explicitly that they independently verify all the conditions given in
Sec. \ref{PA1conditions}, so the final theory will contain 
\begin{equation}
\mathcal{L}_{7}=f_{7}^{\text{Perm,1}}(X)
\mathcal{L}_{7}^{\text{Perm,1}}+f_{7}^{\text{Perm,2}}(X)
\mathcal{L}_{7}^{\text{Perm,2}},
\end{equation}
with $f_{7}^{\text{Perm,1}}$ and $f_{7}^{\text{Perm,2}}$ being two
independent arbitrary functions of $X$.

\subsection{Antisymmetric $\epsilon$ terms.}
\label{PA1sec:eps}

As argued in Sec.~\ref{PA1procedure}, one can also build scalars out of a
vector field by contractions with the completely antisymmetric
Levi-Civita tensor $\epsilon^{\mu\nu\alpha\beta}$. In order for these
new terms to be effectively independent on the previously derived ones,
it is necessary that two such tensors never appear contracted with one
another, since there exists relations such as $\epsilon_{\alpha\beta\mu\nu}
\epsilon^{\alpha\beta\rho\sigma} = 2 \left( \delta_\mu^\rho \delta_\nu^\sigma
-\delta_\mu^\sigma \delta_\nu^\rho \right)$, all such contractions will
give back our previous terms.
This produces only a limited number of
independent new terms at each order.

The first order, quadratic in the derivative terms, contains only one
possible contraction, namely $\mathcal{L}^\epsilon_{4}= \epsilon_{\mu
\nu \rho \sigma} \partial^\mu A^\nu \partial^\rho A^ \sigma  =
\frac{1}{2} F_{\mu \nu}\tilde{F}^{\mu \nu}$, which is not a new term,
being included by construction in $\mathcal{L}_2$. At this order, this is
the only possibility with vanishing Hessian constraints.

Cubic terms produce two test Lagrangians, namely

\begin{equation}
\mathcal{L}^\text{test}_{5,1} =  \epsilon_{\mu \nu \rho
\sigma} \partial^\mu A^\nu \partial^\rho A^ \sigma  \left(\partial \cdot
A \right)= \frac{1}{2} F_{\mu \nu} \tilde{F}^{\mu \nu} \left(\partial
\cdot A \right)
\ \ \ \ \hbox{and} \ \ \ \
\mathcal{L}^\text{test}_{5,2}= \epsilon_{\mu \nu \rho \sigma} 
\partial^\mu A ^\nu \partial ^\rho A_\alpha \partial^\alpha A^\sigma = 
\tilde{F}_{\rho \sigma}   \partial ^\rho A_\alpha \partial^\alpha
A^\sigma,
\end{equation}
which, once combined in such a way as to ensure the vanishing of
the Hessian $\mathcal{H}^{0\mu}$, yield
\begin{equation}
\mathcal{L}_{5}^{\epsilon} = F_{\mu \nu} \tilde{F}^{\mu \nu}
\left(\partial \cdot A \right) - 4  \left( \tilde{F}_{\rho \sigma}  
\partial ^\rho A_\alpha \partial^\alpha A^\sigma \right).
\end{equation}

Increasing the powers of the derivatives, as usual by now, also increases
the complexity of the possible new terms. Quartic test terms are found
to be given by
\begin{equation}
\begin{split}
\mathcal{L}^{\epsilon,\text{test}}_{6,1} & = \left( \epsilon_{\mu \nu \rho \sigma} \partial^\mu A^\nu \partial^\rho A^ \sigma \right) ^2 = \frac{1}{4} \left(F_{\mu \nu} \tilde{F}^{\mu \nu}\right)^2, \\
\mathcal{L}^{\epsilon,\text{test}}_{6,2} & = \left( \epsilon_{\mu \nu \rho \sigma} \partial^\mu A^\nu \partial^\rho A^ \sigma \right) \left(\partial \cdot A \right)^2 =   \frac{1}{2} F_{\mu \nu} \tilde{F}^{\mu \nu} \left(\partial \cdot A \right)^2,  \\
\mathcal{L}^{\epsilon,\text{test}}_{6,3} & = \left( \epsilon_{\mu \nu \rho \sigma} \partial^\mu A^\nu \partial^\rho A^ \sigma \right) \left( \partial ^\alpha A^\beta \partial_\alpha A_\beta \right),   \\
\mathcal{L}^{\epsilon,\text{test}}_{6,4} & = \left( \epsilon_{\mu \nu \rho \sigma} \partial^\mu A^\nu \partial^\rho A ^\sigma \right) \left( \partial ^\alpha A^\beta \partial_\beta A_\alpha \right) , \\
\mathcal{L}^{\epsilon,\text{test}}_{6,5} & = \left( \epsilon_{\mu \nu \rho \sigma}  \partial^\mu A ^\nu \partial ^\rho A_\alpha \partial^\alpha A^\sigma \right) \left(\partial \cdot A \right) =  \left( \tilde{F}_{\rho \sigma}   \partial ^\rho A_\alpha \partial^\alpha A^\sigma \right) \left(\partial \cdot A \right),  \\
\mathcal{L}^{\epsilon,\text{test}}_{6,6} & = \left( \epsilon_{\mu \nu \rho \sigma} \partial^\mu A^\nu \partial^\rho A_\alpha \partial^\sigma A_\beta \partial^\alpha A^\beta \right) = \left(  \tilde{F}_{\rho \sigma} \partial^\rho A_\alpha \partial^\sigma A_\beta \partial^\alpha A^\beta \right), \\
\mathcal{L}^{\epsilon,\text{test}}_{6,7}& = \left( \epsilon_{\mu \nu \rho \sigma} \partial^\mu A^\nu \partial_\alpha A^\rho \partial_\beta A^\sigma \partial^\alpha A^\beta \right) =  \left(  \tilde{F}_{\rho \sigma} \partial_\alpha A^\rho \partial_\beta A^\sigma \partial^\alpha A^\beta \right),\\
\mathcal{L}^{\epsilon,\text{test}}_{6,8} & = \left( \epsilon_{\mu \nu \rho \sigma} \partial^\mu A^\nu \partial^\rho A_\alpha \partial_\beta A^\sigma\partial^\alpha A^\beta \right) =  \left(  \tilde{F}_{\rho \sigma} \partial^\rho A_\alpha \partial_\beta A^\sigma \partial^\alpha A^\beta \right), \\
\mathcal{L}^{\epsilon,\text{test}}_{6,9} & =   \left( \epsilon_{\mu \nu \rho \sigma} \partial^\mu A^\nu \partial^\rho A_\beta \partial_\alpha A^\sigma\partial^\alpha A^\beta \right) =  \left(  \tilde{F}_{\rho \sigma} \partial^\rho A_\beta \partial_\alpha A^\sigma \partial^\alpha A^\beta \right),
\end{split}
\end{equation}
leading to one new solution with vanishing Hessian constraint, namely
\begin{equation}
\mathcal{L}_{6}^{\epsilon}=  \tilde{F}_{\rho \sigma} F^\rho{}_\beta
F^\sigma{}_\alpha \partial^\alpha A^\beta \,,
\end{equation}
together with $\mathcal{L}_{F\tilde{F}\times F\tilde{F}}=\left(F_{\mu \nu}
\tilde{F}^{\mu \nu}\right)^2$ and $\mathcal{L}_{F\tilde{F}\times
FF}=\left(F_{\mu \nu} \tilde{F}^{\mu \nu}\right) \left( F_{\rho \sigma}
F^{\rho\sigma} \right)$, both of which are already part of
$\mathcal{L}_2$. As before, all these extra Lagrangians can be multiplied
by arbitrary functions $f^\epsilon (X)$ without modifying our conclusions.

\subsection{Final Lagrangian in flat spacetime}

Up to this point, we have considered explicit terms involving products
of up to five derivatives of the vector field. Higher-order terms can be
derived by continuing along the same lines of calculation; this would
imply an important number of long terms. We have not found a general
rule allowing to derive a generic action at any given order, so we are
merely led to conjecture, given the above calculations, that there is no
reason the higher-order terms to be vanishing (assuming they do not
lead to trivially vanishing equations of motion \cite{Deffayet:2013tca}).

Considering the result previously derived, we conjecture that terms like
$\mathcal{L}_{n}^{\text{Perm}}$ will continue to show up at higher
order, and in fact, we speculate that the higher order the more numerous
terms one will find. A generic $\mathcal{L}_{n}^{\text{Perm}}$ will
yield vanishing $\mathcal{H}^{00}$ and $\mathcal{H}^{0i}$ and vanish
when going to the scalar limit $A_\mu\to\partial_\mu\pi$ of the theory.
Including the Levi-Civita terms, we propose that the most general
Lagrangian leading to second-order equations of motion for three vector
propagating degrees of freedom contains an infinite number of terms,
whose general formulation takes the form
\begin{equation}
\label{PA1LagFinalSum}
\mathcal{L}_{\text{gen}}\left(A_\mu\right)=-\frac{1}{4}F_{\mu\nu}F^{\mu\nu}+\sum_{n\geq
  2} \mathcal{L}_n + \sum_{n\geq 5}\mathcal{L}^\epsilon_n,
\end{equation}
with
\begin{equation}
\begin{split}
\mathcal{L}_2 = & f_2(A_\mu,F_{\mu\nu},\tilde{F}_{\mu\nu}),\\
\mathcal{L}_3 = & f_3^{\text{Gal}}(X) \mathcal{L}_3^{\text{Gal}},\\
\mathcal{L}_4 = & f_4^{\text{Gal}}(X) \mathcal{L}_4^{\text{Gal}},\\
\mathcal{L}_5 = & f_5^{\text{Gal}}(X) \mathcal{L}_5^{\text{Gal}}+f_5^{\text{Perm}}(X)
\mathcal{L}_5^{\text{Perm}},\\
\mathcal{L}_6 = &f_6^{\text{Perm}}(X) \mathcal{L}_6^{\text{Perm}},\\
\mathcal{L}_7 = & f_7^{\text{Perm,1}}(X) \mathcal{L}_7^{\text{Perm,1}}
+ f_7^{\text{Perm,2}}(X) \mathcal{L}_7^{\text{Perm,2}},\\
\mathcal{L}_{n\geq 8} = & \sum_{i} f_n^{\text{Perm},i}(X) \mathcal{L}_n^{\text{Perm},i},\\
\mathcal{L}^\epsilon_n = & \sum_{i} g_n^{\epsilon,i}(X) \mathcal{L}_n^{\epsilon,i},
\end{split}
\end{equation}
where the terms in $\mathcal{L}_3$ to $\mathcal{L}_5$ are given in
Sec. \ref{PA1L3->5}, those in $\mathcal{L}_6$ and $\mathcal{L}_7$
can be found respectively in Sec. \ref{PA1L6} and \ref{PA1L7}, and those in $\mathcal{L}^\epsilon_n$ are shown in Sec. \ref{PA1sec:eps}. Once again,
$f_2$ is an arbitrary scalar function of
all the possible contractions among $A_\mu$, $F_{\mu\nu}$ and
$\tilde{F}_{\mu\nu}$, and all the other $f_n$ and $g_n^{\epsilon}$
are arbitrary functions of $X$.

\section{Curved space-time}
\label{PA1Covariantization}

Coupling the vector with the metric and rendering the latter dynamical
implies many new possible terms satisfying our conditions. Those are
explored below. Some of the terms have already been discussed in
previous works \cite{Horndeski:1976gi,Heisenberg:2014rta,Hull:2015uwa},
and we introduce new ones below.

\subsection{Covariantization}

In order to take into account the metric $g_{\mu\nu}$ itself as
dynamical in the most general way, one needs to impose its equation of
motion to be also of order two at most. First, we transform partial
derivatives into covariant derivatives. This satisfies the constraints
in the vector sector since the additional terms, coming from the
connection, are only first order in the metric derivatives and thus
ensures that no derivative higher than second order will appear in the
equations of motion of the vector field or the metric.

On the other hand, when going to the scalar sector, one has to pay
special attention to the higher derivative terms which can appear due to
the commutations of derivatives of the scalar field. This can be
problematic only when one considers terms reducing to the scalar
Galileon Lagrangians because the property $\nabla_\mu\nabla_\nu
\pi=\nabla_\nu\nabla_\mu \pi$ remains valid in curved spacetime. As for
the extra terms, called $\mathcal{L}_n^{\text{Perm}}$ above, as well as
the U(1)-invariants, they keep vanishing in the scalar formulation of
the theory. The terms reducing to the scalar Galileon Lagrangians have
already been studied, and their curved space-time extension can be found
in Refs.~\cite{Deffayet:2009wt,Deffayet:2011gz,Heisenberg:2014rta}. In
conclusion, the only terms to be modified in Eq. (\ref{PA1LagFinalSum}) are
\begin{equation}
\begin{split}
\mathcal{L}_4 & = f_4^{\text{Gal}}(X)R -2 f_{4,X}^{\text{Gal}}(X)
\mathcal{L}_4^{\text{Gal}},\\
\mathcal{L}_5 & =f_5^{\text{Gal}}(X)
G_{\mu\nu} \nabla^\mu A^\nu+3 f_{5,X}^{\text{Gal}}(X)
\mathcal{L}_5^{\text{Gal}}+f_5^{\text{Perm}}(X)
\mathcal{L}_5^{\text{Perm}},\\
\end{split}
\end{equation}
where the notation $f_{,X}$ stands for a derivative with respect to $X$,
\emph{i.e.} $f_{,X}\equiv \text{d} f/\text{d} X$.

\subsection{Additional curvature terms}

This last part is dedicated to all the additional terms which can appear from
the coupling contractions of curvature terms with terms implying the vector
field. A similar study was already proposed for the Abelian case, and it was
shown that the only possibility was to contract the field tensor with
divergence-free objects built from curvature
\cite{deRham:2011by,Jimenez:2013qsa}, \emph{i.e.} the Einstein tensor
$G_{\mu\nu}$ and the following fourth-rank divergence-free tensor
\begin{equation}
L_{\mu\nu\rho\sigma}=2R_{\mu\nu\rho\sigma} +
2(R_{\mu\sigma}g_{\rho\nu} + R_{\rho\nu}g_{\mu\sigma} -
R_{\mu\rho}g_{\nu\sigma} - R_{\nu\sigma}g_{\mu\rho}) +
R(g_{\mu\rho}g_{\nu\sigma} - g_{\mu\sigma}g_{\rho\nu}).
\end{equation}
This tensor has the same symmetry properties as the Riemann tensor
$R_{\mu\nu\rho\sigma}$, \emph{i.e.} it is antisymmetric in $(\mu,\nu)$ and
$(\rho,\sigma)$, and symmetric in the exchange of ($\mu\nu)$ and
($\rho\sigma$). The use of divergence-free tensors permits us to avoid higher-order derivatives of the metric in the equations of motion: contracting
such a tensor with another dynamical object of the required derivative
order will naturally lead, in the equations of motion, to the divergence
of this tensor, and hence will vanish in the chosen case.

In order to derive the relevant generalizing terms for the Proca theory,
we proceed in the same way as in the previous section. We first consider
contractions of both $G_{\mu\nu}$ and $L_{\mu\nu\rho\sigma}$ with
$A_\mu$ only, and then these specific contractions with $A_\mu$ and its
first derivative that vanish in the scalar part. We then apply the
Hessian condition given in Sec. \ref{PA1procedure} to obtain the required
terms.

For the contractions with the vector field only, the only term we can
have is
\begin{equation}
\mathcal{L}^\text{Curv}_1=G_{\mu\nu}A^\mu A^\nu,
\end{equation}
with all possible contractions with $L_{\mu\nu\rho\sigma}$ being vanishing due to
its antisymmetry properties. Note that in this case, we cannot multiply
this Lagrangian by a scalar function of the vector field $X$: in the
scalar sector, this would lead to first-order derivatives of the scalar
component $\pi$ which would subsequently yield terms involving
$\nabla_{\alpha} G_{\mu\nu}$ in the equations of motion, terms which
are third order in the metric and thus excluded.

Contractions with the field tensor only have already been studied in
Ref.~\cite{Jimenez:2013qsa}, where it was shown that the only available
term satisfying our requirements can be written as
\begin{equation}
\mathcal{L}^\text{Curv}_2=L_{\mu\nu\rho\sigma} F^{\mu\nu} F^{\rho\sigma},
\label{PA1L2curv}
\end{equation}
the other possible contraction over different indices,
$L_{\mu\nu\rho\sigma}F^{\mu\rho}F^{\nu\sigma}$, being equivalent to
Eq.~\eqref{PA1L2curv} because of the symmetries of $L_{\mu\nu\rho\sigma}$.
Similarly, one could consider the term built out of the dual of the
field tensor;  this actually also reduces to $\mathcal{L}^\text{Curv}_2$
\cite{Jimenez:2013qsa}. Moreover, this last term can be multiplied by
any scalar function of the vector field only without losing its
properties, since it vanishes in the scalar sector.

Last on the list are those non-U(1) invariant terms with first-order
derivatives of the vector field; these must vanish in the scalar sector
in order to comply with our demands. In this case, in order to have at
most a second-order equation of motion of the metric from the equation of
motion of the vector field, it is sufficient to contract only the
indices of the derivatives with the divergence-free curvature tensors.
At this point, we see that we need to introduce a term in
$\nabla^{\mu}A^{\nu}$, if we want to commute derivatives in order to have
a vanishing term in the scalar sector, and that this term will thus
involve the field strength tensor. As a result, we can consider terms
with either $L_{\mu\nu\alpha\beta} F^{\mu\nu}$ or
$L_{\mu\alpha\rho\beta} F^{\mu\rho}$ contracted with $(\nabla^\alpha
A^\beta)$, $(\nabla^\beta A^\alpha)$, $(\nabla^\alpha A^\gamma
\nabla^\beta A_\gamma)$ or $(A^\alpha A^\beta)$. A close examination of
these possibilities reveals that such terms are either vanishing, or
else that they reduce to $\mathcal{L}^\text{Curv}_2$.

We summarize the set of all possible additional terms containing the
curvature in the following Lagrangian
\begin{equation}
\mathcal{L}^\text{Curv}= f^\text{Curv}_1 G_{\mu\nu}A^\mu A^\nu +
f^\text{Curv}_2(X)L_{\mu\nu\rho\sigma} F^{\mu\nu} F^{\rho\sigma},
\end{equation}
where $f^\text{Curv}_1$ is a constant, and $f^\text{Curv}_2$ is an
arbitrary function of $X$ only. We did not consider higher-order
products of curvature terms since our aim is to focus on the the vector
part of the theory.

\subsection{Final Lagrangian in curved spacetime}

We can finally write the complete expression of the generalized
Abelian Proca theory in curved spacetime, which reads
\begin{equation}
\label{PA1LagFinalSumCST}
\mathcal{L}_{\text{gen}}=-\frac{1}{4}F_{\mu\nu}F^{\mu\nu}
+\mathcal{L}^{\text{Curv}}+\sum_{n\geq 2} \mathcal{L}_n
+ \sum_{n\geq 5}\mathcal{L}^\epsilon_n,
\end{equation}
where
\begin{equation}
\label{PA1LagFinalCST}
\begin{split}
\mathcal{L}^\text{Curv} & = f^\text{Curv}_1 G_{\mu\nu}A^\mu A^\nu +
f^\text{Curv}_2(X)L_{\mu\nu\rho\sigma} F^{\mu\nu} F^{\rho\sigma},\\
\mathcal{L}_2 & =f_2(A_\mu, F_{\mu\nu},\tilde{F}_{\mu\nu}),\\
\mathcal{L}_3 & = f_3^{\text{Gal}}(X) \mathcal{L}_3^{\text{Gal}},\\
\mathcal{L}_4 & = f_4^{\text{Gal}}(X)R -2 f_{4,X}^{\text{Gal}}(X)
\mathcal{L}_4^{\text{Gal}},\\
\mathcal{L}_5 & =f_5^{\text{Gal}}(X) G_{\mu\nu} \nabla^\mu A^\nu
+3 f_{5,X}^{\text{Gal}}(X) \mathcal{L}_5^{\text{Gal}}+f_5^{\text{Perm}}(X)
\mathcal{L}_5^{\text{Perm}},\\
\mathcal{L}_6 & =f_6^{\text{Perm}}(X) \mathcal{L}_6^{\text{Perm}},\\
\mathcal{L}_7 & =f_7^{\text{Perm,1}}(X) \mathcal{L}_7^{\text{Perm,1}}+
f_7^{\text{Perm,2}}(X) \mathcal{L}_7^{\text{Perm,2}},\\
\mathcal{L}_{n\geq 8} & =\sum_{i} f_n^{\text{Perm},i}(X) \mathcal{L}_n^{\text{Perm},i},\\
\mathcal{L}^\epsilon_n & = \sum_{i} g_n^{\epsilon,i}(X) \mathcal{L}_n^{\epsilon,i},
\end{split}
\end{equation}
all $f$ and $g$ being arbitrary functions of $X$, except $f^\text{Curv}_1$
which is a constant, and $f_2$ which is an
arbitrary function of $A_\mu$, $F_{\mu\nu}$ and $\tilde{F}_{\mu\nu}$. The complete
expression of the Lagrangians which appears in Eq. (\ref{PA1LagFinalCST})
are given in Sec. \ref{PA1L3->5} to \ref{PA1sec:eps} where
partial derivatives must be replaced by covariant ones.

As in the previous, flat-space case, we conjecture the series to contain an
infinite number of terms.

\section{Conclusion and discussion}

In this paper, we investigated the terms which can be included in the
generalized Proca theory. After obtaining those which contain products
of up to five first derivatives, we proposed a conjecture concerning the
following terms. We discuss a crucial difference between the generalized
scalar and vector theories: in contrast with the Galileon theory, the
generalization of the Proca theory seems to depend on an infinite number
of arbitrary functions, either in front of
$\mathcal{L}^{\text{Perm},i}_n$ or in front of all the possible
contractions among field tensors.

On the other hand, one could ask if this theory is the most general one
can write, in flat as in curved  spacetime\footnote{The most general
multiple-scalar field theory in flat spacetime was given in Ref.
\cite{Sivanesan:2013tba} and covariantized in Ref.
\cite{Padilla:2012dx}.  In the latter paper, it was suggested that the
ensuing theory would be the multi-scalar generalization of Horndeski's
theory.  However, as shown in Ref. \cite{Kobayashi:2013ina}, the
covariantization procedure does not guarantee that the resulting action
is the most general;  indeed, as presented in Ref.
\cite{Ohashi:2015fma}, a better procedure is that devised originally
by Horndeski in Ref. \cite{Horndeski:1974wa}.}. A first extension would,
for instance, consist in including terms containing higher-order
derivatives leading still to second-order equations of motion; this can
be found for instance in Ref.~\cite{Deffayet:2010zh}.  A second
extension would be to investigate the terms additional to those
appearing in the covariantization of the action, through the equation of
motion rather than the Lagrangian, as it is done in Ref.
\cite{Ohashi:2015fma} following the original Horndeski's procedure
\cite{Horndeski:1974wa}; this could be interesting in order to have a
full understanding of the model.

In the future, this work could be completed along many directions. For
instance, a Hamiltonian analysis of the relevant degrees of freedom
would allow us to determine the relevant symmetries. Moreover, one
could study how a general theory such as that defined here can be obtained
from an effective U(1) invariant initial model. Pioneering work on this
direction, along the line of spontaneous symmetry breaking to obtain
extensions of Proca theory, has been done and can be found in, e.g.,
Refs.~\cite{Tasinato:2014mia,Hull:2014bga}. An extension to
the non-Abelian situation is currently undergoing, with a much richer
phenomenology. Finally, one could consider the effects of such a theory in a
cosmological context, be it by means of implementing an inflation model 
\cite{Jimenez:2013qsa,Barrow:2012ay} or by suggesting new solutions to
the cosmological constant problem \cite{Tasinato:2014mia}.

\subsection*{Acknowledgments} Throughout this paper, we have used xTensor,
xCoba and TexAct, all parts of the Mathematica package xAct available in
Ref.~\cite{xAct}. We wish to thank L.~Bernard, C.~Deffayet,
G.~Esposito-Far\`ese, L.~Heisenberg and  G.~Faye for illuminating
discussions.  This work was supported by COLCIENCIAS - ECOS NORD grant
number RC 0899-2012 with the help of ICETEX, and by COLCIENCIAS grant
numbers 110656933958 RC 0384-2013 and 123365843539 RC FP44842-081-2014.

\section{Appendix}
\subsection{Hessian condition for the quartic terms: $\mathcal{L}_6$}
\label{PA1appL6}

The different test Lagrangians that can be written down at the
fourth-order derivative level are
\begin{equation}
\begin{split}
\mathcal{L}^\text{test}_{6,1} & = \left(\partial\cdot A\right)^4,\\
\mathcal{L}^\text{test}_{6,2} & = \left(\partial\cdot A\right)^2
   \left(\partial_ {\sigma}A_ {\rho} \partial^ {\sigma}A^ {\rho}\right),\\
\mathcal{L}^\text{test}_{6,3} & = \left(\partial\cdot A\right)^2
   \left(\partial_ {\rho}A_ {\sigma} \partial^ {\sigma}A^{\rho}\right),\\
\mathcal{L}^\text{test}_{6,4} & = \left(\partial\cdot A\right) \left(\partial_\nu
    A_ {\sigma} \partial^ {\rho}A^{\nu} \partial^{\sigma} A_ {\rho}\right),\\
\mathcal{L}^\text{test}_{6,5} & = \left(\partial\cdot A\right) \left(\partial^ {\rho}A^ {\nu}
   \partial_ {\sigma}A_ {\rho} \partial^ {\sigma}A_\nu\right),\\
\mathcal{L}^\text{test}_{6,6} & = \left(\partial_\mu A_ {\sigma} \partial^{\nu} A^ {\mu}
   \partial^\rho A_ {\nu} \partial^ {\sigma}A_\rho\right),\\
\mathcal{L}^\text{test}_{6,7} & = \left(\partial^\nu A^ {\mu} \partial_{\rho} A_ {\sigma}
   \partial^\rho A_ {\mu} \partial^ {\sigma}A_\nu \right),\\
\mathcal{L}^\text{test}_{6,8} & = \left(\partial_\nu A^ {\sigma} \partial^{\nu} A^{\mu}
   \partial_\rho A_ {\sigma} \partial^ {\rho}A_\mu\right),\\
\mathcal{L}^\text{test}_{6,9} & = \left(\partial^\nu A^ {\mu} \partial^ {\rho}A_ {\mu}
   \partial_\sigma A_ {\rho} \partial^ {\sigma}A_\nu \right),\\
\mathcal{L}^\text{test}_{6,10} & = \left(\partial_\nu A_ {\mu} \partial^ {\nu}A^ {\mu}\right)^2,\\ \mathcal{L}^\text{test}_{6,11} & = \left(\partial_{\mu}A_\nu \partial^ {\nu}A^ {\mu}\right)
   \left(\partial_ {\sigma}A_\rho \partial^ {\sigma}A^\rho \right),\\
\mathcal{L}^\text{test}_{6,12} & = \left(\partial_ {\mu}A_ {\nu} \partial^ {\nu}A^\mu \right)
   \left( \partial_ {\rho}A_{\sigma} \partial^ {\sigma}A^{\rho}\right),
\end{split}
\end{equation}
so that setting
\begin{equation}
\mathcal{L}^\text{test}_{6} = \sum_{k=1}^{12} x_k \, \mathcal{L}^\text{test}_{6,k},
\end{equation}
we obtain for the $(00)$ component of the Hessian
\begin{equation}
\begin{split}
\mathcal{H}_{6}^{00} = & 2 \left(6 x_{1} + x_{2} + x_{3}\right)
\left(\partial \cdot A\right)^2 + 2 \left(2 x_{10} + x_{11} +
x_{2}\right) \left(\partial_ {\mu}A_ {\nu} \partial^ {\mu}A^
{\nu}\right) \\
& + 2 \left(x_{3}+x_{11} + 2 x_{12}\right) \left(\partial^
{\mu}A^ {\nu} \partial_ {\nu}A_ {\mu}\right) - 2 \bigl[4 x_{2} + 4
  x_{3} + 3 \left(x_4 + x_5\right) \bigr] \left(\partial \cdot
A\right) \left(\partial^{0}A^{0}\right) \\ & + 4 \left[ x_6 + x_7 + x_8 + x_9
+2 \left( x_{10} + x_{11} + x_{12}\right) \right]
\left(\partial^{0}A^{0}\right)^2 - 2 \left( x_5 + x_7 + 2
x_8 + x_9\right) \left(\partial_ {\mu}A^{0}\right) \left(\partial^
{\mu}A^{0}\right) \\
& - \left(\partial^{0}A_ {\mu}\right) \left\{2 \left(
x_5 + x_7 + 2 x_8 + x_9\right) \left(\partial^{0}A^
{\mu}\right) + 2 \bigl[3 x_4 + x_5 + 2 \left(2 x_6 + x_7 +
  x_9\right) \bigr] \left(\partial^ {\mu}A^{0}\right)\right\},
\end{split}
\end{equation}
and the $(0i)$ component
\begin{equation}
\begin{split}
\mathcal{H}_{6}^{0i} = & 2 \left(x_7 + 2 x_8 + x_9 + 4 x_{10} + 2 x_{11} \right)
\left(\partial^{0}A^{0}\right)
\left(\partial^{0}A^{i}\right) + 2 \left(2 x_6 + x_7 + x_9 +2 x_{11} + 4 x_{12} \right)
\left(\partial^{0}A^{0}\right)
\left(\partial^{i}A^{0}\right) \\
& - \left(\partial \cdot A\right)
\left[2 \left(2 x_{2} + x_5\right) \left(\partial^{0}A^{i}\right) +
  \left(4 x_{3} + 3 x_4 + x_5\right)
  \left(\partial^{i}A^{0}\right) \right] - \left(3 x_4 + 4 x_6 +
x_7\right) \left(\partial^{i}A_ {\mu}\right)\left( \partial^
{\mu}A^{0}\right) \\
& - \left( x_5 + x_7 + 4 x_8\right)
\left(\partial_ {\mu}A^{i}\right)\left( \partial^ {\mu}A^{0}\right) -
\left(\partial^{0}A_ {\mu}\right) \left(x_5 + x_7 + 2 x_9\right)
\left( \partial^{i}A^ {\mu} + \partial^ {\mu}A^{i}\right).
\end{split}
\end{equation}
Canceling these two functions of $x_k$ provides the solutions 
exhibited in Sec.~\ref{PA1L6}.

\subsection{Hessian condition for the fifth order: $\mathcal{L}_7$}
\label{PA1appL7}

In the last case we considered, for which the highest power in derivatives
is five, we can write the various test Lagrangians as
\begin{equation}
\begin{split}
\mathcal{L}^\text{test}_{7,1} & = \left(\partial\cdot A\right)^5,\\ 
\mathcal{L}^\text{test}_{7,2} & = \left(\partial\cdot A\right) ^3
\left(\partial_ {\gamma}A_ {\sigma} \partial^ {\gamma}A^{\sigma}\right),\\
\mathcal{L}^\text{test}_{7,3} & = \left(\partial\cdot A\right) ^3
   \left(\partial_ {\sigma}A_ {\gamma} \partial^{\gamma} A^{\sigma}\right),\\
\mathcal{L}^\text{test}_{7,4} & = \left(\partial\cdot A\right)^2\left(\partial_ {\rho}A_ {\gamma}
   \partial^ {\sigma}A^ {\rho} \partial^ {\gamma}A_{\sigma}\right),\\
\mathcal{L}^\text{test}_{7,5} & = \left(\partial\cdot A\right) ^2 \left(\partial^{\sigma}
   A^ {\rho} \partial_ {\gamma}A_ {\sigma} \partial^ {\gamma}A_{\rho}\right),\\
\mathcal{L}^\text{test}_{7,6} & = \left(\partial\cdot A\right) \left(\partial_{\nu} A_ {\gamma}
   \partial^ {\rho}A^{\nu} \partial^ {\sigma}A_ {\rho} \partial^ {\gamma}A_{\sigma}\right),\\
\mathcal{L}^\text{test}_{7,7} & = \left(\partial\cdot A\right) \left(\partial^ {\rho}A^ {\nu}
   \partial_ {\sigma}A_ {\gamma}\partial^ {\sigma}A_ {\nu}\partial^ {\gamma}A_{\rho}\right),\\
\mathcal{L}^\text{test}_{7,8} & = \left(\partial\cdot A\right)\left(\partial_{\rho}A^ {\gamma}
   \partial^ {\rho}A^{\nu}\partial_{\sigma}A_{\gamma}\partial^{\sigma}A_{\nu}\right),\\
\mathcal{L}^\text{test}_{7,9} & = \left(\partial\cdot A\right) \left(\partial^ {\rho}A^{\nu}
   \partial^{\sigma}A_{\nu}\partial_{\gamma}A_{\sigma}\partial^{\gamma}A _{\rho}\right),\\
\mathcal{L}^\text{test}_{7,10} & = \left(\partial\cdot A\right)\left(\partial_{\rho}A_{\nu}
   \partial^{\rho}A^{\nu}\right)^2,\\
\mathcal{L}^\text{test}_{7,11} & = \left(\partial\cdot A\right) \left(\partial_
{\nu}A_ {\rho} \partial^ {\rho}A^ {\nu}\right) \left(\partial_
{\gamma}A_ {\sigma} \partial^ {\gamma}A^ {\sigma}\right),\\
\mathcal{L}^\text{test}_{7,12} & = \left(\partial\cdot A\right) \left(\partial_
{\nu}A_ {\rho} \partial^ {\rho}A^ {\nu}\right)\left( \partial_
{\sigma}A_ {\gamma} \partial^ {\gamma}A^ {\sigma}\right),\\
\mathcal{L}^\text{test}_{7,13} & = \left(\partial_ {\mu}A_ {\gamma} \partial^ {\nu}A^
{\mu} \partial^ {\rho}A_ {\nu} \partial^ {\sigma}A_ {\rho} \partial^
{\gamma}A_ {\sigma}\right),\\
\mathcal{L}^\text{test}_{7,14} & = \left(\partial^
{\nu}A^ {\mu} \partial_ {\rho}A_ {\gamma} \partial^ {\rho}A_ {\mu}
\partial^ {\sigma}A_ {\nu} \partial^ {\gamma}A_ {\sigma}\right),\\
\mathcal{L}^\text{test}_{7,15} & = \left(\partial^ {\nu}A^ {\mu} \partial^ {\rho}A_
{\mu} \partial^ {\sigma}A_ {\nu} \partial_ {\gamma}A_ {\sigma} \partial^
{\gamma}A_ {\rho}\right),\\
\mathcal{L}^\text{test}_{7,16} & = \left(\partial^ {\nu}A^
{\mu} \partial_ {\rho}A^ {\gamma} \partial^ {\rho}A_ {\mu} \partial_
{\sigma}A_ {\gamma} \partial^ {\sigma}A_ {\nu}\right),\\
\mathcal{L}^\text{test}_{7,17} & = \left(\partial_ {\nu}A_ {\mu} \partial^ {\nu}A^
{\mu}\right)\left( \partial_ {\rho}A_ {\gamma} \partial^ {\sigma}A^
{\rho} \partial^ {\gamma}A_ {\sigma}\right),\\
\mathcal{L}^\text{test}_{7,18} & =
\left(\partial_ {\nu}A_ {\mu} \partial^ {\nu}A^ {\mu} \right)
\left(\partial^ {\sigma}A^ {\rho} \partial_ {\gamma}A_ {\sigma}
\partial^ {\gamma}A_ {\rho}\right),\\
\mathcal{L}^\text{test}_{7,19} & =
\left(\partial_ {\mu}A_ {\nu} \partial^ {\nu}A^ {\mu}
\right)\left(\partial_ {\rho}A_ {\gamma} \partial^ {\sigma}A^ {\rho}
\partial^ {\gamma}A_ {\sigma}\right),\\
\mathcal{L}^\text{test}_{7,20} & =
\left(\partial_ {\mu}A_ {\nu} \partial^ {\nu}A^ {\mu} \right)
\left(\partial^ {\sigma}A^ {\rho} \partial_ {\gamma}A_ {\sigma}
\partial^ {\gamma}A_ {\rho}\right).
\end{split}
\end{equation}
We again set
\begin{equation}
\mathcal{L}^\text{test}_{6} = \sum_{k=1}^{20} x_k \, \mathcal{L}^\text{test}_{7,k},
\end{equation}
leading to the following Hessian matrix elements: 
\begin{equation}
\begin{split}
\mathcal{H}_{7}^{00} = & 2 \left(10 x_{1} + x_{2} + x_{3}\right)
\left(\partial \cdot A\right)^3 + 2 \left(x_{4} + x_{17} + x_{19}\right)
\left(\partial^ {\mu}A^ {\nu} \partial_ {\nu}A_ {\rho} \partial^
{\rho}A_ {\mu}\right) \\ 
& + 2 \left(x_{5} + x_{18} + x_{20} \right) \left(\partial^
{\mu}A^ {\nu} \partial_ {\rho}A_ {\mu} \partial^ {\rho}A_ {\nu}\right)
 - 6 \left(2 x_{2} + 2 x_{3} + x_{4} + x_{5}\right) \left(\partial \cdot
A\right)^2 \left(\partial^{0}A^{0}\right) \\
& - 2 \bigl[4 x_{10} + 2 x_{11} + 3 \left(x_{17} + x_{18}\right) \bigr] 
\left(\partial_ {\mu}A_ {\nu} \partial^ {\mu}A^ {\nu}\right) \left(\partial^{0}A^{0}\right) \\
& - 2 \bigl[2 x_{11} + 4 x_{12} + 3 \left(x_{20} + x_{19}\right) \bigr] \left(\partial^
{\mu}A^ {\nu} \partial_ {\nu}A_ {\mu}\right) \left(\partial^{0}A^{0}
\right) \\
& + \left(\partial \cdot A\right) \bigl\{2 \left(3 x_{2}+ 2 x_{10} + x_{11} 
\right) \left(\partial_ {\mu}A_ {\nu} \partial^ {\mu}A^
{\nu}\right) + 2 \left(3 x_{3} + x_{11} + 2 x_{12} \right)
\left(\partial^ {\mu}A^ {\nu} \partial_ {\nu}A_ {\mu}\right) \\
& + 4 \left[ x_{6} + x_{7} + x_{8} + x_{9} + 2 \left(
x_{10} + x_{11} + x_{12} \right) \right]
\left(\partial^{0}A^{0}\right)^2 \\
& - 2 \left(2 x_{5} + x_{7} + 2 x_{8}
+ x_{9}\right) \left(\partial_ {\mu}A^{0}\right)\left(\partial^
{\mu}A^{0}\right) - \left(\partial^{0}A_ {\mu}\right) \bigl[2 \left(2
x_{5} + x_{7} + 2 x_{8} + x_{9}\right) \left(\partial^{0}A^ {\mu}
\right)\\
& - 4 \left(3 x_{4} + x_{5} + 2 x_{6} + x_{7} + x_{9}\right)
\left(\partial^ {\mu}A^{0}\right) \bigr]\bigr\} \\
& +
\left(\partial^{0}A^{0} \right)\bigl\{2 \left[ x_{14} + x_{15} + 2 \left(
x_{16} + x_{18} + x_{20} \right)\right] \left[ \left( \partial_ {\mu}A^{0}\right)\left( \partial^
{\mu}A^{0}\right)  + \left(\partial^{0}A_ {\mu}\right) \left(\partial^{0}A^ {\mu}
\right) \right] \\
& + 2 \left(5 x_{13} + 3 x_{14} + 3 x_{15} + x_{16} + 6 x_{17} + 2 x_{18} + 6 x_{19} + 2 x_{20} \right)
\left( \partial^ {\mu}A^{0}\right) \left( \partial^0 A_\mu\right) \bigr]\bigr\} \\
& - \left(\partial^ {\mu}A^ {\nu}\right) \left(\partial^{0}A_ {\mu}
\right)\bigl\{2 \bigl[  x_{7} + x_{14} + x_{16} + 2 \left(x_{9} +
x_{15}\right) \bigr]\left( \partial^{0}A_ {\nu} \right) \\
& - 2 \left(4 x_{6} + x_{7} + 5 x_{13} + 2 x_{14} + x_{15}\right) \left(\partial_
{\nu}A^{0}\right)\bigr\} \\
& -\left(\partial^ {\mu}A^ {\nu}\right) \left(
\partial_ {\mu}A^{0} \right)\bigl\{2 \left(x_{7} + 4 x_{8} + x_{14} + 3 x_{16} \right)
\left(\partial^{0}A_ {\nu} \right) + 2 \bigl[x_{7} + x_{14} + x_{16} + 2 \left(x_{9} + x_{15}
\right) \bigr] \left(\partial_{\nu}A^{0}\right)\bigr\},
\end{split}
\end{equation}
and
\begin{equation}
\begin{split}
\mathcal{H}_{7}^{0i} = &\left(x_{14} + x_{16} + 2 x_{20} \right)
\left(\partial^{0}A_ {\mu}\right)\left( \partial^{0}A^ {\mu}
\right)\left(\partial^{i}A^{0} \right)\\
& - \left(\partial \cdot A\right)^2 \bigl[2 \left(3 x_{2} + x_{5}\right)
  \left(\partial^{0}A^{i} \right) + \left(6 x_{3} + 3 x_{4} + x_{5}\right)
  \left(\partial^{i}A^{0}\right)\bigr] \\ 
& + \left(\partial_ {\mu}A_ {\nu} \partial^ {\mu}A^ {\nu}\right)
   \bigl[2 \left(2 x_{10} + x_{18}\right) \left(\partial^{0}A^{i} \right)
   + \left(2 x_{11} + 3 x_{17} + x_{18}\right) \left(\partial^{i}A^{0}\right) \bigr] \\
&  + \left(\partial^ {\mu}A^ {\nu} \partial_ {\nu}A_ {\mu}\right) \bigl[2
  \left(x_{11} + x_{20}\right) \left(\partial^{0}A^{i} \right)  +
  \left(4 x_{12} + 3 x_{19} + x_{20}\right) \left(\partial^{i}A^{0}\right)
  \bigr] \\
& + \left( 5 x_{13} + x_{14} + 2 x_{15}+ 6 x_{19} + 2 x_{20} \right)
\left(\partial^{0}A_ {\mu}\right)\left( \partial^{i}A^{0}
\right)\left(\partial^ {\mu}A^{0} \right) \\ 
&  + \left(x_{14} + x_{15} + 2 x_{20} \right) \left(\partial^{i}A^{0} \right)\left(\partial_
{\mu}A^{0}\right)\left( \partial^ {\mu}A^{0} \right)+
\left(\partial^{0}A^{i} \right)\bigl[\left(2 x_{16} + 2 x_{18}\right)
  \left(\partial^{0}A_ {\mu} \right)\left(\partial^{0}A^ {\mu} \right)  \\ 
&  + \left(2 x_{14} + x_{15} + x_{16} + 6 x_{17} + 2 x_{18} \right)\left(
  \partial^{0}A_ {\mu} \right)\left(\partial^ {\mu}A^{0} \right)+
  \left(x_{15} + x_{16} + 2 x_{18} \right)\left( \partial_ {\mu}A^{0}
  \right)\left(\partial^ {\mu}A^{0} \right)\bigr] \\ 
&  +  \left(\partial
\cdot A\right) \bigl\{\left( 2 x_{7} + 4 x_{8} + 2 x_{9} + 8 x_{10} + 4 x_{11}\right)
\left(\partial^{0}A^{0} \right)\left(\partial^{0}A^{i}\right) \\
& + \left(4 x_{6} + 2 x_{7} + 2 x_{9} + 4 x_{11} + 8 x_{12} \right)
\left(\partial^{0}A^{0}\right)\left( \partial^{i}A^{0} \right) \\ 
&  - \left(6 x_{4} + 4 x_{6} + x_{7}\right)
\left(\partial^{i}A^ {\mu} \right)\left(\partial_ {\mu}A^{0}\right) -
\left(2 x_{5} + x_{7} + 4 x_{8}\right) \left(\partial_ {\mu}A^{0}
\partial^ {\mu}A^{i} \right)\\
&- \left(\partial^{0}A_ {\mu} \right)
   \bigl[\left(2 x_{5} + x_{7} + 2 x_{9}\right)
  \left(\partial^{i}A^ {\mu} \right)- \left(2 x_{5} + x_{7} + 2 x_{9}\right)
  \left(\partial^ {\mu}A^{i} \right)\bigr]\bigr\} \\
& + \left(\partial^{0}A^{0} \right)\Bigl[\left(5 x_{13} + 2 x_{14} + x_{15}  + 6 x_{17} + 6 x_{19}\right) \left(\partial^{i}A^{\mu}\right)\left( \partial_ {\mu}A^{0} \right)\\
& + \left(x_{14} + 3 x_{16} + 2 x_{18} + 2 x_{20}
\right) \left(\partial_ {\mu}A^{0}\right)
\left(\partial^ {\mu}A^{i} \right)+ \left(\partial^{0}A_ {\mu}
\right)\bigl\{\left(x_{14} + 2 x_{15} + x_{16} + 2 x_{18} + 2 x_{20} \right)
\left(\partial^{i}A^ {\mu}\right) \\
& + \bigl[x_{14} + x_{15} + x_{16} + 2 \left(x_{18} + 2 x_{20} \right) \bigr]
\left(\partial^{\mu}A^{i}\right)\bigr\}\Bigr] - \left( \partial^ {\mu}A^ {\nu}
\right)\bigl\{\left( x_{7} + x_{14} + x_{16} \right)
\left(\partial^{i}A_ {\nu} \right)\left(\partial_ {\mu}A^{0} \right) \\  
& + \left(\partial^{0}A_ {\nu} \right)\bigl[\left( x_{7} + x_{14} + x_{15}\right)
\left(\partial^{i}A_ {\mu} \right)+ 2 \left(2 x_{8} + x_{16}\right)
\left( \partial_ {\mu}A^{i}\right) \bigr] + \left(4 x_{6} + 5 x_{13} + x_{14}\right)
\left(\partial^{i}A_ {\mu}\right)\left(\partial_ {\nu}A^{0} \right) \\  
& + \left( x_{7} + x_{14} + x_{16} \right)
\left(\partial_ {\mu}A^{i}\right)\left( \partial_{\nu}A^{0} \right)
+ \left( 2 x_{9} + x_{15} + x_{16}\right)
\ \left(\partial_ {\mu}A^{0} \right)\left(\partial_ {\nu}A^{i}
\right)\\
& + \left(\partial^{0}A_ {\mu} \right)\bigl[2 \left( x_{9} + x_{15}\right)
\left(\partial^{i}A_ {\nu} \right)+
  \left( x_{7} + x_{14} +  x_{15}\right) \left(\partial_ {\nu}A^{i}
  \right)\bigr]\bigr\},
\end{split}
\end{equation}
whose vanishing leads to the solutions presented in Sec.~\ref{PA1L7}.

\chapter[On the 4D generalized Proca action for an Abelian vector field (article)]{On the 4D generalized Proca action for an Abelian vector field}
\label{PartArticleProca2}

\begin{figure}[h!]
\begin{center}
\includegraphics[scale=0.9]{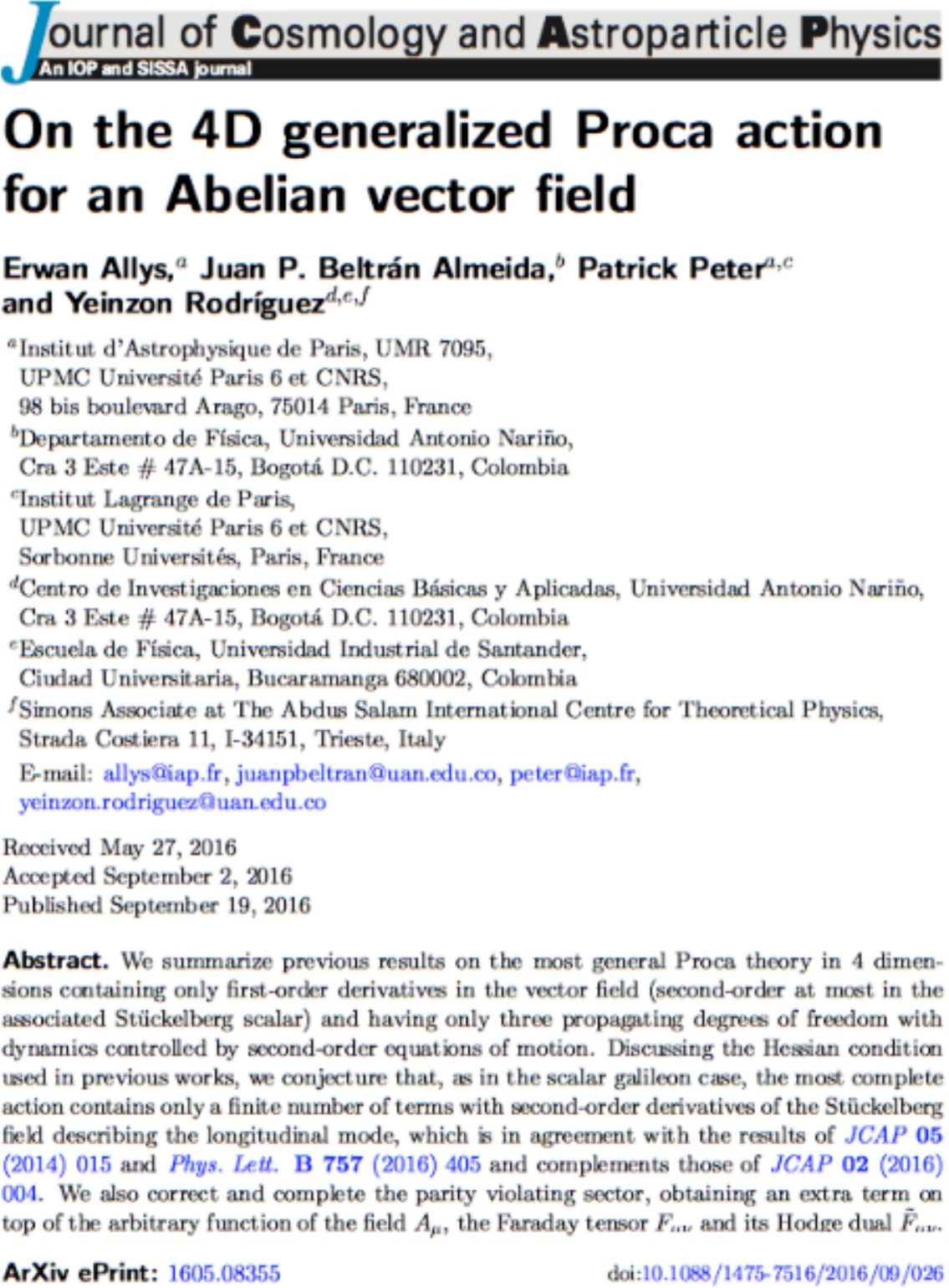}
\end{center}
\end{figure}

\section{Introduction}

Along the line of modifying gravity in a scalar-tensor way, many
proposals have been made to write down theories whose dynamics stem from
second order equations of motion for both the tensor and the scalar
degrees of freedom
\cite{Nicolis:2008in,Deffayet:2009wt,Deffayet:2009mn,Deffayet:2011gz,deRham:2011by},
thus generalizing an old proposal \cite{Horndeski:1974wa}; such theories have been dubbed Galileons. 
The obvious next move consists in obtaining a similar general action for a vector field \cite{Horndeski:1976gi} 
(see also in Refs. \cite{Zumalacarregui:2013pma,Gleyzes:2014dya,Langlois:2015cwa,Langlois:2015skt}), 
thereby forming the vector Galileon case \cite{Heisenberg:2014rta,Tasinato:2014eka}, which 
was investigated thoroughly \cite{EspositoFarese:2009aj,Hull:2014bga,Khosravi:2014mua,Charmchi:2015ggf,Hull:2015uwa,Li:2015vwa,Li:2015fxa,Heisenberg:2016eld}. 
Demanding U(1) invariance led to a no-go theorem \cite{Deffayet:2013tca} which 
can be by-passed essentially by dropping the U(1) invariance hypothesis. Cosmological
implications of such a model can be found e.g. in Refs.~\cite{BeltranJimenez:2010uh,Barrow:2012ay,Jimenez:2013qsa,Tasinato:2013oja,Tasinato:2014mia,Jimenez:2015caa,Gao:2016vtt,Heisenberg:2016dkj,DeFelice:2016yws,Nieto:2016gnp,Chagoya:2016aar,DeFelice:2016uil,Heisenberg:2016wtr}.

Recent papers \cite{Heisenberg:2014rta,Allys:2015sht,Jimenez:2016isa}
have derived the most general action containing a vector field, with
different conclusions as to the number of possible terms given the
underlying hypothesis. In Refs.~\cite{Heisenberg:2014rta,Jimenez:2016isa},
the Lagrangian was built from contractions of derivative terms with
Levi-Civita tensors, whereas Ref.~\cite{Allys:2015sht} used a more systematic
approach based on the Hessian condition. It appears that a consensus has finally been
reached, suggesting only a finite number of terms in the theory, all of
them being given in an explicit form. To describe this consensus and
complete the discussion, we examine in the present paper an alternative
explanation for the presence of, allegedly, only a finite number of
terms in the generalized Proca theory, using the tools developed in
Ref.~\cite{Allys:2015sht} where the infinite series of terms was
conjectured. This discussion also allows us to
compare the systematic procedure used in Ref.~\cite{Allys:2015sht} with the
construction based on Levi-Civita tensors of
Refs.~\cite{Heisenberg:2014rta,Jimenez:2016isa}.  We then summarize these
previously obtained results and settle the whole point in as definite a
manner as possible.

Focusing on the parity violating sector of the model, not thoroughly
investigated in Refs. \cite{Heisenberg:2014rta,Jimenez:2016isa}, 
certain terms obtained in Ref.~\cite{Allys:2015sht}
should not appear according to the abovementioned discussion.
Indeed, we show that, because of an identity not taken into
account in Ref.~\cite{Allys:2015sht}, those unexpected parity-violating terms are
either merely vanishing or can be combined into a simple scalar formed
with the field $A^\mu$, the Faraday tensor $F^{\mu\nu}$ and its Hodge
dual $\tilde{F}^{\mu\nu}$. 
This closes the gap, hence providing an
even firmer footing to the conjecture according to which the most general
theory is in fact given by Eq.~\eqref{PA2TotalLag}, which is, up to a new term uncovered in this paper, 
Eq.~(12) of Ref.~\cite{Jimenez:2016isa} in
Minkowski space, or Eq.~(28) in an arbitrary curved spacetime.

In Sec.~\ref{PA2PartStateOfTheArt}, we summarize the results previously
obtained, together with the associated investigation procedures. We then
present the generic structure in Sec.~\ref{PA2conj}, emphasizing how it
permits an automatic implementation of the Hessian condition, and argue
that the number of acceptable Lagrangian structures satisfying the usual
physical requirements is finite, up to arbitrary functions. Splitting
the possible terms into parity conserving and violating contributions,
we motivate our conclusion in Secs.~\ref{PA2ParCons} and \ref{PA2ParViol},
resolving the apparent disagreement between the present conjecture and
the conclusions of a previous work \cite{Allys:2015sht}; we conclude in
Sec.~\ref{PA2PartFinalModel} by explicitly writing down the final 4D vector
action.

\section{Present status}
\label{PA2PartStateOfTheArt}

Let us first introduce the vector theory, the hypothesis and results
obtained thus far. We assume in what follows the Minkowski metric to
take the form $g^{\mu \nu}=\eta^{\mu \nu}=\rm{diag}(-1,+1,+1,+1)$ and
set $(\partial \cdot A) \equiv \partial_\mu A^\mu$ and $X=A_\mu A^\mu$
for simplification and notational convenience.

\subsection{Generalized Abelian Proca theory}

One seeks to generalize Proca theory, namely that stemming from the action
\begin{equation}
\mathcal{S}_\mathrm{Proca}  = \int \mathcal{L}_\mathrm{Proca}\,\dd^4x 
= \int \left( -\frac14 F_{\mu\nu} F^{\mu\nu} + \frac12 m_A^2 X \right)\,\dd^4x \,,
\end{equation}
with $A^\mu$ being a massive vector field, not subject to satisfy a U(1)
invariance, and $F_{\mu\nu} \equiv \partial_\mu
A_\nu-\partial_{\nu}A_{\mu}$ being the associated Faraday tensor. The
generalization of this action can be made by considering all ``safe''
terms containing the vector field and its first derivative. To explicit
what safe means in this context, one
decomposes the field into a scalar $\pi$ and pure vector $\bar{A}$ parts
according to
\begin{equation}
A_\mu=\partial_\mu \pi + \bar{A}_\mu \ \ \ \hbox{or} \ \ \ A = \dd \pi +
\bar{A} \,,
\end{equation}
where $\pi$ is commonly referred to as the Stückelberg field, and
$\bar{A}^\mu$ is divergence-free. One then demands the equations of
motion for $A^\mu$, and for both $\pi$ and $\bar{A}^\mu$, are
second order, and that the Proca field propagates only three degrees of
freedom \cite{deRham:2010ik}. These conditions are discussed in full depth in
Refs.~\cite{Heisenberg:2014rta,Allys:2015sht,Jimenez:2016isa}. The first
condition ensures that the model can be stable \cite{ostro,Boulware:1973my,Woodard:2006nt}, while
the second stems from the fact that a massive field of spin $s$
propagates $2s+1$ degrees of freedom.

Note that the scalar field will appear in two different parts, one
containing only the Stückelberg field itself, and one containing also the
pure vector contribution $\bar{A}$. Examining the decoupling limit of the
theory, one recovers for the pure scalar part of the Lagrangian the exact
requirements of the Galileon theory
\cite{Nicolis:2008in,Deffayet:2009wt,Deffayet:2009mn,Deffayet:2011gz},
and so this part of the Lagrangian must reduce to this well-studied class
of model.

\subsection{Investigation procedures}
\label{PA2PartInvestigationProcedure}

Two different but equivalent procedures have been devised to write down
the most general theory sought for. The first, originally proposed and explained in
Refs.~\cite{Heisenberg:2014rta,Jimenez:2016isa}, consists in a systematic
construction of scalar Lagrangians in terms of contractions of two Levi-Civita
tensors with derivatives of the vector field. This permits an easy
comparison with the Galileon theory, as the same structure automatically
ensues. The condition that only three degrees of freedom propagate is
then verified on the relevant terms.

The second procedure, put forward in Ref.~\cite{Allys:2015sht}, works somehow
the other way around by systematically constructing all possible scalar
Lagrangians propagating only three degrees of freedom. To achieve this
requirement, a condition on the Hessian of the Lagrangian $\mathcal{L}$
(or each independent such Lagrangian) considered, namely
\begin{equation}
\mathcal{H}^{\mu\nu}=\frac{\partial^2 \mathcal{L}}{\partial(\partial_0
A_\mu)\partial(\partial_0A_\nu)} \,,
\label{PA2Hessian}
\end{equation}
is imposed. As discussed in Ref. \cite{Allys:2015sht}, in order that the
timelike component of the vector field does not propagate in non trivial
theories, the components $\mathcal{H}^{0\mu}$ must vanish. All possible
terms satisfying this constraint are considered at each order.

There are two crucial points concerning the latter method that still need to
be checked once the terms satisfying all other requirements have been
obtained: not only they must reduce to the scalar galileon Lagrangians in
the pure scalar sector, but they must imply second class constraints.
Moreover, given that it is a systematic expansion in terms of scalars
built out of vectors with derivatives, one must make sure they are not
either identically vanishing or mere total derivatives. In other words,
although the method ensures that all possible terms will be found, they
are somehow too numerous and there may remain some redundancy that must
be tracked down and eliminated.


\section{Generic structure}
\label{PA2conj}

As all contractions between vector derivatives and $\delta$ and $\epsilon$
can always be written in terms of $\epsilon$ only, a complete basis for
expanding the general category of Lagrangians of interest is made up with
terms of the form
\begin{equation}
\mathcal{T}^i_N = \underbrace{\epsilon^-{}_- \epsilon^-{}_- \cdots}_N \
 \partial_\centerdot A_\centerdot \partial_\centerdot A_\centerdot\cdots \,,
\label{PA2GenForm}
\end{equation}
where indices appearing in the field derivatives are contracted only
with corresponding indices in the Levi-Civita tensors, the remaining 
indices being contracted possibly in between Levi-Civita tensors in
such a way as to yield a scalar. Each index $i$ reflects the fact that
there can be more than one way to contract the $N$ Levi-Civita 
tensor to form a scalar.
These terms form a complete basis for the Lagrangians containing an arbitrary
number of field derivatives.

The general Lagrangian will then be of the form
\begin{equation}
\mathcal{L} = \sum_{i,N} f^i_N(X) \mathcal{T}^i_N \,,
\label{PA2GenLag}
\end{equation}
where we consider only prefactors that are functions of $X$: one could
envisage contracting a vector field itself with the derivative terms
involved in Eq. \eqref{PA2GenLag}, but that would lead to an equivalent
basis up to integrations by parts \cite{Allys:2015sht}\footnote{We have found
one special case, discussed below Eq.~\eqref{PA2FS}, for which the total
derivative of the integration by parts would actually vanish for
symmetry reasons; we included and discussed this special term in our
final form of the action.}. When written in terms of the Stückelberg
field only, \emph{i.e.} setting $A_\mu \to \partial_\mu\pi$, and restricting
attention to $N=2$, Eq. \eqref{PA2GenLag} automatically yields the subclass
of the generalised galileon theory \cite{Deffayet:2009mn} containing
only derivatives of the scalar field\footnote{The full generalized
galileon theory is recovered if one also makes the replacement $f^i_N(X)
\to f^i_N(\pi,\partial \pi)$.}.

The terms thus built in Eq. \eqref{PA2GenForm} now fall into two distinct
categories, depending on how they behave under a U(1) gauge
transformation. Those invariant under such transformations contracts
all field derivative indices to one and only one Levi-Civita tensor, \emph{i.e.}
they take the form
$$\epsilon^{\mu\nu -} \epsilon^{\rho\sigma -} \cdots \partial_\mu A_\nu
\partial_\rho A_\sigma\cdots \,,$$ which can all be equivalently expressed as functions of
scalar invariants made out of the Faraday tensor  $F_{\mu\nu}$ and its
Hodge dual $\tilde{F}^{\mu\nu}=\frac{1}{2}\epsilon^{\mu\nu\alpha\beta}
F_{\alpha\beta}$. Indeed, written in this form, one can identically replace all
$\partial_\mu A_\nu$ by $\frac12 F_{\mu\nu}$. Conversely, since the following
identities
\begin{equation}
F^{\mu\nu}F_{\mu\nu} = - \epsilon^{\mu\nu\alpha\beta}
\epsilon_{\rho\sigma\alpha\beta} \partial_\mu A_\nu \partial^\rho A^\sigma \,,
\label{PA2EqLFFEps}
\end{equation}
and
\begin{equation}
\tilde{F}^{\mu\nu} F_{\mu\nu} = 2\epsilon^{\mu\nu\rho\sigma} \partial_\mu A_\nu
\partial_\rho A_\sigma \,,
\label{PA2EqLFtildeFEps}
\end{equation}
hold, any function of $F$ and $\tilde F$ can be expressed as a term such as
discussed above.

This leads to the first Lagrangian compatible with our requirements,
namely the so-called $\mathcal{L}_2$, containing all possible scalars
made by contracting $A_\mu$, $F_{\mu\nu}$ and $\tilde F_{\mu\nu}$. Such
terms can always be expressed \cite{Fleury:2014qfa} as functions of the
scalars $X$, $F^2\equiv F_{\mu\nu} F^{\mu\nu}$, $F\cdot \tilde F
\equiv F_{\mu\nu} \tilde F^{\mu\nu}$ and
\begin{equation}
(A\cdot \tilde F)^2 \equiv A_\alpha \tilde F^{\alpha\sigma} A_\beta \tilde
F^\beta{}_\sigma = A_\alpha A_\beta
\epsilon^{\alpha\sigma\mu\nu} \epsilon^\beta{}_{\sigma\kappa\delta}
\partial_\mu A_\nu \partial^\kappa A^\delta \,,
\end{equation}
again up to integrations by part. The Lagrangian $\mathcal{L}_2$ always
satisfies the conditions discussed in the previous section, and in
particular yields a trivially vanishing Hessian condition
$\mathcal{H}^{0\mu}$: varying $\mathcal{L}_2$ with respect to $\partial
_0 A_0$ [see Eq. \eqref{PA2Hessian}] yields a factor containing
$\epsilon^{00-}$, which vanishes identically. It also gives second order
equation of motion both for $\pi$ and $\bar A_\mu$ as it contains
neither $\partial \partial \pi$ nor $\partial \partial \bar A_\mu$
terms. We should emphasize at this point that $\mathcal{L}_2$ contains
parity conserving as well as parity violating terms; we shall not
consider them any more, but they should be assumed always present in the
forthcoming discussion.

All the terms contained in Eq. \eqref{PA2GenLag} but not of the form discussed
in the previous paragraph read
\begin{equation}
\mathcal{L}^i_N = f^i_N \left( X\right)
\underbrace{\epsilon^{\mu-}\cdots\epsilon^{\nu-} }_N \partial_\mu A_\nu
\cdots \,,
\label{PA2Lgen}
\end{equation}
where at least one field derivative has indices contracted with two
distinct Levi-Civita tensors and the $f^i_N \left( X\right)$ are
arbitrary functions of the gauge vector magnitude $X=A^2$. For $N\leq 2$,
the Hessian condition is automatically satisfied: $\mathcal{H}^{0\mu}$
stems for a variation of the Lagrangian with respect to $\partial_0 A_0$
and $\partial_0 A_\mu$. This demands three equal ``0'' indices
distributed on at most two Levi-Civita tensors, resulting in a vanishing
contribution for symmetry reasons. The other requirements, such
as the order of the equations of motion these terms lead to, are
discussed in length in Secs. \ref{PA2ParCons} and \ref{PA2ParViol}.

For $N>2$, the situation is less clear, as the Hessian does not then
identically vanish. Instead, the condition $\mathcal{H}^{0\mu}=0$ then
implies that the coefficients of all the linearly independent terms
stemming from this condition vanish. The number of such linearly
independent terms increases with the number of field derivatives allowed
for in the Lagrangian, and it is therefore to be expected that, above a
given threshold value $N>N_\mathrm{thr}$, up to unforeseeable fortuitous
cancellations, no new term will be obtainable that could possibly
satisfy the requirements of a safe theory. We conjecture that, as in the
scalar galileon case, $N_\mathrm{thr}=2$; the following sections detail
the reasons hinting to such a conjecture, splitting into parity
conserving ($N$ even) and violating ($N$ odd) contributions. Note that
there exists a general argument, based on the fact that the Lagrangian
contains second-order derivatives of the field, for which the scalar galileon
theory automatically stops at $N=2$ \cite{Deffayet:2009mn}, whereas in
the vector case, no such argument can be found, the Lagrangian
containing only first-order derivatives and there could exist terms
which vanish when $A_\mu\to \partial_\mu \pi$ while still satisfying all
other hypothesis. As a result, the arguments below are different from
those needed to show $N_\mathrm{thr}=2$ in the scalar galileon case.

\section{Parity conserving terms}
\label{PA2ParCons}

Previous works discussed parity conserving actions with $N=2$ including
up to 4 field derivatives $\partial A$, the so-called $\mathcal{L}_n$,
with $n=3,\cdots,5$ \cite{Heisenberg:2014rta}, and $n=6$
\cite{Allys:2015sht}, $n$ counting the number of field derivative plus
two (this convention, bizarre in the vector case, is meaningful in the
original galileon construction). Up to $n=5$, the Lagrangians satisfy
the condition that the scalar part of the vector field corresponds only
to non trivial total derivative interactions, a condition which, once
relaxed, yields the extra $n=6$ term: in the latter situation, one can
always factorize the action by some factors involving the Faraday tensor
and its dual, ensuring it vanishes in the pure scalar sector. All these
terms were shown to be of the form presented in Eq. \eqref{PA2Lgen} above
with $N=2$, thus agreeing with our conjecture. They also comply with all
the necessary requirements we asked for the theory to be physically
meaningful, with second-order equations of motion and only three
propagating degrees of freedom \cite{Allys:2015sht,Jimenez:2016isa}.

In Ref.~\cite{Allys:2015sht}, new terms were also suggested which,
similarly to $\mathcal{L}_6$, were of the form $(\partial A)^p F^q
\tilde F^r$ (with $r$ even to ensure parity conservation), and therefore
vanishing in the pure scalar sector. It was even argued that an infinite
tower of such terms could be generated. A further examination of these
terms however revealed a different, and somehow more satisfactory,
picture: some new terms, by virtue of the Cayley-Hamilton theorem,
vanish identically in 4 dimensions, a conclusion that can also be
reached by rewriting the relevant terms in the form presented in Eq.
\eqref{PA2Lgen}, but with Levi-Civita tensors having more than 4 indices
\cite{Jimenez:2016isa}, explaining why the new terms identically
vanish in 4 dimensions to which the present analysis is restricted: in a
way similar to Lovelock theory for a spin 2 field \cite{Lovelock:1971yv},
one can imagine that
for each number of dimensions, a finite number of new terms can be
generated.

In conclusion of this short section, suffices it to say that extra
parity preserving terms involving more fields and not already present
in $\mathcal{L}_2$ have been actively searched for, and never found;
although this does not prove that such terms cannot be found, this
provides a sufficiently solid basis to assume this statement as a
conjecture, which will only make sense provided a similar conclusion can
be reached for the parity-violating terms to which we now turn.

\section{Parity violating terms}
\label{PA2ParViol}

Parity-violating terms can be written as in Eq. \eqref{PA2Lgen} with an odd
number $N$ of Levi-Civita tensors. For $N=1$, it leads to an action
built from Eq. \eqref{PA2EqLFtildeFEps}, and hence is already included in
$\mathcal{L}_2$ discussed above. One thus expects no terms not included
in $\mathcal{L}_2$ since those terms would contain at least three Levi-Civita symbols. In Ref.~\cite{Allys:2015sht} however, two extra such
terms were found to satisfy all the physically motivated requirements,
obtained through the systematic Hessian method. They read
\begin{equation}
\label{PA2EqL5Eps}
\mathcal{L}_{5}^{\epsilon} = F_{\mu \nu} \tilde{F}^{\mu \nu}
\left(\partial \cdot A \right) - 4  \left( \tilde{F}_{\rho \sigma}  
\partial ^\rho A_\alpha \partial^\alpha A^\sigma \right) \,,
\end{equation}
and
\begin{equation}
\label{PA2EqL6Eps}
\mathcal{L}_{6}^{\epsilon}=  \tilde{F}_{\rho \sigma} F^\rho{}_\beta
F^\sigma{}_\alpha \partial^\alpha A^\beta \,.
\end{equation}
According to our conjecture, they should either vanish or be contained
in the previous terms up to a total derivative. We show below that it is
indeed the case, and for that purpose we first recall an identity derived and first reported, to our knowledge, in 
Ref.~\cite{Fleury:2014qfa}; this completes
the proof that the systematic procedure could not find terms having 
up to 4 field derivatives that are not contained in $\mathcal{L}_2$.

\subsection{A useful identity}

Let $A_{\mu\nu}$ and $B_{\mu\nu}$ be two antisymmetric tensors in a
four-dimensional spacetime with mostly positive signature. One has
\begin{equation}
A^{\mu \alpha} \tilde{B}_{\nu \alpha} + B^{\mu \alpha} \tilde{A}_{\nu
\alpha} = \frac{1}{2} (B^{ \alpha \beta} \tilde{A}_{\alpha \beta}
)\delta^{\mu}_{\nu} \,,
\label{PA2EqFP}
\end{equation}
where $\tilde{X}^{\mu\nu}=
\frac{1}{2}\epsilon^{\mu\nu\alpha\beta}X_{\alpha\beta}$ is the Hodge
dual of $X$ \cite{Fleury:2014qfa}. 

In order to prove this identity, one uses the relation (see, e.g., Ref. \cite{Wald:1984rg})
\begin{equation}
\epsilon^{\alpha_1 \dots \alpha_k \delta_{(k+1)}\dots \delta_n}
\epsilon_{\beta_1 \dots \beta_k \delta_{(k+1)}\dots \delta_n} =
(-1)^s (n-k)! k! \delta^{[\alpha_1\dots}_{\beta_1\dots}\delta^{\alpha_k ]}_{\beta_k} \,,
\label{PA2EqContractionEpsGen}
\end{equation}
where $s$ counts the number of minus signs in the signature of the metric
and $n$  the dimension of spacetime. One gets 
\begin{equation}
\varepsilon^{\alpha_1 \alpha_2 \alpha_3 \delta }\varepsilon_{\beta_1
\beta_2 \beta_3 \delta } = - 3! \delta^{[{\alpha_1 }}_{\beta_1 }
\delta^{\alpha_2}_{\beta_2} \delta^{\alpha_3]}_{\beta_3} \,,
\label{PA2EqContractionEps}
\end{equation}
and
\begin{equation}
\varepsilon^{\alpha_1 \alpha_2 \delta_1 \delta_2 }\varepsilon_{\beta_1
\beta_2 \delta_1 \delta_2 } = - 2! 2! \delta^{[{\alpha_1 }}_{\beta_1 }
\delta^{\alpha_2]}_{\beta_2} \,,
\end{equation}
in the $n=4$-dimensional case, leading to
\begin{equation}
X^{\alpha\beta}=-\frac{1}{2}\epsilon^{\mu\nu\alpha\beta}\tilde{X}_{\mu\nu} \,,
\end{equation}
to express a tensor from its Hodge dual.

Beginning from the left-hand side of the identity we wish to prove, we get
\begin{equation}
A^{\mu\alpha} \tilde{B}_{\nu \alpha} = -\frac14
\varepsilon^{\gamma \epsilon \mu \alpha } 
\varepsilon_{\nu  \sigma \rho \alpha} \tilde{A}_{\gamma \epsilon}
B^{\sigma \rho} \,,
\end{equation}
which, upon using Eq. \eqref{PA2EqContractionEps}, yields 
$A^{\mu \alpha} \tilde{B}_{\nu \alpha} 
= \displaystyle \frac32 \delta^{[{\gamma }}_{\nu }
\delta^{\epsilon}_{\sigma} \delta^{\mu]}_{\rho} 
\tilde{A}_{\gamma \epsilon} B^{\sigma \rho}.$
Expanding and simplifying the relevant terms, one finally
obtains Eq. \eqref{PA2EqFP}, as desired.

As a direct consequence, we can easily deduce the identities
\begin{equation}
F^{\mu \alpha } F_{\nu \alpha}  - \tilde{F}^{\mu \alpha } \tilde{F}_{\nu
\alpha} = \frac{1}{2} \left(F^{ \alpha \beta } {F}_{\alpha \beta}\right)
\delta^{\mu}_{\nu} \,,
\label{PA2EqFPExample1}
\end{equation}
and
\begin{equation}
F^{\mu \alpha } \tilde{F}_{\nu \alpha} = \frac{1}{4} \left(F^{ \alpha \beta }
\tilde{F}_{\alpha \beta}\right) \delta^{\mu}_{\nu} \,,
\label{PA2EqFPExample2}
\end{equation}
which follows from substituting  $A_{\mu\nu}=F_{\mu\nu}$, $B_{\mu\nu}
= \tilde{F}_{\mu\nu}$ and $A_{\mu\nu}=B_{\mu\nu}=F_{\mu\nu}$
respectively in Eq. (\ref{PA2EqFP}).

\subsection{Simplification of $\mathcal{L}_5^{\epsilon}$ and
$\mathcal{L}_6^{\epsilon}$}

We now use the above identities to first expand
$\mathcal{L}_5^{\epsilon}$. One has
\begin{equation}
\mathcal{L}_5^{\epsilon} =  F_{\mu \nu}\tilde{F}^{\mu \nu} \partial
\cdot A - 4 \tilde{F}_{\rho \sigma}  \partial^{\rho} A^{\alpha}
\partial_{\alpha} A^{\sigma}= F_{\mu \nu}\tilde{F}^{\mu \nu} \partial
\cdot A - 4 \tilde{F}_{\rho \sigma}  (F^{\rho \alpha} +\partial^{\alpha}
A^{\rho} )\partial_{\alpha} A^{\sigma} \,,
\end{equation}
whose last term can be transformed into 
\begin{equation}
\tilde{F}_{\rho \sigma}  \left(F^{\rho \alpha}
+\partial^{\alpha} A^{\rho} \right)\partial_{\alpha} A^{\sigma} = 
\frac{1}{4}F\tilde{F} \delta_{\sigma}^{\alpha}  \partial_{\alpha} A^{\sigma} 
+ \tilde{F}_{\rho \sigma}  \partial^{\alpha} A^{\rho} \partial_{\alpha}
A^{\sigma} \,,
\end{equation}
so that, finally, one ends up with 
\begin{equation}
\mathcal{L}_5^{\epsilon} = - 4 \tilde{F}_{\rho \sigma} 
\partial^{\alpha} A^{\rho} \partial_{\alpha} A^{\sigma} = 0 \,,
\end{equation}
being a contraction between a fully symmetric and a fully antisymmetric
tensor. 

As for $\mathcal{L}_6^{\epsilon}$, one has
\begin{equation}
 \mathcal{L}_6^{\epsilon} =  \left(\tilde{F}_{\rho \sigma} F^{\sigma
 \alpha} \right) F^{\rho \beta} \partial_{\alpha} A_{\beta}
 =\left(-\frac{1}{4}F\tilde{F} \delta_{\rho}^{\alpha} \right) F^{\rho
 \beta} \partial_{\alpha} A_{\beta} \,.
\end{equation}
A few straightforward manipulations then yield
\begin{equation}
\mathcal{L}_6^{\epsilon} = -\frac{1}{8} \left(F\tilde{F}\right) F^2 \,,
\label{PA2L6e}
\end{equation} 
showing $\mathcal{L}_6^{\epsilon}$ is not in fact a new term but is already
included in the Lagrangian $\mathcal{L}_2$.

\section{Final model}
\label{PA2PartFinalModel}

The two extra parity-violating terms obtained in
Ref. \cite{Allys:2015sht} have been shown here to be either vanishing or
already included in a previous Lagrangian. As mentioned below
Eq.~\eqref{PA2GenLag}, we also found that the term
\begin{equation}
\mathcal{L}^\mathrm{bis}_4 = g_4(X) A^\mu A_\lambda \tilde F_{\mu\nu}
\partial^\lambda A^\nu = g_4(X) A^\mu A_\lambda \epsilon_{\mu\nu\rho\sigma}
\partial^\rho A^\sigma \partial^\lambda A^\nu ,
\label{PA2FS}
\end{equation}
is compatible with all the conditions we demand and could therefore be
included in the general analysis. The corresponding term in
Eq.~\eqref{PA2GenLag} would be proportional to $\tilde
F_{\mu\nu}S^{\mu\nu}$, with $S^{\mu\nu} = \partial^\mu A^\nu
+\partial^\nu A^\mu$ being the symmetric counterpart of the Faraday tensor which
clearly vanishes identically. The line of reasoning leading to
a small number of possible terms of the form \eqref{PA2GenLag} should apply
to higher order terms of the kind \eqref{PA2FS}; we did not find any such
terms.

According to all the above discussions, it seems safe to conjecture that the final
complete action for a Proca vector field involving only first-order
derivatives in 4 dimensions is that given by Eq.~(12) of
Ref. \cite{Jimenez:2016isa}, together with Eq.~\eqref{PA2FS}.
The complete formulation of the
parity-conserving terms were also derived and written in a simple form
in Refs. \cite{Heisenberg:2014rta,Allys:2015sht,Jimenez:2016isa}. We merely
repeat the full action below:
\begin{equation}
\mathcal{S} = \int\dd^4 x\sqrt{-g} \left( -\frac{1}{4} F_{\mu\nu}
F^{\mu\nu} + \sum_{i=2}^6 \mathcal{L}_i + \mathcal{L}^\mathrm{bis}_4  \right),
\label{PA2TotalLag}
\end{equation}
with 
\begin{equation}
\begin{split}
\mathcal{L}_2 &= f_2\left( A_\mu, F_{\mu\nu}, \tilde{F}_{\mu\nu} \right) = f_2\left[
X,F^2,F\cdot \tilde F,\left( A \cdot \tilde F\right)^2\right] \,, \\
\mathcal{L}_3 &= f_3\left( X\right) \partial \cdot A = \frac12 f_3\left( X\right) S_{\mu}{}^{\mu} \,, \\
\mathcal{L}_4 &= f_4\left( X\right) \left[ (\partial \cdot A)^2-\partial_\rho
A_\sigma \partial^\sigma A^\rho \right] = \frac14 f_4 \left( X\right) \left\{ \left[ (S_{\mu}{}^{\mu})^2-S_{\rho}{}^{\sigma} S_{\sigma}{}^{\rho} \right] + F_{\mu\nu}F^{\mu\nu}\right\} \,, \\
\mathcal{L}_5 &= f_5\left( X\right)\left[ (\partial \cdot A)^3-3 (\partial \cdot A)
\partial_\rho A_\sigma \partial^\sigma A^\rho + 2 \partial_\rho A^\sigma
\partial_\gamma A^\rho \partial_\sigma A^\gamma\right] + g_5\left( X\right)
\tilde{F}^{\alpha\mu} \tilde{F}^\beta{}_\mu \partial_\alpha A_\beta \\
& = \frac18 f_5\left( X\right)\left[ (S_{\mu}{}^{\mu})^3-3 (S_{\mu}{}^{\mu}) 
S_{\rho}{}^{\sigma}S_{\sigma}{}^{\rho}  + 2 S_{\rho}{}^{\sigma}S_{\sigma}{}^{\gamma}
S_{\gamma}{}^{\rho}  \right] + \frac14 \left[ 2 g_5\left( X\right) - 3 f_5\left( X\right)\right]
\tilde{F}^{\alpha\mu} \tilde{F}^\beta{}_\mu S_{\alpha \beta } \,, \\
\mathcal{L}_6 &= g_6 \left( X\right) \tilde{F}^{\alpha\beta} \tilde{F}^{\mu\nu} \partial_\alpha 
A_\mu \partial_\beta A_\nu =\frac14 g_6 \left( X\right) \tilde{F}^{\alpha\beta} \tilde{F}^{\mu\nu}
\left( S_{\alpha\mu} S_{\beta\nu} + F_{\alpha\mu} F_{\beta\nu}\right).
\end{split}
\label{PA2Lterms}
\end{equation}
In Eq.~\eqref{PA2Lterms}, $f_2$ is an arbitrary function of all possible scalars made out of
$A_\mu$, $F_{\mu\nu}$ and $\tilde F_{\mu\nu}$, containing both parity
violating and preserving terms, while $f_3$, $f_4$,  $f_5$, $g_5$ and
$g_6$ are arbitrary functions of $X$ only. Note that this dependence
is compatible with our basis choice in Eq.~\eqref{PA2GenLag}, so that any other choice,
for instance $g_k (X, F^2)$, would spoil the Hessian condition. 
We assume also that the standard kinetic term, $-\frac14 F_{\mu\nu} F^{\mu\nu}$, 
does not appear
in $f_2$, in order that the normalization of the vector field follows
that of standard electromagnetism and thus we have pushed it out in
Eq.~\eqref{PA2TotalLag}. The Lagrangians of Eq. \eqref{PA2Lterms} are expressed in
terms of either the ordinary derivatives $\partial_{\mu} A_{\nu}$, or in
terms of its symmetric $S_{\mu\nu}$ and antisymmetric $F_{\mu\nu}$
parts\footnote{The relation between our formulations and those in terms
of the Levi-Civita tensors is given in Ref.~\cite{Jimenez:2016isa}.}: the
second formulation, obtained by setting $\partial_{\mu} A_{\nu}=\frac12
\left( S_{\mu\nu} + F_{\mu\nu}\right)$ and making use, in the case of
$\mathcal{L}_5 $, of Eq.~\eqref{PA2EqFPExample1}, induces extra terms in
$\mathcal{L}_4$ and $\mathcal{L}_6$ which can be absorbed in the
parity-preserving part of $\mathcal{L}_2$, being functions of $A_\mu$
and $F_{\mu\nu}$.

The presence of the new term $\mathcal{L}^\mathrm{bis}_4 $ is not as
surprising as it would appear at first sight when one considers the
generic structure of the terms contained in Eq.~\eqref{PA2Lterms}. For the
dynamics of the Lagrangians to be non trivial, up to terms already
contained in $\mathcal{L}_2$, the functions $f_3$, $f_4$, $f_5$ and
$g_6$ must contain at least one factor of $X=g_{\mu\nu}A^\mu A^\nu$ (see
also Ref.~\cite{Jimenez:2016isa}). Assuming $2g_5 - 3 f_5$ to also
contain such a factor (generic situation, no fine-tuning of the
arbitrary functions), we conclude that each term can be written in the
form\footnote{Sometimes also up to terms included in $\mathcal{L}_2$.}
\begin{equation}
\mathcal{L}_i = A^2 h(X) \langle \mathcal{O}_i\rangle = \tilde h(X) \langle A
\tilde{\mathcal{O}}_i A \rangle + \partial_\mu J_i^\mu,
\label{PA2Li}
\end{equation}
$A^2 h$ standing for the relevant $f$ or $g$ function (this
transformation is indeed not possible for $\mathcal{L}^\mathrm{bis}_4$). 
In Eq.~\eqref{PA2Li},
the brackets indicate a trace over spacetime indices,
$\mathcal{O}_i$ and $\tilde{\mathcal{O}}_i$ are operators constructed
from $\tilde F$'s (possibly none) and at least one $S$, and $J_i^\mu$ is the relevant
current to make the identity true\footnote{See
Ref.~\cite{Deffayet:2011gz} for an extensive discussion of these
equivalent formulations in the scalar galileon case.}. So, the terms
vanishing in the purely scalar case, \emph{i.e.} those for which
$\mathcal{O}_i$ contains at least one factor of $\tilde F$, take the
form $\langle AS\tilde FA\rangle$, $\langle AS\tilde F\tilde F A\rangle$
and $\langle AS\tilde FS\tilde F A\rangle$. The first such term, which
is nothing but our $\mathcal{L}^\mathrm{bis}_4 $, is then seen to appear
in a totally natural way.

Our final action is, up to the new term $\mathcal{L}^\mathrm{bis}_4$,
exactly the same as that of Ref.~\cite{Jimenez:2016isa}. There is
however a subtle difference in the fact that all possible
parity-violating terms are also written, being included in $f_2$ and
$\mathcal{L}^\mathrm{bis}_4$. Note that the curved space-time
generalization of this action is also given in
Ref.~\cite{Jimenez:2016isa}, the covariantization of
$\mathcal{L}^\mathrm{bis}_4$ being obtained by a trivial replacement
$\partial \to \nabla$.

A legitimate question to ask is whether Eq. \eqref{PA2TotalLag} is indeed
the most general theory that can be written involving a vector field
with three propagating degrees of freedom and second-order equations of
motion. This has already been conjectured in
Refs.~\cite{Heisenberg:2014rta,Jimenez:2016isa}. Now, the discussion and
calculations of the present article correct the conjecture made in Ref.
\cite{Allys:2015sht} about an infinite tower of terms, and also suggests
a finite number of terms, even in the parity violating sector. So, there
is finally a complete agreement on this point.

An additional indication of the correctness of this conjecture is that
the systematic investigation procedure of Ref. \cite{Allys:2015sht}
completed by the calculation of Ref. \cite{Jimenez:2016isa} for the
parity-conserving sector, and by the present paper in the
parity-violating sector, did not find any term other than those shown
above up to the orders of $\mathcal{L}_6$ (parity violating) and
$\mathcal{L}_7$ (parity conserving). However, if there were an infinite
tower of possible Lagrangians, one would expect such Lagrangians to
appear in our systematic procedure, which is not the case. Note
especially that these works show that the parity violating sector
contains no other terms than $\mathcal{L}^\mathrm{bis}_4$ and those
contained in $f_2$, which is a very strong constraint, and greatly
strengthens the conjecture we have made.

Finally, this work heavily relies on the postulate that spacetime is 4
dimensional. Relaxing this assumption permits to include the extra terms
proposed in Ref. \cite{Allys:2015sht} which, as shown in Ref.
\cite{Jimenez:2016isa}, can be expressed with higher dimensional
Levi-Civita tensors. For a given spacetime dimensionality, one thus
expects, just like in the Lovelock case for a spin 2 field \cite{Lovelock:1971yv}, a finite
number of new terms to appear: in practice, in $D$ dimensions, one
expects terms containing up to $D$ first order derivatives of the 
vector field.

\subsection*{Acknowledgments}

We acknowledge P. Fleury and C. Pitrou for pointing out the importance
of the identity in Eq. (\ref{PA2EqFP}). We also  wish to thank C.~Deffayet,
G.~Esposito-Farese, L.~Heisenberg and J.~Beltr\'an Jim\'enez for useful
clarifications and enlightening discussions. This work was supported by
COLCIENCIAS grant numbers 110656933958 RC 0384-2013 and 123365843539 RC
FP44842-081-2014. P.P. would like to thank the Labex Institut Lagrange
de Paris (reference ANR-10-LABX-63) part of the Idex SUPER, within which
this work has been partly done.

\chapter[Generalized SU(2) Proca Theory (article)]{Generalized SU(2) Proca Theory}
\label{PartArticleProcaSU2}

\begin{figure}[h!]
\begin{center}
\includegraphics[scale=1.4]{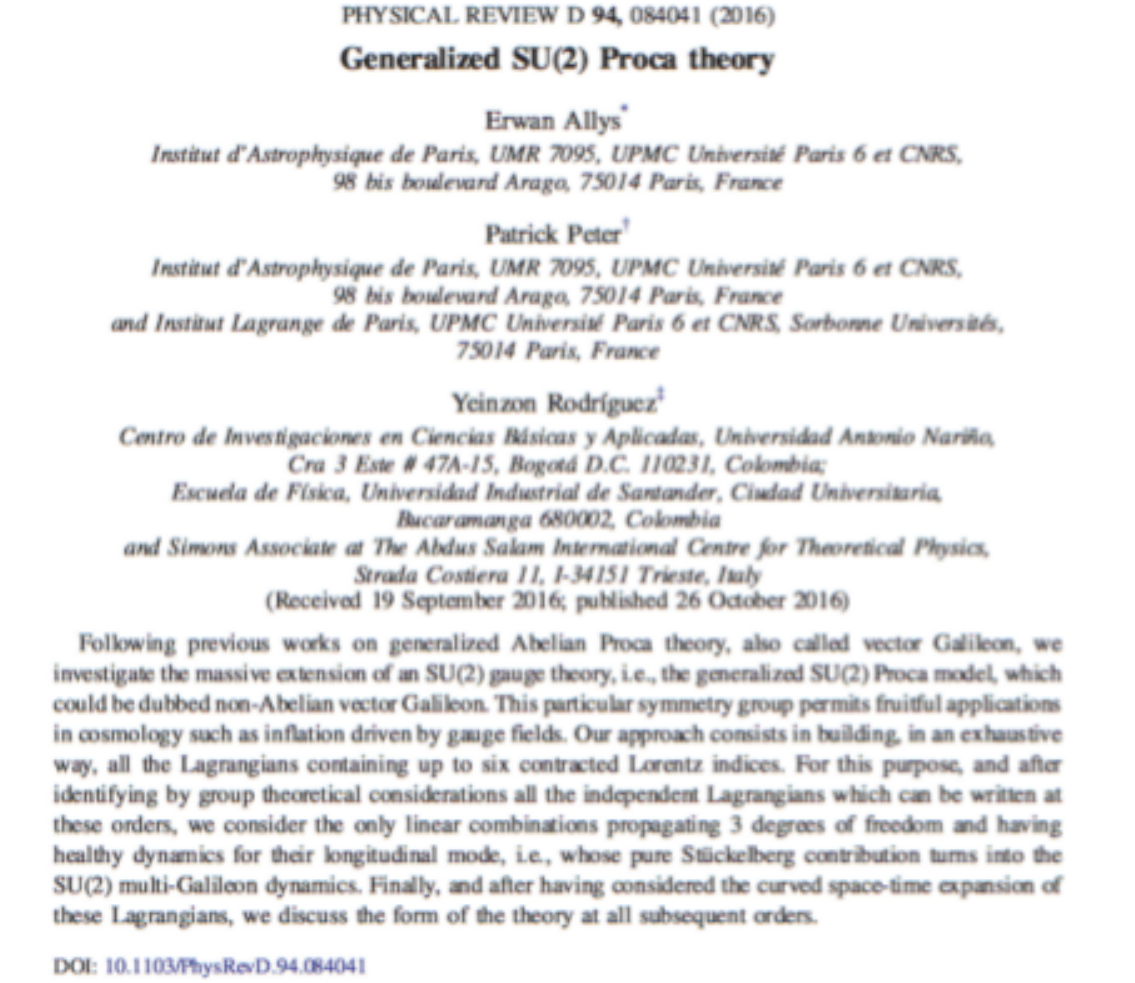}
\end{center}
\end{figure}

\section{Introduction}

In the search for well-motivated theories that describe the primordial
universe, several attempts have been made to obtain inflationary
descriptions from particle physics (the Standard Model, Supersymmetry,
Grand Unified Theories, etc.; see, e.g., Refs.~\cite{Bezrukov:2012sa,Bezrukov:2007ep,Mazumdar:2010sa,Lyth:1998xn,Hertzberg:2014sza}),
 or from quantum theories of gravity such as Supergravity, String
 Theory, and Loop Quantum Gravity (see, e.g.,
 Refs.~\cite{Ferrara:2013rsa,Olive:1989nu,Baumann:2014nda,Ashtekar:2009mm,Barrau:2010nd}).
This top-down approach has been very fruitful, providing new ways to
understand the structure of the high energy theories necessary to
reproduce the observable properties of the Universe, ranging from the
Cosmic Microwave Background Radiation (CMB) to the Large-Scale Structure
(LSS). However, little is known from the observational point of view for
many of these theories (those whose characteristic energy scale is
much higher than the electroweak one), the CMB and LSS being, at present, the
only situations in which they would have had observable consequences and
would thus leave testable signatures. Since the power of the current and
proposed accelerators is not going to increase as much as would be
needed to directly test these theories in the foreseeable future, we
need to devise another approach to the fundamental theory that describes
nature.

Such an approach already exists, and it boils down to the question of
whether there is any choice in formulating the fundamental theory. This
bottom-up approach consists in finding an action completely free of
pathologies, the first of them being the Ostrogradski instability
\cite{ostro} (the Hamiltonian could be unbounded from below), and
satisfying a given set of assumptions, e.g., symmetry requirements.  One
then needs to define the material content of the universe (scalar
fields, vector fields, ...), although, in principle, the construction itself
and the stability requirements constrain some content and allow others
so that the material content is, once the conditions are applied,
somehow redefined. This very ambitious program is just beginning to be
implemented, and interesting works have been carried out in which the
extra material content (on top of gravity) is composed of one or many
scalar fields. It was Horndeski \cite{Horndeski:1974wa} who found, for
the first time, the most general action for a scalar field and gravity
that produces second-order equations of motion. In general, if the
Lagrangian is nondegenerate, having equations of motion of second order
at most is a necessary requirement to avoid the Ostrogradsky instability
\cite{Woodard:2006nt,Woodard:2015zca}. By pursuing this goal, an action
is found that, however, still requires a Hamiltonian analysis in order
to guarantee that the instability is not present.

Horndeski's construction was rediscovered in the context of what is
nowadays called Galileons \cite{Nicolis:2008in}.  The Galileons are the
scalar fields whose action, in flat spacetime, leads to equations of
motion that involve only second-order derivatives.  The idea has been
extended by finding the so-called Generalized Galileons, by allowing for
lower-order derivatives in the equations of motion
\cite{Deffayet:2011gz,Deffayet:2013lga}.  The background space-time
geometry where these Generalized Galileons live can be promoted to a
curved one by replacing the ordinary derivatives with covariant ones and
adding some counterterms that involve nonminimal couplings to the
curvature \cite{Deffayet:2009wt,Deffayet:2009mn}.  The latter guarantees
the equations of motion for both geometry and matter are still
second order, so the Galileons, both Generalized and Covariantized,
are found.  This procedure is equivalent to that proposed by Horndeski
for one scalar field \cite{Kobayashi:2011nu}, but it loses some interesting
terms when more than one scalar field is present
\cite{Kobayashi:2013ina}. The Galileon approach for scalar fields has
found multiple applications in cosmology, ranging from inflation (see,
e.g., Refs.
\cite{Creminelli:2010ba,Kobayashi:2010cm,Mizuno:2010ag,Burrage:2010cu,Creminelli:2010qf,Kamada:2010qe,Libanov:2016kfc,Banerjee:2016hom,Hirano:2016gmv,Brandenberger:2016vhg,Nishi:2016wty}) to dark energy (see, e.g., Refs. \cite{Chow:2009fm,Silva:2009km,Kobayashi:2010wa,Gannouji:2010au,Tsujikawa:2010zza,DeFelice:2010pv,Ali:2010gr,Padilla:2010tj,DeFelice:2010nf,Mota:2010bs,Nesseris:2010pc,Gabadadze:2016llq,Neveu:2016gxp,Salvatelli:2016mgy,Shahalam:2016kkg,Minamitsuji:2016qyc,Saridakis:2016ahq,Biswas:2016bwq}).

The original proposal was based on the requirement of second-order
equations of motion for all the additional degrees of freedom to gravity, all of them therefore being 
dynamical so that the system is nondegenerate. The generalization
to the so-called extended Horndeski theories also includes nonphysical
degrees of freedom and thus considers degenerate theories
\cite{Gleyzes:2014dya,Langlois:2015cwa,Langlois:2015skt,Motohashi:2016ftl,Crisostomi:2016czh}.
%
Such a construction is by now well understood, and some cosmological
applications have also been considered 
\cite{Harko:2016xip,Babichev:2016rlq,Sakstein:2016ggl,Kobayashi:2016xpl,Lagos:2016wyv,Crisostomi:2016czh,Frusciante:2016xoj,Qiu:2015aha,Akita:2015mho}.

However, scalar fields are not the only possibilities as the matter
content of the universe.  Horndeski indeed wondered some fourty years
ago what the action would be for an Abelian vector field in curved
spacetime \cite{Horndeski:1976gi}.  Working with curvature is a way to
bypass the no-go theorem presented in Ref. \cite{Deffayet:2013tca}, which
states that the only possible action for an Abelian vector field in flat
spacetime that leads to second-order equations of motion is the
Maxwell-type one.  Relaxing the gauge invariance allows for a nontrivial action in flat spacetime, in this way generalizing the Proca action
\cite{Heisenberg:2014rta,Tasinato:2014eka}.  The construction of the
resulting vector Galileon action has been well investigated and discussed, so
there is already a consensus about the number and type of terms in the
action, even in the covariantized version
\cite{Allys:2015sht,Jimenez:2016isa,Allys:2016jaq}.  Moreover, the
analogous extended Horndeski theories have been built for a vector field
\cite{Heisenberg:2016eld,Kimura:2016rzw}, and the corresponding
cosmological applications have been explored
\cite{Tasinato:2014eka,Tasinato:2014mia,Hull:2014bga,DeFelice:2016cri,DeFelice:2016yws,DeFelice:2016uil,Heisenberg:2016wtr}.

Some cosmological applications of vector fields have been investigated,
and interesting scenarios, such as the $fF^2$ model
\cite{Watanabe:2009ct} and the vector curvaton
\cite{Dimopoulos:2006ms,Dimopoulos:2009am}, have been devised. There is,
however, an obstacle when dealing with vector fields in cosmology:  they
produce too much anisotropy, both at the background and at the perturbation
levels, well above the observable limits, unless one implements some
dilution mechanism or considers only the
temporal component of the vector field (which is, however, usually nondynamical). In the $fF^2$ model, the potentially huge anisotropy is
addressed by coupling the vector field to a scalar that dominates the
energy density of the universe and, therefore, dilutes the anisotropy;
in contrast, in the vector curvaton scenario, the anisotropy is diluted
by the very rapid oscillations of the vector curvaton around the minimum
of its potential.  Another dilution mechanism is to consider many
randomly oriented vector fields \cite{Golovnev:2008cf}; however, this requires
a large number of them, indeed hundreds, so it is difficult to
justify it from a particle physics point of view.  There is,
nevertheless, another possibility, the so-called ``cosmic triad''
\cite{Golovnev:2008cf,ArmendarizPicon:2004pm}, a situation in which three
vector fields orthogonal to each other and of the same norm can give
rise to a rich phenomenology while making the background and
perturbations completely isotropic \cite{Rodriguez:2015xra}.  A couple
of very interesting models, gauge-flation
\cite{Maleknejad:2011jw,Maleknejad:2011sq} and chromo-natural inflation
\cite{Adshead:2012kp}, have implemented this idea by embedding it in a
non-Abelian framework and exploiting the local isomorphism between the
SO(3) and SU(2) groups of transformations.  At first sight, the cosmic
triad configuration looks very unnatural, but dynamical system studies
have shown that it represents an attractor configuration
\cite{Maleknejad:2011jr}.  Unfortunately, although the background
dynamics of these two models is successful, their perturbative dynamics
makes them incompatible with the latest Planck observations
\cite{Namba:2013kia,Adshead:2013nka}.  Despite this failure, such models
have shown the applicability that non-Abelian gauge fields can have in
cosmological scenarios.

Having in mind the above motivations, the purpose of this paper is to
build the first-order terms of the generalized SU(2) Proca theory and
to discuss the general form of the complete theory. For the most part,
we focus on those Lagrangians containing up to six contracted Lorentz indices,
which we obtain exhaustively. To ensure that we do not forget some terms, we first
construct from group theoretical considerations all possible Lagrangians
at these orders, before imposing the standard dynamical condition,
\emph{i.e.}, that only three degrees of freedom propagate. Then, after
identifying all the Lagrangians that imply the same dynamics, e.g.,
those related by a conserved current, we verify that the pure
Stückelberg part of the Lagrangians is healthy, \emph{i.e.}, that it implies the SU(2)
multi-Galileon dynamics. To this end, it is useful to derive all the
equivalent formulations of the SU(2) adjoint multi-Galileon model, which
we provide in the Appendix. Then, after computing the relevant
curved space-time extension of our Lagrangians, we conclude about the status
of the complete formulation of the theory, \emph{i.e.}, that containing the higher
order terms we did not consider in this work.

The layout of this paper is the following.  In Section  \ref{PSU2gnapc}, the
generalized non-Abelian Proca theory is introduced, and some technical
aspects needed for later sections are laid out;  the procedure to build
the theory is also described.  In Section \ref{PSU2PartSinglet}, the
building blocks of the Lagrangian are systematically obtained.  Section
\ref{PSU2cht} deals with the right number of propagating degrees of freedom
and the consistency of the obtained Lagrangian with the scalar Galileon
nature of its longitudinal part.  The covariantization of the theory is
performed in Section \ref{PSU2covproc} and the final model, together with a
discussion and comparison with the Abelian case, is presented in Section
\ref{PSU2FinalModel}.  The appendix presents the construction of the
multi-Galileon scalar Lagrangian in the 3-dimensional representation of
SU(2) and its equivalent formulations.  Throughout this paper, we have
employed the mostly plus signature, \emph{i.e.}, $\eta_{\mu\nu} =
\rm{diag}\,\left( -, +,+,+\right)$, and set $\hbar=c=1$.

\section{Generalized non-Abelian Proca Theory} \label{PSU2gnapc}

Our aim is to generalize the non-Abelian Proca theory, described below,
to include all possible second-order ghost-free terms propagating only
three degrees of freedom. After discussing the general symmetry case, we
concentrate on the SU(2) symmetry, which is particularly interesting in
a cosmological perspective, as discussed in the Introduction, and we
roughly present the procedure, which will be thoroughly explained below.

\subsection{Non-Abelian Proca Theory}

Let us first present the nowadays standard non-Abelian Proca theory.
Also called a massive Yang-Mills model, this theory had been extensively
studied in the past, such as, e.g., in
Refs.~\cite{Boulware:1970zc,Shizuya:1975ek,GrosseKnetter:1993nu,Banerjee:1994pp,Su:2002qj},
with a Hamiltonian formulation detailed in
Refs.~\cite{Senjanovic:1976br,Banerjee:1997sf}. Our starting point
Lagrangian, including the mass term, reads
\begin{equation}
\label{PSU2EqLagProcaNonAbelian}
\mathcal{L} = -\frac{1}{4} F^{\mu\nu}_{\a} F^{\a}_{\mu\nu} +\frac{1}{2}
m^2 A^{\mu}_{\a} A_{\mu}^{\a},
\end{equation}
with the non-Abelian Faraday tensor given by
\begin{equation}
\label{PSU2EqFaradayNonAbelian}
F_{\mu \nu}^{\a} = \partial_\mu A_\nu^{\a}  - \partial_\nu A_\nu^{\a} +
g f^{\a}{}_{\b \EAc} A_\mu^{\b} A_\nu ^{\EAc},
\end{equation}
with $g$ being the coupling constant and $f^{\a}{}_{\b \EAc}$ the structure
constants of the symmetry group under consideration. This can be
considered as the limit of a valid particle physics model based on a
Higgs condensate whose corresponding degree of freedom is assumed
to be frozen, hence breaking the relevant
symmetry~\cite{Hull:2014bga,Tasinato:2014mia}.

Let us emphasize a technical point at this stage: one could work with
the vector field assumed as an operator, namely,
\begin{equation}
A_\mu (x) = A_\mu ^{\a} (x) T_{\a},
\end{equation}
with $T_{\a}$ representing the operators associated with corresponding
elements of the underlying group in a given representation. We then
have, by definition of the algebra, the commutation relations
\begin{equation}
\left[T_{\a}, T_{\b} \right] = i f_{\a\b}{}^{\EAc} T_{\EAc}.
\label{PSU2TaTb}
\end{equation}
Since this work concentrates on the vector fields
themselves and not on their action on other fields, it is simpler to
restrict our attention to the fields themselves, \emph{i.e.},
\begin{equation}
A_\mu (x)=\left\{ A_\mu^{\a} (x) \right\},
\end{equation}
which are in the Lie algebra of the symmetry group under consideration.
These two ways of writing the field operators are, of course, strictly
equivalent, but the latter formalism, with group indices attached to the
vectors themselves, merely does not need the introduction of the algebra
operators themselves and is thus more appropriate for our purpose.

Any action needs to be a scalar, and this includes not only the Lorentz
group but also any internal symmetry, such as that stemming from the
algebra in Eq. \eqref{PSU2TaTb}. If the relevant symmetry is of the local type,
and for an infinitesimal transformation,
the vectors transform through
\begin{equation}
\delta A_\mu ^{\a} = -\frac{1}{g}  \partial_\mu \alpha^{\a}(x)+
f^{\a}{}_{\b\EAc}\alpha^{\b}(x) A_\mu ^{\EAc},
\label{PSU2dAloc}
\end{equation}
which leaves invariant only the kinetic term $F_{\mu\nu}^{\a} F^{\mu\nu}_{\a}$,
but of course not even a mass term $A_{\mu}^{\a} A^{\mu}_{\a}$, much less any
extension such as those we want to consider below. This is merely a restatement of
the well-known fact that mass breaks gauge symmetry. We therefore restrict our
attention to global transformations of the kind
\begin{equation}
\partial_\mu \alpha^{\a}=0\ \ \ \ \Longrightarrow \ \ \ \ \
\delta A_\mu ^{\a} = f^{\a}{}_{\b\EAc}\alpha^{\b} A_\mu ^{\EAc};
\label{PSU2dAglo}
\end{equation}
\emph{i.e.}, we assume the vector field itself transforms as the adjoint
representation, with dimension equal to that of the symmetry group
itself. It is also profitable, and maybe more enlightening, to look at
the effect of a finite local transformation of the group, still
described by a set of parameters $\alpha^{\a}(x)$. Under this
transformation, the vector field transforms as
\begin{equation}
\label{PSU2EqGaugeTransfoInt}
A_\mu (x)=A_\mu^\a (x) T_\a \mapsto U\left[\alpha^{\a}(x)\right]
\left[-\frac1{g}\partial_\mu + A_\mu (x) \right]
U^{-1}\left[\alpha^{\a}(x)\right],
\end{equation}
where $U\left[\alpha^{\a}(x)\right]$ describes the action of the group
element labeled by $\alpha^{\a}(x)$. This allows us to emphasize that in the
case where the symmetry becomes global, \emph{i.e.}, where $\alpha^{\a}(x)$ no longer depends on the space-time point, the vector field transforms
exactly as the adjoint representation of the symmetry group. This is
indeed the symmetry assumed for the non-Abelian Proca (massive
Yang-Mills) field. In the Abelian case, this transformation is trivial
because the action of the group commutes with the vector field, and the
transformation in Eq. \eqref{PSU2EqGaugeTransfoInt} thus reduces to the identity in the
global symmetry case. In the non-Abelian case, however, one needs to
specify how the extra indices are to be summed over in order to produce
a singlet with respect to this global symmetry transformation. To relate
the set of theories under considerations here with the more usual ones
in particle physics involving a local symmetry broken by means of a
Higgs field, one can envisage our transformation in Eq. \eqref{PSU2dAglo} as the
limit of that in Eq. \eqref{PSU2dAloc}.

With these motivating considerations, we now move on to evaluating the
most general theory with a massive vector field transforming according
to the adjoint representation of a given global symmetry group.

\subsection{Restricting Attention to the SU(2) Case}

In view of the potentially relevant cosmological consequences, from now on we restrict our attention to the case for which the relevant
symmetry group is SU(2), with dimension equal to $3$, and therefore
consider a vector field also of dimension $3$. Since SU(2) is locally
isomorphic to SO(3), one can then simply use a vector representation
with group indices varying from $1$ to $3$ in $A_{\mu}^{\a}$; \emph{i.e.}, we
restrict our attention to the fundamental representation of SO(3).

The set of SU(2) structure constants is identical to the 3-dimensional Levi-Civita
tensor $\epsilon^{\a}_{\ \b\EAc}$, whereas the group metric $g_{\a\b}$, given by
$g_{\a\b} = -f_{\a\d}{}^{\e} f_{\b\e}{}^{\d}$, is simply the flat metric
$2 \delta_{\a\b}$. The only primitive invariants are $\epsilon_{\a\b\EAc}$
and $\delta_{\a\b}$ \cite{Slansky:1981yr,Fuchs:1997jv,Padilla:2010ir},
and one can therefore write all possible contractions by merely
contracting fields with contravariant indices with all appropriate
combinations of those two primitive invariants written with covariant indices.
Recall also the further simplification induced by the fact that
contractions among structure constants (Levi-Civita symbols in the
case at hand) leaving one, two or three free indices will, respectively,
lead to a vanishing result, or terms proportional to $\delta_{\a\b}$ and
$\epsilon_{\a\b\EAc}$ \cite{Metha:1983mng}; it is therefore often
unnecessary to use multiple contractions.

As already alluded to earlier, choosing SU(2) is not innocuous as we aim
at cosmological applications, in view, in particular, of implementing
inflation driven by gauge fields (see, e.g., Refs.
\cite{Maleknejad:2011jw,Maleknejad:2011sq,Namba:2013kia,Maeda:2012eg,Adshead:2012qe,Nieto:2016gnp,Davydov:2015epx,Alexander:2014uza,Sharif:2014aaa,Darabi:2014pua,Maleknejad:2013npa,Maeda:2013daa,Setare:2013dra,Maleknejad:2011jr,Maleknejad:2012fw,Cembranos:2012ng,Adshead:2012kp,Dimastrogiovanni:2012ew,Adshead:2013nka,Ghalee:2012gg}):
since its adjoint representation is 3-dimensional, SU(2) permits us to
generate configurations for which all three vectors are nonvanishing
while ensuring isotropy.

\subsection{Generalization}

What follows is very similar to the generalized Abelian Proca case as
discussed, e.g., in
Refs.~\cite{Heisenberg:2014rta,Tasinato:2014eka,Li:2015vwa,Allys:2015sht,Jimenez:2016isa,Allys:2016jaq} 
(see also Refs.~\cite{deRham:2011by,Jimenez:2013qsa,Hull:2015uwa}
for the equivalent curved space-time construction). In brief, we 
construct the most general action generalizing that of Proca for a
massive SU(2) vector field, \emph{i.e.},
\begin{equation}
\mathcal{S}_\mathrm{Proca}  = \int \mathcal{L}_\mathrm{Proca}\,\dd^4x =
\int \left( -\frac14 F^{\a}_{\mu\nu} F_{\a}^{\mu\nu} + \frac12 m_A^2 X
\right)\,\dd^4x,
\label{PSU2Proca}
\end{equation}
where $X\equiv A_\mu^{\a} A^\mu_{\a}$. To the above action
(\ref{PSU2Proca}), we add all possible terms containing not only
functions of $X$ but also derivative self-interactions. These terms will
have to fulfill some conditions for the corresponding theory to make
sense. We first split the vector into a scalar-pure vector decomposition
\begin{equation}
A_\mu^{\a}=\partial_\mu \pi^\a + \bar{A}_\mu^\a,
\end{equation}
 where $\pi^\a$ is a scalar multiplet in the $\bm{3}$ representation of
 SU(2), \emph{i.e.}, the St\"uckelberg field generalized to the non-Abelian
 case, and $\bar{A}_\mu^\a$ is a divergence-free vector ($\partial_\mu
 \bar{A}^{\mu\a}=0$), containing the curl part of the field, \emph{i.e.}, that
 for which the Abelian form of the Faraday tensor is nonvanishing. The conditions one then
 must  impose on the theory in order for it to make (classical) sense are
\begin{itemize}
\item[a)] the equations of motion for all physical degrees of freedom,
\emph{i.e.},, for both $\bar{A}_\mu^\a$ and $\pi^\a$, and hence $A_\mu^\a$ and
$\pi^\a$, must be at most second order, thus ensuring stability
\cite{ostro,Woodard:2006nt,Woodard:2015zca},
\item[b)] the action may contain at most second-order derivative terms
in $\pi^\a$ and first-order derivatives for $A_\mu^\a$,
\item[c)] each component of the SU(2) multiplet propagates only three
degrees of freedom, the zeroth component being nondynamical.
\end{itemize} 
In what follows, we apply these conditions and restrict our attention
to the theories involving terms with up to six Lorentz indices
contracted. From the cosmological perspective, such theories are
expected to allow for a richer phenomenology since this is what happens
for the Abelian Proca case \cite{Tasinato:2014eka,Tasinato:2014mia,Hull:2014bga,DeFelice:2016cri,DeFelice:2016yws,DeFelice:2016uil,Heisenberg:2016wtr}

\subsection{Procedure}
\label{PSU2PartProcedure}

We now proceed along the lines of Ref.~\cite{Allys:2015sht}; \emph{i.e.},
we build, in Sec.~ \ref{PSU2PartSinglet}, a complete basis of linearly
independent test Lagrangians describing all possible Lagrangians
containing a given number of vector fields and their derivatives; the
detailed prescription is given in Sec.~\ref{PSU2PartProcedureSinglet}.
Next we demand only three degrees of freedom per multiplet component of
the vector field, which translates into a condition on the
Hessian~\cite{Heisenberg:2014rta,Allys:2015sht}, the latter being
defined by
\begin{equation}
\label{PSU2EqDefHessian}
\mathcal{H}^{\mu\nu \d \e} = \frac{\partial}{\partial (\partial_0
A_{\mu\d}) } \frac{\partial}{\partial (\partial_0 A_{\nu \e} )}
\mathcal{L},
\end{equation}
for a given Lagrangian $\mathcal{L}$. This functional over the fields is
symmetric under the index exchange $\left(\mu,\d \right) \leftrightarrow
\left( \nu,\e \right)$.

In order for $\mathcal{H}^{\mu\nu \d \e}$ to have three vanishing
eigenvalues, one for each timelike component of the three vectors
$A_{\mu\d}$, and since all the terms it is built of are \emph{a priori}
independent (up to symmetries), a necessary condition is that we demand
$\mathcal{H}^{0 \nu \d \e}=0$; this requirement will be explicitly
checked in Sec.~\ref{PSU2PartHessianCondition} for each test Lagrangian.

The above condition is, however, not sufficient, for it does not exhaust
all the constraints and thus does not count the effectively propagating
degrees of freedom. For instance, some terms inducing no dynamics for
the time components of the vector fields may also yield no dynamics for
some other component, or even for the overall vector field. The required
analysis is tedious and must be followed step by
step~\cite{Senjanovic:1976br,Banerjee:1997sf}.

As the final step of the above analysis, we consider the scalar
part associated with those linear combinations of test Lagrangians
verifying the Hessian condition. One must check, which is done in
Sec.~\ref{PSU2PartScalarContribution}, that they are of two kinds: either
they have no dynamics at all, being vanishing or given by a total
derivative, or their dynamics is second order in the equations of motion
of the scalar field; \emph{i.e.}, they belong to the class of generalized
Galileons
\cite{Deffayet:2010zh,Padilla:2010ir,Padilla:2010de,Hinterbichler:2010xn,Padilla:2012dx,Sivanesan:2013tba}. 
This will provide the most general terms that verify the requirements we
demand, formulated in terms of the non-Abelian Faraday tensor; see
Sec.~\ref{PSU2PartSwitchFaradayYM}.

Before moving on, we mention that even though the
procedure discussed above and applied below is allegedly tedious, it
guarantees an exhaustive list of all possible terms at each
order, and, in particular, all those specific to the non-Abelian case.
Those terms might have been obtained by some quicker method, but we prefer to
be able to produce all the theoretically acceptable terms rather than
constructing a few. In view of possible cosmological applications, there
is indeed no way to say which terms will be relevant and which ones will not.

\section{Construction of the test Lagrangians}
\label{PSU2PartSinglet}

As anticipated above, our method relies heavily on the construction of a
basis of test Lagrangians satisfying the symmetry requirement, on which
we later apply the Hessian condition. This is the purpose of this
section.

\subsection{Description of the Procedure}
\label{PSU2PartProcedureSinglet}

We now proceed to build the complete basis, in the sense of linear
algebra, of test Lagrangians, for a given number of fields and their
first derivatives. Since they are linearly independent, we will then be able to
write down the most general theory at the given order as a linear
combination of these Lagrangians.

In order to construct Lagrangians, \emph{i.e.}, scalars, we need to consider the
Lorentz and group indices. The former spacetime indices run from $0$
to $3$ and are denoted by small Greek letters, while the latter group
indices run from $1$ to $3$, since we assume the adjoint 3-dimensional
SU(2) representation, and are represented by small Latin letters from
the beginning of the alphabet. We first write down all the Lorentz
scalar quantities that may be formed with a given number of fields and
first derivatives, and then consider all the SU(2) index combinations
leading to SU(2) scalars of these Lorentz scalars.

For the sake of simplicity, beginning with the Lorentz sector, we 
dismiss the group indices altogether, keeping in mind, however, that their
presence might spoil some symmetry properties: contractions between
symmetric and anti-symmetric (with respect to Lorentz indices only)
tensors will not necessarily vanish when group indices are included,
as exemplified by the starred equations in the next section. The Lorentz scalars,
once formed, will then subsequently be assigned SU(2) indices following
simple alphabetical order, leaving as many free SU(2) indices as there
are fields in the term, to then be contracted with a relevant pure SU(2)
tensor. For instance, a term like $A^{\mu}A^{\nu}\left(\partial_{\mu}
A^{\nu} \right)$ will be indexed as
$A^{\mu\a}A^{\nu\b}\left(\partial_{\mu} A^{\nu\EAc} \right)$, demanding
contraction with a structure constant $\epsilon_{\a\b\EAc}$ to form a
Lorentz and SU(2) scalar.
This procedure can seem rather tedious, and it most
definitely is, but it ensures that we construct a complete basis.

For simplicity, we restrict our attention to those Lagrangians
containing up to 6 Lorentz indices contracted as they should be to form a
scalar.

\subsection{Lorentz Sector}  \label{PSU2LSSection}

An easy way to classify the Lorentz scalars that one can form with a given
number of 4-vectors consists in using the local equivalence, at the
Lie-algebraic level, between SO(3,1) and SU(2)$\times$SU(2) (see, e.g.,
Ref. \cite{Ramond:2010zz}). One obtains the following
table~\cite{Slansky:1981yr,Feger:2012bs}:

\begin{center}
\begin{tabular}{|l|c|c|c|c|c|c|c|c|}
\hline
$\#$ of vector fields $A^{\mu} $ & 1 & 2 & 3 & 4 & 5 & 6 & 7 & 8 \\
\hline
$\# $ of Lorentz scalars & 0  & 1 & 0 & 4 & 0 & 25 & 0 & 196 \\
\hline
\end{tabular}
\end{center}

These scalars can be written in terms of the primitive invariants,
namely, $g_{\mu\nu}$ and $\epsilon_{\mu\nu\rho\sigma}$. As shown in the
table, an odd number of vector fields is impossible, as is obvious from
the fact that one cannot form primitive Lorentz invariants with an odd number of
indices. For two fields, the only contracting possibility is $g_{\mu\nu}$,
while for four free Lorentz indices, the contractions 
with a term of the form $A^\mu B^\nu C^\rho D^\sigma$ 
can be performed with any member of
the list
\begin{equation}
\left\{
\begin{array}{l}
g_{\mu\nu} g_{\rho\sigma},\\
g_{\mu\rho}g_{\nu\sigma},\\
g_{\mu\sigma}g_{\nu\rho},\\
\epsilon_{\mu\nu\rho\sigma}.
\end{array}
\right.
\end{equation}
For the case with six free indices 
of the form $A^\mu B^\nu C^\rho D^\sigma E^\delta F^\epsilon$, 
one finds the fifteen independent
possibilities of combining three metrics, \emph{i.e.}, $g_{\mu\nu}g_{\rho\sigma}
g_{\delta\epsilon}$ and the nonequivalent permutations of indices, as
well as fifteen combinations of a metric and a Levi-Civita tensor, of which
only ten are independent, which we choose to be
\begin{equation}
\left\{
\begin{array}{l}
g_{\nu\rho} \epsilon_{\mu\sigma\delta\epsilon},\\
g_{\nu\sigma} \epsilon_{\mu\rho\delta\epsilon},\\
g_{\nu\delta} \epsilon_{\mu\rho\sigma\epsilon},\\
g_{\nu\epsilon} \epsilon_{\mu\rho\sigma\delta},\\
g_{\rho\sigma} \epsilon_{\mu\nu\delta\epsilon},\\
g_{\rho\delta}\epsilon_{\mu\nu\sigma\epsilon},\\
g_{\rho\epsilon}\epsilon_{\mu\nu\sigma\delta},\\
g_{\sigma\delta}\epsilon_{\mu\nu\rho\epsilon},\\
g_{\sigma\epsilon}\epsilon_{\mu\nu\rho\delta},\\
g_{\epsilon\delta}\epsilon_{\mu\nu\rho\sigma}.
\end{array}
\right.
\end{equation}

Now, one needs to take into account that when only one vector $A^\mu$
and its gradient are plugged into these expressions, some terms are
identical and can thus be simplified. The following table sums up the
number of independent terms that can be built for a given product of vectors and gradients. Numbers in parentheses indicate those terms that would vanish
if it were not for the group index; in our listings of all available
Lagrangians below, we indicate these contractions with a star.
Given the above discussion, we are sure that all the possible terms have
been found, and they are all linearly independent.
\begin{center}
\begin{tabular}{|c|c|c|c|}
\hline
\backslashbox{$\# (\partial^\mu A^\nu)$}{$\# A^\rho A^\sigma$} & 0 & 1 & 2 \\
\hline
1 & 1 (0)  & 3 (1) & 6 (4)  \\
\hline
2 & 4 (0)  & 13 (3)  & 34 (23) \\
\hline
3 & 9 (2)  & 52 (22)  &  \\
\hline
\end{tabular}
\end{center}
We now discuss each case separately.

For a single derivative and no additional field, one gets the simplest
combination, namely, $\left(\partial \cdot A \right)$. With two
additional fields, one gets
\begin{equation}
\left\{ \begin{array}{l}
\left(\partial \cdot A \right) \left(A \cdot A \right),\\
\left[\left(\partial^\mu A^\nu \right) A_\mu A_\nu \right],\\
\left[ \epsilon_{\mu \nu \rho \sigma} \left(\partial^\mu A^\nu \right) 
A^\rho A^\sigma \right], ~~~~ ~~~~ (*)
\end{array} \right.
\end{equation}
and with four additional fields, one obtains
\begin{equation}
\left\{ \begin{array}{l}
\left(\partial \cdot A \right) \left(A \cdot A \right)\left(A \cdot A
\right),\\
\left[\left(\partial^\mu A^\nu \right) A_\mu A_\nu \right]\left(A \cdot
A \right),\\
\left[ \epsilon_{\mu \nu \rho \sigma} \left(\partial^\mu A^\nu \right) 
A^\rho A^\sigma \right]\left(A \cdot A \right), ~~~~ ~~~~ (*) \\
\left[\epsilon_{\mu \nu \rho \sigma}\left(\partial^\mu A^\alpha \right)
A^\nu A^\rho A^\sigma A_\alpha    \right], ~~~~ ~~~~ (*)  \\
\left[\epsilon_{\mu \nu \rho \sigma}\left(\partial^\alpha A^\mu \right)
A^\nu A^\rho A^\sigma A_\alpha    \right],  ~~~~ ~~~~ (*) \\
\left(\partial \cdot A \right) \left[\epsilon_{\mu \nu \rho \sigma}
A^\mu A^\nu A^\rho A^\sigma  \right]. ~~~~ ~~~~ (*)
\end{array} \right.
\end{equation}

With two derivatives and no additional field, one then finds
\begin{equation}
\left\{ \begin{array}{l}
\left(\partial \cdot A \right)\left(\partial \cdot A \right),\\
\left[\left(\partial^\mu A^\nu \right) \left(\partial_\mu A_\nu \right)
\right],\\
\left[\left(\partial^\mu A^\nu \right) \left(\partial_\nu A_\mu \right)
\right],\\
\left[ \epsilon_{\mu \nu \rho \sigma} \left(\partial^\mu A^\nu \right)
\left(\partial^\rho A^\sigma \right)  \right],
\end{array} \right.
\end{equation}
whereas with two additional fields, one finds\footnote{As an example of the fact that not
every reshuffling of indices is independent, let us consider the term $
\epsilon_{\mu \nu \rho \sigma} A^\mu A^\nu \left(\partial^\alpha A^\rho
\right)   \left(\partial_\alpha A^\sigma \right)$, which could, in
principle, have appeared in the list in Eq. \eqref{PSU217}. It is indeed not
necessary because the property
\begin{equation}
g_{\mu\rho} \epsilon_{\nu\sigma\delta\epsilon} = g_{\nu\rho}
\epsilon_{\mu\sigma\delta\epsilon} - g_{\rho\sigma}
\epsilon_{\mu\nu\delta\epsilon} + g_{\rho\delta}
\epsilon_{\mu\nu\sigma\epsilon} -
g_{\rho\epsilon}\epsilon_{\mu\nu\sigma\delta}
\end{equation}
allows us to write it as a linear combination of the terms in
Eq.~\eqref{PSU217}.
}

\begin{equation}
\left\{ \begin{array}{l}
\left(\partial \cdot A \right)\left(\partial \cdot A \right)\left(A
\cdot A \right),\\
\left[\left(\partial^\mu A^\nu \right) \left(\partial_\mu A_\nu \right)
\right]\left(A \cdot A \right),\\
\left[\left(\partial^\mu A^\nu \right) \left(\partial_\nu A_\mu \right)
\right]\left(A \cdot A \right),\\
\left[ \epsilon_{\mu \nu \rho \sigma} \left(\partial^\mu A^\nu \right)
\left(\partial^\rho A^\sigma \right)  \right]\left(A \cdot A \right),\\
\left[\left(\partial^\mu A^\nu \right) A_\mu A_\nu \right]\left(\partial
\cdot A \right),\\
\left[ \epsilon_{\mu \nu \rho \sigma} \left(\partial^\mu A^\nu \right) 
A^\rho A^\sigma \right]\left(\partial \cdot A \right),  ~~~~ ~~~~ (*) \\
\left[ A_\mu A_\nu  \left(\partial^\mu A^\alpha \right)
\left(\partial^\nu A_\alpha \right) \right],\\
\left[ A_\mu A_\nu  \left(\partial^\mu A^\alpha \right)
\left(\partial_\alpha A^\nu \right) \right],\\
\left[ A_\mu A_\nu  \left(\partial^\alpha A^\mu \right)
\left(\partial_\alpha A^\nu \right) \right],\\
\left[ \epsilon_{\mu \nu \rho \sigma} A^\mu A^\nu  \left(\partial^\rho
A^\alpha \right)  \left(\partial^\sigma A_\alpha \right)\right],  ~~~~
~~~~ (*) \\
\left[ \epsilon_{\mu \nu \rho \sigma} A^\mu A^\nu  \left(\partial^\rho
A^\alpha \right)  \left(\partial_\alpha A^\sigma \right)\right],  ~~~~
~~~~ (*) \\
\left[ \epsilon_{\mu \nu \rho \sigma} A^\mu A^\alpha  \left(\partial^\nu
A^\rho \right)  \left(\partial^\sigma A_\alpha \right)\right], \\
\left[ \epsilon_{\mu \nu \rho \sigma} A^\mu A^\alpha  \left(\partial^\nu
A^\rho \right)  \left(\partial_\alpha A^\sigma \right)\right].
\end{array} \right.
\label{PSU217}
\end{equation}

Finally, demanding three gradients of the vector field and no vector
field itself, one obtains
\begin{equation}
\left\{ \begin{array}{l}
\left(\partial \cdot A \right)\left(\partial \cdot A \right)\left(
\partial \cdot A \right),\\
\left[\left(\partial^\mu A^\nu \right) \left(\partial_\mu A_\nu \right)
\right]\left( \partial \cdot A \right),\\
\left[\left(\partial^\mu A^\nu \right) \left(\partial_\nu A_\mu \right)
\right]\left( \partial \cdot A \right),\\
\left[ \epsilon_{\mu \nu \rho \sigma} \left(\partial^\mu A^\nu \right) 
\left(\partial^\rho A^\sigma \right)  \right] \left( \partial \cdot A
\right),  \\
\left[ \left(\partial^\mu A_\nu \right) \left(\partial^\nu A_\rho
\right) \left(\partial^\rho A_\mu \right) \right],\\
\left[ \left(\partial^\mu A_\nu \right) \left(\partial^\nu A_\rho
\right) \left(\partial_\mu A^\rho \right) \right],\\
\left[ \epsilon_{\mu \nu \rho \sigma} \left(\partial^\mu A^\alpha
\right) \left(\partial^\nu A_\alpha \right) \left(\partial^\rho A^\sigma
\right) \right], ~~~~ ~~~~ (*) \\
\left[ \epsilon_{\mu \nu \rho \sigma} \left(\partial^\mu A^\alpha
\right) \left(\partial_\alpha A^\nu \right) \left(\partial^\rho A^\sigma
\right) \right],\\
\left[ \epsilon_{\mu \nu \rho \sigma} \left(\partial^\alpha A^\mu
\right)\left(\partial_\alpha A^\nu \right)\left(\partial^\rho A^\sigma
\right) \right]. ~~~~ ~~~~ (*)
\end{array} \right.
\end{equation}

\subsection{Group Sector}

Let us now proceed with the similar procedure but now in the group sector.
Since we assumed that the vector fields transform according to the
representation of dimension $3$ of SU(2), one can safely use known
results from representation theory of compact Lie groups. The table
below summarizes the different possibilities to obtain an SU(2) singlet
as a function of the number of fields belonging to the $\mathbf{3}$
representation of SU(2) \cite{Slansky:1981yr,Feger:2012bs}:

\begin{center}
\begin{tabular}{|l|c|c|c|c|c|c|c|}
\hline
$\#$ vector fields in the $\mathbf{3}$ of SU(2) & 1 & 2 & 3 & 4 & 5 & 6
& 7\\
\hline
$\#$ of SU(2) singlets & 0  & 1 & 1 & 3 & 6 & 15 & 36 \\
\hline
\end{tabular}
\end{center}

We reproduce below the procedure explained in
Sec.~\ref{PSU2PartProcedureSinglet}, whereby one constructs the necessary
products of group metric coefficients $\delta_{ab}$ and structure
constants $\epsilon_{abc}$. Getting as many independent terms as
predicted by the representation theory (table above) ensures completeness of
the basis. Similar to the Lorentz invariance discussed in the previous
section, these two tensors are the only
primitive invariants of the group
\cite{Slansky:1981yr,Fuchs:1997jv,Padilla:2010ir}.

To contract with two or three free SU(2) indices, the only possible
choices are, respectively, $\delta_{ab}$ and $\epsilon_{abc}$. With four fields,
one can make use of the three combinations
\begin{equation}
\label{PSU2DabDcd}
\left\{ \begin{array}{l}
\delta_{ab}\delta_{cd},\\
\delta_{ac}\delta_{bd},\\
\delta_{ad}\delta_{bc},
\end{array} \right.
\end{equation}
while five fields demand the following six possibilities, namely
\begin{equation}
 \left\{ \begin{array}{l}
\delta_{ab}\epsilon_{cde},\\
\delta_{ac}\epsilon_{bde},\\
\delta_{ad}\epsilon_{bce},\\
\delta_{bc}\epsilon_{ade},\\
\delta_{bd}\epsilon_{ace},\\
\delta_{cd}\epsilon_{abe}.
\end{array} \right. 
\end{equation}

As in Sec.~\ref{PSU2LSSection}, one can devise other possible
formulations that apply, but they will always be expressible as linear
combinations of the above. For instance, relations between the structure
constants, such as
\begin{equation}
\epsilon_{ab}{}^{e}\epsilon_{cde}=\delta_{ac}\delta_{bd} -
\delta_{ad}\delta_{bc},
\end{equation}
imply that contracting a four-index term with two structure
constants is equivalent to a linear combination of the terms given in
Eq.~\eqref{PSU2DabDcd}.

\subsection{Final Test Lagrangians}

Gathering the results and applying the procedure of
Sec.~\ref{PSU2PartProcedureSinglet}, we are now in a position to write down
our test Lagrangians, scalars under both Lorentz and SU(2)
transformations. Some of these terms simplify through contractions, e.g.,
$\epsilon_{\a\b\EAc}\left(A^{\a} \cdot A^{\b} \right)\left(\partial \cdot
A^{\EAc}\right) = 0$, and we are left with fewer terms than the naive
multiplication of all singlet possibilities of each sector would have
otherwise suggested. This is fortunate because the number of terms to be
considered \emph{a priori} is quickly increasing with the number of fields
involved, as shown in the table below:
\begin{center}
\begin{tabular}{|c|c|c|c|}
\hline
\backslashbox{$\# \partial^\mu A^{\nu\a}$}{$\# A^{\rho \b}$} & 0 & 2 & 4
\\
\hline
1 & 0  & 3 & 36  \\
\hline
2 & 4  & 42  & 510 \\
\hline
3 & 9  & 312  &  \\
\hline
\end{tabular}
\end{center}					

After simplifications, we find two terms (instead of three according to the
table) containing a single derivative term and two additional vector fields,
\begin{equation}
\left\{ \begin{array}{l}
\mathcal{L}_{1}=\epsilon_{\a \b \EAc}\left[\left(\partial^\mu A^{\a\nu}
\right) A^{\b}_\mu A^{\EAc}_\nu \right],\\
\mathcal{L}_{2}=\epsilon_{\a\b\EAc}\left[ \epsilon_{\mu \nu \rho \sigma}
\left(\partial^\mu A^{\a\nu} \right)  A^{\b\rho} A^{\EAc\sigma} \right]  ,
\end{array} \right.
\end{equation}
and eight with four such fields, namely,
\begin{equation}
\left\{ \begin{array}{l}
\mathcal{L}_{1}=\epsilon_{\a\b\EAc}\left[\left(\partial^\mu A^{\d\nu}
\right) A^{\a}_\mu A^{\b}_\nu \right]\left(A^{\EAc} \cdot A_{\d}
\right),\\
\mathcal{L}_{2}=\epsilon_{\a\b\EAc}\left[\left(\partial^\mu A^{\a\nu}
\right) A^{\d}_\mu A^{\b}_\nu \right]\left(A^{\EAc} \cdot A_{\d}
\right),\\
\mathcal{L}_{3}=\epsilon_{\a\b\EAc}\left[\left(\partial^\mu A^{\a\nu}
\right) A^{\b}_\mu A^{\d}_\nu \right]\left(A^{\EAc} \cdot A_{\d}
\right),\\
\mathcal{L}_{4}=\epsilon_{\a\b\EAc}\left[ \epsilon_{\mu \nu \rho \sigma}
\left(\partial^\mu A^{\d\nu} \right)  A^{\a\rho} A^{\b\sigma}
\right]\left(A^{\EAc} \cdot A_{\d} \right),  \\
\mathcal{L}_{5}=\epsilon_{\a\b\EAc}\left[ \epsilon_{\mu \nu \rho \sigma}
\left(\partial^\mu A^{\a\nu} \right)  A^{\d\rho} A^{\b\sigma}
\right]\left(A^{\EAc} \cdot A_{\d} \right),  \\
\mathcal{L}_{6}=\epsilon_{\a\b\EAc}\left[\epsilon_{\mu \nu \rho
\sigma}\left(\partial^\mu A^{\d\alpha} \right) A_{\d}^\nu A^{\a\rho}
A^{\b\sigma} A^{\EAc}_\alpha    \right],\\
\mathcal{L}_{7}=\epsilon_{\a\b\EAc}\left[\epsilon_{\mu \nu \rho
\sigma}\left(\partial^\alpha A^{\d\mu} \right) A_{\d}^\nu A^{\a\rho}
A^{\b\sigma} A^{\EAc}_\alpha    \right] ,\\
\mathcal{L}_{8}=\epsilon_{\a\b\EAc}\left[\epsilon_{\mu \nu \rho
\sigma}\left(\partial \cdot A_{\d} \right) A^{\d \mu} A^{\a\nu}
A^{\b\rho} A^{\EAc \sigma}    \right] .\\
\end{array} \right.
\end{equation}
Note that one cannot build a single derivative term without an additional
field, as it would otherwise belong to the $\mathbf{3}$ representation of SU(2).

For two first-order vector field derivatives without additional fields, one gets
\begin{equation}
\left\{ \begin{array}{l}
\mathcal{L}_1=\left(\partial \cdot A^{\a} \right)\left(\partial \cdot
A_{\a} \right),\\ \mathcal{L}_2=\left[\left(\partial^\mu A^\nu_{\a}
\right) \left(\partial_\mu A_\nu^{\a} \right) \right],\\
\mathcal{L}_3=\left[\left(\partial^\mu A^\nu_{\a} \right)
\left(\partial_\nu A_\mu^{\a} \right) \right],\\
\mathcal{L}_4=\left[ \epsilon_{\mu \nu \rho \sigma} \left(\partial^\mu
A^{\a\nu} \right)  \left(\partial^\rho A^\sigma_{\a} \right)  \right],
\end{array} \right.
\end{equation}
whereas with two additional vector fields, one gets 
\begin{equation}
\label{PSU2EqLagBeforeHessian}
\left\{ \begin{array}{l}
\mathcal{L}_1=\left(\partial \cdot A^{\a}\right)\left(\partial \cdot
A_{\a} \right)\left(A^{\b} \cdot A_{\b} \right),\\
\mathcal{L}_2=\left(\partial \cdot A^{\a} \right)\left(\partial \cdot
A^{\b} \right)\left(A_{\a} \cdot A_{\b} \right),\\
\mathcal{L}_3=\left[\left(\partial^\mu A^\nu_{\a} \right)
\left(\partial_\mu A_\nu^{\a} \right) \right]\left(A^{\b} \cdot A_{\b}
\right),\\
\mathcal{L}_4=\left[\left(\partial^\mu A^\nu_{\a} \right)
\left(\partial_\mu A_\nu^{\b} \right) \right]\left(A^{\a} \cdot A_{\b}
\right),\\
\mathcal{L}_5=\left[\left(\partial^\mu A^\nu_{\a} \right)
\left(\partial_\nu A_\mu^{\a} \right) \right]\left(A^{\b} \cdot A_{\b}
\right),\\
\mathcal{L}_6=\left[\left(\partial^\mu A^\nu_{\a} \right)
\left(\partial_\nu A_\mu^{\b} \right) \right]\left(A^{\a} \cdot A_{\b}
\right),\\
\mathcal{L}_7=\left[ \epsilon_{\mu \nu \rho \sigma} \left(\partial^\mu
A^{\a\nu} \right)  \left(\partial^\rho A^\sigma_{\a} \right) 
\right]\left(A^{\b} \cdot A_{\b} \right),\\
\mathcal{L}_8=\left[ \epsilon_{\mu \nu \rho \sigma} \left(\partial^\mu
A^{\a\nu} \right)  \left(\partial^\rho A^\sigma_{\b} \right) 
\right]\left(A_{\a} \cdot A^{\b} \right),\\
\mathcal{L}_9=\left[\left(\partial^\mu A^\nu_{\a} \right) A_\mu^{\a}
A_\nu^{\b} \right]\left(\partial \cdot A_{\b} \right),\\
\mathcal{L}_{10}=\left[\left(\partial^\mu A^\nu_{\a} \right) A_\mu^{\b}
A_\nu^{\a} \right]\left(\partial \cdot A_{\b} \right),\\
\mathcal{L}_{11}=\left[\left(\partial^\mu A^\nu_{\a} \right) A_\mu^{\b}
A_{\b\nu} \right]\left(\partial \cdot A^{\a} \right),\\
\mathcal{L}_{12}=\left[ \epsilon_{\mu \nu \rho \sigma}
\left(\partial^\mu A^{\a\nu} \right)  A^\rho_{\a} A^\sigma_{\b}
\right]\left(\partial \cdot A^{\b} \right), \\
\mathcal{L}_{13}=\left[ A_\mu^{\a} A_{\a\nu}  \left(\partial^\mu
A^\alpha_{\b} \right) \left(\partial^\nu A_{\alpha}^{\b} \right)
\right],\\
\mathcal{L}_{14}=\left[ A_\mu^{\a} A_{\nu}^{\b}  \left(\partial^\mu
A^\alpha_{\a} \right) \left(\partial^\nu A_{\b\alpha} \right) \right],\\
\mathcal{L}_{15}=\left[ A_\mu^{\a} A_{\a\nu}  \left(\partial^\mu
A^{\b\alpha} \right) \left(\partial_\alpha A^\nu_{\b} \right) \right],\\
\mathcal{L}_{16}=\left[ A_\mu^{\a} A_{\nu}^{\b}  \left(\partial^\mu
A^{\alpha}_{\a} \right) \left(\partial_\alpha A^\nu_{\b} \right)
\right],\\
\mathcal{L}_{17}=\left[ A_\mu^{\a} A_{\nu}^{\b}  \left(\partial^\mu
A^{\alpha}_{\b} \right) \left(\partial_\alpha A^\nu_{\a} \right)
\right],\\
\mathcal{L}_{18}=\left[ \epsilon_{\mu \nu \rho \sigma} A^{\a\mu}
A^{\b\nu}  \left(\partial^\rho A^\alpha_{\a} \right) 
\left(\partial^\sigma A_{\b\alpha} \right)\right], \\
\mathcal{L}_{19}=\left[ \epsilon_{\mu \nu \rho \sigma} A^{\a\mu}
A^{\b\nu}  \left(\partial^\rho A^\alpha_{\a} \right) 
\left(\partial_\alpha A^\sigma_{\b} \right)\right] ,\\
\mathcal{L}_{20}=\left[ \epsilon_{\mu \nu \rho \sigma} A^{\a\mu}
A^{\b\nu}  \left(\partial^\alpha A^\rho_{\a} \right)  
\left(\partial_\alpha A^\sigma_{\b} \right)\right]  ,\\
\mathcal{L}_{21}=\left[ \epsilon_{\mu \nu \rho \sigma} A^{\a\mu}
A^\alpha_{\a}  \left(\partial^\nu A^{\rho}_{\b} \right) 
\left(\partial^\sigma A_\alpha ^{\b}\right)\right], \\
\mathcal{L}_{22}=\left[ \epsilon_{\mu \nu \rho \sigma} A^{\a\mu}
A^\alpha_{\b}  \left(\partial^\nu A^{\rho}_{\a} \right) 
\left(\partial^\sigma A_\alpha ^{\b}\right)\right], \\
\mathcal{L}_{23}=\left[ \epsilon_{\mu \nu \rho \sigma} A^{\mu}_{\a}
A^\alpha_{\b}  \left(\partial^\nu A^{\b\rho} \right) 
\left(\partial^\sigma A_\alpha ^{\a}\right)\right] ,\\
\mathcal{L}_{24}=\left[ \epsilon_{\mu \nu \rho \sigma} A^{\a\mu}
A^\alpha_{\a}  \left(\partial^\nu A^{\b\rho} \right) 
\left(\partial_\alpha A^\sigma_{\b} \right)\right] ,\\
\mathcal{L}_{25}=\left[ \epsilon_{\mu \nu \rho \sigma} A^{\a\mu}
A^\alpha_{\b}  \left(\partial^\nu A^{\rho}_{\a} \right) 
\left(\partial_\alpha A^{\b\sigma} \right)\right], \\
\mathcal{L}_{26}=\left[ \epsilon_{\mu \nu \rho \sigma} A^{\mu}_{\a}
A^\alpha_{\b}  \left(\partial^\nu A^{\b\rho} \right) 
\left(\partial_\alpha A^{\a\sigma} \right)\right] ,\\
\mathcal{L}_{27}=\left[ A_\mu^{\a} A_{\nu}^{\b}  \left(\partial^\mu
A^\alpha_{\b} \right) \left(\partial^\nu A_{\a \alpha} \right)
\right],\\
\mathcal{L}_{28}=\left[ A_\mu^{\a} A_{\nu}^{\b}  \left(\partial^\alpha
A^\mu_{\b} \right) \left(\partial_\alpha A^\nu_{\a} \right) \right].
\end{array} \right.
\end{equation}

Finally, with three derivatives, one finds
\begin{equation}
\left\{ \begin{array}{l}
\mathcal{L}_{1}=  \epsilon_{\a \b \EAc} \left[\left(\partial^\mu
A_\nu^{\a} \right) \left(\partial^\nu A_\rho^{\b} \right)
\left(\partial^\rho A_\mu^{\EAc} \right) \right], \\
\mathcal{L}_{2}=  \epsilon_{\a \b \EAc} \left[ \left(\partial^\mu
A_\nu^{\a} \right) \left(\partial^\nu A_\rho^{\b} \right)
\left(\partial_\mu A^{\EAc \rho} \right) \right], \\
\mathcal{L}_{3}=  \epsilon_{\a \b \EAc} \left[ \epsilon_{\mu \nu \rho
\sigma} \left(\partial^\mu A^{\a \alpha} \right) \left(\partial^\nu
A_\alpha^{\b} \right) \left(\partial^\rho A^{\EAc \sigma} \right) \right],
\\
\mathcal{L}_{4}=  \epsilon_{\a \b \EAc} \left[ \epsilon_{\mu \nu \rho
\sigma} \left(\partial^\mu A^{\a \alpha} \right) \left(\partial_\alpha
A^{\b \nu} \right) \left(\partial^\rho A^{\EAc \sigma} \right) \right], \\
\mathcal{L}_{5}=  \epsilon_{\a \b \EAc} \left[ \epsilon_{\mu \nu \rho
\sigma} \left(\partial^\alpha A^{\a \mu} \right)\left(\partial_\alpha
A^{\b \nu} \right)\left(\partial^\rho A^{\EAc \sigma} \right) \right],
\end{array} \right.
\end{equation}
completing our list of test Lagrangians.

\section{Construction of the healthy terms} \label{PSU2cht}
\subsection{Hessian Condition}
\label{PSU2PartHessianCondition}

Let us now apply the Hessian condition, as discussed in
Sec.~\ref{PSU2PartProcedure}. The first step is to calculate the Hessians
associated with the various test Lagrangians, defined by
Eq.~(\ref{PSU2EqDefHessian}). One sees that only those terms containing at
least two first-order derivatives of the vector field yield a nonvanishing value. In practice, one gets
\begin{equation}
\left\{ \begin{array}{l}
\mathcal{H}^{\mu\nu \al\be}_1= 2 g ^{0 \mu} g ^{0 \nu} g^{\al
\be},\\
\mathcal{H}^{\mu\nu \al\be}_2=-2 g^{\mu\nu} g^{\al \be},\\
\mathcal{H}^{\mu\nu \al\be}_3=2  g ^{0 \mu} g^{0 \nu} g^{\al \be},\\
\mathcal{H}^{\mu\nu \al\be}_4=0,
\end{array} \right.
\end{equation}
for the terms with two first-order derivatives and no additional fields, and
\begin{equation}
\left\{ \begin{array}{l}
\mathcal{H}^{\mu\nu \al\be}_1=2 g ^{0 \mu} g ^{0 \nu} g^{\al
\be}\left(A^{\b}\cdot A_{\b} \right),\\
\mathcal{H}^{\mu\nu \al\be}_2=2 g ^{0 \mu} g ^{0 \nu} \left(A^{\al}\cdot
A^{\be} \right),\\
\mathcal{H}^{\mu\nu \al\be}_3= -2g^{\mu\nu}g^{\al \be} \left(A^{\b}\cdot
A_{\b} \right),\\
\mathcal{H}^{\mu\nu \al\be}_4= -2g^{\mu\nu} \left(A^{\al}\cdot A^{\be}
\right),\\ \mathcal{H}^{\mu\nu \al\be}_5=2 g ^{0 \mu} g ^{0 \nu}g^{\al
\be} \left(A^{\b}\cdot A_{\b} \right),\\
\mathcal{H}^{\mu\nu \al\be}_6=2 g ^{0 \mu} g ^{0 \nu} \left(A^{\al}\cdot
A^{\be} \right),\\
\mathcal{H}^{\mu\nu \al\be}_7=0,\\
\mathcal{H}^{\mu\nu \al\be}_8=0,\\
\mathcal{H}^{\mu\nu \al\be}_9=A ^{0 \al} A ^{\mu \be} g^{0\nu}+A ^{0
\be} A ^{\nu \al} g^{0\mu},\\
\mathcal{H}^{\mu\nu \al\be}_{10}=A ^{0 \be} A ^{\mu \al} g^{0\nu}+A ^{0
\al} A ^{\nu \be} g^{0\mu},\\
\mathcal{H}^{\mu\nu \al\be}_{11}=A ^{0 \b} A ^{\mu}_{\b} g^{0\nu}g^{\al
\be}+A ^{0 \b} A ^{\nu}_{\b} g^{0\mu}g^{\al \be},\\
\mathcal{H}^{\mu\nu \al\be}_{12}=\epsilon^{0\mu}{}_{\rho\sigma} A^{\rho
\al} A^{\sigma \be} g^{0\nu}+\epsilon^{0\nu}{}_{\rho\sigma} A^{\rho \be}
A^{\sigma \al} g^{0\mu},\\
\mathcal{H}^{\mu\nu \al\be}_{13}=2 A^{0 \b} A^{0}_{\b} g^{\mu\nu} g^{\al
\be},\\
\mathcal{H}^{\mu\nu \al\be}_{14}=2 A^{0 \al} A^{0 \be} g^{\mu\nu} ,\\
\mathcal{H}^{\mu\nu \al\be}_{15}=A^{0 \b} A^{\nu}_{\b} g^{\mu 0} g^{\al
\be}+A^{0 \b} A^{\mu}_{\b} g^{\nu 0} g^{\al \be},\\
\mathcal{H}^{\mu\nu \al\be}_{16}=A^{0 \al} A^{\nu \be} g^{\mu 0}+A^{0
\be} A^{\mu \al} g^{\nu 0},\\
\mathcal{H}^{\mu\nu \al\be}_{17}=A^{0 \be} A^{\nu \al} g^{\mu 0}+A^{0
\al} A^{\mu \be} g^{\nu 0},\\
\mathcal{H}^{\mu\nu \al\be}_{18}=0,\\
\mathcal{H}^{\mu\nu \al\be}_{19}=\epsilon _{\rho \sigma}{}^{0 \nu}
A^{\rho \al}\ A^{\sigma \be} g^{\mu 0} + \epsilon _{\rho \sigma}{}^{0
\mu} A^{\rho \be}\ A^{\sigma \al} g^{\nu 0} ,\\
\mathcal{H}^{\mu\nu \al\be}_{20}= -2 \epsilon_{\rho \sigma}{}^{\mu\nu}
A^{\rho \al}A^{\sigma \be},\\
\mathcal{H}^{\mu\nu \al\be}_{21}=0,\\
\mathcal{H}^{\mu\nu \al\be}_{22}=0,\\
\mathcal{H}^{\mu\nu \al\be}_{23}=0,\\
\mathcal{H}^{\mu\nu \al\be}_{24}=0,\\
\mathcal{H}^{\mu\nu \al\be}_{25}=\epsilon_{\rho}{}^{0\mu\nu} A^{\rho
\al} A^{0\be} + \epsilon_{\rho}{}^{0\nu\mu} A^{\rho \be} A^{0\al} ,\\
\mathcal{H}^{\mu\nu \al\be}_{26}=\epsilon_{\rho}{}^{0\mu\nu} A^{\rho
\be} A^{0\al} + \epsilon_{\rho}{}^{0\nu\mu} A^{\rho \al} A^{0\be} ,\\
\mathcal{H}^{\mu\nu \al\be}_{27}=2 A^{0 \al} A^{0 \be} g^{\mu\nu} ,\\
\mathcal{H}^{\mu\nu \al\be}_{28}= -2 A^{\nu \al} A^{\mu\be},
\end{array} \right.
\end{equation}
for those with two first-order derivatives and two additional vector fields.

For the terms with three first-order derivatives, we have
\begin{equation}
\left\{ \begin{array}{l}
\mathcal{H}^{\mu\nu \al\be}_1=3 \epsilon^{\al \be}{}_{\EAc} (g^{0 \mu}
\partial^\nu A^{\EAc 0} - g^{0 \nu} \partial^\mu A^{\EAc 0}), \\
\mathcal{H}^{\mu\nu \al\be}_2=\epsilon^{\al \be}{}_{\EAc} (g^{0 \mu}
\partial^0 A^{\EAc \nu} - g^{0 \nu} \partial^0 A^{\EAc \mu}) + \epsilon^{\al
\be}{}_{\EAc} (\partial^\mu A^{\EAc \nu} - \partial^\nu A^{\EAc \mu}), \\
\mathcal{H}^{\mu\nu \al\be}_3=0,\\
\mathcal{H}^{\mu\nu \al\be}_4= \epsilon^{\al \be}{}_{\EAc} (\epsilon^{0
\nu \rho \sigma} g^{0 \mu} \partial_\rho A^{\EAc}_\sigma - \epsilon^{0 \mu
\rho \sigma} g^{0 \nu} \partial_\rho A^{\EAc}_\sigma) + 2 \epsilon^{\al
\be}{}_{\EAc} \epsilon^{\rho \mu 0 \nu} \partial_\rho A^{\EAc 0} , \\
\mathcal{H}^{\mu\nu \al\be}_5= -2 \epsilon^{\al \be}{}_{\EAc}
\epsilon^{\mu \nu \rho \sigma} \partial_\rho A^{\EAc}_\sigma + 4
\epsilon^{\al \be}{}_{\EAc} \epsilon^{\rho \mu 0 \nu} \partial^0
A^{\EAc}_\rho .
\end{array} \right.
\end{equation}

With these partial Hessians, we now construct a basis of terms
fulfilling the condition discussed above, \emph{i.e.}, such that
$\mathcal{H}^{0\mu\al\be} = 0$ for all values of $\mu$, $\al$ and $\be$; see
Sec.~\ref{PSU2PartProcedure}. To reach this goal, using notations already
introduced in Ref.~\cite{Allys:2015sht}, we produce a Lagrangian by
means of a linear combination of our test ones, namely,
\begin{equation}
\mathcal{L}_{\text{test}}=\sum_i x_i \mathcal{L}_i,
\end{equation}
for a yet-unknown set of constant parameters $x_i$. The Hessian is then
calculated for this Lagrangian, leading to algebraic equations for the
$x_i$ whose roots provide the required actions. It turns out to be
easier to separately compute the cases $\mu=0$ and $\mu=i$, as well as
$\al=\be$ and $\al\neq\be$.

Let us begin with the case $\al=\be$. Test Lagrangians with two
derivatives and no additional fields have only one Hessian component not
identically vanishing, namely,
\begin{equation}
\mathcal{H}^{00\al \al} =  4(x_1 + x_2 + x_3), 
\end{equation}
while for two additional vector fields, there are four independent Hessian
conditions, given by
\begin{align}
\mathcal{H}^{00\al\al} = & ~  4\left(x_1 + x_3 + x_5 \right) \left(
A^{\b} \cdot A_{\b}\right) + 2 \left( x_2 + x_4 + x_6 \right)
\left(A^{\al} \cdot A^{\al} \right) \nonumber\\
&- 2 \left( x_9 + x_{10} + x_{14} + x_{16} + x_{17}+ x_{27} + x_{28}
\right) \left( A^{0\al} A^{0\al} \right) \nonumber \\
& - 4\left( x_{11} + x_{13} + x_{15}  \right) \left( A^{0 \b} A^{0}_{\b}
\right),
\end{align}
\begin{align}
\mathcal{H}^{0i\al\al} =&  -\left( x_9 + x_{10} + x_{16} + x_{17}  +
2x_{28} \right) \left( A^{0\al} A^{i \al} \right) - 2 \left(
x_{11} + x_{15} \right) \left(A^{0 \b} A^{i}_{\b} \right) \nonumber\\
& -\left(x_{12} + x_{19} + 2 x_{20} \right) \left( \epsilon_{\rho
\sigma}{}^{0i}A^{\rho \al}A^{\sigma \al} \right).
\end{align}
On the other hand, the case $\al \not= \be$ implies 
\begin{equation}
\mathcal{H}^{00\al\be} = 2 \left(x_2 + x_4 + x_6 \right) \left( A^{\al}
\cdot A^{\be} \right) - 2 \left( x_9 + x_{10} +x_{14} + x_{16} + x_{17}
+ x_{27} + x_{28} \right) \left( A^{0\al} A^{0\be} \right),
\end{equation}
\begin{align}
\mathcal{H}^{0i \al\be} = -\left(x_9 + x_{17} + 2x_{28} \right) \left(
A^{0\be} A^{i\al} \right) - \left(x_{10} + x_{16} \right) \left(A^{0\al}
A^{i\be} \right) -\left(- x_{12} + x_{19} + 2 x_{20} \right)
\left(\epsilon^{0i}{}_{\rho\sigma} A^{\rho \al}A^{\sigma \be} \right).
\end{align}
Making these four terms vanish can be done, without loss of
generality (since all linear combinations of the resulting terms are all
also acceptable):
\begin{equation}
\begin{array}{l}
x_3 = -x_1 - x_5, \\
x_4 = - x_2 - x_6, \\
x_{12} = 0, \\
x_{13} = 0,\\
x_{14} = - x_{27} + x_{28},  \\
x_{15} = - x_{11} , \\
x_{16} = -x_{10} , \\
x_{17} = -x_9 - 2 x_{28},\\
x_{19} =  -2 x_{20}.
\end{array}
\end{equation}

With three derivatives, one finds that $\mathcal{H}^{00\al\al} $,
$\mathcal{H}^{00\al\be}$ ($\al\neq\be$) and $\mathcal{H}^{0i\al\al}$
identically vanish, whereas for $\al \neq \be$, we have
\begin{equation}
\mathcal{H}^{0i\al\be} = \epsilon_{\EAc}{}^{\al \be} \left[(-3x_1 - x_2)
\partial^i A^{0 \EAc} - (x_4 + 2x_5) \epsilon^{0 i \rho \sigma}
\partial_\rho A_\sigma^{\EAc}\right],
\end{equation}
thus leading to the conditions
\begin{equation}
\begin{array}{l}
x_2 = - 3x_1, \\
x_4 = - 2x_5.
\end{array}
\end{equation}

\subsection{Simplification of the Lagrangian}

For one gradient and two vector fields, we can define the current
\begin{equation}
J^\mu = \epsilon_{\a\b\EAc} \epsilon^{\mu\nu\rho\sigma} A_\nu^{\a}
A_\rho^{\b} A_\sigma^{\EAc},
\end{equation}
showing that $\mathcal{L}_2$ is a total derivative, namely,
\begin{equation}
\partial_\mu J^\mu = 3 \mathcal{L}_2.
\end{equation}

A similar technique applies for one derivative term and 4 additional
vector fields: in this case, one forms the following two currents,
\begin{equation}
\begin{array}{l}
J^{\mu}_1 = \epsilon^{\mu}{}_{\nu\rho\sigma} A^{\nu\a} A^{\rho\b} A
^{\sigma\EAc} A^{\alpha \d} A_{\alpha\d} \epsilon_{\a\b\EAc},\\
J^{\alpha}_{2} = \epsilon_{\mu\nu\rho\sigma}
A^{\mu\a}A^{\nu\b}A^{\rho\EAc} A^{\sigma\d} A^{\alpha}_{\d}
\epsilon_{\a\b\EAc},
\end{array}
\end{equation}
yielding
\begin{equation}
\begin{array}{l}
\partial_{\mu} J^{\mu}_1 = 3 \left( \mathcal{L}_{3}- 2 \mathcal{L}_{5} +
2 \mathcal{L}_{6} \right),\\
\partial_{\alpha} J^{\alpha}_2 = - \mathcal{L}_{8}.
\end{array}
\end{equation}
Finally, some terms involving two first-order derivatives can
be described by 
\begin{eqnarray}
J^{\mu_1} &=& \delta^{\mu_1\mu_2}_{\nu_1\nu_2} A^{\nu_1 \a}
\partial_{\mu_2} A^{\nu_2}_{\a}, \\
J^{\mu}_\epsilon &=& \epsilon^{\mu\nu\rho\sigma} A_\nu^{\a}
(\partial_\rho A_{\sigma \a}),
\end{eqnarray}
where we have used the definition $\delta^{\mu_1 \mu_2}_{\nu_1 \nu_2} \equiv
\delta^{\mu_1}_{\nu_1}\delta^{\mu_2}_{\nu_2}
-\delta^{\mu_1}_{\nu_2}\delta^{\mu_2}_{\nu_1}$
stemming from Eq.~\eqref{PSU2deltaMult}, leading to
\begin{eqnarray}
\partial_{\mu_1} J^{\mu_1} &=& \mathcal{L}_{1} - \mathcal{L}_{3}, \\
\partial_\mu J^\mu_\epsilon &=& \mathcal{L}_4.
\end{eqnarray}

Terms containing two derivatives and two fields are
slightly more involved. We first make use of the identity
\cite{Allys:2016jaq,Fleury:2014qfa}
\begin{equation}
\label{PSU2EqFP}
A^{\mu \alpha} \tilde{B}_{\nu \alpha} + B^{\mu \alpha} \tilde{A}_{\nu
\alpha} = \frac{1}{2} (B^{ \alpha \beta} \tilde{A}_{\alpha \beta}
)\delta^{\mu}_{\nu},
\end{equation}
valid for all antisymmetric tensors $A$ and $B$. This provides the relations
\begin{equation}
\left( G^{\mu\alpha\a}\tilde{G}_{\nu\alpha}^{\b} +
G^{\mu\alpha\b}\tilde{G}_{\nu\alpha}{}^{\a}\right) A_{\mu\a}
A^{\nu}_{\b} = \frac{1}{2} \left( G^{\alpha\beta \a}
\tilde{G}_{\alpha\beta}^{\b} \right) \left(A_{\a}\cdot A_{\b} \right)
\end{equation}
and
\begin{equation}
G^{\mu\alpha\a}\tilde{G}_{\nu\alpha\a} A_{\mu}^{\b}A^{\nu}_{\b} =
\frac{1}{4}\left( G^{\alpha\beta \a} \tilde{G}_{\alpha\beta \a}\right)
\left(A^{\b}\cdot A_{\b} \right),
\end{equation}
where $G^{\mu\alpha\a}$ is the Abelian form of the Faraday tensor
as defined below in Eq.~\eqref{PSU2EqFaradayAbelian}.
From these, one then derives the following two identities relating
the Lagrangians in Eq.~\eqref{PSU2EqLagBeforeHessian}:
\begin{equation}
\mathcal{L}_{25}+\mathcal{L}_{26}-\mathcal{L}_{22}-\mathcal{L}_{23}
=\mathcal{L}_{8}
\end{equation}
and
\begin{equation}
2\left(\mathcal{L}_{24}-\mathcal{L}_{21} \right) = \mathcal{L}_{7}.
\end{equation}

It is also possible to find total derivatives to reduce the number of
independent terms. First, one can use the fact that $\tilde{G}$ is
divergence-free, introducing the currents
\begin{equation}
\begin{array}{l}
J^{\mu}_{G,1}=\tilde{G}^{\mu\nu}_{\a}A_{\nu}^{\a} \left(A^{\b}\cdot
A_{\b} \right),\\
J^{\mu}_{G,2}=\tilde{G}^{\mu\nu}_{\a}A_{\nu\b} \left(A^{\a}\cdot A^{\b}
\right),
\end{array}
\end{equation}
providing
\begin{equation}
\begin{array}{l}
\partial_{\mu}J^{\mu}_{G,1}= \mathcal{L}_{7} - 2 \mathcal{L}_{22},\\
\partial_{\mu}J^{\mu}_{G,2}= \mathcal{L}_{8} - \mathcal{L}_{21} -
\mathcal{L}_{23}.
\end{array}
\end{equation}
One can subsequently use the antisymmetric forms written
from $\delta^{\mu_1 \mu_2}_{\nu_1 \nu_2}$: 
\begin{equation}
\begin{array}{l}
J^{\mu}_{\delta,1}=\delta^{\mu_1 \mu_2}_{\nu_1 \nu_2}
A_{\b}^{\lambda}A_{\lambda}^{\b} A^{\nu_1}_{\a} \partial_{\mu_2}
A^{\nu_2\a},\\
J^{\mu}_{\delta,2}=\delta^{\mu_1 \mu_2}_{\nu_1 \nu_2}
A_{\a}^{\lambda}A_{\lambda}^{\b} A^{\nu_1\a} \partial_{\mu_2}
A^{\nu_2}_{\b},\\
J^{\mu}_{\delta,3}=\delta^{\mu_1 \mu_2}_{\nu_1 \nu_2}
A^{\lambda\a}A^{\nu_1}_{\a} A^{\nu_2}_{\b}
\partial_{\mu_2}A_{\lambda}^{\b},
\end{array}
\end{equation}
resulting in
\begin{equation}
\begin{array}{l}
\partial_{\mu}J^{\mu}_{\delta,1} = \mathcal{L}_{1}-\mathcal{L}_{5}+2
\mathcal{L}_{10} - 2 \mathcal{L}_{16},\\
\partial_{\mu}J^{\mu}_{\delta,2} = \mathcal{L}_{2}-\mathcal{L}_{6}+ 
\mathcal{L}_{9}+ \mathcal{L}_{11} -  \mathcal{L}_{15} - 
\mathcal{L}_{17},\\
\partial_{\mu}J^{\mu}_{\delta,3} = \mathcal{L}_{14}+\mathcal{L}_{9}+
\mathcal{L}_{15}- \mathcal{L}_{27} -  \mathcal{L}_{17} - 
\mathcal{L}_{11}.
\end{array}
\end{equation}
Finally, we can write
\begin{equation}
J^{\mu}_{\epsilon,1}= \epsilon^{\mu}{}_{\nu\rho\sigma} A^{\nu \a}
A^{\rho\b}A^{\alpha}_{\a} \partial^{\sigma}A_{\alpha \b},
\end{equation}
implying
\begin{equation}
\partial_{\mu}J^{\mu}_{\epsilon,1}=\mathcal{L}_{18} + 
\mathcal{L}_{23} - \mathcal{L}_{21}.
\end{equation}

All the above conditions are linearly independent. They allow us to write
Lagrangians $\mathcal{L}_{9}$, $\mathcal{L}_{10}- \mathcal{L}_{16}$,
$\mathcal{L}_{11}- \mathcal{L}_{15}$, $\mathcal{L}_{18}$,
$\mathcal{L}_{21}$, $\mathcal{L}_{22}$, $\mathcal{L}_{24}$, and
$\mathcal{L}_{25}$ as functions of the other Lagrangians. Note, however,
that one can always add these to other terms of the final basis for
simplification purposes.

Lastly, the current
$J^{\mu} = \epsilon^{\mu}{}_{\nu\rho\sigma} \partial^{\nu} A^{\alpha \a}
\partial^{\rho} A_{\alpha}^{\b} A^{\sigma \EAc} \epsilon_{\a\b\EAc}$
permits us to simplify one of the terms containing three first-order
derivatives by making use of
$\partial_{\mu} J^{\mu} = \mathcal{L}_3$.

\subsection{A New Basis}

One can now rewrite our basis of Lagrangians satisfying the Hessian
condition, taking into account the extra relations stemming from the
total derivatives and the identity of Ref. \cite{Fleury:2014qfa}. We
group our terms to produce a new and more convenient basis, and for that
purpose, we use the Abelian form
of the Faraday tensor, namely,
\begin{equation}
\label{PSU2EqFaradayAbelian}
G_{\mu\nu}^{\a} = \partial_\mu A_\nu^{\a} - \partial_\nu A_\mu^{\a},
\end{equation}
as well as its Hodge dual $\tilde{G}_{\mu\nu}^{\a}=\frac12 \epsilon_{\mu
\nu\rho\sigma} G^{\rho\sigma\a}$, also defined in the usual way. Using
the Abelian form of the Faraday tensor to describe a non-Abelian vector field theory
may seem a bit unusual, but it considerably simplifies our
forthcoming considerations since this term naturally appears from the
first-order derivatives of the vector field and cancels in the scalar
sector. We later move on to a formulation using the actual non-Abelian 
Faraday tensor, as is given by Eq.~(\ref{PSU2EqFaradayNonAbelian}).
We also make use of the symmetric counterpart of the Abelian
Faraday tensor, namely, 
\begin{equation}
S_{\mu\nu}^{\a} = \partial_{\mu}A_{\nu}^{\a} +
\partial_{\nu}A_{\mu}^{\a}.
\label{PSU2S}
\end{equation}

For one first-order derivative of the vector field and two additional vector fields, we obtain
\begin{equation}
\tilde{\mathcal{L}}_1= 2 \mathcal{L}_{1} = \epsilon_{\a \b
\EAc}\left[G^{\mu\nu \a} A^{\b}_\mu A^{\EAc}_\nu \right],
\end{equation}
and with four additional vector fields, we obtain
\begin{equation}
\left\{ \begin{array}{l}
\tilde{\mathcal{L}}_1=  2 \mathcal{L}_{1}
=\epsilon_{\a\b\EAc}\left[G^{\mu\nu \d} A^{\a}_\mu A^{\b}_\nu
\right]\left(A^{\EAc} \cdot A_{\d} \right),\\
\tilde{\mathcal{L}}_2= \mathcal{L}_{2} + \mathcal{L}_{3} = 
\epsilon_{\a\b\EAc}\left[S^{\mu\nu\a}A^{\b}_\mu A^{\d}_\nu
\right]\left(A^{\EAc} \cdot A_{\d} \right),\\
\tilde{\mathcal{L}}_3= \mathcal{L}_{3} - \mathcal{L}_{2} =
\epsilon_{\a\b\EAc}\left[ G^{\mu \nu \a} A^{\b}_{\mu} A^{\d}_{\nu}
\right]\left(A^{\EAc} \cdot A_{\d} \right) , \\
\tilde{\mathcal{L}}_4= \mathcal{L}_{4} =
\epsilon_{\a\b\EAc}\left[\tilde{G}^{\mu\nu\d} A^{\a}_\mu A^{\b}_\nu
\right]\left(A^{\EAc} \cdot A_{\d} \right),\\
\tilde{\mathcal{L}}_5= \mathcal{L}_{5} =
\epsilon_{\a\b\EAc}\left[\tilde{G}^{\mu\nu\a}  A^{\d}_{\mu} A^{\b}_{\nu}
\right]\left(A^{\EAc} \cdot A_{\d} \right),  \\
\tilde{\mathcal{L}}_6=\mathcal{L}_{6} - \mathcal{L}_{7} =
\epsilon_{\a\b\EAc}\left[\epsilon_{\mu \nu \rho \sigma} G^{\mu \alpha \d}
A_{\d}^\nu A^{\a\rho} A^{\b\sigma} A^{\EAc}_\alpha    \right].
\end{array} \right.
\end{equation}

Terms with two first-order derivatives and no additional fields can be written as
\begin{equation}
\mathcal{L}_1=2 \left( \mathcal{L}_{2} - \mathcal{L}_{3} \right) =
G^{\mu\nu}_{\a}G_{\mu\nu}^{\a},
\end{equation}
and with two additional fields, they are given by
\begin{equation}
\label{PSU2EqLagPostHessian}
\left\{
\begin{array}{l}
\tilde{\mathcal{L}}_1 =\mathcal{L}_{1} - \mathcal{L}_{5} = \delta^{\mu_1
\mu_2}_{\nu_1 \nu_2} A^{\lambda}_{\b} A^{\b}_{\lambda}
\left(\partial_{\mu_1} A^{\nu_1}_{\a} \right) \left(\partial_{\mu_2}
A^{\nu_2 \a} \right),\\
\tilde{\mathcal{L}}_2 =  2 \left(\mathcal{L}_{3} - \mathcal{L}_{5}
\right)=G^{\mu\nu}_{\a}G_{\mu\nu}^{\a} \left(A^{\b} \cdot A_{\b}
\right), \\
\tilde{\mathcal{L}}_3 = \mathcal{L}_{2} - \mathcal{L}_{6} =\delta^{\mu_1
\mu_2}_{\nu_1 \nu_2} A^{\lambda}_{\a} A_{\lambda\b}
\left(\partial_{\mu_1} A^{\nu_1\a} \right) \left(\partial_{\mu_2}
A^{\nu_2 \b} \right),\\
\tilde{\mathcal{L}}_4 = 2 \left(\mathcal{L}_{4} - \mathcal{L}_{6}
\right)=G^{\mu\nu}_{\a}G_{\mu\nu\b} \left(A^{\a} \cdot A^{\b} \right) \\
\tilde{\mathcal{L}}_5 = 2 \mathcal{L}_{7} =\tilde{G}_{\mu\nu}^{\a}
G^{\mu\nu}_{\a}\left(A^{\b} \cdot A_{\b} \right) ,\\
\tilde{\mathcal{L}}_6 =  2 \mathcal{L}_{8}  = \tilde{G}_{\mu\nu\a}
G^{\mu\nu}_{\b}\left(A^{\a} \cdot A^{\b} \right),\\
\tilde{\mathcal{L}}_{7} =\mathcal{L}_{18} + \mathcal{L}_{20} - 2
\mathcal{L}_{19} =   \left[ \epsilon_{\mu \nu \rho \sigma} A^{\a\mu}
A^{\b\nu}  G^{\rho \alpha}_{\a} G^\sigma_{\, \, \, \alpha \b}\right], \\
\tilde{\mathcal{L}}_{8} =  \mathcal{L}_{26} + \mathcal{L}_{23} =
\tilde{G}_{\mu\sigma}^{\b} A^{\mu}_{\a} A_{\alpha\b} S^{\alpha \sigma
\a} , \\
\tilde{\mathcal{L}}_{9} = \mathcal{L}_{26} - \mathcal{L}_{23} =
\tilde{G}_{\mu\sigma}^{\b}A^{\mu}_{\a} A_{\alpha\b}   G^{\alpha \sigma
\a} ,\\
\tilde{\mathcal{L}}_{10} = \mathcal{L}_{14} - \mathcal{L}_{27} =
\delta^{\mu_1 \mu_2}_{\nu_1\nu_2} A_{\mu_1}^{\a} A_{\mu_2}^{\b}
\left(\partial^{\mu_1}A^{\alpha}_{\a} \right) \left(\partial^{\mu_2}
A_{\alpha \b} \right),\\
\tilde{\mathcal{L}}_{11} = \mathcal{L}_{27} + \mathcal{L}_{28} - 2
\mathcal{L}_{17} = A^{\a}_\mu A^{\b}_\nu G^\mu_{\;\; \alpha \b}  G^{\nu
\alpha}_{\; \; \; \; \a} .
\end{array}
\right.
\end{equation}
As anticipated, we obtain 11
independent terms, which correspond to 28 terms to begin with, with 8
constraints and 9 Hessian conditions.

Finally, the three-gradient case yields
\begin{equation}
\left\{ \begin{array}{l}
\tilde{\mathcal{L}}_{1}= 2\left( \mathcal{L}_1 -3\mathcal{L}_2\right) = 
\epsilon_{\a\b\EAc} G^{\mu}{}_{\nu}{}^{\a} G^{\nu}{}_{\rho}{}^{\b}
G^{\rho}{}_{\mu}{}^{\EAc} ,\\
\tilde{\mathcal{L}}_{2} =  2 \mathcal{L}_4 -\mathcal{L}_3 -
\mathcal{L}_5 =   \epsilon_{\a\b\EAc} G^{\mu\alpha\a} G_{\alpha}{}^{\nu\b}
\tilde{G}_{\mu\nu}{}^{\EAc}.
\end{array} \right.
\end{equation}

\subsection{Scalar Contribution}
\label{PSU2PartScalarContribution}

Let us now consider the scalar part of the previously developed
Lagrangian, as explained in Sec.~\ref{PSU2PartProcedure}, making the
substitution $A_\mu^{\a} \rightarrow \partial_\mu\pi^{\a}$ and writing
only those terms that do not identically vanish, using the results of
the Appendix, where the useful Galileon
Lagrangians are provided (Sec.~\ref{PSU2PartAlternativeGalileonLag}), as
well as the linear combinations leading to second-order equations
(Sec.~\ref{PSU2PartResultLagGal}).

With one derivative and four vector fields, the only remaining term of
the scalar sector out of the original three is
\begin{equation}
\tilde{\mathcal{L}}_2 =  \epsilon_{\a\b\EAc} S^{\mu\nu\a}A^{\b}_\mu
A^{\d}_\nu \left(A^{\EAc} \cdot A_{\d} \right),
\label{PSU2limit}
\end{equation}
which does not yield second-order equations in the scalar limit. 

Lagrangians involving two derivatives of the vector fields provide
\begin{equation}
\left\{
\begin{array}{l}
\tilde{\mathcal{L}}_1 = \delta^{\mu_1 \mu_2}_{\nu_1 \nu_2}
A^{\lambda}_{\b} A^{\b}_{\lambda} \left(\partial_{\mu_1} A^{\nu_1}_{\a}
\right) \left(\partial_{\mu_2} A^{\nu_2 \a} \right),\\
\tilde{\mathcal{L}}_3 = \delta^{\mu_1 \mu_2}_{\nu_1 \nu_2}
A^{\lambda}_{\a} A_{\lambda\b} \left(\partial_{\mu_1} A^{\nu_1\a}
\right) \left(\partial_{\mu_2} A^{\nu_2 \b} \right),\\
\tilde{\mathcal{L}}_{10} = \delta^{\mu_1 \mu_2}_{\nu_1\nu_2}
A_{\mu_1}^{\a} A_{\mu_2}^{\b} \left(\partial^{\mu_1}A^{\alpha}_{\a}
\right) \left(\partial^{\mu_2} A_{\alpha \b} \right),
\end{array}
\right.
\end{equation}
leading to the corresponding scalar terms
\begin{equation}
\left\{
\begin{array}{l}
\left.\tilde{\mathcal{L}}_1\right|_{\pi}
=\mathcal{L}^{\text{Gal},3}_{4,\mathrm{I}},\\
\left.\tilde{\mathcal{L}}_3
\right|_{\pi}=\mathcal{L}^{\text{Gal},3}_{4,\mathrm{II}},\\
\left.\tilde{\mathcal{L}}_{10} \right|_{\pi} =
\mathcal{L}^{\text{Gal},2}_{4,\mathrm{II}} -
\mathcal{L}^{\text{Gal},2}_{4,\mathrm{III}}.
\end{array}
\right.
\end{equation}
One can derive two linear combinations having second-order equations,
namely,
\begin{equation}
 \left.\tilde{\mathcal{L}}_1\right|_{\pi} +2
 \left.\tilde{\mathcal{L}}_3\right|_{\pi} =
 \mathcal{L}^{\text{Gal},3}_{4,\mathrm{I}} +
 2\mathcal{L}^{\text{Gal},3}_{4,\mathrm{II}} \\ \end{equation}
 [see Eq.  \eqref{PSU2EqGal1}] and
 \begin{equation}
\left.\tilde{\mathcal{L}}_{10}\right|_{\pi} +
\left.\tilde{\mathcal{L}}_3\right|_{\pi}   =
\mathcal{L}^{\text{Gal},2}_{4,\mathrm{II}} - \frac12
\left(2\mathcal{L}^{\text{Gal},2}_{4,\mathrm{III}} +
\mathcal{L}^{\text{Gal},3}_{4,\mathrm{I}}  \right) + \frac12 \left(
\mathcal{L}^{\text{Gal},3}_{4,\mathrm{I}} +
2\mathcal{L}^{\text{Gal},3}_{4,\mathrm{II}}\right),
\label{PSU2L10pi}
\end{equation}
yielding second-order equations, as each of the three terms on the right-hand side
of Eq.  \eqref{PSU2L10pi} does so, as shown in the Appendix [see
Eqs.~\eqref{PSU2EqGal3}, \eqref{PSU2EqGal2} and \eqref{PSU2EqGal1}].


\subsection{Final Flat Spacetime Model}
\label{PSU2PartSwitchFaradayYM}

Let us regroup the results of the above sections to produce the final
theory in flat spacetime with the Minkowskian metric. We first gather
most of the new terms induced by the nonlinear contributions into an
arbitrary function $f(A_{\mu}^{\a}, G_{\mu\nu}^\a,
\tilde{G}_{\mu\nu}^\a)$. Indeed, this is possible because they not only
appear in the systematic procedure we have exposed, but they also
satisfy all our conditions; this is equivalent to the general proof
discussed in Ref.~\cite{Allys:2016jaq}, where the typical term is built
out of Levi-Civita tensors, necessarily inducing terms proportional to
$\epsilon^{00\dots}$ in the Hessian, and hence vanishing contributions.

Up to now, we have used the Abelian form
of the Faraday tensor to express the
relevant Lagrangians, although there can be situations in which working
with the non-Abelian counterpart in Eq. (\ref{PSU2EqFaradayNonAbelian}) can be more
convenient, in particular, in view of the fact that this is the relevant
tensor that appears naturally when one extends the theory to its gauged
version. This is quite simple since the arbitrary function
$f(A_{\mu}^{\a}, G_{\mu\nu}^\a, \tilde{G}_{\mu\nu}^\a)$ can be
equivalently written as a new function $\tilde{f}(A_{\mu}^{\a},
F_{\mu\nu}^{\a}, \tilde{F}_{\mu\nu}^\a)$ using
Eq.~(\ref{PSU2EqFaradayNonAbelian}). It is worth noting that such a change
of variable implies no other terms than those already included in the
original function.

Gathering the above considerations into a compact form, we obtain
a first generic term, reminiscent of the Abelian case, namely,
\begin{equation}
\mathcal{L}_2 = f(A_{\mu}^{\a}, G_{\mu\nu}^{\a}, \tilde{G}_{\mu\nu} ^{\a}) =
\tilde{f}(A_{\mu}^{\a}, F_{\mu\nu}^{\a}, \tilde{F}_{\mu\nu} ^{\a}).
\end{equation}

In addition to this term, all the remaining previously derived terms involving
contractions with up to six Lorentz indices are
\begin{equation}
\left\{
\begin{array}{l}
\hat{\mathcal{L}}_1 = \delta^{\mu_1 \mu_2}_{\nu_1 \nu_2}
A^{\lambda}_{\b} A^{\b}_{\lambda} \left(\partial_{\mu_1} A^{\nu_1}_{\a}
\right) \left(\partial_{\mu_2} A^{\nu_2 \a} \right) + 2  \delta^{\mu_1
\mu_2}_{\nu_1 \nu_2} A^{\lambda}_{\a} A_{\lambda\b}
\left(\partial_{\mu_1} A^{\nu_1\a} \right) \left(\partial_{\mu_2}
A^{\nu_2 \b} \right) , \\
\hat{\mathcal{L}}_2 =  \delta^{\mu_1 \mu_2}_{\nu_1 \nu_2}
A^{\lambda}_{\a} A_{\lambda\b} \left(\partial_{\mu_1} A^{\nu_1\a}
\right) \left(\partial_{\mu_2} A^{\nu_2 \b} \right) + \delta^{\mu_1
\mu_2}_{\nu_1\nu_2} A_{\mu_1}^{\a} A_{\mu_2}^{\b}
\left(\partial^{\mu_1}A^{\alpha}_{\a} \right) \left(\partial^{\mu_2}
A_{\alpha \b} \right),\\
\hat{\mathcal{L}}_3 = \tilde{G}_{\mu\sigma}^{\b}A^\mu_{\a} A_{\alpha\b} 
 S^{\alpha\sigma \a},\\
\end{array}
\right.
\end{equation}
the first two actually being equivalent in the pure scalar sector since they
lead to the same equations of motion, \emph{i.e.}, those stemming from the
Galileon Lagrangian containing four scalar fields in the ${\bm 3}$
representation of SU(2). Note that there is no term containing only one
gradient.

With this general basis, which we expand upon in the final
discussion section, we can now turn to the covariantization required to
apply this category of theories to cosmologically relevant situations.

\section{Covariantization}  \label{PSU2covproc}
\subsection{Procedure}

Below we follow a procedure similar to that proposed for the Galileon
case~\cite{Deffayet:2009mn,Deffayet:2009wt,deRham:2011by}, the
generalized Proca model
\cite{Jimenez:2013qsa,Gleyzes:2014dya,Heisenberg:2014rta,Allys:2015sht},
and the
multi-Galileon situation~\cite{Deffayet:2010zh,Padilla:2012dx,Sivanesan:2013tba}.
The principle is simple: one first transforms all partial derivatives into
covariant ones and then checks that only those terms leading to at most second-order
equations of motion are kept.

The pure vector part now contains $A$ and $\nabla A$ terms, which
translate into $A$, $\partial A$, $g$ and $\partial g$ terms. None of
these terms could lead to any derivative of order higher than two in the
equations of motion. On the other hand, the Faraday tensor terms do not
yield metric derivatives since partial derivatives can be replaced by
covariant ones by virtue of the antisymmetry of these terms. We also
leave these terms aside.

As for the scalar part, derivatives of order three or more could appear
for the curvature. To fix this potential problem, we write the equations of
motion in terms of covariant derivatives and commute them in order to
generate the curvature tensor, which contains only second-order
derivatives of the metric: the problem is with the derivatives of
the curvature terms. As these particular contributions stem from terms
implying at least fourth-order derivatives of the scalar field, it is
easy to identify them and to write down the required counterterms.

In practice, this does not show that the resulting equations of motion of
the metric do not involve higher-order derivatives of the scalar field.
We merely apply the results of Ref.~\cite{Jimenez:2013qsa}, where it was
shown that if the equations of motion for the scalar field are safe,
then so are those for the metric. This result translates directly to
our case.

For many of the terms discussed below, it turns out to be
easier to write the Lagrangian as a function of the vector
field rather than of its scalar part, even though we are ultimately
interested in the latter. Indeed, the scalar Euler-Lagrange
equation
\begin{equation}
0=\frac{\partial\mathcal{L}}{\partial \pi_{\al}} - \nabla_\nu
\frac{\partial \mathcal{L}}{\partial (\nabla_\nu \pi_{\al})} +
\nabla_\nu \nabla_\mu \frac{\partial \mathcal{L}}{\partial (\nabla_\mu
\nabla_\nu \pi_{\al})} 
\end{equation}
can be written as
\begin{equation}
0=- \nabla_\nu \frac{\partial \mathcal{L}}{\partial (\nabla_\nu
\pi_{\al})} + \nabla_\nu \nabla_\mu \frac{\partial \mathcal{L}}{\partial
(\nabla_\mu \nabla_\nu \pi_{\al})} = - \nabla_\nu \left( \frac{\partial
\mathcal{L}}{\partial A_{\nu\al}} -  \nabla_\mu
\frac{\partial\mathcal{L}}{\partial (\nabla_\mu A_{\nu\al})} \right)
\end{equation}
since the action is assumed to be local in $A_\mu$ and therefore cannot
contain terms involving nonderivative functions of the scalar field
$\pi$.

In the following sections, we write those terms containing only the
curvature and its derivative, or only its derivative, by the respective
notation $\left.\mathcal{F}\right|_{R}$ or
$\left.\mathcal{F}\right|_{\nabla R}$, where $\mathcal{F}$ is the term
whose restriction is being considered. We concentrate on terms
which are nonvanishing in the scalar sector only.

\subsection{Terms in $\mathcal{L}^{\text{Gal}}$}

The Lagrangians we consider give, in the scalar sector, 
\begin{equation}
\left\{
\begin{array}{l}
\left.\hat{\mathcal{L}}_1\right|_{\pi}  = 
\mathcal{L}^{\text{Gal},3}_{4,\mathrm{I}} + 2
\mathcal{L}^{\text{Gal},3}_{4,\mathrm{II}} , \\
\left.\hat{\mathcal{L}}_2\right|_{\pi}  =
\mathcal{L}^{\text{Gal},3}_{4,\mathrm{II}} +
\mathcal{L}^{\text{Gal},2}_{4,\mathrm{II}}  -
\mathcal{L}^{\text{Gal},2}_{4,\mathrm{III}},
\end{array}
\right.
\label{PSU2EqLagFinGal}
\end{equation}
where we use the Galileon Lagrangians of the Appendix. In the following, working in the vector sector,
we substitute $\partial_\mu\pi^\a\to A_\mu^\a$.
Equation~\eqref{PSU2EqLagFinGal} implies that only three independent
counterterms are needed, \emph{i.e.}, those associated with
$\mathcal{L}^{\text{Gal},3}_{4,\mathrm{I}}$,
$\mathcal{L}^{\text{Gal},3}_{4,\mathrm{II}}$ and
$(\mathcal{L}^{\text{Gal},2}_{4,\mathrm{II}}  -
\mathcal{L}^{\text{Gal},2}_{4,\mathrm{III}})$. We now proceed to find
these counterterms.

First, we have
\begin{equation}
\left.\left\{ \nabla_{\nu} \nabla_{\mu} \left[\frac{\partial
\mathcal{L}^{\text{Gal},3}_{4,\mathrm{I}}}{\partial \left( \nabla_{\mu}
A_{\nu\al} \right)}\right]\right\}\right\vert_{R} = - 2 A^{\lambda}_{\b}
A^{\b}_{\lambda}  R_{\mu\nu} \nabla^{\nu}A^{\mu\al} - 2 A_{\b}^{\lambda}
A^{\b}_{\lambda} A^{\mu\al} \nabla^{\nu}R_{\mu\nu}.
\end{equation}
Introducing
\begin{equation}
\mathcal{L}^{\text{Gal},3}_{4,\mathrm{I},\mathrm{CT}} = \frac{1}{4}
A^{\lambda}_{\b} A ^{\b}_{\lambda} A^{\mu}_{\a} A^{\a}_{\mu} R,
\end{equation}
we find that
\begin{equation}
\left.\left\{ \nabla_{\nu} \left[\frac{\partial
\mathcal{L}^{\text{Gal},3}_{4,\mathrm{I},\mathrm{CT}}}{\partial \left(
A_{\nu\al} \right)} \right] \right\}\right|_{\nabla R}= 
A_{\b}^{\lambda}A^{\b}_{\lambda} A^{\mu\al} \nabla^{\nu}\left(g_{\mu\nu}
R \right),
\end{equation}
which finally implies the equation of motion (EOM)
\begin{equation}
EOM_\pi\left.\left(\mathcal{L}^{\text{Gal},3}_{4,\mathrm{I}} +
\mathcal{L}^{\text{Gal},3}_{4,\mathrm{I},CT}  \right)\right|_{\nabla R}
=- 2A_{\b}^{\lambda}A^{\b}_{\lambda} A^{\mu\alpha}
\nabla^{\nu}\left(R_{\mu\nu} - \frac{1}{2}g_{\mu\nu} R \right) = 0,
\end{equation}
vanishing by virtue of the properties of the Einstein tensor.

Similarly, for $\mathcal{L}^{\text{Gal},3}_{4,\mathrm{II}}$, we have
\begin{equation}
\left.\left\{\nabla_{\nu} \nabla_{\mu} \left[\frac{\partial
\mathcal{L}^{\text{Gal},3}_{4,\mathrm{II}}}{\partial \left( \nabla_{\mu}
A_{\nu\al} \right)} \right] \right\}\right|_{R} = - 2 A^{\lambda}_{\b}
A^{\al}_{\lambda}  R_{\mu\nu}\nabla^{\nu}A^{\mu\b} - 2 A_{\b}^{\lambda}
A^{\al}_{\lambda} A^{\mu\b} \nabla^{\nu}R_{\mu\nu}.
\end{equation}
Introducing
\begin{equation}
\mathcal{L}^{\text{Gal},3}_{4,\mathrm{II},\mathrm{CT}} = \frac{1}{4}
A^{\lambda}_{\b} A _{\lambda\a} A^{\mu\b} A^{\a}_{\mu} R,
\end{equation}
which verifies
\begin{equation}
\left.\left\{ \nabla_{\nu} \left[\frac{\partial
\mathcal{L}^{\text{Gal},3}_{4,\mathrm{II},\mathrm{CT}}}{\partial \left( 
A_{\nu\al} \right)} \right] \right\}\right|_{\nabla R}  = 
A_{\b}^{\lambda}A^{\al}_{\lambda} A^{\mu\b} \nabla^{\nu}\left(g_{\mu\nu}
R \right),
\end{equation}
we obtain
\begin{equation}
EOM_\pi\left.\left(\mathcal{L}^{\text{Gal},3}_{4,\mathrm{II}} +
\mathcal{L}^{\text{Gal},3}_{4,\mathrm{II},CT}  \right)\right|_{\nabla
R}=- 2A_{\b}^{\lambda}A^{\al}_{\lambda} A^{\mu\b}
\nabla^{\nu}\left(R_{\mu\nu} - \frac{1}{2}g_{\mu\nu} R \right) = 0.
\end{equation}

Finally, using the previous notation
\begin{equation}
\tilde{\mathcal{L}}_{10} =\mathcal{L}^{\text{Gal},2}_{4,\mathrm{II}} -
\mathcal{L}^{\text{Gal},2}_{4,\mathrm{III}},
\end{equation}
we have
\begin{equation}
\left.\left\{\nabla_{\nu} \nabla_{\mu} \left[\frac{\partial
\tilde{\mathcal{L}}_{10} }{\partial \left( \nabla_{\mu} A_{\nu\al}
\right)} \right] \right\}\right|_{R} =-2 A^{\mu\al}A^{\lambda\b}
R^{\nu}{}_{\rho\lambda\mu} \nabla_\nu A^{\rho}_{\b}-2
A^{\mu\al}A^{\lambda\b}A^{\rho}_{\b}\nabla_\nu R^{\nu}{}_{\rho\lambda\mu}.
\end{equation}
We introduce the counterterm
\begin{equation}
\mathcal{L}_{10,\mathrm{CT}} = -\frac12 A^{\mu\a} A^{\nu \b}
A^{\rho}_{\a} A^{\sigma}_{\b} R_{\mu\nu\rho \sigma},
\end{equation}
giving
\begin{equation}
\nabla_{\rho}\left. \left( \frac{\partial
\mathcal{L}_{10,\mathrm{CT}}}{\partial \left(A_{\rho \al} \right)}
\right)\right|_{\nabla R} = -2 A^{\mu\a} A_{\lambda \a} A^{\rho \al}
\nabla^{\nu} R^{\lambda}{}_{\rho \mu \nu} = 2 A^{\mu\al}
A^{\lambda\b}A^{\rho}_{\b} \nabla_\nu R^{\nu}{}_{\rho\lambda\mu},
\end{equation}
which, as expected, results in
\begin{equation}
EOM_\pi\left.\left(\tilde{\mathcal{L}}_{10}  +
\mathcal{L}_{10,\mathrm{CT}} \right)\right|_{\nabla R}= 0.
\end{equation}

Then, to obtain the covariantized form of the action, it is sufficient
to add the counterterms obtained in this part to the action given
previously in flat spacetime. The result is summarized in
Sec.~\ref{PSU2FinalModel}.

\subsection{Coupling with Curvature}

Once the derivatives have been covariantized, one must also include
possible direct coupling terms between the vector field and the
curvature tensors, which we do below in a way entirely similar to that
of Ref.~\cite{Allys:2015sht}. First, we demand contractions with tensors
whose divergences vanish on all indices (to ensure that integration by parts provides no
higher-order contributions in the equations of motion)
\cite{deRham:2011by,Jimenez:2013qsa}: this means the Einstein tensor as
well as
\begin{equation}
L_{\mu\nu\rho\sigma}=2R_{\mu\nu\rho\sigma} + 2(R_{\mu\sigma}g_{\rho\nu}
+ R_{\rho\nu}g_{\mu\sigma} - R_{\mu\rho}g_{\nu\sigma} -
R_{\nu\sigma}g_{\mu\rho}) + R(g_{\mu\rho}g_{\nu\sigma} -
g_{\mu\sigma}g_{\rho\nu}),
\end{equation}
whose symmetries are those of the Riemann tensor, to which it is dual in
the sense that it can be written as
\begin{equation}
L^{\alpha\beta\gamma\delta}= - \frac12
\epsilon^{\alpha\beta\mu\nu}\epsilon^{\gamma\delta
\rho\sigma}R_{\mu\nu\rho\sigma}.
\end{equation}

Even limiting ourselves to the same number of fields as in the flat
spacetime situation, many terms are \emph{a priori} possible. To begin with,
all contractions involving a single vector field are impossible. With two
such fields, the reasoning is exactly equivalent to the Abelian case,
which means the Lagrangians
\begin{equation}
\mathcal{L}^\mathrm{curv}_{1} = G_{\mu\nu}A^{\mu\a}A^{\nu}_{\a}
\end{equation}
and
\begin{equation}
\mathcal{L}^\mathrm{curv}_{2} = L_{\mu\nu\rho\sigma} G^{\mu\nu\a}
G^{\rho\sigma}_{\a}
\end{equation}
are acceptable.

Terms in which at least one of the Abelian-like Faraday tensors is replaced by its Hodge
dual can always be rewritten as a contraction between the Riemann tensor
and two Abelian-like Faraday tensors, which cannot give second-order equations of
motion~\cite{Jimenez:2013qsa}. One could envisage a contraction with
a term like $G^{\mu\rho\a} G^{\nu\sigma}_{\a}$, but which is proportional
to $\mathcal{L}^\mathrm{curv}_{2}$: to show this, one needs to use the
following identity,
\begin{equation}
\epsilon^{\alpha\beta\gamma\delta} \epsilon^{\rho\sigma\mu\nu} -
\epsilon^{\alpha\rho\sigma\mu} \epsilon^{\beta\gamma\delta\nu} +
\epsilon^{\alpha\gamma\delta\nu} \epsilon^{\beta\rho\sigma\mu} +
\epsilon^{\alpha\beta\delta\nu} \epsilon^{\rho\gamma\sigma\mu} -
\epsilon^{\alpha\beta\gamma\nu} \epsilon^{\rho\delta\sigma\mu}=0,
\end{equation}
and the first Bianchi identity.

With three fields, one can obtain a new nonvanishing term, in
contrast to the Abelian case. This is mostly due to the fact
that it is possible to have an antisymmetry in the exchange of two
underived vector fields. We get
\begin{equation}
\mathcal{L}^\mathrm{curv}_{3} = L_{\mu\nu\rho\sigma} \epsilon_{\a\b\EAc}
G^{\mu \nu \a} A^{\rho\b} A^{\sigma\EAc},
\end{equation}
which is shown to be proportional to $L_{\mu\nu\rho\sigma}
\epsilon_{\a\b\EAc} G^{\mu\rho\a} A^{\nu\b} A^{\sigma\EAc}$, by making use
of the previous identity on the Levi-Civita tensor.

Four fields provide, again in contrast to the Abelian situation,
the extra contribution
\begin{equation}
\mathcal{L}^\mathrm{curv}_{4} = L_{\mu\nu\rho\sigma}  A^{\mu\a} A^{\nu
\b} A^{\rho}_{\a}A^{\sigma}_{\b}.
\end{equation}
It is worth noticing at this point that it is possible to go from the
expression of $\mathcal{L}^\mathrm{curv}_{2}$ and
$\mathcal{L}^\mathrm{curv}_{3}$ using $G_{\mu\nu}^\a$ (the Abelian form of the
Faraday tensor) to that using $F_{\mu\nu}^\a$ (the non-Abelian one),
both of which are equal in an Abelian theory: it is sufficient for
this purpose to include the terms $\mathcal{L}^\mathrm{curv}_{3}$ and
$\mathcal{L}^\mathrm{curv}_{4}$ only (they are generated by the
transformation from $G_{\mu\nu}^\a$ to $F_{\mu\nu}^\a$).

\section{Final model, discussion}  \label{PSU2FinalModel}

Let us summarize the results obtained for the generalized SU(2) Proca
theory. First, we showed that any function of the vector field, Faraday
tensor, and its Hodge dual (either in their Abelian or non-Abelian
formulation) was possible, \emph{i.e.},
\begin{equation}
\mathcal{L}_2 = f(A_{\mu}^{\a}, G_{\mu\nu}^\a, \tilde{G}_{\mu\nu} ^\a) =
\tilde{f}(A_{\mu}^{\a}, F_{\mu\nu}^\a, \tilde{F}_{\mu\nu} ^\a).
\end{equation}
Such a  general $\mathcal{L}_2$ term involving only gauge-invariant
quantities for the derivatives is also present in the Abelian case; we
will not discuss it any further since it appears similarly (and for the
same reasons) in both the Abelian and non-Abelian theories.

Before presenting the other terms contained in the non-Abelian action,
let us pursue the summary of what was found for its Abelian
counterpart, as worked out in
Refs.~\cite{Heisenberg:2014rta,Allys:2015sht,Jimenez:2016isa,Allys:2016jaq}; 
as usual, we denote $\mathcal{L}_{n+2}$ the Lagrangians containing $n\geq
1$ first-order derivatives of the vector field.
First, the relation between the more general scalar and vector theories, \emph{i.e.},
the Galileon and generalized Proca models, provide, in this case, a deeper
understanding through the use of the Stückelberg trick to go from one sector to
another (\emph{i.e.}, switching between $\partial_\mu \pi$ and $A_\mu$). In the scalar
Galileon theory, only one term exists in the Lagrangians $\mathcal{L}_3$ to
$\mathcal{L}_5$, each of which generates a contribution to the vector
sector by the Stückelberg trick, \emph{i.e.}, those with a prefactor $f_i(X)$ in the
conclusion of Ref.~\cite{Allys:2016jaq}. An additional freedom
stems from the fact that a given scalar Lagrangian can give different
vector Lagrangians when permuting the second-order derivatives before
introducing the vector field: although $\partial_\mu\partial_\nu \pi =
\partial_\nu\partial_\mu \pi$, this symmetry is absent in the pure
vector case since $\partial_\mu A_\nu \not= \partial_\nu A_\mu$. This
property led to one additional contribution to the vector sector of each
$\mathcal{L}_4$ to $\mathcal{L}_6$. These contributions appear with the 
prefactor $g_i(X)$ in Ref.~\cite{Allys:2016jaq}; they vanish in the
pure scalar sector.

Coming back to the non-Abelian situation, and in addition to
$\mathcal{L}_2$, we derived those relevant Lagrangians implying up to 6
contracted Lorentz indices and being nontrivial in flat spacetime.
Contrary to the Abelian case, we found no such Lagrangian for $n=1$. For
$n=2$, there are three possible terms; \emph{i.e.}, $\mathcal{L}_4$ contains
\begin{equation}
\left\{
\begin{array}{l}
\mathcal{L}_4^1 = \delta^{\mu_1 \mu_2}_{\nu_1 \nu_2} A^{\lambda}_{\b}
A^{\b}_{\lambda} \left(\nabla_{\mu_1} A^{\nu_1}_{\a} \right)
\left(\nabla_{\mu_2} A^{\nu_2 \a} \right) + \frac14 A^{\lambda}_{\b} A
^{\b}_{\lambda} A^{\mu}_{\a} A^{\a}_{\mu} R \\
~~~~~~~ + 2 \delta^{\mu_1 \mu_2}_{\nu_1 \nu_2} A^{\lambda}_{\a} A_{\lambda\b} \left(\nabla_{\mu_1} A^{\nu_1\a} \right) \left(\nabla_{\mu_2} A^{\nu_2 \b} \right) + \frac12 A^{\lambda}_{\b} A _{\lambda\a} A^{\mu\b} A^{\a}_{\mu} R , \\
\mathcal{L}_4^2 = \delta^{\mu_1 \mu_2}_{\nu_1 \nu_2} A^{\lambda}_{\a} A_{\lambda\b} \left(\nabla_{\mu_1} A^{\nu_1\a} \right) \left(\nabla_{\mu_2} A^{\nu_2 \b} \right) + \frac{1}{4} A^{\lambda}_{\b} A _{\lambda\a} A^{\mu\b} A^{\a}_{\mu} R \\
~~~~~~~ + \delta^{\mu_1 \mu_2}_{\nu_1\nu_2} A_{\mu_1}^{\a} A_{\mu_2}^{\b} \left(\nabla^{\nu_1}A^{\alpha}_{\a} \right) \left(\nabla^{\nu_2} A_{\alpha \b} \right) -\frac12 A^{\mu\a} A^{\nu \b} A^{\rho}_{\a} A^{\sigma}_{\b} R_{\mu\nu\rho \sigma},\\
\mathcal{L}_4^3 = \tilde{G}_{\mu\sigma}^{\b}A^\mu_{\a} A_{\alpha\b}  S^{\alpha\sigma \a},\\
\end{array}
\right.
\end{equation}
the first two terms giving, once developed, the following forms:
\begin{equation}
\left\{
\begin{array}{l}
\mathcal{L}_4^1 = (A_{\b} \cdot A^{\b}) \left[ \left(\nabla\cdot A_\a \right)\left(\nabla\cdot A^\a \right) - (\nabla_\mu A^\nu_{\a})(\nabla^\mu A_\nu^{\a}) +\frac14
A_{\a} \cdot A^{\a} R \right] \\ 
\hspace{1cm} + 2 (A_{\a} \cdot A_{\b}) \left[\left(\nabla\cdot A^\a \right)\left(\nabla\cdot A^\b \right) - (\nabla_\mu A^{\nu \a})(\nabla^\mu A_\nu^{\b}) +\frac12 A^{\a}
\cdot A^{\b} R\right], \\ 
\mathcal{L}_4^2 = (A_{\a} \cdot A_{\b}) \left[\left(\nabla\cdot A^\a \right)\left(\nabla\cdot A^\b \right) - (\nabla_\mu A^{\nu \a})(\nabla^\mu A_\nu^{\b}) +\frac14 A^{\a}
\cdot A^{\b} R\right] \\
\hspace{1cm} + (A^{\mu \a} A^{\nu\b}) \left[\left(\nabla_\mu A^\alpha_\a \right)\left(\nabla_\nu A_{\alpha\b} \right) - \left(\nabla_\nu A^\alpha_\a \right)\left(\nabla_\mu A_{\alpha\b} \right) 
-\frac12 A^\rho_{\b} A^{\sigma \b} 
R_{\mu\nu\rho\sigma}\right],
\end{array}
\right.
\end{equation}
which are more easily compared with the equivalent results for the Abelian case.
Finally, we also found four extra possibilities for the Lagrangians,
implying a coupling with the curvature
\begin{equation}
\begin{array}{l}
\mathcal{L}^\mathrm{curv}_{1} = G_{\mu\nu}A^{\mu\a}A^{\nu}_{\a},\\
\mathcal{L}^\mathrm{curv}_{2} = L_{\mu\nu\rho\sigma} F^{\mu\nu}_{\a}
F_{\mu\nu}^{\a},\\
\mathcal{L}^\mathrm{curv}_{3} = L_{\mu\nu\rho\sigma} \epsilon_{\a\b\EAc}
F^{\mu \nu \a} A^{\rho\b} A^{\sigma\EAc},\\
\mathcal{L}^\mathrm{curv}_{4} = L_{\mu\nu\rho\sigma} A^{\mu\a} A^{\nu
\b} A^{\rho}_{\a}A^{\sigma}_{\b},
\end{array}
\end{equation}
thereby completing the full action at that order.

Let us first consider the actions whose equations of motion involve only
second-order derivatives for the scalar (not first-order ones), which is
equivalent to having only two vector fields together with the relevant 
gradients in the action. The
multi-Galileon SU(2) model in the adjoint representation has been
considered in \cite{Padilla:2010ir}, where it was shown that building a
Lagrangian is only possible at the order of $\mathcal{L}_4$ (not to
mention the order $\mathcal{L}_2$ already discussed above). The
equivalent formulations of this Lagrangian are detailed in
Appendix~\ref{PSU2AppendixGalileon}. Following the previous considerations,
no Lagrangian in the vector sector should appear at the order of
$\mathcal{L}_3$ since there is no such associated Lagrangian for the
multi-Galileon at that order; we explicitly confirmed this expectation.
In addition, two Lagrangians should appear at the order of
$\mathcal{L}_4$, one associated with the multi-Galileon dynamics and one
associated with the commutation of second-order derivatives of the scalar
field. In fact, three Lagrangians have been found, two of them giving
the multi-Galileon dynamics in the scalar sector. We then interpret
these two previous terms as contributions which are equivalent in the
scalar case but not in the vector case. The fact that there are two nonvanishing
Lagrangians in the scalar sector is also due to
a commutation of the second-order derivatives of the scalar fields but
in a current term, which implies that it is not possible to describe
this commutation with a Lagrangian vanishing in the pure scalar
sector. This additional term is specific to the non-Abelian case:
the term in $\delta^{\mu_1 \mu_2}_{\nu_1\nu_2} A_{\mu_1}^{\a}
A_{\mu_2}^{\b} \left(\nabla^{\nu_1}A^{\alpha}_{\a} \right)
\left(\nabla^{\nu_2} A_{\alpha \b} \right)$ vanishes in the Abelian
case, while $\mathcal{L}_4^1$ and $\mathcal{L}_4^2$ both reduce to
$\mathcal{L}_4^{\text{Abelian}} = \delta^{\mu_1 \mu_2}_{\nu_1 \nu_2}
A^{\lambda}A_{\lambda} \left(\nabla_{\mu_1} A^{\nu_1}\right)
\left(\nabla_{\mu_2} A^{\nu_2} \right)$.

To go further, let us first consider terms implying more derivatives,
\emph{i.e.}, having $n\geq 3$. At the order of $\mathcal{L}_5$, and since there
is no possible dynamics for the SU(2) adjoint multi-Galileon, we expect that
no term having a nonvanishing pure scalar contribution is possible.
This suggests that the only possible term is
\begin{equation}
\mathcal{L}_5 = \epsilon_{\a\b\EAc} \left(A^\a \cdot A^\d\right)
\tilde{G}^{\alpha\mu}_\d\tilde{G}^{\beta}{}_\mu^\b S_{\alpha\beta}^\EAc,
\end{equation}
with the other SU(2) index contractions giving a vanishing result. At the
order of $\mathcal{L}_6$, the only possibility seems to be the
independent possible contractions of SU(2) indices on
$\mathcal{L}_6^{\text{Abelian}}=\left(A\cdot A
\right)\tilde{G}^{\alpha\beta}\tilde{G}^{\mu\nu}
S_{\alpha\mu}S_{\beta\nu}$, since there is no possibility of having a term that
does not vanish in the pure scalar sector. However, one
should verify that there is no other term vanishing in the pure scalar
sector, not included in $\mathcal{L}_2$, and whose dynamics is not
described by the previous ones. This kind of terms would be
specific to a non-Abelian theory, as is the second term of $\mathcal{L}_4^2$, 
and they would vanish for a vector field in a trivial group representation.

Concerning the Lagrangians with more than two vector fields together
with the relevant gradients, one has to pay attention to the fact that
fully factorizing an $f\left(A_\mu^\a\right)$ as in the Abelian case
is not guaranteed to lead to a valid procedure, although factorizing
such an arbitrary function in front of any valid contribution also leads
to another valid contribution. In addition, one could think that if there
is no valid Lagrangian with only a few nongradient vector fields at
a given derivative order, it is fairly probable that there is also
no such valid Lagrangian at all at this order. For instance, we showed
explicitly that terms at the order of $\mathcal{L}_3$ are not possible
with up to $4$ vector fields, and this questions the possibility of having
such a term even with a higher number of vector fields. An interesting
point is that if a Lagrangian is allowed which does not vanish in the
pure scalar sector, it corresponds to a possible term in the
multi-Galileon action, which shows that both theories are closely
related.

To conclude, this discussion showed that even if the full action of the
model has not been obtained yet, discussing the low order terms permits us
to identify and understand the whole Lagrangian structure. The above
discussion is not specific to the SU(2) case and therefore can be
extended to other group representations. For a theory with a vector
field transforming under any representation of any group, a systematic
study of all possible terms in the action should be performed in
parallel with the corresponding multi-Galileon theory.

\subsection*{Acknowledgments}  This work was supported by COLCIENCIAS - ECOS NORD grant number RC 0899-2012 with the help of ICETEX, and by
COLCIENCIAS grant numbers 110656933958 RC 0384-2013 and 123365843539 RC
FP44842-081-2014. P.P. would like to thank the Labex Institut Lagrange
de Paris (reference ANR-10-LABX-63) part of the Idex SUPER, within which part of
this work has been completed.

\section{Appendix: SU(2) Galileon Lagrangian Equivalent Formulations}
\label{PSU2AppendixGalileon}
\subsection{Introduction}

The purpose of this appendix is to write explicitly all the Lagrangians
describing the multi-Galileon dynamics in the 3-dimensional
representation of SU(2), focusing on the Lagrangians containing only four
Galileon fields, \emph{i.e.}, those which are useful in this article. A
Lagrangian describing this dynamics is given in Ref. \cite{Padilla:2012dx},
namely,
\begin{equation}
\label{PSU2EqLagIniPSZ}
\mathcal{L}^{\pi}_{m} = \alpha^{i_1 \cdots i_m} \delta^{\mu_2 \cdots
\mu_m}_{\left[ \nu_2\cdots \nu_m\right]} \pi_{i_1}
\partial_{\mu_2}\partial^{\nu_2}\pi_{i_2}\cdots
\partial_{\mu_m}\partial^{\nu_m}\pi_{i_m},
\end{equation}
with $m$ running from 1 to 5, and with the notation 
\begin{equation}
\frac{1}{(D-n)!}\epsilon^{i_1\cdots i_n \sigma_1 \cdots \sigma_{D-n}}
\epsilon_{j_1\cdots j_n \sigma_1 \cdots \sigma_{D-n}} =  n! \delta^{j_1
\cdots j_n}_{[i_1 \cdots i_n]} = \delta^{j_1 \cdots j_n}_{i_1 \cdots
i_n} = \delta^{j_1}_{i_1} \cdots \delta^{j_n}_{i_n} \pm \cdots,
\label{PSU2deltaMult}
\end{equation}
for $n$ running from 1 to 4 (in a four-dimensional spacetime). Other
equivalent formulations are possible, which is the purpose of
this appendix.

This investigation is necessary for two reasons. First, the
formulation given in Eq. \eqref{PSU2EqLagIniPSZ} cannot be obtained from a
vector Lagrangian using the switch $A_{\mu}^{\a} \rightarrow
\partial_{\mu} \pi^{\a}$ since a scalar field without derivatives is
present. Second, if different Lagrangians are equivalent in the scalar
sector, they could give Lagrangians that are not equivalent in the vector sector.
We thus expect that different Lagrangians valid in the vector sector
become different but equivalent formulations of the multi-Galileon
dynamics when considering the pure scalar part of the action.

For this purpose, we use the results of Ref.~\cite{Deffayet:2011gz},
which describe equivalent formulations of the Galileon theory in the Abelian
case, introducing a Lagrangian similar to that in Eq.
\eqref{PSU2EqLagIniPSZ}, together with the following Lagrangians:
\begin{equation}
\label{PSU2EqGalAbelien1}
\mathcal{L}^{\text{Gal},1}_{m}= \delta^{\mu_1 \cdots \mu_{m-1}}_{\left[
\nu_1\cdots \nu_{m-1}\right]} \partial_{\mu_{1}}\pi
\partial^{\nu_{1}}\pi \partial_{\mu_2}\partial^{\nu_2} \pi \cdots
\partial_{\mu_{m-1}}\partial^{\nu_{m-1}} \pi,
\end{equation}
\begin{equation}
\mathcal{L}^{\text{Gal},2}_{m}= \delta^{\mu_1 \cdots \mu_{m-2}}_{\left[
\nu_1\cdots \nu_{m-2}\right]} \partial_{\mu_{1}}\pi
\partial_{\lambda}\pi \partial^{\nu_1}\partial^{\lambda} \pi \cdots
\partial_{\mu_{m-2}}\partial^{\nu_{m-2}} \pi,
\end{equation}
\begin{equation}
\label{PSU2EqGalAbelien3}
\mathcal{L}^{\text{Gal},3}_{m}= \delta^{\mu_1 \cdots \mu_{m-2}}_{\left[
\nu_1\cdots \nu_{m-2}\right]} \partial_{\lambda}\pi
\partial^{\lambda}\pi \partial_{\mu_1}\partial^{\nu_1} \pi \cdots
\partial_{\mu_{m-2}}\partial^{\nu_{m-2}} \pi,
\end{equation}
for $m\geq2$, the case $m=1$ giving $\mathcal{L}=\pi$. These Lagrangians all give
second-order equations of motion.

\subsection{Lagrangians}
\label{PSU2PartAlternativeGalileonLag}

We first write all possible Lagrangians appearing when we add the
group indices to the previous Lagrangians, restricting ourselves to the
case $m=4$. They are more numerous than in the multi-Galileon case since we
have an additional freedom when choosing the group index contractions.

The only possible Lagrangian associated with the formulation of
Ref.~\cite{Padilla:2012dx} is
\begin{equation}
\mathcal{L}^{\text{PSZ}}_{4} = \delta^{\mu_1 \cdots \mu_3}_{\nu_1 \cdots
\nu_3} \pi_{\a} \partial_{\mu_1} \partial^{\nu_1} \pi^{\a}
\partial_{\mu_2} \partial^{\nu_2} \pi_{\b} \partial_{\mu_3}
\partial^{\nu_3} \pi^{\b} .
\end{equation}
The Lagrangians appearing in Ref.~\cite{Deffayet:2011gz}, given in
Eqs.~\eqref{PSU2EqGalAbelien1} to~\eqref{PSU2EqGalAbelien3}, can be endowed with SU(2)
indices in several ways, namely, two possibilities for
$\mathcal{L}^{\mathrm{Gal},1}_{4}$:
\begin{equation}
\mathcal{L}^{\text{Gal},1}_{4,\mathrm{I}} = \delta^{\mu_1 \cdots
\mu_3}_{\nu_1 \cdots \nu_3} \partial _{\mu_1}\pi_{\a} \partial^{\nu_1}
\pi^{\a} \partial_{\mu_2} \partial^{\nu_2} \pi_{\b} \partial_{\mu_3}
\partial^{\nu_3} \pi^{\b}
\end{equation}
and
\begin{equation}
\mathcal{L}^{\text{Gal},1}_{4,\mathrm{II}} = \delta^{\mu_1 \cdots
\mu_3}_{\nu_1 \cdots \nu_3} \partial _{\mu_1}\pi_{\a} \partial^{\nu_1}
\pi_{\b} \partial_{\mu_2} \partial^{\nu_2} \pi^{\a} \partial_{\mu_3}
\partial^{\nu_3} \pi^{\b};
\end{equation}
three possibilities for $\mathcal{L}^{\text{Gal},2}_{4}$:
\begin{equation}
\mathcal{L}^{\text{Gal},2}_{4,\mathrm{I}} = \delta^{\mu_1 \mu_2}_{\nu_1
\nu_2}  \partial_{\mu_1} \pi_{\a} \partial_{\lambda} \pi^{\a}
\partial^{\lambda}\partial^{\nu_1} \pi_{\b} \partial_{\mu_2}
\partial^{\nu_2} \pi^{\b},
\end{equation}
\begin{equation}
\mathcal{L}^{\text{Gal},2}_{4,\mathrm{II}} = \delta^{\mu_1 \mu_2}_{\nu_1
\nu_2}  \partial_{\mu_1} \pi_{\a} \partial_{\lambda} \pi_{\b}
\partial^{\lambda}\partial^{\nu_1} \pi^{\a} \partial_{\mu_2}
\partial^{\nu_2} \pi^{\b},
\end{equation}
and
\begin{equation}
\mathcal{L}^{\text{Gal},2}_{4,\mathrm{III}} = \delta^{\mu_1
\mu_2}_{\nu_1 \nu_2}  \partial_{\mu_1} \pi_{\a} \partial_{\lambda}
\pi_{\b} \partial^{\lambda}\partial^{\nu_1} \pi^{\b} \partial_{\mu_2}
\partial^{\nu_2} \pi^{\a};
\end{equation}
and finally two possibilities for $\mathcal{L}^{\text{Gal},3}_{4}$:
\begin{equation}
\mathcal{L}^{\text{Gal},3}_{4,\mathrm{I}} = \partial_{\lambda} \pi_{\a}
\partial^{\lambda} \pi^{\a} \delta ^{\mu_1 \mu_2} _{\nu_1 \nu_2}
\partial_{\mu_1} \partial^{\nu_1} \pi_{\b} \partial_{\mu_2} \partial
^{\nu_2} \pi^{\b}
\end{equation}
and
\begin{equation}
\mathcal{L}^{\text{Gal},3}_{4,\mathrm{II}} = \partial_{\lambda} \pi_{\a}
\partial^{\lambda} \pi_{\b} \delta ^{\mu_1 \mu_2} _{\nu_1 \nu_2}
\partial_{\mu_1} \partial^{\nu_1} \pi^{\a} \partial_{\mu_2} \partial
^{\nu_2} \pi^{\b}.
\end{equation}

Looking for the Lagrangians implying second-order equations of motion,
one can quickly verify that $\mathcal{L}^{\text{PSZ}}_{4}$,
$\mathcal{L}^{\text{Gal},1}_{4,\mathrm{I}}$ and
$\mathcal{L}^{\text{Gal},1}_{4,\mathrm{II}}$ have this property due to
the symmetry properties of $\delta^{\mu_1 \cdots \mu_3}_{\nu_1 \cdots
\nu_3}$. However, the other Lagrangians do not give \emph{a priori} second-order 
equations of motion\footnote{The automatic cancellation between
third-order derivatives discussed in Ref.~\cite{Deffayet:2011gz} is not
valid anymore since this cancellation can be spoiled by the group
indices.}. We then investigate, in the following, the relations among
the different Lagrangians.

\subsection{Relations among the Lagrangians}
\paragraph{Between PSZ and Gal,1.}

We first relate $\mathcal{L}^{\text{PSZ}}_{4}$ and the Lagrangians
$\mathcal{L}^{\text{Gal},1}_{4}$ by means of conserved currents. Indeed,
\begin{equation}
J^{\mu_1}_{0,\mathrm{I}}=J^{\text{PSZ-Gal},\mu_1} _{4,\mathrm{I}} =
\delta^{\mu_1 \cdots \mu_3}_{\nu_1 \cdots \nu_3} \pi_{\a}
\partial^{\nu_1} \pi^{\a} \partial_{\mu_2} \partial^{\nu_2} \pi_{\b}
\partial_{\mu_3} \partial^{\nu_3} \pi^{\b} 
\end{equation}
gives
\begin{equation}
\partial_{\mu_1} J^{\mu_1}_{0,\mathrm{I}} =\partial_{\mu_1}
J^{\text{PSZ-Gal},\mu_1} _{4,\mathrm{I}} = \mathcal{L}^{\text{PSZ}}_{4} 
+ \mathcal{L}^{\text{Gal},1}_{4,\mathrm{I}},
\end{equation}
and
\begin{equation}
J^{\mu_1}_{0,\mathrm{II}}=J^{\text{PSZ-Gal},\mu_1} _{4,\mathrm{II}} =
\delta^{\mu_1 \cdots \mu_3}_{\nu_1 \cdots \nu_3} \pi_{\a} 
\partial^{\nu_1} \pi_{\b} \partial_{\mu_2} \partial^{\nu_2} \pi^{\a}
\partial_{\mu_3} \partial^{\nu_3} \pi^{\b}
\end{equation}
gives
\begin{equation}
\partial_{\mu_1} J^{\mu_1}_{0,\mathrm{II}} = \partial_{\mu_1}
J^{\text{PSZ-Gal},\mu_1} _{4,\mathrm{II}} = \mathcal{L}^{\text{PSZ}}_{4}  +
\mathcal{L}^{\text{Gal},1}_{4,\mathrm{II}} .
\end{equation}

It is also possible to make a direct correspondence between
$\mathcal{L}^{\text{Gal},1}_{4,\mathrm{I}}$ and $
\mathcal{L}^{\text{Gal},1}_{4,\mathrm{II}}$ with the current
\begin{equation}
J^{\mu_2}_{0,\mathrm{I\to II}}=J_{4,\mathrm{I\to
II}}^{\text{Gal},1,\mu_2} = \delta^{\mu_1 \cdots \mu_3}_{\nu_1 \cdots
\nu_3} \partial_{\mu_1} \pi_{\a} \partial^{\nu_1} \pi^{\a}
\partial^{\nu_2} \pi_{\b} \partial_{\mu_3} \partial ^{\nu_3} \pi^{\b},
\end{equation} 
yielding
\begin{equation}
\partial_{\mu_2} J^{\mu_2}_{0,\mathrm{I\to II}} = \partial_{\mu_2}
J_{4,\mathrm{I\to II}}^{\text{Gal},1,\mu_2}  =
\mathcal{L}^{\text{Gal},1}_{4,\mathrm{I}} -
\mathcal{L}^{\text{Gal},1}_{4,\mathrm{II}}.
\end{equation}

\paragraph{Between Gal,2 and Gal,3}

Introducing
\begin{equation}
J^{\mu_1}_{1} = J^{\text{Gal},2-3,\mu_1}_{4,\mathrm{I}} = \partial_{\lambda}
\pi_{\a} \partial^{\lambda} \pi ^{\a} \delta ^{\mu_1 \mu_2} _{\nu_1
\nu_2} \partial^{\nu_1} \pi_{\b} \partial_{\mu_2} \partial^{\nu_2}
\pi^{\b},
\end{equation}
we get
\begin{equation}
\label{PSU2EqGal3}
\partial_{\mu_1} J^{\mu_1}_{1}  = \partial_{\mu_1}
J^{\text{Gal},2-3,\mu_1}_{4,\mathrm{I}} = 2
\mathcal{L}^{\text{Gal},2}_{4,\mathrm{III}} +
\mathcal{L}^{\text{Gal},3}_{4,\mathrm{I}}.
\end{equation}

In a similar way, from
\begin{equation}
J^{\mu_1}_{2} = J^{\text{Gal},2-3,\mu_1}_{4,\mathrm{II}} =
\partial_{\lambda} \pi_{\a} \partial^{\lambda} \pi_{\b} \delta ^{\mu_1
\mu_2} _{\nu_1 \nu_2} \partial^{\nu_1} \pi^{\a} \partial_{\mu_2}
\partial^{\nu_2} \pi^{\b},
\end{equation}
we obtain
\begin{equation}
\partial_{\mu_1} J^{\mu_1}_{2}  =\partial_{\mu_1}
J^{\text{Gal},2-3,\mu_1}_{4,\mathrm{II}} = 
\mathcal{L}^{\text{Gal},2}_{4,\mathrm{I}} +
\mathcal{L}^{\text{Gal},2}_{4,\mathrm{II}} +
\mathcal{L}^{\text{Gal},3}_{4,\mathrm{II}}.
\end{equation}

\paragraph{Between Gal,1, Gal,2 and Gal,3 through Kronecker properties}

We use the following identity given in Ref.~\cite{Deffayet:2011gz}:
\begin{equation}
\label{PSU2EqDevAntiSym}
\delta^{\mu_1 \cdots \mu_n} _{\nu_1 \cdots \nu_n} = \delta^{\mu_1}
_{\nu_1} \delta^{\mu_2 \cdots \mu_n} _{\nu_2 \cdots \nu_n} +
\sum_{i=2}^{n} \left(-1 \right)^{i-1} \delta^{\mu_1}_{\nu_i} \delta
^{\mu_2 \cdots \mu_n} _{\nu_1 \nu_2 \cdots \nu_{i-1} \nu_{i+1} \cdots
\nu_n},
\end{equation}
which gives, for $n=3$,
\begin{equation}
\label{PSU2EqAlgebraicAntisym}
\delta^{\mu_1 \cdots \mu_3} _{\nu_1 \cdots \nu_3} =  \delta^{\mu_1}
_{\nu_1}  \delta^{\mu_2 \mu_3} _{\nu_2 \nu_3} - \delta^{\mu_1} _{\nu_2}
\delta^{\mu_2 \mu_3}_{\nu_1 \nu_3} + \delta ^{\mu_1} _{\nu_3} \delta
^{\mu_2 \mu_3} _{\nu_1 \nu_2}.
\end{equation}
It is then possible to obtain two additional relations among the
different Lagrangians. Indeed, applying this identity to
$\mathcal{L}^{\text{Gal},1}_{4,\mathrm{I}}$ and
$\mathcal{L}^{\text{Gal},1}_{4,\mathrm{II}}$, we get
\begin{equation}
\mathcal{L}^{\text{Gal},1}_{4,\mathrm{I}} = -2
\mathcal{L}^{\text{Gal},2}_{4,\mathrm{I}} +
\mathcal{L}^{\text{Gal},3}_{4,\mathrm{I}}
\end{equation}
and
\begin{equation}
\mathcal{L}^{\text{Gal},1}_{4,\mathrm{II}} =
-\mathcal{L}^{\text{Gal},2}_{4,\mathrm{II}}  -
\mathcal{L}^{\text{Gal},2}_{4,\mathrm{III}} +
\mathcal{L}^{\text{Gal},3}_{4,\mathrm{II}}.
\end{equation}

\subsection{Lagrangians with Second-Order Equations of Motion}
\label{PSU2PartResultLagGal}

Using the results of the previous subsections, we can summarize the Lagrangians
that give second-order equations of motion:
\begin{equation}
\mathcal{L}^{\text{PSZ}}_{4},
\end{equation}
\begin{equation}
\mathcal{L}^{\text{Gal},1}_{4,\mathrm{I}}  = -2
\mathcal{L}^{\text{Gal},2}_{4,\mathrm{I}} +
\mathcal{L}^{\text{Gal},3}_{4,\mathrm{I}}= -
\mathcal{L}^{\text{PSZ}}_{4} - \partial_{\mu} J^{\mu}_{0,\mathrm{I}},
\end{equation}
\begin{equation}
\mathcal{L}^{\text{Gal},1}_{4,\mathrm{II}} = 
-\mathcal{L}^{\text{Gal},2}_{4,\mathrm{II}}  -
\mathcal{L}^{\text{Gal},2}_{4,\mathrm{III}} +
\mathcal{L}^{\text{Gal},3}_{4,\mathrm{II}} = -
\mathcal{L}^{\text{PSZ}}_{4} - \partial_{\mu} J^{\mu}_{0,\mathrm{II}},
\end{equation}
\begin{equation}
\label{PSU2EqGal2}
\mathcal{L}^{\text{Gal},2}_{4,\mathrm{II}} =
\frac{1}{4}\mathcal{L}^{\text{Gal},1}_{4,\mathrm{I}}
-\frac{1}{2}\mathcal{L}^{\text{Gal},1}_{4,\mathrm{II}} -\frac{1}{4}\partial_\mu
J^{\mu}_1 +\frac{1}{2}\partial_\mu J^{\mu}_2,
\end{equation}
\begin{equation}
\mathcal{L}^{\text{Gal},2}_{4,\mathrm{I}} +
\mathcal{L}^{\text{Gal},2}_{4,\mathrm{III}}  =
-\frac{1}{2}\mathcal{L}^{\text{Gal},1}_{4,\mathrm{I}}
+\frac{1}{2}\partial_\mu J^{\mu}_1,
\end{equation}
and
\begin{equation}
\label{PSU2EqGal1}
\mathcal{L}^{\text{Gal},3}_{4,\mathrm{I}} +2
\mathcal{L}^{\text{Gal},3}_{4,\mathrm{II}} =
\frac{1}{2}\mathcal{L}^{\text{Gal},1}_{4,\mathrm{I}}
+\mathcal{L}^{\text{Gal},1}_{4,\mathrm{II}} +\frac{1}{2}\partial_\mu
J^{\mu}_1 +\partial_\mu J^{\mu}_2.
\end{equation}



\chapter*{Conclusion\markboth{Conclusion}{Conclusion}} 
\addcontentsline{toc}{chapter}{\protect\numberline{}Conclusion} 
\label{ConclusionThese}

Au cours de cette thèse, j'ai présenté les concepts centraux dans la construction de théories physiques, et comment appliquer ces concepts au développement de théories allant au-delà du Modèle Standard de la physique des particules -- et plus particulièrement les théories de grande unification --, ainsi qu'à l'examen des théories allant au-delà de la relativité générale -- en prenant l'exemple des théories de gravité modifiée de Galiléons. Tout au long de cette discussion, j'ai mis l'accent sur le rôle des symétries dans la construction de modèles physiques, ainsi que sur la grande polyvalence des outils de la théorie des groupes permettant de décrire ces symétries. J'ai également discuté certaines conséquences de ces théories au-delà des modèles standards dans un cadre cosmologique, et tout particulièrement la formation de cordes cosmiques lors de la brisure des symétries de grande unification. Je synthétise ici les principaux résultats qui ont été obtenus, ainsi que les différentes perspectives qu'ils offrent.

Dans les articles~\cite{Allys:2015yda} et~\cite{Allys:2015kge} reproduits chapitre~\ref{PartArticleCordes1} et~\ref{PartArticleCorde2}, les structures microscopiques réalistes des cordes cosmiques ont été étudiées, en considérant une implémentation complète d'un schéma de brisure de théorie de grande unification. Ce travail a été fait à la fois dans le cadre d'un GUT général, puis pour un modèle de GUT spécifique basé sur le groupe SO(10). Ces investigations ont montré qu'il est nécessaire de décrire la structure réaliste des cordes cosmiques non pas en ne considérant qu'une étape de brisure de symétrie, mais en considérant toutes les étapes successives d'un schéma de brisure\footnote[1]{Ces résultats semblent cependant montrer qu'une limite de découplage est atteinte lorsque les échelles d'énergie caractéristiques des brisures de symétrie sont très éloignées.}, et que l'intégralité des champs de Higgs à l'origine de ces étapes successives de brisure de symétrie apparaissent dans ces structures. Des ansatz permettant de décrire les structures réalistes de cordes cosmiques ont également été détaillés, et l'ordre de grandeur des modifications de l'énergie par unité de longueur de ces structures par rapport à celle des modèles jouets comme le modèle de Higgs abélien a été calculé et confirmé par des résolutions numériques.

Ces investigations ont permis d'évaluer perturbativement les modifications de l'énergie par unité de longueur des cordes réalistes par rapport aux modèles jouets, et ont montré que celles-ci n'étaient importantes que dans une certaine limite de l'espace des paramètres pour les schémas de brisure les plus communs, donnant des \og single-field strings \fg{}. Ces résultats ne sont cependant pas surprenants si la présence de champs supplémentaires à ceux des modèles jouets ne changent pas la structure de soliton des cordes, comme c'est le cas dans les premières investigations qui ont été faites dans les articles~\cite{Allys:2015yda} et~\cite{Allys:2015kge}. La prochaine étape de l'étude des cordes cosmiques réalistes serait alors de comprendre si la présence de ces champs supplémentaires peut permettre de former des structures non perturbatives ne pouvant pas être approximées par des vortex de Nielsen-Olesen du modèle de Higgs abélien\footnote[2]{C'est exactement ce type de structure qui apparaît pour les \og many-fields strings \fg{}, qui se forment dans le cas de schémas de brisure peu courants.}. La possibilité de former de telles structures stables aurait des conséquences importantes dans l'étude des cordes cosmiques, puisqu'elle signifierait que les vortex étudiés dans le cadre de modèles jouets ne sont pas forcément une description effective correcte des cordes cosmiques apparaissant dans des schémas de brisure réalistes.

Une autre perspective permise par les structures réalistes de cordes cosmiques provient de l'identification des différents champs et symétries apparaissant dans la description des cordes vis-à-vis des théories de grande unification. Un telle identification avait déjà été effectuée pour les champs apparaissant dans l'étape de brisure de symétrie formant la corde, par exemple dans les références~\cite{Ma:1992ky,Davis:1996sp}, sans que cela permette d'étudier la potentielle apparition de courants, nécessitant la présence de champs supplémentaires se couplant avec la corde. Dans le cadre de la structure réaliste des cordes cosmiques, les champs de Higgs participant aux étapes de brisure de symétrie antérieures à la formation des cordes et apparaissant dans la structure de celles-ci forment cependant des candidats naturels pour former des courants. L'étude de courants portés par ces champs n'est alors pas du tout arbitraire, puisque ceux-ci sont identifiés dans la théorie de grande unification qui a mené à la formation des cordes. Une telle investigation permettrait donc de justifier sur des bases de physique des particules l'apparition possible de courants dans les cordes.

Dans les articles \cite{Allys:2016hfl,Allys:2015sht,Allys:2016jaq,Allys:2016kbq}, présentés chapitres~\ref{PartArticleMultigal},~\ref{PartArticleProca1},~\ref{PartArticleProca2} et~\ref{PartArticleProcaSU2}, une investigation de différents modèles de gravité modifiée de Galiléons scalaires et vectoriels a été effectuée. Ces travaux ont notamment contribué à construire et à justifier les théories considérées comme les plus générales à l'heure actuelle pour les Galiléons vectoriels et multi-scalaires. Les premiers termes de théories de multi-Galiléons vectoriels ont également été obtenus par une méthode d'investigation exhaustive basée sur des considérations de théorie des groupes. Ces différents travaux ont fait apparaître un lien manifeste entre les théories de Galiléons décrivant des champs scalaires et vectoriels. Dans le contexte de la construction de théories décrivant toutes les dynamiques possibles pour un jeu d'hypothèses donné, comme le fait d'avoir des équations du mouvement au plus du second ordre, il serait intéressant d'étudier plus en détails ce lien entre les différentes théories de Galiléons. Il pourrait notamment être possible de transposer directement une part importante des contraintes et des propriétés de structure des modèles de scalaires sur les modèles de vecteurs, voire à d'autres théories plus complexes. Il est également nécessaire de continuer à étudier les conséquences phénoménologiques des théories de Galiléons, notamment en ce qui concerne la cosmologie et la dynamique d'objets compacts. 

Cette discussion nous ramène finalement à la question posée en ouverture de ce manuscrit~: Pourrons-nous aller au-delà des modèles standards actuels de la physique ? J'espère avoir montré tout au long de ce document que la construction de tels modèles a à présent atteint une certaine maturité dans le domaine de la gravitation comme de la physique des particules, et que la cosmologie peut fournir un laboratoire très prometteur pour tester de telles théories. Avec le développement actuel de la cosmologie de précision et l'avènement de l'astronomie gravitationnelle, des tests directs des théories au-delà des modèles standards auront même peut-être lieu dans un futur proche.

\appendix
\addtocontents{toc}{\protect\setcounter{tocdepth}{2}}

\chapter{Éléments de théorie des groupes}
\label{ChapterTdG}

\section{Préambule}

\noindent
Cet appendice récapitule les principales définitions et propriétés des groupes et algèbres de Lie utilisées tout au long de ce document. Les propriétés des représentations des groupes et annexes de Lie sont entre autres discutées. La rédaction de cette partie est notamment basée sur les notes de cours~\cite{ZuberGroupe} et les livres~\cite{Ramond:2010zz,Fuchs:1997jv}.

\section{Groupes et algèbres de Lie}
\subsection{Généralités sur les groupes}

\noindent
$\bullet$ \emph{Groupe} : Ensemble $G$ d'élements $g$, muni d'une loi de composition interne notée \og $\cdot$ \fg{}. Le groupe contient un élément neutre $e$ et un inverse $g^{-1}$ pour chaque élément $g$. Il est abélien si l'opération est commutative. S'il est fini, son ordre est son cardinal, noté $\# G$ ou $\left|G\right|$. 
\newline
$\bullet$ \emph{Classes de conjugaison} : On considère une relation d'équivalence $\sim$ sur le groupe :
\begin{equation}
a \sim b \Longleftrightarrow \exists g\in G ~~ : ~~ a = g \cdot b \cdot g^{-1}.
\end{equation}
Les éléments $a$ et $b$ sont alors dits conjugués. Les classes d'équivalence résultantes de cette relation forment une partition de $G$.
\newline
$\bullet$ \emph{Remarque} : Les classes sont \emph{a priori} de tailles différentes. En particulier, la classe de l'élément neutre $e$ est toujours réduite à $\{e\}$.
\newline
$\bullet$ \emph{Sous-groupe} : Sous-ensemble $H$ de $G$ qui est un groupe pour la restriction de la loi de composition \og $\cdot$ \fg{} à $H$. Il est propre si non identique à $G$. Si $H$ est un sous-groupe, $a^{-1}Ha$ pour $a \in G$ forme aussi un sous groupe, dit sous-groupe conjugé de $H$.
\newline
$\bullet$ \emph{Centre d'un groupe} : Ensemble des éléments qui commutent avec tous les éléments de $G$. C'est un sous-groupe de $G$, qui est propre si $G$ n'est pas abélien.
\newline
$\bullet$ \emph{Homéomorphisme} : Application $\rho$ de $G$ dans un autre groupe $G^\prime$ qui vérifie
\begin{equation}
\forall g,h\in G ~~ \rho(g \cdot h) = \rho(g) \cdot \rho(h).
\end{equation}
En particulier, on a $\rho(e) = e^\prime$ et $\rho(g^{-1})=\rho(g)^{-1}$. Le noyau de l'homéomorphisme, antécédent de $e^\prime \in G^\prime$, est un sous-groupe de $G$.
\newline 
$\bullet$ \emph{Automorphisme} : Un automorphisme est un homéomorphisme bijectif d'un groupe sur lui-même.
\newline
$\bullet$ \emph{Classes par rapport à un sous-groupe $H$} : On définit la relation d'équivalence 
\begin{equation}
g \underset{~\scriptstyle{R}}{\sim} g^\prime \Longleftrightarrow g \cdot {g^\prime}^{-1} \in H 
\Longleftrightarrow \exists h \in H ~~: ~~g = h \cdot g^\prime 
\Longleftrightarrow g \in H g^\prime.
\end{equation}
Cette relation est appelée équivalence à droite. On définit de même l'équivalence à gauche $g \underset{~\scriptstyle{L}}{\sim} g^\prime$.
\newline
$\bullet$ \emph{Propriété} : Les classes d'équivalence pour l'équivalence à droite forment une partition de $G$. Si $g_j$ est un représentant d'une classe donnée, celle-ci peut se noter $H\cdot g_j$., et on a $\left| H\cdot g_j\right| = \left| H \right|$.
\newline
$\bullet$ \emph{Remarque} : Toutes les classes d'équivalence à gauche ont les mêmes propriétés, mais avec des classes \emph{a priori} différentes.
\newline
$\bullet$ \emph{Sous-groupe invariant} : Le sous-groupe $H$ est invariant si
\begin{equation}
\forall g\in G, ~ \forall h\in H, ~ g h g^{-1}\in H 
~~ \Longleftrightarrow ~~ \underset{~\scriptstyle{R}}{\sim} \simeq  \underset{~\scriptstyle{L}}{\sim}
~~ \Longleftrightarrow ~~ H \text{ est égal à ses conjugés, \emph{i.e.} } g H g^{-1} = H ~\forall g.
\end{equation}
$\bullet$ \emph{Propriété} : Si $H$ est un sous-groupe invariant de $G$, $G/H$ peut être muni d'une structure de groupe. Le cardinal de $G/H$ est alors l'indice de $H$ dans $G$. 
\newline
$\bullet$ \emph{Remarque} : Le noyau d'un homéomorphisme de $G$ dans $G^\prime$ est un sous-groupe invariant.
\newline
$\bullet$ \emph{Groupes simples, semi-simples} : Un groupe est simple s'il n'a pas de sous-groupe invariant non trivial, semi-simple s'il n'a pas de sous-groupe invariant abélien non-trivial.
\newline
$\bullet$ \emph{Remarque} : Un groupe simple est semi-simple, l'implication réciproque est fausse.

\subsection{Groupes continus, propriétés topologiques}
\label{AnnexePartTopologie}

\noindent
$\bullet$ \emph{Groupe continu (topologique)} : Espace topologique avec une structure de groupe, et tel que $(g,h) \mapsto g\cdot h$ et $g \mapsto g^{-1}$ sont des fonctions continues.
\newline
$\bullet$ \emph{Connexité} : Si $G$ est non connexe, la composante connexe de l'identité est un sous-groupe invariant.
\newline
$\bullet$ \emph{Homotopie} : Deux chemins dans la même classe d'homotopie peuvent être déformés continument l'un en l'autre ; on dit qu'ils sont homotopes.
\newline
$\bullet$ \emph{Premier groupe d'homotopie} : Grâce à la composition des chemins, on a une structure de groupe sur les classes d'homotopie, qui définit le premier groupe d'homotopie $\pi_1(G)$
\newline
$\bullet$ \emph{Simple connexité} : Si $\pi_1(G)=1$, le groupe est simplement connexe.
\newline
$\bullet$ \emph{Recouvrement universel} : Si un groupe $G$ n'est pas simplement connexe, on peut construire un groupe simplement connexe $\tilde{G}$ tel que $G$ et $\tilde{G}$ sont localement homéomorphes : $\tilde{G}$ est le recouvrement universel de $G$.
\newline
$\bullet$ \emph{Propriété} : Il y a unicité du recouvrement universel à un homéomorphisme près.
\newline
$\bullet$ \emph{Propriété} : Le noyau de l'homéomorphisme $\rho$ de $\tilde{G}$ dans $G$ est $\pi_1(G)$, et on a donc que 
\begin{equation}
G \simeq \tilde{G}/\pi_1(G).
\end{equation}
$\bullet$ \emph{Mesure invariante de Haar} : Une mesure de Haar $\text{d}\mu(g)$ définie sur un groupe continu est telle que : $\text{d}\mu (g) =\text{d}\mu (g\cdot g^{\prime})=\text{d}\mu (g^{\prime}\cdot g)=\text{d}\mu (g^{-1})~$. C'est une mesure invariante par composition sur le groupe.
\newline
$\bullet$ \emph{Propriété} : Si $G$ est compact, une telle mesure existe et est unique à une normalisation près.
\newline
$\bullet$ \emph{Propriété} : Pour une telle mesure, l'intégration d'une fonction continue sur le groupe donne un résultat fini.
\newline
$\bullet$ \emph{Groupe de Lie} : Groupe topologique sur une variété différentiable, et où les lois de groupe sont infiniment différentiables. Lorsque le groupe est de dimension finie d, $g$ dépend de façon continue de $d$ paramètres réels $(\xi^1,\cdots,\xi^d) \in D \subset \mathbb{R}^d$.

\subsection{Écriture locale d'un groupe de Lie, algèbre de Lie}

\noindent
$\bullet$ \emph{Algèbre de Lie} : Algèbre dont le produit noté $[X,Y]$ (crochet de Lie) est antisymétrique et satisfait l'identité de Jacobi :
\begin{equation}
[X_1,[X_2,X_3]] + [X_3,[X_1,X_2]] + [X_2,[X_3,X_1]] =0.
\end{equation}
$\bullet$ \emph{Remarque} : Toute algèbre associative de loi de composition \og $*$ \fg{} est une algèbre de Lie pour le produit $[X,Y] = X*Y - Y*X$.
\newline
$\bullet$ \emph{Espace tangent d'un groupe de Lie} : On considère $g(t)$ sous-groupe à un paramètre de $G$ tel que $g(0)=e(=\mathds{1})$. Le générateur $X$ de ce sous-groupe est tel que 
\begin{equation}
\label{EqPassageGroupeAlgebre1}
g(\delta t) = \mathds{1} + \delta t X + \cdots ~~ \Longleftrightarrow ~~ X = \left.\frac{\text{d}}{\text{d}t}g(t)\right|_{t=0} ~~ \Longleftrightarrow ~~ \frac{\text{d}}{\text{d}t}g(t) = g(t) X.
\end{equation}
L'espace tangent en $e$ à $G$, qu'on notera $\mathfrak{g}$, est l'espace engendré par tous les générateurs $X$ des sous-groupes $g(t)$. On a dim$(G) =$ dim$(\mathfrak{g})$.
\newline
$\bullet$ \emph{Propriété} : Si $G$ est décrit avec des coordonnées $\xi^\alpha$, ces vecteurs tangent sont les opérateurs différentiels $X = X^\alpha \left( \partial / \partial \xi^\alpha \right)$.
\newline
$\bullet$ \emph{Propriété} : On peut aussi écrire les éléments du sous-groupe défini précédemment via
\begin{equation}
\label{EqPassageGroupeAlgebre2}
g(t) = \exp (tX) = \sum_n \frac{t^n}{n!} X^n.
\end{equation}
$\bullet$ \emph{Propriété} : L'application $X\in \mathfrak{g} \mapsto e^X\in G$ est bijective au voisinage de $e$. Elle est surjective si $G$ est connexe et compact, et injective si et seulement si $G$ est simplement connexe.
\newline 
$\bullet$ \emph{Exemple} :  Si $G=\text{GL}(n,\mathbb{R})$, $\mathfrak{g}=\text{M}(n,\mathbb{R})$. Si $G=\text{U}(n)$, $\mathfrak{g}$ est l'ensemble des matrices antihermitiennes [sans traces si $G=\text{SU}(n)$]. Si $G=\text{O}(n)$, $\mathfrak{g}$ est constituée des matrices antisymétriques (et donc de trace nulle).
\newline
$\bullet$ \emph{Propriété} : Le commutateur de deux éléments de l'espace tangent d'un groupe de Lie $G$ reste dans l'espace tangent. L'algèbre $\mathfrak{g}$ est donc une algèbre de Lie avec cette opération comme crochet de Lie.
\newline
$\bullet$ \emph{Application adjointe} : On définit la fonction $\text{ad}X : \mathfrak{g} \mapsto \mathfrak{g}$, telle que 
\begin{equation}
\label{EqApplicationAdjointe}
Y \mapsto (\text{ad} X) Y = [X,Y].
\end{equation}

\subsection{Relation entre les propriétés de $G$ et $\mathfrak{g}$}

\noindent
$\bullet$ \emph{Idéal d'une algèbre de Lie} : Sous espace $J$ de $\mathfrak{g}$, stable par multiplication par un élément quelconque de $\mathfrak{g}$, \emph{i.e.} $[\mathfrak{g},J]\subset J$. L'idéal est abélien si $[J,J]={0}$.
\newline
$\bullet$ \emph{Simplicité} : Une algèbre de Lie est simple si seul $\{0\}$ en est un idéal, semi-simple si seul $\{0\}$ en est un idéal abélien.
\newline
$\bullet$ \emph{Propriété} : $G$ simple $\Rightarrow$ $\mathfrak{g}$ simple, et $G$ semi-simple $\Rightarrow$ $\mathfrak{g}$ semi-simple.
\newline
$\bullet$ \emph{Propriété} : Une algèbre de Lie semi-simple est compacte si $G$ est compact.
\newline
$\bullet$ \emph{Remarque} : L'algèbre de Lie d'un groupe de Lie $G$ récupère toutes ses propriétés locales, mais pas ses propriétés globales.
\newline
$\bullet$ \emph{Complexification} : On autorise les paramètres réels à être complexes.
\newline
$\bullet$ \emph{Remarque} : Des algèbres différentes peuvent devenir isomorphes une fois complexifiées.
\newline
$\bullet$ \emph{Propriété} : Toute algèbre semi-simple a une unique forme réelle compacte.
\newline
$\bullet$ \emph{Propriété} : Si $G$ n'est pas connexe, et si $G^\prime\subset G$ est la composante connexe de l'identité, alors les algèbres de $G$ et $G^\prime$ coïncident ; par exemple, o(3)=so(3).
\newline
$\bullet$ \emph{Propriété} : Si $G$ est non simplement connexe, comme $G$ et son recouvrement universel $\tilde{G}$ sont localement homéomorphes, ils ont la même algèbre de Lie.
\newline
$\bullet$ \emph{Propriété} : À tout algèbre $\mathfrak{g}$ correspond un unique $G$ simplement connexe dont $\mathfrak{g}$ est l'algèbre de Lie. Les $G^\prime$ donnant $\mathfrak{g}$ vérifient $G= G^\prime/H$, avec $H$ un sous groupe invariant de $G^\prime$

\subsection{Structure des algèbres de Lie}
\label{TheorèmeCartan}

\noindent
$\bullet$ \emph{Constante de structure} : On dénote $\{t_\alpha\}$ une base de $\mathfrak{g}$ de dimension $D$. Les constantes de structures $f_{\alpha\beta}{}^{\gamma}$ sont telles que 
\begin{equation}
[t_\alpha,t_\beta] = f_{\alpha\beta}{}^\gamma t_\gamma,
\end{equation}
et sont donc antisymétriques sur les deux indices bas, \emph{i.e.} $f_{\alpha\beta}{}^\gamma = - f_{\beta\alpha}{}^\gamma$. On peut aussi écrire l'opérateur adjoint $(\text{ad}X) Y = \sum x^\alpha y^\beta f_{\alpha \beta}{}^{\gamma} t_{\gamma}$, avec $X=x^\alpha t_\alpha$ et $Y=y^\beta t_\beta$.
\newline
$\bullet$ \emph{Identité de Jacobi} : On peut l'écrire $[\text{ad}X,\text{ad}Y]Z = \text{ad}[X,Y]Z$, ce qui donne en terme des constantes de structures : 
\begin{equation}
f_{\alpha\delta}{}^{\epsilon}f_{\beta\gamma}{}^{\delta}
+f_{\beta\delta}{}^{\epsilon}f_{\gamma\alpha}{}^{\delta}
+f_{\gamma\delta}{}^{\epsilon}f_{\alpha\beta}{}^{\delta} =0.
\end{equation}
$\bullet$ \emph{Forme de Killing} : Forme bilinéaire symétrique sur l'algèbre de Lie, qui s'écrit
\[
(X,Y) = \text{Tr}(\text{ad}X \cdot \text{ad}Y) = f_{\alpha\delta}{}^\gamma f_{\beta\gamma}{}^{\delta} x^\alpha x^\beta \equiv g_{\alpha\beta} x^\alpha x^\beta.
\]
$\bullet$ \emph{Tenseur métrique} : Il est défini via la forme de Killing,
\begin{equation}
g_{\alpha\beta} = f_{\alpha\delta}{}^\gamma f_{\beta\gamma}{}^{\delta} = \text{Tr}(\text{ad}t_\alpha \cdot \text{ad}t_\beta).
\end{equation}
$\bullet$ \emph{Propriété} : La forme de Killing est invariante par l'action de tout $\text{ad} Z$, à savoir que $([Z,X],Y) + (X,[Z,Y]) = 0$. Dans une algèbre de Lie simple, toute forme bilinéaire invariante est multiple de la forme de Killing.
\newline
$\bullet$ \emph{Propriété} : Le tenseur métrique $g_{\alpha\beta}$ peut être utilisé pour monter et descendre les indices de l'algèbre (on a $g^{\alpha\beta}$ tel que $g^{\alpha\gamma}g_{\gamma\beta} = \delta^\alpha_\beta$), et $f_{\alpha\beta\gamma}$ est antisymétrique sur tout ses indices.
\newline
$\bullet$ \emph{Théorèmes de Cartan} :
\newline
$~~$ - Une algèbre de Lie est semi-simple si et seulement si la forme de Killing est non dégénérée, \emph{i.e.} $\text{det} (g_{\alpha\beta}) \neq 0$.
\newline
$~~$ - Une algèbre de Lie semi-simple est compacte si et seulement si $g_{\alpha\beta}$ est définie négative.
\newline
$~~$ - Toute algèbre de Lie semi-simple est somme directe d'algèbres de Lie simples $\mathfrak{g}_i$.
\newline
$\bullet$ \emph{Opérateur de Casimir quadratique} : Si $\mathfrak{g}$ est semi-simple, elle a une forme de Killing inversible, et on peut définir l'opérateur de Casimir quadratique
\begin{equation}
C_2 = \sum_{\alpha\beta} g^{\alpha\beta} t_\alpha t_\beta.
\end{equation}
Il commute avec tout les éléments de l'algèbre, \emph{i.e.} $[C_2,t_\gamma]=0$. 
\newline
$\bullet$ \emph{Propriété} : Toute expression quadratique en $t_\alpha$ qui commute avec tous les éléments de l'algèbre est proportionnelle à $C_2$.
\newline
$\bullet$ \emph{Remarque} : Dans une algèbre non simple, il y a autant d'opérateurs de Casimir quadratiques que de composantes simples.
\newline
$\bullet$ \emph{Sous-algèbre de Cartan, rang} : Une sous-algèbre de Cartan est une sous-algèbre maximale telle que tout ses éléments sont simultanément diagonalisables. Dans la représentation adjointe cela signifie que le crochet de Lie entre deux de ses éléments est nul, impliquant donc que ceux-ci commutent (d'où la diagonalisation simultanée). Elle n'est pas unique, mais deux telles sous-algèbres peuvent être reliées par un automorphisme. Sa dimension est le rang $n$ de l'algèbre.
\newline
$\bullet$ \emph{Propriété} : En général, une algèbre de rang $n$ possède $n$ invariants de Casimir, qui sont polynomiaux en les éléments de l'algèbre\footnote{Par exemple pour su($n$) ils sont d'ordre 2, 3, $\cdots$, n, et pour so($2n$) ils sont d'ordre 2, 4, 6, $\cdots$, $2 n-4$, $n-2$.}.
\newline
$\bullet$ \emph{Propriété} : Pour les algèbres de Lie simples compactes de dimension finie, les contractions arbitraires de constantes de structures sont toujours proportionnelles à zéro, au tenseur métrique, et la constante de structure selon qu'on laisse respectivement 1, 2 ou 3 indices libres~\cite{Metha:1983mng}. Pour su($n$) avec $n\geq 3$, ce résultat est aussi valable en effectuant des contractions avec l'invariant primitif symétrique $d_{\alpha\beta\gamma}$, et la contraction avec trois indices libres peut alors être proportionnelles à celui-ci\footnote{Le résultat est proportionnel à $f$ si la contraction implique un nombre impair de $f$ et un nombre pair de $d$, et proportionnel à $d$ pour un nombre pair de $f$ et impair de $d$. Le cas où les nombres de $f$ et $d$ ont la même parité est impossible dans les contractions à trois indices libres.}.
\newline
$\bullet$ \emph{Remarque} : Sur le groupe de Lie $G$, les Casimirs correspondent à des opérateurs différentielles invariants. 

\section{Représentation linéaire des groupes}
\subsection{Représentation d'un groupe}

\noindent
$\bullet$ \emph{Représentation d'un groupe} : $G$ est représenté dans un espace vectoriel $E$ (on dit aussi que $E$ porte une représentation de $G$) si il existe un homéomorphisme $D$ de $G$ dans GL($E$). Cette application $g \mapsto D(g) \in \text{GL}(E)$ vérifie donc
\[
D(g\cdot g^\prime) = D(g) \cdot D(g^\prime) ~~,~~ D(e)=\mathds{1} ~~,~~ D(g)^{-1} = D(g^{-1}) ~~,~~\forall g, g^\prime \in G.
\]
Si $E$ est de dimension $p$, la représentation est de dimension $p$. La représentation identité $D(g)=\mathds{1}$ pour tout $g$ est dite triviale.
\newline
$\bullet$ \emph{Représentation fidèle (faithful)} : Une représentation est dite fidèle si elle est injective ; son noyau ne contient que l'élément neutre de $G$ et $D(g) = D(g^\prime) \Longleftrightarrow g= g^\prime$.
\newline
$\bullet$ \emph{Propriété} : Si une représentation n'est pas fidèle, son noyau $H$ est un sous-groupe invariant, et la représentation de $G/H$ dans $E$ est fidèle. Toute représentation d'un groupe simple est donc fidèle.
\newline
$\bullet$ \emph{Écriture sur une base} : Soit $\{e_i\}$ une base de $E$ de dimension finie, on peut écrire la représentation sur cette base, et on a 
\[
D(g) e_j = D_{ij}(g)e_i ~~,~~ D_{ij}(g\cdot g^\prime) = D_{i\alpha}(g) D_{\alpha j}(g^\prime) ~~,~~
D_{ij}(g^{-1}) = (D_{ij}(g))^{-1}~~,~~\forall g, g^\prime.
\]
$\bullet$ \emph{Opérateur d'entrelacement (interwining opérator)} : Soit $D$ et $D^\prime$ deux représentations de $G$ sur $E$ et $E^\prime$, un opérateur d'entrelacement $V$ est un opérateur linéaire vérifiant $V D(g) = D^\prime (g) V$ pour tout élément du groupe.
\newline
$\bullet$ \emph{Représentations équivalentes} : Si l'opérateur d'entrelacement $V$ entre deux représentations est inversible (ce qui implique que $E$ et $E^\prime$ sont de même dimension), ces représentations sont équivalentes.
\newline
$\bullet$ \emph{Caractère} : On définit le caractère d'une représentation par $\chi (g) = \text{Tr}\left[ D(g) \right]$. 
\newline
$\bullet$ \emph{Propriétés} : Le caractère est indépendant du choix de base.  Deux représentations équivalentes ont le même caractère. Le caractère prend la même valeur pour deux éléments d'une même classe de conjugaison~: c'est une fonction de classe. On a $\chi (e) = dim(D)$.

\subsection{Réductibilité, représentations irréductibles}

\noindent
$\bullet$ \emph{Représentations réductibles et irréductibles} : Une représentation réductible laisse invariant un sous-espace de $E$. Le cas contraire, elle est irréductible. Une représentation réductible qui laisse un sous-espace $E_1$ et son complémentaire $E_2$ invariants est complètement réductible ; on peut alors l'écrire comme une somme de représentations sur les espaces $E_1$ et $E_2$.
\newline
$\bullet$ \emph{Remarque} : Si une représentation laisse $E_1$ invariant, il existe une représentation dans le complémentaire $E_2$.
\newline
$\bullet$ \emph{Remarque} : La notion de réductibilité dépend de si on considère la représentation sur $\mathbb{R}$ ou sur $\mathbb{C}$.
\newline
$\bullet$ \emph{Représentation conjuguée} : $D^*$ est la représentation conjuguée de $D$, formée des matrices complexes conjuguées. Celles-ci vérifient bien l'homéomorphisme de groupe, par exemple $D_{ij}^*(g\cdot g^\prime) = D_{i\alpha}^* (g) D_{\alpha j}^* (g^\prime)$.
\newline
$\bullet$ \emph{Représentations réelles, pseudo-réelles} : Une représentation est réelle si il existe une base où $D=D^*$, ou de façon équivalente si $\chi$ est une fonction réelle sur le groupe. Les deux représentations sont alors équivalentes. Si les deux représentations complexes conjuguées sont équivalentes mais qu'il n'existe pas une telle base, les représentations sont pseudo-réelles.
\newline
$\bullet$ \emph{Représentation contragrédiente (duale)} : Elle vérifie à partir d'une représentation initiale $\bar{D} = (D^{-1})^{T}$, ou de façon équivalente $\bar{D}_{ij}(g) = D_{ji}(g^{-1})$.  
\newline
$\bullet$ \emph{Propriété} : Si $D$ est unitaire, alors $\bar{D} = D^{*}$.
\newline
$\bullet$ \emph{Propriété} : Les représentations $D$, $D^{*}$ et $\bar D$ sont simultanément réductibles ou irréductibles.
\newline
$\bullet$ \emph{Représentations unitaires} : L'espace vectoriel $E$ doit être pré-hilbertien, et donc porter un produit scalaire. Une représentation $D$ est unitaire si $D(g)$ est unitaire pour tout élément du groupe. Elle vérifie donc les propriétés $\braket{D(g) x| D(g) y} = \braket{x|y}$ et $D(g^{-1})=D(g)^{-1} = D(g)^\dagger$.
\newline
$\bullet$ \emph{Propriétés} : Toute représentation d'un groupe fini ou compact sur un espace pré-hilbertien est unitarisable, et toute représentation unitaire réductible est complément réductible. 
\newline
$\Rightarrow$ Toute représentation réductible d'un groupe fini ou compact sur un espace pré-hilbertien est équivalente à une représentation unitaire complètement réductible.
\newline
$\bullet$ \emph{Lemme de Schur} : Soient $D$ et $D^{\prime}$ représentations irréductibles sur $E$ et $E^{\prime}$, et $V$ un opérateur d'entrelacement entre les deux représentations. Soit $V=0$, soit $V$ est une bijection et les représentations sont équivalentes.
\newline
$\bullet$ \emph{Corollaires} : Un opérateur d'entrelacement d'une représentation irréductible de $\mathbb{C}$ avec elle-même, et donc qui commute avec tout les représentants du groupe, est un multiple de l'identité. Une représentation irréductible sur $\mathbb{C}$ d'un groupe abélien est donc de dimension 1.

\subsection{Produit tensoriel de représentations, représentations restreintes}

\noindent
$\bullet$ \emph{Produit tensoriel de représentations} : Les espaces vectoriels $E_1$ et $E_2$ portent deux représentations de $G$ $D_1$ et $D_2$. On définit une représentation de $G$ sur l'espace $E_1 \otimes E_2$ (de base $z_{i,j}=x_i\otimes y_j$), écrivant que 
\begin{equation}
D(g) z = D_1(g) x \otimes D_2(g) y.
\end{equation}
$\bullet$ \emph{Propriétés} : On a alors $\chi(g) = \chi_1(g) \cdot \chi_2(g)$ et $dim (D) = dim (D_1) \cdot  dim (D_2)$.
\newline
$\bullet$ \emph{Décomposition de Clebsch-Gordan} : Le produit de représentations irréductibles donne \emph{a priori} une représentation réductible, qu'on peut décomposer en représentations irréductibles (quand c'est possible). On écrit alors $D \otimes D^{\prime} = \oplus_p m_p D_p$, où les coefficients $m_p$ donnent la multiplicité de la représentation irréductible $D_p$ dans le produit tensoriel $D\otimes D^{\prime}$.
\newline
$\bullet$ \emph{Propriété} : On a $\chi_D \cdot \chi_{D^\prime} = \sum_p m_p \chi_p$ et $dim(D) \cdot dim(D^\prime) = \sum_p m_p dim(D_p)$.
\newline
$\bullet$ \emph{Coefficients de Clebsch-Gordan} : La donnée de la décomposition en représentations irréductibles revient à écrire la base de $E\otimes E^\prime$ ($e^{E}_\alpha \otimes e^{E^\prime}_\beta$) en fonction de celle de $\oplus_p m_p E_p$ (écrite sous la forme $e^{p,i}_\gamma$ pour l'espace $E_p$, l'indice $i \in \{1,\cdots m_p\}$ explicitant la multiplicité de cette espace), à savoir 
\[
e_\alpha^{E} \otimes e_\beta^{E^\prime} = \bigoplus_{p,i,\gamma} C_{(E,\alpha;E^\prime,\beta | p,i,\gamma)} e_\gamma^{p,i}
\]
Les $C_{(E,\alpha;E^\prime,\beta | p,i,\gamma)}$ sont les coefficients de Clebsch-Gordan\footnote{
On trouve souvent les coefficients de Clebsch-Gordan écrits dans la littérature avec les notations spécifiques à SU(2), où les bases des espaces vectoriels associés à une représentation irréductible donnée sont notées sous la forme $\ket{l,m}$. On évitera cette notation ici, car elle est peu adaptée aux représentations des groupes de rang supérieur à 1, où plusieurs entiers sont nécessaires pour identifier une représentation irréductible et décrire ses éléments.
}, ils encodent complètement la décomposition en représentation irréductible de $D\otimes D^\prime$.
\newline
$\bullet$ \emph{Propriété} : Avec des représentations unitaires et des bases orthonormales, ces coefficients encodent un changement de bases orthonormales, et vérifient donc des relations d'orthogonalité et de complétude, du type 
\[
\sum_{p,i,\gamma} C_{(~~)} C_{(~~)}^* = \delta_{\alpha,\alpha^\prime} \delta_{\beta,\beta^\prime},
\]
et
\[
\sum_{\alpha,\beta}  C_{(~~)} C_{(~~)}^* = \delta_{p,p^\prime}\delta_{i,i^\prime}\delta_{\gamma,\gamma^\prime}.
\]
$\bullet$ \emph{Propriété} : Les coefficients de Clebsch-Gordan permettant de passer d'une base à l'autre, ils permettent aussi de passer des formules matricielles de $D\otimes D^\prime$ à celles de sa décomposition en représentations irréductible.
\newline
$\bullet$ \emph{Représentations restreintes} : Si $H$ est un sous-groupe de $G$, on peut restreindre une représentation $D$ de $G$ sur $H$, obtenant une nouvelle représentation $D^\prime$ sur $H$.
\newline
$\bullet$ \emph{Règles de branchement} : Une représentation restreinte d'une représentation irréductible ne l'est \emph{a priori} pas. On peut alors écrire sa décomposition en représentations irréductibles, de façon similaire à la décomposition de Clebsch-Gordan pour les produits tensoriels de représentations. Cela constitue les règles de branchement.

\subsection{Produits direct et semi-direct de groupes, représentations associées}
\label{ProduitsGroupes}

\noindent
$\bullet$ \emph{Produit direct de groupes} : Soient deux groupes $G_1$ et $G_2$ d'éléments $g_1$ et $g_2$. Le produit direct $G=G_1 \times G_2$ constitué des éléments $(g_1 ,g_2 )$ forme un groupe de dimension $dim(G)=dim(G_1)\cdot dim (G_2)$ pour la loi 
\begin{equation}
(g_1 ,g_2 )\cdot(g_1^\prime,g_2^\prime)=(g_1 \cdot g_1^\prime,g_2\cdot g_2^\prime).
\end{equation}
$\bullet$ \emph{Propriété} : $G_1 \simeq {(g_1,e_2)}$ et $G_2 \simeq {(e_1 ,g_2 )}$ sont des sous-groupes invariants de $G=G_1 \times G_2$, et $G$ n'est donc pas simple.
\newline
$\bullet$ \emph{Représentation des produits directs} : Si $E_1$ porte une représentation $D_1$ de $G_1$ et $E_2$ une représentation $D_2$ de $G_2$, alors $G=G_1 \times G_2$ a une représentation $D$ de dimension $dim(D) = dim(D_1) \cdot dim(D_2)$ sur $E_1 \otimes E_2$ définie par
\begin{equation}
D(g_1,g_2)(x_1,x_2)=(D_1(g_1)x_1,D_2(g_2)x_2).
\end{equation}
$\bullet$ \emph{Propriété} : La représentation $D$ ainsi définie est réductible si $D_1$ ou $D_2$ le sont, et irréductible si $D_1$ et $D_2$ le sont.
\newline
$\bullet$ \emph{Produit semi-direct de groupes} : Soient $N$ et $H$ deux groupes d'éléments $n$ et $h$, et $\varphi$ un homéomorphisme de $H$ dans les automorphismes de $N$. Le produit semi-direct $G=N \rtimes_\varphi H$ d'éléments $(n,h)$ est un groupe pour la loi
\begin{equation}
(n_1 , h_1 )\cdot(n_2 ,h_2)=(n_1 \cdot \varphi(h_1)[n_2] ,h_1\cdot h_2).
\end{equation}
Chaque homéomorphisme $\varphi$ décrit \emph{a priori} une structure de groupe particulière. L'indice $\varphi$ est cependant omis quand il n'y a pas de confusion possible.
\newline
$\bullet$ \emph{Propriété} : $N\simeq {(n,e_H)}$ et $H \simeq{(e_N,h)}$ sont des sous-groupes de de $G=N \rtimes_\varphi H$, dont seul $N\simeq {(n,e_H)}$ est invariant ; $G=N \rtimes_\varphi H$ n'est donc pas simple.
\newline
$\bullet$ \emph{Remarque} : Pour les produits semi-directs de groupes, et contrairement aux produits directs, il n'y a pas de lien simple entre les représentations des groupes initiaux et celles de leur produit semi-direct.

\subsection{Représentation des algèbres de Lie}

$\bullet$ \emph{Représentation d'une algèbre de Lie} : Homéomorphisme $d$ d'une algèbre de Lie $\mathfrak{g}$ dans GL($E$) qui respecte la linéarité et le crochet de Lie, \emph{i.e.}
\[
\forall X,Y \in g ~~,~~[X,Y] \rightarrow d([X,Y]) = [d(X),d(Y)] \in \text{GL}(E).
\]
$\bullet$ \emph{Corollaire} : Les représentant des générateurs de l'algèbre vérifient les mêmes relations de commutation quelque soit la représentation, soit
\begin{equation}
T_i = d(t_i) ~~ \Longrightarrow ~~ [T_i,T_j] = f_{ij}{}^{k} T_k.
\end{equation}
Toutes les propriétés dépendant seulement des crochets de Lie sont donc conservées.
\newline
$\bullet$ \emph{Remarque} : Les autres grandeurs, par exemple les opérateurs de Casimir, peuvent prendre des valeurs différentes\footnote{C'est d'ailleurs une façon de repérer les représentations.}.
\newline
$\bullet$ \emph{Représentation fidèle} : Pour une telle représentation, le noyau de $d$ se réduit à $\{0\}\subset \mathfrak{g}$.
\newline
$\bullet$ \emph{Lien entre les représentations de $G$ et $\mathfrak{g}$} : On peut obtenir une représentation $d$ de $\mathfrak{g}$ à partir d'une représentation $D$ de $G$, en utilisant les liens entre un groupe et l'algèbre associée donnés dans les équations~\eqref{EqPassageGroupeAlgebre1} et~\eqref{EqPassageGroupeAlgebre2} :
\[
d(X) = \left. \frac{\text{d}}{\text{d}t}\right|_{t=0} D[g(t)] ~~\Longleftrightarrow ~~ D\left(e^{t X}\right) = e^{t d(X)}.
\]
\newline
$\bullet$ \emph{Représentation adjointe de $G$ dans $\mathfrak{g}$} : Représentation définie par 
\begin{equation}
D^{\text{adj}}(g) \left(X\right) = g X g^{-1} = \text{Ad}(g) X.
\end{equation}
$\bullet$ \emph{Représentation adjointe de $\mathfrak{g}$ dans $\mathfrak{g}$} : Elle s'obtient à partir de la représentation adjointe de $G$ dans $\mathfrak{g}$, et donne [en nommant $Y$ le générateur associé à $g(t)$]
\begin{equation}
\left.\frac{\text{d}}{\text{d}t} \text{Ad}[g(t)] X \right|_{t=0} = [Y,X] = (\text{ad}Y) X,
\end{equation}
où on retrouve l'application \og ad \fg{} de $\mathfrak{g}$ dans $\mathfrak{g}$ définie en~\eqref{EqApplicationAdjointe}.
\newline
$\bullet$ \emph{Propriété} : Les constantes de structure forment une base de la représentation adjointe de $\mathfrak{g}$ dans $\mathfrak{g}$, \emph{i.e.} à l'élément $t_i$ est associé la matrice ${\left(T_i\right)}_j^k=f_{ij}{}^k$. En effet, agissant sur un élément $X=x^i t_i$, on a 
\begin{equation}
(\text{ad}t_i) X = [t_i,x^j t_j] = f_{ij}{}^{k} x^j t_k = \left(T_i\right)_j^k x^j t_k.
\end{equation}
$\bullet$ \emph{Propriété} : À partir de la représentation adjointe de $\mathfrak{g}$ sur $\mathfrak{g}$, on peut obtenir de façon unique la représentation adjointe associée au groupe $G$ simplement connexe associé à $\mathfrak{g}$, puis à tous les $G^\prime$ associés à $\mathfrak{g}$.  La description des représentations d'une algèbre de Lie suffit donc pour décrire toutes les représentations des groupes de Lie associées.



\bibliographystyle{unsrt}
{\footnotesize \bibliography{TheseEA} }


\newpage
\thispagestyle{empty}
~ 

\newpage
\thispagestyle{empty}

~
\vfill

\begin{center}
\textbf{{\Large Résumé}}
\end{center}

\noindent
La description actuelle des interactions fondamentales repose sur deux théories ayant le statut de modèle standard. Les interactions électromagnétiques
et nucléaires sont décrites à un niveau quantique par le Modèle Standard de la physique des particules, alliant théories de jauge et brisures spontanées de symétrie par le mécanisme
de Higgs. À l'opposé, l'interaction gravitationnelle est décrite par la relativité générale,
basée sur une description dynamique de l'espace-temps dans un cadre classique.

Bien que ces deux modèles soient vérifiés avec une grande précision dans
le système solaire, ils sont affligés d’un certain nombre de problèmes théoriques et
manquent de force prédictive aussi bien à l'échelle de Planck qu’à l'échelle cosmologique ; 
ils ne sont par conséquent plus perçus comme fondamentaux. La cosmologie, dont la phénoménologie fait apparaitre ces deux échelles extrêmes, apparaît alors comme un laboratoire privilégié pour tester les théories au delà de ces modèles standards.

La première partie de cette thèse concerne l'étude des cordes cosmiques, défauts topologiques se formant lors de la brisure spontanée de théories de grande unification dans l'univers primordial.
J’y montre notamment comment étudier ces défauts en prenant en compte la structure
complète des théories de physique des particules dont ils sont issus, ce qui représente
une avancée importante par rapport à la description courante en termes de modèles ”jouets”
très simplifiés. La deuxième partie de cette thèse consiste en la construction et l'étude de
différentes théories de gravité modifiée liées au modèle de Galiléon, un modèle tentant notamment
d'expliquer la phénoménologie liée à l'énergie noire.

\vspace{0.5cm}

\textbf{Mots-clés~: } Cosmologie, Cordes cosmiques, Défauts topologiques, Galiléons, Grande unification, Gravité modifiée, Physique des particules

\vspace{0.5cm}

\begin{center}
\rule{0.6\linewidth}{.5pt}
\end{center}

\vspace{0.5cm}

\begin{center}
\textbf{{\Large Abstract}}
\end{center}

The current description of fundamental interactions is based on two theories with the status of standard models. The electromagnetic and nuclear interactions are described at a quantum level by the Standard Model of particle physics, using tools like gauge theories and spontaneous symmetry breaking by the Higgs mechanism. The gravitational interaction is described on the other hand by general relativity, based on a dynamical description of space-time at a classical level.

Although these models are verified to high precision in the solar system experiments, they suffer from several theoretical weaknesses and a lack of predictive power at the Planck scale as well as at cosmological scales; they are thus not viewed anymore as fundamental theories. As its phenomenology involves both these extreme scales, cosmology is then a good laboratory to probe theories going beyond these standard models.

The first part of this thesis focus on cosmic strings, topological defects forming during the spontaneous symmetry breaking of grand unified theories in the early universe. I show especially how to study these defects while taking into account the complete structure of the particles physics models leading to their formation, going beyond the standard descriptions in terms of simplified toy-models. The second part is devoted to the construction and the examination of different theories of modified gravity related to the Galileon model, a model which tries in particular to explain the dark energy phenomenology.

\vspace{0.5cm}

\textbf{Keywords: } Cosmology, Cosmic strings, Galileons, Grand unification theories, Modified gravity, Particle physics.

\vfill
~


\end{document}